%% file: main.tex
\documentclass[aps,rmp,twocolumn,longbibliography,unsortedaddress]{revtex4-1}
\usepackage{amsmath,amssymb,graphicx}
\usepackage{bm}
\usepackage{color}
\usepackage[colorlinks]{hyperref}
\hypersetup{
	linkcolor = {blue},
    citecolor = {blue},
    urlcolor = {blue}
}

\newcommand{\sub}[1]{_{\mathrm{#1}}}

\newcommand{\eq}[1]{Eq.~\eqref{#1}}

\begin{document}

\title{Topological Photonics}

\date{\today}

\author{Tomoki Ozawa}
\affiliation{Interdisciplinary Theoretical and Mathematical Sciences Program (iTHEMS), RIKEN, Wako, Saitama 351-0198, Japan}
\affiliation{Center for Nonlinear Phenomena and Complex Systems, Universit\'e Libre de Bruxelles, CP 231, Campus Plaine, B-1050 Brussels, Belgium}
\affiliation{INO-CNR BEC Center and Dipartimento di Fisica, Universit\`a di Trento, I-38123 Povo, Italy}

\author{Hannah M. Price}
\affiliation{School of Physics and Astronomy, University of Birmingham, Edgbaston, Birmingham B15 2TT, United Kingdom}
\affiliation{INO-CNR BEC Center and Dipartimento di Fisica, Universit\`a di Trento, I-38123 Povo, Italy}

\author{Alberto Amo}
\affiliation{Universit\'e de Lille, CNRS, UMR 8523 - PhLAM - Laboratoire de Physique des Lasers Atomes et Mol\'ecules, F-59000 Lille, France}

\author{Nathan Goldman}
\affiliation{Center for Nonlinear Phenomena and Complex Systems, Universit\'e Libre de Bruxelles, CP 231, Campus Plaine, B-1050 Brussels, Belgium}

\author{Mohammad Hafezi}
\affiliation{Joint Quantum Institute, Institute for Research in Electronics and Applied Physics, Department of Electrical and Computer Engineering, Department of Physics, University of Maryland, College Park, Maryland 20742, USA}

\author{Ling Lu}
\affiliation{Institute of Physics, Chinese Academy of Sciences/Beijing National Laboratory for Condensed Matter
Physics, Beijing 100190, China}
\affiliation{Songshan Lake Materials Laboratory, Dongguan, Guangdong 523808, China}

\author{Mikael C. Rechtsman}
\affiliation{Department of Physics, The Pennsylvania State University, University Park, Pennsylvania 16802, USA}

\author{David Schuster}
\affiliation{The James Franck Institute and Department of Physics, University of Chicago, Chicago, Illinois 60637, USA}

\author{Jonathan Simon}
\affiliation{The James Franck Institute and Department of Physics, University of Chicago, Chicago, Illinois 60637, USA}

\author{Oded Zilberberg}
\affiliation{Institute for Theoretical Physics, ETH Zurich, 8093 Zurich, Switzerland}

\author{Iacopo Carusotto}
\affiliation{INO-CNR BEC Center and Dipartimento di Fisica, Universit\`a di Trento, I-38123 Povo, Italy}

\begin{abstract}
Topological photonics is a rapidly emerging field of research in which geometrical and topological ideas are exploited to design and control the behavior of light. Drawing inspiration from the discovery of the quantum Hall effects and topological insulators in condensed matter, recent advances have shown how to engineer analogous effects also for photons, leading to remarkable phenomena such as the robust unidirectional propagation of light, which hold great promise for applications. Thanks to the flexibility and diversity of photonics systems, this field is also opening up new opportunities to realize exotic topological models and to probe and exploit topological effects in new ways. This article reviews experimental and theoretical developments in topological photonics across a wide range of experimental platforms, including photonic crystals, waveguides, metamaterials, cavities, optomechanics, silicon photonics, and circuit QED. A discussion of how changing the dimensionality and symmetries of photonics systems has allowed for the realization of different topological phases is offered, and progress in understanding the interplay of topology with non-Hermitian effects, such as dissipation, is reviewed. As an exciting perspective, topological photonics can be combined with optical nonlinearities, leading toward new collective phenomena and novel strongly correlated states of light, such as an analog of the fractional quantum Hall effect. 
\end{abstract}

\maketitle

\tableofcontents

\input{RMP_I_Intro}

\input{RMP_IIA_Review}
\input{RMP_II_Pump}
\input{RMP_IIC_Floquet}
\input{RMP_IIB_DriveDissipation}

\input{RMP_IIIA1_Gyro}
\input{RMP_IIIA2_Propagating}
\input{RMP_IIIA3_Optomechanics}

\input{RMP_IIIA4_QHEOther}
\input{RMP_IIIB1_SiliconRing}
\input{RMP_IIIB2_Chicago}
\input{RMP_IIIB4_new}
\input{RMP_IIIB5_Metamaterials}

\input{RMP_IIIB_Wu}
\input{RMP_IIIB6_QSHOther}
\input{RMP_IIIC_Anomalous}
\input{RMP_IIID_Gapless}
\input{RMP_IVA_1DChiral}
\input{RMP_IVB_Pump}%
\input{RMP_VA_3D}
\input{RMP_VB_Synthetic}

\input{RMP_VIA_NonHermitian-aug1}
\input{RMP_VIB_Bogoliubov}
\input{RMP_VIII_Interaction}
\input{RMP_IX_Conclusion}

\begin{acknowledgements}
T.O. was supported by the EU-FET Proactive grant AQuS (Project No. 640800), the ERC Starting Grant TopoCold, and the Interdisciplinary Theoretical and Mathematical Sciences Program (iTHEMS) at RIKEN.

H.M.P. received funding from the Royal Society and from the European Unions Horizon 2020 research and innovation programme under the Marie Sklodowska-Curie grant agreement No 656093: ``SynOptic."

A.A. was supported by the ERC grant Honeypol, the EU-FET Proactive grant AQUS (Project No. 640800), the French National Research Agency (ANR) project Quantum Fluids of Light (ANR-16-CE30-0021) and the program Labex CEMPI (ANR-11-LABX-0007), the CPER Photonics for Society P4S, and the M\'{e}tropole Europ\'{e}enne de Lille.

N.G. was supported by the FRS-FNRS (Belgium) and by the ERC Starting Grant TopoCold.

M.H. acknowledges Sunil Mittal, and was supported by AFOSR MURI Grant No. FA95501610323, the Sloan Foundation, and the Physics Frontier Center at the Joint Quantum Institute.

L.L. was supported by the National key R\&D Program of China under Grants No. 2017YFA0303800 and No. 2016YFA0302400 and the NSFC under Project No.11721404.

M.C.R. acknowledges the National Science Foundation under Awards No. ECCS-1509546 and No. DMS-1620422, the David and Lucile Packard Foundation, the Charles E. Kaufman Foundation, a supporting organization of the Pittsburgh Foundation, the Office of Naval Research under the YIP program, Grant No. N00014-18-1-2595, and the Alfred P. Sloan Foundation under fellowship No. FG-2016-6418.

D.S. and J.S. were supported by the University of Chicago Materials Research Science and Engineering Center, which is funded by the National Science Foundation under Award No. DMR-1420709. This work was supported by ARO Grant No. W911NF-15-1-0397. D.S. acknowledges support from the David and Lucile Packard Foundation. This work was supported by DOE Grant No. DE-SC0010267 and AFOSR Grant No. FA9550-16-1-0323.

O.Z. was supported by the Swiss National Science Foundation (SNSF).

I.C. acknowledges funding from Provincia Autonoma di Trento, partly through the SiQuro project (``On Silicon Chip Quantum Optics for Quantum Computing and Secure Communications"), from ERC through the QGBE grant and from the EU-FET Proactive grant AQuS, Project No.640800 and EU-FET-Open grant MIR-BOSE Project No.737017.
\end{acknowledgements}

\bibliography{combined_biblio}

\end{document}

%% file: RMP_I_Intro.tex
\section{Introduction}
\label{sec:intro}

Over the last decade, {\it topological photonics} appeared as a rapidly growing field of study. This field aims to explore the physics of topological phases of matter, originally discovered in solid-state electron systems, in a novel optical context. In this review, we attempt to cover the main achievements of topological photonics, beginning from the basic concepts of topological phases of matter and photonics, so that readers can follow our discussion independently of their background.

Topological photonics is rooted in ideas that were first developed to understand topological phases of matter in solid-state physics. This field of research began with the discovery of the integer quantum Hall effect in 1980~\cite{Klitzing:1980PRL,VonKlitzing:RMP1986}. In this effect, a two-dimensional electron gas in the presence of a strong perpendicular magnetic field was found to exhibit robust plateaus in the Hall conductance as a function of the magnetic field at values equal to integer multiples of the fundamental constant $e^2/h$. The far-reaching conceptual consequences of this integer quantum Hall effect were soon highlighted by~\textcite{Thouless:1982PRL, Kohmoto:1985AnnPhys}. These works related the integer appearing in the Hall conductance to a {\it topological invariant} of the system, the Chern number, that is an integer-valued quantity which describes the global structure of the wavefunction in momentum space over the Brillouin zone.

An important insight into the physical meaning of the topological invariant is given by the {\it bulk-edge correspondence}~\cite{Jackiw:1976PRD,Hatsugai:1993PRL,Hatsugai:1993PRB,Qi:2006PRB}: when two materials with different topological invariants are put in contact, there must exist edge states that are spatially localized at the interface at energies that lie within the energy gap of the surrounding bulk materials. 

The bulk-edge correspondence can be heuristically understood in the following way: an integer topological invariant of a gapped system cannot change its value under perturbations or deformations of the system, unless the energy gap to excited states is somewhere closed. This implies that when two materials with different topological invariants are put in contact, the energy gap must close somewhere in the interface region, which leads to the appearance in this region of localized states. In a finite-size sample of a topologically nontrivial material, the physical edge of the sample can be considered as an interface between a region with a nonzero topological invariant and the topologically trivial vacuum, guaranteeing the existence of localized states at the system boundary. 

In the quantum Hall effect, these edge modes display {\it chiral} properties, in the sense that they can propagate only in one direction along the sample boundary but not in the opposite direction. The number of such chiral edge modes that are available at the Fermi energy for electric conduction is proportional to the Hall conductance. Because of the unidirectional nature of the edge states, the edge currents are immune to backscattering, resulting in the precise and robust quantization of the measured Hall conductance~\cite{Halperin1982,MacDonald:1984PRB,Buttiker:1988PRB}.

Interest in the topological physics of electronic systems surged further when a different class of topological phases of matter, now known as the quantum spin-Hall systems or $\mathbb{Z}_2$ topological insulators, was discovered in 2005~\cite{Kane:2005PRLa,Kane:2005PRLb,Bernevig:2006PRL,Bernevig:2006Science,Konig:2007Science}. In these systems, the Chern number is zero but the wave function is characterized by a binary ($\mathbb{Z}_2$) topological invariant, that can be nonzero and robust in the presence of time-reversal symmetry. Since then, there has been intense investigation in condensed-matter physics into what different topological phases of matter are possible under various symmetries, and what are the physical consequences of this physics~\cite{BernevigBook,Chiu:2016RMP,Hasan:2010RMP,Qi:2011RMP}. Besides electronic systems in solid-state materials, topological phases of matter are also being actively studied in other quantum many-body systems, in particular liquid helium~\cite{Volovik:2009Book} and ultracold atomic gases~\cite{Cooper:AdvPhys2008, Dalibard:RMP2011,Goldman:2014ROPP,Goldman:2016NatPhys}.

\begin{figure}
\includegraphics[trim=2.5cm 5.5cm 2.cm 6.5cm,width=0.95\columnwidth]{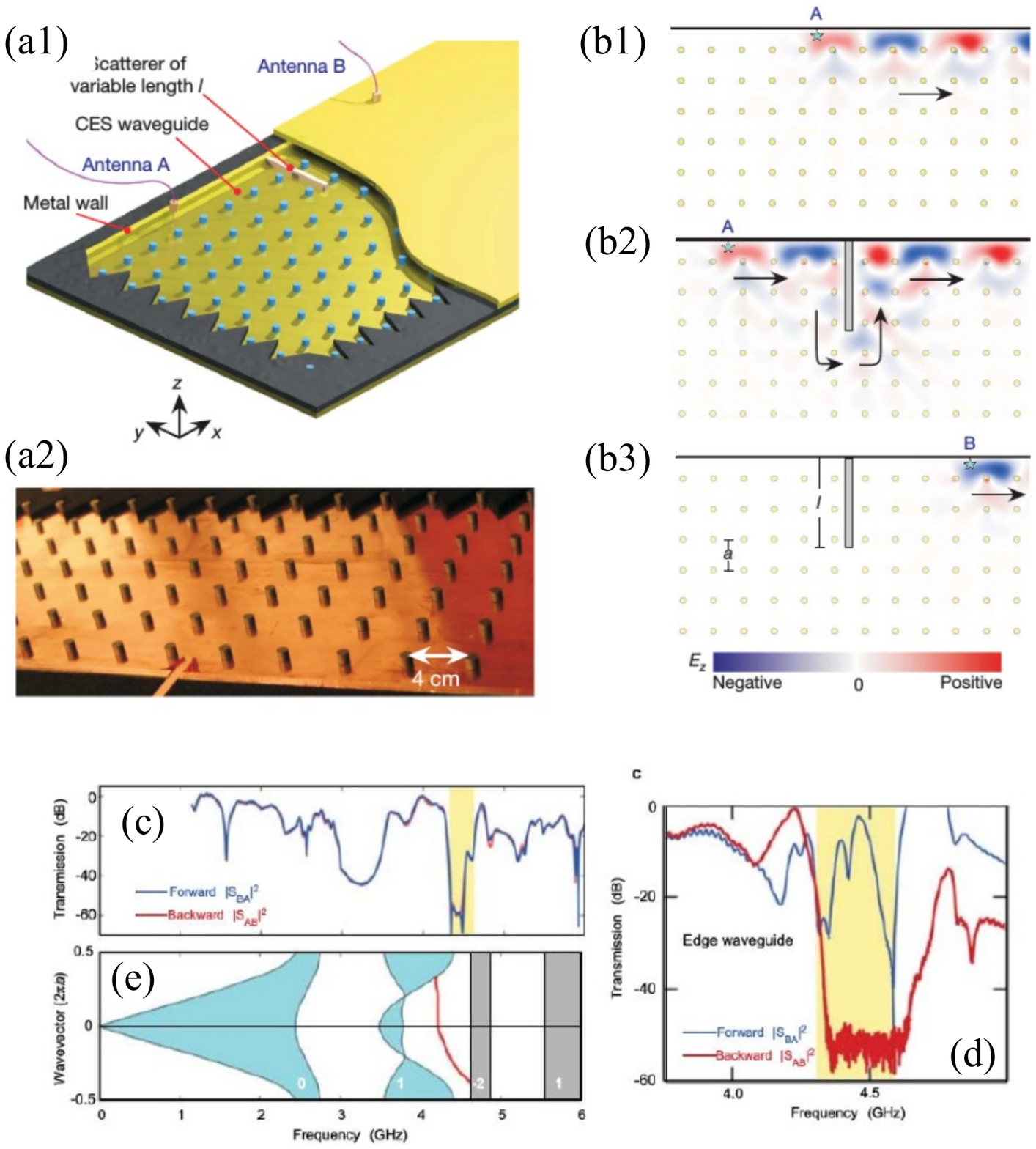}
\caption{(a1)  Sketch of the gyromagnetic photonic crystal slab used in the experiments of~\textcite{Wang:2009Nature}. The blue dots indicate the ferrite rods which are organized in a two-dimensional square lattice along the $x$-$y$ plane and are subject to a magnetic field of $0.2\,$T. The structure is sandwiched between two parallel copper plates providing confinement along $z$.  The chiral edge state is located at the boundary of the photonic crystal next to the metal wall. Two dipole antennas A and B serve as feeds and/or probes. Backscattering is investigated by inserting a variable-length metal obstacle between the antennas. (a2) A top view photograph of the actual waveguide with the top plate removed.
(b) Theoretical calculations of light propagation on edge states: (b1), (b3) unidirectional, non-reciprocal propagation from the antennas A and B, respectively. (b2) The immunity to backscattering against a defect. (c) Reciprocal transmission when the two antennas are located in the bulk. (d) Non-reciprocal transmission via the chiral edge state. Blue and red curves refer to transmission from antenna A to antenna B and viceversa. (e) Projected dispersion of the allowed photonic bands in the bulk (blue and gray) and the chiral edge state (red). The white numbers indicate the Chern number of each band. Adapted from~\textcite{Wang:2009Nature}.
}\label{fig:wangintro}
\end{figure}

Parallel to the growth in the study of topological phases of matter in condensed-matter systems, Haldane and Raghu made the crucial observation that topological band structures are, in fact, a ubiquitous property of waves inside a periodic medium, regardless of the classical or quantum nature of the waves. In their seminal works \cite{Haldane:2008PRL,Raghu:2008PRA}, they considered electromagnetic waves in two-dimensional spatially periodic devices embedding time-reversal-breaking magneto-optical elements and showed that the resulting photonic bands would have nontrivial topological invariants. Consequently, they predicted that such photonic systems would support robust chiral states propagating along the edge of the system at frequencies inside the photonic band gap. 

Shortly afterward, following a realistic proposal from~\textcite{Wang:2008PRL}, the idea of Haldane and Raghu was experimentally implemented using the two-dimensional magneto-optical photonic crystal structure in the microwave domain sketched in Figs.~\ref{fig:wangintro}(a1) and~\ref{fig:wangintro}(a2)~\cite{Wang:2009Nature}: a clear signature of the nontrivial band topology was indeed found in the unidirectionally propagating edge states and in the corresponding nonreciprocal behavior, as illustrated in the simulations of Figs.~\ref{fig:wangintro}(b) and in the experimental data of Figs.~\ref{fig:wangintro}(c) and~\ref{fig:wangintro}(d). More details on this and related following experiments are given in Sec.\ref{sec:analogIQHE}.

Further progress toward the implementation of such a model in the optical domain and the exploration of other topological models remained however elusive. One major challenge was the absence of a large magneto-optical response in the optical domain. One way to overcome this difficulty is to consider internal degrees of freedom of photons as pseudospins and look for an analogy of quantum spin Hall systems, namely, where the overall time-reversal symmetry is not broken but each pseudospin feels an artificial magnetic field~\cite{Hafezi:2011NatPhys,Umucalilar:2011PRA,Khanikaev:2013NatMat}. 
A second way is to use ideas from the Floquet topological insulators~\cite{Oka:2009PRB, Kitagawa:2010PRB,Lindner:2011NatPhys} known in condensed-matter physics, where temporal modulation is applied to the system to generate an effective time-independent Hamiltonian which breaks time-reversal symmetry~\cite{Fang:2012NatPhot}. A third way is to employ time-dependent modulation to implement a ``topological pump"~\cite{Thouless:1983PRB}; this last approach was realized experimentally in photonics in 2012~\cite{Kraus:2012a}, while the previous two ideas were realized in 2013 by two concurrent experiments~\cite{Rechtsman:2013Nature,Hafezi:2013NatPhot}.

Since then there has been great activity in the study of a variety of photonic systems realizing band structures with nontrivial topological invariants, leading to the emerging research field of {\em topological photonics}~\cite{lu2014topological, Lu:2016NatPhys2, khanikaev2017two, Sun:2017PQE}. Along similar lines, intense theoretical and experimental work has also been devoted to related topological effects in other areas of classical physics, such as in mechanical and acoustic systems. Reviews of the advances of these other fields have been given by~\textcite{Fleury:Acoustics2015,Huber:2016NatPhys}. 

This review is focused on the recent developments in the study of topological phases of matter in the photonics context. As we shall see in the following, in the last decade, topological ideas have successfully permeated the field of photonics, having been applied to a wide range of different material platforms, arranged in lattices of various dimensionalities, and operating in different regions of the electromagnetic spectrum, from radio waves and microwaves up to visible light.
One long term goal of topological photonics is to achieve and control strongly correlated states of photons with topological features such as fractional quantum Hall states. In addition to opening up perspectives for exploring the fundamental physics of topological phases of matter beyond solid-state systems, topological photonics also offers rich potential applications of these concepts to a novel generation of optoelectronic devices, such as optical isolators and topological lasers. 

The structure of this review article is the following. In Sec.\ref{sec:topologybasic}, we offer a general review of the main geometrical and topological concepts that have been developed in the study of solid-state electronic systems and that are commonly used in topological photonics. The following Sec.\ref{sec:drivedissipation} gives a general overview of the specific features that characterize photonic systems in contrast to electronic topological insulators.

In Sec.~\ref{sec:2D}, we discuss two-dimensional photonic systems which show topological features. This section is divided into four subsections. In Sec.~\ref{sec:analogIQHE}, we discuss two-dimensional photonic structures which break time-reversal symmetry and hence display physics analogous to integer quantum Hall systems. Relevant photonic systems include gyromagnetic photonic crystals, waveguide arrays, optomechanics, cavity QED, and circuit QED. The following Sec.~\ref{sec:analogQSHE} deals with two-dimensional photonic systems which do not break time-reversal symmetry and hence can be considered as analogs of the quantum spin Hall systems. Systems discussed in this section include silicon ring resonator arrays, radio-frequency circuits, twisted optical resonators, and photonic metamaterials. In Sec.~\ref{section_anomalous_Floquet}, we review photonic realizations of anomalous Floquet topological systems with waveguide arrays, namely, temporally modulated systems displaying topological features that do not have an analog in static Hamiltonians. Section~\ref{subs:gapless} discusses two-dimensional gapless systems such as honeycomb lattices, whose features can also be understood from topological considerations. Systems discussed in this section include microwave resonators, photorefractive crystals, coupled microlasers, exciton-polariton lattices, and waveguide arrays.

Section~\ref{sec:1D} is devoted to photonic realizations of one-dimensional topological systems. In Sec.\ref{sec:1Dchiral} we concentrate on systems with chiral symmetry, such as the Su-Schrieffer-Heeger (SSH) model. In Sec.\ref{sec:pump} we review photonic realizations of topological pumping.
Systems of higher dimensionality are then considered in Sec.~\ref{sec:higher}.
Three-dimensional gapless phases with features originating from topological charges in momentum space, such as Weyl points, are discussed in Sec.\ref{sec:3Dgapless}. In Sec.\ref{sec:3dgapped}, we discuss gapped three-dimensional phases and their topological interface states. The following Sec.\ref{sec:4D} presents the concept of synthetic dimensions, which could be used to realize models with an effective spatial dimensionality higher than three, e.g., four-dimensional quantum Hall systems.

In Sec.~\ref{sec:gainloss}, we discuss photonic systems where gain and loss play an essential role. Such systems are described by non-Hermitian Hamiltonians and do not find a direct counterpart in electronic topological insulators. Section~\ref{sec:nonhermitian} focuses on the interplay of gain and loss, while the following Sec.~\ref{sec:bogoliubov} reviews the emergent topology of Bogoliubov modes that arises from parametric down conversion processes.

Section~\ref{sec:interaction} is devoted to an overview of the interplay between topology and optical nonlinearities. Theoretical work on nonlinear effects stemming from weak nonlinearities is reviewed in Sec.\ref{subsec:weak}, while the following Sec.\ref{subsec:strong} highlights the prospect of strong photon-photon interactions mediated by strong nonlinearities to realize topologically nontrivial strongly correlated states of photons. Some of the future perspectives of the field of topological photonics are finally illustrated in Sec.~\ref{sec:conclusion}.

Compared to earlier reviews on topological photonics~\cite{lu2014topological, Lu:2016NatPhys2, khanikaev2017two, Sun:2017PQE}, this review aims to be comprehensive, starting from the very basics of topological phases of matter and trying to cover most of the works that have appeared in the last years in relation to topological phases of matter in optical systems in any dimensionality.
However, we need to warn the readers that space restrictions force us to leave out many other fields of the optical sciences that relate to topological concepts, e.g. the rich dynamics of optical vortices in singular optics~\cite{Dennis:2009ProgOpt,Gbur:Book}, the topology underlying knots in complex electromagnetic fields~\cite{Arrayas:2017PhysRep}, and the topological ideas underlying bound states in the continuum~\cite{Hsu:2016NatRev}. For all these advances, we refer the readers to the rich specific literature that is available on each of them.

%% file: RMP_IIA_Review.tex
\section{Basic concepts}
\label{sec:basicconcepts}

In this section, we introduce general concepts of topological phases of matter and optical and photonic systems that are needed in the following sections. In Sec.~\ref{sec:topologybasic}, we briefly review the paradigm of topological phases of matter, as it was originally developed in the context of electronic systems in solid-state materials, and illustrate the basic technical and mathematical tools to describe them. Then in Sec.~\ref{sec:drivedissipation}, we review the principal features of photonic systems used for topological photonics, with a special emphasis on their differences and peculiarities  as compared to electronic systems.

\subsection{Topological phases of matter}
\label{sec:topologybasic}

According to Bloch's theorem, the eigenstates of a quantum particle in a periodic potential are organized into energy bands separated by energy gaps. This band structure determines the metallic or insulating nature of different solid-state materials~\cite{AshcroftMermin}. Besides the energy dispersion of the bands, the geometrical structure of the Bloch eigenstates in momentum space can also have an impact on the electronic properties of materials as first discovered by~\onlinecite{Karplus:1954PR} and~\onlinecite{Adams:1959JPCS}. This geometrical structure is reflected also in integer-valued global topological invariants associated with each band, as we see later. In spite of their seemingly abstract nature~\cite{Simon:1983PRL},
nontrivial values of these topological invariants have observable consequences such as the quantized bulk conductance in the quantum Hall effect and in the emergence of topologically protected edge states located on the physical boundary of the system~\cite{Qi:2011RMP,Hasan:2010RMP,BernevigBook,Volovik:2009Book}. 

This section is devoted to an introduction to the basic concepts of such {\em topological phases of matter}. 

\subsubsection{Integer quantum Hall effect}
\label{sec:iqhe}

The quantum Hall effect is historically the first phenomenon where momentum-space topology was recognized to lead to observable physical phenomena.
The integer quantum Hall effect was discovered in a two-dimensional electron gas subject to a strong perpendicular magnetic field by~\onlinecite{Klitzing:1980PRL}, who observed a robust quantization of the Hall conductance in units of $e^2/h$, where $e$ is the charge of an electron and $h$ is Planck's constant. Soon after, the extremely robust quantization of the Hall conductance was related to the topology of bands in momentum space by Thouless, Kohmoto, Nightingale, and den Nijs (TKNN)~\cite{Thouless:1982PRL}.

In order to review this landmark result, we first need to introduce the basic geometrical and topological properties of eigenstates in momentum space, such as the local Berry connection and Berry curvature and the global Chern number, respectively.
We consider a single-particle Hamiltonian $\hat{H}(\hat{\mathbf{r}}, \hat{\mathbf{p}})$ in generic dimension $d$, where $\hat{\mathbf{r}}$ and $\hat{\mathbf{p}}$ are, respectively, the position and momentum operators. We assume that the Hamiltonian obeys the spatial periodicity condition $\hat{H}(\hat{\mathbf{r}} + \mathbf{a}_i, \hat{\mathbf{p}}) = \hat{H}(\hat{\mathbf{r}}, \hat{\mathbf{p}})$, where $\{ \mathbf{a}_i \}$ are a set of $d$ lattice vectors. Thanks to the spatial periodicity, one can invoke Bloch's theorem to write the eigenstates as
\begin{equation}
\psi_{n,\mathbf{k}}(\mathbf{r}) = e^{i\mathbf{k}\cdot \mathbf{r}} u_{n,\mathbf{k}} (\mathbf{r}), 
\end{equation}
where $n$ is the band index and $\mathbf{k}$ is the crystal momentum defined within the first Brillouin zone. The Bloch state $u_{n,\mathbf{k}} (\mathbf{r})$ obeys the same periodicity as the original Hamiltonian $u_{n,\mathbf{k}} (\mathbf{r} + \mathbf{a}_i) = u_{n,\mathbf{k}} (\mathbf{r})$ and is an eigenstate of the Bloch Hamiltonian 
\begin{equation}
 \hat{H}_\mathbf{k} \equiv e^{-i\mathbf{k}\cdot \hat{\mathbf{r}}}\hat{H}(\hat{\mathbf{r}},\hat{\mathbf{p}}) e^{i\mathbf{k}\cdot \hat{\mathbf{r}}},
\end{equation}
namely 
\begin{equation}
\hat{H}_\mathbf{k} u_{n,\mathbf{k}}(\mathbf{r}) = E_n (\mathbf{k}) u_{n,\mathbf{k}} (\mathbf{r}),
\label{eigenvalueHk}
\end{equation}
where $E_n (\mathbf{k})$ is the energy dispersion of the $n$th band~\cite{AshcroftMermin}. 

The physics of an energy band is captured in part by its dispersion relation $E_n (\mathbf{k})$, but also by the {\it geometrical} properties of how its eigenstates $u_{n,\mathbf{k}}(\mathbf{r})$ vary as a function of $\mathbf{k}$~\cite{Karplus:1954PR, Adams:1959JPCS, Resta:1994RMP, Resta:2011EPJB, Nagaosa:2010RMP}. This geometry of the eigenstates is encoded by the Berry phase~\cite{Pancharatnam:1956PIAS,Berry:1984PRSLA,Hannay:1985JPA}, which is defined in the following.
Whereas the Berry phase can be defined for a very general parameter space, in our discussion of topological phases of matter we restrict ourselves to the case where the parameters are the crystal momentum $\mathbf{k} = (k_x,k_y,k_z)$ varying over the first Brillouin zone. Then if one prepares a localized wave packet from states in band $n$ and makes it adiabatically move along a closed path in momentum space, it will acquire a dynamical phase, determined by the time integral of its $\mathbf{k}$-dependent energy, but also a Berry phase~\cite{Xiao:2010RMP}
\begin{align}
	\gamma
	=
	\oint \boldsymbol{\mathcal{A}}_n (\mathbf{k})\cdot d\mathbf{k}, \label{beryphasedef}
\end{align}
that is geometrically determined by an integral over the same momentum-space path, of the Berry connection, defined as
\begin{align}
	\boldsymbol{\mathcal{A}}_n (\mathbf{k})
	\equiv
	i\langle u_{n,\mathbf{k}} | \boldsymbol\nabla_\mathbf{k} | u_{n,\mathbf{k}} \rangle. \label{eq:berryconnection}
\end{align}
Note that the definition of the Bloch states via \eq{eigenvalueHk} does not specify the overall phase of $|u_{n,\mathbf{k}}\rangle$, so one can freely choose the phase of the Bloch states. 
Under a gauge transformation $|u_{n,\mathbf{k}}\rangle \to e^{i\chi(\mathbf{k})} |u_{n,\mathbf{k}}\rangle$, the Berry connection is not gauge invariant and transforms as $\boldsymbol{\mathcal{A}}_n (\mathbf{k}) \to \boldsymbol{\mathcal{A}}_n (\mathbf{k}) - \boldsymbol\nabla_\mathbf{k} \chi(\mathbf{k})$. However, the single valuedness of $e^{i\chi(\mathbf{k})}$ at the beginning and end of the path imposes that the Berry phase~(\ref{beryphasedef}) for a given closed path is gauge invariant modulo $2\pi$. Additionally, from the gauge-dependent Berry connection $\boldsymbol{\mathcal{A}}_n (\mathbf{k})$ one can construct a gauge invariant Berry curvature, which in three dimensions takes the following form:
\begin{align}
	\boldsymbol{\Omega}_n (\mathbf{k})
	= \boldsymbol\nabla_\mathbf{k} \times \boldsymbol{\mathcal{A}}_n (\mathbf{k}), \label{II_omega}
\end{align}
and which encodes the geometrical properties of the $n$th band.
In two dimensions, the Berry curvature has only one component:
\begin{align}
	\Omega_{n} (\mathbf{k})
	=
	i\left(
	\langle \partial_{k_x} u_{n,\mathbf{k}}|\partial_{k_y} u_{n,\mathbf{k}} \rangle
	-
	\langle \partial_{k_y} u_{n,\mathbf{k}}|\partial_{k_x} u_{n,\mathbf{k}} \rangle
	\right).
\end{align}
Importantly, although the Berry curvature is a gauge invariant quantity that is continuously defined over the whole Brillouin zone, the phase of the Bloch states themselves cannot always be chosen to be continuous. Whether this is possible or not depends on the value of a topological invariant of the band, the Chern number, defined as the integral
\begin{align}
	C_n
	=
	\frac{1}{2\pi}\int_{\mathrm{BZ}} d^2 k\, \Omega_n (k_x, k_y), \label{II_chern}
\end{align}
over the whole first Brillouin zone. If one can define the phase of the Bloch state, and hence the Berry connection $\boldsymbol{\mathcal{A}}_n (\mathbf{k})$, continuously over the whole Brillouin zone, a direct consequence of the definition $\boldsymbol\Omega_n (\mathbf{k}) = \boldsymbol\nabla_\mathbf{k} \times \boldsymbol{\mathcal{A}}_n (\mathbf{k})$ and of Stokes's theorem is that the Chern number must necessarily be zero. Conversely, having a nonzero Chern number implies that the Bloch state and hence the Berry connection $\boldsymbol{\mathcal{A}}_n (\mathbf{k})$ cannot be continuously defined.

\begin{figure}
\resizebox{0.15 \textwidth}{!}{\includegraphics*{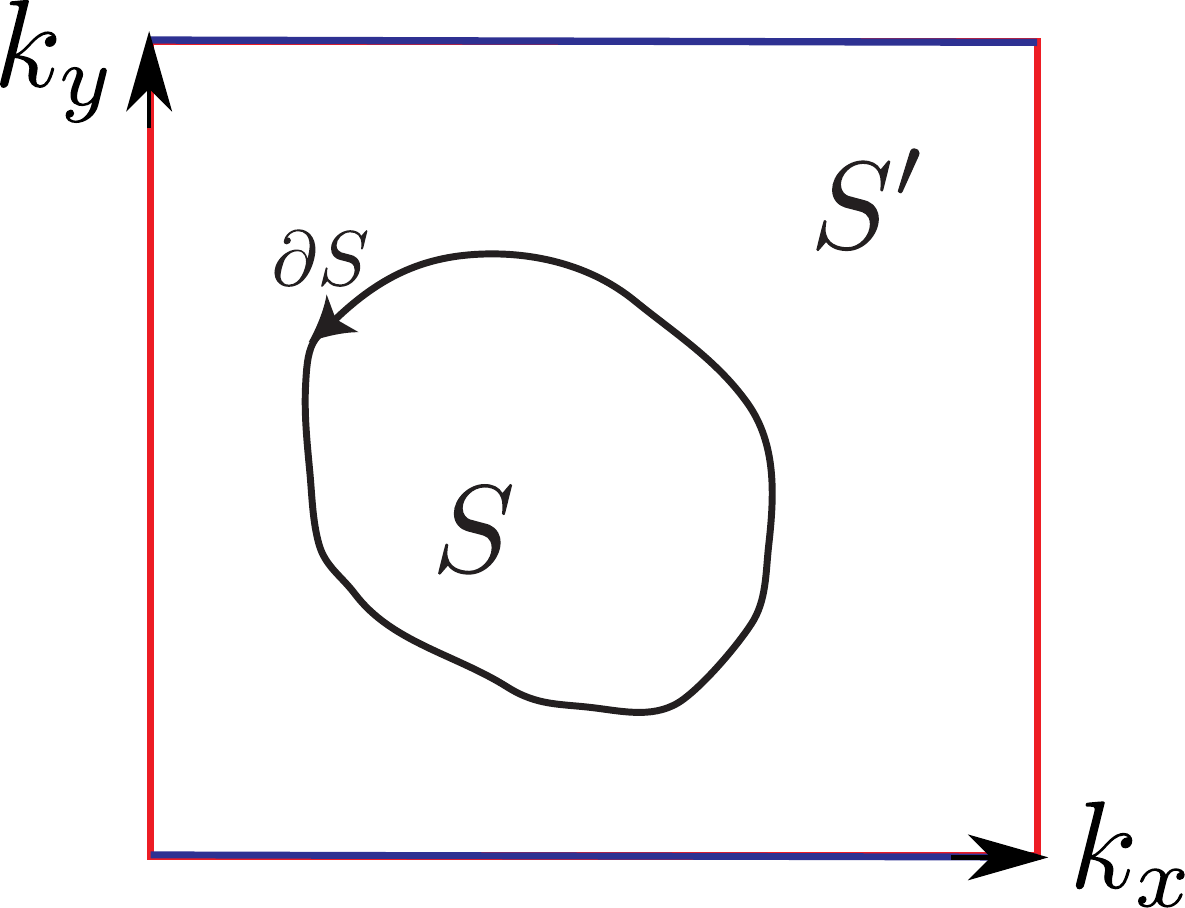}}
\caption{A schematic illustration of how the Brillouin zone is divided into two parts, $S$ and $S^\prime$. Thanks to the periodicity of the quasimomentum, the two-dimensional Brillouin zone has a torus-like structure, in which top-bottom and left-right edges (purple and red) should be identified.} \label{II_divide}
\end{figure}

It is not difficult to see that the Chern number always takes an integer value~\cite{Kohmoto:1985AnnPhys}. To this purpose, we divide the integration domain of Eq.~(\ref{II_chern}) into two regions $S$ and $S^\prime$ as sketched in Fig.~\ref{II_divide}. Within $S$, we choose a continuous gauge for the Bloch state, giving the Berry connection $\boldsymbol{\mathcal{A}}_n (\mathbf{k})$. Similarly, within $S^\prime$,  we choose a continuous gauge, which yields the Berry connection $\boldsymbol{\mathcal{A}}^\prime_n (\mathbf{k})$. Keeping in mind that the first Brillouin zone can be thought of as a torus, thanks to the periodicity of the quasimomentum, we can use Stokes's theorem within $S$ and $S^\prime$ and rewrite the Chern number in terms of the line integral along the common boundary $\partial S = -\partial S^\prime$,
\begin{align}
	C_n
	&=
	\frac{1}{2\pi}\int_{S} d^2 k\, \Omega_n (k_x, k_y)
	+
	\frac{1}{2\pi}\int_{S^\prime} d^2 k\, \Omega_n (k_x, k_y)
	\notag \\
	&=
	\frac{1}{2\pi}\oint_{\partial S} d \mathbf{r} \cdot \boldsymbol{\mathcal{A}}_n (\mathbf{k})
	-
	\frac{1}{2\pi}\oint_{\partial S} d \mathbf{r} \cdot \boldsymbol{\mathcal{A}}^\prime_n (\mathbf{k})
	\notag \\
	&=
	\frac{1}{2\pi}\left( \gamma - \gamma^\prime \right),
\end{align}
where $\gamma$ and $\gamma^\prime$ are the Berry phases along the contour $\partial S$ calculated using $\boldsymbol{\mathcal{A}}_n (\mathbf{k})$ and $\boldsymbol{\mathcal{A}}_n (\mathbf{k})^\prime$, respectively.
As the Berry phases are calculated along the same path, they must be equal up to multiples of $2\pi$. This in turn implies that the Chern number $C_n$ must be an integer. Importantly, this integer-valued quantity has a profound topological origin~\cite{Avron:1983PRL,Simon:1983PRL,Niu:1985PRB}, which indicates that its value must remain strictly constant under smooth perturbations that preserve the band gaps separating the band $n$ to neighboring bands~\cite{Avron:1983PRL}. Fermionic systems in which the fermions completely fill Bloch bands with nonzero Chern numbers are generically termed {\it Chern insulators}.

Within linear response theory and ignoring interparticle interactions, one can show that the Hall conductance $\sigma_{xy}$ of a two-dimensional insulator is~\cite{Thouless:1982PRL}
\begin{align}
	\sigma_{xy}
	&=
	-\frac{e^2}{h}
	\sum_{n}
	C_n, \label{II_TKNNformula}
\end{align}
where the Chern numbers are summed over the $n$ occupied bands. Since the Chern numbers can take only integers, it follows that the Hall conductance is quantized in units of $e^2/h$. As we shall see shortly, in the simplest case of a uniform two-dimensional electron gas under a strong magnetic field, the energy levels form flat Landau levels, and all the Landau levels have the same Chern number. Therefore the Hall conductance of the integer quantum Hall effect is proportional to the number of occupied Landau levels.

The quantization of the Hall conductance can also be related to the number of modes that propagate unidirectionally around the system, the so-called chiral edge modes. Indeed, each of such edge modes contributes $-e^2/h$ to the measured Hall conductance~\cite{Halperin1982,MacDonald:1984PRB,Buttiker:1988PRB}. The existence of such current-carrying edge modes is also constrained by topology, in the sense that the sum of the Chern numbers associated with the occupied bulk bands is equal to the number of edge modes contributing to the edge current~\cite{Hatsugai:1993PRL, Hatsugai:1993PRB, Qi:2006PRB}. This relationship between a bulk topological invariant, such as the Chern number, and the number of localized edge modes is an example of the bulk-edge correspondence, i.e., a matching between the topological properties defined in the bulk of a material with its boundary phenomena~\cite{BernevigBook}.

While the previous discussions are based on single-particle energy bands in a perfect crystal, the definition of the Chern number can also be generalized to include the effects of interactions and disorder~\cite{Niu:1985PRB}. When the interparticle interactions become very strong, the Hall conductance can become quantized at {\it fractional values} of $e^2 / h$~\cite{Tsui:1982PRL}. This is known as the ``fractional quantum Hall effect" in which the quantum many-body ground state is strongly correlated and topological. Remarkably, the excitations of such a fractional quantum Hall state can have a fractional charge and possibly even fractional statistics~\cite{Laughlin:1983PRL,Arovas:1984PRL}. Progress toward realizing analog fractional quantum Hall states of light is reviewed in Sec.~\ref{subsec:strong}.

In the rest of this section, we proceed with a detailed discussion of a few important models for integer quantum Hall systems and Chern insulators.
We start by considering a two-dimensional electron gas under a strong and uniform magnetic field, which gives rise to Landau levels and to the integer quantum Hall effect~\cite{PrangeBook,YoshiokaBook}.
The second example is the Harper-Hofstadter model that is a tight-binding lattice model in a uniform magnetic field~\cite{Harper:1955PPSA, Hofstadter:1976PRB,Azbel:1964JETP}.
The third one is the Haldane model~\cite{Haldane:1988PRL}, which is the first example of a Chern insulator model with alternating magnetic flux patterns. We then conclude by illustrating the bulk-edge correspondence on a simple Jackiw-Rebbi model~\cite{Jackiw:1976PRD}.

{\it Landau levels:}
The quantum Hall effect was originally found in a semiconductor hetero-junction where electrons are confined to move in a two-dimensional plane~\cite{Klitzing:1980PRL}. This system can be modeled, to a first approximation, as a two-dimensional electron gas in free space under a constant magnetic field. The single-particle Hamiltonian is
\begin{align}
	\hat{H} = \frac{[\hat{p}_x - e A_x (\hat{\mathbf{r}})]^2 + [\hat{p}_y - e A_y (\hat{\mathbf{r}})]^2}{2m}, \label{II_LLHam}
\end{align}
where $\mathbf{A}(\mathbf{r}) = (A_x (\mathbf{r}), A_y (\mathbf{r}),0)$ is the magnetic vector potential, associated with the magnetic field $\mathbf{B} = \boldsymbol\nabla_\mathbf{r} \times \mathbf{A}(\mathbf{r})$.

For a given magnetic field, different forms of the vector potential $\mathbf{A}(\mathbf{r})$ can be chosen. For our case of a constant magnetic field along the $z$ direction, $\mathbf{B} = (0,0,B)$, the two most common choices are the Landau gauge, which keeps translational symmetry along one direction as $\mathbf{A}(\mathbf{r}) = (-yB,0,0)$ or $(0,xB,0)$, and the symmetric gauge, $\mathbf{A}(\mathbf{r}) = (-yB/2,xB/2,0)$, which keeps instead rotational symmetry. Physical observables such as the energy spectrum and the Hall conductance do not depend on the choice of gauge.

The single-particle energy spectrum of this system consists of equally spaced {\it Landau levels} of energy
\begin{align}
	E_n = \hbar \omega_c \left( n + 1/2\right),
\end{align}
where $\omega_c \equiv |e|B/m$ is the cyclotron frequency and the integer $n\geq 0$.
For a large system, each Landau level is highly degenerate with a degeneracy equal to the number of unit magnetic fluxes $\phi_0=h/|e|$ piercing the system.

Regarding each Landau level as an energy band, the system can be considered as a Chern insulator when the Fermi energy lies within an energy gap.
The Chern number is one for any Landau level. Then, from the TKNN formula~(\ref{II_TKNNformula}), the Hall conductance is thus proportional to the number of occupied Landau levels, which explains the basic phenomenology of the integer quantum Hall effect. 

{\it Harper-Hofstadter model:}
The next model we consider is the discrete lattice version of the Landau level problem, the Harper-Hofstadter model~\cite{Harper:1955PPSA, Hofstadter:1976PRB,Azbel:1964JETP}. In tight-binding models, the magnetic vector potential $\mathbf{A} (\mathbf{r})$ enters as a nontrivial phase of the hopping amplitude between neighboring sites, called the \textit{Peierls phase}. In the simplest cases, the phase accumulated when hopping from a site at $\mathbf{r}_1$ to a site at $\mathbf{r}_2$ can be written in terms of the vector potential as
\begin{equation}
\Phi_{\mathbf{r}_1\rightarrow \mathbf{r}_2}= \frac{e}{\hbar}\int_{\mathbf{r}_1}^{\mathbf{r}_2} \mathbf{A}(\mathbf{r}) \cdot d\mathbf{r},
\end{equation}
where the integral is taken along a straight line connecting the two points~\cite{Peierls:1933ZPhys,Luttinger:1951PR}.

\begin{figure}
\resizebox{0.45 \textwidth}{!}{\includegraphics*{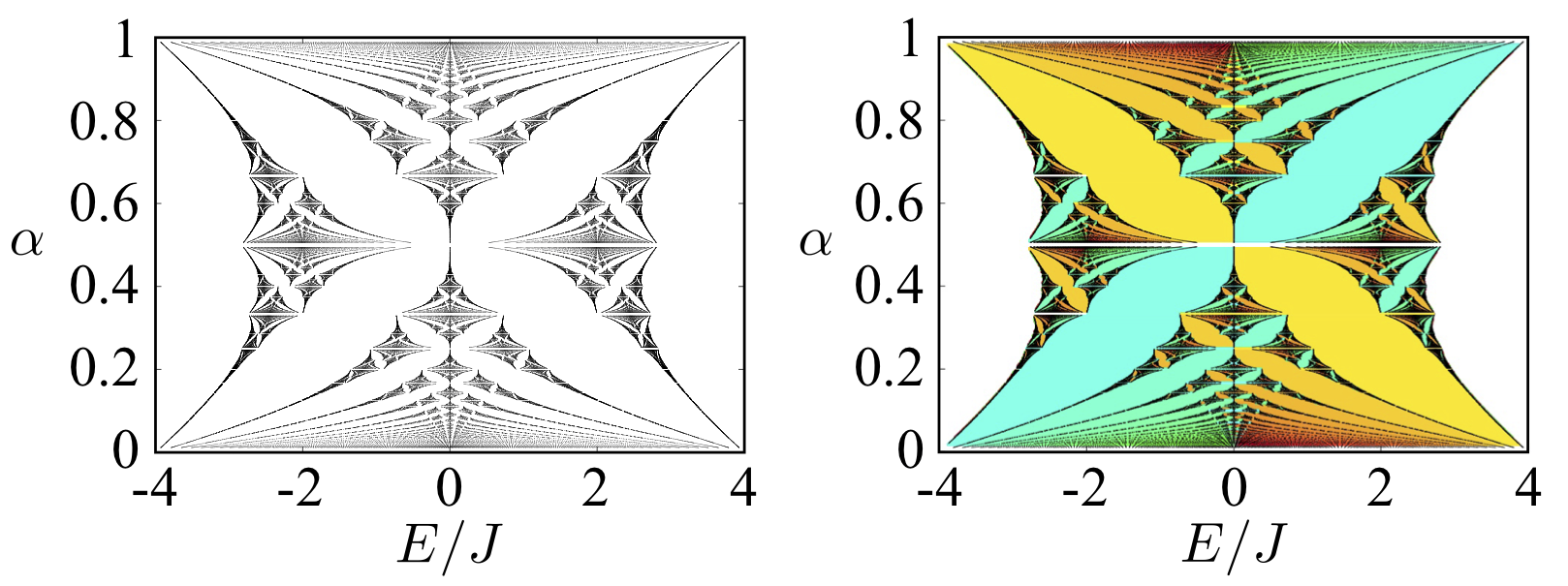}}
\caption{(Left) Energy spectrum of the Harper-Hofstadter model, which is called the Hofstadter butterfly. (Right) The colored Hofstadter butterfly in which the color of each band gap indicates the topological invariant of the gap, given by the sum of the Chern numbers of all bands below. Warm colors indicate a positive topological invariant, whereas the cool colors indicate a negative topological invariant. The horizontal axes are the energies and the vertical axes are the flux $\alpha$.} \label{II_HHspectrum}
\end{figure}

Choosing for definiteness the Landau gauge along the $y$ direction, $\mathbf{A} (\mathbf{r}) = (0, Bx, 0)$, the Hamiltonian of the Harper-Hofstadter model on a square lattice is
\begin{align}
	\hat{H}
	=
	-J
	\sum_{x,y}
	\left(
	\hat{a}^\dagger_{x+a,y}\hat{a}_{x,y}
	+
	e^{i2\pi \alpha x/a}\hat{a}^\dagger_{x,y+a}\hat{a}_{x,y}
	+
	\mathrm{H.c.}
	\right), \label{II_HHHamiltonian}
\end{align}
where $\hat{a}_{x,y}$ is the annihilation operator of a particle at site $(x,y)$, $J$ is the magnitude of the (isotropic) hopping amplitude, and $a$ is the lattice spacing. The intensity of the magnetic field in the system is quantified by the parameter $\alpha$, obeying $\alpha \phi_0 = Ba^2$, which identifies the magnetic flux per plaquette of the lattice in units of the magnetic flux quantum $\phi_0$.
The main distinction from the Landau level case previously discussed is that in the Hofstadter model there are two competing length scales: the lattice spacing and the magnetic length. As a result, the electron paths interfere to give the fractal energy spectrum as a function of $\alpha$, which is widely known as the {\it Hofstadter butterfly} and which is plotted in the left panel of Fig.~\ref{II_HHspectrum}.
The first experimental demonstration of the Hofstadter butterfly was performed in a microwave waveguide, exploiting the analogy between the transfer matrix governing the transmission of microwaves and the eigenvalue equation of the Harper-Hofstadter model~\cite{Kuhl:1998PRL}.

To get more insight into this spectrum, it is useful to concentrate on cases where $\alpha$ is a rational number, $\alpha = p/q$ with $p$ and $q$ being co-prime integers.
Because of the spatially varying hopping phase, the Hamiltonian breaks the basic periodicity of the square lattice. Periodicity is, however, recovered if we consider a larger unit cell of $q\times 1$ plaquettes: this is called the {\it magnetic unit cell}~\cite{Zak:1964PR,Dana:1985JPC}. As the number of bands in lattice models is equal to the number of lattice sites per magnetic unit cell, the Harper-Hofstadter model with $\alpha = p/q$ has $q$ bands.

To find the geometrical and topological properties of the model, one can diagonalize the momentum-space Hamiltonian~\cite{Harper:1955PPSA, Hofstadter:1976PRB,Azbel:1964JETP}
\begin{align}
	\hat{H}_\mathbf{k}
	=
	-J
	\sum_{i = 0}^q
	&\left[
	\cos \left( k_y - 2\pi \alpha\right) \hat{a}_i(\mathbf{k})^\dagger \hat{a}_i(\mathbf{k})
	\right.	
	\notag \\
	&\left.
	+ e^{-ik_x} \hat{a}_{i+1} (\mathbf{k})^\dagger \hat{a}_i(\mathbf{k})
	+
	\mathrm{H.c.}
	\right],
\end{align}
where $i$, defined mod $q$, indicates the site within a magnetic unit cell, and the momentum $\mathbf{k}$ is chosen within the magnetic Brillouin zone: $-\pi/q \le k_x \le \pi/q$ and $-\pi \le k_y \le \pi$. An explicit calculation shows that the Chern numbers of all isolated bands of the Harper-Hofstadter model are nonzero and can be found as solutions of a simple Diophantine equation~\cite{Thouless:1982PRL}. As shown in the right panel of Fig.~\ref{II_HHspectrum}, this model exhibits a rich structure of positive and negative Chern numbers depending on the magnetic flux. 

{\it Haldane model:}
The Haldane model~\cite{Haldane:1988PRL} is the first model system that exhibits a nonzero quantized Hall conductance in a nonuniform magnetic field with a vanishing average flux per plaquette. This model demonstrated that, to obtain the quantum Hall effect, the essential feature required is, in fact, not a net magnetic field but the breaking of time-reversal symmetry. As the Haldane model consists of a honeycomb lattice with suitable hopping amplitudes, it is useful to start by briefly reviewing the physics of a tight-binding model on a honeycomb lattice, which is often used to describe electrons in graphene~\cite{CastroNeto:2009RMP} and which has recently been widely implemented in photonics, as we will review in Sec.~\ref{subs:gapless}.

\begin{figure}
\resizebox{0.49 \textwidth}{!}{\includegraphics*{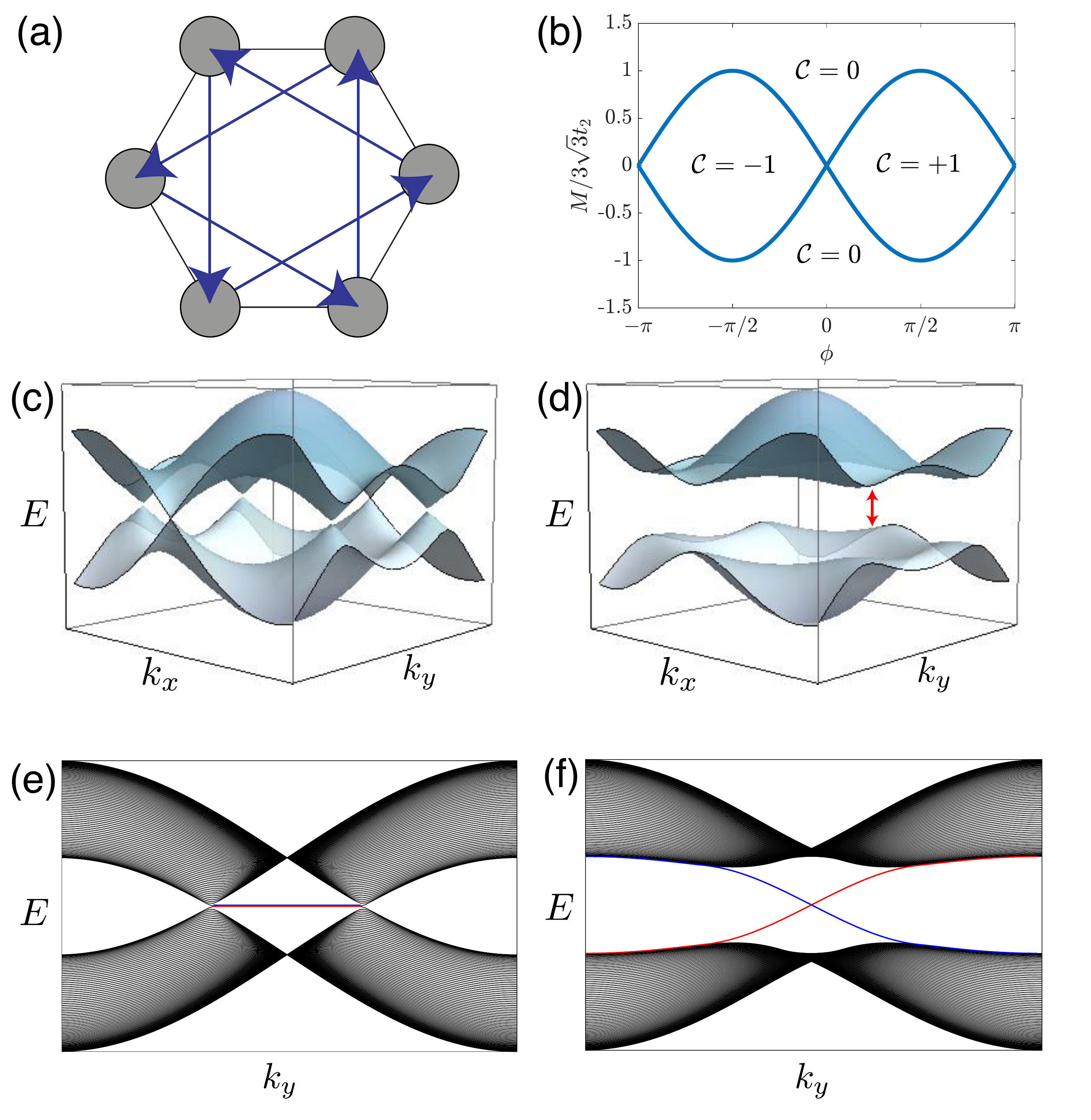}}
\caption{(a) A plaquette of the Haldane model. In addition to the usual nearest-neighbor hoppings, there are also complex next-nearest-neighbor hoppings. For the latter, hopping along the arrows carries a phase of $\phi$, whereas the hopping opposite to the arrows carries the opposite phase of $-\phi$. (b) The phase diagram of the Haldane model as a function of the next-nearest-neighbor hopping phase $\phi$ and the sublattice energy difference $2M$. (c) Bulk band structure of the honeycomb lattice with nearest- neighbor hopping only. Conical touchings of the bands are Dirac points. (d) Typical bulk band structure of the Haldane model in the presence of a band gap. (e) Typical band structure for a gapless honeycomb lattice with zigzag edges. Flat-band localized edge states (indicated in color) connect between the Dirac cones, with one per edge. (f) Typical band structure with edges on both sides of the system when a topological gap opens. Red and blue lines indicate edge states on left and right edges, respectively.} \label{II_Haldane}
\end{figure}

The honeycomb lattice with nearest-neighbor hopping is one of the simplest examples of a system which exhibits {\it Dirac cones} in the band structure, namely linear crossings of the energy dispersion of two neighboring bands. A honeycomb lattice has two lattice sites per unit cell, which gives two bands. These are degenerate at two isolated and inequivalent points in the Brillouin zone, called Dirac points. Around the Dirac points, the effective Hamiltonian of the two bands takes the following form in the sublattice basis:
\begin{align}
	\hat{H}_\mathrm{D} \approx \hbar v_D \left( q_x \sigma_x + q_y \sigma_y\right), \label{II_DiracHam}
\end{align}
where $v_D \equiv 3t_1/2$ is called the Dirac velocity, with $t_1$ being the nearest-neighbor hopping amplitude, and $(q_x, q_y)$ is the momentum measured from a Dirac point. As a result, the dispersion around the Dirac point is linear, $E = \pm \hbar v_D \sqrt{q_x^2 + q_y^2}$, and is referred to as the Dirac cone. A complete plot of the band dispersion is shown in Fig.~\ref{II_Haldane}(c).

In order to open an energy gap at Dirac points, one needs to add a term proportional to $\sigma_z$ in $\hat{H}_\mathrm{D}$. This can be achieved by either breaking time-reversal symmetry or inversion symmetry, which implies that as long as both time-reversal symmetry and inversion symmetry are preserved the gapless Dirac points are protected~\cite{BernevigBook}.

The key novelty of the Haldane model is to add two more sets of terms to the nearest-neighbor honeycomb lattice model, which open energy gaps at the Dirac cones in complementary ways, by breaking inversion symmetry or time-reversal symmetry. The first set of terms is a constant energy difference $2M$ between two sublattices, which breaks inversion symmetry. The second set of terms are next-nearest-neighbor hoppings with magnitude $t_2$ and complex hopping phases breaking time-reversal symmetry, i.e., hopping along the arrows in Fig.~\ref{II_Haldane}(a) carries a phase of $\phi$, whereas the hopping along the opposite direction carries the opposite phase of $-\phi$.
While adding the energy difference between sublattices opens a trivial gap, in the sense that the resulting bands are topologically trivial, adding complex next-nearest-neighbor hoppings results in opening a gap with topologically nontrivial bands. This different behavior is due to the breaking of time-reversal symmetry by the complex hopping phases, a necessary condition to obtain a nonzero Chern number~\cite{BernevigBook}. The full topological phase diagram of the Haldane model as a function of the magnitude of the two gap-opening terms is summarized in Fig.~\ref{II_Haldane}(b), while an example of the energy band dispersion for a bulk system is shown in Fig.~\ref{II_Haldane}(d).

The nontrivial topology of the Haldane model can also be seen from the appearance of chiral-propagating edge states. 
The gapless dispersion of a finite slab of honeycomb lattice is displayed in Figs.~\ref{II_Haldane}(e), to be contrasted with the topologically nontrivial gap of the Haldane model displayed in Fig.~\ref{II_Haldane}(f).

The dispersion of a finite slab of the Haldane model is displayed in Figs.~\ref{II_Haldane}(e) and~\ref{II_Haldane}(f) for the gapless case and the gapped case with a topologically nontrivial gap, respectively. Along the $y$ direction the system is taken to be periodic, so the momentum $k_y$ is a good quantum number. In the other $x$ direction, we have a large but finite-size system with a sharp edge. While the edge states of the unperturbed honeycomb lattice are flat at the energy of the Dirac points [Fig.~\ref{II_Haldane}(e)], two propagating edge states appear when a bulk topological gap opens, which traverse the energy gap in opposite directions along opposite edges of the system [Fig.~\ref{II_Haldane}(f)]. Propagation of each of these states is therefore unidirectional and is protected by the fact that these edge states are spatially separated, suppressing scattering processes from one edge state into the other.

The Haldane model, and its generalization, 
has been realized in solid-state systems by~\onlinecite{Chang:2013Science}, as well as in photonics~\cite{Rechtsman:2013Nature} and ultracold atomic gases~\cite{Jotzu:2014Nat}.

\begin{figure}
\resizebox{0.35 \textwidth}{!}{\includegraphics*{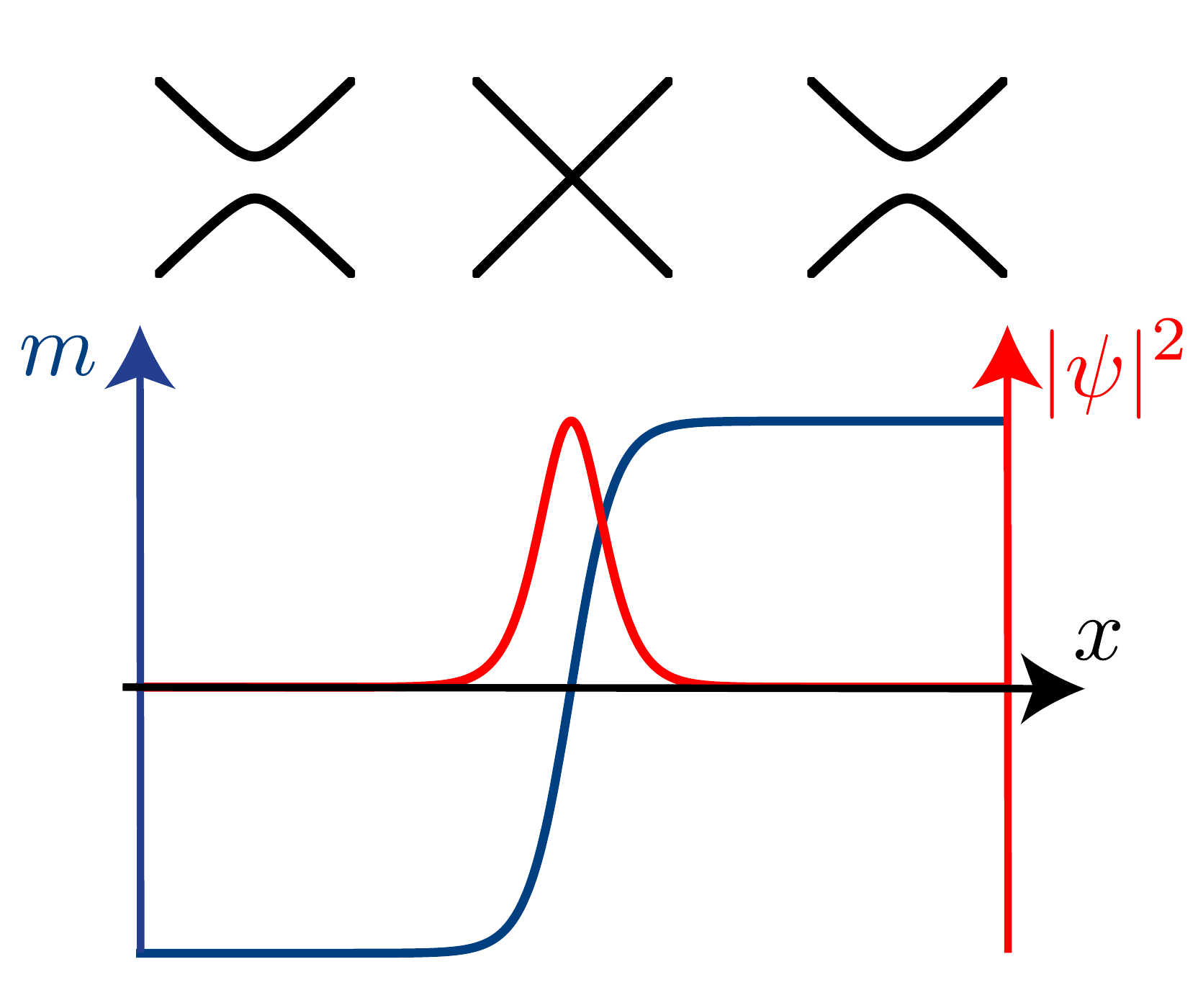}}
\caption{A schematic illustration of how the interface of two regions with different topological numbers can host a localized state. The spatial dependence of the mass term $m(x)$ and the wave function residing at the interface ($x=0$) are schematically plotted. Sketched on top is the schematic dispersion for values of $m$ at corresponding positions, showing that the bulk band gap closes when $m(0)=0$.} \label{II_JR}
\end{figure}

{\it Bulk-edge correspondence:}
The relationship between a topologically nontrivial band structure and the presence of topologically protected edge states is a very general phenomenon known as the bulk-edge correspondence~\cite{Hatsugai:1993PRL, Hatsugai:1993PRB, Qi:2006PRB}.
We now illustrate the bulk-edge correspondence through a simple model. Note that a solid mathematical formulation of this bulk-edge correspondence has been developed based on the index theorem~\cite{Callias:1978CMP,Chiu:2016RMP}.

As we have seen in the Haldane model, one can obtain topological bands by adding proper gap-opening terms, proportional to $\sigma_z$, to the gapless unperturbed Hamiltonian (\ref{II_DiracHam}). Looking at the phase diagram of the Haldane model, Fig.~\ref{II_Haldane}(b), a topological phase transition can be induced by changing $M/t_2$ or $\phi$. In the Dirac Hamiltonian description, this topological phase transition corresponds to adding a $\sigma_z$ term and changing the sign of its coefficient~\cite{Haldane:1988PRL,BernevigBook}. We can therefore model the interface of two Haldane models with different Chern numbers by considering the following Jackiw-Rebbi Hamiltonian~\cite{Jackiw:1976PRD}:
\begin{align}
	\hat{H} = \hbar v_D \left( q_x \sigma_x + q_y \sigma_y\right) + m(x)\sigma_z, \label{eq:jackiwrebbi}
\end{align}
where the mass term $m(x)$ varies along the $x$ direction, obeying $m(x) < 0$ at $x < 0$, $m(0) = 0$, and $m(x) > 0$ at $x > 0$. The gap closes at $x = 0$, and the system is divided into two parts with $m < 0$ and $m > 0$ with different Chern numbers, which are sketched in Fig.~\ref{II_JR} for illustration. Writing the momentum operators in terms of a spatial derivative ($q_{x,y} = -i\nabla_{x,y}$), it is straightforward to see that the wave function of the form
\begin{align}
	\psi (x) \propto 
	e^{ik_y y}
	\exp \left(-\frac{1}{\hbar v_D}\int_0^{x}m(x^\prime) dx^\prime \right) 
	\begin{pmatrix}
	1 \\ i
	\end{pmatrix}
\end{align}
is an eigenstate of the Hamiltonian with the energy $\hbar v_D k_y$. This state $\psi(x)$ is localized around $x = 0$,
and has a positive group velocity along the $y$ direction. As there is no other normalizable state around $x=0$ with a negative group velocity, this state is chiral and robust against backscattering. Analogously, edge states around a generic topologically nontrivial system can be understood as interface states between the system and the topologically trivial vacuum~\cite{Hasan:2010RMP}. 

\subsubsection{Quantum spin Hall system}
\label{se:QSH}

In all the models that we have seen so far, time-reversal symmetry was explicitly broken through either the applied magnetic field or the complex hopping phase. When time-reversal symmetry is present, the Berry curvature obeys $\Omega_n (-\mathbf{k}) = -\Omega_n (\mathbf{k})$ for nondegenerate bands, implying that the Chern number, which is an integral of the Berry curvature over the Brillouin zone, is necessarily zero. A similar argument also holds for degenerate bands: no Chern bands can be found in two-dimensional systems preserving time-reversal symmetry~\cite{BernevigBook}.

In 2005, new classes of two-dimensional topological phases were proposed~\cite{Kane:2005PRLa, Kane:2005PRLb,Bernevig:2006PRL,Bernevig:2006Science}.
These models consist of two copies of a Chern insulator, one for each spin, where the magnetic fields acting on two spins are opposite and hence the Chern number for spin up $C_\mathrm{up}$ is opposite to that for spin down $C_\mathrm{down} = -C_\mathrm{up}$. Since the magnetic fields for two spins are opposite, time-reversal-symmetry is preserved, and the sum of the Chern numbers for the two spins is zero. In this model, as long as there are no spin-flip processes, the two spin components are uncoupled and independently behave as Chern insulators with opposite Chern numbers $C_\mathrm{up,down}$. As a result, there are the same number of edge states in the two spin states, but with opposite propagation direction; instead of chiral, the edge states are then called {\it helical}.

Even when the spin is not conserved, at least one pair of edge states survives and is topologically protected, as long as time-reversal symmetry holds. This feature is a consequence of Kramers's theorem, which holds in fermionic systems in the presence of time-reversal symmetry. The theorem tells us that if there is a state with energy $E$ and momentum $\mathbf{k}$, there must exist another distinct state with the same energy but with the opposite momentum $-\mathbf{k}$. In particular, at time-reversal-symmetric momenta such as $\mathbf{k} = 0$, states should be doubly degenerate. As a consequence, when there is a pair of edge states with spin-up and spin-down crossings at $\mathbf{k} = 0$, 
the edge states cannot open a gap, and hence there are topologically protected helical edge states. This is clearly different from a trivial insulator where there are no robust edge states traversing the gap.
The topological invariant characterizing the system is given by $C_\mathrm{up}$ mod 2 ($= C_\mathrm{down}$ mod 2), which takes values of either 0 (trivial) or 1 (nontrivial), and thus is called the $\mathbb{Z}_2$ topological invariant. This $\mathbb{Z}_2$ topological invariant keeps the same value even when spin-mixing terms are added, provided time-reversal symmetry is maintained~\cite{Kane:2005PRLa,Fu:2006PRB,Sheng:2006PRL,Roy:2009PRB}. Such topological phases are called the quantum spin Hall insulators or the $\mathbb{Z}_2$ topological insulators.
The $\mathbb{Z}_2$ topological insulator has been experimentally realized in HgTe quantum wells~\cite{Konig:2007Science} following the theoretical proposal~\cite{Bernevig:2006Science}.
Shortly afterward, $\mathbb{Z}_2$ topological insulators were found to exist also in three dimensions~\cite{Fu:2007PRL, Moore:2007PRB, Roy:2009PRB,Qi:2008PRB}.

One may also envisage an analog of the quantum spin Hall insulators for photons by using, for example, the polarization degree of freedom as pseudospins. However, the bosonic nature of photons forbids the existence of direct photonic analogs of the quantum spin-Hall insulators~\cite{DeNittis:2017arXiv}.  For Kramers's theorem to hold, one needs for the time-reversal operator $\mathcal{T}$ to be fermionic, which satisfies $\mathcal{T}^2 = -1$, while the bosonic time-reversal operator obeys $\mathcal{T}^2 = +1$. 

However, if there is no coupling between pseudospins, i.e., if there is an extra symmetry in the system, then each pseudospin component can independently behave as a model with nonzero Chern number and hence shows helical edge states in the presence of the bosonic time-reversal symmetry~\cite{Hafezi:2011NatPhys, albert2015topological}.
Note however that topological edge states of such systems are not robust against terms coupling different pseudospin states, which would be the photonic analog of time-reversal-symmetry breaking magnetic impurities for electronic $\mathbb{Z}_2$ topological insulators. Photonic models showing analogs of quantum spin Hall systems with no (or little) couplings between different pseudospin states are reviewed in Sec.\ref{sec:analogQSHE}.

\subsubsection{Topological phases in other dimensions}
\label{sec:otherdimensions}

We have so far reviewed the topological phases of matter in two dimensions with and without time-reversal symmetries.
Generally speaking, in the presence of a given symmetry, one can consider topological phases which are protected as long as the symmetry is preserved, which lead to the concept of the {\it symmetry-protected topological phases.}
A complete classification of noninteracting fermionic topological phases in any dimension based on the time-reversal, particle-hole, and chiral symmetries is known in the literature~\cite{Ryu:2010NJP, Schnyder:2009AIP, Kitaev:2009AIP,Teo:2010PRB,Chiu:2016RMP}.

As the topological band structure is a single-particle property and does not depend on the statistics of the particles, this classification, originally derived for fermionic systems, directly applies to bosonic systems as well, provided the Hamiltonian conserves the number of particles. The situation is in fact different when the number of particles is allowed to change, e.g., in Bogoliubov-de Gennes Hamiltonians of superconductors; in this case the bosonic and fermionic band structures are different. The fermionic case is included in the above-mentioned classification, while the bosonic case is reviewed in Sec.~\ref{sec:bogoliubov}. We now focus on three specific examples of topological phases of matter in dimensions other than two: one-dimensional Hamiltonians with chiral symmetry, three-dimensional Weyl points, and higher-dimensional quantum Hall systems.

{\it One-dimensional chiral Hamiltonian:}
One-dimensional Hamiltonians with chiral symmetry can have topologically nontrivial phases characterized by an integer-valued {\em winding number}. In noninteracting tight-binding models, chiral symmetry is equivalent to having a bipartite lattice, i.e. a lattice that can be divided into two sublattices with hopping occurring only between different sublattices. When a discrete translational symmetry is present, the Hamiltonian in momentum space can be written in the following generic form~\cite{Ryu:2010NJP}:
\begin{align}
	\hat{H}_k
	=
	\begin{pmatrix}
	\mathbf{0} & Q(k)^\dagger \\ Q(k) & \mathbf{0}
	\end{pmatrix}, \label{IIIA_kspaceform}
\end{align}
where $Q(k)$ is an $n\times n$ matrix satisfying $Q(k) = Q(k + 2\pi/a)$, each unit cell consists of $2 n$ sites, and $a$ is again the lattice spacing.

When there is a gap at zero energy, namely $\det Q(k) \neq 0$ for any $k$, the topology of this Hamiltonian is characterized by the winding number defined through the phase of $\det Q(k) \equiv |\det Q(k)|e^{i\theta (k)}$ as~\cite{Zak:1989PRL,Kane:2014NatPhys}
\begin{align}
	\mathcal{W}
	=
	\frac{1}{2\pi}\int_0^{2\pi/a} dk \frac{d\theta(k)}{dk}. \label{IIAwinding}
\end{align}
The winding number tells us the number of times $\det Q(k)$ wraps around the origin when plotted in a complex plane as one varies $k$ along the Brillouin zone.
The bulk-boundary correspondence states that the number of zero-energy edge modes on one side of the one-dimensional chain is given by the absolute value of the winding number $|\mathcal{W}|$~\cite{Ryu:2002PRL,Delplace:2011PRB}. Such zero-energy edge modes are topologically protected in the sense that they remain locked at zero energy even in the presence of chiral-symmetry-preserving perturbations provided the gap remains open. 

\begin{figure}
\resizebox{0.45 \textwidth}{!}{\includegraphics*{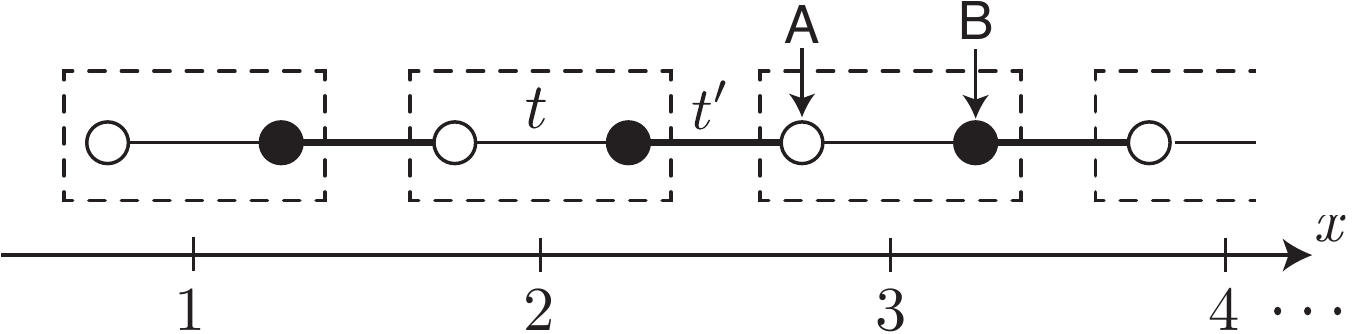}}
\caption{Schematic illustration of the Su-Schrieffer-Heeger model. Dashed squares indicate the unit cell of the lattice; each unit cell contains two lattice sites, one belonging to the $A$ sublattice and the other to the $B$ sublattice. The lattice terminates on the left-hand side with a complete unit cell.} \label{II_sshfigure}
\end{figure}

The prototypical example of a one-dimensional chiral Hamiltonian with nontrivial topology is the Su-Schrieffer-Heeger (SSH) model~\cite{Su:1979}, which is a chain with two alternating hopping strengths as sketched in Fig.~\ref{II_sshfigure}.
The system can be divided into two sublattices $A$ and $B$, and the tight-binding Hamiltonian can be written as
\begin{align}
	\hat{H}_\mathrm{SSH}
	=
	\sum_{x} \left( t \hat{b}^\dagger_x \hat{a}_x  + t^\prime \hat{a}_{x+1}^\dagger \hat{b}_x + \mathrm{H.c.}\right), \label{IIIA_sshham}
\end{align}
where $a_x$ and $b_x$ are annihilation operators for $A$ and $B$ sublattice sites at position $x$. The intracell and intercell hoppings are described by the (real) hopping amplitudes $t$ and $t^\prime$, respectively.
After a Fourier transformation, one sees that the momentum-space Hamiltonian has the form of Eq.~(\ref{IIIA_kspaceform}) with $n=1$ and $Q(k) = t + t^\prime e^{ik}$. The system is gapped as long as $t \neq t^\prime$, and the corresponding winding numbers are $\mathcal{W} = 0$ for $t > t^\prime$ and $\mathcal{W} = 1$ for $t < t^\prime$. Therefore, when the chain is terminated at one end with the final hopping of $t$, there exists a zero-energy topological edge state if $t < t^\prime$~\cite{Ryu:2002PRL}.

One-dimensional photonic structures with nontrivial topology will be discussed in Sec.~\ref{sec:1D}.

{\it Three-dimensional Weyl points:} As briefly introduced, two-dimensional band structures can host Dirac cones, corresponding to gapless points around which bands disperse linearly with respect to the two quasimomenta (\ref{II_DiracHam}). In three dimensions, the analog of a Dirac point is a Weyl point~\cite{Wan:2011PRB,Lu:2013NatPhot,Armitage:2017arXiv}: a point degeneracy between two bands which have a linear dispersion in all three directions in momentum space at low energy, as described by the Weyl Hamiltonian:
\begin{align}
	H_W = \hbar v(q_x \sigma_x + q_y\sigma_y+q_z\sigma_z), \label{IIAweyl}
\end{align}
where $\mathbf{q} = (q_x,q_y,q_z)$ is the momentum measured relative to the degenerate point and $v$ is the group velocity, taken here to be locally isotropic.

Close to a Weyl point, the resulting Berry curvature~\eqref{II_omega} is reminiscent of the magnetic field around a magnetic monopole, where the field can either point outward or inward toward the Weyl point. In analogy with magnetic monopoles, a quantized ``charge" can be associated with a Weyl point; this is nothing other than a Chern number calculated by integrating the Berry curvature over a two-dimensional surface enclosing the Weyl point, generalizing Eq.~\eqref{II_chern}. It can also be shown that Weyl points generate only nonzero Berry curvature when either $\mathcal{P}$ (parity) and/or $\mathcal{T}$ (time-reversal symmetry) is broken. Consequently, to get Weyl points in a band structure, one can break $\mathcal{P}$, $\mathcal{T}$, or both symmetries. The photonic realization of Weyl points and other three-dimensional gapless phases will be discussed in Sec.~\ref{sec:3Dgapless}.

{\it Higher-dimensional quantum Hall systems:} The two-dimensional integer quantum Hall effect, previously introduced, is just the first in a family of related quantum Hall phenomena that can exist in higher dimensions~\cite{Chiu:2016RMP}. Here we briefly review what happens in three or four spatial dimensions, highlighting key differences that emerge due to having odd or even numbers of spatial dimensions.  

First, in an odd number of spatial dimensions, the possible quantum Hall responses are closely related to those of a lower-dimensional system, as they are only associated with lower-dimensional topological invariants. In three dimensions, for example, a gapped band can be labeled by a triad of Chern numbers, e.g.~[$\mathbf{C^{(1)}}\equiv(C^{(1)}_x, C^{(1)}_y, C^{(1)}_z)$], where each Chern number is calculated from \eq{II_chern} by integrating the Berry curvature over a 2D momentum plane orthogonal to the quasimomentum direction indicated~\cite{Avron:1983PRL}. This triad of two-dimensional topological invariants leads to a linear three-dimensional quantum Hall effect, calculated by extending \eq{II_TKNNformula} to three spatial dimensions~\cite{Halperin:1987JJAP,montambaux:1990PRB,Kohmoto:1992PRB}. In the simplest case, such a three-dimensional quantum Hall system could be constructed by stacking 2D quantum Hall systems along a third direction and weakly coupling the layers together~\cite{Stormer:1986PRL,Chalker:1995PRL,bernevig:2018PRL,Balents:1996PRL}. In Sec.~\ref{sec:3dgapped}, we discuss three-dimensional quantum Hall systems and other gapped 3D topological phases in the context of photonics. 

Second, in even numbers of spatial dimensions, such as four dimensions, new types of quantum Hall effect can emerge that are intrinsically associated with that higher dimensionality~\cite{Meng:2003JPA,prodan2016bulk}. These quantum Hall effects can be distinguished from their lower-dimensional cousins as they appear as a higher-order response to perturbing electromagnetic fields and because they are related to topological invariants that vanish in fewer spatial dimensions. 

The four-dimensional quantum Hall effect, in particular, was first discussed by~\onlinecite{Frohlich:2000,Zhang:2001Science}, and later played an important role in understanding time-reversal invariant topological insulators in two and three dimensions through a dimensional-reduction procedure from four dimensions~\cite{Qi:2008PRB}. This quantum Hall effect consists of a nonlinear quantized current response in one direction, when perturbative electric and magnetic fields are applied in the other three directions. We assume for simplicity that only one nondegenerate band is occupied and an electromagnetic gauge potential $\mathbf{A} = (A_0,A_1,A_2,A_3,A_4)$ is applied as a perturbation, where $A_0$ is the electric potential and $(A_1,A_2,A_3,A_4)$ is the magnetic vector potential in four dimensions.
The current in response to the applied electric field $E_\nu \equiv \partial_0 A_\nu - \partial_\nu A_0$ and the applied magnetic field $B_{\rho \sigma} \equiv \partial_{\rho} A_\sigma - \partial_\sigma A_\rho$ is then given by~\cite{Price:2015PRL}
\begin{align}
	j^\mu
	=
	-\frac{e^2}{h} E_\nu \int_{\mathrm{BZ}} \Omega^{\mu \nu} \frac{d^4 k}{(2\pi)^3}
	+
	\frac{e^3}{2 h^2} \varepsilon^{\mu \nu \rho \sigma} E_{\nu} B_{\rho \sigma}
	\,C^{(2)},
	\label{VB_4DQH}
\end{align}
where $\mu$, $\nu$, $\rho$, and $\sigma$ run through spatial indices. The Berry curvature along the $\mu$-$\nu$ plane is defined as $\Omega^{\mu \nu} = \partial_{k_\mu} \mathcal{A}_{\nu} - \partial_{k_\nu} \mathcal{A}_{\mu}$ in terms of the usual Berry connection $\mathcal{A}_\mu = i \langle u_{k}|\partial_{k_\mu} | u_{k}\rangle$ and the integral is now over the four-dimensional Brillouin zone. Here, $\varepsilon^{\mu \nu \rho \sigma}$ is the 4D Levi-Civita symbol.

The second term in Eq.~(\ref{VB_4DQH}) vanishes in fewer than four spatial dimensions, and so this is the new quantum Hall effect that can emerge in a 4D system. It depends also on a four-dimensional topological invariant~\cite{NakaharaBook}
\begin{align}
	C^{(2)}
	=
	\frac{1}{32\pi^2}\int_{\mathrm{BZ}}\varepsilon_{\alpha \beta \gamma \delta} \Omega^{\alpha \beta} \Omega^{\gamma \delta} d^4 k, \label{VB_2CN}
\end{align}
which is known as the second Chern number. In contrast, the Chern number (\ref{II_chern}) appearing in the two-dimensional quantum Hall effect is sometimes called the first Chern number. Note that both first and second Chern numbers can be extended to the case of degenerate bands by taking a trace of the integrand over the degenerate bands. The first term in Eq.~(\ref{VB_4DQH}) is a contribution to the current that is reminiscent of the two-dimensional quantum Hall response, where only two directions are involved, and is characterized by the first Chern number in the $\mu$-$\nu$ plane; physically, this is similar to the quantum Hall physics of odd dimensions introduced above. Note that when the system possesses time-reversal symmetry, the first term of Eq.~(\ref{VB_4DQH}) vanishes and only the second term remains~\cite{Zhang:2001Science}.
Experimental observations of the four-dimensional quantum Hall effect through topological pumping are discussed in Sec.\ref{sec:pump}, and proposals for directly observing the four-dimensional quantum Hall effect using synthetic dimensions are discussed in Sec.\ref{sec:4D}.

%% file: RMP_II_Pump.tex
\subsubsection{Topological pumping}
\label{sec:pumpintro}

Topological invariants, such as the first and second Chern numbers defined above, can also lead to a quantization of particle transport in systems which are ``pumped" periodically and adiabatically in time. In this section, we introduce the concept of topological pumping by reviewing connections between the Archimedes screwpump and a topological pump in the semiclassical limit. We then discuss how a 1D topological pump can be related to the 2D quantum Hall effect, and how the topological framework developed so far leads to a simple lattice model for a topological pump. 

\begin{figure}[t!]
\includegraphics[width=\columnwidth]{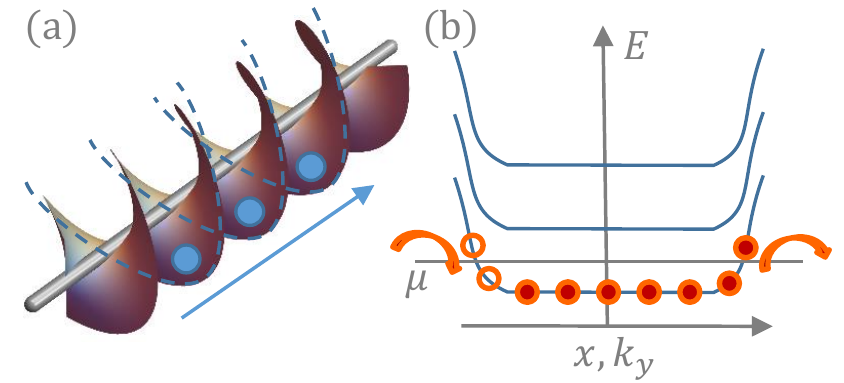}
\caption{ (a) Schematic of an Archimedes screw pump, which mechanically transports fluid in the direction of the blue arrow as the helicoidal surface is rotated. In a quantum mechanical treatment, the screw pump can be approximated by a series of parabolic potentials, as indicated by the blue dashed lines. (b) Illustration of the quantum Hall effect for Landau levels; the spectrum is sketched in the Landau gauge on a cylinder, with open boundary conditions along $x$ and periodic boundary conditions along $y$. Each state in a given Landau level is indexed by the transverse momentum $k_y$, setting the center of the state along $x$. As magnetic flux is threaded through the cylinder, an electric field is generated along $y$ and there is a spectral flow of states along $x$, corresponding to the quantum Hall transport. At the system boundaries, the Landau levels bend up in energy, forming chiral-propagating edge states, as expected from the bulk-boundary correspondence. At the left-hand side, the threading of flux means empty states will move down in energy and be filled up from an external reservoir as they cross the Fermi level $\mu$. At the right-hand side, states will flow up in energy and particles will be ejected into a reservoir. Signatures of these chiral edge states persist in a 1D topological pump as edge modes that cross the bulk energy gap as the pump parameter is tuned. 
\label{fig1_OZ}}
\end{figure}

{\it Archimedes screw pump:}
A pump is a device that moves fluids by mechanical action, i.e., it consumes energy to perform mechanical work by moving the fluid.
One of the oldest pumps known to man is the so-called Archimedes screw pump, which is still in use today. In this device, a fluid is pumped by turning a screw-shaped surface inside a cylindrical shaft; see Fig.~\ref{fig1_OZ}(a). As the screw-shaped surface is made to rotate around its axis, a volume of fluid is collected at one end of the device. It is then pushed along the tube by the rotating helicoid until it pours out at the other end of the device. Ideally, at each full turn of the pump, the collected volume is identical and the fluid is homogeneously transported a unit of distance along the device. Consequently, the screwpump is used as a variable rate feeder to deliver a measured rate or quantity of material in industrial processes.

Let us now adopt a quantum mechanical description of the screw pump. At any given time, the fluid is approximately confined within a series of parabolic potentials; see Fig.~\ref{fig1_OZ}(a). 
Assuming that the fluid is noninteracting, it suffices to write the Hamiltonian for a single particle of mass $m$ in the resulting chain of parabolic potentials
\begin{align}
\label{eq:ArchiHam}
\hat{H} = &\frac{\hat{p}_x^2}{2 m} + V \sum_j [\hat{x} - x_j(t)]^2\,\theta\left\{\left(x_{j-1}(t)+x_j(t)\right)/2\right\}\notag \\
&\theta\left\{-\left(x_{j+1}(t)+x_j(t)\right)/2\right\}\,,
\end{align}
where $\hat{x}$  and $\hat{p}_x$ are the position operator and its conjugate, respectively. The parabolic potentials are characterized by time-dependent minima located at points $x_m(t)$, with a potential amplitude $V$, and equidistant separation $\Delta x=x_{m+1}-x_{m}$. The Heaviside function $\theta[x]$ cuts the influence of neighboring sites on one another. Neglecting the coupling between neighboring wells, each well hosts the standard harmonic oscillator states of a single particle.
Within a pump cycle, the minima can be parameterized as $x_m(t)=x_{m,0}+U(\varphi(t))\Delta x$, where $x_{m,0}$ are some initial positions, $\varphi(t)=2\pi t/T_p$ is the periodic pumping parameter, and the displacement satisfies $U(\varphi+2\pi) = U(\varphi)+1$. After a full period $T_p$ of the pumping, each minimum moves by one site, and thus the Hamiltonian is invariant. Therefore, the eigenstates of the Hamiltonian are periodic as $\varphi \to \varphi + 2\pi$.

Turning on weak tunnel coupling between the localized states leads to one-dimensional Bloch bands that span the device, with time-dependent Bloch states $u_{n,{k}_x,\varphi} ({x})$ and Bloch energies $\omega_{n,k_x,\varphi}$. In the semiclassical limit, the transport of particles by pumping can generally be captured using the semiclassical equations of motion for a wavepacket that is prepared with a well-defined center-of-mass position $x$ and momentum $k_x$ in a given instantaneous Bloch band $n$. Under adiabatic modulation of the pumping parameter, the wave packet remains in the instantaneous band and evolves with a velocity  
\begin{align}
\dot{x}_n = \frac{\partial \omega_{n,k_x,\varphi}}{\partial k_x}+\Omega_n \partial_t \varphi
\end{align}
resulting from the sum of the usual group velocity plus an anomalous velocity term. The latter is
determined by the Berry curvature  $\Omega_n (k_x, \varphi) = i \left( \langle \partial_{\varphi} u_n | \partial_{k_x} u_n \rangle  - \langle \partial_{k_x} u_n | \partial_{\varphi}  u_n \rangle \right)$ 
associated with the instantaneous Bloch states \cite{Karplus:1954PR, Xiao:2010RMP}.

Hence, changing the pump parameter $\varphi$ induces an extra motion of the particle which, depending on the sign of $\Omega_n$, can be either in the same or in the opposite direction as the motion of the lattice. The resulting displacement of the wavepacket after one pump cycle is obtained by integrating $\dot{x}_n$ and can in principle be arbitrary. As this anomalous particle transport depends on the geometrical Berry curvature, it is often referred to as ``geometrical pumping." Experiments along these lines have been performed with cold atoms by~\textcite{LuSpielman:2016} and in photonics by~\textcite{Wimmer:2017NatPhys}.

For a filled or homogeneously populated bulk band $n$, however, the periodicity of the Bloch energy $\omega_{n,k_x,\varphi}$ in $k_x$ around the Brillouin zone guarantees that the group velocity contribution integrates to zero. The displacement per cycle can then be related to the 2D topological first Chern number~$C$ of the pumping process:
\begin{equation}
\label{eq:ChernNumber}
C = \frac{1}{2 \pi} \int\sub{BZ} \int_0^{2 \pi} \Omega_n (k_x, \varphi) d \varphi dk_x
\end{equation}
whose expression closely resembles Eq.~(\ref{II_chern}) for the Chern number of a band. As the displacement is proportional to this topological invariant, it neither depends on the pumping speed, provided adiabaticity still holds, nor on the specific lattice parameters as long as band crossings do not occur. Hence, the transport is highly robust against perturbations such as interaction effects or disorder \cite{Niu:1984}. For the screw pump potential \eqref{eq:ArchiHam}, one intuitively expects the fluid to move along with the moving potential. This is in agreement with our effective Bloch-bands description, where all lowest bands of the system appear with $C=1$ corresponding to the displacement by $\Delta x$ per unit cycle. The situation is of course much richer in the wave mechanics case where the displacement can also be in the opposite direction depending on the sign of $C$.

There is a deep connection between the screw pump and the physics of the quantum Hall effect, as suggested by the quantization of the pumped particles per cycle in terms of the topological first Chern number. To explore this further, we start from the Landau-level Hamiltonian introduced above in Eq.~(\ref{II_LLHam}) in the Landau gauge ${\mathbf{A}} = (0, xB, 0)$. Then the transverse momentum $\hbar k_y$ is a good quantum number, and, for a given state, the Hamiltonian reduces to that of a shifted 1D harmonic oscillator
\begin{equation}
\label{eq:ReducedLandauHamiltonian}
\hat{H}\sub{LL} = \frac{\hat{p}_x^2}{2m_a} +  \frac{1}{2} m \omega^2_c \left( \hat{x} - \frac{\hbar k_{y}}{m \omega_{c}}\right)^2
\end{equation}
\noindent where $\omega_{c}$ is the cyclotron frequency, and $x_c={\hbar k_{y}}/{m \omega_c}$ is the shifted center of the harmonic potential. Comparing this expression with \eq{eq:ArchiHam}, one can see that a state in a given potential minimum $j$ in the 1D screw pump is analogous to a state with a given $k_y$ in a 2D Landau level. 

In the Landau gauge, we can imagine putting the system on a cylinder, i.e., applying periodic boundary conditions in the $y$ direction, and threading a magnetic flux through the cylinder. [Note that applying periodic boundary conditions puts $k_y$ on a lattice of discretely allowed values given by the Born-von Karman boundary conditions, analogously to Eq.~\eqref{eq:ArchiHam}.] This threading of flux generates an electromotive electric field $E_y$ in the $y$ direction, leading to a Hall response in the $x$ direction. An adiabatic threading of flux causes continuous evolution of eigenvalues known as the spectral flow~\cite{phillips1996self,Waterstraat}. The number of states which cross the gap in the positive direction over the flux-threading cycle is equal to the Chern number of the spectral gap $\sigma_n$. The central position $x_c$ of each eigenstate shifts to that associated with the next allowed value of $k_y$, similar to how minima move in the screwpump, and in agreement with Laughlin's pumping argument~\cite{Laughlin:1981}. Indeed, Landau levels have first Chern numbers $|C|=1$ in agreement with the intuition developed from the screw pump.  

This analogy shows that
the pump parameter $\varphi$ of the 1D screw pump can be thought of as a threaded magnetic flux and hence a momentum in a perpendicular fictitious dimension; this correspondence can be exploited  in a procedure called ``dimensional extension" to derive the appropriate higher-dimensional model beginning from the lower-dimensional pump~\cite{Qi:2008PRB, Kraus:2012a,Kraus:2012b,Verbin:2013,Kraus:2013,Verbin:2015,prodan2016bulk}. Finally, as sketched in Fig.~\ref{fig1_OZ}, the bulk-edge correspondence of the quantum Hall effect can be used to explain how a topological pump functions with open boundary conditions and couples to particle reservoirs.

{\it Thouless pump and the Hofstadter model:} Turning the above analogy around, we can start with a 2D quantum Hall system, such as the Hofstadter model~\eq{II_HHHamiltonian}, and obtain the corresponding topological pump. For the Hofstadter model in the Landau gauge, we proceed by Fourier-transforming the model only in the $y$ direction to obtain 
\begin{align} \label{Eq:H1D}
    \hat{H} = - J \sum_{x,k_y} &\left[ \hat{a}_{x,k_y}^\dag \hat{a}_{x + a,k_y} + h.c. 
    \right. \notag \\
      &\left.+ 2 \cos\left(2\pi \alpha x/a - k_y a\right) \hat{a}_{x,k_y}^\dag \hat{a}_{x,k_y} \right ]\,.
\end{align}
Then applying the procedure of ``dimensional reduction," the momentum along $y$ is now considered as an external parameter $\varphi=k_y$; this reduces the dimensionality of the Hamiltonian by one dimension by removing the sum over $k_y$ and making the operators $k_y$ independent~\cite{Thouless:1983PRB}. This 1D model \eq{Eq:H1D} is then commonly known as the Harper model~\cite{Harper:1955PPSA} and will be discussed in detail in Sec.~\ref{sec:pump}. This system can be adiabatically pumped by slowly changing the external parameter $\varphi$ and hence modulating the on-site energy periodically in time. For a filled band insulator, this is known as a Thouless pump~\cite{Thouless:1983PRB}, as there is a quantization of particle transport over each pump cycle, as set by the sum over the first Chern numbers of the filled bands. 

The main distinction from the Landau-level case discussed above is that in the Hofstadter model there are two competing length scales: the lattice spacing and the magnetic length. As a result, the electron paths interfere to give the fractal Hofstadter butterfly, with a resulting band structure composed of bands with positive and negative Chern numbers. Similarly, in the 1D Thouless pump, the on-site potential imposes a new length scale, and in the resulting band structure, we encounter bands that will pump against the direction in which the potential is moved. This is a purely wave-physics interference effect in contrast to the classical particle picture used to understand the screw pump.

In the lattice configuration realized in the ultracold atomic experiment~\cite{Lohse:2016NatPhys,Nakajima:2016NatPhys}, the lowest energy band had a positive $C > 0$. Observing pumping in the opposite direction then required working with higher bands, which was achieved using the atomic gas in strongly nonequilibrium conditions. As reviewed in Sec.~\ref{sec:pump}, photonic systems proved to be an ideal platform for realizing topological pumps in more complex and also higher-dimensional geometries.

%% file: RMP_IIC_Floquet.tex
\subsubsection{Floquet engineering: Topology through time-periodic modulations}\label{section_floquet}

After having introduced in the previous section the general concepts of topological bands, we now briefly review the manner by which topological band properties can be generated by subjecting a static system to an external time-periodic drive, an approach which is generally referred to as \emph{Floquet engineering}, due to its direct relation to the Floquet theorem~\cite{Oka:2009PRB,Kitagawa:2010PRB,Lindner:2011NatPhys,Kitagawa:2011PRB,Cayssol:2013PS,Goldman:PRX2014,Bukov:2015AdPhys,Eckardt:2016Review}. We will describe how topological properties emerge in this general context and will then briefly discuss how Floquet engineering can be exploited in photonics.

Let us first consider a generic quantum system described by the static Hamiltonian $\hat H_0$. The aim of Floquet engineering is to modify the band structure of this Hamiltonian, effectively, by subjecting the system to a time-periodic modulation $\hat V (t+T)\!=\!\hat V(t)$, where $T\!=\!2\pi/\Omega$ denotes the period of the drive. In the nontrivial case where $[\hat H_0,\hat V(t)]\!\ne\!0$, the time-evolution operator $\hat U(t,t_0)$, which is associated with the total time-dependent Hamiltonian $\hat H(t)\!=\!\hat H_0\!+\!\hat V(t)$, forms an intricate object that one can formally write as a time-ordered integral $\hat U(t,t_0)\!=\! \mathcal{T} \mathrm{exp}\left[-(i/\hbar)\int^t_{t_0}\mathrm{d}t' \hat H(t')\right]$. However, due to the time-periodicity inherent to the system $\hat H (t+T)\!=\!\hat H(t)$, this time-evolution operator can be factorized, leading to the more suggestive form~\cite{Kiss:1994PRA,Rahav:2003PRA,Goldman:PRX2014,Bukov:2015AdPhys}
		\begin{align}
			\label{eq:Floquet_thm}
			\hat U(t,t_0) =e^{-i \hat K_\mathrm{kick}(t)}e^{-i (t-t_0) \hat H_\mathrm{eff}/\hbar}e^{i \hat K_\mathrm{kick}(t_0)},
		\end{align}  
where the operator $\hat H_\mathrm{eff}$ is time independent, and where $\hat K_\mathrm{kick}(t+T)\!=\!\hat K_\mathrm{kick}(t)$ has zero average over a period of the drive. The latter expression~\eqref{eq:Floquet_thm} indicates that the dynamics of the periodically driven system is essentially ruled by an effective Hamiltonian $\hat H_\mathrm{eff}$, whose properties are potentially distinct from those associated with the initial static Hamiltonian $\hat H_0$. In addition, the final ``kick" $e^{- i \hat K_\mathrm{kick}(t)}$ in Eq.~\eqref{eq:Floquet_thm} reflects the micromotion, namely, the dynamics taking place within each period of the drive. Both the effective Hamiltonian $\hat H_\mathrm{eff}$ and the kick operator $\hat K_\mathrm{kick}(t)$ result from a rich interplay between the static Hamiltonian $\hat H_0$ and the drive operator $\hat V (t)$;  these two operators, and hence the time-evolution operator in Eq.~\eqref{eq:Floquet_thm}, can be systematically computed using various perturbative treatments~\cite{Goldman:PRX2014,Goldman:2015PRA,Bukov:2015AdPhys,Eckardt:2015NJP,Mikami:2015PRB}.

In traditional realizations, Floquet engineering operates in the so-called ``high-frequency" regime of the drive ($\Omega\!\rightarrow\!\infty$); physically, this corresponds to situations where the period $T$ sets the shortest time scale in the system~\cite{Kitagawa:2010PRB}, and hence, where the micromotion is typically irrelevant. In this regime of the drive, it is instructive to probe the dynamics stroboscopically, namely, by considering discrete observation times $t_N\!=\!NT$, where $N$ is an arbitrary integer and $t_0\!=\!0$. Up to a unitary (gauge) transformation, the long-time dynamics is then captured by the stroboscopic time-evolution operator [Eq.~\eqref{eq:Floquet_thm}]
\begin{equation}
\hat{\mathcal U}(t_N)\!=\!e^{-i t_N \hat H_\mathrm{eff}/\hbar}=\left [ e^{-i T \hat H_\mathrm{eff}/
\hbar} \right ]^N =\left [\hat{\mathcal U}(T) \right ]^N. 
\end{equation}
Hence, in this framework, the relevant dynamics is governed by the Floquet operator $\hat{\mathcal U}(T)$, or equally, by the effective Hamiltonian $\hat H_\mathrm{eff}\!=\!(i\hbar/T)\log \hat{\mathcal U}(T)$. As an important corollary, the topological properties of the system are then entirely dictated by the band structure of the effective Hamiltonian:~for a proper choice of the drive protocol [$\hat V(t)$], the effective Hamiltonian $\hat H_\mathrm{eff}$ can host topological properties, even when the underlying static system ($\hat H_0$) is trivial. Consequently, in the high-frequency regime ($\Omega\!\rightarrow\!\infty$), the topological classification of periodically driven systems is strictly equivalent to that of static systems~\cite{Kitagawa:2010PRB,Lindner:2011NatPhys,Cayssol:2013PS}: topological band theory~\cite{Qi:2011RMP,Hasan:2010RMP,Bansil:2016RMP} directly applies to the Bloch bands associated with the effective Hamiltonian $\hat H_\mathrm{eff}$, i.e., the so-called ``Floquet spectrum."

A simple but important example of such driven-induced topological states is found when analyzing the properties of a particle hopping on a 2D circularly shaken honeycomb lattice~\cite{Oka:2009PRB,Rechtsman:2013Nature,Jotzu:2014Nat,Eckardt:2015NJP,Zheng:2014PRA,plekhanov2017floquet}. In a frame moving with the shaken lattice, the drive takes the form of a time-periodic inertial force ${\bf F} (t)$, so that the time-dependent Hamiltonian can be written in the form
\begin{equation}
\hat H (t) = - J \sum_{\langle j, k \rangle} \hat a^{\dagger}_j \hat a_k - \sum_j {\bf F} (t) \cdot {\bf r}_j \, \hat a^{\dagger}_j \hat a_j ,  \label{honey_shake}
\end{equation}
where the first term describes hopping between nearest-neighboring sites of the honeycomb lattice, with hopping amplitude $J$, where ${\bf F} (t)\!=\! F \left [\cos (\Omega t) {\bf e}_x + \sin (\Omega t) {\bf e}_y \right ]$ reflects the circular shaking, and where ${\bf r}_j$ denotes the position of site $j$. Interestingly, we note that the time-dependent Hamiltonian in Eq.~\eqref{honey_shake} is equivalent to that describing electrons in graphene when the latter is irradiated by a circularly polarized light~\cite{Oka:2009PRB,Cayssol:2013PS}; in that case, the force ${\bf F} (t)$ is then directly related to the AC electric field of the radiation. The effective Hamiltonian $\hat H_\mathrm{eff}$ associated with the time-dependent Hamiltonian in Eq.~\eqref{honey_shake} can be evaluated using the so-called $1/\Omega$ expansion~\cite{Goldman:PRX2014,Goldman:2015PRA,Bukov:2015AdPhys,Eckardt:2015NJP,Mikami:2015PRB}, which yields
\begin{equation}
\hat H_\mathrm{eff} \approx - J_\mathrm{eff} \sum_{\langle j, k \rangle} \hat a^{\dagger}_j \hat a_k - J_\mathrm{eff}^{\text{NNN}} \sum_{\langle \langle m, n \rangle \rangle}  i^{\circlearrowright} \hat a^{\dagger}_m \hat a_n  ,  \label{honey_shake_2}
\end{equation}
where the first term describes the renormalization of the nearest neighbor hopping term in Eq.~\eqref{honey_shake}, with $J_\mathrm{eff}\!=\!J \mathcal{J}_0 (F d/\Omega)$, and where the second term corresponds to a next-nearest-neighbor hopping term, with effective hopping matrix elements $J_\mathrm{eff}^{\text{NNN}} i^{\circlearrowright}\!=\! \pm i (\sqrt{3} J^2/\Omega) \mathcal{J}_1^2 (F d/\Omega)$ whose sign depends on the orientation of the hopping event. Here $\mathcal{J}_{0,1}$ denote Bessel functions of the first kind and $d$ is the lattice spacing. Importantly, the effective Hamiltonian in Eq.~\eqref{honey_shake_2} is equivalent to the Haldane model~\cite{Haldane:1988PRL} introduced in Sec.\ref{sec:iqhe}.

In direct analogy with the original Haldane model, the spectrum associated with the effective Hamiltonian~\eqref{honey_shake_2} displays two Bloch bands with nonzero Chern numbers and chiral edge states. In this Floquet-engineered context, we note that the time-reversal-symmetry-breaking nature of the Chern bands  naturally stems from the chirality of the circular drive. In condensed matter, a driven system exhibiting effective Bloch bands with nonzero Chern numbers is generally called a {\em Floquet Chern insulator}. Such a strategy was considered in various physical contexts, ranging from irradiated materials~\cite{Oka:2009PRB,Lindner:2011NatPhys,Cayssol:2013PS} to ultracold atoms in shaken optical lattices~\cite{Jotzu:2014Nat,Eckardt:2016Review,Flaschner:2016arxiv}, but it was in fact pioneered in photonic experiments using femtosecond-laser-written lattices~\cite{Rechtsman:2013Nature} as reviewed in detail in Sec.\ref{sec:propagating}. 

This Floquet-engineering strategy is not restricted to the Haldane model, but can indeed be extended to engineer other toy models of topological matter. For instance, the Harper-Hofstadter model~\cite{Hofstadter:1976PRB} discussed in Sec.~\ref{sec:iqhe} [Eq.~\eqref{II_HHHamiltonian}] can be designed by subjecting a 2D square lattice to a proper time-periodic modulation; various theoretical proposals can be found in different physical contexts, including cold atoms~\cite{Kolovsky:2011EPL,Creffield:2016NJP}, ion traps~\cite{Bermudez:2011PRL}, and photonics~\cite{Fang:2012NatPhot}. As further discussed in Sec.~\ref{sec:qheothertata}, these schemes build on the concept of resonant ``modulation-assisted tunneling"~\cite{Goldman:2015PRA}, where a time-periodic modulation is set on resonance with respect to an intrinsic energy offset between neighboring sites. Cold-atom realizations of the Harper-Hofstadter model, using resonant moving optical potentials, were reported by~\textcite{Aidelsburger:2013PRL,Miyake:2013PRL,Aidelsburger:2014NatPhys,tai2017microscopy}. More recently, a photonics implementation based on a circuit-QED architecture was described by~\textcite{Roushan:2016NatPhys}; see Fig.~\ref{fig:Roushan} and Sec.~\ref{subsec:strong}.

The simple topological characterization presented above for the high-frequency regime ($\Omega\!\rightarrow\!\infty$) breaks down when the period of the drive becomes comparable to other time scales in the problem (e.g., when $\hbar \Omega$ becomes comparable to the bandwidth of the effective spectrum~\cite{Kitagawa:2010PRB}). Indeed, in that situation, the micromotion becomes crucial and the topological classification based on $\hat H_\mathrm{eff}$ only must be revised~\cite{Kitagawa:2010PRB,Rudner:2013PRX,Nathan:2015,Carpentier:2015}. In particular, away from the high-frequency regime, topologically protected edge modes are shown to exist even when the topological invariants (e.g., Chern numbers) associated with $\hat H_\mathrm{eff}$ are all trivial~\cite{Kitagawa:2010PRB,Rudner:2013PRX}. The discovery of such {\em anomalous Floquet topological  phases} suggested that novel types of topological invariants had to be introduced in order to accurately recover the bulk-edge correspondence in this regime. Such topological invariants (winding numbers) were identified by~\textcite{Kitagawa:2010PRB,Asboth:2012PRB,Rudner:2013PRX,Nathan:2015,Carpentier:2015,Yao:2017PRB, Bi:2017PRB} and were indeed shown to depend on the complete time-evolution operator $\hat U(t,t_0)$. The crucial role played by the micromotion in this topological characterization~\cite{Nathan:2015} indicates a shift of paradigm with respect to  the standard topological classification of static systems. 

As a final remark, we note that a fruitful approach to Floquet topological physics is offered by the so-called {\em quantum walks}~\cite{Broome:2010PRL,Kitagawa:2010PRA,Kitagawa:2012NatComm}, where the time-evolution operator of a system is digitally built by repeatedly applying a series of unitary operations $\hat{\mathcal U} (T)\!=\! \hat U_M \hat U_{M-1} \cdots \hat U_2 \hat U_1$. Because of the $T$ periodicity of such quantum walks, their topological classification is equivalent to that of Floquet-engineered systems discussed above~\cite{Kitagawa:2010PRA,Kitagawa:2012NatComm}. 

As we shall see at multiple places in the course of this review, photonic systems have shown a great potential to implement Floquet techniques in different platforms. This has led to the experimental realization of intriguing Floquet phases~\cite{Rechtsman:2013Nature,Noh:2017NatPhys,Bandres:2016PRX,Roushan:2016NatPhys,Bellec:2017EPL}, in particular, anomalous Floquet topological states~\cite{Gao:2016NatCom,Maczewsky:2017NatCom,Mukherjee:2017NatCom} and topologically protected states in quantum walks~\cite{Kitagawa:2012NatComm,Cardano:2017NatComm}. 

%% file: RMP_IIB_DriveDissipation.tex
\subsection{Features of photonic systems}
\label{sec:drivedissipation}
Historically, the study of topological effects in quantum condensed-matter systems originated from electric conduction experiments measuring the current versus voltage characteristics of two-dimensional electron gases. In these systems, the basic constituents, the electrons, obey Fermi statistics; the topologically nontrivial states arise as the equilibrium state for sufficiently low temperature; and the electric conductivity is measured under weak or moderate external fields that do not dramatically affect the underlying many-body state. 

This equilibrium or quasiequilibrium condition is shared by almost all condensed-matter experiments, with a few remarkable exceptions such as Floquet topological insulators~\cite{Inoue:PRL2010,Lindner:2011NatPhys,Cayssol:2013PS,Oka:2009PRB} and light-induced superconductivity~\cite{Fausti:Science2011}. In these systems, new phases of matter are induced by intentionally keeping the system far away from equilibrium by means of some incident electromagnetic radiation.

The situation is fundamentally different in photonic systems for at least two fundamental reasons: (i) the basic constituent, the photon, possibly dressed by some matter excitation into a polariton, obeys bosonic statistics; and (ii) photons can reside in any realistic device only for a finite time and some external driving is needed to inject them into the system. In the next two sections we review the consequences of these remarkable differences. In the last section we review some basic concepts of nonlinear optics and illustrate how a third-order $\chi^{(3)}$ nonlinearity can be seen as an effective binary interaction between photons, therefore opening the way toward many-body physics using gases of interacting photons~\cite{carusotto:2013}.

\subsubsection{Bosonic nature}
\label{sec:bosonicnature}

As previously introduced, the first key difference between quantum condensed-matter systems, based on electron gases, and photonic systems is that the basic constituents of the latter are bosonic. In Sec.~\ref{se:QSH}, we have seen how this difference can have an impact already at the single-particle level as Kramer's theorem, which underlies quantum spin Hall physics, holds only in the presence of fermionic time-reversal operators satisfying $\mathcal{T}^2=-1$.

At the many-body level, the difference between bosons and fermions is even more apparent as quantum statistics impose a specific symmetry on the many-body wavefunction under particle exchange: because of the Pauli exclusion principle, a noninteracting gas of fermions at low temperatures fills all states below the Fermi level with just one particle per state, while leaving all higher states empty. In the particular case of insulators, where the Fermi level lies within an energy gap, all valence (conduction) bands are filled (empty), so that integrals over the Brillouin zone naturally appear in the calculations. In contrast, a weakly interacting bosonic system at low temperatures consists of a Bose-Einstein condensate with a macroscopic population of particles accumulated in the single lowest-energy state~\cite{Huang,BECbook}. As we see in the course of this section, the picture is made different in optical systems by the presence of losses and/or the propagating nature of photons, so that the ground state of the system is typically a trivial vacuum state: generating and maintaining the photon gas in a state with interesting topological properties then requires injecting light from some external source.

\subsubsection{Nonequilibrium nature}
\label{subsec:noneq}

\label{subsubsec:noneq}

In any realistic optical device, photons are typically subject to a variety of processes that tend to reduce the number of photons present in the device. Depending on the geometry and the materials used, these processes can range from absorptive losses in the underlying material medium, which make the photons simply disappear, to radiative losses and/or a finite propagation time through the system, which result in the emission of light in the surrounding space as propagating radiation. As a consequence, some external pumping mechanism must be introduced to inject and sustain the light field in the device. Except for very specific cases~\cite{Klaers:Nature2010,Hafezi:PRB2015,Lebreuilly:PRA2017,Silberberg:PRL2009,Rasmussen:PRL2000}, the resulting state of the light field is thus quite distinct from a thermal equilibrium state. 

Given this intrinsically nonequilibrium nature of photonic systems, the ways in which topological effects manifest, as well as the experimental probes and diagnostics that are available, can be completely different in photonic systems compared to condensed-matter setups. For example, the light emitted by the device carries out detailed information on the field distribution and the photon statistics inside the device. Depending on the specific setup, this information can be extracted by imaging the emission in free space, as is typical for planar microcavities, and/or by collecting the emission with local probes such as antennas or waveguides. Different ways of injecting light into the system have also been experimentally used to highlight different properties of topological states. 

We now review the main such pumping schemes, highlighting the key general features of each. The impact of the nonequilibrium condition on specific topological effects is then discussed throughout the following sections, e.g., on non-Hermitian topological photonics in Sec.\ref{sec:gainloss}, on the stabilization of many-body phases in Sec.\ref{subsec:strong}, and on the measurement of topological invariants in Sec.\ref{subsec:quantummagn}.

\paragraph{Coherent pumping}

In a typical coherent pumping scheme, the system is illuminated with an externally incident laser beam or by placing an antenna or an external waveguide in the system's vicinity, so as to inject coherent light at specific spatial locations. Light propagation through the system is then monitored by collecting transmitted and/or scattered light with a second antenna or a detector. 

The conceptually simplest theoretical description of such an approach consists of solving Maxwell's equations, including the specific geometrical arrangement of dielectric and magnetic elements and suitable source terms to describe the emitting antenna. Since a complete analytical solution is typically beyond reach, a number of numerical methods have been developed for this task, ranging from the same plane wave expansions used to obtain the band structure, to finite element methods for the time-evolution~\cite{Joannopoulos:2011book}. 
Such techniques were for instance used in the theoretical calculations presented by~\textcite{Wang:2009Nature} and shown in Figs.~\ref{fig:wangintro}(b),~\ref{fig:wangintro}(c), and~\ref{fig:wangintro}(e). 

For nonbianisotropic materials with no magnetoelectric coupling, one can eliminate the magnetic field and write Maxwell's equation for a field oscillating at frequency $\omega$ in the compact form~\cite{Joannopoulos:2011book,Wang:2008PRL}
\begin{align}
	\nabla_\mathbf{r} \times \left[ \boldsymbol{\mu}^{-1}(\mathbf{r}) \nabla_\mathbf{r}\times \mathbf{E}(\mathbf{r})\right] = \omega^2 \boldsymbol{\epsilon}(\mathbf{r})\mathbf{E}(\mathbf{r}), \label{eq:maxwellreduced}
\end{align}
where $\mathbf{E}(\mathbf{r})$ is the electric field, and $\boldsymbol{\mu}(\mathbf{r})$ and $\boldsymbol{\epsilon}(\mathbf{r})$ are the magnetic permeability and dielectric permittivity tensors, respectively. The gyromagnetic photonic crystals used in the experiment~\cite{Wang:2009Nature} and illustrated in Figs.~\ref{fig:wangintro}(a1) and~\ref{fig:wangintro}(a2) were characterized by spatially periodic $\boldsymbol{\mu}(\mathbf{r})$ and $\boldsymbol{\epsilon}(\mathbf{r})$, as well as nonvanishing nondiagonal elements of $\boldsymbol{\mu}$, which quantify the strength of the gyro-magnetic effect. Emission by an antenna can be straightforwardly included in the model just by adding source terms on the right-hand side (rhs) of Eq.~(\ref{eq:maxwellreduced}).

In spatially periodic systems, a basis of solutions of Eq.~(\ref{eq:maxwellreduced}) can be found which satisfies the Bloch theorem just like the Schr\"odinger equation for electrons in a periodic potential. In particular, these solutions are labeled by the crystal momentum $\mathbf{k}$ and the band index $n$ and the role of the Bloch wavefunction is played here by the electric field $\mathbf{E}_{n,\mathbf{k}}(\mathbf{r})$. 
For a given configuration, the specific form of the photonic Bloch bands and the Bloch wave functions can be obtained with standard photonic techniques reviewed by~\textcite{Joannopoulos:2011book}. 

Given the different form of the wave equation (\ref{eq:maxwellreduced}) compared to the Schr\"odinger equation, calculation of the topological invariants for photonic crystals requires some specific work beyond the picture presented in Sec.\ref{sec:topologybasic} for electronic systems. 
While the operator acting on $\mathbf{E}(\mathbf{r})$ on the left-hand side (lhs) of Eq.~ (\ref{eq:maxwellreduced}) is still a Hermitian operator, the eigenvalue problem involves the dielectric permittivity $\boldsymbol{\epsilon}(\mathbf{r})$  multiplying on the right-hand side.
It is then useful to consider the modified scalar product, 
\begin{equation}
\langle \mathbf{E}_1 \vert \mathbf{E}_2 \rangle = 
\int d^2\mathbf{r}\,\sum_{i,j}\, {E}_{1,i}^*(\mathbf{r})\,{\epsilon}_{i,j}(\mathbf{r})\,{E}_{2,j}(\mathbf{r}),
\end{equation}
where $i,j=\{x,y,z\}$.
The Berry connection introduced in \eq{eq:berryconnection} then takes the form~\cite{Wang:2008PRL}
\begin{multline}
\boldsymbol{\mathcal{A}}_n (\mathbf{k})=
i\langle \mathbf{E}_{n,\mathbf{k}} \vert \nabla_\mathbf{k} \mathbf{E}_{n,\mathbf{k}} \rangle = \\
= i\int d^2\mathbf{r}\,\sum_{ij}\,{E}^*_{n,\mathbf{k},i}(\mathbf{r})\,{\epsilon}_{ij}(\mathbf{r})\,\nabla_\mathbf{k} {E}_{n,\mathbf{k},j}(\mathbf{r})
\label{eq:Berry_E}
\end{multline}
in terms of the normalized Bloch eigenfunctions $\mathbf{E}_{n,\mathbf{k}}(\mathbf{r})$ of the photon state of crystal momentum $\mathbf{k}$ and band index $n$.
Starting from \eq{eq:Berry_E} for the Berry connection, the geometrical and topological invariants such as the Berry curvature, the Chern number, and the bulk-boundary correspondence display the usual features as reviewed in Sec.~\ref{sec:topologybasic}; in particular, as before, the Chern number can become nonzero only when the time-reversal symmetry is broken.
Of course, equivalent geometrical and topological properties would be found if the Bloch wavefunction was written in terms of magnetic field after having eliminated the electric one.

While this {\em ab initio} technique can be applied in full generality to any photonic structure, a different strategy consists of developing simplified models that are able to capture the main physics, while, at the same time, providing some analytical insight as well as the possibility of extending to quantum optical features. The most celebrated such model, most suitable for discrete systems of coupled resonators, is inspired by the tight-binding picture of solid-state physics~\cite{AshcroftMermin} and is naturally expressed in a quantum language. In the classical optics and photonics literature, it often goes under the name of coupled mode theory, as reviewed by~\textcite{Haus:1991IEEE}.

The starting point is an expansion of the electromagnetic field
\begin{equation}
 \mathbf{E}(\mathbf{r})=\sum_j \mathbf{E}_j(\mathbf{r})\,\hat{a}_j+ \mathbf{E}^*_j(\mathbf{r})\,\hat{a}_j^\dagger 
\end{equation}
onto a basis of localized quantized modes labeled (in the simplest case of single-mode resonators) by the site index $j$. The suitably normalized mode profiles $\mathbf{E}_j(\mathbf{r})$ are obtained as the eigenmodes of Maxwell's equations of eigenfrequency $\omega_j$, and the quantum $\hat{a}_j$ and $\hat{a}^\dagger_j$ operators, respectively, destroy or create a photon in each resonator $j$ and satisfy Bose statistics, that is, $[\hat{a}_j,\hat{a}_{j'}]=0$ and $[\hat{a}_j,\hat{a}^\dagger_{j'}]=\delta{j,j'}$.

The corresponding Hamiltonian has the form of a collection of independent harmonic oscillators in which tunneling between neighboring sites $j'\longrightarrow j$ can be straightforwardly included as hopping terms of amplitude $J_{j,j'}$,
\begin{equation}
 H_{\rm res}=\sum_j \hbar \omega_j\, \left[\hat{a}^\dagger_j\,\hat{a}_j+\frac{1}{2}\right] -\sum_{j,j'}J_{j,j'}\hat{a}^\dagger_j\,\hat{a}_{j'}. 
 \label{eq:phot_modes}
\end{equation}
Pioneering examples of the application of this tight-binding formalism are found in~\textcite{Yariv:OptLett1999,Bayindir:PRL2000}. If the hopping amplitudes $J_{j,j'}$ can be made complex, photons behave as if they are experiencing a synthetic magnetic vector potential. The site dependence of the resonator frequency $\omega_j$ can model an external potential acting on photons.

The main difference between photonic systems and the usual solid-state ones is that photons can radiate away from the resonators into the surrounding empty space, e.g. through the nonperfectly reflecting cavity mirrors. At the level of the Hamiltonian (\ref{eq:phot_modes}), this requires the inclusion of a continuum of radiative modes $\hat{A}_\eta$, labeled by the index $\eta$ and satisfying Bose commutation relations. These modes are linearly coupled to the localized resonator modes via terms of the form
\begin{multline}
H= H_{\rm res}+\int \!d\eta\,\hbar \omega_\eta \hat{A}^\dagger_\eta\,\hat{A}_\eta + \\
- \sum_j \int\!d\eta\, \left[\hbar g_{j,\eta} \hat{A}^\dagger_\eta \hat{a}_j + \textrm{h.c.}\right],
\label{eq:phot_modes2}
\end{multline}
where $\omega_\eta$ is the frequency of a given radiative mode $\eta$ and $g_{j,\eta} $ is the coupling between that radiative mode and the resonator mode $j$. As discussed in full detail in quantum optics textbooks~\cite{QuantumOptics}, this Hamiltonian is the starting point of the so-called input-output formulation of the cavity field dynamics in terms of a quantum Langevin equation~\cite{Gardiner:PRA1985}. Under the simplifying assumptions that the different resonators are coupled to independent continua of radiative modes with an approximately constant spectral weight within the frequency range of interest, one can write
\begin{equation}
 i\frac{d\hat{a}_j}{dt}=\omega_j\hat{a}_j-\sum_{j^\prime}J_{j,j'}\,\hat{a}_{j'} -\frac{i \gamma_j}{2}\hat{a}_j + \hat{A}_j^{\rm in}(t),
 \label{eq:qlang}
\end{equation}
where the radiative damping rate 
\begin{equation}
\gamma_j= 2\pi \left| g_{j,\eta}\right|^2 \, \left|\frac{d \omega_\eta}{d\eta}\right|^{-1}
\end{equation}
has to be evaluated for the resonant radiative mode such that $\omega_\eta=\omega_j$ and the bosonic input operators 
\begin{equation}
\hat{A}_j^{\rm in}(t)= - \int\!d\eta\, g^*_{j,\eta} \hat{A}_\eta
\end{equation}
include the zero-point quantum noise as well as the incident radiation. 

The model can of course be straightforwardly extended to account for loss channels of nonradiative origin. More complex configurations may require including more field components on each site to describe multimode cavities: dissipative terms of different forms, e.g., a nondiagonal damping matrix $\gamma_{j,j'}$ to describe simultaneous coupling of several sites to the same continuum~\cite{Harris:PRL1989,Chen:PRL1990,Ghulinyan:PRA2014,CohenTannoudji4}, and/or light amplification processes by population-inverted emitters~\cite{QuantumOptics,QuantumNoise}. 

In the most relevant case of a coherent incident field and a quadratic resonator Hamiltonian $H_{\rm res}$, we can replace the operators with $\mathbb{C}$-number-valued expectation values $\alpha_j$ that evolve according to the ordinary differential equations
\begin{equation}
 i\frac{d\alpha_j}{dt}=\omega_j\alpha_j-\sum_{j^\prime}J_{j,j'}\alpha_{j'} -\frac{i \gamma_j}{2}\alpha_j + F_j(t),
 \label{eq:dalpha/dt}
\end{equation}
where the source term $F_j(t)=\langle \hat{A}_j^{\rm in}(t) \rangle$ corresponds to the classical amplitude of the incident field. Techniques to efficiently evaluate the tight-binding parameters from classical transmission and reflection calculations are discussed, e.g., by~\textcite{Hafezi:2011NatPhys} for coupled ring resonator arrays. More details on other specific systems can be found, e.g., in~\textcite{Bellec:2013PRB} for microwave resonators,~\textcite{Kruk:2017Small} for dielectric nanoparticles,~\textcite{Poddubny:2014ACS,Downing:PRB2017} for plasmonic chains.
Generalization to spatially continuous systems such as planar microcavities is reviewed by~\textcite{carusotto:2013}.

Apart from the last two terms describing driving and dissipation, this equation (\ref{eq:dalpha/dt}) has exactly the same form as the Schr\"odinger equation for noninteracting tight-binding electrons, where the field amplitude $\alpha_j$ plays the role of the discrete electron wave function. This formal equivalence between equations allows one to simulate the single-particle properties of tight-binding models using photonics.

Depending on the specific spatiotemporal shape of the coherent drive $F_j(t)$, the motion equation (\ref{eq:dalpha/dt}) can be used to describe various different phenomena such as the time-dependent response to a pulsed excitation or the nonequilibrium steady state under a monochromatic excitation. As a general rule, a coherent drive selectively excites only those modes that are on or close to resonance with the pump frequency spectrum and that have a significant overlap with the pump profile. 

For instance, when a monochromatic pump is shone on the bulk of the system, the injected light intensity dramatically depends on whether its frequency corresponds to an allowed energy band or to a band gap. On resonance with a band, a spatially extended and periodic pump can selectively excite Bloch states with specific $k$~\cite{Bardyn:2014NJP}, while a spatially localized pump generates an outward propagating field up to distances roughly proportional to the group velocity of the excited modes over the total photon decay rate~\cite{Ozawa:2014PRL}. Within a forbidden gap, no propagating state is instead available and the spatial light intensity profile will show a sharp exponential decay, typically determined by the distance to the nearest band edge. 

When the pump is focused on a system edge, pumping in an energy band will result in light penetrating into the bulk, while pumping in a band gap will concentrate the excitation on edge states, if available. Of course, different modes can be selectively excited by playing with the spatial and/or polarization shape and symmetry of the pump spot.

In more complex geometries, the frequency selectivity of a coherent pump has been used to selectively excite specific localized modes, ranging from the Landau levels in the nonplanar ring cavity of~\textcite{Schine:2016Nature} to complex Penrose tiling geometries~\cite{Vignolo:PRB2016}, an approach which may be extended to explore a variety of other interesting cases, including, e.g., the relativistic Landau levels of strained honeycomb lattices and the momentum-space Landau levels that appear under a harmonic confinement, as theoretically studied by, respectively,~\textcite{Salerno:20152DMat} and~\textcite{Berceanu:2016PRA}. 

\paragraph{Incoherent pumping}
\label{sec:IncohPump}

Photoluminescence experiments using incoherent pumping are a straightforward but powerful tool to visualize the energy distribution and the structure of the eigenstates of a system. In particular, this approach is routinely used in planar microcavity devices. Using this technique, states in a specific energy range can be isolated by spectrally resolving the emission, and then near- or far-field images recover the spatial profile or the $k$-space momentum distribution of modes. In the topological photonics context, this technique was used, e.g., to visualize the relativistic Dirac-like dispersion in honeycomb lattices~\cite{Jacqmin:2014PRL} and the corresponding edge states~\cite{Milicevic:20152DMat,Milicevic:2017PRL}.

Typical photoluminescence experiments are performed in a low pump power regime where the emission occurs spontaneously and is distributed fairly uniformly across all modes. Ramping up the pump power, experiments can enter a regime where bosonic stimulation and then mode-competition effects conspire to concentrate the emission into a reduced number of modes. For high enough pump power, above the so-called lasing threshold, stimulated emission exceeds losses and a new kind of steady state is achieved: in this state, a strong and coherent light intensity is concentrated into a single mode, which absorbs all pump power and which has an emission linewidth that is dramatically reduced~\cite{QuantumOptics,QuantumNoise}. 

As a rule of thumb, the lasing mode is typically selected by the largest gain condition; attention must however be paid to complex spatial mode deformation effects due to interplay of gain with the nonlinearity and the gain saturation~\cite{Tureci:PRA2007}, such as ballistic outward flows~\cite{Richard:PRL2005,Wouters:PRB2008,Wertz:NatPhys2010} or solitonic-like self-bound modes~\cite{Tanese:NatComm2013,Jacqmin:2014PRL}. In the topological photonics context, in spite of these complications, ramping the pump power above lasing threshold has been instrumental in resolving small spin-orbit coupling effects in hexagonal chains of micropillar resonators~\cite{Sala:PRX2015,Zambon:2018arXiv}. On-going advances in topological lasing will be outlined in the concluding Sec.~\ref{sec:conclusion}.  

\paragraph{Propagating geometries}
\label{sec:IIpropagating}

While both previous schemes are based on a driven-dissipative evolution of the light field, the conservative dynamics of light flowing through so-called "propagating geometries" has been exploited in a number of recent breakthrough experiments in topological photonics, starting from the realization of topological quantum walks in~\textcite{Kitagawa:2012NatComm} to Floquet topological insulators in coupled waveguide systems in~\textcite{Rechtsman:2013Nature} and Berry-phase-induced anomalous transport in~\textcite{Wimmer:2017NatPhys}. Well before these advances, such propagating geometries had already been used to realize a variety of novel phenomena, e.g., spatial lattice solitons \cite{christodoulides2003discretizing, efremidis2002discrete, fleischer2003observation}, two-dimensional Anderson localization \cite{schwartz2007transport, lahini2008anderson}, and wave dynamics in quasicrystals \cite{freedman2006wave, freedman2007phason, levi2011disorder, Verbin:2013}. All these experiments are based on purely classical properties of light and do not involve any quantum optical feature. 

A convenient theoretical description of classical monochromatic light in these systems is based on paraxial diffraction theory~\cite{yariv1976introduction,Boyd,Rosanov}. We start from classical Maxwell's wave equation for light propagating in a source-free, nonmagnetic material of (spatially dependent) dielectric constant $\varepsilon=\varepsilon(x,y,z)$:
\begin{equation}
\nabla \times \nabla \times {\bf E} = \varepsilon\,\left( \frac{\omega}{c} \right)^2 {\bf E},
\end{equation}
where ${\bf E}$ is the electric field, $\omega$ is the light frequency, and $c$ is the speed of light in vacuum.  Using a standard vector identity and the fact that $\nabla \cdot \left( \varepsilon {\bf E}\right) = 0$, we arrive at
\begin{equation}
- \nabla^2 {\bf E}= \varepsilon\left( \frac{\omega}{c} \right)^2 {\bf E}+\nabla \left( {\bf E} \cdot \frac{\nabla \varepsilon}{\varepsilon} \right) .\label{eq2}
\end{equation}
Assuming that the dielectric constant $\varepsilon(x,y,z)=\varepsilon_0+\Delta\varepsilon(x,y,z)$ displays relatively small variations from the background value $\varepsilon_0=n_0^2$, and that the length scale of the spatial variation is large compared to the wavelength $\lambda_0=2\pi/k_0=2\pi c/(n_0\omega)$ in the medium, we can neglect the last  term coupling the polarization and the orbital degrees of freedom and assume that the two polarizations evolve independently according to a scalar equation. In the experiment of \textcite{Rechtsman:2013Nature}, $\varepsilon$ varies by $\Delta\varepsilon\sim 10^{-3}$ on a length scale $\sim 10\lambda_0$.

We further assume that light is made to propagate along {\em paraxial} directions close to the axis of the waveguides (denoted as the positive $z$ direction). This guarantees that the wave vector component in the $z$ direction must be much greater than those in the $x$ and $y$ directions ($k_0\simeq k_z \gg k_{x,y}$).  This suggests to write the electric field in the carrier-envelope form 
\begin{equation}
 {\bf E}(x,y,z)=\hat{{\bf e}}\,\tilde{E}(x,y,z)\, \exp\left[ i k_0 z \right ],
\end{equation}
where $\hat{{\bf e}}$ is the unit vector in the direction of polarization and $\tilde{E}$ is a slowly varying function satisfying $| \nabla \tilde{ E}| \ll | k_0 \tilde{ E}|$.  
Plugging this expression into the propagation equation (\ref{eq2}), we find that
\begin{equation}
-\partial_z^2\tilde{ E}  - 2 i k_0 \partial_z \tilde{ E} - \nabla^2_\perp\tilde{ E} + k_0^2 \tilde{ E} =  \varepsilon\,\left(\frac{\omega}{c}\right)^2 \tilde{ E},\label{Eq3}
\end{equation}
where the $\nabla_\perp$ operator acts on the transverse $(x,y)$ plane.  We now use the fact that $\tilde{ E}$ varies slowly in the $z$ direction to neglect the $\partial_z^2\tilde{ E}$ term from Eq. (\ref{Eq3}). Using for convenience the refractive index $n=\sqrt{\varepsilon}=n_0 + \Delta n\simeq n_0 + \Delta \varepsilon /(2 n_0)$, 
we arrive at the paraxial equation for the diffraction of light through the structure:
\begin{equation}
i \partial_z \tilde{ E} = -\frac{1}{2k_0}\nabla_\perp^2 \tilde{ E} - \frac{k_0 \Delta n}{n_0}\tilde{ E}.\label{paraxial}
\end{equation}
For a strong enough confinement within the waveguides, the usual tight-binding approximation~\cite{AshcroftMermin} can be performed on the paraxial equation (\ref{paraxial}), which leads to evolution equations for the field amplitude $\alpha_j$ in each waveguide of the form
\begin{equation}
 i\frac{d\alpha_j}{dz}=k^z_j\alpha_j-J_{j,j'}\alpha_{j'},
 \label{eq:dalpha/dz}
\end{equation}
where the wave vector $k_j^z$ of light propagating inside the waveguide $j$ is determined by the background index $n_0$ as well as by the lateral confinement length, while the tunneling matrix $J_{j,j'}$ depends on the overlap of the evanescent tails of $j,j'$ modes.

The formal similarity of the paraxial propagation equation (\ref{paraxial}) with the Schr\"odinger equation of quantum mechanics establishes a close analogy between the diffraction of classical light and the motion of a spinless massive quantum particle, where the diffraction along the $x$-$y$ plane sets the particle mass and the refractive index modulation gives the external potential. Note, however, that the role of the temporal coordinate is played here by the propagation direction $z$, whose spatial derivative replaces on the lhs of \eq{paraxial} the usual temporal $t$ derivative of the Schr\"odinger equation. As a result, the Floquet approach discussed in Sec.\ref{section_floquet} can be implemented in propagating geometries by spatially modulating system properties along the $z$ direction, as experimentally pioneered by~\textcite{Rechtsman:2013Nature} and reviewed here in Sec.\ref{sec:propagating}.

Going beyond the assumption of monochromaticity, the propagating light can have nontrivial temporal dynamics, such that the physical time $t$ becomes like a third spatial coordinate, in addition to the $x,y$ transverse coordinates, and the propagation equation for the slowly varying field $\tilde{E}(\mathbf{r},t;z)$ must include an additional kinetic energy term with respect to the temporal direction,
\begin{equation}
i \frac{\partial \tilde{E}}{\partial z} = -\frac{1}{2k_0}\nabla_\perp^2 \tilde{E}- \frac{k_0\,\Delta n}{n_0}\tilde{E}-\frac{1}{2m_t}\frac{\partial \tilde{E}}{\partial t^2}
\label{paraxial_t}
\end{equation}
whose masslike coefficient $m_t^{-1} \equiv - d^2 k/d\omega^2$ [with $k(\omega)=\omega \, n(\omega)/c$] is proportional to the group velocity dispersion in the medium of frequency-dependent refractive index $n(\omega)$~\cite{Boyd}.

In order to go beyond the classical light case considered in the above equations, theoretical works~\cite{Lai:PRA1989,Lai:PRA1989b,Larre:PRA2015} have developed a full quantum theory to map the light propagation of quasimonochromatic quantum light onto a standard many-body theory of interacting bosons. In the near future, this reformulation may be useful in studying the interplay of quantum fluctuations, strong nonlinearities, and topological effects.

As a consequence of the exchanged roles of the $z$ coordinate and the physical time $t$, radiation that propagates through such a device does not provide a real-time monitoring of the field evolution, as is usual for the cavity systems discussed above, but instead only offers access to the field state at the end point $z=z_{fin}$ of the evolution. Correspondingly, extracting spectral information about eigenmodes requires the inclusion of additional elements, such as an extra one-dimensional chain of waveguides that can be used to selectively inject light at one side of the system, according to a $k_z$ resonance condition~\cite{Noh:arx2017}. Analogously, the role of the incident field on the front interface of the device is only to set the initial condition of the evolution at $z=z_{in}$. The counterpart of these limitations is that such propagating geometries allow us to study time-dependent problems with conservative dynamics, extending in the long run even to quantum many-body physics~\cite{Polkovnikov:RMP2011,Larre:EPJD2016}. 

From an experimental point of view, focusing on works related to topological photonics, it is worth noting that a discrete version of the position-to-time mapping was exploited by~\textcite{Schreiber:Science2012} to obtain a 2D analog quantum walk by encoding two extra spatial variables in the arrival time of optical pulses. A related quantum walk was used in the experiment~\cite{Wimmer:2017NatPhys} to reconstruct the geometrical Berry curvature of a photonic lattice model from anomalous transport features.

\subsubsection{Basics of nonlinear optics}
\label{subsec:nlo}

We conclude this section by reviewing the main aspects of photonic systems with a brief summary of the key concepts of nonlinear optics. In view of the ongoing developments toward the realization of strongly correlated many-photon states in strongly nonlinear systems, as reviewed in Sec.\ref{subsec:strong}, we will pay special attention to the reformulation of $\chi^{(3)}$ Kerr nonlinearities in terms of a binary interaction between photons.

The standard semiclassical description of nonlinear optical processes is based on Maxwell's equations, including nonlinear terms resulting from the nonlinear dependence of the dielectric polarization on the applied field~\cite{Boyd,Butcher}. In the most naive form, this reads
\begin{equation}
 P = \chi^{(1)}\,E+\chi^{(2)}\,E^2+\chi^{(3)}\,E^3+\cdots , \label{polarization}
\end{equation}
where the linear electric susceptibility $\chi^{(1)}$ is responsible for the usual refractive index; the second-order susceptibility $\chi^{(2)}$ gives rise, e.g., to second harmonic generation, optical rectification, and parametric downconversion processes; and the third-order susceptibility $\chi^{(3)}$ leads to four-wave mixing processes as well as an intensity-dependent refractive index.  In the simplest form, this last effect can be reformulated as 
\begin{equation}
 n(\mathbf{r})=n_0+n_{\mathrm{nl}}|\mathbf{E}(\mathbf{r})|^2,
 \label{nnl}
\end{equation}
where $n_0$ is the refractive index in the linear regime, and the $n_{\mathrm{nl}}$ coefficient, proportional to the $\chi^{(3)}$ susceptibility, quantifies the refractive index dependence on the local light intensity.

\begin{figure}
\includegraphics[width=0.4\columnwidth,angle=0,clip]{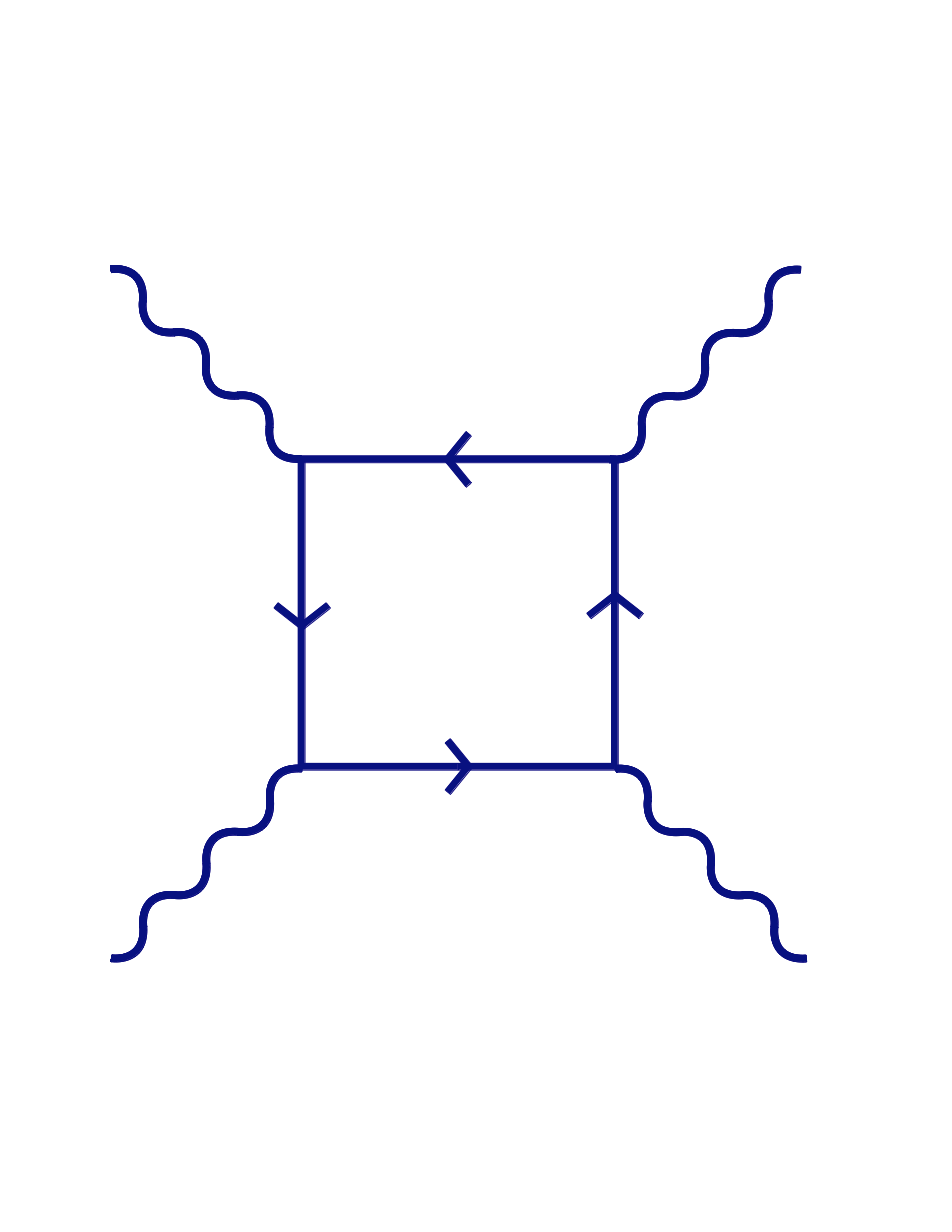}
\caption{QED Feynman diagram contributing to photon-photon scattering in vacuum via creation of a virtual electron-positron pair. Wavy lines represent photons and directed arrows represent electrons and positrons.
\label{fig:HeisenbergEuler}}
\end{figure} 

From the point of view of photons as quantum mechanical particles, such an intensity-dependent refractive index can be reinterpreted in terms of binary interactions between photons, which are mediated by the nonlinear polarization of the underlying medium. This picture of interacting photons was pioneered in the calculation of the effective $\chi^{(3)}$ third-order nonlinear polarizability of the vacuum arising from the exchange of a virtual electron-positron pair~\cite{Heisenberg:Zeit1936,Karplus:PR1951}, as sketched in the Feynman diagram for photon-photon scattering shown in Fig.\ref{fig:HeisenbergEuler}. Given the large mass $m_e$ of electrons and positrons as compared to the optical energies, the low-energy cross section of such processes
\begin{equation}
\sigma\propto \alpha^4\left(\frac{\hbar}{m_e c}\right)^2 \left[\frac{\hbar\omega}{m_e c^2}\right]^6
\label{eq:Hei-Eu}
\end{equation}
is very low \textit{in vacuo}: in the $\sigma\approx 10^{-68}\,\textrm{m}^2$ range for $1$~eV optical photons. This has made the experimental observation of this physics \textit{in vacuo} extremely challenging. Given the $(\hbar\omega)^6$ dependence of the scattering cross section, the most natural strategy is to use high energy photons, e.g., in the $\gamma$-ray range. The first experimental observation of photon-photon scattering using the electromagnetic fields surrounding ultrarelativistic colliding ions was recently reported by~\textcite{ATLAS:NatPhys2017}.

As compared to the vacuum, condensed-matter media offer the much more accessible option of replacing electron-positron pairs of MeV-ranged mass $m_e$ with electron-hole pairs of eV-ranged rest mass (set by the band gap of the material). According to Eq.~(\ref{eq:Hei-Eu}) the corresponding reduction of the intermediate-state detuning provides a dramatic reinforcement of the cross section by $\approx 36$ orders of magnitude. This corresponds to a significant value of the nonlinear $\chi^{(3)}$ polarizability in \eq{polarization}, which leads to many nonlinear optical phenomena, including the intensity-dependent refractive index of \eq{nnl}, two-photon absorption, and parametric amplification or oscillation, etc.~\cite{Butcher,Boyd}. 

Beyond the basic Feynman diagram sketched in Fig.\ref{fig:HeisenbergEuler}, more complex configurations arising in specific materials may offer interesting advantages to experiment, e.g.,  when photons are dressed by optical transitions in discrete emitters~\cite{Noh:ROPP2017} or excitons in solid-state materials~\cite{Ciuti:PRB1998}, when the collision process occurs via a long-lived biexcitonic bound state~\cite{Wouters:PRB2007,Carusotto:EPL2010,Takemura:NatPhys2014}, or when the nonlinearity inherits the long-range character of the interactions between Rydberg states in solid-state~\cite{Kazimierczuk:Nature2014} or gaseous media~\cite{Saffman:RMP2010,Chang:NatPhot2014}. 

In most media, the third-order $\chi^{(3)}$, introduced in \eq{polarization}, can however be viewed in the many-body language as the result of simple two-photon collisions. This alternative perspective shines new light on basic nonlinear optical phenomena from a novel angle and allows to take advantage of the artillery of many-body techniques originally developed in the context of condensed-matter and nuclear physics to predict new optical effects. In the last decade, the resulting concept of {\em quantum fluids of light}~\cite{carusotto:2013} has led to the experimental observation of a remarkable number of many-body effects in weakly interacting gases of photons, such as Bose-Einstein condensation, superfluidity and quantum hydrodynamics, etc.

Under the simplifying assumptions that the frequency dependence of $\chi^{(3)}$ is negligible and that the rotating-wave approximation can be performed, the real part of $\chi^{(3)}$ provides an interaction Hamiltonian
\begin{equation}
 \hat{H}_{\rm int}=\frac{g^{(3)}}{2} \int\!d^3{\bf r}\, \hat{E}^\dagger({\mathbf r}) \, \hat{E}^\dagger({\mathbf r})\, \hat{E}({\mathbf r})\,\hat{E}({\mathbf r}),
\label{eq:H_int}
\end{equation}
which is quartic in the electric field operators $\hat{E}({\mathbf r})$ and where the amplitude $g^{(3)}$ is proportional to the real part of the $\chi^{(3)}$ nonlinearity. The imaginary part gives instead two-body losses from two-photon absorption. 
The assumption of a spatially and temporally local polarization response of the medium to the applied electric field medium that underlies Eq.~(\ref{eq:H_int}) is valid in most common silicon- or silica-based materials used for topological photonics in the infrared and visible range~\cite{Hafezi:2013NatPhot,Rechtsman:2013Nature}, as well as in Jaynes-Cummings-type lattices embedding strongly detuned emitters~\cite{Hoffman:PRL2011,Fitzpatrick:PRX2017}.
Nonetheless, one must not forget that several other experiments make use of thermal~\cite{Vocke:Optica2015}, photorefractive~\cite{fleischer2003observation,Jia:PRA2009}, or Rydberg-mediated nonlinearities~\cite{Saffman:RMP2010,Chang:NatPhot2014}, whose response may be slow in time and/or long range in space due to heat and charge diffusion effects and to the inherently long-range nature of dipole interactions.

From a quantitative point of view, it is crucial to keep in mind that the optical nonlinearity of most commonly used materials results in very weak interactions between {\em single} photons, so that an accurate theoretical description is provided by a mean-field approach. Under this approximation, high-order averages are split into products of the mean field, e.g.,
\begin{equation}
 \langle \hat{E}^\dagger (\mathbf{r})\,\hat{E}(\mathbf{r})\,\hat{E}(\mathbf{r})\rangle \simeq \langle \hat{E}^\dagger (\mathbf{r})\rangle \langle \hat{E}(\mathbf{r})\rangle \langle \hat{E}(\mathbf{r})\rangle,
\label{eq:MF}
\end{equation}
and the Heisenberg equation of motion for the field expectation value $E(\mathbf{r})=\langle \hat{E}(\mathbf{r})\rangle$ recovers the classical Maxwell's equations including a nonlinear polarization term \eq{polarization}. Of course, the extremely small intensity of single-photon nonlinear effects does not preclude that a huge number of photons can collectively have a dramatic impact on the macroscopic optical response to a strong light beam.

Within the mean-field approximation \eq{eq:MF}, an intensity-dependent refractive index can be included in the classical equations of motion (\ref{eq:dalpha/dt}) of the tight-binding formalism described in the previous Sec.\ref{subsubsec:noneq}, by simply adding to the rhs of the motion equation for $\alpha_j$ an additional term of the form 
\begin{equation}
+\omega_{\mathrm{nl}}\,|\alpha_j|^2\,\alpha_j,
\label{eq:alpha_nl}
\end{equation}
where the nonlinearity parameter $\omega_{\mathrm{nl}}$ is proportional (with an opposite sign) to the real part of the $\chi^{(3)}$ nonlinearity and, typically, inversely proportional to the spatial volume of the optical mode under consideration~\cite{carusotto:2013}. In propagating geometries, instead, an interaction term of the form 
\begin{equation}
-\frac{k_0 n_{\mathrm{nl}}}{n_0}|\tilde{E}(\mathbf{r})|^2\tilde{E}(\mathbf{r})  \label{eq:prop_nl}
\end{equation}
has to be added to the right-hand side of the paraxial propagation equation (\ref{paraxial}) for monochromatic light, which then takes the form of a Gross-Pitaveskii equation of dilute Bose-Einstein condensates~\cite{BECbook}. In both cases, a nonvanishing imaginary part of $\chi^{(3)}$ and $n_{\mathrm{nl}}$ in Eqs.~(\ref{eq:alpha_nl}) and~(\ref{eq:prop_nl}) can be included to model saturable absorption, two-photon absorption, or gain saturation effects.

Going beyond the mean-field regime and realizing strongly correlated photon states requires very special materials with extremely strong nonlinearities. Finding such materials is one of the most active research lines in modern nonlinear optics~\cite{carusotto:2013,Chang:NatPhot2014,Roy:RMP2017}. So far, the most exciting results have been obtained using polaritons in gases of coherently driven atoms with electromagnetically induced transparency (EIT) in Rydberg states~\cite{Gorshkov:2011PRL,Peyronel:Nature2012,Firstenberg:Nature2013} -- the so-called Rydberg polaritons-- or circuit-QED devices, where microwave cavity photons are strongly coupled to a superconducting qubit element~\cite{Schoelkopf:2008Nature, You:2011Nature,Houck:NatPHys2012,Gu:2017PhysRep}. Even though a complete and quantitative account of the complex features of these optical nonlinearities calls for a more sophisticated theoretical description of the interactions between Rydberg polaritons~\cite{Bienias:PRA2014,Jachymski:PRL2016} and of the Josephson dynamics in circuit-QED devices~\cite{Bourassa:PRA2012}, the simplest form (\ref{eq:H_int}) of the interaction Hamiltonian is typically sufficient to capture the main physics. 

The quantum Langevin equation for the cavity field dynamics \eq{eq:qlang} is also straightforwardly extended to interacting regimes by adding a two-photon interaction term
\begin{equation}
 H_{\rm nl} = \sum_j \frac{\hbar \omega_{\mathrm{nl}}}{2}  \hat{a}^\dagger_j \hat{a}^\dagger_j \hat{a}_j \hat{a}_j,
\label{eq:H_omega_nl}
\end{equation}
to the resonator Hamiltonian $H_{\rm res}$.
Actual calculations are often simpler to perform by recasting the input-output formalism in terms of a master equation for the density matrix $\hat{\rho}$. For a coherent drive, the driving and dissipation terms have the form
\begin{multline}
  \frac{d\hat{\rho}}{dt}= -\frac{i}{\hbar} \left[\hat{H}_{\rm res}+H_{\mathrm{nl}}+\sum_j F_j(t) \hat{a}^\dagger_j + F^*_j(t) \hat{a}_j,\hat{\rho}\right] \\ + \sum_j \frac{\gamma_j}{2} [2 \hat{a}_j \hat{\rho} \hat{a}^\dagger_j -  \hat{a}^\dagger_j \hat{a}_j \hat{\rho} -  \hat{\rho} \hat{a}^\dagger_j \hat{a}_j ].
\end{multline}
Generalization of this approach to incoherent pumps, as discussed in Sec.\ref{subsubsec:noneq}, can be found in quantum optics textbooks~\cite{QuantumOptics,QuantumNoise}.

As introduced in Sec.\ref{subsubsec:noneq}, a quantum description of light propagation in the propagating geometries crucially requires going beyond the monochromatic light assumption. Among many recent developments in this direction~\cite{Petrosyan:PRL2011,Gorshkov:2011PRL,Bienias:PRA2014,Moos:PRA2015,Maghrebi:PRL2015,Gullans:PRL2016}, a particularly promising theoretical approach is based on a quantum version of the $t-z$ mapping. As discussed by~\textcite{Lai:PRA1989,Lai:PRA1989b,Larre:PRA2015}, this reformulation leads to a model of interacting bosons, again with the physical roles of time $t$ and propagation coordinate $z$ exchanged.

%% file: RMP_IIIA1_Gyro.tex
\section{Topological photonics in two dimensions}
\label{sec:2D}

Having reviewed basic ideas of topological physics and optical and photonic systems in the previous section, we are now in the position to dive into the exciting recent advances of topological photonics. Taking inspiration from the well-known classification of electronic topological insulators~\cite{Chiu:2016RMP}, the next sections are organized according to the dimensionality and symmetry class to which each topological system belongs. 

The present section is focussed on two-dimensional systems, starting from the analog quantum Hall systems that sparked the whole field of topological photonics. In the following sections we then move to quantum spin Hall systems, anomalous Floquet insulators and, finally, gapless systems such as honeycomb lattices. For each class, we present the principal material platforms that have been developed and the main topological effects that each system has been able to observe.

\subsection{Analog quantum Hall systems in photonics}
\label{sec:analogIQHE}

In this first section we review two-dimensional photonic systems in which time-reversal symmetry is explicitly broken, so that the topology can be classified in terms of the integer-valued (first) Chern number. These systems can be considered as direct photonic analogs of integer quantum Hall states of a two-dimensional electron gas in the presence of an out-of-plane magnetic field.

Our attention is focused on those optical platforms that have led to major experimental advances in the field, namely, gyromagnetic photonic crystals and arrays of coupled waveguides, but we briefly discuss also other platforms that have been theoretically proposed and are presently under experimental investigation. While most of the experiments focused on the  chiral edge modes and the resulting topologically protected one-way propagation, we conclude by briefly reviewing theoretical work in the direction of measuring geometrical and topological properties of the bulk. 

\subsubsection{Gyromagnetic photonic crystals }

\begin{figure}[t]
\includegraphics[width=0.99\columnwidth]{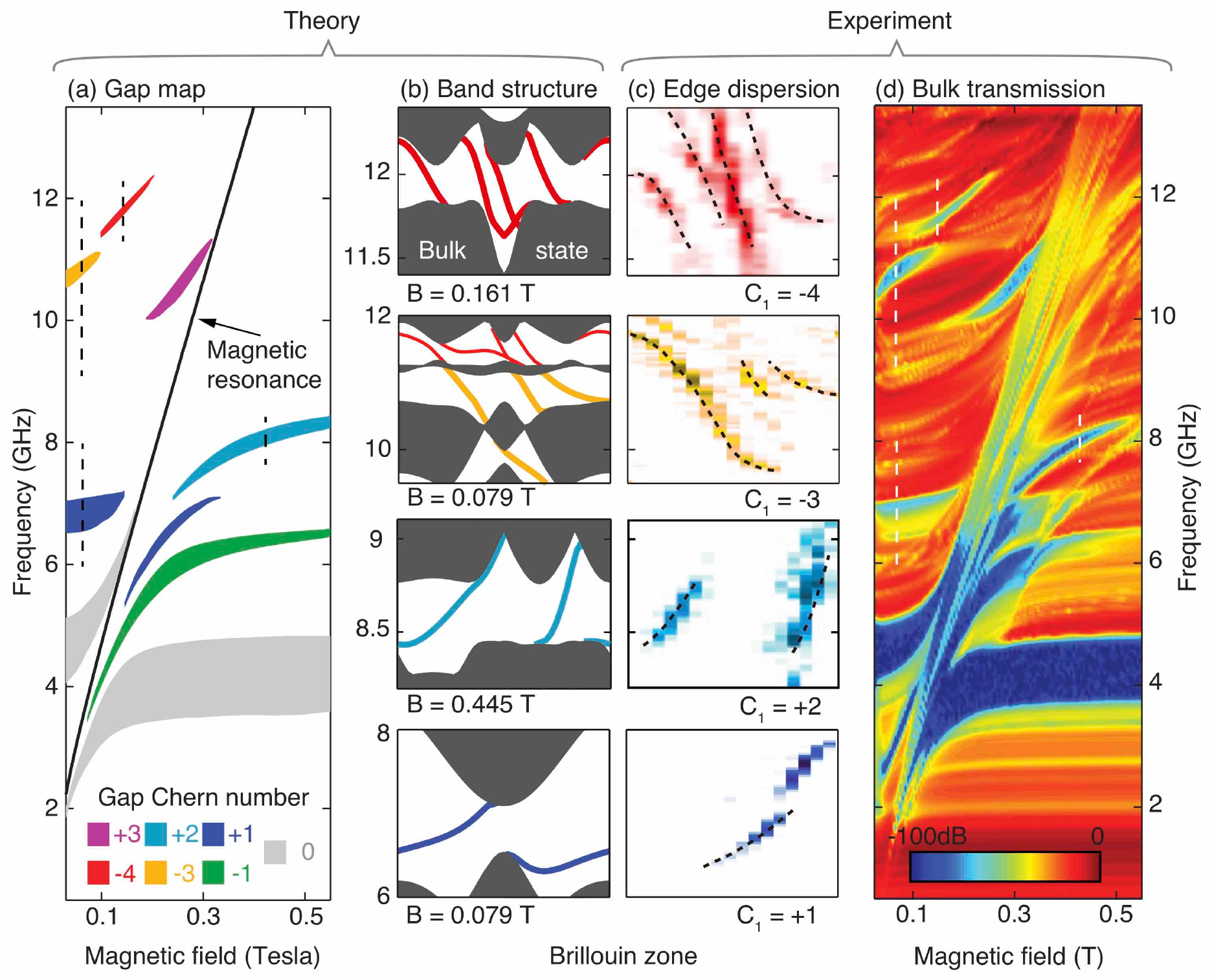}
\caption{(Color online)
Comparison of theoretical and experimental results for the setup shown in Fig.~\ref{fig:wangintro}(a) for varying magnetic fields.
(a) Theoretical topological gap map as a function of the static magnetic field applied. Each band gap is labeled by its gap Chern number, the sum of Chern numbers of the bulk bands below the gap.
(b) Calculations of the band and edge state dispersion for a few different values of the magnetic field.
(c) Experimental edge dispersions obtained by Fourier transforming the edge-mode profiles containing both intensity and phase information.
(d) Experimental bulk transmission as a function of magnetic field and frequency, in agreement with the gap map in (a).
Adapted from~\textcite{Skirlo:2015PRL}. 
}
\label{Gyromagnetic}
\end{figure}

In a nutshell, photonic crystals~\cite{Joannopoulos:2011book} consist of a spatially periodic arrangement of material elements giving spatially periodic dielectric permittivity $\epsilon_{ij}(\mathbf{r})$ and magnetic permeability $\mu_{ij}(\mathbf{r})$ tensors. In such a geometry, one can apply to photons the Bloch theorem originally developed in solid-state physics for electrons in crystalline solids~\cite{AshcroftMermin}: photon states organize themselves in allowed bands separated by forbidden gaps and are labeled by their wave vector defined within the first Brillouin zone of the periodic lattice. 

Most of the early literature on photonic crystals focused on the possibility of realizing a complete photonic band gap~\cite{Yablonovitch:1987PRL,John:1987PRL} that could, e.g., suppress spontaneous emission on embedded emitters and, in the presence of defects in the otherwise crystalline order, create strongly localized states. The study of such in-gap states gave rise to a number of exciting developments in view of photonic applications such as high-$Q$ photonic crystal cavities~\cite{Noda:OFCC16}, high-performance, low-noise semiconductor lasers~\cite{Painter:Science1999}, and low-loss-waveguides insensitive to bends~\cite{Mekis:PRL1996,Lin:Science1998,Yariv:OptLett1999,Bayindir:PRL2000}. On the other hand, propagating band states regained major attention when it was realized that photonic bands are not completely characterized by their energy dispersion, but also encode geometrical and topological features, which could lead to photonic analogs of the electronic quantum Hall effect~\cite{Haldane:2008PRL}.

The seminal proposal of Haldane and Raghu focused on the case of a gyroelectric photonic crystal where a pair of Dirac cones are gapped under a static magnetic field which breaks the time-reversal symmetry~\cite{Haldane:2008PRL,Raghu:2008PRA}.
Soon after, this was followed by a realistic design of a gyromagnetic photonic crystal operating at microwave frequencies and displaying a Chern number of 1~\cite{Wang:2008PRL,Chong:2008PRB}.
The basic formalism, such as how to define the Berry curvature of photonic bands, was reviewed in Sec.~\ref{subsubsec:noneq}.

Experiments~\cite{Wang:2009Nature} were performed using the material platform sketched in Fig.~\ref{fig:wangintro}(a), namely, a periodic array of ferrite rods of vanadium-doped calcium–iron garnet, a material that under a static magnetic field shows strong gyro-magnetic properties encoded in the nondiagonal matrix elements of the magnetic tensor $\mu_{ij}$. The theoretical dispersion of photonic energy bands for such a system is displayed in Fig.\ref{fig:wangintro}(e), which includes labels indicating the Chern number of the different bands.

In the experiment, the photonic crystal slab had of course a finite spatial size and was bounded by a metal wall on one side: since the reflecting gap of the surrounding metal has a topologically trivial nature, the bulk-boundary correspondence predicts that a topologically protected chiral edge mode appears within the bulk energy gaps of the photonic crystal, as indicated by the red line in the dispersion plot in Fig.\ref{fig:wangintro}(e). Since time-reversal symmetry is broken by the external magnetic field, the number of edge states in each gap and their direction of propagation is determined by the sum of the Chern numbers of all lower-lying bands. 

The main phenomenological consequence of such chiral edge states is that they support propagation in one direction only, so that backscattering from defects and scatterers is completely suppressed independently of their nature and strength. Since no state is available at the same frequency that propagates in the opposite direction, any wave incident on the defect can only circumnavigate it and then recover its original path along the edge of the system,  at most accumulating some phase shift. This remarkable feature is apparent in the numerical simulation shown in Fig.\ref{fig:wangintro}(b) and is in stark contrast with standard waveguides where generic defects are responsible for a strong backscattering of light and, therefore, a significantly suppressed transmission. 

In contrast to the reciprocal behavior of the bulk visible in Fig.\ref{fig:wangintro}(c), the large nonreciprocity of the transmission between a pair of antennas located on the edge shown in Fig.\ref{fig:wangintro}(d), as well as its insensitivity to the presence of a metallic scatterer located in between them, was the smoking gun of the nontrivial topology.

Similar experiments were soon performed by other groups~\cite{Fu:2010APL,Fu2010:2010EPL,Fu:2011APL,Fu:2011EPL,Lian:2012CPL,Lian:2012PRB,Li:2014IJMPB,Poo:2011PRL,Poo:2012APL,Yang:2013APL,Li:2014APA,Poo:2016SR}. 
The many possibilities of these systems to engineer a variety of different band topologies were then explored. For instance, photonic bands with large Chern number bands were theoretically identified by~\textcite{Skirlo:2014PRL} by simultaneously gapping multiple pairs of Dirac cones. This prediction was confirmed by the experiment~\cite{Skirlo:2015PRL}, where the edge-mode profiles were directly scanned and Fourier transformed, so to observe chiral dispersions of Chern numbers ranging from 1 to 4, as illustrated in Fig. \ref{Gyromagnetic}. 

As further features of topological edge states, it was later shown that they can self-guide in air~\cite{Ao:2009PRB,Poo:2011PRL,Liu:2012OL,Lu:2013AO,Li:2014JO,Li:2015AO}, appear in coupled defect cavities~\cite{Fang:2011PRB}, have robust local density of states~\cite{Asatryan:2013PRB,Asatryan:2014JOSAA}, be modeled in time domain~\cite{Li:2013CPL}, self-collimate unidirectionally~\cite{Li:2015APL}, be realized in materials of Tellegen magnetoelectric couplings~\cite{Ochiai:2015JPSJ,He:2016PNAS,Jacobs:2015NJP,Sun:2017PQE}, form bulk flat bands~\cite{Yang:2017JOSAB,Yang:2017APL}, and be immune to disorder in the bulk~\cite{Mansha:2017arXiv,Xiao:2017arXiv}. Remarkably, while it was previously known that one-way modes can exist on the surfaces of continuous magnetic media~\cite{Hartstein:1973JOPC,Yu:2008PRL,Zhang:2012APL,Yu:2014EML,Deng:2015AO,Shen:2015OE,Ochiai:2015STAM,Gangaraj:2017IEEE}, the topological origin of these modes was only recently unveiled~\cite{Silveirinha:2015PRB,Silveirinha:2016PRB-Z2,Silveirinha:2016PRB}. Topologically robust, zero-dimensional defect states localized on a dislocation in a two-dimensional topological photonic crystal were theoretically predicted and experimentally observed by~\textcite{Li:2018NatComm}.

From the application point of view, these one-way edge waveguides inspired novel device designs for tunable delays and phase shifts with unity transmission~\cite{Wang:2008PRL}, reflectionless waveguide bends and splitters~\cite{He:2010JAP,He:2010APL,Liu:2010APL,Wang:2013JOSAB}, signal switches~\cite{Zang:2011JOSAB}, directional filters~\cite{Fu:2010APL} and couplers~\cite{Zhu:2011JLT,Wang:2011JAP}, broadband circulators~\cite{Qiu:2011OE,Zhang:2013OLT}, slow-light waveguides~\cite{Yang:2013APL}, terahertz circuits~\cite{Bahari:2016topological}, photonic pulling force~\cite{Wang:2015CLEO}, and other functions~\cite{Wu:2017AOM}.

Whereas all these experiments were carried out using magneto-optic photonic crystals in the microwave domain, there is a strong push toward extending these ideas toward optical frequencies. In this domain, the magneto-optical effects are typically weaker by at least 2 to 3 orders of magnitude, but the material is continuously being improved~\cite{Onbasli:2016SR} and enhanced~\cite{Luo:2016APL}. Even though the resulting topological band gap is correspondingly smaller than that in the microwave range, such a small bandwidth is still enough to provide topological features in narrow-band phenomena such as topological laser operation~\cite{Bahari:Science2017}. More details on these very recent developments are given in the outlook Sec.\ref{sec:conclusion}.

%% file: RMP_IIIA2_Propagating.tex
\subsubsection{Propagating geometries}
\label{sec:propagating}

In this section, we will discuss an alternative approach that does not make use of magneto-optical effects, thus enabling the realization of a photonic topological system in the optical frequency range: by using three-dimensional systems (in particular, arrays of optical waveguides)~\cite{Szameit:2010JPB}, it is possible to use one of the dimensions as a temporal coordinate and thus observe topological protection in the orthogonal plane \cite{Rechtsman:2013Nature}.   For waveguide arrays, this means breaking inversion symmetry along the direction of propagation ($z$) in order to observe chiral edge states in the $(x,y)$ plane.  This is akin to substituting {\it optical activity} (i.e., circular birefringence) for the Faraday effect required for the Haldane-Raghu mechanism.  In this section, we start by introducing the fundamentals of waveguide arrays and how they may be described as a (2+1)-dimensional system (two spatial dimensions, and one temporal) in which diffraction of optical wave packets substitutes for temporal evolution of quantum mechanical particles.  We then describe how such arrays may break $z$-reversal symmetry and realize topological protection through the Floquet-engineering approach introduced in Sec.~\ref{section_floquet}.  The waveguide array geometry (also called ``photonic lattices'') has previously been used across different physical systems to realize novel phenomena such as spatial lattice solitons \cite{christodoulides2003discretizing, efremidis2002discrete, fleischer2003observation}, two-dimensional Anderson localization \cite{schwartz2007transport, lahini2008anderson}, and wave dynamics in quasicrystals \cite{freedman2006wave, freedman2007phason, levi2011disorder, Verbin:2013}, among many others.  
There is also a proposal for obtaining a non-Abelian braiding phase for photons~\cite{Iadecola:2016PRL} by guiding light through zero-energy midgap modes in a two-dimensional host lattice (although braiding of photons would not be of use for quantum computing in the same spirit as the braiding of Majorana fermions).

\begin{figure}
	\includegraphics[width=7cm]{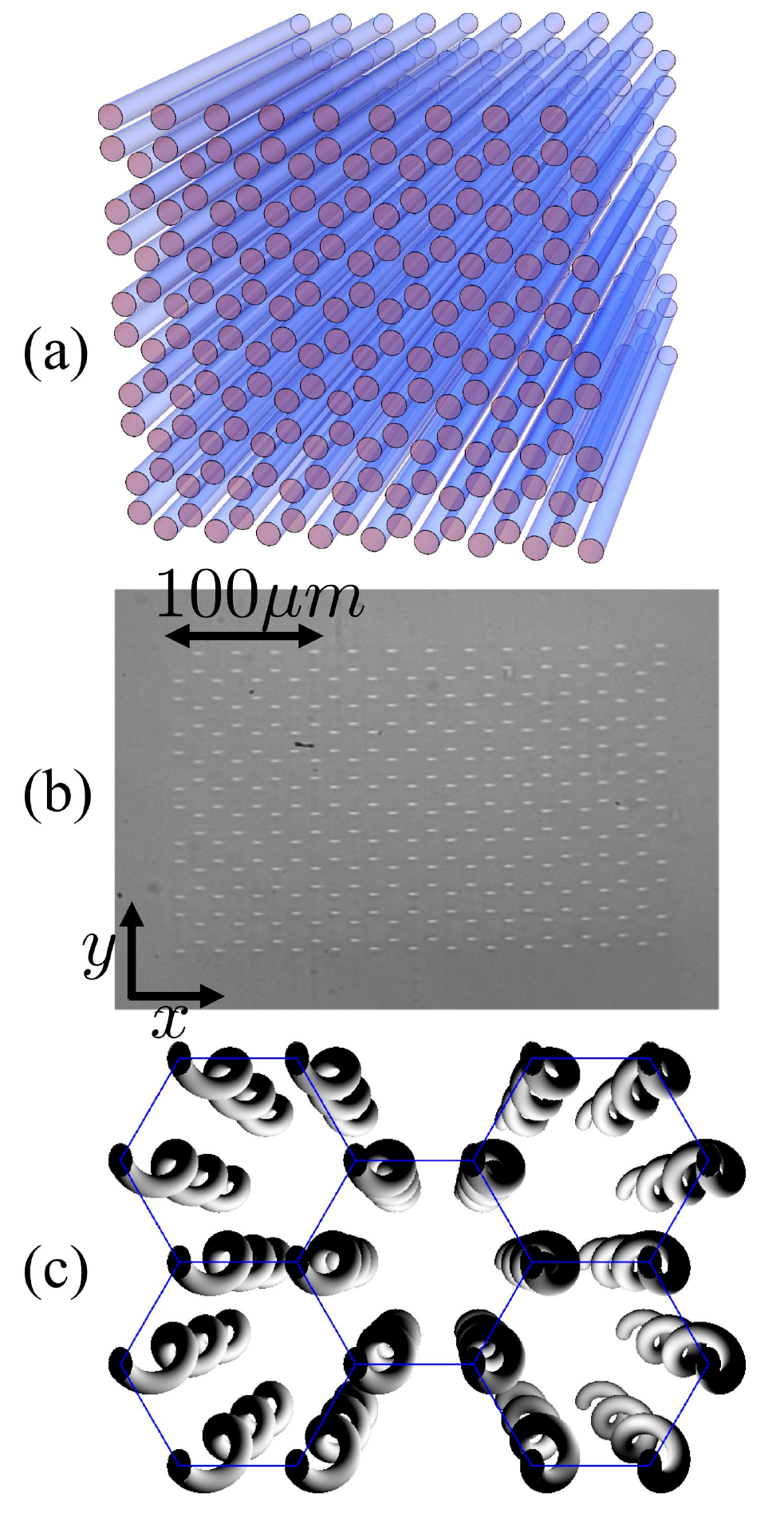}
	\caption{(a) Schematic diagram of a honeycomb lattice waveguide array with straight waveguides. Adapted from~\textcite{Rechtsman:2013NatPhot}. (b) Microscope image of input facet of a honeycomb lattice waveguide array. Adapted from~\textcite{Rechtsman:2013PRL}. (c) Schematic of an array of helical waveguides; each waveguide helix has the same radius, period, and phase (i.e., they are all moving in concert). Adapted from~\textcite{Rechtsman:2013Nature}.   }
	\label{fig1}
\end{figure}

A typical waveguide array geometry is depicted schematically in Fig. \ref{fig1}(a).  The waveguides are fabricated using the direct laser writing method, and their properties are highly dependent on the host material [Fig. \ref{fig1}(b) shows a microscope image of a transverse cross section of the structure].  Below, we will be describing experiments performed in fused silica glass (refractive index $n=1.46$ at wavelength $\lambda = 633nm$).  The specifics of the fabrication procedure are described in detail elsewhere \cite{Szameit:2010JPB, szameit2007control}.  Typical parameters that describe waveguide properties are refractive index increase from the background $\Delta n\sim 1.0\times 10^{-3}$, waveguide radii in the $x$ and $y$ directions are $r_x=2\mu m$ and $r_y=5\mu m$, and their shapes may be described with a hyper-Gaussian functional form $\Delta n(x,y)=\Delta n_0\exp \left\{ -\left[(x/r_x)^2 + (y/r_y)^2 \right]^\alpha  \right\}$, with the exponent $\alpha=3$ in fused silica glass.  In a typical waveguide array experiment, a beam of light is injected at the input facet of the array ($z$=0) and allowed to propagate through until it exits the array, at which point it is imaged onto a charge-coupled device (CCD) camera.  As discussed in Sec. \ref{sec:IIpropagating}, it is described by the paraxial equation for the diffraction of light:
\begin{equation}
i \partial_z \tilde{E} = -\frac{1}{2k_0}\nabla_\perp^2 \tilde{E} - \frac{k_0 \Delta n}{n_1}\tilde{E}, \label{Eq4}
\end{equation}
where $\tilde{E}$ represents the envelope function of the electric field.  Note that the paraxial equation takes the form of a Schr\"odinger equation, even though it describes the diffraction of classical light rather than the motion of a massive quantum particle.  However, in the usual Schr\"odinger equation of quantum mechanics, the left side of the equation has a time derivative; here it is a derivative in $z$, the spatial coordinate in the propagation direction.  Therefore, $z$ takes the role of a temporal coordinate and the transverse $(x,y)$ plane takes the role of an artificial two-dimensional material.  The diffraction of static continuous wave light therefore emulates the evolution of the wave function of a single quantum mechanical particle.  

The structure shown schematically in Fig. \ref{fig1}(a) and in an experimental image in Fig. \ref{fig1}(b) is a honeycomb lattice of waveguides.  Each waveguide is single mode, meaning it can be thought of as a potential well with a single bound state.  The waveguides are placed at a distance from one another such that the modes of neighboring waveguides can evanescently couple (i.e., ``tunnel'') between neighbors (typical spacing $d=15$ $\mu$m).  This results in a typical hopping parameter (aka, coupling constant) between waveguides of $J\sim 1$ cm$^{-1}$, but this can be tuned by changing the wavelength, waveguide refractive index, and/or spacing between the waveguides.  The length of the sample (which corresponds to the amount of ``time,'' i.e., propagation distance, that the optical wave function can propagate) is typically taken to be on the order $Z\sim 10$ cm.  Given that this particular array is a honeycomb lattice, the diffraction of photons therefore has a perfect correspondence with the motion of noninteracting electrons in graphene. Honeycomb waveguide arrays were first used to demonstrate optical Dirac physics via the observation of conical diffraction \cite{Peleg:2007PRL}.

Since each waveguide acts as an ``artificial atom'' in the analogy between waveguide arrays and two-dimensional materials, it is possible to employ the tight-binding approximation to Eq. (\ref{Eq4}), as described in Sec. \ref{sec:IIpropagating}.  In this approximation, the wave function $\tilde{E}$ is expanded in a subspace composed of bound modes of each waveguide.  Thus, we can write the paraxial equation as
\begin{equation}
i\partial_z \alpha_m = -\sum_{\langle m,n \rangle} J_{mn} \alpha_n, \label{tight-binding}
\end{equation}
where $\alpha_m$ is the amplitude of the mode in waveguide $m$, $J_{mn}$ is the hopping strength between waveguides $m$ and $n$, and the summation is taken over neighboring waveguides (nearest, next-nearest, and so on, as necessary).  Here we henceforth assume the tight-binding description due to its ubiquity across other experimental platforms (condensed matter, ultracold atoms, coupled resonators, among others).  The bulk and edge band structures of the honeycomb lattice of Fig.~\ref{fig1}(a) are shown in Figs.~\ref{II_Haldane}(c) and~\ref{II_Haldane}(e), assuming nearest-neighbor hopping only.  For a discussion of edge band structure, see Sec. \ref{subs:gapless}.  This is exactly the band structure of graphene \cite{Wallace:1947PR,neto2009electronic} exhibiting Dirac cones - conical touchings between bands at the Brillouin zone corners.  We note here this system may be explored beyond the tight-binding limit by performing photonic \textit{ab initio} simulations by diagonalizing the full continuum Schr\"odinger equation~(\ref{Eq4}).  Most often, continuum simulations yield only minor quantitative corrections to tight binding, but in special cases can reveal profound qualitative differences, including the presence of edge states in regions of the edge Brillouin zone where tight binding predicts none \cite{Plotnik:2014NatMat}.  A rigorous description of continuous topological systems can be found in a series of works by~\textcite{fefferman2012honeycomb, fefferman2014topologically, fefferman2014wave, lee2016photonic}.   

A key requirement of realizing topologically protected chiral edge states is breaking time-reversal symmetry, as previously described.  Since $z$ acts as a temporal coordinate in waveguide arrays, breaking $z$-reversal symmetry can allow for topologically protected edge states in the transverse $(x,y)$ plane.  This is accomplished by using helical, instead of straight, waveguides in a honeycomb waveguide array, as depicted in Fig. \ref{fig1}(c).  Similar helical waveguide arrays have been used to demonstrate dynamical localization \cite{crespi2013dynamic}.  To describe the diffraction of light through the helical array, we move into a coordinate frame comoving with the helices: $x \rightarrow x + R \cos\Omega z$, $y \rightarrow y + R \sin\Omega z$, and $z \rightarrow z$.  In the new coordinate system, the Laplacian remains unchanged, but the $z$ derivative transforms as
\begin{equation}
\partial_z \rightarrow R\Omega \left[- \sin \left(\Omega z\right) \partial_x  + \cos \left( \Omega z \right) \partial_y  \right] + \partial_z.
\end{equation}

We now rewrite the $z$ derivative of Eq. (\ref{Eq4}) in this new coordinate system and find
\begin{equation}
i \partial_z \tilde{E} = \left[i \nabla_\perp - {\bf A}(z)\right]^2\tilde{E} - \frac{k_0 \Delta n}{n_1}\tilde{E} + \frac{k_0}{2} R^2\Omega^2\tilde{E} , \label{paraxial_gauge}
\end{equation}
where ${\bf A}(z)=k_0 R \Omega (-\sin \Omega z, \cos \Omega z)$ is the vector potential induced by the helical rotation, and the final term in Eq. (\ref{paraxial_gauge}) can be ignored because it is simply proportional to the identity.  This vector potential ${\bf A}$ corresponds to a circularly rotating electric field (note that it is curl free, so the corresponding magnetic field is zero).  It can be incorporated into Eq. (\ref{tight-binding}) simply by including the appropriate Peierls phase factors in the hopping, namely $J_{mn}\rightarrow J_{mn}\exp \left[ -i {\bf A(z)}\cdot {\bf r}_{mn}\right]$, where ${\bf r}_{mn}$ is the vector that connects site $m$ to site $n$.  Therefore, the Schr\"odinger equation~(\ref{tight-binding}) is time periodic and must be solved using the machinery of Floquet engineering; see Sec.~\ref{section_floquet}. This equation is perfectly analogous to the Schr\"odinger equation that describes the motion of electrons in graphene under the influence of circularly polarized classical light; see also Eq.~\eqref{honey_shake} and paragraph below.  The rotating field acts to break time-reversal symmetry (actually $z$-reversal symmetry, $z$ being the temporal coordinate), without requiring the presence of a magnetic field (even a fictitious one).  This Floquet system precisely maps to the Haldane model \cite{Haldane:1988PRL} in the high-frequency driving limit; see Eq.~\eqref{honey_shake_2} for a description of the effective Haldane-like Hamiltonian.  The bulk and edge band structures for the Haldane model are shown in Figs.~\ref{II_Haldane}(d) and~\ref{II_Haldane}(f), respectively; these are qualitatively akin to those of the waveguide array described here.  Comparing Figs.~\ref{II_Haldane}(c) and~\ref{II_Haldane}(e) with~\ref{II_Haldane}(d) and~\ref{II_Haldane}(f), it can be seen that the helicity acts to break the degeneracy at the Dirac points and open a bulk band gap.  In the edge band structure [Fig.~\ref{II_Haldane}(f)], edge states are present (these are the two bands crossing the gap), with one localized to the top of the structure and the other localized to the bottom.  They are part of a single chiral edge state that flows around the edges.       

\begin{figure}
	\includegraphics[width=7cm]{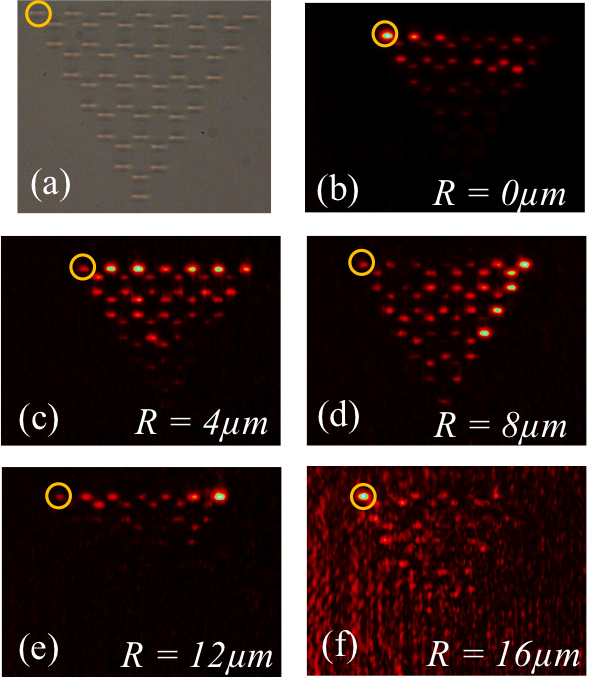}
	\caption{(a) Microscope image of input facet of waveguide array \cite{Rechtsman:2013Nature}.  The yellow circle indicates the point of injection of input light. (b)-(f) Output facet of waveguide array after 10 cm of propagation, for different devices with increasing helix radius.  The edge states do not backscatter as they turn around the top-right corner, i.e., in (d).  Their group velocity increases and then decreases with $R$, as predicted from the Floquet lattice model.  Significant bending loss can be observed in the background for $R=16$ $\mu$m.  Helix period is 1 cm in all cases.}
	\label{mikael-fig2}
\end{figure}

The topologically protected chiral edge states can be directly observed experimentally.  A series of such experiments were performed by~\textcite{Rechtsman:2013Nature}, observing perfect transmission around corners and past defects; we highlight one of these in Fig. \ref{mikael-fig2}.  Here light is injected in the top-left corner of a honeycomb lattice arranged in an equilateral triangle geometry, for a series of samples of increasing helix radius.  The optical wave function travels clockwise around the structure, with increasing group velocity as a function of the helix radius (i.e., gauge field strength) until about $R\sim 8$ $\mu$m, whereupon the group velocity decreases and goes to zero at $R\sim 16$ $\mu$m.  This is quantitatively consistent with the lattice model, which predicts that the band gap opens, and then closes, as a function of increased helix radius \cite{Rechtsman:2013Nature}.  The high background signal that is observable at $R=16$ $\mu$m arises from the bending loss associated with the helicity of the waveguides (this corresponds to the phenomenon of ``heating'' in condensed-matter Floquet systems).  Note that in each case, some bulk modes are excited (i.e., not all light is localized to the edge of the structure).  This is attributable to the fact that the bulk modes have some support at the corner of the structure and are therefore excited along with the edge states when light is injected at the corner waveguide.

Since the advent of topological phenomena in the paraxial geometry, there has been significant progress in this direction.  To highlight a few examples, waveguide array geometries have been used either theoretically or experimentally to demonstrate the optical Rashba effect~\cite{plotnik2016analogue}, the photonic anomalous Floquet topological insulator state \cite{Rudner:2013PRX, Maczewsky:2017NatCom, Mukherjee:2017NatCom, Bellec:2017EPL}, topological transitions \cite{Leykam:2016PRLanomalous, guglielmon2017prediction}, chiral edge states in quasicrystals (in particular, Penrose tilings) \cite{Bandres:2016PRX}, protected zero-dimensional cavity modes in two-dimensional lattices \cite{Noh:2018NatPhot} (protected modes two dimensions lower than host lattice), type-II Weyl points in three dimensions \cite{Noh:2017NatPhys}, and the topological Anderson insulator~\cite{Stutzer:2018Naure}, a one-dimensional version of which was realized in an ultracold atomic gas~\cite{Meier:2018arXiv}. Furthermore, Floquet topological insulators were realized in waveguide arrays fabricated with two-photon polymerization in photoresist materials; these were used to demonstrate protection against time-dependent defects \cite{jorg2017dynamic}.  Because of the difficulty of breaking time-reversal symmetry in a planar geometry, the paraxial platform described here provides a rich methodology for photonic topological phenomena, including those that incorporate non-Hermiticity, nonlinearity and other effects that go beyond solid-state physics. Recently, a state-recycling technique was developed to significantly enhance the effective timescales over which dynamics take place within arrays of coupled optical waveguides~\cite{mukherjee2017state}. This scheme consists of placing the photonic lattice into a cavity, which allows the optical state to be reinjected many times into the lattice. This approach also allows one to image real-time (stroboscopic) evolution in photonic lattices, by recording the state of the system after each round-trip.

%% file: RMP_IIIA3_Optomechanics.tex
\subsubsection{Optomechanics}

There have also been interesting developments in breaking time-reversal symmetry and implementing synthetic gauge fields in optomechanical systems. In general, the field of optomechanics deals with the coherent interaction between photons and acoustic phonons confined in a cavity or an array, which can be controlled at the single phonon level~\cite{Aspelmeyer2014}. This field has generated a lot of excitement due to its potential applications, ranging from sensing to quantum information processing.

In 2012, it was proposed that time-reversal symmetry could be broken in optomechanical resonators, and therefore, one could use them as an optical isolator and nonreciprocal phase shifter~\cite{Hafezi:2012OPtExp}. Specifically, a directional laser pump was used to select one circulation direction, such that, consequently, the manifestation of time-reversal breaking could be observed in the nonreciprocal optical response~\cite{shen2016experimental,alu2016nonreciprocity}. One can also switch the role of phonon and photons and study nonreciprocal transport of phonons~\cite{stannigel2012optomechanical,habraken2012continuous,Alu2014sound,bahl2016dynamically}.

More recently, there has been an intriguing proposal for the implementation of synthetic gauge fields in optomechanical crystals \cite{Schmidt:2015Opt}. Thanks to the uniformity in their fabrication, optomechanical crystals can form a 1D or 2D array of resonators, with a significant degree of controllability. Initial experimental demonstration of such an approach for a few sites was recently reported by~\textcite{fang2016generalized}. These advances could lead to the realization of various topological phases in optomechanical crystals~\cite{Schmidt:2015Opt} and nonreciprocal baths and amplifiers~\cite{Clerk2015}. Other optomechanical systems to implement synthetic gauge fields include quantum wells~\cite{Sasha2017} and superconducting circuits\cite{Cleland2010,Teufel2017,Delsing2014,Schoelkopf2017}.

%% file: RMP_IIIA4_QHEOther.tex
\subsubsection{Cavity- and circuit-QED systems}
\label{sec:circuitQED}

Arrays of cavity- and circuit-QED devices can offer several advantages for exploring topological states of light, including highly controllable geometry and connectivity, as well as the possibility of reaching the regime of strong effective photon-photon interactions. Depending on the material platform and the frequency domain considered, these interactions can be obtained by coupling photons to a variety of emitters, e.g., atoms, chemical impurities in a solid-state material, artificial atoms such as quantum dots, or even superconducting circuits~\cite{Gu:2017PhysRep}. While progress in interacting topological photonics will be discussed in Sec.~\ref{sec:interaction}, we introduce here experimental and theoretical advances toward realizing noninteracting analog quantum Hall states in cavity- and circuit-QED related systems. 

\begin{figure}[thbp]
	\centering
	\includegraphics[width=0.9\columnwidth]{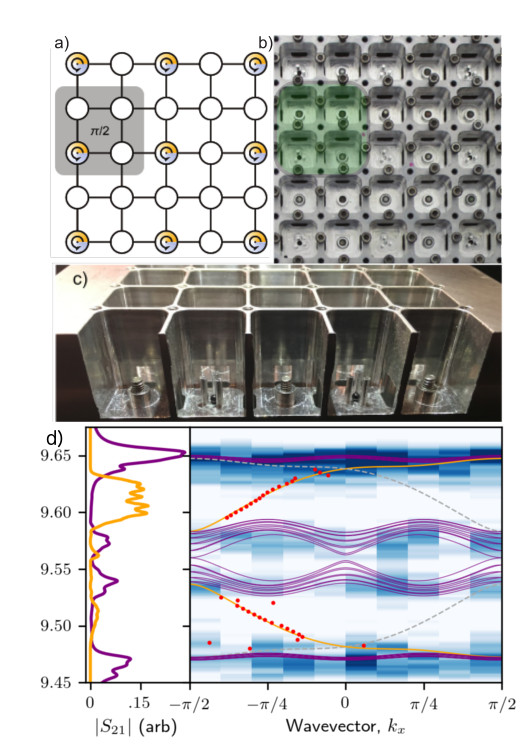}
	\caption{(a) Connectivity of a microwave cavity model engineered to realize a Harper-Hofstadter model [Eq.~\ref{II_HHHamiltonian}]. Open circles are fundamental-mode resonators exhibiting an $s$-like on-site orbital. Blue $\rightarrow$ orange circles are chiral resonators with a single isolated $p_x+i p_y$ orbital, employed to induce an effective magnetic flux per plaquette of $\alpha=1/4$. (b) A photograph of the lattice from overhead, with a single four-site magnetic unit cell highlighted in green. Fundamental-mode resonators consist of a single-post coaxial cavity; chiral resonators consist of a three-post cavity where time-reversal symmetry is broken using a yttrium-iron-garnet sphere in a magnetic field. Tunneling between cavities is achieved via a slot cut between them. (c) A cutout side view of the same structure: now evident are aluminum screws threaded into the center post of the fundamental-mode resonators, used to tune their frequencies to degeneracy. (d) Site-resolved measurement of the band structure and edge dispersion of the lattice. Frequencies on the vertical axis are in GHz. From~\textcite{Owens:2018PRA}.\label{Fig:TopoCircuitUWave}}
\end{figure}

One approach in this direction relies upon engineering arrays of tunnel-coupled reentrant coaxial microwave cavities, as outlined theoretically by~\textcite{anderson2016engineering} and demonstrated experimentally by~\textcite{Owens:2018PRA}. In this setup, time-reversal symmetry is broken through the chiral on-site spatial wave function of every fourth resonator, as depicted in Fig.~\ref{Fig:TopoCircuitUWave}. This chiral on-site wave function is engineered through the coupling of a ferrimagnetic yttrium-iron-garnet (YIG) crystal in a magnetic field to the near-degenerate $P$ modes of a triple-coax cavity. The resulting model corresponds to a Harper-Hofstadter model [Eq.~\ref{II_HHHamiltonian}] with an effective magnetic flux per plaquette of $\alpha=1/4$. By using a real magnetic field to break time-reversal symmetry, this approach is akin to earlier work~\cite{Wang:2009Nature} at room temperature, but allows one to substantially suppress losses by focusing on those topological bands that are dark to the YIG crystals. Measuring and compensating disorder in this system relies upon Hamiltonian tomography techniques based upon one- and two-point network analysis of the lattice~\cite{ma2016hamiltonian}. Forthcoming challenges include extensions of these ideas to higher-dimensional topological circuits and then to  strongly interacting systems.  For the latter, quality factors need to be significantly higher as could likely be realized using superconducting cavities, and the system will need to be operated in a cryogenic environment where $\hbar \omega_0 \ll k_B T$. Specific issues related to the population of strongly correlated states are addressed in Sec.\ref{subsec:strong}.

A different strategy to induce an artificial gauge field in a coupled cavity array was proposed by~\textcite{Cho:2008PRL}. The idea is to trap in each cavity a single atom with two ground states that can be described as a lattice of effective spin $1/2$ and can thus be mapped onto a lattice of impenetrable bosons. Hopping between neighboring sites is provided by the effective spin-exchange coupling that results from intercavity hopping of virtually excited photons when atoms are optically dressed. The artificial magnetic field is imposed by making the optical dressing field spatially dependent.
While this first proposal focused on the fractional quantum Hall regime, similar ideas have since been applied to noninteracting polaritons in a hybrid circuit-QED system~\cite{Yang:2012PRA}, where a superconducting resonator at each lattice site is coupled to a nitrogen-vacancy center ensemble, whose internal states are dressed by spatially dependent microwave sources. 

Another proposal to realize a synthetic magnetic field in a circuit-QED architecture by means of passive circulator elements was proposed by~\textcite{Koch:2010PRA}: the circulators mediating the tunnel coupling between resonators are designed to break time-reversal symmetry and can be implemented with simple superconducting circuits, such as three-junction Josephson rings. Extended to a lattice of resonators, this approach could be used to realize a Chern insulator, for example, in a kagome geometry as also further studied by~\textcite{Petrescu:2012PRA}. Along these lines, a general strategy to break time-reversal symmetry in a resonator lattice by exploiting the magneto-optical effect in the waveguide or resonator elements mediating intersite coupling was discussed for photonic crystal resonator lattices by~\textcite{Fang:2013PRA}.

\subsubsection{Other proposals}
\label{sec:qheother}

Beyond current experiments, there have been many proposals for alternative routes to breaking time-reversal symmetry in photonics by exploiting light-matter coupling, dynamical modulation, and/or novel lattice elements. As we now briefly review, these developments may lead to analog quantum Hall effects in a variety of new systems, including microcavity polaritons and different realizations of resonator lattices. 

\paragraph{Topolaritons} 
Microcavity polaritons provide a particularly suitable photonic platform to address the physics of lattices with broken time-reversal symmetry. Polaritons are mixed light-matter quasiparticles arising from the strong coupling between photons and excitons, electron-hole bound pairs, confined in a semiconductor microcavity~\cite{carusotto:2013}. While they are neutral particles, excitons possess a non-negligible magnetic moment arising from the spin of the electron and hole in the pair. Thus, polaritons, via their excitonic component, show significant Zeeman splittings ($\Delta_Z$) when subject to an external magnetic field~\cite{Mirek:2017PRB}. In this situation, the lowest-energy mode of polaritons confined in a single resonator splits into two states of different emission energy characterized by opposite circular polarizations.

This feature has been exploited in a number of theoretical works to propose a Chern insulator based on a polariton lattice in an external magnetic field~\cite{Nalitov:2015PRL, Karzig:2015PRX, Bardyn:2015PRB,Yi:2016PRB}. The combination of the polariton Zeeman splitting and the transverse-electric--transverse-magnetic (TE-TM) splitting ($\Delta_{\mathrm{TE-TM}}$) characteristic of the photonic part of polaritons~\cite{Kavokin:2005PRL} results in the opening of a topological gap whenever band crossings are present in the spectrum of the lattice. Simultaneously, in finite-size samples, protected chiral edge states emerge at the boundaries, with a chiral direction determined by the sign of the external magnetic field. A prominent example of these \textit{topolaritons} is a honeycomb lattice of semiconductor micropillars in the presence of an external magnetic field~\cite{Nalitov:2015PRL}. In this case, the external magnetic field is expected to open a gap at the Dirac cones with a magnitude given by $\Delta_{Z}$ and $\Delta_{\mathrm{TE-TM}}$; the resulting bands acquire a Chern number of $\mp 2$ or $\pm 1$ depending on the ratio of these splittings to the nearest-neighbor hopping~\cite{Bleu:2016PRB, Bleu:2017PRB}.  One of the most attractive features of polaritons is the possibility of combining these topological properties with significant Kerr nonlinearities; more discussion on interacting topological systems is given in Sec.~\ref{sec:interaction}.
Recently, topolaritons have been experimentally realized in the context of topological lasers~\cite{Klembt:arxiv2018}, which are discussed in more detail in Sec.~\ref{sec:conclusion}.

Besides exciton polaritons in suitable semiconductor devices,~\textcite{Jin:2016NatComm} recognized that magnetoplasmons arising from the coupling of the electromagnetic field with electron-hole excitations in electronic quantum Hall systems also have nontrivial topological properties. In particular, since magnetoplasmons possess a particle-hole symmetry while breaking time-reversal symmetry, they are a unique example of class D 2D topological systems~\cite{Ryu:2010NJP}.

\paragraph{Dynamical modulation}
\label{sec:qheothertata}

A flexible and powerful way to break time-reversal symmetry in both the microwave and optical domain is offered by the dynamical modulation of the properties of a resonator array. Typically, one dynamically tunes either the resonance frequencies of different cavities~\cite{Hayward:2012PRL,Minkov:2016Optica} or hopping amplitudes~\cite{Fang:2012NatPhot}. In this approach of {\it modulation assisted tunneling}, a large difference $\Delta \omega$ in the on-site resonance frequencies between neighboring tight-binding lattice sites initially suppresses particle tunneling. This tunneling can be restored by applying a suitable resonant time-dependent modulation at frequency $\Delta \omega$. The phase of the external modulation then appears in the phase of the tunneling amplitudes, simulating the effects of a gauge field on a charged particle. The idea was first introduced by~\textcite{Jaksch:2003NJP} and experimentally implemented by~\textcite{Miyake:2013PRL,Aidelsburger:2013PRL} for ultracold gases. A related idea was proposed in a classical mechanical framework by~\textcite{Salerno:2016PRB}.

For a suitable spatial dependence of phases in a square geometry, the dynamical resonator array maps directly onto the Harper-Hofstadter model [Eq.~\ref{II_HHHamiltonian}]. Although this mapping relies on the rotating wave approximation, nontrivial topological features persist even if the inter-resonator coupling is ultrastrong and the rotating wave approximation breaks down~\cite{Yuan:2015PRA}. Other choices of the modulation phases can be used to engineer spatially inhomogeneous effective gauge fields~\cite{Fang:2013PRL,Lin:2014PRX} or effective electric fields~\cite{Yuan:2016PRB, Yuan:2015PRL} that may also be useful in controlling light. Complex modulations were implemented by~\textcite{Lumer:2018arXiv} to realize novel light guiding effects by a
spatially dependent gauge field, as proposed by~\textcite{Lin:2014PRX}. Going up in geometrical complexity, nonreciprocal propagation based on interband photonic transitions induced in a waveguide by means of a running-wave-shaped, electrically driven modulation of the refractive index was experimentally demonstrated by~\textcite{Lira:2012PRL}. In general, note that dynamical modulation as well as magneto-optical effects could also be exploited to imprint a synthetic magnetic field for photons in a resonator-free implementation, based on a waveguide network~\cite{Lin:2015NJP}.

The actual implementation of the general temporal modulation idea depends on the specific photonic system under consideration. In the infrared and optical domains, the main limitation is posed by the rate at which photonic structures can actually be modulated (e.g., by carrier injection or optomechanics), which is typically significantly lower than the optical frequency itself (GHz versus 100THz). A resonator lattice with hoppings on the order of the modulation frequency (or lower) is thus required, which in turn imposes stringent bounds on the quality factor of the resonators. In circuit-QED architectures, schemes to dynamically modulate the superconducting circuit elements, e.g., superconducting quantum interference devices (SQUIDs), used to couple the lattice of resonators together were proposed by~\textcite{Peropadre:2013PRB,Wang:2015Scientificreports,Wang:2016NPJ}. As we shall see in more detail in Sec.~\ref{subsec:strong}, this strategy was employed for a pioneering demonstration of the interplay of magnetic and interaction effects by~\textcite{Roushan:2016NatPhys}. In propagating geometries based on waveguide arrays, the role of time and the propagation direction are exchanged (see Sec.~\ref{sec:propagating}), and so a suitable ``time-dependent" modulation may be realized by spatially varying the refractive index of the medium~\cite{Longhi:2013OptLett, Dubcek:2015NJP}.  Pioneering experiments along these lines were reported by~\textcite{Rechtsman:2013Nature,Mukherjee:2015NJP} and are reviewed in Sec.~\ref{sec:propagating}.

%% file: RMP_IIIB1_SiliconRing.tex
\subsection{Analog quantum spin Hall systems in photonics}
\label{sec:analogQSHE}

The second main class of topological photonics systems in 2D are those which preserve time-reversal symmetry for photons and which are analogous to quantum spin Hall systems in condensed matter. In this class of systems, the one-way edge modes are only topologically protected if spin-changing scattering processes can be neglected. Here we provide an in-depth discussion of key systems that have been experimentally realized, and also briefly review some promising theoretical proposals.

\subsubsection{Silicon ring resonator arrays }
\label{sec:siliconring}

This first section is devoted to a review of the experiments by~\textcite{Hafezi:2013NatPhot,Mittal:2014PRL,Mittal:2016NatPhot} using arrays of silicon ring resonators. In these systems, a synthetic magnetic field for photons can be engineered by controlling the differential optical paths, when photons  hop between neighboring sites in two opposite directions. A pseudospin-$1/2$ degree of freedom naturally arises as ring resonators support a pair of degenerate whispering gallerylike modes, propagating in opposite clockwise and counterclockwise directions. As the device does not contain any real magnetic element, time-reversal symmetry imposes the fact that the two spin states experience opposite synthetic magnetic fields, and therefore, the corresponding edge states have opposite chiralities, in direct analogy to quantum spin Hall systems.

We start by explaining how a nonsymmetric phase hopping can be engineered between a pair of optical resonators. By arranging these pairs into a square lattice, one can synthesize a uniform magnetic field. Furthermore, we review the experimental progress in this platform, from imaging topological edge states,  demonstration of their robustness in transport observables such as transmission and delay time, and finally measurement of the associated topological invariants.  We also discuss the prospect of extension of these ideas in the quantum regime, in particular, in generation of photon pairs and quantum transport of two-photons.

\begin{figure}
	\includegraphics[width=7cm]{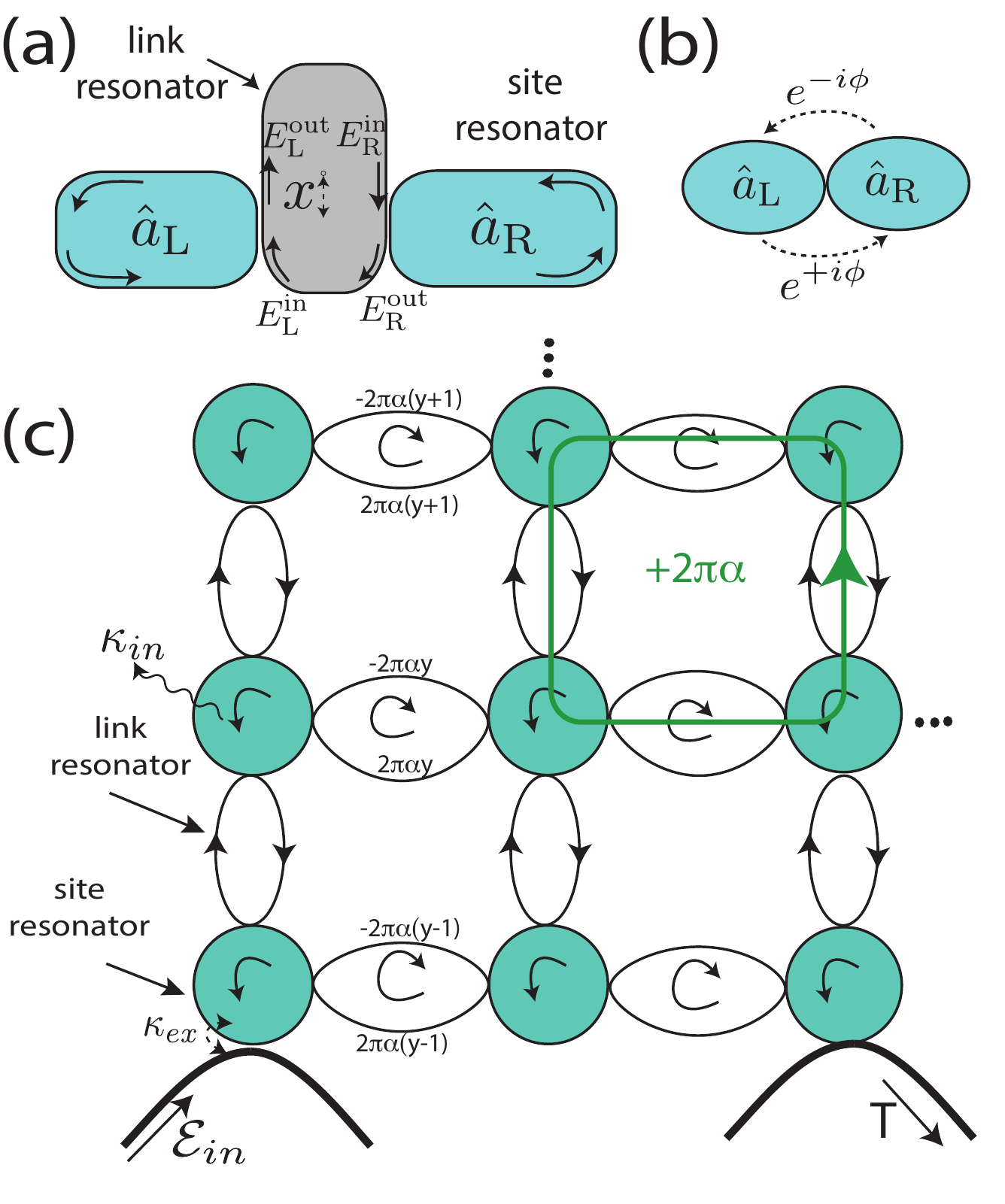}
	\caption{(a) Schematic of the coupling mechanism between two site resonators via an off-resonant link ring. (b) After integrating out the off-resonant link, its asymmetric location leads a nonzero hopping phase $\phi$ between site resonators. (c) A two-dimensional array of rings realizing the Harper-Hofstadter model for photons.
	From~\textcite{Hafezi:2011NatPhys}.}
	\label{fig:MH_lattice}
	\end{figure}

\textit{Two-resonator case:} 
As the first step, we show a nonzero hopping phase can be obtained for photons hopping back and forth between a pair of coupled site resonators. As originally proposed by~\textcite{Hafezi:2011NatPhys}, this can be obtained when two neighboring resonators are coupled through a third off-resonant link ring, as sketched in Fig.~\ref{fig:MH_lattice}(a).  We consider the length of resonators to be $m\lambda$, where $2 m$ is an even integer and $\lambda$ is the resonant wavelength. The off-resonant ring link length is set to $2 m\lambda+3\lambda/2$ to guarantee that it remains antiresonant. The physics would be analogous for an odd integer, except for the sign of tunneling. Now, if the off-resonant ring link is vertically shifted by $x$ between site resonators, as shown in Fig.~\ref{fig:MH_lattice}(a), the accumulated optical phase during the hopping from one site to the other is asymmetrical, and we achieve the desired hopping phase between two sites, given by $\phi=4\pi x/\lambda$,  as shown in Fig.~\ref{fig:MH_lattice}(b). 

The formal derivation of this hopping phase was presented by~\textcite{Hafezi:2011NatPhys}  using the input-output formalism~\cite{Gardiner:PRA1985}, which, as discussed in Sec.\ref{subsec:noneq}, is equivalent to coupled mode theory in the noninteracting case. An alternative approach using transfer matrices is presented in the supplemental material of~\textcite{Hafezi:2013NatPhot}. We briefly recapitulate the input-output formalism approach. Specifically, the boundary condition for the electric field is given by $\hat{E}_{\rm i}^{\rm out}=\hat{E}_{\rm i}^{\rm in}+\sqrt{2\kappa}\hat{a}_{\rm i}$, where $\kappa$ is the coupling efficiency and the site index can be left or right, $\rm {i=R,L}$. Assuming no intrinsic loss in waveguides, the dynamics of each optical resonator is given by $\dot{\hat {a}}_{\rm i}=-\kappa \hat{a}_{\rm i }-\sqrt{2 \kappa} \hat{E}_{\rm i}^{\rm in} $. The input of each resonator is related to the output of the other resonator, by a propagation phase which depends on the optical length. For our set of parameters, $\hat{E}_{\rm R(L)}^{\rm in}=-i \hat{E}_{\rm L(R)}^{\rm out} \exp({\mp2 \pi i \phi})$. For counter-clockwise modes of the side resonators, as shown in Fig.~\ref{fig:MH_lattice}(a), photons acquire a phase proportional to the upper or lower connecting optical lengths, when they hop from left to right and vice versa. If the link ring is positioned asymmetrically, this hopping phase can be different for forward or backward hopping. Specifically, by eliminating the fields in the link rings $(\hat{E}_{\rm R}^{\rm in},\hat{E}_{\rm L}^{\rm in},\hat{E}_{\rm R}^{\rm out},\hat{E}_{\rm L}^{\rm out})$, the photon dynamics is given by $\dot{\hat{a}}_{\rm R(L)}=i\kappa \exp({\mp 2\pi i \phi})\hat{a}_{\rm L(R)}$ which is equivalent to having a Hamiltonian \begin{equation} 
H=-\kappa \hat{a}_{\rm R}^\dagger \hat{a}_{\rm L} \exp({-2\pi i \phi}) +\rm{h.c.} \end{equation}

Note that since the overall length of the link resonators is fixed, to maintain the antiresonance condition,  the acquired phases for forward and backward hops are not independent from each other. Specifically, they are equal in magnitude and opposite in sign. In the general case of an arbitrary off-resonant link ring, the nonzero hopping phases in the forward and backward directions remain opposite in sign to guarantee the Hermiticity of the Hamiltonian. However, if the antiresonancy condition is not met,  the resonant frequency of the resonators gets shifted; see the discussion in the supplementary material of \textcite{Hafezi:2013NatPhot}.

\textit{2D lattice:}  By arranging the resonator in a lattice structure, one can implement an effective uniform magnetic field in the form of the Harper-Hofstadter model, which was discussed in Sec.\ref{sec:iqhe}. In particular, we implement the Landau gauge, as in Eq.~\ref{II_HHHamiltonian}, to obtain the same phase $\alpha$ over each square plaquette, except that here the $x$-$y$ directions are swapped.  The phase pattern can be obtained by arranging the positioning of the link rings according to the Landau gauge. Specifically, the link rings responsible for tunneling in the $y$ direction are symmetrically positioned to get zero hopping phase. In contrast, the link rings for tunneling in the $x$ direction are asymmetrically positioned, in an increasing order in the $y$ direction, giving rise to a $e^{i\ 2 \pi \alpha y}$ hopping phase. The scheme is illustrated in Fig.~\ref{fig:MH_lattice}(c). As discussed in Sec.~\ref{sec:iqhe}, for a fixed magnetic field, which is set by the parameter $\alpha$, any finite lattice displays bulk and edge bands, the former being organized in the celebrated Hofstadter butterfly, and the latter being located in the gaps of the bulk dispersion. 

\begin{figure}
	\includegraphics[width=7cm]{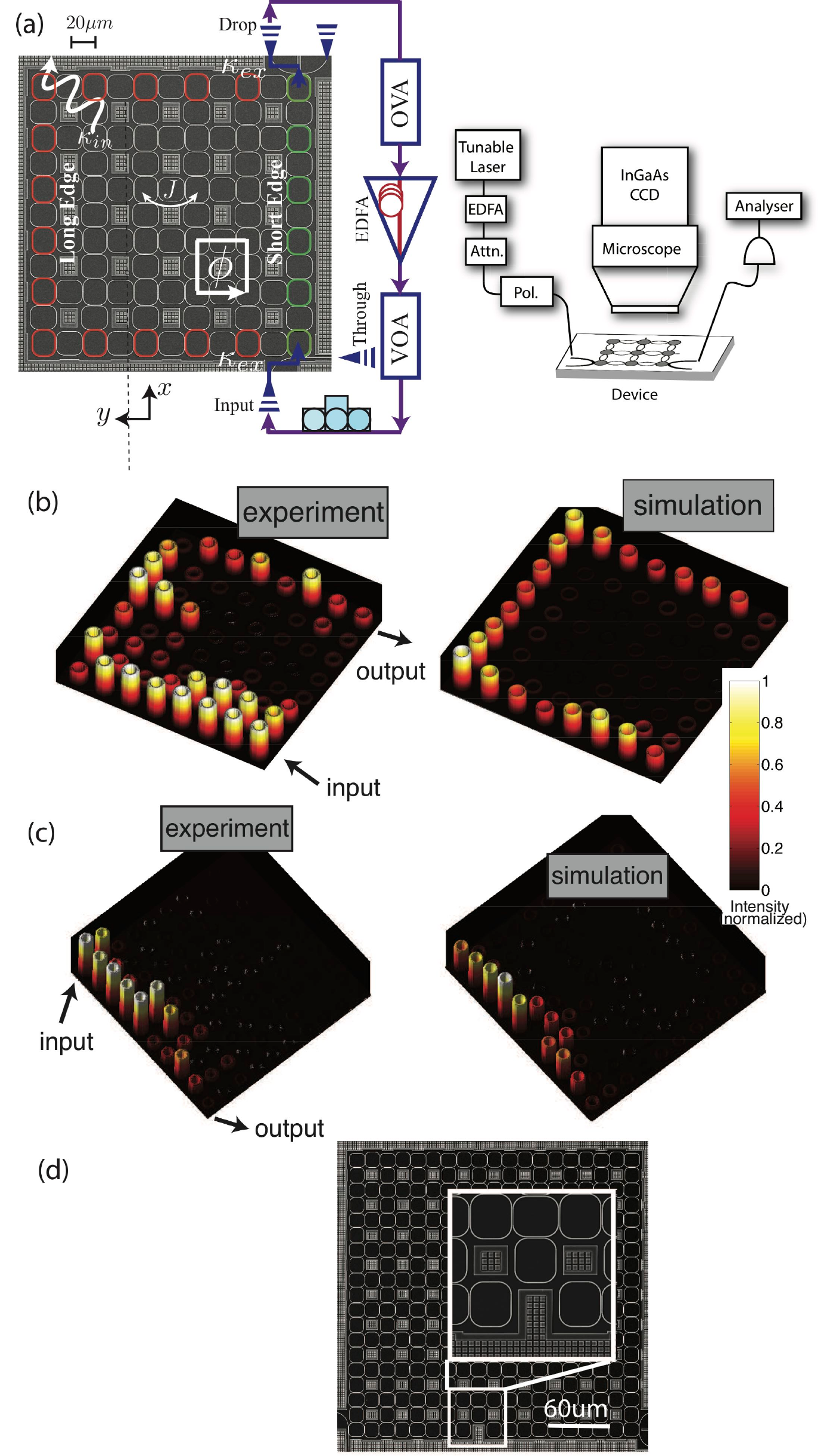}
	\caption{(a) Sketch of the experimental setup. (b) Spatial intensity profile showing light propagation along the edge of the system. (c) Spatial intensity profile showing routing of light along the edge and around a missing resonator on the edge. (d) SEM image of the system with a missing resonator, intentionally removed in the design. Adapted from~\textcite{Hafezi:2013NatPhot}}
	\label{fig:MH_imaging}
	\end{figure}
	
\textit{Experimental setup:} The lattice model illustrated in Fig.~\ref{fig:MH_lattice}(c) was implemented by~\textcite{Hafezi:2013NatPhot} using standard silicon-on-insulator technology working in the telecom range at $\lambda \simeq 1.55\,\mu$m. As illustrated in Fig.~\ref{fig:MH_imaging}, high-quality silicon ring resonators with a $Q$ factor exceeding $10^4$ were
fabricated on top of an oxide substrate using deep-UV projection photolithography. The
cross section of the waveguides, which form the site resonators and off-resonant link rings,
was designed to ensure single-mode propagation of the TE mode. The evanescent coupling between the site resonators and the link rings was controlled by the thickness of the air gaps separating the elements. Because of the surface roughness of the waveguides, a fraction of the light in the resonators scatters orthogonally to the plane and can be captured by a microscope. In addition to transmission measurement through the input and output waveguides coupled to specific site resonators, imaging this scattered light on a CCD camera gives direct information on the spatial profile of the photonic modes. The good quality of the resonators guarantees that the spin-flip-like couplings between clockwise and counterclockwise propagating modes are effectively negligible.

\textit{Topologically protected edge states:}
When the system is excited via the input waveguide with a laser field at a frequency resonant with one of the edge modes, the photons are guided through the edge and exit the system from the output waveguide. Figure~\ref{fig:MH_imaging}(b) shows the light propagation clockwise along the edge of the system. The transverse width of the edge state was about one to two resonators, as observed in both experiment and simulation. As a direct manifestation of the topological protection of edge states, when a resonator is removed from the path of an edge state, the photons route around the missing resonator and then continue their path to the output coupler without being backreflected [Fig.~\ref{fig:MH_imaging}(c)]. 

\begin{figure}
	\includegraphics[width=7cm]{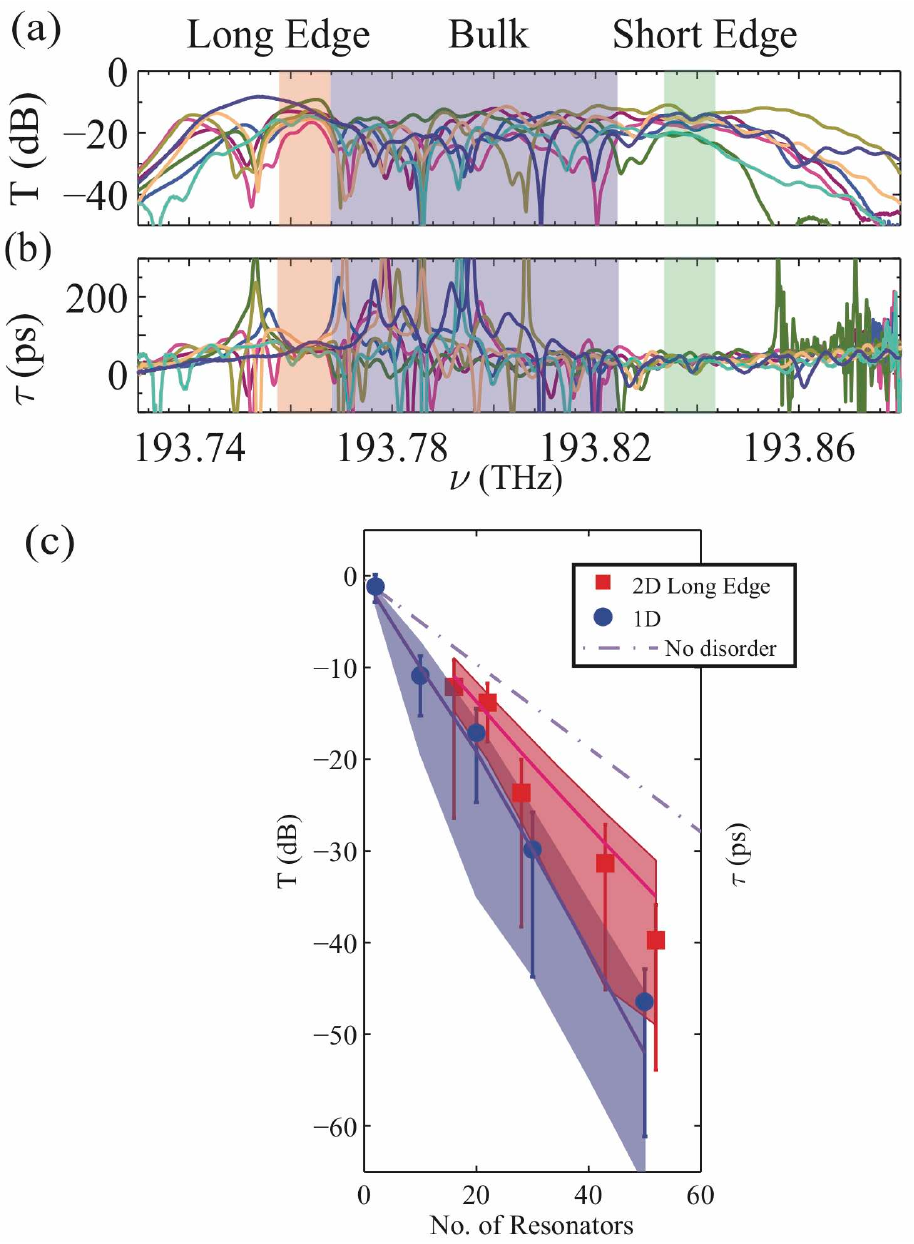}
	\caption{(a) Measured transmission and (b) delay-time spectra for eight different 8$\times$8 lattice devices. The two regions with reduced variance in both the transmission and the delay time are indicated by the red and green shading. A noisy region of propagation through bulk states is indicated by the blue shading. (c) Scaling of the transmission as a function of system size for 2D and 1D devices. Solid markers with error bars are the measured average and standard deviation (65\% confidence band) values. Solid lines with shaded areas are the simulated average and standard deviation. Adapted from~\textcite{Mittal:2014PRL}.}
	\label{fig:MH_robustness}
	\end{figure}

Beside the spatial imaging the edge states and the qualitative study of their robustness, a more quantitative experimental demonstration of their robustness was reported by~\textcite{Mittal:2014PRL} using the structure described above and sketched in Fig.~\ref{fig:MH_imaging}(a). Figure~\ref{fig:MH_robustness} shows the transmission and delay spectra interferometrically measured at the output port for eight different $8\times 8$ lattice devices. Overall, the transmission spectra of the different devices show significant fluctuations because of intrinsic fabrication variations in the dimensions of the site resonators and link rings. On closer inspection, one can see that the fluctuations in both the transmission and the  delay time are suppressed in the two regions indicated by the red and green shading in Figs.~\ref{fig:MH_robustness}(a) and~\ref{fig:MH_robustness}(b) corresponding to propagation via edge states in the counterclockwise and clockwise directions along the short and long edges, respectively. 

More insight on this physics can be seen in Fig.~\ref{fig:MH_robustness}(c), which shows the measured average transmission and its standard deviation for a sample of 95 devices. The transmission through the topological edge states of a 2D lattice and the one through the nontopological band of an analogous 1D array (a so-called coupled resonator optical waveguides) are plotted as a function of system size, i.e., the number of resonators traveled from input to output. In both cases, the transmission decays exponentially with system size. However, the decay rate is slower for the topological 2D system compared to the 1D system. The shaded regions are simulation results, using the experimentally estimated parameters, which agree with the experimental observation. In order to differentiate the decay of transmission stemming from resonator losses from the one due to disorder, both resulting in exponential attenuation, the simulated result in the absence of disorder is presented as a dashed line: while losses affect both 2D and 1D systems in the same way, transport through topological edge states of a 2D system appears to be much less disturbed by disorder than the 1D counterpart.

As a further feature of topologically protected edge states,~\textcite{Mittal:2014PRL} experimentally demonstrated how transport in lossy edge states was unambiguously distinguished from tunneling through localized bulk states by considering the statistical distribution of the delay time during propagation. Specifically, the delay distribution for edge states is
approximately Gaussian with a Gaussian width independent of system size, as typical of diffusive transport in one-dimensional systems~\cite{Cooper:2010}. On the other hand, the distribution for bulk states is asymmetric with the most probable value being less than the average, as typical of transport governed by localization and earlier observed in one-dimensional systems in the microwave domain~\cite{Chabanov:2001}.

\begin{figure}
	\includegraphics[width=.5\textwidth]{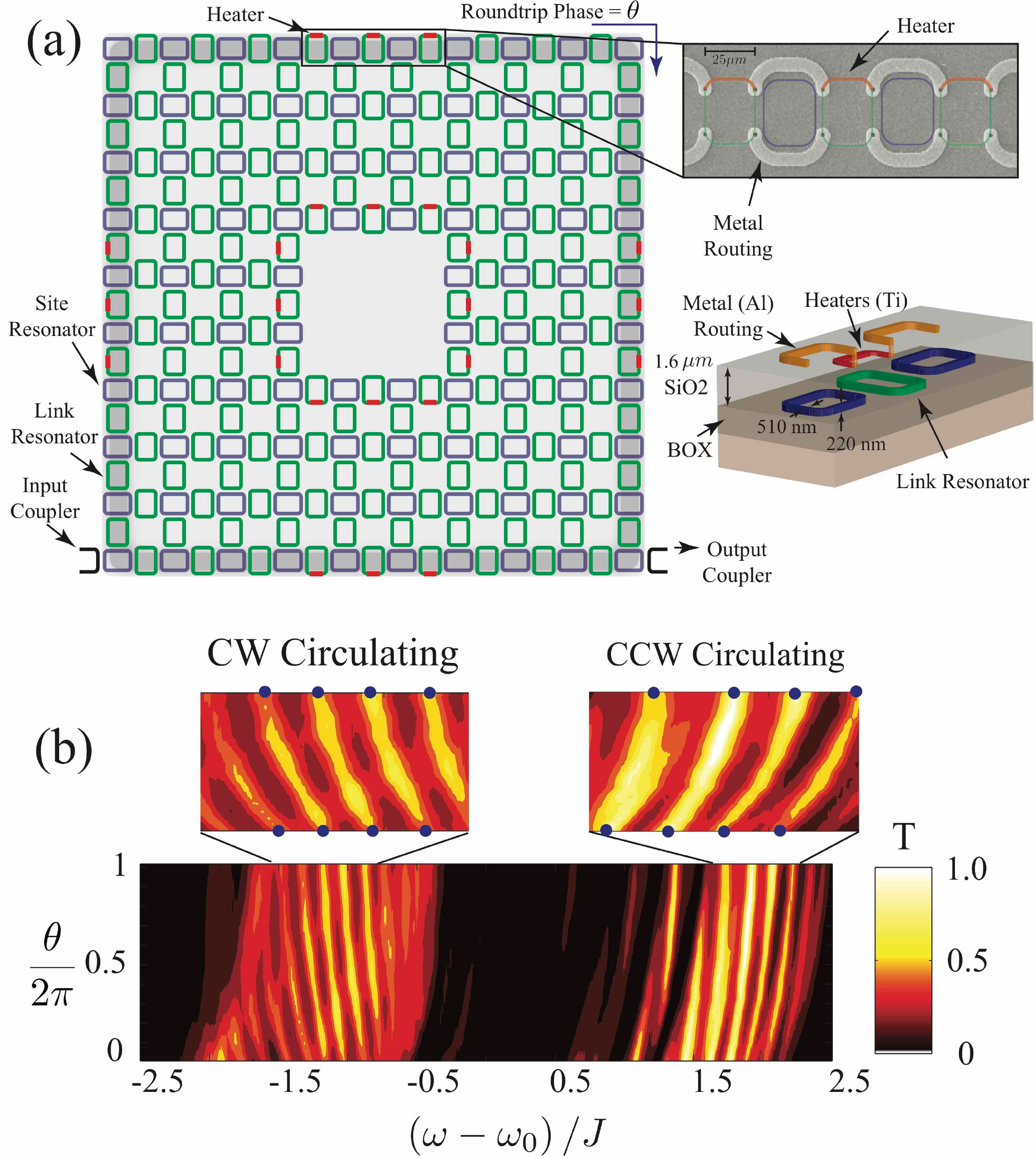}
	\caption{(a) Sketch of the experimental device (left). The edge states considered in the experiment lie on the outer edge. The tunable gauge field coupled only to the edge states is introduced by fabricating heaters on link rings situated on the lattice edges. SEM image showing heaters fabricated on top of link rings (top right) and schematic of the waveguide cross section showing the ring resonators, the metal heaters, and the metal routing layer  (bottom right). (b) Measured transmission as a function of the coupled flux $\theta$ and the incident laser frequency $\omega$. Insets: zoom-in of the edge state bands. 
	Adapted from~\textcite{Mittal:2016NatPhot}.}
	\label{fig:MH_invariant}
	\end{figure}
	
\textit{Invariant measurement:} The hallmark feature of topological physics is the presence of one-way propagating modes at the system boundary, whose chirality is a consequence of topological character of the bulk. Specifically, the bulk-boundary correspondence dictates that the number of chiral edge modes is completely determined by the bulk topological invariant, the Chern number. Following a proposal by~\textcite{Hafezi:2014PRL}, the winding number associated with a spectral flow of edge states (see Sec.~\ref{sec:pumpintro}) was experimentally studied by~\textcite{Mittal:2016NatPhot}. 

While the transverse conductance experiment usually performed in electronic systems is not applicable to photonic systems, the general spectral flow argument~\cite{Laughlin:1981, Halperin1982} is in fact applicable also to this case. To model the spectral flow of a quantum Hall edge of winding number $k=1$, one can consider a linear edge dispersion $E_p = v p$, where $E_p$ is the energy, $v$ is the group velocity, and $p$ is the momentum along the edge.  When a gauge flux ($\theta$) is coupled to the edge, the momentum is replaced by the covariant momentum, i.e., $ E_p=v\left(p-q\theta / L\right),$ where $L$ is the length of the edge and $q$ is the charge of the edge excitations. For noninteracting photons, the synthetic charge can be set to $q=1$ so that the corresponding vector potential is simply $\theta/L$. For a finite system, quantization of momentum on the edge results in
\begin{equation}
E_n= \frac{2\pi v}{L} \left(n-\frac{\theta}{2\pi}\right),
\end{equation}
where $n$ is an integer. Therefore, the insertion of $\theta=2\pi$ flux shifts $E_n\to E_{n-1}$, resulting  in a spectral flow.

To experimentally observe and measure this spectral flow, the synthetic gauge field system described above should be supplemented with an extra tunable gauge flux \cite{Mittal:2016NatPhot}. To couple a tunable gauge field to the edges, metallic heaters were fabricated above the link ring waveguides on the lattice edge, as shown in Fig.~\ref{fig:MH_invariant}(a). These heaters use the thermo-optic effect to modify the accumulated phase of light propagating through the waveguides and hence result in a gauge flux. 

Figure~\ref{fig:MH_invariant}(b) shows the measured  transmission spectrum as a function of the coupled flux $\theta$. Edge states of the outer edge and the bulk states are easily identifiable as bright and dark regions, respectively. As the coupled flux $\theta$ increases, the energy of the clockwise edge states decreases, whereas the energy of counterclockwise edge states increases. For a $2\pi$ increase in  flux, the edge state resonances move by one resonance to replace the position once held by its neighbor. This flow indicates that the measured winding number is $k = +1.0 (1)$ for the clockwise circulating edge states, and $k=-1.0 (2)$ for the counterclockwise circulating edge states.

Specific theoretical proposals to directly detect the bulk topological invariants without using the edge physics have been suggested by many authors. A short review of the main ones can be found in Sec.\ref{subsec:quantummagn}. 

%% file: RMP_IIIB2_Chicago.tex
\subsubsection{Topological rf circuits}
\label{sec:circuits}

Radio-frequency (rf) circuits are an excellent substrate for the study of topological band structures and eventually strongly interacting topological phases of matter.  Their connectivity can literally be wired in an arbitrary manner, with arbitrary numbers of connections per node and long-range connections, allowing a wide range of exotic band structures to be realized.  Furthermore, the physical size of the components is macroscopically large compared to atoms, phonons, and optical photons, enabling easy access for site- and time-resolved measurements. Moreover, topological properties in linear lattice models can be observed at room temperature and in the presence of significant dissipation, despite the fact that there are many thermal photons present. This is a vivid demonstration that topological band structures are the property of waves rather than of quantum mechanical interactions.

Two types of rf topological circuits have been experimentally demonstrated so far; the first type was discussed in Sec.\ref{sec:circuitQED} and was used to realize a topological model with time-reversal symmetry breaking and nonzero Chern numbers. The second type of circuit, which is the focus of this section, employed kHz frequency lumped-element inductors and capacitors to implement a time-reversal invariant topological system~\cite{ningyuan2015time}. In this system, it was possible to observe many signature effects of a topological insulator, as well as several features that are difficult or impossible to find in a solid-state material system. 

\begin{figure}[t]
	\centering
	\includegraphics[width=0.9\columnwidth]{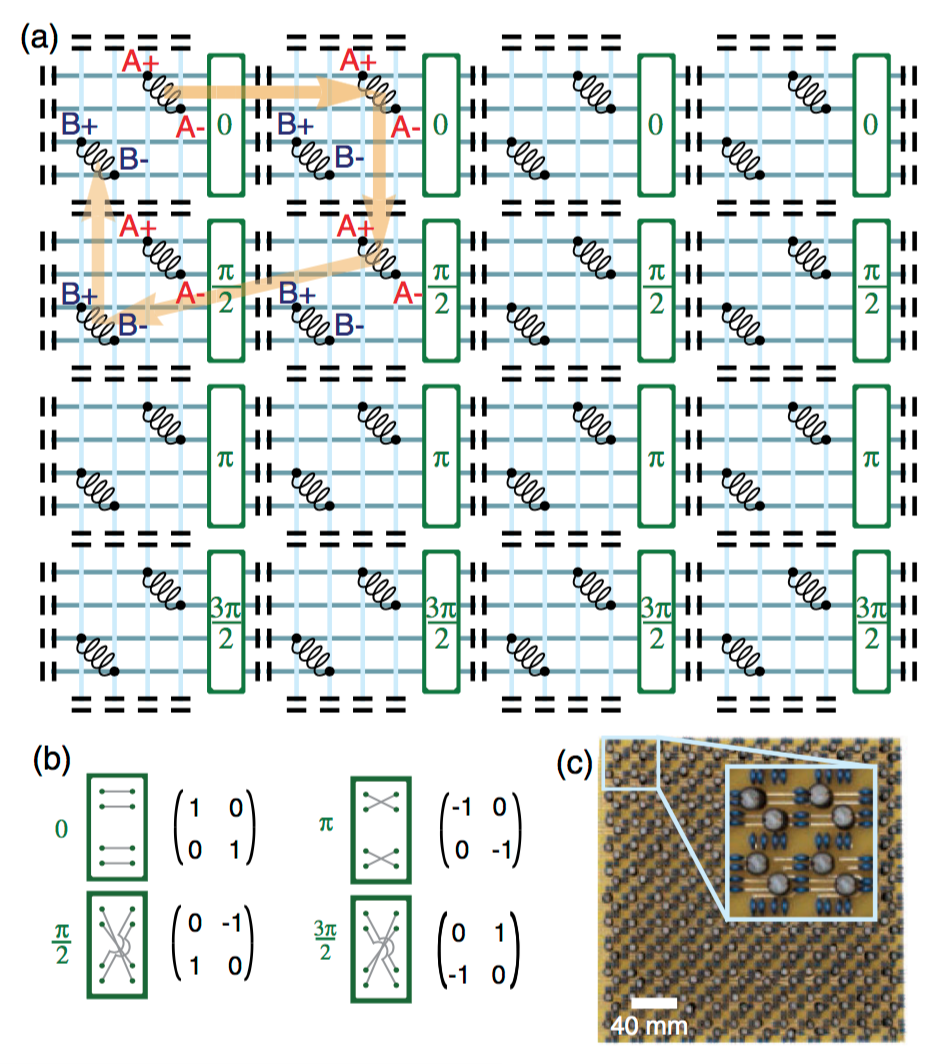}	
	\caption{(a) Schematic of the experimental circuit used to realize a time-reversal-invariant quantum spin Hall system. In this setup, the periodic structure is composed of inductors and coupling capacitors (black) that are connected by wires (light and dark blue lines). Each lattice site consists of two inductors, labeled $A$ and $B$, corresponding to right and left circularly polarized spins. As a photon hops around a single plaquette (indicated in orange) it accumulates a Berry phase of $\pi / 2$; this is engineered by braiding the capacitive couplings, as detailed in (b) and indicated here in green. (b) The synthetic spin-orbit coupling is engineered through the structure of the coupling elements between lattice sites. Each of the four tunneling phases implemented (indicated by the left column of each set of figures) is induced by a particular coupling between inductors (middle column), as described in the corresponding rotation matrix (right column); the signs of the couplings are controlled by whether the $+$ end of one inductor is coupled to the $+$ or $-$ end of the adjacent inductor. (c) The experimental circuit, in which the inductors (black cylinders) are coupled via the capacitors (blue). Inset: Zoom-in on a single plaquette that consists of four adjacent lattice sites. From~\textcite{ningyuan2015time}. 
\label{Fig:TopoCircuitRF}}
\end{figure}

In what follows, we briefly describe the experimental realization of a quantum spin Hall version of the Harper-Hofstadter model~\cite{Goldman:2010PRL} with a quarter flux per plaquette that uses only inductors and capacitors; it was subsequently pointed out~\cite{albert2015topological} that a minimal circuit model could be realized at a flux per plaquette of $1/3$. As shown in Fig.~\ref{Fig:TopoCircuitRF}, each lattice site in the experimental circuit consisted of two inductors (labeled $A$ and $B$), allowing for the representation of two pseudospin states. Tunneling between lattice sites was achieved by capacitive coupling, with the sign of the coupling reflected in which ends of the inductors are coupled to one another. Finally, the spin-orbit coupling was implemented by changing, on a site-by-site basis, whether $A$ was coupled to $A$, $-A$, $B$, or $-B$. 

In more detail, a localized excitation on a single lattice site was represented by rf fields in a $\frac{1}{\sqrt{2}}(A\pm i B)$ superposition of the two inductors on the lattice site, where $+$($-$) corresponded to spin up (down). Under these conditions, for a spin-up excitation, coupling $(A,B) \rightarrow (A,B)$ between adjacent lattice sites corresponded to implementing a tunneling phase of $0^\circ$, while connecting $(A,B) \rightarrow (-A,-B)$ corresponded to a tunneling phase of $180^\circ$, and $(A,B) \rightarrow \pm(B,-A)$ to a tunneling phase of $\pm90^\circ$. Repeating $(0^\circ,90^\circ,180^\circ,-90^\circ)$ horizontal tunnel couplers every four lattice sites, with all vertical tunnel couplers having a phase of $0^\circ$, resulted in a Harper-Hofstadter model with a quarter flux per plaquette. Spin-down excitations experienced the opposite Peierls phase and hence the opposite effective flux per plaquette.

It is important to note that this approach transcends the tight-binding regime: instead of a full $LC$ \emph{resonator} on each $A$ and $B$ site, only an inductor was included. When combined with coupling capacitors, this resulted in a Harper-Hofstadter spin-band structure exhibiting nonzero spin Chern numbers identical to those observed in the analogous tight-binding model; the lack of on-site inductors changes the bandwidths and gaps, but, perhaps surprisingly, does \emph{not} change the topology of the spin bands. An equivalent construction swaps inductors and capacitors, sending $\omega\rightarrow \omega_0^2 / \omega$ (with $\omega_0\equiv 1/\sqrt{LC}$), in much the same way that swapping inductors and capacitors  in the (topologically trivial) 1D lumped-element transmission line converts the transmission line between left and right handed. 

By forgoing on-site resonators and using on-site inductors with \emph{coupling} capacitors, this circuit operates in what can be referred to as a ``massless, left-handed'' configuration. This requires fewer inductors, which are the primary source of loss and disorder, and provides band gaps of order $\omega_0$, making the system less susceptible to disorder than tight-binding approaches with an on-site resonator, where the band gaps are all reduced by the ratio of the on-site capacitance to the coupling capacitance.

\begin{figure}[h!]
\includegraphics[width=3.4in]{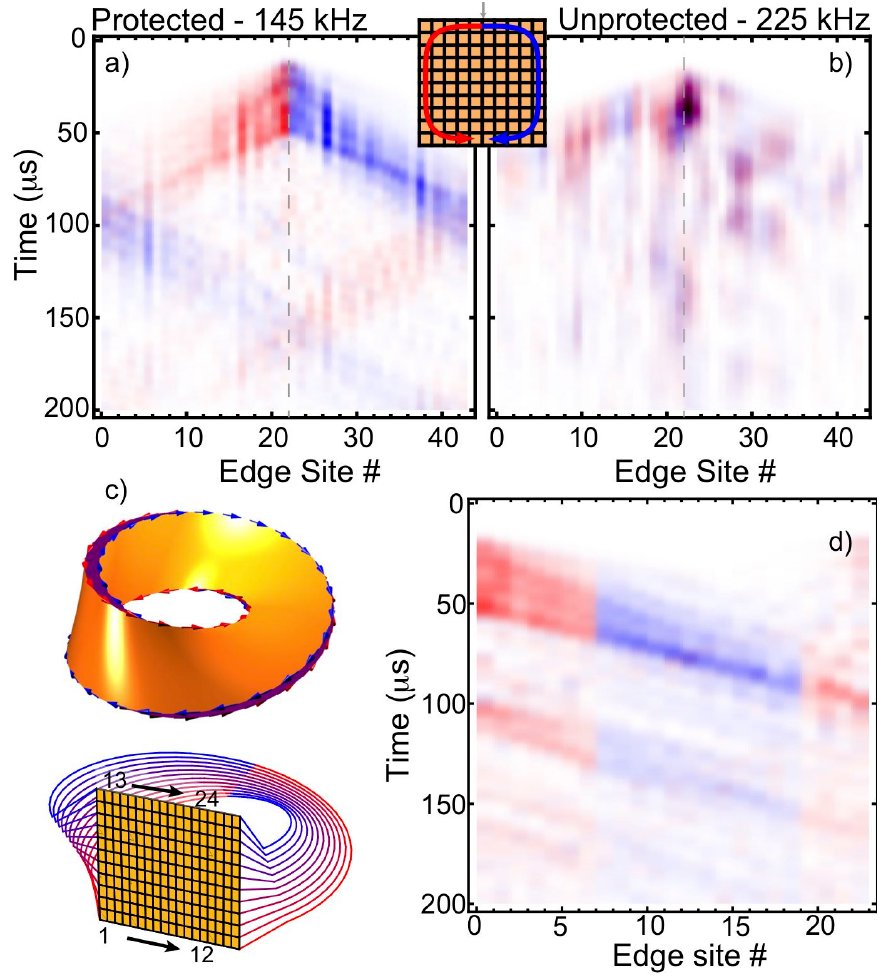}
\caption{\label{fig:timeresolved}
Time-resolved transport dynamics of the edge modes in the topological circuit, shown in Fig.~\ref{Fig:TopoCircuitRF}. (a), (b) The spin-resolved time-evolution of sites around the edge of a system, following the initial excitation of an $A$ inductor at the edge. This corresponds, in the pseudospin basis, to an initial localized excitation of $(\uparrow + \downarrow)/\sqrt{2}$ at a single lattice site. (a) When exciting a topological edge state (at 145 kHz),  the $\uparrow$ (red) and $\downarrow$ (blue) signals propagate around the edge in opposite directions, demonstrating the expected spin-momentum locking. Despite the presence of disorder, two round-trips are visible (as sites 0 and 42 are equivalent). (b) When exciting a nontopological edge state (at 225 kHz), disorder immediately leads to backscattering. Inset: The gray arrow indicates the initially-excited lattice site, with the edge site numbering convention indicated by the red and blue arrows. (c) Illustration of quantum spin Hall edge states on a M\"{o}bius strip (top panel), with arrows indicating the edge propagation direction, and colors representing the spin states. Experimentally, this was realized by imposing the connectivity of a M\"{o}bius strip on the circuit (bottom panel), with the printed circuit board shown in orange and the additional external connections colored according to the spin. (d) Spin-resolved detection of edge-transport after the excitation of $\uparrow$; the $\uparrow$ (red) and $\downarrow$ (blue) signals show the conversion from $\uparrow$ to $\downarrow$ as the excitation moves from one edge to the other, with three round-trips being visible. From~\textcite{ningyuan2015time}. }
\end{figure}

Because this approach exhibits tunneling energies comparable to the photon energy $\omega_0$, it can be operated with effective quality factors (resulting from inductor loss $R$) of order $\sim 100$ (characteristic of off-the-shelf room-temperature electronic components), with tunneling still observed over $\sim 30$ lattice sites within the photon $1/e$ lifetime; furthermore, a few percent disorder in the components, typical of off-the-shelf electronics, do not induce noticeable backscattering within the photon lifetime. Figures~\ref{fig:timeresolved}(a) and~\ref{fig:timeresolved}(b) show the dynamical evolution of a spin-mixed pulse injected on the edge, as it splits into spin components which move with opposite chiralities. 

Because this system is a \emph{circuit}, its global connectivity is easily modified. Previous work has seen the exploration of M\"{o}bius topologies [Figs.~\ref{fig:timeresolved}(c) and~\ref{fig:timeresolved}(d)], with only a single edge~\cite{ningyuan2015time}; by connecting the left and right edges with a twist and spin flip (c), a M\"{o}bius strip geometry is realized, where spinful excitations propagate to the right along the top edge, undergo a spin flip, propagate to the left along the bottom edge, and repeat. Related prospects include for creation of conical defects to explore inter-Landau-level states~\cite{biswas2016fractional}. 

A topoelectrical circuit~\cite{Lee:2018CommPhys} displaying zero-dimensional topological corner midgap states, protected by the bulk spectral gap, reflection symmetries, and a spectral symmetry has also experimentally been realized by~\textcite{Imhof:2018NatPhys}. A three-dimensional circuit displaying Weyl points was realized by~\textcite{Lu:arxiv2018}, providing access to the Weyl points and the Fermi arcs as well as to the Berry curvature distribution underlying the quantized chiral charge of the Weyl points. Finally, a microwave network was used by~\textcite{Hu:2015PRX} to measure a topological edge invariant. 

\subsubsection{Twisted optical resonators}
\label{sec:twisted}

Exploring Landau-level physics with charge-neutral particles is a persistent goal of the synthetic matter and metamaterial communities, both because learning to create ``effective magnetic fields'' for charge-neutral particles illuminates the meaning of a magnetic field, and because interacting topological matter in the continuum (fractional quantum Hall phases, for example) admits simpler theoretical description than lattice analogs (fractional Chern insulators).

Proposals to explore Landau-level physics with light rely upon coupling an optical field to a rotating atomic medium~\cite{otterbach2010effective} or phase plate~\cite{longhi2015synthetic} to inject angular momentum. The connection to synthetic magnetic fields may be understood by realizing that sending light through a rotating medium induces an image rotation~\cite{franke2011rotary}, thus turning the laboratory frame into a rotating frame. When viewed in a constantly rotating frame, massive particles experience fictitious Coriolis and centrifugal forces $2 m\vec{\Omega}\times \vec{v}$ and $m\vec{\Omega}\times\vec{\Omega}\times\vec{r}=m|\Omega|^2 \vec{r}_\perp$, respectively, where $\vec{r}$ and $\vec{v}$ are the particle position and velocity, $\vec{\Omega}$ is the angular velocity of the rotating frame, and $m$ is the particle mass. The Coriolis force has the same form as the Lorentz force $q\vec{v}\times\vec{B}$, with the identification $q \vec{B}\equiv m\vec{\Omega}$. Thus, to investigate Landau-level physics of light in the \emph{lab} frame, it is only necessary to induce a continuous image rotation on the light, and to somehow compensate for the centrifugal force that this rotation applies to the light.

\begin{figure}[t]
	\centering
	\includegraphics[width=0.9\columnwidth]{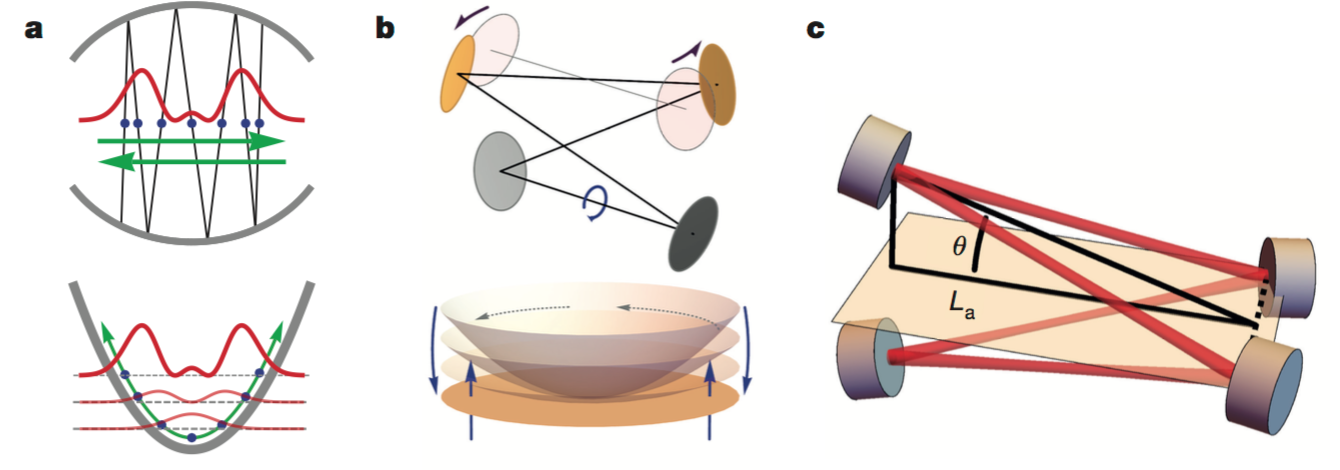}
	\caption{Engineering Landau levels for optical photons. Light in an optical resonator behaves as 2D particles in a harmonic trap. (a) When the ray trajectory is followed over many round-trips through a two-mirror resonator, its ``hit pattern'' in the central plane of the resonator corresponds to the stroboscopic evolution of a classical massive trapped particle. The projection of the eigenmodes of the resonator onto the central plane of the resonator are Hermite Gauss, akin to a quantum harmonic oscillator. (b) When a four-mirror resonator is twisted out of the plane, the ray trajectories undergo a round-trip rotation about the resonator axis, equivalent to Lorentz (Coriolis) and centrifugal forces; when the centrifugal force precisely cancels the harmonic confinement induced by the mirror curvature, the resulting Lorentz force produces Landau levels. (c) A realistic rendering of the four-mirror resonators. From~\textcite{Schine:2016Nature}. \label{Fig:TwistedCav}}
\end{figure}

The combination of image rotation and centrifugal-force cancellation was first realized by~\textcite{Schine:2016Nature}, using techniques proposed by~\textcite{sommer2016engineering} and depicted in Fig.~\ref{Fig:TwistedCav}. The image rotation is achieved by directing the light repeatedly through a nonplanar path using a four-mirror optical resonator: akin to a pair of back-to-back dove prisms, this extremely low-loss construction is able to send the light through the nonplanar path \emph{thousands} of times, inducing an image rotation on each round-trip set by the nonplanarity of the resonator.

To achieve the centrifugal-force cancellation, the mirrors that produce the image rotation are curved to repeatedly focus the light toward the resonator axis. The interplay of wave propagation, reflection off of curved surfaces, and image rotation results in a complex mode structure for such a resonator. Qualitatively, wave-propagation in the paraxial limit is equivalent to evolution of a free massive particle, and thus imbues the photon with mass; reflection off of the curved mirror surface provides a radial impulse which is proportional to the distance from the resonator axis, providing harmonic trapping, while the resonator twist (nonplanarity) induces the Lorentz and centrifugal forces.

This analogy may be sharpened through 2D $ABCD$ matrices~\cite{siegman1986lasers}, or, more intuitively, by treating the repeated passage of the optical field through the resonator as a periodically driven system~\cite{sommer2016engineering}, resulting in an effective ``Floquet'' Hamiltonian for the optical field in a particular plane of the resonator. The final result is manifolds of degenerate resonator eigen-modes with energies
\begin{align}
\frac{E_{npq}}{\hbar}=\omega_{npq}=n\frac{2\pi c}{L}+p\omega_{cyc},
\end{align}
where $c$ is the speed of light in vacuum, $L$ is the resonator round-trip length, and $\omega_{cyc}$ is the effective cyclotron frequency which determines the energy-gap between Landau levels. The $n$ quantum number determines how many wavelengths fit within the resonator longitudinally, or equivalently which Floquet copy we are referring to. Then $p$ is the Landau-level index, and $q$ is the angular momentum index of the eigenstate within the Landau level.

\textit{Stabilizing the Landau levels against astigmatism:}
In practice, such resonators are sensitive to mirror astigmatism (due to off-axis incidence of the optical field on the mirrors), which results in different harmonic confinement along the $x$ and $y$ axes, and a consequent destabilization of the Landau level. This destabilization may be understood in various ways: (1) astigmatism means that centrifugal force cannot be simultaneously canceled along both the $x$ and $y$ axes, and because particles in magnetic fields move along equipotentials, the residual confinement along a single axis guides the particles off to infinity; (2) when the astigmatism is optimally canceled, the residual confining potential takes the form $x^2-y^2=r^2 \left(e^{i 2 \theta}+e^{-i 2 \theta}\right)$ ---a potential that drives $\Delta l=\pm 2$ transitions within the degenerate manifold of states comprising the Landau level. This effect has also been observed in rotating atomic gases~\cite{Cooper:AdvPhys2008}.

\begin{figure}[t]
	\centering
	\includegraphics[width=0.9\columnwidth]{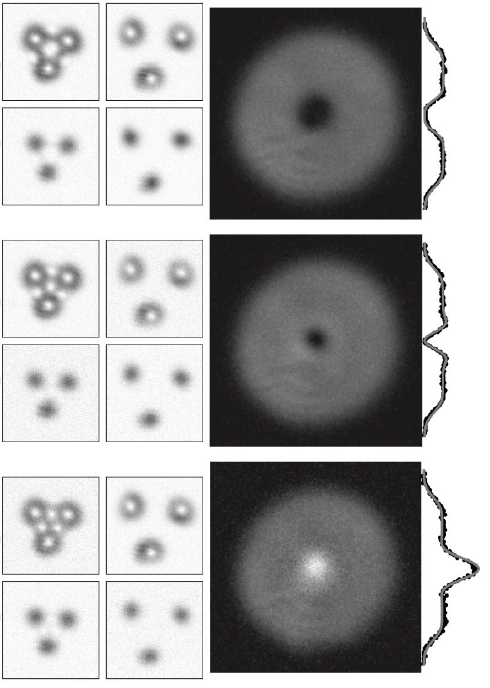}
	\caption{Translational invariance of Landau levels, and topological dynamics on a conical manifold. The astigmatism-robust Landau levels exhibit a three-fold symmetry, manifest in their spectrum by only every third angular momentum state being degenerate $q=0,3,6,\dots$, or $q=1,4,7,\dots$ or $q=2,5,8,\dots$. Panels in the top, center, and bottom rows correspond to starting with $q=0$, 1, and 2, respectively. Away from the cone tip, the system supports translational invariance, as well as the possibility of generating states with arbitrary angular momentum. Left column: Injection of the two lowest Landau levels centered away from the cone tip: within each group of images, the lower subpanels are for the lowest Landau level, while the upper ones are for the first excited one. In each case, the resonator responds with three copies of the Landau level under consideration: far enough away from the cone tip (right subpanels) the copies are independent of one another. When they are closer (left subpanels), the form of the interference of the copies near the cone tip is sensitive to the initial $q=0$, 1, and 2, corresponding to flux threading the tip itself. Right panels: The breaking of translational invariance near the cone tip reflects the Wen-Zee coupling of the lowest Landau level to the geometric curvature of the cone tip; the excess density for the unthreaded cone reflects the mean-orbital spin of the integer quantum Hall state. From~\cite{Schine:2016Nature}.  \label{Fig:MeanOrbitalSpin}}
\end{figure}

To realize a degenerate Landau level, this astigmatism issue must be addressed. One approach is to engineer a Landau-level-like scenario which does not exhibit states whose angular momenta are separated by 2$\hbar$; this is possible by further twisting the resonator, resulting in a situation where the new ``lowest'' manifold of degenerate states exhibits only every third unit of angular momentum $q=0,3,6,\dots$. This new, demonstrably stable system corresponds to a Landau level on a cone with opening angle $\alpha=\arcsin(1/3)$, which may be understood by realizing that the allowed values of $q$ support only threefold symmetric light patterns. When a photon leaves a particular third, it reenters that third from the opposite edge, but the dynamics are otherwise that of a planar Landau level. This is precisely how a charged particle behaves when constrained to the surface of a cone, in a $B$ field normal to the cone's surface. This platform has enabled the first direct measurement of the mean-orbital spin (of the lowest Landau level, see Fig.~\ref{Fig:MeanOrbitalSpin}), a topological quantum number which quantifies the coupling of density to manifold curvature through the Wen-Zee action~\cite{wen1992shift}.

Time-reversal symmetry breaking is particularly important in interacting systems, as collisions between particles in a quantum spin Hall system can induce back-scattering which would otherwise be symmetry protected. The twisted optical resonator does not break time-reversal symmetry, and so it must have a hidden spin degree of freedom which when flipped induces the photons to experience the opposite magnetic field. This turns out to arise from the order in which the photon traverses the mirrors of the running-wave twisted resonator. Forward and backward modes exhibit opposite synthetic magnetic fields. To break the degeneracy between them, it is sufficient to differentially couple to the polarization of the modes, which are opposite relative to a fixed axis (although they are the same relative to the direction of propagation). This Faraday-type coupling was recently demonstrated with an optically pumped ensemble in a weak magnetic field~\cite{ningyuan2017photons}.

%% file: RMP_IIIB4_new.tex
\subsubsection{Intrinsic spin-orbit interactions for light}
\label{sec:soc}
The fundamental properties of Maxwell's equations lead to an ``intrinsic" spin-orbit coupling for light, in contrast to ``extrinsic" spin-orbit effects, engineered by the design of photonic materials as discussed in the previous sections. Intrinsic spin-orbit coupling plays an important role on length scales comparable to the wavelength of light and so has attracted attention across photonics, nano-optics, and plasmonics~\cite{bliokh:2015NatPhot}. It can also lead to an analog of the quantum spin Hall effect for light~\cite{Bliokh:2015Science}, as we briefly introduce. 

In free space, a propagating plane wave has two spin states, given by the left- and right-handed circular polarizations, which have opposite helicities $\sigma=\pm 1$. The corresponding spin vector is ${\bf S} = \sigma {\bf k} / |{\bf k}|$ in units of $\hbar$ and is aligned with the propagation vector ${\bf k}$. This is an intrinsic coupling between the orbital and spin degrees of freedom for light, underlying a wide range of phenomena~\cite{bliokh:2015NatPhot}. 

When light is strongly confined transverse to its propagation direction, intrinsic spin-orbit coupling can lead to behavior reminiscent of the electronic quantum spin Hall effect~\cite{Kavokin:2005PRL,Leyder:NatPhys2007,Bliokh:2015Science,VanMechelen:2016Optica}. In a nutshell, the idea is the following. As a general consequence of Gauss's law in free space, $\nabla \cdot {\bf E} =0$, the transversality condition (${\bf k} \cdot {\bf E} = 0$) implies that the electric field polarization directly depends on the wave vector ${\bf k}$. Since the wave vector ${\bf k}$ in the evanescent tail of a confined optical mode has an imaginary component orthogonal to the surface, the polarization acquires a circular component corresponding to a nonvanishing transverse spin component, whose sign changes with the propagation direction~\cite{Bliokh:2012PRA, Bliokh:2014Naturecommunications}.

An analogy can thus be drawn between this so-called ``spin-direction" (or ``spin-momentum") locking of confined optical modes and the spin-momentum locked edge states of a 2D quantum spin Hall system or
surface states of a 3D topological insulator. However, a
key difference is that the optical modes are bosonic and
so not topologically protected by time-reversal symmetry~\cite{Bliokh:2015Science}. Formally, this can be seen by noting that, although a nonzero topological Chern number can be found for each uncoupled helicity state in free space, these Chern numbers are even and so the $\mathbb{Z}_2$ topological invariant, introduced in Sec.~\ref{se:QSH}, vanishes.

Nevertheless, previous experiments had observed how spin-polarized emitters give rise to a spin-controlled unidirectional excitation of surface or guided modes in a wide-range of systems, including metal surfaces~\cite{Lee:2012PRL, Rodriguez:2013science, oconnor:2014naturecommunications}, optical nanofibers~\cite{Petersen:2014science,Mitsch:2014naturecommunications,Sayrin:2015PRX}, and waveguides~\cite{lefeber:2015naturecommunications, Sollner:2015naturenanotech}. These experiments also show that, despite the lack of topological protection, the spin-direction locking and this optical analog of the quantum spin Hall effect is very robust to system details~\cite{bliokh:2015NatPhot}.  

%% file: RMP_IIIB5_Metamaterials.tex
\subsubsection{Bianisotropic metamaterials}
\label{sec:metamaterial}

In this section, we review how a quantum spin Hall effect for photons can be realized in electromagnetic metamaterials, namely, artificial composite materials containing subwavelength dielectric and/or metallic structures~\cite{liu2011metamaterials}. The key advantage of such materials is the great flexibility that they offer for engineering the effective dielectric permittivity $\varepsilon$, magnetic permeability $\mu$, and bianisotropy or magnetoelectric coupling $\chi$ that appear in the electric and magnetic response to long-wavelength fields,
\begin{eqnarray}
\left(\begin{array}{c}
{\bf D} \\
{\bf B}
\end{array} \right)
= \left(\begin{array}{cc} 
\varepsilon & i  \chi \\
-i  \chi^T & \mu \end{array}  \right)
\left(\begin{array}{c}
{\bf E} \\
{\bf H}
\end{array} \right).
\end{eqnarray}
As first proposed by~\textcite{Khanikaev:2013NatMat}, a properly designed metamaterial structure can exhibit a quantum spin Hall effect for light, as experimentally evidenced by the presence of topologically robust spin-dependent edge states, analogous to the helical edge states of an electronic topological insulator.

Following the initial theoretical proposal, the first step in this direction is to construct photonic modes which mimic the electron spin-$1/2$ eigenstates. This can be achieved, for example, through an enforced matching of $\varepsilon=\mu$ in a metamaterial, ensuring that, in the absence of bianisotropy, the TE- and TM-polarized modes propagate along a 2D metamaterial slab with the same wavenumbers, restoring the $({\bf E},{\bf H})\rightarrow (-{\bf H},{\bf E})$ duality of the electromagnetic field in free space. For a given wavevector $\bf{k}$, linear combinations $\psi^\pm_{\bf{k}}$  of these degenerate TE and TM modes can be constructed with the special property that $\psi^+_{\bf{k}}$ can be transformed into $\psi^-_{\bf{-k}}$ by a suitable symmetry operation $\hat{D}$, similarly to how an electron's spin is flipped by time-reversal symmetry. Since the square of the transformation satisfies the usual $\hat{D}^2=-1$ condition of electronic time reversal, these states can be identified as a pair of photonic pseudospin-$1/2$ states and show symmetry-protected Kramers doublets for time-reversal symmetric momenta.

The second step required is then to engineer an appropriate bianisotropy or magneto-electric coupling $\chi$~\cite{serdyukov2001electromagnetics} that generates a strong spin-orbit coupling between the pseudospin states, mimicking that found in a topological insulator. To first order, the effect of a finite $\chi$ can in fact be recast in reciprocal space as an explicit coupling between photon momentum and polarization ${\bf D} = \epsilon {\bf E} + (i c/\omega) \chi \mu^{-1} {\bf k} \times {\bf E}$ and ${\bf B} = \mu {\bf H} + (i c/\omega) \chi^T \epsilon^{-1} {\bf k} \times {\bf H}$. While the bianisotropy $\chi$ of materials found in nature, such as optically active media containing chiral molecules, is typically small, a very large value can be obtained in metamaterial structures containing, for example, split-ring resonators~\cite{pendry:1999IEEE, Marques:2002PRB,Shelby:2001Sci,Li:2009PRE}.

This approach was exploited by~\textcite{Khanikaev:2013NatMat} in the design of $\varepsilon=\mu$ matched metamaterial rods arranged into a ``metacrystal" in the form of a hexagonal lattice, which had a significant value of the bianisotropy $\chi_{xy}=-\chi_{yx}$ terms. In such a hexagonal geometry, the photonic bands host doubly degenerate Dirac points, which are gapped out by the bianisotropy $\chi$. Around the gapped Dirac points, the effective low-energy model can then be mapped to the Kane-Mele model for the quantum spin Hall effect (see Sec.~\ref{se:QSH}), where the topological states are protected by the engineered symmetry $\hat{D}$ of the electromagnetic field in the metamaterial. 

Experimentally, the key signature of this photonic quantum spin Hall effect is the appearance of robust polarization-dependent edge states. Such ``spin"-polarized transport has been observed for microwave photons by~\textcite{Chen:2014NatComm} for a uniaxial hexagonal metacrystal of nonresonant meta-atoms sandwiched between two metallic plates, where the effective bianisotropy arises from field inhomogenities. The topological robustness of such edge modes was also further demonstrated by~\textcite{slobozhanyuk:2016SciRep}, who used near-field imaging in a square lattice of bianisotropic meta-molecules to directly show the lack of backscattering around sharp corners. 

\begin{figure}
\resizebox{0.5 \textwidth}{!}{\includegraphics*{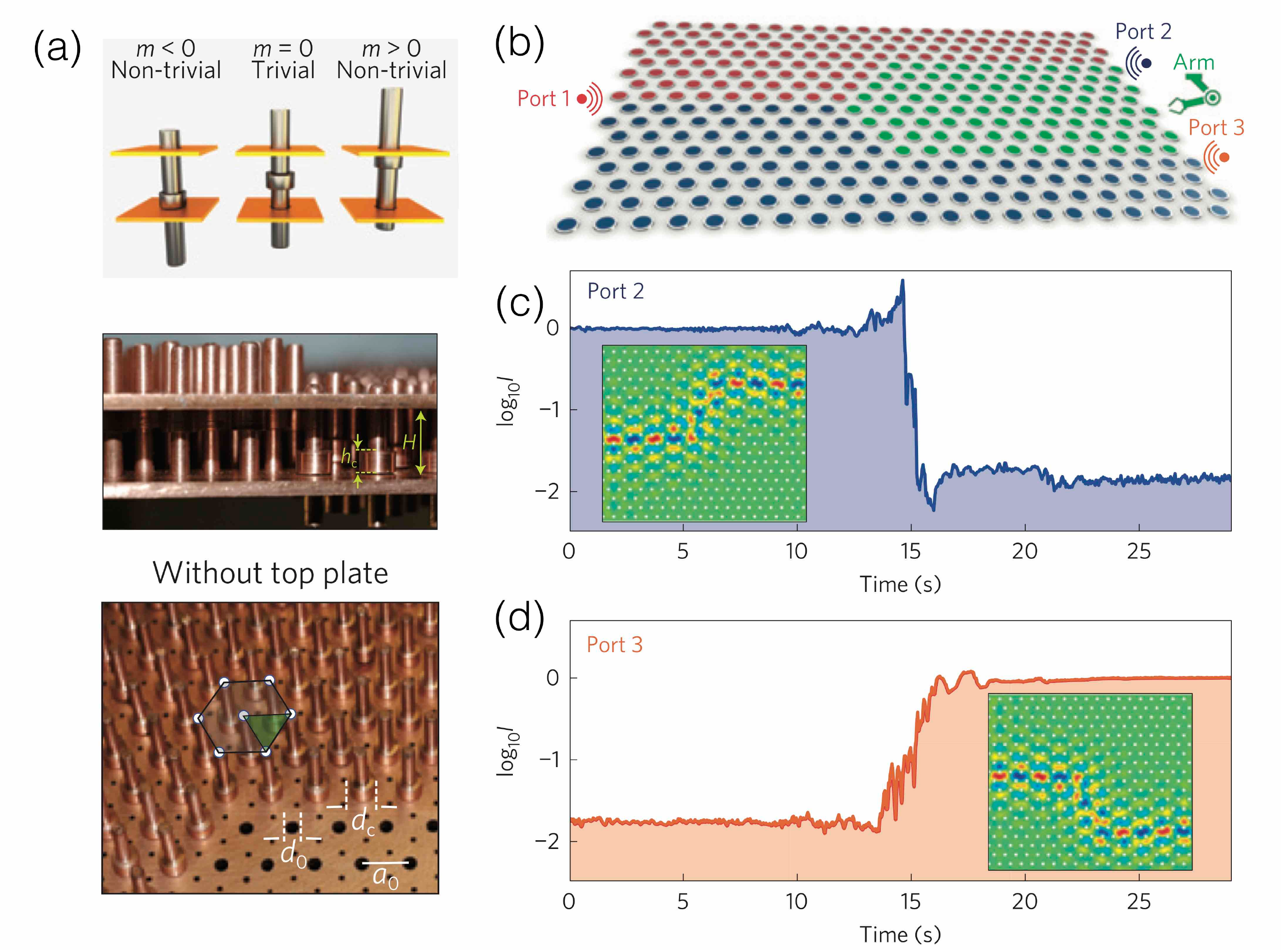}}
\caption{(a) (Top) Displacing a metallic collar relative to the two metal plates of a waveguide introduces an effective bianisotropy, coupling the electric and magnetic fields. Shifting a collar from the ``down" to ``up" position reverses the sign of this term and changes the sign of the ``mass" term $m$ that opens a gap at the Dirac cone, once these rods are arranged into a hexagonal or triangular lattice. (Middle and bottom panels) Photograph of the experiment, in which collars in a triangular array can be moved up and down to create arbitrary and reconfigurable topological domain walls. (b) Configuration of the topological switch where in the red (blue) regions, the metallic collars are always up (down), and where in the green region, the collars are dynamically moved from down to up to change the location of the domain wall. (c), (d) The time-resolved switching of transmission via the edge states through (c) Port 2 and (d) Port 3, with insets illustrating the power flow. Adapted from~\textcite{cheng:2016NatMat}.} \label{meta_copper}
\end{figure}

The following works have shown that topological photonic states displaying the photonic analog of the quantum spin Hall effect can also be realized in an even more simple structure, as recently proposed~\cite{Ma:2015PRL} and experimentally realized~\cite{cheng:2016NatMat,lai:2016SciRep,Xiao2016prb}. In this ``metawaveguide" setup, a parallel-plate metal waveguide is filled with a hexagonal or triangular lattice of metallic cylinders, which are connected at the top and bottom to the metal plates [see Fig.~\ref{meta_copper}(a)]. The geometry of the cylinders and plates is carefully optimized such that the spectrum contains an overlapping degenerate pair of Dirac cones for the TE and TM modes, allowing for the introduction of photonic pseudospin states. The bianisotropy is then introduced either by allowing a finite air gap between a cylinder and one of the metal plates or by adding an asymmetrically placed collar to the cylinder~\cite{Ma:2015PRL}. Moving such a metallic collar relative to the metal plates can be understood as changing the sign of the ``mass" term that opens up the gaps at the Dirac cones [see Eq.~(\ref{eq:jackiwrebbi})] allowing for the straightforward and reconfigurable engineering of arbitrary topological domain walls. As proposed by~\textcite{cheng:2016NatMat}, this could form the basis of a topological switch, in which the movement of metallic collars is used to switch the propagation path from one port to another port [see Figs.~\ref{meta_copper}(b) -~\ref{meta_copper}(d)]. 

Recently, these ideas have also been extended to all-dielectric bianisotropic metamaterials~\cite{Slobozhanyuk:2016NatPhot,slobozhanyuk:2017arxiv}, which may offer advantages for reducing Ohmic losses, scaling up to optical frequencies, and increasing compatibility with on-chip integration. In these systems, arrays of dielectric disks are carefully designed such that the electric and magnetic dipole modes are degenerate and electromagnetic duality is restored~\cite{Slobozhanyuk:2016NatPhot}. The bianisotropy term is then introduced by adding a raised circular notch on one of the flat faces of the disk;  when the disks are arranged into a hexagonal or triangular metacrystal, this gaps out the Dirac points, leading to a photonic topological insulator. This has been experimentally realized in the microwave regime for a 2D array of ceramic disks~\cite{slobozhanyuk:2017arxiv}. Such arrays could also be layered to make a 3D system, analogous to a ``weak" 3D topological insulator~\cite{slobozhanyuk:2017arxiv}, as discussed further in Sec.~\ref{sec:3dgapped}.

%% file: RMP_IIIB_Wu.tex
\subsubsection{Photonic structures with crystalline symmetries}
\label{sec:wuhu}

As seen, the coupling of electric and magnetic fields in bianisotropic materials requires subwavelength structures that are asymmetric in the direction perpendicular to the plane of the topological metacrystal. Because of the small size of the features required for their fabrication, their implementation remains challenging for visible and near-infrared wavelengths. A different method to implement an analog of the quantum spin Hall effect for photons in two dimensions was proposed by~\textcite{Wu:2015PRL}. The configuration is based on subwavelength dielectric structures with inversion symmetry with respect to the 2D plane, which is in principle easier to implement experimentally.

\begin{figure}[t]
		\includegraphics[width=\columnwidth]{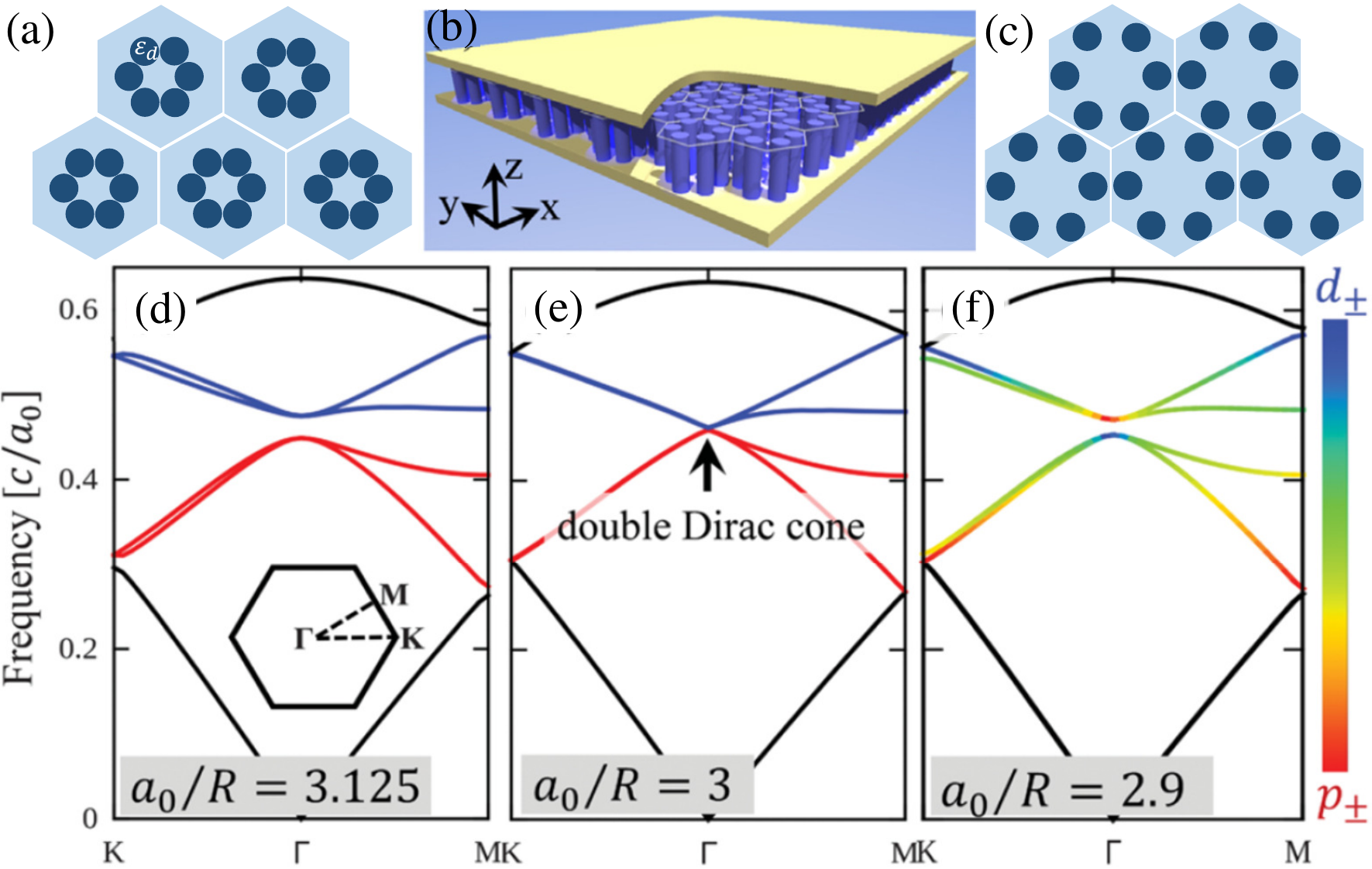}
		\caption{(a)-(c) Scheme of the lattice of hexagonal clusters of rods with dielectric constant $\epsilon_d$. (d)-(f) Energy-momentum dispersion corresponding to (d) contracted, (e) centered-honeycomb, and (f) expanded clusters, showing in the latter case the opening of a topological gap. (b), (d)-(f) From~\onlinecite{Wu:2015PRL}.}
		\label{fig_WuHu}
	\end{figure}

The idea is to consider the lowest TM mode of a slab of cylindrical subwavelength dielectric rods surrounded by air and confined between two metallic plates [see Fig.~\ref{fig_WuHu}(b)]. The rods are arranged in a honeycomb geometry resulting in a photonic Dirac dispersion at $K$ and $K'$ points. If instead of taking the usual two-sites unit cell of the honeycomb lattice, we consider a hexagonal cluster of rods as the unit cell, the additional band folding translates the Dirac points to the center of the Brillouin zone, resulting in a doubly degenerate Dirac crossing at the $\Gamma$ point [see Fig.~\ref{fig_WuHu}(e)]. Solving Maxwell's equations for the lowest TM mode of the slab shows that the in-plane magnetic field distributions of the doubly degenerate lower Dirac bands present a $p_x$, $p_y$ character, while the upper Dirac bands contain states with $d_{xy}$, $d_{x^2-y^2}$ symmetry. Their symmetric and antisymmetric combinations $p_{\pm}$ and $d_{\pm}$ constitute a pseudospin basis [Fig.~\ref{fig_WuHu}(e), red and blue lines, respectively].

When shifting the rods toward the center of the hexagonal clusters, the dispersion shows the opening of a trivial gap at the Dirac point, with the upper and lower bands preserving their symmetry [Figs.~\ref{fig_WuHu}(a) and~\ref{fig_WuHu}(d)]. If the rods are pushed away from the center of each cluster a gap opens again, but this time it is the result of a band inversion at the Dirac point, with the new states having $d$/$p$ character for the lower and upper bands [Fig.~\ref{fig_WuHu}(c) and~\ref{fig_WuHu}(f)]~\cite{Wu:2015PRL,Xu:OptExp2016}. The bands now possess nonzero pseudospin Chern numbers, an analog to the case of an electronic $\mathbb{Z}_2$ topological insulator. If we now consider a ribbon of the topologically nontrivial photonic crystal, on a given edge, two bands of edge states with opposite group velocities traverse the gap around the $\Gamma$ point. The wavefunctions of each edge band are associated with a specific pseudospin. A similar situation takes place at the interface between two photonic crystals with trivial or nontrivial arrangements.

We can relate these topological features to a photonic \textit{pseudo} quantum spin Hall effect. Indeed, while Maxwell's equation respect bosonic time-reversal-symmetry operations ($\mathcal{T}^2=+1$), in the presence of the $C_{6 \nu}$ symmetry, we can construct an antiunitary operator $\mathcal{\tilde{T}}^{2}=(\mathcal{K}\mathcal{T})^2=-1$ in which $\mathcal{K}$ respects the parity of $p$ and $d$ eigenstates with respect to $\pi/2$ and $\pi/4$ rotations, respectively~\cite{Wu:2015PRL}. This pseudo-time-reversal symmetry provides a Kramers doublet at the expense of keeping the $C_{6 \nu}$ geometry of the lattice. In particular, the breaking of the crystalline order at the edge of the ribbon couples the two pseudospins and results in the opening of a very small gap at the crossing of the edge state bands at the time-reversal symmetric point at the $\Gamma$ point.

The simplicity of this proposal has triggered its implementation in a wide variety of systems, ranging from acoustics~\cite{Simon:NJP2017} to phononics~\cite{He:NatPhys2016, Xia:PRB2017,Brendel:arx2017}. In optics it was implemented for microwaves~\cite{Yves:NatComm2017,Yang:2018PRL}, infrared photons in a photonic crystal slab~\cite{Barik:NJP2016, Barik:Science2018, Anderson:OE2017, Zhu:PRB2018}, and metasurfaces~\cite{Gorlach:NatCom2018}. The experiments of \onlinecite{Barik:Science2018} showed that the two pseudospins characterizing each interface state band correspond to opposite circular polarizations of the confined photons. A semiconductor quantum dot embedded in the photonic crystal was used to couple single photons to the topological interface states. In a different configuration, using coupled waveguides, it was shown that the lattice configuration with the nontrivial gap can hold localized zero-dimensional defect states~\cite{Noh:2018NatPhot}. All these features open interesting perspectives in views of engineering novel whispering-gallery-mode geometries~\cite{Siroki:PRB2017} or exploring quantum chiral optics~\cite{Lodahl:Nat2017}. 

%% file: RMP_IIIB6_QSHOther.tex
\subsubsection{Other proposals}
\label{sec:QSHEOther}

Two alternative routes to simulating an artificial gauge field for a coupled optical cavity array were proposed by~\textcite{Umucalilar:2011PRA}, by harnessing the polarization degree of freedom for light. In the first scheme, a nontrivial tunneling phase between lattice sites is generated by coupling together the orbital and polarization degrees of freedom, through, for example, a suitable embedding of either birefrigent slabs or optically active layers within an array of distributed Bragg reflector microcavities. In the second scheme, photons move in a single planar microcavity, where a suitable periodic lateral patterning of the cavity generates a confining lattice potential but also a position-dependent polarization mixing. As a result, the polarization of a photon then traces a closed loop in polarization space when it evanescently tunnels between lattice wells, and gains a geometrical Berry phase. This effect corresponds to a generalization for evanescent waves of how propagating photons can be imprinted with geometrical Pancharatnam~\cite{Pancharatnam:1956PIAS} or Berry phases~\cite{Tomita:1986PRL,Chiao:1988PRL}. This second scheme, in particular, offers the potential for reaching the strongly interacting photon regime (see Sec.~\ref{subsec:strong}), if the lateral patterning is scaled down to the micrometer scale, where the tighter confinement of light within the lattice wells will lead to greatly enhanced photon-photon interactions. 

%% file: RMP_IIIC_Anomalous.tex
\subsection{Anomalous Floquet topological insulators }\label{section_anomalous_Floquet}
The third main class of 2D topological photonic systems is the so-called anomalous Floquet topological insulators, where unusual topological properties can emerge due to periodically driving the system in time. In this section, we set the focus on the two recent experiments of~\textcite{Maczewsky:2017NatCom,Mukherjee:2017NatCom}, which realized such topological properties experimentally by designing suitable propagating waveguide arrays for photons. 
 
	\begin{figure*}[t]
		\includegraphics[width=\textwidth]{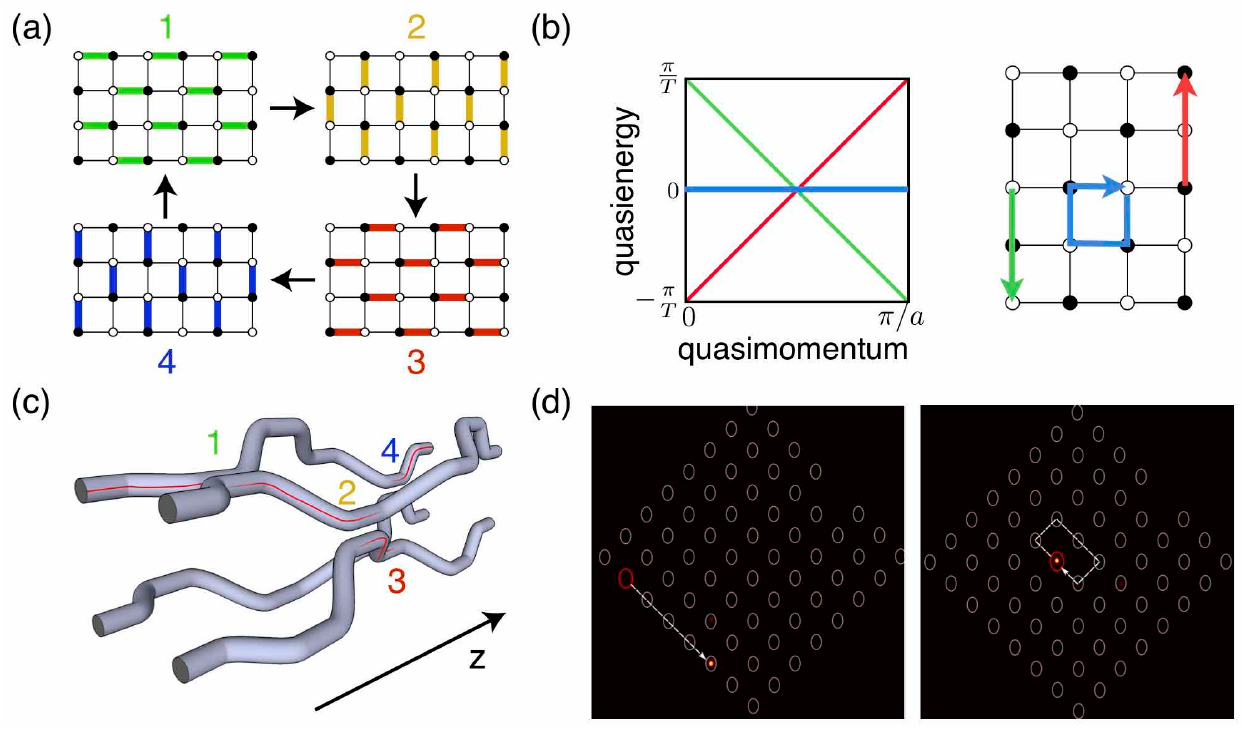}
		\caption{(a) Sketch of the Rudner \emph{et al.} toy model for anomalous Floquet topological phases~\cite{Rudner:2013PRX}, where hopping on a square lattice is activated in a cyclic and time-periodic manner. (b) Energy bands and sketch of the corresponding states in the ideal case where the activated-hopping amplitude $J$ is exactly set on resonance $J\!=\!\hbar \Omega$, where $\Omega$ is the drive frequency. The dispersionless bulk band (blue) corresponds to cyclotron-localized states and is associated with a zero Chern number; dispersive branches (green and red) correspond to propagating edge states (skipped-cyclotron orbits). These edge states are topologically protected by a winding number that takes the full-time-evolution (including the micromotion) into account~\cite{Nathan:2015}. (c) The optical waveguide implementation of~\textcite{Mukherjee:2017NatCom}, where the sequential activation of neighboring coupling was finely engineered. (d) Experimental evidence for the simultaneous existence of robust propagating edge states and (quasi) localized bulk states. (a), (b) Adapted from~\textcite{Rudner:2013PRX}. (c), (d) From~\textcite{Mukherjee:2017NatCom}.}
		\label{fig_anomalous}
	\end{figure*} 

As already discussed in Sec.~\ref{section_floquet}, topological band structures can be engineered through Floquet engineering, namely, by subjecting a system to a time-periodic modulation. In two dimensions, an emblematic example is provided by the ``Floquet Chern insulator," which can be realized by subjecting a honeycomb (graphenelike) lattice to a circular shaking, either produced by a mechanical modulation of the lattice~\cite{Rechtsman:2013Nature,Jotzu:2014Nat} or by irradiation~\cite{Oka:2009PRB,Lindner:2011NatPhys}. In the high-frequency regime of the drive, i.e., when $\hbar \Omega$ is much larger than any other energy scale in the system, the dynamics is well captured by an effective Hamiltonian, whose band structure (the Floquet spectrum) exhibits Bloch bands with nonzero Chern numbers (Sec.~\ref{section_floquet}). In this high-frequency regime, the topological characterization of the driven system reduces to applying the standard topological band theory to the effective Hamiltonian (i.e., to the Floquet band structure);~ in particular, the usual bulk-edge correspondence guarantees that a Floquet Chern insulator exhibits chiral edge modes at its boundaries. This approach was pioneered in photonics~\cite{Rechtsman:2013Nature}, where a 2D honeycomb-shaped array of optical waveguides was circularly modulated along a third spatial direction playing the role of ``time."

This simple topological characterization breaks down when the bandwidth of a Floquet Chern insulator approaches $\hbar \Omega$ (i.e., when the system no longer operates in the high-frequency regime $\Omega\!\rightarrow\!\infty$). This can be understood in two ways. First, let us recall that the Floquet spectrum, which is associated with the Floquet operator $\hat{\mathcal{U}} (T)\!=\!e^{-(i/\hbar) T \hat H_{\text{eff}}}$, is only uniquely defined within a Floquet-Brillouin zone $[-\hbar \Omega/2; \hbar \Omega/2]$. Consequently, when the effective bandwidth approaches $\hbar \Omega$, gap-closing events are possible at the Floquet-Brillouin zone's boundaries~\cite{Kitagawa:2010PRB}. This can produce a cancellation of the Chern numbers associated with the Floquet bands, however, and this is a crucial observation, topological chiral edge states can still be present in the inner part of the spectrum. Such a coexistence of chiral edge modes with \emph{seemingly trivial} Bloch bands, which is in apparent contradiction with the bulk-edge correspondence, is referred to as ``anomalous Floquet topological phases"~\cite{Rudner:2013PRX}, as opposed to the standard Floquet Chern insulators previously discussed. A second, and more fundamental, reason for this breakdown of the usual topological characterization stems from the fact that the micromotion plays a crucial role as soon as the period of the drive $T$ no longer sets the shortest time scale in the problem:~in the ``low-frequency" regime, the dynamics (including the topological nature of the system) cannot possibly be ruled by $\hat H_{\text{eff}}$ only; see Sec.~\ref{section_floquet}. All together, this indicates that the anomalous Floquet topological phase cannot be accurately characterized by the Chern number related to the effective Hamiltonian, but rather by a distinct topological invariant: a ``winding number," which fully takes the micromotion into account~\cite{Kitagawa:2010PRB,Rudner:2013PRX,Nathan:2015,Carpentier:2015}.

An instructive toy model leading to a dramatic instance of an anomalous Floquet topological phase was introduced by~\textcite{Rudner:2013PRX}. In a certain parameter regime, the system simultaneously exhibits a \emph{single completely flat} Bloch band (with a trivial Chern number) together with a topologically protected chiral edge mode. As represented in Fig.~\ref{fig_anomalous}(a), the driven system consists of a 2D square lattice, whose allowed hopping events are activated in a sequential and $T$-periodic manner:~the time-evolution operator describes a four-step quantum walk. One should note that the circular nature of the sequence imprints a chirality to the system, similarly to the circular shaking leading to Floquet Chern insulators in graphenelike lattices~\cite{Rechtsman:2013Nature,Jotzu:2014Nat,Oka:2009PRB,Lindner:2011NatPhys}; such a chirality is crucial for the emergence of chiral edge modes in the system. The effective (Floquet) band structure of this model is shown in Fig.~\ref{fig_anomalous}(b) for the extreme case where the activated-hopping amplitude $J$ is exactly set on resonance $J\!=\!\hbar \Omega$. The resulting unique (two-degenerate) Bloch band is completely flat, which reflects the fact that the stroboscopic motion, i.e., the motion over each period of the drive $T$, is necessarily restricted to closed loops in any region of the bulk. However, as in the quantum Hall effect, such ``orbits" reduce to skipped orbits at the boundaries, hence resulting in a chiral edge mode [see Fig.~\ref{fig_anomalous}(b)]. This simple picture is confirmed by the topological analysis of the model, which associates a nontrivial winding number to the Floquet Bloch band~\cite{Rudner:2013PRX}. 

Recently,~two independent experimental teams~\cite{Maczewsky:2017NatCom,Mukherjee:2017NatCom} reported on the realization of this intriguing ``Rudner-toy" model, using an array of laser-inscribed coupled waveguides using the technique described by~\textcite{Szameit:2010JPB}. Both teams considered a 2D array of optical waveguides, defined in the $x$-$y$ plane, with a propagation direction $z$ playing the role of time. As illustrated in Fig.~\ref{fig_anomalous}(a), the model relies on a sequential activation of neighboring coupling within this 2D square lattice.~\textcite{Maczewsky:2017NatCom,Mukherjee:2017NatCom} achieved this goal by spatially modulating the waveguides along the $z$ direction, in such a way that different pairs of waveguides are locally moved together (to activate the coupling) and then apart (to switch it off); see Fig.~\ref{fig_anomalous}(c) for a sketch of this protocol. These independent teams implemented two distinct configurations of the model, the activated couplings being homogeneous in~\textcite{Maczewsky:2017NatCom}, while these were chosen to be inhomogeneous by~\textcite{Mukherjee:2017NatCom}. However, both experiments reached the aforementioned ``anomalous" regime of the topological phase diagram, where chiral edge modes are uniquely determined by the nontrivial winding number~\cite{Kitagawa:2010PRB,Rudner:2013PRX,Nathan:2015}.~\textcite{Mukherjee:2017NatCom} demonstrated the realization of the anomalous Floquet topological phase through a thoughtful study of the activated-coupling strength (which uniquely identified the realized anomalous phase within the topological phase diagram) combined with direct observations of the chiral edge states propagation and bulk localization [see Fig.~\ref{fig_anomalous}(d)]. This analysis was further validated through numerical simulations based on the theoretical model.~\textcite{Maczewsky:2017NatCom} also signaled the anomalous phase by demonstrating the dispersionless nature of the bulk (i.e., the existence of a flat band) and the chiral nature of the edge mode. This latter work also analyzed a topological transition from the anomalous topological phase to a trivial phase (characterized by the absence of edge mode) by designing lattices with decreasing coupling strengths. These two experiments demonstrated the high tunability offered by laser-inscribed photonic crystals in view of designing intriguing toy models and simulating exotic phases of matter.

Various other schemes have been proposed to reach the anomalous regime of Floquet topological systems, which can be applied to a variety of physical platforms~\cite{Kitagawa:2010PRB,Kitagawa:2010PRA,Reichl:2014PRA,Leykam:2016PRLanomalous,Quelle:2017arxiv}. In this broader context, robust localized states, associated with nontrivial winding numbers~\cite{Kitagawa:2010PRA}, were first demonstrated in a photonic setup realizing a 1D quantum walk~\cite{Kitagawa:2012NatComm}; a similar setup was recently explored in view of directly extracting winding numbers through Zak-phase measurements performed in the bulk~\cite{Cardano:2017NatComm}. Besides, the winding number of 2D anomalous Floquet topological insulators was also measured in a microwave network, using a dimensional-reduction (topological pump) approach~\cite{Hu:2015PRX}. Finally, the existence of anomalous Floquet edge modes was also shown in a designer surface plasmon structure operating in the microwave regime~\cite{Gao:2016NatCom}.

%% file: RMP_IIID_Gapless.tex
\subsection{Topology in gapless photonic systems }\label{subs:gapless}

\begin{figure*}[t]
		\includegraphics[width=\textwidth]{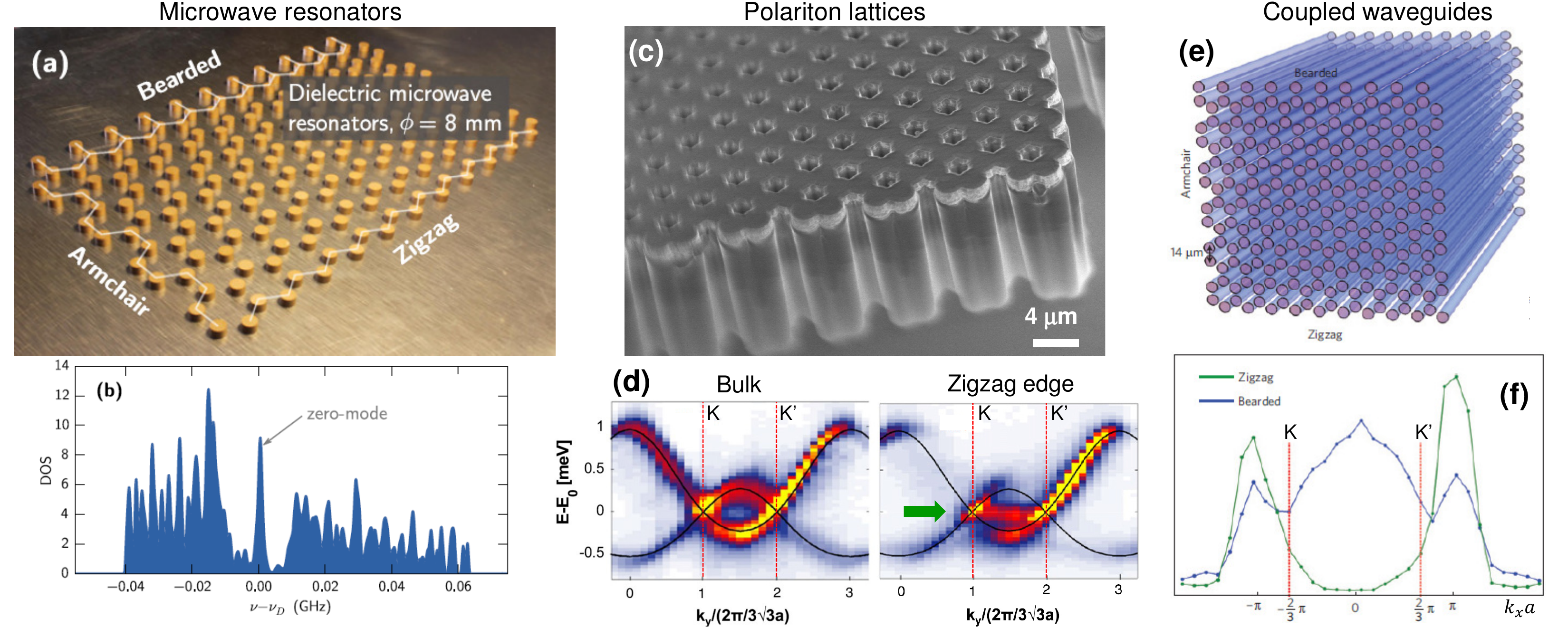}
		\caption{(a) Image of a honeycomb lattice of microwave resonators with armchair, bearded, and zigzag edges. (b) Measured density of states as a function of microwave frequency $\nu$. The peak at the Dirac energy $\nu_d$ indicates the presence of zero-energy states. From~\onlinecite{Bellec:2014NJP}. (c) Scanning electron microscope image of a honeycomb lattice of coupled polariton micropillars. (d) Photoluminescence spectrum measured at the center of the lattice showing Dirac crossings (left), and at the zigzag edge (right), showing a flat band of edge states (pointed by the arrow). From~\onlinecite{Milicevic:20152DMat}. (e) Representation of a lattice of coupled waveguides and (f) measured momentum-space distribution of the transmission through the zigzag and bearded edges. Red lines show the position of the Dirac points. From~\onlinecite{Plotnik:2014NatMat}.}
		\label{fig_honeycomb_photons}
	\end{figure*}

Another important class of topological systems is gapless photonic lattices with Dirac points. The primary example of this kind is the honeycomb lattice of coupled photonic resonators or waveguides. The Hamiltonian describing the dynamics of photons in these lattices is equivalent to that of $p_z$ electrons in graphene, giving rise to two bands with linear crossings, at the Dirac points, as illustrated in Fig.~\ref{fig_honeycomb_photons}(d) for a lattice of polariton resonators. In the absence of external fields, spin-orbit coupling, or temporal modulation the system remains ungapped. Nevertheless, this type of lattice presents features that are topological in the sense that they can be related to certain topological invariants or geometrical properties of the system. These features include topological edge states, topological phase transitions, and the emergence of synthetic gauge fields when suitably deforming the lattice.

\begin{figure}[t]
		\includegraphics[width=\columnwidth]{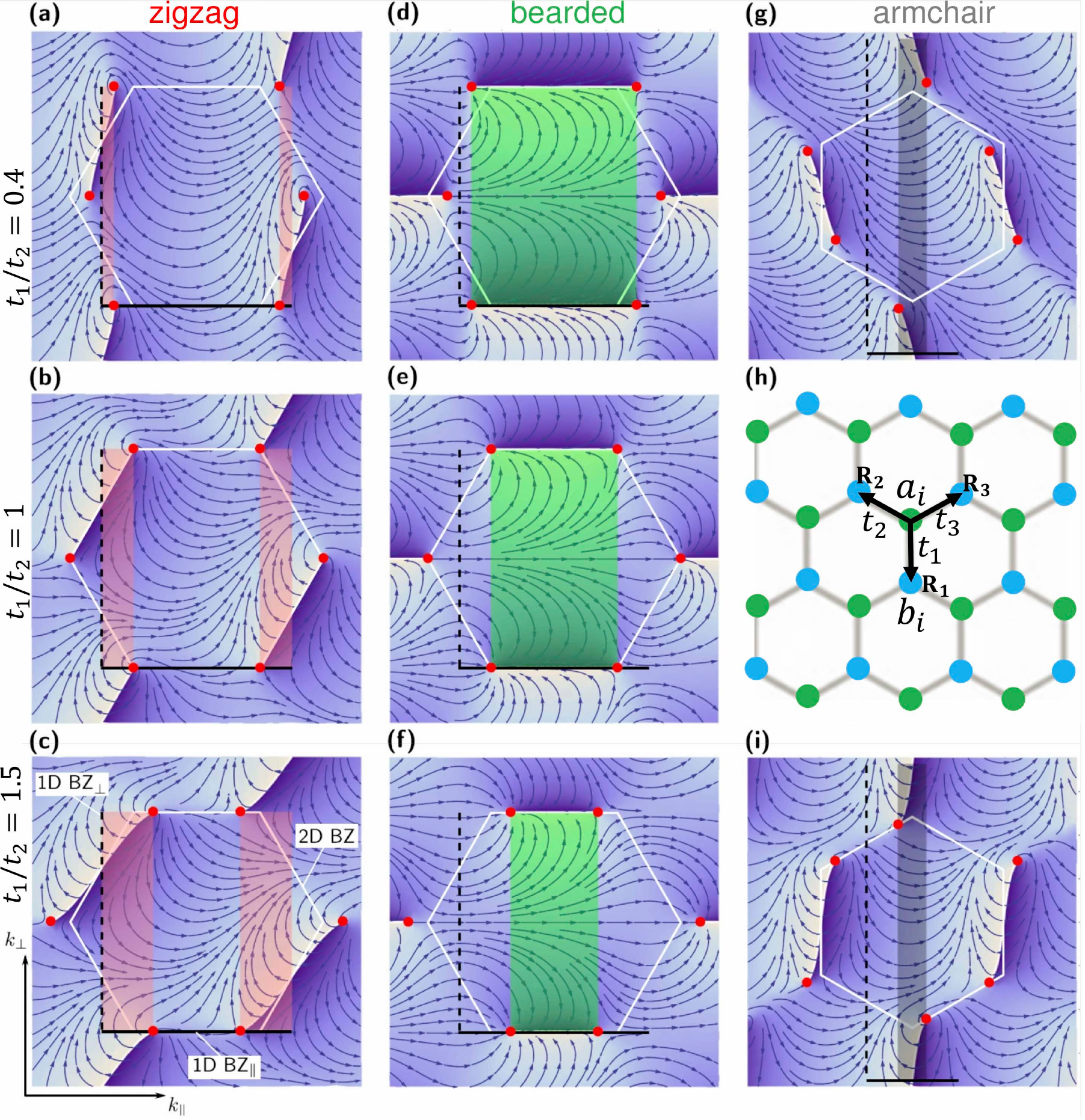}
		\caption{Stream plots of $\theta \left(k_{\perp}, k_{\parallel} \right)$ in momentum space for zigzag, bearded, and armchair edges, and for different values of the homogeneous strain $t_{1}/t_{2}$ ($t_{2}=t_{3}$). The colored areas show the regions of $k_{\parallel}$ in which $\mathcal{W}=1$, corresponding to the presence of edge states. Black solid and dashed lines represent the 1D Brillouin zones in $k_{\parallel}$ and $k_{\perp}$, respectively. From~\onlinecite{Bellec:2014NJP}. Red points show the position of the Dirac points. (h) Scheme of the honeycomb lattice and nearest-neighbor hoppings.}
		\label{fig_edge_strain}
	\end{figure}

To analyze these topological properties, let us consider the hopping of the lowest photonic mode of each individual resonator to its nearest neighbor. In the tight-binding approximation with nearest-neighbor hoppings, the Hamiltonian in momentum space is chiral and takes the form of Eq.~\ref{IIIA_kspaceform} 
in the $A$, $B$ sublattice basis depicted in Fig.~\ref{fig_edge_strain}(h) by green and blue dots. In this case, $Q(\mathbf{k})$ in Eq.~\ref{IIIA_kspaceform} takes the form $Q(\mathbf{k})=-t_1 e^{i\mathbf{k}\cdot \mathbf{R_1}}-t_2 e^{i\mathbf{k}\cdot \mathbf{R_2}} -t_3 e^{i\mathbf{k}\cdot \mathbf{R_3}} \equiv \left|Q (\mathbf{k}) \right| e^{-i\theta(\mathbf{k})}$, where $t_{1,2,3}$ are the nearest-neighbor hopping amplitudes and $\mathbf{R_{1,2,3}}$ are the vectors connecting a site to its three nearest neighbors [see Fig.~\ref{fig_edge_strain}(h)]. For equal hoppings ($t_{1}=t_{2}=t_{3}\equiv t$) the two bands of eigenvalues of the honeycomb Hamiltonian are $\epsilon (\mathbf{k}) = \pm t \left|Q (\mathbf{k}) \right|$~\cite{Wallace:1947PR}, resulting in Dirac-like crossings at the $K$, $K^\prime$ points in the first Brillouin zone. The eigenfunctions are $\left|u_{k,\pm}\right\rangle=\left(1/\sqrt{2}\right)\left(e^{-i \theta(\mathbf{k})}, \pm 1 \right)^{\dagger}$.

The honeycomb Hamiltonian has been implemented in a number of photonic systems including microwave resonators~\cite{Bittner:2010PRB, Bellec:2013PRB}, photorefractive crystals~\cite{Peleg:2007PRL, Song:2015NatComm, Sun:2018PRL}, atomic vapor cells~\cite{zhang:2018arxiv}, coupled microlasers~\cite{Nixon:2013PRL}, polariton lattices~\cite{Kusudo:2013PRB, Jacqmin:2014PRL}, and propagating waveguides~\cite{Rechtsman:2013NatPhot, Plotnik:2014NatMat}. Some examples are shown in Fig.~\ref{fig_honeycomb_photons}.

The first noticeable feature of the eigenstates is their pseudospin structure, with two components that reflect the underlying $A$, $B$ interlaced triangular sublattices. The pseudospin structure results in particular scattering properties close to the Dirac cones: using a honeycomb lattice imprinted in a photorefractive medium it was shown that the conical diffraction characteristic of Dirac crossings~\cite{Peleg:2007PRL} 
presents an orbital angular momentum $l=\pm 1$, depending on whether sublattice $A$ or $B$ is excited~\cite{Song:2015NatComm}.

A second relevant feature of the eigenfunctions is their nontrivial Berry phase: if the eigenfunctions are transported adiabatically on a closed loop in momentum space around one of the Dirac points, the eigenfunctions change sign~\cite{CastroNeto:2009RMP}. In other words, they get a Berry phase of $\pm \pi$, the sign being opposite for the $K$ and $K^\prime$ points. This effect is present even when a gap is opened at the Dirac cones, which can be induced by introducing an on-site energy difference $\Delta$ between the $A$ and $B$ sublattices. In this case, a nonzero Berry curvature extends around the Dirac points. Therefore, if a wave packet is created close to one of the Dirac points and subject to acceleration, the Berry curvature results in an anomalous velocity whose sign depends on the Dirac point around which the wave packet is created~\cite{Ozawa:2014PRL}. An efficient way of accelerating the photon wavepacket is to design a lattice whose resonators continuously increase in size from one lattice site to the next: the decreasing photon confinement results in an on-site energy gradient. If the photon lifetime is long enough, the force can induce magnetic Bloch oscillations with a displacement perpendicular to the gradient and a direction determined by the Dirac point around which oscillations take place~\cite{Cominotti:EPL2013}.

Photonic simulators can also be used to explore other properties of propagating wave packets in a honeycomb lattice. For instance, the chiral symmetry of the honeycomb Hamiltonian is preserved in the presence of a potential step, resulting in phenomena such as Veselago lensing~\cite{Cheianov:2007Science}, the Goos-H\"anchen effect~\cite{Grosche:2016PRA}, or Klein tunneling~\cite{Dreisow:2012EPL, Solnyshkov:2016prb}. Dissipation present, for example, in polariton lattices, does not significantly affect these phenomena~\cite{Ozawa:2017PRA}.
Dirac cones can also be used to tailor the dispersion of photonic structures, even producing ``epsilon-near-zero" materials~\cite{Huang:2011NatMat,Moitra:2013NatPhot}.

\noindent \textit{Edge states in Dirac systems:} One of the characteristics of the honeycomb lattice is the existence of zero-energy edge states in finite-size ribbons~\cite{Klein:1994ChemPhysLett, Nakada:1996PRB}. These edge states are topological in the sense that they are related to the winding properties of the bulk Hamiltonian~\cite{Ryu:2002PRL}. To understand this bulk-edge relation, let us consider a graphene ribbon of finite size in the direction perpendicular to the edge ($\perp$), and infinite in the parallel direction ($\parallel$). By Fourier transforming the real-space Hamiltonian along this axis, we can reduce it to an effective 1D Hamiltonian for each value of $k_{\parallel}$ \cite{CastroNeto:2009RMP,Delplace:2011PRB}. This effective 1D chiral Hamiltonian has the same form as the SSH Hamiltonian discussed in Sec.~\ref{sec:otherdimensions}, characterized by the complex function $Q (k_{\perp}, k_{\parallel})=\left| Q \left(k_{\perp}, k_{\parallel} \right) \right|e^{i \theta \left(k_{\perp}, k_{\parallel} \right)}$. We can then apply the topological arguments discussed in Sec.~\ref{sec:otherdimensions}. The number of pairs of zero-energy edge states, appearing at the two edges of the ribbon, is thus determined by the winding of $\theta \left(k_{\perp}, k_{\parallel} \right)$, Eq.~\ref{IIAwinding}, along the $k_{\perp}$ direction over the first Brillouin zone~\cite{Ryu:2002PRL, Mong:2011PRB, Delplace:2011PRB}:

\begin{align}
\mathcal{W}(k_\parallel) = \frac{1}{2\pi} \int_{BZ} d k_\perp{\frac{d \theta\left(k_{\perp}, k_{\parallel} \right) }{d k_\perp}} .
\label{eq:winding_honeycomb}
\end{align}

\noindent The information about the specific type of edge is contained in the choice of unit cell dimer and unit vectors when writing down the honeycomb Hamiltonian, such that they allow for the full reconstruction of the lattice (including the edges). Therefore, the information on the edges is reflected in the specific form of $\theta \left(k_{\perp}, k_{\parallel} \right)$ when writing down Eq.~\ref{IIIA_kspaceform} \cite{Delplace:2011PRB}. Figures~\ref{fig_edge_strain}(b) and~\ref{fig_edge_strain}(e) show $\theta \left(k_{\perp}, k_{\parallel} \right)$ for zigzag and bearded edges, respectively. The colored areas indicate the values of $k_\parallel$ for which $\mathcal{W}=1$, corresponding to the existence of edge states.

The direct access to the wave functions in photonic lattices has been employed to study the local properties of these edge states. Their existence has been evidenced experimentally in lattices of coupled waveguides~\cite{Plotnik:2014NatMat}, microwave resonators~\cite{Bittner:2012PRB, Bellec:2014NJP}, and polaritons~\cite{Milicevic:20152DMat}, showing that edge states for zigzag and bearded terminations connect Dirac cones in complementary regions in momentum space [Figs.~\ref{fig_edge_strain}(b) and~\ref{fig_edge_strain}(e)], while armchair terminations do not possess any edge state ($\mathcal{W}=0$ for any $k_{\parallel}$).

The topological arguments used to predict the existence of edge states in a honeycomb lattice can be extended to other Hamiltonians, for instance, when more than one mode per site is involved. As long as the system possesses the chiral symmetry, they can be written in the form of Hamiltonian~\ref{IIIA_kspaceform}. With more than one mode per site, $Q (\mathbf{k})$ takes the form of a $n \times n$ matrix whose determinant can be written as $\det Q(k) \equiv |\det Q(k)|e^{i\theta (k)}$. The existence of pairs of zero-energy edge states is again given by the winding of $\theta (\mathbf{k})$ along the momentum direction perpendicular to the edge~\cite{Kane:2014NatPhys}. An example of this kind of chiral Hamiltonian is the $p$-orbital version of graphene, in which orbitals with $p_{x}$, $p_{y}$ geometry are considered at each lattice site~\cite{Wu2007:PRL}. This $4\times 4$ orbital Hamiltonian has been implemented in a polariton honeycomb lattice when considering the doubly degenerate first excited states of each coupled micropillar. The spectrum consists of four bands with Dirac crossings, and the presence of zero-energy edge states for different kinds of terminations is well accounted for by the analysis of the winding of $\theta (\mathbf{k})$ just presented~\cite{Milicevic:2017PRL}.

Similar arguments can be applied to Dirac Hamiltonians without chiral symmetry, for instance with next-nearest-neighbor hopping or with a staggered potential in the $A$, $B$ sublattices, both effects giving rise to nonzero diagonal terms in Eq.~\ref{IIIA_kspaceform}. The presence of edge states can also be determined via winding arguments, but their energy is not necessarily zero~\cite{Mong:2011PRB}.\\

\noindent \textit{Valley Hall edge states:} Propagating edge states with weak topological protection can be engineered in lattices with appropriately staggered potentials. The staggered potential between the $A$, $B$ sublattices $\Delta$ breaks the inversion symmetry and opens a gap at the Dirac points. By integrating the Berry curvature around each Dirac point, we can define a \textit{valley} Chern number whose sign is opposite for $K$ and $K^\prime$ points (the total Chern number of the band still being zero). If the staggered potential is changed to $-\Delta$, the gap is still open but the signs of the \textit{valley} Chern numbers switch between $K$ and $K^\prime$ points. When joining two honeycomb semi-infinite ribbons subject to opposite staggered potentials $\Delta$ and $-\Delta$, it has been shown that interface states appear in two bands that traverse the gap~\cite{Zhang:PRL2011,Ma:NJP2016,Weinstein:2DMat2016,Chen:arx2016,Goldman:2016PRA,Bleu:PRB2017,Qiu:OE2017}. Indeed, if the $K$ and $K^\prime$ valleys of the two lattices have opposite valley Chern numbers, the gap needs to close locally at those points at the interface between the two lattices~\cite{Ma:NJP2016}. This situation is restricted to very specific conditions, for instance, it applies to zigzag interfaces. For this configuration, the propagation of a wave packet in an interface state with a given valley polarization is protected against any perturbation that does not mix the two valleys, i.e., bends of the interface of $120^\circ$, which preserve the zigzag character of the interface~\cite{Chen:arx2016,Wu:arx2017,Ma:NJP2016}.
Interface states between photonic crystals which exhibit valley Hall and quantum spin Hall effects have also been experimentally demonstrated~\cite{Kang:2018NatComm}.

For armchair interfaces, the breaking of translational symmetry in the direction perpendicular to the interface mixes the $K$ and $K^\prime$ points of each lattice, and the valley Chern number is not well defined. Similarly, if the gap becomes too large ($\left| \Delta \right| \gtrsim t $) in zigzag interfaces, the Berry curvatures associated with both valleys overlap. The valley Chern numbers are not well defined anymore and a gap opens in the edge state bands~\cite{Noh:arx2017}.

Despite the fact that valley Hall edge states are not robust against arbitrary spatial disorder at the interface, they can be used to route photons in photonic structures~\cite{Wu:arx2017,Noh:arx2017} and to design delay cavities in Si photonics technologies~\cite{Ma:NJP2016}. A route to recover topological protection is to use chiral nonlinear excitations (i.e., vortices) as proposed recently for a polariton fluid in a honeycomb lattice~\cite{Bleu:arxiv2017}.

\noindent \textit{Effect of strain:} The application of strain to an ungapped honeycomb lattice ($\Delta=0$) strongly modifies its spectrum and eigenfunctions and, consequently, its topological properties. We consider two main classes of strain: (i) homogeneous strain, in which hopping is different along different spatial directions ($t_{1}\neq t_{2}\neq t_{3}$), and (ii) inhomogeneous strain, in which hopping takes different values at different positions [$t_{i}(\mathbf{r})$]. The first case was theoretically studied by~\textcite{Montambaux:2009PRB}. They predicted a topological phase transition occurring when one of the three nearest-neighbor hopping amplitudes ($t_{1}$) is twice as large as the other two ($t_{2}=t_{3}$). At the transition point ($t_{1}=2 t_{2}$), the two nonequivalent Dirac cones merge and disappear resulting in the opening of a full gap. This transition was first experimentally observed in cold atoms~\cite{Tarruell:2012Nature}, but it is in photonic lattices where its effect on the existence of edge states has been studied (see Fig.~\ref{fig_edge_strain}). Experiments in lattices of microwave resonators and coupled waveguides~\cite{Rechtsman:2013PRL, Bellec:2014NJP} have shown that above the transition point ($t_{1} > 2 t_{2}$), ribbons with zigzag terminations contain a flat energy band of edge states covering the whole momentum space, while for bearded boundaries, edge states disappear. For armchair terminations, edge states appear for any value of the unidirectional strain as long as the anisotropy axis is not parallel to the edge, as in Figs.~\ref{fig_edge_strain}(g) and~\ref{fig_edge_strain}(i)~\cite{Bellec:2013PRL, Bellec:2014NJP}. The existence of edge states in homogeneously strained honeycomb lattices can also be predicted from topological arguments. The Hamiltonian including this kind of strain still possesses the chiral symmetry, and the number of zero-energy edge states is governed by the winding of $\theta (\mathbf{k})$.
Interestingly, homogeneous strain has been employed to synthesize critically tilted Dirac cones in the orbital $p$ bands of lattices of coupled micropillars~\cite{Milicevic:2018arXiv}.

	\begin{figure}[b]
		\includegraphics[width=\columnwidth]{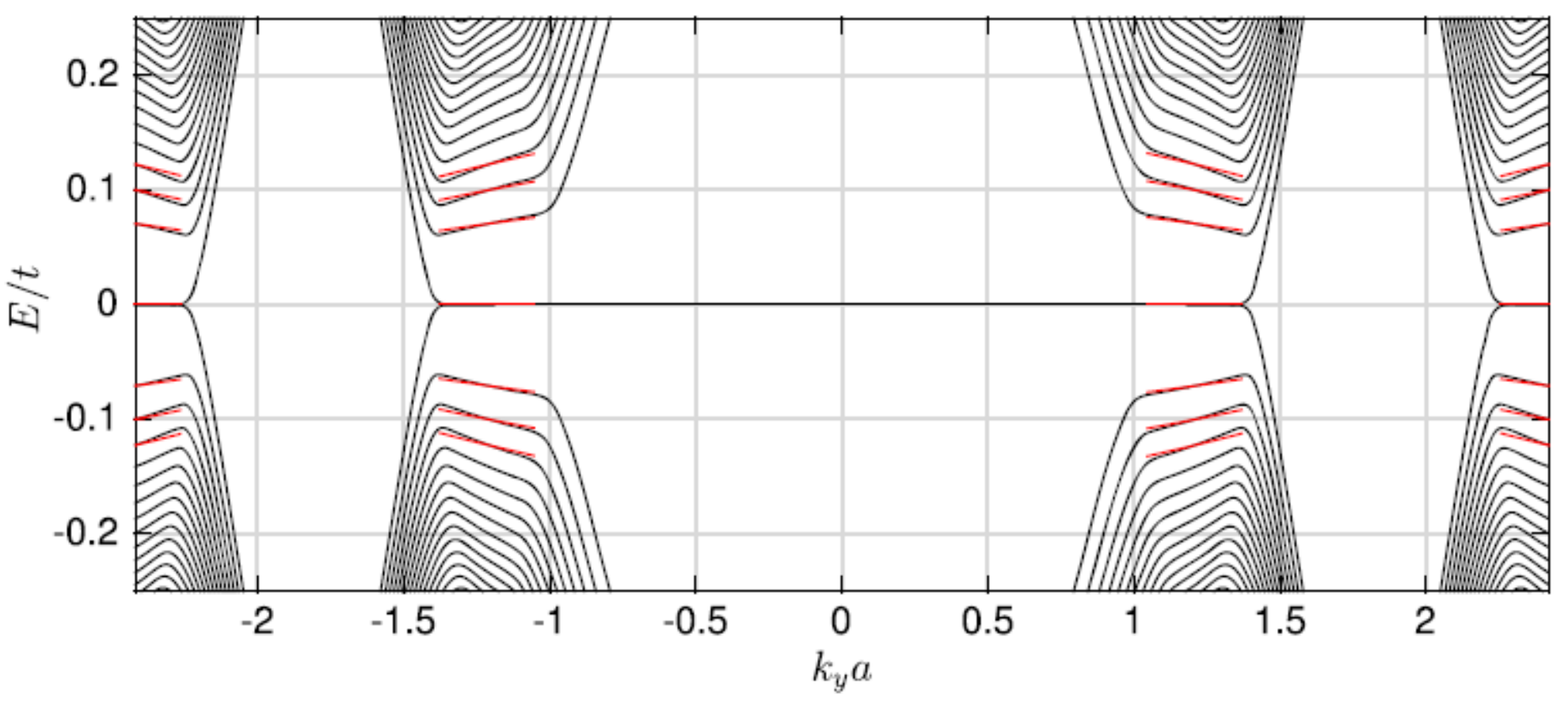}
		\caption{Tight-binding calculation of the level structure of a uniaxially strained photonic graphene ribbon as a function of $k_y$, perpendicular to the strain direction. The hopping strength along the narrow dimension of the ribbon increases linearly from one edge to the other, resulting in a gauge field for photons. The Dirac points, located at $k_y=1.2/a$ with $a$ being the nearest-neighbor distance, split into Landau levels (red lines). The lowest one, $n_L=0$ at $E=0$, is flat, while the others have a nonzero group velocity. From~\onlinecite{Salerno:20152DMat}.}
		\label{fig_salerno}
	\end{figure}

The second kind of strain concerns the continuous variation of the hopping over the lattice. 
Originally studied in the context of electronic graphene, this kind of inhomogeneous strain has been shown to induce a gauge field~\cite{Kane:1997PRL}. If the strain takes the specific trigonal shape $(u_{r}, u_{\theta}) =\beta r^2 (\sin 3\theta, \cos 3\theta$), where $u_{r}, u_{\theta}$ are the real-space displacements of the positions of the carbon atoms in polar coordinates, the modified hopping induces a gauge field for electrons at the Dirac cones corresponding to a homogeneous pseudomagnetic field perpendicular to the graphene sheet~\cite{Guinea:2010NatPhys}. In other words, the Hamiltonian describing the time evolution of electrons in the strained lattice can be cast in the form of an unstrained honeycomb Hamiltonian subject to a homogeneous pseudomagnetic field perpendicular to the graphene plane. Note, however, that this field does not break time-reversal symmetry: it has opposite signs at the two Dirac points. Therefore, edge states appearing in the gaps between consecutive Landau levels are associated with propagation on both directions along the edge~\cite{Gopalakrishnan:2012PRB, Salerno:2017PRB} and backscattering-protected transport is not expected.

Translated to the photonic realm, this configuration provides an efficient way of inducing a pseudomagnetic field acting on photons as if they were charged particles. This precise idea was implemented by~\onlinecite{Rechtsman:2013NatPhot} in a lattice of coupled waveguides by continuously varying their separation (i.e., the nearest-neighbor photon hopping) following the above-mentioned strain configuration. The value of the valley dependent pseudomagnetic field acting on the propagating photons can take effective values of several thousand tesla, much higher than strengths of real magnetic field currently realizable in the laboratory. The main consequence is the emergence of Landau levels in the vicinity of the Dirac points~\cite{CastroNeto:2009RMP}. Analogously to the effect of a real magnetic field in graphene, the energy of the photonic Landau levels $n_{L}$ scales as $\pm \sqrt{n_{L}}$. This was observed in numerical tight-binding calculations and experimentally via the localization of a wave packet on the edge of a strained lattice~\cite{Rechtsman:2013NatPhot}, attesting to both the presence of flat Landau bands and states localized at the edge emerging from the gauge field.

The trigonal strain discussed so far is not the only way of inducing a homogeneous magnetic field.~\textcite{Salerno:20152DMat} showed that linear uniaxial strain along one of the crystallographic directions results in a homogeneous pseudomagnetic field similar to that emerging from trigonal strain. Figure~\ref{fig_salerno} shows the emergence of Landau levels associated with pseudomagnetic fields of opposite sign at the $K$, $K^\prime$ points. The most noticeable feature in this tight-binding calculation is that, except for $n_{L}=0$, the Landau levels are not completely flat. The reason is that the position-dependent hopping results in a local Dirac velocity that varies along the lattice~\cite{DeJuan:2012PRL}.

The engineering of Landau levels in photonic structures is particularly interesting in the quest for confined lasing geometries. The possibility of introducing flat gapped bands in the bulk of a photonic lattice, with a high density of states, could be used to fabricate low threshold on-chip lasers. Moreover, the combination of gauge fields with gain and losses provides exciting perspectives on the study of the parity anomaly and sublattice selective lasing~\cite{Schomerus:2013PRL}.

%% file: RMP_IVA_1DChiral.tex
\section{Topological photonics in one dimension}
\label{sec:1D}
The previous section was devoted to a review of two-dimensional photonics systems where topological concepts were first investigated. In the present section, we now turn our attention to one-dimensional models in Sec.~\ref{sec:1Dchiral} and then, in Sec.~\ref{sec:pump}, to the topological pumping effects that such systems have been used to study.

\subsection{Topology in 1D chiral systems}
\label{sec:1Dchiral}

In one dimension, topological phases of matter cannot exist without imposing symmetries on the system~\cite{Schnyder:2009AIP, Kitaev:2009AIP}. An important symmetry in one dimension, which can lead to topological phases, is chiral symmetry, for which the representative one-dimensional topological model is the Su-Schrieffer-Heeger (SSH) model as introduced in Sec.~\ref{sec:otherdimensions}. 

The first experimental realization of the SSH model and its associated topological edge state in a photonics context was in a photonic superlattice~\cite{Malkova:2009OptLett}. Since then, the SSH model or related models have been discussed and realized in photonic crystals~\cite{Keil:2013NatComm, Xiao:2014PRX}, electromagnetic metamaterials~\cite{Tan:2014SciRep, Yannopapas:2014IJMPB}, plasmonic and dielectric nanoparticles~\cite{Poddubny:2014ACS, Ling:2015OptExp, Slobozhanyuk:2015PRL, Sinev:2015Nanoscale, Slobozhanyuk:2016LaserPhotRev, Kruk:2017Small}, plasmonic waveguide arrays~\cite{Bleckmann:2017PRB, Cherpakova:2018arXiv},  polariton micropillars~\cite{Solnyshkov:2016PRL}, and coupled optical waveguides~\cite{Naz:2017arXiv}.
Lasing in the edge state of the SSH model was recently experimentally observed by~\onlinecite{St-Jean:2017NatPhot,Parto:PRL2018,Zhao:2018NatComm}. These experiments constitute the first realizations of topological lasers, i.e., lasers which make use of topological edge states as we further discuss in Sec.~\ref{sec:topolaser}.
The interplay between the SSH model and the radiative loss of photons have also been discussed in photonic crystals~\cite{Schomerus:2013OptLett, Poshakinskiy:2014PRL}, microwave cavity arrays~\cite{Poli:2015NatComm}, and coupled waveguide arrays~\cite{Zeuner:2015PRL}.
The SSH model of optical waveguides with periodic modulation and its edge states have been studied with Floquet theory~\cite{Zhu:2018PRA}.
The SSH model has also been realized with electric circuits~\cite{Lee:2018CommPhys}, and the Zak phase (Berry phase across the Brillouin zone) has been measured to show its topological nature~\cite{Goren:2018PRB}.
By properly adding loss, the parity-time ($\mathcal{PT}$) symmetric version of the SSH model was realized by~\textcite{Weimann:2017NatMat}. A one-dimensional electric circuit has also been shown to exhibit quantized energy transport due to the topological nature of the system in the presence of dissipation~\cite{Rosenthal:2018PRB}. More details on non-Hermitian topological models are given in Sec.~\ref{sec:nonhermitian}.
There is also a proposal to realize the one-dimensional Jackiw-Rebbi model, introduced in Sec.\ref{sec:iqhe}, in a driven slow-light setup~\cite{Angelakis:2014SciRep}, where the zero-energy bound mode can be probed through the transmission spectrum.

Another strategy to realize 1D chiral Hamiltonians with nontrivial topology involves discrete-time quantum walks~\cite{Kitagawa:2010PRA, Kitagawa:2012QIP}.
As discussed in Sec.~\ref{section_floquet}, a discrete-time quantum walk consists of a repeated application of a set of operations represented by a unitary matrix $\hat{\mathcal{U}}$. Topological properties of such unitary evolution can be understood analogously to Floquet topological phases by defining an effective Hamiltonian $\hat{H}_\mathrm{eff}$ through $\hat{\mathcal{U}} = e^{-i\hat{H}_{\mathrm{eff}}T/\hbar}$. If $\hat{H}_\mathrm{eff}$ has nontrivial topology, its effect can be detected through the discrete-time quantum walk. The great flexibility in choosing a set of operations to realize $\hat{\mathcal{U}}$ and hence $\hat{H}_\mathrm{eff}$ makes the discrete-time quantum walk a powerful platform to explore topological phases of matter.

A topologically nontrivial discrete-time quantum walk was first experimentally realized by~\onlinecite{Kitagawa:2012NatComm} using single photons going through a series of polarization rotations. In the experiment, a topological bound state between the interface of regions with different winding numbers was observed. Subsequently,~\onlinecite{Cardano:2015SciAdv,Cardano:2016NatComm, Cardano:2017NatComm} studied the topological invariant of a quantum walk in the orbital angular momentum space and detected the topological transition between different phases. Following the proposal~\cite{Tarasinski:2014PRA}, the topological invariant of a one-dimensional quantum walk was measured and its robustness to disorder was assessed, using a fiber loop architecture based on the time-multiplexing technique~\cite{Barkhofen:2017PRA}.

%% file: RMP_IVB_Pump.tex
\subsection{Topological pumps}
\label{sec:pump}

\begin{figure}[t!]
\includegraphics[width=\columnwidth]{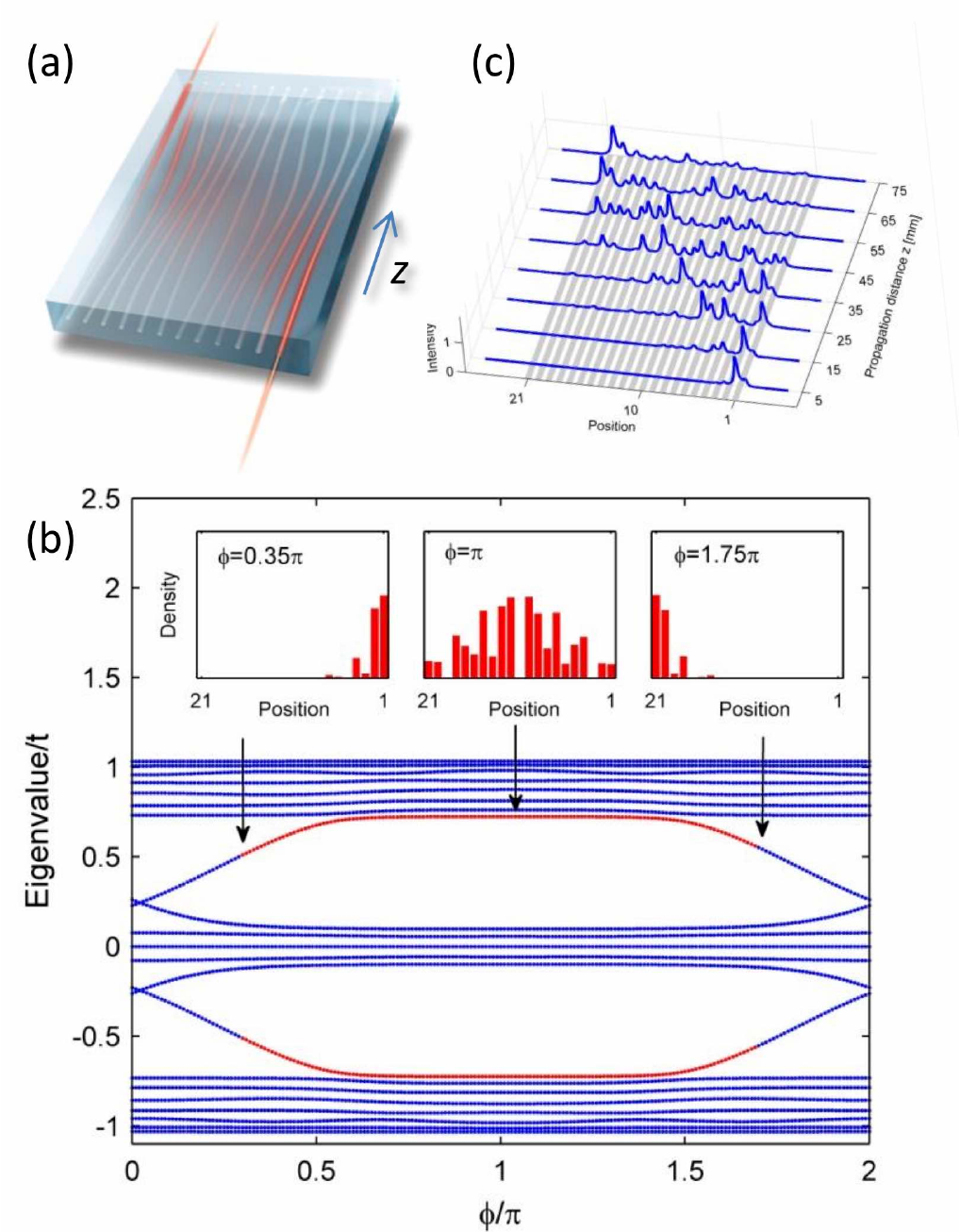}
\caption{Experimental observation of adiabatic pumping via topologically protected boundary states in a photonic waveguide array. (a) An illustration of the adiabatically modulated photonic waveguide array, constructed by slowly varying the spacing between the waveguides along the propagation axis $z$. Consequently, the injected light experiences an adiabatically modulated Hamiltonian $H_\text{off}(\phi(z))$ as it propagates and is pumped across the sample. (b) The spectrum of the model \eq{Eq:HHH1D} as a function of the phase
$\phi$ for $t=40/75$, $2t_{xy}=0.6$, and $b=\left(1+\sqrt{5}\right)/2$.
In the experiment, a 21-site lattice was used and $\phi$ was scanned between $0.35\pi$ and $1.75\pi$, marked by arrows (and red dots). The insets depict the spatial density of a boundary eigenstate as a function of the position at three different stages
of the evolution: At $\phi=0.35\pi$, the eigenstate is localized on the right boundary. At $\phi=\pi$, it is delocalized across the system, while at $\phi=1.75\pi$ the state is
again localized, but on the left boundary. (c) Experimental results: Light was injected into the rightmost waveguide at $z=0$ with $\phi=0.35\pi$. The measured intensity distributions as a function of the position are presented at different stages of the adiabatic evolution, i.e., different propagation distances. It is evident that during the adiabatic evolution, the light initially injected on the rightmost waveguide crosses the lattice from right to left and is finally concentrated on the leftmost waveguides. From~\textcite{Kraus:2012a}.  
\label{fig2_OZ}}
\end{figure}

Electric currents are usually generated by applying a voltage across a material, inducing longitudinal charge transport. Using Faraday's induction law, we can similarly generate electric currents via a time-dependent variation of a magnetic flux. In both cases, the longitudinal conductivity is determined by the microscopic details of the material and can take arbitrary values. As seen in Sec.~\ref{sec:basicconcepts}, the situation is completely different in topological systems, where currents can show quantization effects.

In particular, in topological charge pumps~\cite{Thouless:1983PRB}, Faraday's induction law is encoded in an adiabatic cyclic variation of a Hamiltonian potential that mimics the magnetic flux threading in a higher-dimensional topological Chern insulator. Consequently, the charge transport across the system per unit cycle of the pump parameter turns out to be quantized in much the same way that quantized Hall conductance appears in the higher-dimensional static Chern insulator. This feature has drawn much attention for controlled low-current nanoscale device applications~\cite{Kouwenhoven:1991}.

The first experiments toward the implementation of a topological charge pump were
conducted in solid-state systems with demonstrations of quantized charge transport using single electron pumps~\cite{Geerligs:1990,Kouwenhoven:1991,Pothier:1991,Pothier:1992}. The quantization, however, of the pumped charge in these devices relied on simple Coulomb blockade rather than on topological concepts. Later attempts using open mesoscopic systems incorporated geometrical ideas to generate a quasiadiabatic, nonquantized current that is proportional to the area enclosed in parameter space~\cite{Spivak:1995,Brouwer:1998, Zhou:1999,Switkes:1999,Pekola:2008}. The challenge of realizing quantized topological pumping was only recently accomplished using cold atoms in optical superlattices~\cite{Lohse:2016NatPhys,Nakajima:2016NatPhys}.

In this section, we introduce realizations of topological pumps in photonic systems.
Unitary photonic topological pumps have been realized using coupled waveguide arrays~\cite{Kraus:2012a,Verbin:2015}, whereas non-Hermitian pumps with exceptional points were realized using microwave cavities~\cite{Hu:2015PRX}. In both realizations, the experiments focused on states localized at the system's boundary which were directly excited by the incident light. Similar to the aforementioned cold-atom experiments~\cite{Lohse:2016NatPhys,Nakajima:2016NatPhys} theoretical proposals have addressed the possibility of studying quantized bulk pumping in photonic systems~\cite{Mei:2015,Ke:2016}. Geometric pumping has been experimentally realized using a fiber loop architecture~\cite{Wimmer:2017NatPhys}.

In the following, we focus on the first photonic realizations of topological pumps using waveguide arrays~\cite{Kraus:2012a}. While it is technologically challenging to modulate the waveguide profile in a way to precisely realize the on-site modulation involved in Thouless' original topological pump model \eqref{Eq:H1D}, the excellent control in the waveguide spacing achievable with femtosecond laser microfabrication technology~\cite{szameit2007control,Szameit:2010JPB} allowed for the emulation of an off-diagonal pump model where the interwaveguide hopping amplitudes are slowly modified along the propagation axis $z$ and the light evolves according to the Hamiltonian
\begin{align} \label{Eq:HHH1D}
    \hat{H}_\text{off} = - J \sum_{x} &\left[ \left(1+ \frac{2J_{xy}}{J} \cos\left[2\pi \alpha x/a + \phi(z)\right]\right)\hat{a}_{x}^\dagger \hat{a}_{x + a}\right.
    \notag \\
    &\left.+\mathrm{H.c.} \right]\,.
\end{align}
where $J$ is the bare hopping amplitude from waveguide $n$ to waveguide $n-1$, $2J_{xy}$ is the amplitude of its $z$-dependent modulation (which is equivalent to time-dependent modulation in propagating geometries, cf.~Sec.~\ref{sec:propagating}), and $\alpha$ is a spatial modulation frequency; see Fig.~\ref{fig2_OZ}(a) for an illustration. 

The mapping between the 2D quantum Hall effect on a lattice and the 1D pump discussed in Sec.~\ref{sec:pumpintro} and Eqs.~\eqref{II_HHHamiltonian} and \eqref{Eq:H1D} can be extended beyond the Harper-Hofstadter model~\cite{Kraus:2012b}. Performing dimensional extension on Eq.~\eqref{Eq:HHH1D}, we obtain a 2D tight-binding model where motion along $x$ occurs via a standard nearest-neighbor hopping in the $x$ direction and motion along $y$ occurs only via diagonal hoppings to next-nearest neighbors~\cite{Hatsugai:1990}
\begin{multline}
	\hat{H}
	=
	-J
	\sum_{x,y}
	\left(
	\hat{a}^\dagger_{x+a,y}\hat{a}_{x,y}
	+
	\frac{J_{xy}}{J} e^{i2\pi \alpha x/a}\hat{a}^\dagger_{x+a,y+a}\hat{a}_{x,y}
	\right.\\ \left.+
	\mathrm{H.c.}
	\right). \label{II_HHHHamiltonian}
\end{multline}
Each plaquette in the model is threaded by $2\pi \alpha$ flux as in the Harper-Hofstadter model, cf.~Eq.~\eqref{II_HHHamiltonian}.

The model~\eqref{Eq:HHH1D} is commonly known as the off-diagonal Harper model~\cite{Ketoja:1997,Jitomirskaya:2012}. Figure \ref{fig2_OZ}(b) depicts its spectrum as a function of $\phi$. We observe a characteristic gapped structure with topological modes crossing the gaps as a function of $\phi$. These modes are localized at the system's boundary when they are well within the energy gap, while they become spatially extended when they spectrally approach the bulk modes; see the insets of Fig.~\ref{fig2_OZ}(b).

In the experiment [Fig.~\ref{fig2_OZ}(c)], light is injected via fiber coupling directly into the left end. The initial $\phi$ is chosen in a way to support a localized state on this boundary, so that the injected light can directly excite this state.
The value of $\phi$ is then scanned along the propagation axis by correspondingly varying the interwaveguide distances. Depending on the final value of $\phi$, the spatial intensity distribution at the output will recover the spatial shape of the wave function of the topological state in different regimes, e.g., either extended over the bulk or even completely localized at the other boundary of the system.
Such pumping through the boundary states of the pump highlights the existence of localized states on both ends of the system for suitable values of $\phi$, in agreement with the pump's bulk-edge correspondence.

Moreover, it also illustrates the fact that the topological boundary mode on one end of the system can be connected to the state localized on the other end through semiadiabatic scanning of $\phi$ in sufficiently short systems. The latter is an interesting effect that is not implied by the bulk topology of the system: the bulk topology implies in fact that a quantized number of boundary states will cross the gap on each side of the sample as a function of $\phi$, but does not necessarily imply that these states have to be directly connected.

Using the same technology and extending this idea further, a topological pump was realized for an off-diagonal Fibonacci chain. This realization relied on a mapping between quasiperiodic chains and topological pumps~\cite{Kraus:2012a,Kraus:2012b,Verbin:2015}. Thus, using a two-parameter pumping, the Fibonacci chain was deformed into an off-diagonal Harper model, pumped as above, and deformed back into a Fibonacci chain~\cite{Verbin:2015}. Furthermore, topological phase transitions between quasiperiodic chains with smooth boundaries were studied using photonic waveguide arrays~\cite{Verbin:2013}. A study of the spectral flow of edge states across the energy gaps of a Fibonacci quasicrystal was reported by~\textcite{Baboux:2017PRB} by scanning a suitable structural parameter through many copies of a polariton lattice device.

Interestingly, simultaneous realizations of atomic and photonic 2D topological pumps were recently reported. Such pumps were shown to be directly mappable to a 4D quantum Hall system~\cite{Kraus:2013}. While the atomic experiment~\cite{Lohse:2018} performed a direct mapping of the Berry curvature by looking at anomalous transport in the bulk of the system, the photonic experiment~\cite{Zilberberg:2018Nature} studied the boundary states associated with a second Chern number response using similar methods to those previously discussed.

%% file: RMP_VA_3D.tex
\section{Topological photonics in higher dimensions}
\label{sec:higher}

Having reviewed the photonic realizations of two- and one-dimensional topological models, we now briefly highlight recent and on-going works in higher-dimensional topological systems. In Secs.~\ref{sec:3Dgapless} and~\ref{sec:3dgapped}, we focus on the study of three-dimensional topological photonics, for which macroscopic photonic crystals and metamaterials operating in the microwave domain have provided the main experimental platform. Then, in Sec.~\ref{sec:4D}, we discuss topological physics in even higher spatial dimensions, including perspectives in this direction opened up through the concept of ``synthetic dimensions."

\subsection{Three-dimensional gapless phases}
\label{sec:3Dgapless}

\subsubsection{Weyl points and helicoid surface states}

\begin{figure}[t]
\includegraphics[width=0.5\textwidth]{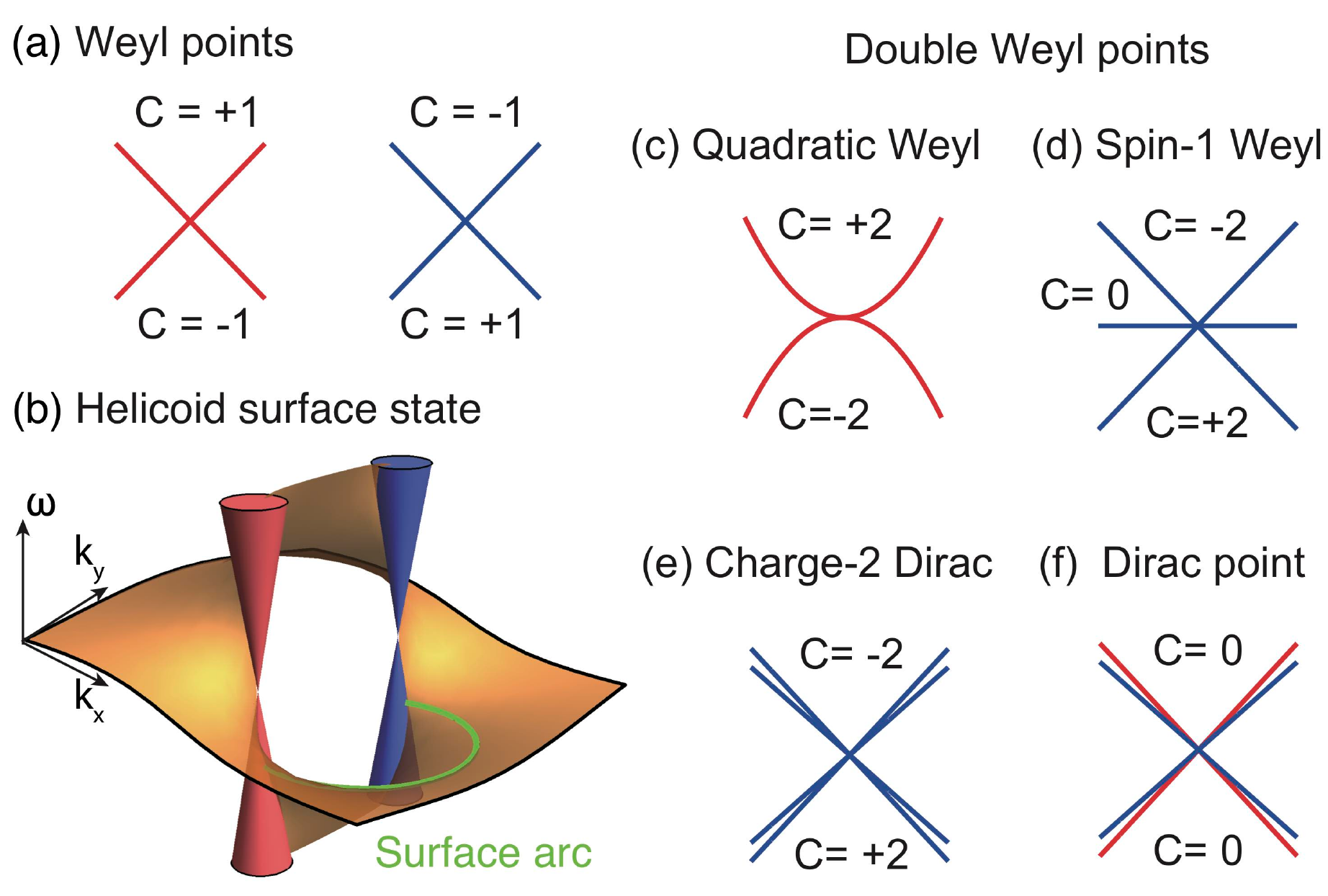}
\caption{(Color online)
(a) Weyl points of opposite Chern numbers~($C$).
(b) Surface dispersion near the projections of a pair of Weyl points with opposite Chern numbers, where the red and blue cones represent the bulk states
projection at $k_+$ and $k_-$. The surface state is plotted using the Riemann sheet of $\textnormal{Im}(log[(k-k_+)/(k-k_-)])$ in the complex plane of $k$. The green surface arc is an isofrequency contour.
(c)--(e) Double-Weyl points of two bands, three bands and four bands. (f) A Dirac point consists of two Weyl points of opposite Chern numbers.
(b) From~\textcite{Fang:2016NatPhys}. Other panels adapted from~\textcite{Zhang:2018PRL}.
}
\label{3Dpoints}
\end{figure}

As briefly reviewed in Sec.~\ref{sec:otherdimensions}, three-dimensional band-structures can host Weyl points, corresponding to gapless points around which bands disperse linearly with respect to the three quasimomenta [\eq{IIAweyl}]. Each Weyl point can be associated with a topological Chern number, calculated by integrating the Berry curvature over a two-dimensional surface enclosing the Weyl point. It can be shown that Weyl points have Chern numbers of $\pm1$~[Fig. \ref{3Dpoints}(a)] and require the breaking of either $\mathcal{P}$ (parity) and/or $\mathcal{T}$ (time-reversal) symmetry. If $\mathcal{T}$ is broken, the minimum number of Weyl points in the band structure is two, whereas, if only $\mathcal{P}$ is broken, as is typically much easier to implement in experiments, then the minimum number of Weyl points is four. For a strong enough tilting of the Weyl cone, the group velocities of the two crossing bands can have the same sign along one direction, in which case one speaks of a type-II Weyl point~\cite{Soluyanov:2015Nature}.

The topological character of Weyl points is reflected in the appearance of topologically protected states on the surface of the three-dimensional system. These surface states are topologically equivalent to helicoid Riemann surfaces~\cite{Fang:2016NatPhys,Zhang:2018PRL} defined with the two-dimensional surface Brillouin zone as the complex plane, shown in Fig. \ref{3Dpoints}(b). A helicoid surface is a noncompact Riemann surface, which is unbounded in the frequency axis, corresponding to the gapless nature of the Weyl surface state.
Locally around each Weyl cone, the surface states can be expressed as $\omega\propto\textnormal{Im}[log(k^C)]$, where $C$ is the Chern number of the Weyl point.
The bulk Weyl points project onto the surface Brillouin zone as poles~($C>0$) and zeros~($C<0$) of the multivalued helicoid surface sheets winding around these singularities.
Their winding direction is determined by the sign of $C$, while the order of the pole or zero is given by $|C|$.
As shown in Fig. \ref{3Dpoints}(b), the isofrequency contours of the helicoid surface are always open arcs connecting the surface projections of the positive and negative bulk Weyl points. These open surface arcs are known as ``Fermi arcs'' in Weyl semimetals.

Theoretically, Weyl points were first proposed to appear in double-gyroid photonic crystals, with a breaking of $\mathcal{P}$ or $\mathcal{T}$~\cite{Lu:2013NatPhot}. Since then, theoretical studies have shown that Weyl points could be realized in optical lattices~\cite{Dubvcek:2015PRL,Roy:2017arXiv}, photonic superlattices~\cite{Bravo:20152DM}, magnetized plasmas~\cite{Gao:2016NC}, chiral metamaterials~\cite{Gao:2015PRL,Xiao:2016PRL,Liu:2017PRL}, Floquet networks~\cite{Wang:2016PRBa,Ochiai:2016JOP}, chiral woodpile crystals~\cite{Chang:2017PRB}, magnetic tetrahedral crystals~\cite{Yang:2017OE}, and classical~\cite{Luo:2018arXiv} and quantum~\cite{Mei:2016QST} circuits. 
In an ideal Weyl system, all Weyl points would be frequency isolated and symmetry related at the exact same frequency~\cite{Wang:2016PRA}.
We also note that ideal Weyl points move the classical free-space scattering laws from DC to the Weyl frequency by design~\cite{Zhou:2017arXiv}, and that, after including losses, Weyl points evolve into exceptional lines~\cite{Xu:2017PRL}.

Experimentally, Weyl points were demonstrated at microwave frequencies in a double-gyroid photonic crystal~\cite{Lu:2015Science}, metallic photonic crystals with multi-Weyl points and surface transport~\cite{Chen:2016NatComm}, photonic metamaterials with type-II Weyl points and surface arcs~\cite{Yang:2017NatComm}, in lumped-element circuits~\cite{Lu:arxiv2018}, and at optical frequencies in coupled waveguides with type-II Weyl points and surface states~\cite{Noh:2017NatPhys}. Synthetic Weyl points in the parameter space of 1D dielectric stacks were also observed by~\onlinecite{Wang:2017PRX}, and ideal Weyl points have been found in a metallic design~\cite{yang2018ideal}.
In this latter platform, the helicoid surface states of the four Weyl points were experimentally mapped out and were topologically equivalent to a Riemann sheet defined by the Jacobi elliptical function, analytical in the whole double-periodic surface Brillouin zone.

\subsubsection{Multi-Weyl and Dirac points}
A Weyl point of nonzero Chern number does not require any symmetry for protection, other than translational symmetry. With an increase of symmetry, multi-Weyl points  can stabilize at high-symmetry momenta~\cite{Xu:2011PRL,Fang:2012PRL,Chen:2016NatComm,Chang:2017PRB}. For example, double-Weyl points~\cite{Zhang:2018PRL} of Chern number of $\pm2$ can form between two bands as a quadratic Weyl point, between three bands as a spin-1 Weyl point, or between four bands as a charge-2 Dirac point, as shown in Figs.~\ref{3Dpoints}(c)-\ref{3Dpoints}(e). In the latter case, charge 2 refers to the Berry charge~(Chern number) of 2, corresponding to the overlapping of Weyl points of the same Chern number. The double-Weyl surface states can be mapped, in the entire Brillouin zone, to the double-periodic Weierstrass elliptic functions: a type of Riemann surface with second-order poles and zeros~\cite{Zhang:2018PRL}.

More generally, a Dirac point in 3D refers to the overlapping of any two Weyl points of opposite Chern numbers, as shown, for example, in Fig. \ref{3Dpoints}(f). Such 3D Dirac points were discussed by~\onlinecite{Lu:2016NatPhys,Slobozhanyuk:2016NatPhot,Wang:2016PRBb,Wang:2017arXiv,Guo:2017arXiv,guo2018observation}. Since a 3D Dirac point has zero Chern number, it does not require the breaking of either $\mathcal{P}$ or $\mathcal{T}$.

\subsubsection{Nodal lines and surface}

As well as the above point degeneracies~(nodal points), line degeneracies are also important in 3D. Such nodal lines~\cite{Fang:2016CPB} can be protected by $\mathcal{PT}$ symmetry with $\pi$ Berry phase, same as the 2D Dirac cones. The nodal lines known so far can be classified into several families, namely, nodal rings~\cite{Burkov:2011PRB}, nodal chains~\cite{Bzdusek:2016Nature}, nodal links~\cite{Yan:2017PRB} and nodal knots~\cite{Bi:2017arXiv}. In photonics, a nodal ring was proposed in gyroid photonic crystals~\cite{Lu:2013NatPhot} and nodal chains were proposed in a face-centered cubic lattice~\cite{Kawakami:2016arXiv} and discovered experimentally in a simple-cubic metallic photonic crystal~\cite{Yan:2018NatPhys}. Nodal lines can also exist in two-dimensional photonic crystals~\cite{Lin:2017PRB}, such as at the zone boundary of two-dimensional lattices with glide reflection symmetry and $\mathcal{T}$. 
Nodal lines can also carry a $Z_2$ charge~\cite{Fang:2015PRB}.

Nodal surfaces can be protected by screw rotations and $\mathcal{T}$. It can even carry nonzero Chern numbers~\cite{Xiao:2017arXiv2}.

\subsection{Three-dimensional gapped phases}
\label{sec:3dgapped}

\begin{figure}[b!]
\includegraphics[width=0.5\textwidth]{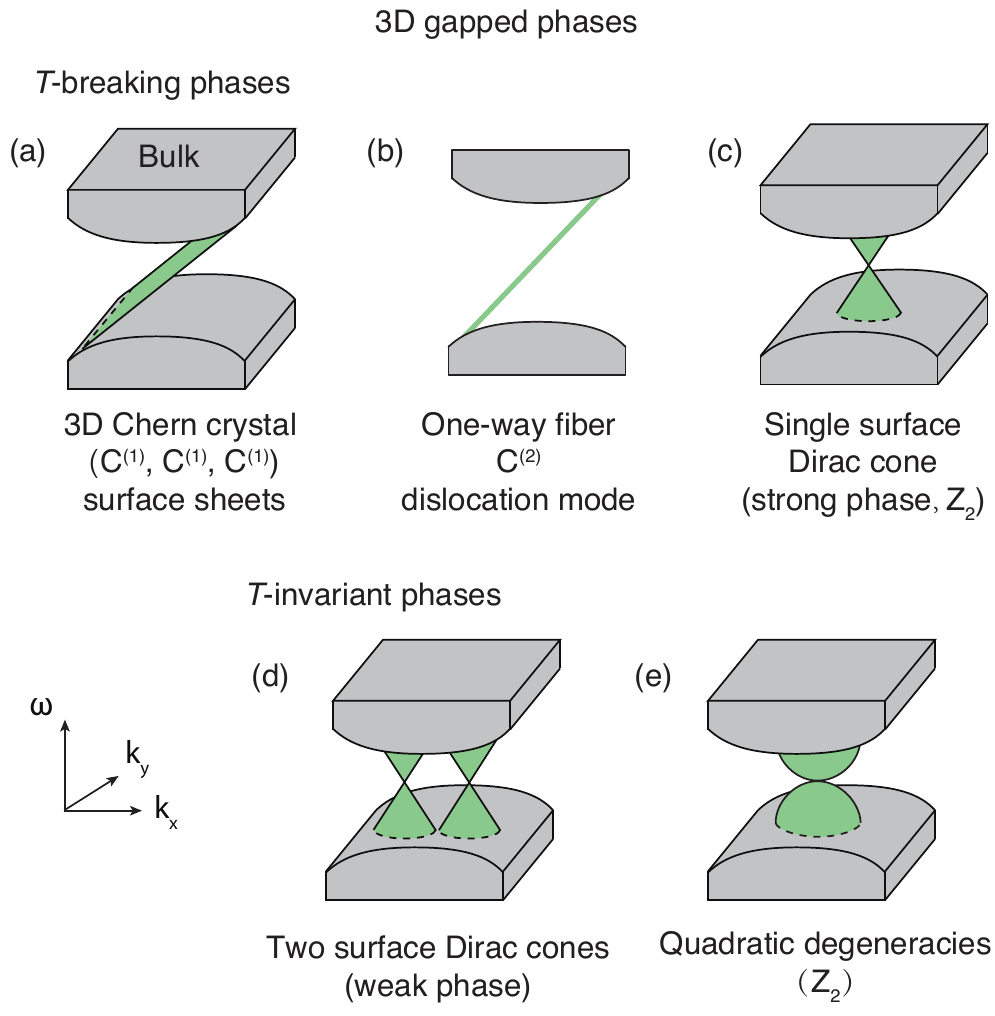}
\caption{(Color online) List of 3D gapped phases in photonics, in which (a)--(c) require $\mathcal{T}$ breaking while (d) and (e) do not.
(a) Analog of the 3D Chern insulator labeled by the three first Chern numbers.
(b) One-way fiber of second Chern number.
(c) Single-Dirac cone surface state with a $\mathbb{Z}_2$ invariant.
(d) Two surface Dirac cones similar to those of weak 3D topological insulators.
(e) Spatial-symmetry-protected gapless quadratic touchings with a $\mathbb{Z}_2$ invariant.
}
\label{3Dgaps}
\end{figure}

Gapping topological degeneracies such as Weyl and Dirac points is the most effective way to obtain 3D band gaps supporting various topological interfacial states.
The space-group analysis of photonic bands~\cite{Watanabe:2018PRL} is helpful for such designs.

\subsubsection{3D Chern crystal}

As introduced in Sec.~\ref{sec:otherdimensions}, a three-dimensional quantum Hall system is characterized by a triad of first Chern numbers~[$\mathbf{C^{(1)}}\equiv(C^{(1)}_x, C^{(1)}_y, C^{(1)}_z)$], as indicated in Fig. \ref{3Dgaps}(a). In such systems, the gapless surface states are unidirectional sheets, whose number and directionality equals the magnitude and sign of the Chern numbers for that surface~\cite{Koshino:2002PRB}.In the context of photonics, the elementary case of $\mathbf{C^{(1)}}=(0,0,1)$ was proposed by annihilating a single pair of Weyl points, by supercell coupling, in the magnetic gyroid photonic crystals~\cite{Lu:2016arXiv}. Note that as first Chern numbers vanish in the presence of time-reversal symmetry, the breaking of $\mathcal{T}$ is required to realize such gapped topological phases. 

\subsubsection{One-way fiber}

One-way fiber modes can form along topological line defects in 3D magnetic photonic crystals, illustrated in Fig. \ref{3Dgaps}(b). This was proposed by~\onlinecite{Bi:2015PRB} by Dirac mass engineering and designed in the gyroid photonic crystal by~\onlinecite{Lu:2016arXiv}. 
The coupling of two Weyl points of opposite Chern numbers makes a 3D Dirac point, as already introduced. The resulting Dirac Hamiltonian
\begin{equation}
H_{D} = \hbar v(q_x \sigma_x + q_y \sigma_y + q_z \sigma_z \tau_z) + m \tau_+ + m^* \tau_-
\end{equation}
has a complex mass term $m$, where $\tau_z$ and $\tau_\pm \equiv (\tau_x \pm i\tau_y)/2$ are Pauli matrices acting on the valley degrees of freedom. In-plane winding of its argument $\textrm{Arg}[m]$ generates a vortex line in 3D supporting a zero mode at the vortex core, topologically protected by the second Chern number $C^{(2)}$ in the 4D parameter space ($k_x,k_y,k_z,\theta$), where $\theta$ is the winding angle of $m$.

In the photonic context, such a topological defect line inside an otherwise fully gapped gyroid photonic crystal can be obtained by means of a helical winding of the supercell modulation coupling the two Weyl points. Depending on the spatial pitch and the handedness of the helical winding, one-way fiber modes of arbitrary $C^{(2)}$ can then be readily designed with arbitrary number of one-way modes.

This is in direct contrast with the one-way edge mode in 2D Chern crystals where high Chern numbers are difficult to obtain~\cite{Skirlo:2014PRL,Skirlo:2015PRL}. Another advantage of the one-way fiber design is that all one-way modes have almost identical group and phase velocities, due to the absence of sharp boundaries.

\subsubsection{Single surface Dirac cone}
A single-Dirac-cone surface state, the hallmark of 3D topological insulators~\cite{Fu:2007PRL}, can also be realized on the surface of magnetic photonic crystals, as shown by~\onlinecite{Lu:2016NatPhys} and illustrated in Fig. \ref{3Dgaps}(c). Instead of the Kramers' degeneracy of electrons due to $\mathcal{T}$, the double degeneracy in photonics can be replaced by a crystalline symmetry --- glide reflection. On the other hand, $\mathcal{T}$ has to be broken to split the dispersions in all surface directions away from degeneracy. 

The starting point to construct this phase is a pair of 3D Dirac points pinned at the high-symmetry points of the bulk Brillouin zone. By breaking $\mathcal{T}$ using magnetic materials, the authors gapped the 3D Dirac points and obtained the gapless single-Dirac cone surface states.
The topological invariant is $\mathbb{Z}_2$~\cite{Fu:2007PRL, Moore:2007PRB, Roy:2009PRB}. When the glide reflection symmetry is broken uniformly on the surface, the single-Dirac cone opens a frequency gap. Other than that, the surface state is robust against arbitrary random disorder, i.e., when the glide symmetry is preserved on average on the surface~\cite{Lu:2016NatPhys2}.

\subsubsection{Nonmagnetic designs}
The above topological phases with 3D bulk gaps require $\mathcal{T}$ breaking for highly robust interfacial states. However, the magnetic response is extremely weak toward optical frequencies, which has motivated the search for $\mathcal{T}$-invariant designs. One approach has been to search for nonmagnetic designs that mimic weak 3D topological insulators, which can be viewed as stacks of 2D topological insulators. In photonics, this was first proposed for all-dielectric bianisotropic metamaterials~\cite{Slobozhanyuk:2016NatPhot}, as introduced in Sec.~\ref{sec:metamaterial}, and has been experimentally realized very recently in a 3D array of bianisotropic metallic split-ring resonators~\cite{yang:2018arxiv}. 

A second approach to $\mathcal{T}$-invariant designs has been to search for systems, where the topological properties are protected instead by crystalline symmetries. The first proposal of such a topological crystalline insulator~\cite{Fu:2011PRL} (see Sec.~\ref{sec:wuhu} for the discussion of such states in two dimensions) utilized $C_4$ rotation symmetry and exhibited double degeneracies at two surface momentum points. This defined a $\mathbb{Z}_2$ invariant for its surface states to connect gaplessly.  At the degenerate point, the dispersion was quadratic due to the $C_4$ symmetry and $\mathcal{T}$. As illustrated in Fig. \ref{3Dgaps}(e), this phase is realizable in photonic crystals~\cite{Yannopapas:2011PRB,Alexandradinata:2014PPL}.
A concrete design for a tetragonal photonic crystal was also proposed by~\onlinecite{Ochiai:2017arXiv}.

%% file: RMP_VB_Synthetic.tex
\subsection{Toward even higher dimensions}
\label{sec:4D}

Topological phases of matter with spatial dimensions of 4 or higher can also be of experimental relevance in photonics. As already introduced in Sec.~\ref{sec:pump}, a recent experiment has used topological pumping to probe a four-dimensional quantum Hall system~\cite{Zilberberg:2018Nature}. In this approach, some of the dimensions are replaced by externally tuned parameters, effectively freezing out the dynamics along these directions. An alternative approach, which could also offer access to the dynamics of particles moving in effectively spatial four dimensions, is based on so-called \textit{synthetic dimensions}. In this, the key concept is to reinterpret internal degrees of freedom as spanning additional spatial dimensions, so that higher-dimensional lattice models are simulated in lower-dimensional systems. In this section, we first review the development of synthetic dimensions in general, before discussing progress in the exploration of four-dimensional topological systems with photons.

\subsubsection{Synthetic dimensions}
\label{sec:synd}

There are several different ways to make the effective spatial dimensionality of a lattice system larger than the physical dimensionality of the real space in which the lattice is located. One natural idea for this purpose is to increase the connectivity of the lattice, as proposed by~\onlinecite{Tsomokos:2010PRA} for superconducting qubit circuits, by~\onlinecite{Jukic:2013PRA} for photonic lattices, by~\onlinecite{Schwartz:2013OptExp} for multidimensional laser-mode lattices,
and by~\onlinecite{Grass:2015PRA} for trapped ions. Another strategy that can allow for even greater flexibility is to use the internal degrees of freedom, reinterpreting these as if they label different sites along an additional synthetic dimension in the system, as originally proposed by~\onlinecite{Boada:2012PRL} in the context of ultracold atomic gases, and later extended by~\onlinecite{Celi:2012PRL} to allow for complex hoppings along the synthetic direction, and so to realize quantum Hall systems. 
Methods to create lattice structures more complex than just a square lattice were proposed by~\onlinecite{Boada:2015NJP,Suszalski:2016PRA,Anisimovas:2016PRA}.

The idea of synthetic dimensions was soon experimentally realized in the context of cold atoms by two groups~\cite{Mancini:2015Science, Stuhl:2015Science}, in which a two-dimensional ladder with a magnetic field was simulated using a one-dimensional chain of atoms. The following experiments have then extended the synthetic dimension idea by using the different electronic states of atoms~\cite{Livi:2016PRL,Kolkowitz:2017Nature} and discrete states in momentum space~\cite{An:2017SciAdv}. Furthermore, there are theoretical proposals to use harmonic oscillator eigenstates~\cite{Price:2017PRA} and orbital angular momentum states~\cite{Pelegri:2017PRA} as synthetic dimensions. Typically, the interparticle interaction along the synthetic direction is very long ranged, resulting in a variety of interesting phenomena~\cite{Grass:2014PRA,Zeng:2015PRL,Lacki:2016PRA,Bilitewski:2016PRA,Barbarino:2016NJP,Taddia:2017PRL,Strinati:PRX2017,Junemann:2017PRX}.

In photonics the first proposal for how to implement a synthetic dimension was made by~\onlinecite{Luo:2015NatComm}, and later extended by~\onlinecite{Zhou:2017PRL,Luo:2017NatComm}, in which different orbital angular momentum states of light, coupled via spatial light modulators, were regarded as the synthetic dimension. This was followed by a proposal in optomechanics~\cite{Schmidt:2015Opt}, in which photon and phonon degrees of freedom were considered as two lattice sites along the synthetic dimension.~\onlinecite{Ozawa:2016PRA,Yuan:2016OptLett} have proposed to use different frequency modes of a multimode ring resonator, coupled via external modulation of refractive index, as a synthetic dimension. By modulating a resonator with multiple frequencies, models with any dimensions can also be simulated~\cite{Yuan:2017arXiv}.
A synthetic frequency dimension could also be realized in a Raman medium, where the synthetic magnetic field is controlled by the alignment of the two Raman  beams~\cite{Yuan:2017PRA}.
Instead of different frequency modes, the angular coordinate within a ring resonator may be used as a synthetic dimension~\cite{Ozawa:2017PRL}, in which the interphoton interaction is local along the synthetic direction, in contrast to extremely long-ranged interactions in other proposals.

There have also been many theoretical proposals for the different physics that could be accessed with synthetic dimensions. In a single resonator with a synthetic dimension, it may be possible to study the edge state of the SSH model~\cite{Zhou:2017PRL} and Bloch oscillations along the synthetic direction~\cite{Yuan:2016Optica}.
In a one-dimensional array of optical cavities with one synthetic dimension, the effect of topological edge states of two-dimensional Chern insulators may be observed~\cite{Luo:2015NatComm}.
Such a topological edge mode can be useful for high-efficiency frequency conversion if there is an edge along the synthetic direction made of frequency modes~\cite{Ozawa:2016PRA, Yuan:2016OptLett} and for realizing an optical isolator if an edge is along the spatial direction~\cite{Ozawa:2016PRA}.
A two-dimensional array of resonators could also be augmented by one frequency dimension to realize photonic Weyl points~\cite{Lin:2016NatComm,Sun:2017PRA}  or a weak 3D topological insulator~\cite{Lin:2018arxiv}. In the long run, one may expect that the idea of synthetic dimensions could find applications in increasing the complexity of optical networks in photonic devices, also in connection with frequency-multiplexing~\cite{saleh_book} and optical comb~\cite{cai2017multimode,Schwartz:2013OptExp} techniques. 

The experimental observation of topologically protected edge states in a photonic lattice with a synthetic dimension was recently pioneered by~\textcite{Lustig:arxiv2018}. A two-dimensional array of single-mode waveguides was used, with the synthetic dimension being encoded in the modal space of coupled waveguides. This allowed for the generation of a synthetic magnetic field and, thus, of a topologically nontrivial two-dimensional model supporting chiral edge states.

\subsubsection{Four-dimensional quantum Hall effect}
\label{sec:4DQHE}

As mentioned in Sec.~\ref{sec:pump}, a recent photonic experiment has used topological pumping to probe the edge states of a four-dimensional quantum Hall system~\cite{Zilberberg:2018Nature}, based on the proposal of~\onlinecite{Kraus:2013}. Concurrently with this experiment, the hallmarks of the quantized bulk response of the 4D quantum Hall effect \eqref{VB_4DQH}, including the second Chern number, were measured through the topological pumping of a two-dimensional ultracold atomic system~\cite{Lohse:2018}. By defining the second Chern number in a general parameter space, this topological invariant has also been shown to be experimentally relevant in helically modulated fibers~\cite{Lu:2016arXiv}, as discussed in Sec.~\ref{sec:3dgapped}, where the angular coordinate in the cross section of a fiber acts as the fourth parameter, and in ultracold gases, where it was measured over a parameter space spanned by properties of two Raman lasers applied to the system~\cite{Sugawa:2016arXiv}.

The first proposal for observing the full dynamics of a four-dimensional system was presented by~\onlinecite{Jukic:2013PRA}, based on using photonic lattices with high connectivity to study four-dimensional solitons. Using a synthetic dimension to directly observe the four-dimensional quantum Hall was then originally proposed in ultracold atomic gases~\cite{Price:2015PRL} and soon after extended to photonics~\onlinecite{Ozawa:2016PRA}. These proposals focused on 
the four-dimensional tight-binding model~\cite{Kraus:2013}:
\begin{align}
	\hat{H}
	&=
	-J\sum_{\mathbf{r}}
	\left(
	\hat{a}^\dagger_{\mathbf{r}+a\hat{e}_x} \hat{a}_{\mathbf{r}}
	+
	\hat{a}^\dagger_{\mathbf{r}+a\hat{e}_y} \hat{a}_{\mathbf{r}}
	\right.
	\notag \\
	&	
	\left.
	+
	e^{i2\pi \Phi_1 x/a} \hat{a}^\dagger_{\mathbf{r}+a\hat{e}_z} \hat{a}_{\mathbf{r}}
	+
	e^{i2\pi \Phi_2 y/a} \hat{a}^\dagger_{\mathbf{r}+a\hat{e}_w} \hat{a}_{\mathbf{r}}
	+\mathrm{H.c.}
	\right),
\end{align}
where $a_{\mathbf{r}}$ is the annihilation operator of a particle at position specified by a four-dimensional vector $\mathbf{r} = (x,y,z,w)$ with $w$ being the synthetic direction, and $a$ being the lattice spacing. The fluxes $\Phi_1$ and $\Phi_2$ pierce the $x$-$z$ and $y$-$w$ planes, respectively.
This is a generalization of the two-dimensional Harper-Hofstadter Hamiltonian to a 4D model with magnetic fields applied in two orthogonal planes, breaking time-reversal symmetry. However, unlike in 2D, time-reversal symmetry breaking is not the only route to observing a quantum Hall effect. Instead, 4D topological lattice models have also been proposed that exploit time-reversal-invariant spin-dependent gauge fields~\cite{Qi:2008PRB}, or lattice connectivity for spinless particles~\cite{Price:2018arxiv}.

Under the addition of weak electromagnetic perturbations, the filling of a 4D topological energy band would lead to a quantized nonlinear Hall current proportional to the second Chern number of the band\eqref{VB_4DQH}. However, in contrast to the atomic case, in photonic systems with loss, this current $j^\mu$ is not a direct observable. Instead, it was proposed to extract this topological response from the shift of the center of mass of the photonic steady-state intensity distribution under a monochromatic pump~\cite{Ozawa:2014PRL,Ozawa:2016PRA}.

%% file: RMP_VIA_NonHermitian-aug1.tex
\section{Gain and loss in topological photonics}
\label{sec:gainloss}
In this section, we discuss the interplay of gain and loss with topology in photonics. We divide this section into two main parts; in the first, we discuss non-Hermitian topological models with gain and loss, while in the second, we focus on recent works concerning topology in Bogoliubov systems. 

\subsection{Non-Hermitian topological photonics}
\label{sec:nonhermitian}

The study of topological physics with photons allows for the exploration of phenomena inaccessible in the context of condensed matter. A case in point is {\it non-Hermiticity} in the form of optical gain and loss. In photonics, gain and loss is much more common than in electrons in solids: gain media are the basis for lasers, and loss of photons is ubiquitous in every photonic device (loss is associated with absorption and surface roughness of a waveguide, for example).

There have thus been a series of works delving into the interplay of non-Hermiticity and topology with a number of disparate aims [including a recent review on the topic by~\textcite{alvarez2018topological}].  Inspired by a model proposed by~\textcite{rudner2009topological},~\textcite{Zeuner:2015PRL} used an optical waveguide array to demonstrate that the winding number of a one-dimensional topological system could be extracted from a non-Hermitian quantum walk.  In that work, it was precisely the finite lifetime (induced by optical loss) of the ``quantum walker" that allowed for the observation of a topological transition.  This amounts to the extraction of a topological number of a Hermitian system using non-Hermiticity, rather than exploring the topological invariants and edge states of non-Hermitian systems \textit{per se}.  

Another direction of non-Hermitian topological photonics is parity-time ($\mathcal{PT}$) symmetric \cite{makris2008beam, ruter2010observation} topological systems.  These are systems with balanced gain and loss such that the Hamiltonian commutes with the $\mathcal{PT}$ operator (where $\mathcal{P}$ represents parity and $\mathcal{T}$ represents time reversal).  It was shown~\cite{bender1998real} that such systems may exhibit real eigenvalue spectra despite their non-Hermiticity; they have been the basis of a major research effort in photonics due to the possibility of overcoming parasitic loss and absorption in optical devices using gain.  Thus, $\mathcal{PT}$-symmetric systems allow for the possibility of well-defined bands and gaps and are thus a natural place to start in studying non-Hermitian topological effects.  That said, it was shown~\cite{hu2011absence, esaki2011edge} that a large class of systems that are $\mathcal{PT}$ symmetric in the bulk must have edge states that ``break" $\mathcal{PT}$; namely, they have complex eigenvalues.  Fortunately, under certain conditions, topological edge states with real eigenvalues can be found and have been demonstrated \cite{Weimann:2017NatMat}.  Topological edge states in $\mathcal{PT}$-symmetric quantum walks have also been experimentally observed~\cite{Xiao:2017NatPhys}.  Recent work has examined $\mathcal{PT}$-symmetric Hamiltonians in the context of topological superconductors~\cite{kawabata2018parity}.

A number of other unconventional phenomena arise when non-Hermiticity and topology are combined.   One example is photonic ``tachyonlike" dispersion \cite{szameit2011p} that was demonstrated in the form of exceptional rings in photonic crystals \cite{zhen2015spawning} as well as ``Fermi arc"-type states that connect between exceptional points in two-dimensional systems~\cite{kozii2017non, Zhou:2018Science, malzard2018bulk}; a related phenomenon was examined in the context of heavy-fermion systems~\cite{yoshida2018non}.  ``Exceptional rings" arise in three-dimensional topological systems exhibiting Weyl points \cite{szameit2011p, Xu:2017PRL}, although to date these have not been experimentally realized.  The enhancement of topological interface states in one-dimensional systems was proposed \cite{Schomerus:2013OptLett} and demonstrated in the microwave regime \cite{Poli:2015NatComm} (along with a related work in optics~\cite{pan2018photonic}). Furthermore, it was shown that topological states absent when the system is Hermitian can be induced by adding losses~\cite{malzard2015topologically}. The interplay of non-Hermiticity and flat bands has been shown to result in a photonic analog of Aharonov-Bohm caging \cite{leykam2017flat}.  Beyond photonic systems, the interplay between non-Hermiticity or dissipation and topology has been explored in a number of theoretical works in varying contexts~\cite{diehl2011topology, PhysRevLett.109.130402, 1367-2630-15-8-085001, PhysRevA.91.042117, PhysRevB.91.165140}.  Finally, two-dimensional topological edge state lasers \cite{Bahari:Science2017, Harari:Science2018, Bandres:Science2018}, intrinsically non-Hermitian systems, have been observed recently; thus, it will become increasingly necessary to put on firm ground the complex interplay of non-Hermiticity and topological protection.  More details on topological lasers are given in Sec.~\ref{sec:topolaser}.

Despite this progress, the major challenge of non-Hermitian topological photonics remains the formulation of a general framework akin to that which exists for Hermitian systems.  In particular, perhaps the key contemporary question is: what is the right topological invariant to consider for a given non-Hermitian Hamiltonian~\cite{esaki2011edge,hu2011absence}, and what is its relevance to bulk-edge correspondence (some recent progress has been made in this direction \cite{lee2016anomalous, Weimann:2017NatMat, leykam2017edge,shen2018topological, yao2018edge, yin2018geometrical, yao2018non, kawabata2018non})?  In this spirit, there has been a recent effort to classify non-Hermitian topological systems~\cite{gong2018topological} in the same vein as Hermitian ones.  This is certainly not a direct generalization however: it may be the case that the symmetries conventionally associated with the Hermitian case may not be well suited to classify Hamiltonians in the non-Hermitian case.  Specifically, recent work has shown that particle-hole and time-reversal symmetries are aspects of the same larger symmetry when Hermiticity is relaxed~\cite{kawabata2018topological}.    

%% file: RMP_VIB_Bogoliubov.tex
\subsection{Emergent topology of Bogoliubov modes}
\label{sec:bogoliubov}

Photons under a parametric driving can be described by a Hamiltonian with terms that do not conserve the number of photons. Such number nonconserving bosonic systems can have topological features which are qualitatively different from fermionic topological systems.
To understand the origin of the number nonconserving terms, let us consider a photonic cavity whose resonant frequency is $\omega$ and assume that the cavity is made of optically nonlinear material with a second-order nonlinear susceptibility $\chi^{(2)}$. When one pumps the system with frequency $2\omega$, the nonlinearity converts the pumped photon into two photons with frequency $\omega$ in the cavity. Assuming that the pump beam is sufficiently strong and can be treated classically, the effective Hamiltonian describing the cavity takes the following form~\cite{Berry:2005Book}:
\begin{align}
	\hat{H}_\mathrm{cavity} = i\hbar \chi^{(2)} \left( \beta^* \hat{a}^2 - \beta \hat{a}^{\dagger 2} \right),
	\label{VIB_hamcav}
\end{align}
which does not conserve the number of photons, where $\hat{a}$ is the annihilation operator of a photon in the cavity and a $\mathbb{C}$-number $\beta$ characterizes the pumping field.
Such a cavity can be aligned to form a periodic lattice. The second-quantized momentum-space Hamiltonian of the lattice system can be written in the following form:
\begin{align}
	\hat{H}_\mathrm{lattice}
	&=
	\frac{1}{2}
	\sum_\mathbf{k}
	\begin{pmatrix} \hat{\Psi}_\mathbf{k}^\dagger & \hat{\Psi}_{-\mathbf{k}} \end{pmatrix}
	H_\mathbf{k}
	\begin{pmatrix} \hat{\Psi}_\mathbf{k} \\ \hat{\Psi}_{-\mathbf{k}}^\dagger \end{pmatrix},
	\notag \\
	H_\mathbf{k}
	&=
	\begin{pmatrix}
	A(\mathbf{k}) & B(\mathbf{k}) \\ B(\mathbf{-k})^* & A(-\mathbf{k})^t
	\end{pmatrix},
	\label{VIB_hampara}
\end{align}
where $\hat{\Psi}_\mathbf{k}$ is an $N$-component vector of annihilation operators with crystal momentum $\mathbf{k}$, and $N$ is the number of lattice sites per unit cell.
The $N$-by-$N$ matrix $A(\mathbf{k})$ is Hermitian and $B(\mathbf{k})^t = B(-\mathbf{k})$. The terms due to $B(\mathbf{k})$ do not conserve the number of photons.
At first glance, the Hamiltonian (\ref{VIB_hampara}) is similar to the Bogoliubov--de Gennes Hamiltonian of superconducting electronic systems. In fact, the Hamiltonian (\ref{VIB_hampara}) has particle-hole symmetry as in the fermionic Bogoliubov--de Gennes Hamiltonian, and the spectrum is symmetric with respect to the zero of the energy.
However, the transformation needed to diagonalize the Hamiltonian is drastically different between bosons and fermions.
The fermionic counterpart of $H_\mathbf{k}$ for the Bogoliubov--de Gennes Hamiltonian can be diagonalized by a unitary matrix to obtain eigenenergies of the systems which are guaranteed to be all real. On the other hand, in order to preserve the bosonic commutation relations, the bosonic Bogoliubov Hamiltonian $H_\mathbf{k}$ should be diagonalized by Bogoliubov transformations which are not a unitary matrix but a paraunitary matrix $\hat{U}$ obeying $\hat{U}^\dagger \left( \sigma_z \otimes \mathbb{I}_N \right) \hat{U} = \sigma_z \otimes \mathbb{I}_N$. The associated eigenenergies can be complex. These differences imply that the standard wisdom on topological phases of matter known for fermions may not hold for bosonic Bogoliubov Hamiltonians. Because of the possibility of having complex eigenvalues, we also need to pay attention to the possibility of instability.

Bosonic Bogoliubov Hamiltonians appear not only in photonic systems. In fact, the topological properties of such Hamiltonians were first discussed in the context of magnons in ferromagnetic crystals~\cite{Shindou:2013PRBa, Shindou:2013PRBb, Shindou:2014PRB}, where analogs of the Chern insulators in bosonic Bogoliubov Hamiltonians were discussed. It was found that the relevant Berry connection of the $n$th band of the Bogoliubov Hamiltonian is 
\begin{align}
	\boldsymbol{\mathcal{A}}_n (\mathbf{k}) = i\langle u_{n,\mathbf{k}} | \sigma_z \nabla_\mathbf{k} | u_{n,\mathbf{k}}\rangle,
\end{align}
where $|u_{n,\mathbf{k}}\rangle$ is the Bloch state of the $n$th band.
Note the additional $\sigma_z$ in the definition of the Berry connection.
The Berry curvature is then defined as $\boldsymbol{\Omega}_n = \nabla_\mathbf{k} \times \boldsymbol{\mathcal{A}}_n (\mathbf{k})$. The Chern number calculated by integrating this Berry curvature over the Brillouin zone is guaranteed to be integer and is related to the number of chiral edge modes.

Bogoliubov excitations are typically gapless at zero energy, but there can be gaps between higher energy bands. In exciton polaritons,~\onlinecite{Bardyn:2016PRB, Bleu:2016PRB} analyzed the Bogoliubov modes of exciton-polariton condensates and proposed models which have topological edge states at the gaps with nonzero excitation energy. The topological edge states at higher energy gaps of Bogoliubov excitations were also discussed in ultracold atomic gases~\cite{Engelhardt:2015PRA, Li:2015PRA, Furukawa:2015NJP, DiLiberto:2016PRL}.
In a lattice of photonic cavities under parametric driving,~\onlinecite{Peano:2016NatComm} proposed a model which has a nonzero gap at zero energy. In order to have a stable system, the gap at zero energy cannot have an edge state, so the sum of the Chern numbers of bands at the negative energy is zero, but gaps between higher energy bands can have topological edge states.

A distinctive feature of the bosonic Bogoliubov Hamiltonian (\ref{VIB_hampara}) is that the eigenenergies can become complex, hence triggering parametric instabilities~\cite{Shi:2017PNAS}.~\onlinecite{Peano:2016PRX} proposed a model where the topological edge states become unstable, even though all the bulk modes are stable. Such an unstable edge mode could be used as a traveling wave parametric amplifier. Instability caused by the topological edge modes was also analyzed in the context of ultracold atomic gases~\cite{Barnett:2013PRA, Galilo:2015PRL, Engelhardt:2016PRL}. The interplay between the topology and the parametric instability was also discussed in classical harmonic oscillators under periodic driving~\cite{Salerno:2016PRB}.

Finally, combined with strong optical nonlinearities (see Sec.~\ref{subsec:strong}), a $p$-wave version of parametric driving underlies the proposal in~\onlinecite{Bardyn:2012PRL} to obtain Majorana modes in a one-dimensional system of strongly interacting, fermionized photons.

%% file: RMP_VIII_Interaction.tex
\section{Topological effects for interacting photons}
\label{sec:interaction}

Most of the discussion of the previous sections concerned linear optical systems whose physics can be accurately described in terms of the standard Maxwell's equations including suitable linear dielectric and magnetic susceptibilities. In this regime, photons behave as independent particles. In this section, we focus our attention on the novel features that originate from the interplay of the topology with nonlinear optical effects. 

Basing ourselves on the general introduction to the basic nonlinear optics concepts of Sec.\ref{subsec:nlo}, the next two Secs.~\ref{subsec:weak} and~\ref{subsec:strong} summarize the main effects of an intensity-dependent refractive index in, respectively, the cases of weak and strong nonlinearity. In the former, a classical mean-field description based on Maxwell's equations with a nonlinear polarization term is accurate, while, in the latter, the physics is dominated by quantum optical effects due to the discreteness of the photon. We also note that Sec.\ref{sec:bogoliubov} provides a brief review of how parametric processes generated by a $\chi^{(2)}$ optical nonlinearity can give rise to a rich emergent topological structure for linear Bogoliubov modes. 

\subsection{Weak nonlinearities}
\label{subsec:weak}

For sufficiently weak values of the optical nonlinearities, one can legitimately perform the mean-field approximation of \eq{eq:MF}, in which the photons lose their particle character and collectively behave as a macroscopic wave, experiencing effective material properties that depend on the local amplitude of the light field according to the model of classical nonlinear polarization \eq{polarization}. While some parametric processes generated by a $\chi^{(2)}$ optical nonlinearity in the topological photoncis context were reviewed in Sec.~\ref{sec:bogoliubov}, in this section we focus on the case of an intensity-dependent refractive index \eq{nnl}: As reviewed in the following, theoretical works have anticipated that the modification of the refractive index induced by the nonlinearity may have dramatic observable consequences such as modifying the effective topology experienced by the wave. 

The first and most natural question of nonlinear topological physics was to understand how solitons~\cite{Segev:PRL1992,Eisenberg:PRL1998,fleischer2003observation} or vortices~\cite{Kivshar:2003book} are affected by the underlying geometry and topology of the band. This physics has attracted great interest in many fields such as ultracold atomic gases, where relativistic solitons and vortices in honeycomb geometries have been studied by~\textcite{Haddad:EPL2011,Haddad:NJP2015,Haddad2:NJP2015}.

Focusing on optical systems, intense research has been devoted to solitons in either the bulk or the edges of the Floquet photonic topological insulators of~\textcite{Rechtsman:2013Nature}, that were introduced in Sec.~\ref{sec:propagating}. The first work in this direction~\cite{Lumer:PRL2013} highlighted different families of long-lived, self-localized wave packets residing in the bulk of the system. Depending on their size, the wave packets may rotate in opposite directions, either following or opposite to the global Floquet modulation of the lattice. Most interestingly, the current profile of a rotationally symmetric six-site wide wave packet can be understood as an edge state residing on the inner boundary of a self-induced effective hole due to the nonlinearity.

The study of the effect of nonlinearity on topological edge states was pioneered by~\textcite{Ablowitz:2013PRA,Ablowitz:PRA2014,Ablowitz:20152DMat}, where the linear edge states of the Floquet bands of the experiment~\cite{Rechtsman:2013Nature} were classified as a function of the Floquet modulation parameters and their nonlinear evolution recast in terms of an effective one-dimensional nonlinear Schr\"odinger-like equation, possibly including higher-order derivative terms. Based on this equation, unidirectionally propagating edge soliton states have been identified: the topological robustness of linear edge states to backscattering translates into an enhanced robustness of edge solitons against higher-order terms. Following works have then analyzed the topological robustness of edge solitons traveling around sharp corners~\cite{Ablowitz:2015OptLett} and developed a general methodology to understand the tight-binding approximation in the context of nonlinear Floquet systems~\cite{Ablowitz:2017PRA}.

Dynamical modulational instabilities of edge states under the effect of nonlinearity and their eventual breakup into a train of solitons has been explored by several authors.~\textcite{Lumer:PRA2016} showed that nonlinear extended edge states of the Floquet system of~\textcite{Rechtsman:2013Nature} are always unstable independently of the sign of their linear dispersion and the actual strength of the nonlinearity. This modulational instability eventually leads to the break up of the extended wave into solitonlike localized wave packets.
Depending on the strength of the nonlinearity, such solitons can extend over many sites along the edge or localize to a single site. For polariton honeycomb lattices, the modulational instability of edge states and the consequent appearance of long-lived quasisoliton edge states were studied by~\textcite{Kartashov:Optica2016}, while, for kagome-shaped polariton lattices, topological edge solitons were studied by~\textcite{Gulevich:SciRep2017}. This latter work also highlighted the wide tunability of the edge soliton group velocity from positive to negative values as well as the robustness of topological edge solitons upon intersoliton collisions. 

The idea of nonlinear effects inducing transitions between states with different symmetries was pioneered by~\textcite{Lumer:PRL2013b} with a theoretical study of nonlinearity-induced transitions between $\mathcal{PT}$-broken and $\mathcal{PT}$-symmetric states in a non-Hermitian system and, then, in~\textcite{Katan:CLEO2016} with a theoretical study of the effect of long-range nonlinearities on topological transport. Along these lines, a novel kind of topological solitons was investigated by~\textcite{Leykam:PRL2016}: nonlinear effects locally induce a topological transition in an otherwise topologically trivial lattice and solitons naturally arise as the edge states at the topological interface. Possible applications of such nonlinearly induced topological transition to optical isolation were explored in different geometries by~\textcite{Xin:NJP2017}.
A related study in a nonlinear but conservative one-dimensional SSH model was reported by~\textcite{Hadad:PRB2016}.

\subsection{Strong nonlinearities}
\label{subsec:strong}

When nonlinearities are large, the discrete nature of the photons constituting the field starts being important and one has to resort to a fully quantum description. Correspondingly, the physics of these systems is qualitatively different, as they are expected to support strongly correlated states of light that closely resemble their electronic counterparts, e.g., fractional quantum Hall liquids~\cite{carusotto:2013}.

The simplest example of a quantum nonlinear effect is the so-called {\em photon blockade} phenomenon~\cite{Imamoglu:PRL97} that occurs in single-mode nonlinear cavities when the single-photon nonlinearity $\omega_{nl}$ [i.e., the frequency shift experienced by the mode for a single-photon occupation, as introduced in \eq{eq:H_omega_nl}] exceeds the damping rate $\gamma$ of the cavity mode. For an incident beam on resonance with the empty cavity mode, a first photon can freely enter the cavity, but a second one will find the effective resonance shifted by $\omega_{nl}$ and can not enter until the first has left. In analogy with the Coulomb blockade of electronics, one can think of the first photon blocking the entrance of the second: hence the term {\em photon blockade}. 

In the last decade or so, photon blockade has been observed in a variety of cavity configurations using different optically nonlinear elements~\cite{Birnbaum:Nature2005,Faraon:NatPhys2008,Reinhard:NatPhot2011,Lang:PRL2011}. 
In relation to topological photonics, the most promising platforms to combine photon blockade with synthetic gauge fields and/or nontrivial band topologies are the nonplanar cavities containing coherently dressed atomic gases in a Rydberg-EIT configuration and circuit-QED devices embedding strongly nonlinear superconducting elements, as recently pioneered by~\textcite{Jia:arXiv2017,ningyuan2017photons} and~\textcite{Roushan:2016NatPhys}, respectively.

The extension of photon blockade to many-cavity configurations so as to obtain complex strongly correlated many-photon states started attracting the interest of researchers in the mid-2000s with several proposals for how to realize Mott-insulator states of light~\cite{Angelakis:PRA2007,Greentree:2006,Hartmann:NatPhys2006}. Along these lines, the first proposal of a quantum Hall effect for light appeared by~\textcite{Cho:2008PRL}. While all these pioneering works made the quite strong assumption of a quasiequilibrium photon gas, which is able to equilibrate and/or be adiabatically manipulated before disappearing due to losses, specific studies of the consequences of the intrinsically driven-dissipative nature of photon systems appeared just a few years later~\cite{Gerace:NatPhys2009,Carusotto:PRL2009}.

The first theoretical study of the interplay of strong interactions with a synthetic gauge field in a driven-dissipative context appeared by~\textcite{Nunnenkamp:NJP2011}, where strongly correlated states of photons in few-sites lattices were highlighted, together with the signatures of such states in the transmission properties of a device. Soon after, a proposal to generate fractional quantum Hall states of light as a driven-dissipative steady state of a lossy Bose-Hubbard model with nontrivial hopping phases under a coherent pump was reported by~\textcite{Umucalilar:2012PRL}. The selective excitation of the desired many-body state can be obtained via the same multiphoton frequency selection mechanism first introduced by~\textcite{Carusotto:PRL2009} for Tonks-Girardeau gases of light. Signatures of the strongly correlated nature of quantum Hall states are then anticipated to transfer to the quantum statistical properties of the emitted light. While the coherent pumping scheme considered by~\onlinecite{Carusotto:PRL2009} and further investigated by~\onlinecite{Hafezi:NJP2013} is promising for the generation of few-photon quantum Hall states, its performance does not scale up favorably to larger numbers of photons: the frequency selection mechanism loses efficiency as many-photon peaks get spectrally closer, and the effective matrix element of the many-photon transition to the quantum Hall state may be quickly decreasing. 

Soon afterward, the coherent pumping scheme was extended to single cylindrical cavity geometries by~\textcite{Umucalilar:PLA2013}. In analogy with related research in rotating atomic gases~\cite{Cooper:AdvPhys2008}, one can take advantage of the formal similarity between the magnetic Lorentz force and the Coriolis one to study quantum Hall physics in the rotating fluids of light that are generated by a Laguerre-Gauss shaped coherent drive. Serious difficulties of this scheme were quickly pointed out~\cite{Fleischhauer:private}: any deviation from the perfect rotational symmetry of the cavity would result in a quick spin down of the rotating photon gas, while the spectral detuning between Landau levels prevents the use of narrow-band nonlinear elements such as Rydberg-EIT atoms.

These difficulties have been solved by the twisted optical resonators used in the experiments of~\textcite{Schine:2016Nature} reviewed in Sec.\ref{sec:twisted}. Replacing the mechanical rotation of the fluid of light with a synthetic magnetic field recovers the degeneracy between states in the lowest Landau level and, at the same time, introduces a sizable detuning between states of opposite angular momentum, which prevents the cloud from spinning down. Crucial experimental steps toward the generation of quantum Hall fluids of light in these systems have been recently reported, including the observation of Rydberg cavity polaritons in running-wave resonators~\cite{Ningyuan:PRA2016}, the use of optically pumped atomic samples to break time reversal and thus suppressing backscattering processes that would otherwise effectively reverse the synthetic gauge field~\cite{ningyuan2017photons}, and the use of the ultrastrong photon-photon interactions of Rydberg EIT to get photon blockade in the Gaussian-shaped fundamental mode of a cavity~\cite{Jia:arXiv2017}.

In the meantime, further theoretical work has anticipated an exotic phase diagram resulting from the short-distance saturation of the Rydberg-Rydberg interaction~\cite{grusdt2013fractional}. On the other hand, since these experiments are exploring quantum Hall physics on the surface of a cone, direct measurement of the properties of anyonic excitations appears to be possible through the density distribution in the vicinity of the cone tip, which will reflect the central charge of the topological fluid~\cite{can2016emergent}. Additional theoretical work on the many-body aspects of this challenge was presented by~\textcite{Sommer:arxiv2015,Dutta:2017arXiv}. 

Along related directions, fractional quantum Hall states in Jaynes-Cummings-Hubbard lattices were theoretically explored by~\textcite{Hayward:2012PRL}. The possibility of realizing topological models based on the spin dynamics of Rydberg polaritons confined in a widely spaced microcavity array was explored by~\textcite{Maghrebi:2015PRA}. Here hopping between sites does not occur by photon tunneling between neighboring cavities, but rather by dipolar interactions which exchange the spin state of neighboring dark polaritons.

\begin{figure}
\includegraphics[width=0.95\columnwidth,angle=0,clip]{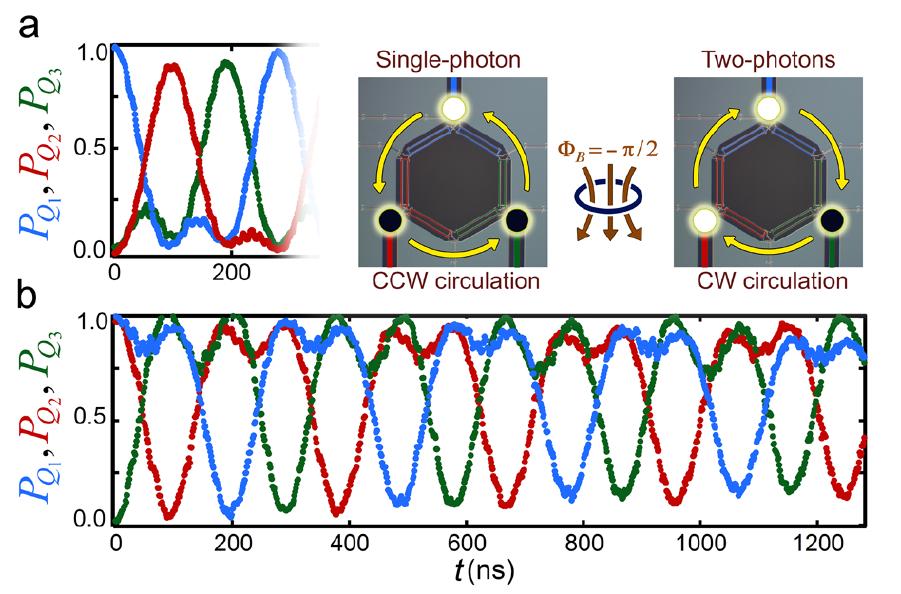}
\caption{Upper cener and right panels: Sketches of the circulation dynamics of a single- (center) and two-photon (right) state superimposed on the circuit-QED system under consideration. (a) Upper left: Time evolution of the excitation probability in the three $Q_{1,2,3}$ qubits for a single-photon state at magnetic flux $\Phi_B=-\pi/2$. (b) The same for a two-photon case. From~\textcite{Roushan:2016NatPhys}.
\label{fig:Roushan}}
\end{figure}

In parallel to these advances with macroscopic optical cavities embedding atoms, the first experimental studies of magnetic effects in a strongly interacting gas of photons were reported by~\textcite{Roushan:2016NatPhys} using a circuit-QED architecture with a closed loop three-site geometry. The synthetic magnetic fields are obtained following the theoretical proposal by~\textcite{Fang:2012NatPhot}, where the hopping phase is determined by the oscillation phase of the temporal modulation needed to compensate the frequency mismatch of neighboring sites. This technique falls within the class of Floquet techniques discussed in Sec.\ref{section_floquet} and was pioneered in the cold-atom context by~\textcite{Aidelsburger:2013PRL,Miyake:2013PRL}.

As illustrated in Fig.\ref{fig:Roushan}, a directional circulation of photons is the signature of broken time-reversal symmetry. Whereas noninteracting photons would independently rotate as single photons in a direction fixed by the synthetic magnetic field, the effect of strong interactions is manifested in the opposite circulation direction of photon vacancies. A similar inversion of the rotation direction as a consequence of strong interactions was recently reported also in the cold-atom context by~\textcite{Tai:arxiv2016}.

Thanks to the relatively long lifetime of photons, the experiment~\cite{Roushan:2016NatPhys} could be performed by initializing the system in a suitable one- or two-photon Fock state and then following its dynamics in the absence of any pump during the evolution: such an approach is able to directly probe the coherent quantum dynamics, but it is strongly limited by the decay rate of the full many-photon state, that typically scales proportionally to the photon number~\cite{Milman:PRA2000}. As a further important result of~\textcite{Roushan:2016NatPhys}, the adiabatic generation of a few-body interacting ground state for arbitrary synthetic magnetic flux $\Phi_B$ was also reported by slowly ramping up $\Phi_B$ to the desired final value.~\textcite{Roushan:Science2017} used this same platform in the presence of a quasiperiodic potential for photons to observe the Hofstadter butterfly of one-photon states and, then, localization effects in the two-photon sector.

This experimental result is all the more promising as several adiabatic protocols to create strongly correlated macroscopic topological states have been theoretically investigated, based on either the melting of Mott-insulator states~\cite{Cho:2008PRL} or by a sequence of flux-insertion and then quasihole refilling processes~\cite{Grusdt:PRL2014,Letscher:PRB2015,Ivanov:arxiv2018}.  An interesting proposal to manipulate quantum states of light by means of a generalized Thouless pumping in strongly nonlinear arrays was suggested by~\textcite{Tangpanitanon:PRL2016}. One has however to keep in mind that all these adiabatic schemes typically require that the process must be completed in a time scale shorter than the lifetime of the quantum many-body state of interest, a condition that may become extremely demanding for macroscopic photon fluids.

An alternative approach to dynamically stabilize topological many-body states of light against losses without any active intervention from external observers was started by~\textcite{Kapit:2014PRX}. In a circuit-QED context, this can be achieved using, e.g., the frequency-selective parametric emission from a shadow lattice: refilling of holes occurs at a fast rate as long as it spectrally coincides with the emission bandwidth, while the injection of extra photons on top of the topological state is suppressed via a generalized blockade phenomenon due to the many-body energy gap. In a related proposal~\cite{Hafezi:PRB2015}, parametric coupling to a thermal bath was proposed to generate a tunable chemical potential for effectively thermalized light. A conceptually similar idea based on population-inverted two-level systems as frequency-dependent light emitters was investigated by~\textcite{Lebreuilly:CRAS2016,Biella:PRA2017} in the context of Mott-insulator states of light. The significant advantages of replacing the Lorentzian emission spectrum with a more sophisticated square spectrum were pointed out by~\textcite{Lebreuilly:PRA2017}. The study of frequency-dependent incoherent pumping schemes applied to the generation of quantum Hall states in the twisted optical resonators of~\textcite{Schine:2016Nature} was discussed by~\textcite{Umucalilar:arXiv2017}. An experimental realization of a dissipatively stabilized Mott insulator of photons was very recently reported in~\textcite{Ma:2019Nature}.

In the long run, the application of photonic systems as useful platforms for topological quantum computation crucially requires strongly correlated topological states that support excitations with non-Abelian braiding statistics~\cite{Nayak:RMP2008}. A promising candidate for this purpose is the so-called Pfaffian states. Originally predicted in the context of the quantum Hall effect of electrons~\cite{Moore:NPB1991}, they can be obtained as the ground state in the presence of suitably engineered three-body interactions. A proposal to put this strategy into practice in a circuit-QED context was presented by~\textcite{Hafezi:PRB2014}. 

Before concluding, it is worth mentioning a completely different approach to the quantum dynamics of systems of few interacting particles~\cite{Krimer:PRA2011,Longhi:OL2011}: in the simplest formulation, the idea is to map the $x_{1,2}$ spatial coordinates of two quantum particles onto the $x,y$ spatial coordinates of a single particle moving in two dimensions. The two-body interactions are then modeled by letting the $x_1=x_2$ line of sites have slightly different linear optical properties. As the dynamics remains at a fully single-particle level, it can be simulated in any of the linear optical devices discussed in the first sections of this review.

Along these lines, physically two-dimensional arrays of waveguides were used by~\textcite{Mukherjee:PRA2016} to show evidence of tunneling processes for two-particle bound states in regimes where single-particle tunneling is instead strongly suppressed. The experiment~\cite{Schreiber:Science2012} demonstrated encoding of the spatial coordinates of a two-dimensional quantum walk into the arrival time of a single optical pulse, with a potential generalization to higher-dimensional configurations. 

Recently, the application of this mapping to study the interplay of interactions and topology in the two-particle dynamics of a one-dimensional SSH model was simultaneously proposed in the theoretical works by~\textcite{DiLiberto:PRA2016} and~\textcite{Gorlach:PRA2017}. An extension of this work exploiting driven-dissipative effects in resonator arrays was proposed by~\textcite{Gorlach:arxiv2018}. While these schemes are naturally transferred to the fermionic sector of the two-particle dynamics by means of the suitably antisymmetric initialization, a proposal to study supersymmetry (SUSY) effects in a light-matter context appeared by~\textcite{Tomka:SciRep2015}.

%% file: RMP_IX_Conclusion.tex
\section{Conclusion and perspectives}
\label{sec:conclusion}

In the previous sections we have seen how topological photonics has grown from its first proposal~
\cite{Raghu:2008PRA,Haldane:2008PRL} into a wide and mature field of research with a number of active exciting directions. In this final section of this review, we aim at summarizing those future developments that are promising for the next future. 

\subsection{Optical isolation and robust transport}
\label{sec:isolation}

Topological photonics may have short-term and midterm technological impact. The most straightforward applications involve the use of topologically protected unidirectional edge states as robust optical delay lines or optically isolating elements as originally discussed by~\textcite{Hafezi:2011NatPhys}. While firm experimental evidence of robust unidirectional propagation is currently available in a number of systems, practical application of these ideas into actual devices is still a subject of active investigation.

In particular, the stringent conditions that an optical isolator device must fulfill to be of practical utility were discussed by~\textcite{Jalas:NatPhot2013}. Different strategies to match the requirements are presently being actively explored using either magnetic elements~\cite{Bahari:Science2017,Solnyshkov:APL2018}, optical nonlinearities~\cite{Khanikaev:NatPhot2015,Shi:NatPhot2015}, or externally modulated elements~\cite{Yu:NatPhot2009,Hua:NatComm2015,fang2016generalized}.

Another exciting direction in view of applications in quantum information processing is to extend these results from classical light fields to quantum optical ones and link with on-going developments in chiral quantum optics~\cite{Lodahl:Nat2017}. The first step in this direction is to show that the dynamics of externally generated entangled photon pairs inherits the topological protection of single-photon states, as theoretically studied by~\textcite{Mittal:2016OptEx,Rechtsman:2016Optica}. The emission and robust propagation of single-photon states into chiral edge states was experimentally demonstrated by~\textcite{Barik:Science2018}. A proposal to generate quantum states of light by coupling an array of two-level emitters to a chiral edge state was reported by~\textcite{Ringel:NJP2014}: interesting correlations between photons originate when an externally generated few-photon wave packet scatters off the strongly nonlinear emitters.

\subsection{Quantum emitters and topological laser}
\label{sec:topolaser}

One of the most active directions of development to date is the study of the interplay of topology with light emitters and with optical gain, which is expected to offer novel features to be exploited in light sources, amplifiers, and laser devices. Strong motivations supported this study in view of optoelectronic and photonics applications, to improve the performance of topological devices compared to their trivial counterparts.

A detailed study of parametric amplification on the chiral edge states of a two-dimensional Harper-Hofstadter model in the presence of parametric downconversion or spontaneous four-wave mixing emission processes was reported by~\textcite{Peano:2016PRX}. With a suitably chosen pump frequency, amplification can be restricted to the edge states, while bulk modes remain quickly damped as in the passive system. The unidirectional nature of the edge state guarantees that amplification is not only quantum noise limited as in standard parametric amplifiers, but also nonreciprocal and almost perfectly insensitive to disorder. Interesting consequences of the topology on the zero-point quantum fluctuations and on the emission of entangled photon pairs are also pointed out. 

The robustness of the generated entangled photon pairs against disorder was specifically studied by~\textcite{Mittal2017}. In particular, it was shown how the generation of entangled photons using spontaneous four-wave mixing into topological states can outperform their topologically trivial counterparts.

\subsubsection{Topological lasers: Theory}

Strong activity is presently being devoted to the theoretical and experimental study of laser oscillation in topological systems, the so-called {\em topological lasers}. For one-dimensional systems, theoretical explorations of laser operation in topological states have appeared by~\textcite{Pilozzi:2016PRB}, which focused on a 1D Aubry-Andr\'e-Harper bichromatic photonic crystal, and by~\textcite{malzard:2018NJP}, which studied a 1D SSH chain of resonators, showing that the topological mode selection can persist even in the presence of nonlinearities. 

In two dimensions, topological lasing has been shown to lead to a highly efficient laser operation that remains monomode even well above the threshold and is robust against disorder~\cite{Harari:Science2018}. Simulations of the nonlinear wave equation with on-site saturable gain terms of the form
\begin{equation}
 \frac{i}{2}\frac{P_j}{1+|\alpha_j|^2/n_{\rm sat}}\alpha_j
\end{equation}
included into the lattice model of \eq{eq:dalpha/dt} without coherent pumping term ($F_j=0$) were performed for both trivial and topological lattice models with a pump profile $P_j$ concentrated on the edge. Gain saturation at high power is modeled by the saturation density $n_{\rm sat}$. In the topologically trivial case, laser operation is not able to exhaust all available gain, because many other modes can easily go above threshold for increasing power, leading to a complex multimode operation. This problem is particularly serious in the presence of disorder, which further suppresses the mode competition effect by spatially localizing modes.

As shown by~\textcite{Harari:Science2018}, all these problems turn out to be no longer relevant in topological lattices: the single lasing mode is extended around the whole system perimeter, maintaining a unidirectional flow and a spatially very uniform intensity profile even in the presence of disorder and for high pumping levels well above the threshold. In laser terms, this means a robust monomode operation with a high slope efficiency. 

Further theoretical work explicitly including the carrier dynamics in the amplifying solid-state material has pointed out the possibility that the topological laser operation is made dynamically unstable by the interplay of nonlinear frequency shifts and a slow population dynamics~\cite{Longhi:2018EPL}. General consequences of the chirality of the lasing mode of topological lasers were investigated in~\cite{Secli:MSc}, in particular the appearance of slow relaxation dynamics and the need to distinguish convective and absolute instabilities when studying the laser threshold in different geometries.

\subsubsection{Topological lasers: Experiments}

From the experimental point of view, the study of the interplay of lasing with geometrical and topological features was initiated by~\textcite{Sala:PRX2015}. In this experiment, the linewidth narrowing effect associated with laser operation was exploited to spectrally resolve the effect of the spin-orbit coupling terms in a hexagonal chain of pillar microcavities.
The first examples of a lasing operation in a topological nontrivial system were reported in one-dimensional chains of pillar microcavities~\cite{St-Jean:2017NatPhot}, ring resonators~\cite{Parto:PRL2018,Zhao:2018NatComm}, or photonics crystal nano cavity~\cite{Ota:2018NatComm}. In all the cases, for suitable pump geometries, a single-mode laser emission occurred into the edge states of the chain.

Recently, lasing in the topological edge states of a time-reversal-breaking two-dimensional photonic crystal embedding magnetic YIG elements was reported by~\textcite{Bahari:Science2017}. This work made use of a photonic crystal structure where gain is provided by quantum well emitters and time-reversal symmetry is broken by bonding the photonic crystal to a magnetic YIG material. In spite of the small width of the magneto-optically induced band gap, and the comparatively large linewidth of the bulk modes, lasing of modes on the edge was observed.

Soon after, another experiment~\cite{Bandres:Science2018} reported topological laser operation in a topological ring resonator array (as discussed in Sec.\ref{sec:siliconring}) embedding, in addition, quantum well emitters that provide optical gain. This setup does not require magnetic elements. By selectively pumping the edge resonators, a highly efficient single-mode emission into the topologically protected edge state was obtained even for gain values high above threshold. The performances of the novel device and the robustness against disorder were benchmarked with an extensive comparison to a topologically trivial laser device.
A technique to break the spinlike symmetry between the clockwise and counterclockwise modes of the rings by adding S-bend elements into the resonators was also demonstrated, further reinforcing the unidirectional properties of the emission. 

A third experiment reported lasing in chiral edge modes using a hexagonal lattice of microcavity polariton resonators~\cite{Klembt:arxiv2018}. The scheme relies on the sensitivity of exciton polaritons to an external magnetic field to break time-reversal symmetry in the configuration described in Sec.~\ref{sec:qheother}.

\subsection{Measurement of bulk topological and geometrical properties}
\label{subsec:quantummagn}

Topological photonics are also opening new perspectives in the study of topological effects of wide interest for quantum condensed-matter physics. As we have reviewed, a variety of new, possibly high-dimensional lattice configurations are becoming available thanks to the advances in photonic fabrication and manipulation. Furthermore, thanks to the high flexibility of the optical excitation and diagnostic schemes, an intense study is being devoted to the observable consequences of the geometrical and topological concepts: while the topologically protected chiral edge states have been the smoking gun of a nontrivial topology starting from the pioneering work of~\textcite{Wang:2009Nature}, the geometrical quantities characterizing the bulk bands are nowadays the subject of active study.

From early studies of transport in electronic systems~\cite{Xiao:2010RMP}, it is well known that the Berry curvature enters the semiclassical equations of motion for electrons as a sort of momentum-space magnetic field. The corresponding Lorentz-like force in reciprocal space provides an anomalous velocity term which is responsible, e.g., for the anomalous and integer~\cite{Thouless:1982PRL} quantum Hall effects. 
In the topological photonics context, this idea underlies the experimental reconstruction by~\textcite{Wimmer:2017NatPhys} of the $k$-space distribution of the bulk Berry curvature from the anomalous velocity of a wavepacket performing Bloch oscillations under an external force as theoretically proposed by~\textcite{Dudarev:2004PRL,Price:2012PRA,Cominotti:EPL2013}.

Application of the anomalous velocity idea to the coherently pumped systems discussed in Sec.~\ref{subsec:noneq} was theoretically proposed by~\textcite{Ozawa:2014PRL}. In addition to putting forward driven-dissipative versions of the anomalous Hall effects, specific pumping schemes able to equally distribute photons among the different momentum states were identified. In analogy to the classical theory of the integer quantum Hall effect in electronic systems~\cite{Thouless:1982PRL,Xiao:2010RMP} and to recent experiments with ultracold atoms~\cite{Aidelsburger:2014NatPhys}, the spatial displacement of the center of mass of the light intensity distribution then provides information on the Chern number of the band.
Later works~\cite{Price:2015PRL,Price:2016PRA,Ozawa:2016PRA} have generalized this idea to the second Chern number of four-dimensional lattice models and to the measurement of other geometrical quantities such as the Fubini-Study metric tensor~\cite{Ozawa:2018PRB}.
On a similar basis, the concept of mean chiral displacement was introduced to measure the winding of a chiral Hamiltonian from real-space measurements~\cite{Mondragon:PRL2014, Cardano:2016NatComm, Maffei:2017NJP,Meier:2018arXiv}. One of the most attractive aspects of this proposal is the possibility of measuring topological invariants in the presence of disorder, which breaks translational symmetry. An alternative scheme to extract the Chern number from the steady-state field amplitude of a small photonic lattice with twisted boundary conditions was discussed by~\textcite{Bardyn:2014NJP}. It was also suggested to calculate the full quantum geometric tensor, including the Berry curvature and the Fubini-Study metric, from direct measurements of the photon wave function in radiative photonic systems~\cite{Bleu:2018PRB}. Furthermore, dissipation has been used in the context of non-Hermitian systems to probe topological invariants in 1D photonic systems~\cite{rudner2009topological,Zeuner:2015PRL}, as discussed in Sec.~\ref{sec:nonhermitian}.

Building on top of the anomalous velocity concept, recent works~\cite{Bliokh:AnnPhys2005,Price:2014PRL} undertook the challenge of upgrading the semiclassical equations of motion to a full quantum mechanical theory of quantum particles in the presence of a nonvanishing momentum-space Berry curvature. As an example of application of this theory, the eigenstates of a topologically nontrivial lattice model subject to an additional external harmonic potential can be physically understood as momentum-space Landau levels on the torus-shaped first Brillouin zone, with a degeneracy set by the Chern number of the band~\cite{Price:2014PRL}. A proposal to investigate this physics in a driven-dissipative array of optical cavities with site-dependent resonance frequencies appeared by~\textcite{Berceanu:2016PRA}. 

Further on-going work in this direction is investigating how the momentum-space magnetic field may lead to momentum-space analogs of the quantum Hall effects~\cite{Ozawa:2015PRA,Ozawa:2016PRB,Claassen:PRL2015}. In these last works, the minima of the $k$-space dispersion play the role of lattice sites, the harmonic trapping provides the momentum-space analog of the kinetic energy, and the Berry curvature of the band plays the role of the magnetic field. Periodic boundary conditions are automatically inherited from the topology of the first Brillouin zone and the phase twist can be adjusted via the position of the harmonic trap minimum within the unit cell.

\subsection{Topological quantum computing}
\label{sec:qip}

A most exciting long term perspective is to use topological photonics devices as a platform for novel quantum information storage and processing tasks that exploit topological effects to protect their operation from external disturbances. A crucial requirement for such {\em topological quantum computing} with light appears to be the availability of strongly nonlinear elements to generate and manipulate strongly correlated states of light. The key experimental issues that researchers are facing along this route have been discussed in Sec.\ref{sec:interaction}, together with the promising results that were reported in the last few years. Here we highlight a possible strategy along which this research may develop in the future.

In the presence of suitably tailored optical nonlinearities, the fluid of light can form quantum states, e.g., Pfaffian ones~\cite{Hafezi:PRB2014}, that are anticipated to display a manifold of topologically degenerate ground states protected by a finite energy gap and elementary excitations with non-Abelian anyonic statistics. In systems with such properties~\cite{Nayak:RMP2008}, quantum information can be encoded in the ground state manifold of states and the unitary transforms corresponding to quantum logical operations can be performed by braiding quasiholes around each other. 
With respect to standard quantum information protocols, quantum computing based on these topological states of matter has the advantage that the states within the topologically degenerate manifold cannot be coupled nor mixed with each other by local disturbances, at least as long as their amplitude is not able to cross the energy gap.

To date, several condensed-matter systems such as quantum Hall states of two-dimensional electrons under strong magnetic fields~\cite{Tong:QHbook} or Majorana fermions in suitable superconductor-based solid-state nanostructures\cite{Elliott:RMP2015} are being seriously considered for this purpose, but to the best of our knowledge no experimental evidence is yet available of anyonic braiding statistics. First proposals taking advantage of the peculiarities of the optical systems to observe anyonic statistics have been recently brought forth~\cite{Umucalilar:PLA2013,Dutta:2017arXiv}, but a key question that remains open is the degree of robustness of the topologically encoded quantum information against the typical dissipative processes of optical systems. On the other hand, using an all-optical platform will be extremely favorable in view of integration of the quantum processing unit into an optical communication network.

%% file: main.bbl
\begin{thebibliography}{780}%
\makeatletter
\providecommand \@ifxundefined [1]{%
 \@ifx{#1\undefined}
}%
\providecommand \@ifnum [1]{%
 \ifnum #1\expandafter \@firstoftwo
 \else \expandafter \@secondoftwo
 \fi
}%
\providecommand \@ifx [1]{%
 \ifx #1\expandafter \@firstoftwo
 \else \expandafter \@secondoftwo
 \fi
}%
\providecommand \natexlab [1]{#1}%
\providecommand \enquote  [1]{``#1''}%
\providecommand \bibnamefont  [1]{#1}%
\providecommand \bibfnamefont [1]{#1}%
\providecommand \citenamefont [1]{#1}%
\providecommand \href@noop [0]{\@secondoftwo}%
\providecommand \href [0]{\begingroup \@sanitize@url \@href}%
\providecommand \@href[1]{\@@startlink{#1}\@@href}%
\providecommand \@@href[1]{\endgroup#1\@@endlink}%
\providecommand \@sanitize@url [0]{\catcode `\\12\catcode `\$12\catcode
  `\&12\catcode `\#12\catcode `\^12\catcode `\_12\catcode `\%12\relax}%
\providecommand \@@startlink[1]{}%
\providecommand \@@endlink[0]{}%
\providecommand \url  [0]{\begingroup\@sanitize@url \@url }%
\providecommand \@url [1]{\endgroup\@href {#1}{\urlprefix }}%
\providecommand \urlprefix  [0]{URL }%
\providecommand \Eprint [0]{\href }%
\providecommand \doibase [0]{http://dx.doi.org/}%
\providecommand \selectlanguage [0]{\@gobble}%
\providecommand \bibinfo  [0]{\@secondoftwo}%
\providecommand \bibfield  [0]{\@secondoftwo}%
\providecommand \translation [1]{[#1]}%
\providecommand \BibitemOpen [0]{}%
\providecommand \bibitemStop [0]{}%
\providecommand \bibitemNoStop [0]{.\EOS\space}%
\providecommand \EOS [0]{\spacefactor3000\relax}%
\providecommand \BibitemShut  [1]{\csname bibitem#1\endcsname}%
\let\auto@bib@innerbib\@empty
\bibitem [{\citenamefont {Aaboud}\ \emph {et~al.}(2017)\citenamefont {Aaboud}
  \emph {et~al.}}]{ATLAS:NatPhys2017}%
  \BibitemOpen
  \bibfield  {author} {\bibinfo {author} {\bibnamefont {Aaboud}, \bibfnamefont
  {Morad}},  \emph {et~al.} (\bibinfo {collaboration} {ATLAS Collaboration})}
  (\bibinfo {year} {2017}),\ \bibfield  {title} {\enquote {\bibinfo {title}
  {{Evidence for light-by-light scattering in heavy-ion collisions with the
  ATLAS detector at the LHC}},}\ }\href
  {https://www.nature.com/nphys/journal/v13/n9/full/nphys4208.html} {\bibfield
  {journal} {\bibinfo  {journal} {Nat. Phys.}\ }\textbf {\bibinfo {volume}
  {13}}~(\bibinfo {number} {9}),\ \bibinfo {pages} {852--858}}\BibitemShut
  {NoStop}%
\bibitem [{\citenamefont {Ablowitz}\ and\ \citenamefont
  {Cole}(2017)}]{Ablowitz:2017PRA}%
  \BibitemOpen
  \bibfield  {author} {\bibinfo {author} {\bibnamefont {Ablowitz},
  \bibfnamefont {Mark~J}}, \ and\ \bibinfo {author} {\bibfnamefont {Justin~T.}\
  \bibnamefont {Cole}}} (\bibinfo {year} {2017}),\ \bibfield  {title} {\enquote
  {\bibinfo {title} {Tight-binding methods for general longitudinally driven
  photonic lattices: Edge states and solitons},}\ }\href
  {https://link.aps.org/doi/10.1103/PhysRevA.96.043868} {\bibfield  {journal}
  {\bibinfo  {journal} {Phys. Rev. A}\ }\textbf {\bibinfo {volume} {96}},\
  \bibinfo {pages} {043868}}\BibitemShut {NoStop}%
\bibitem [{\citenamefont {Ablowitz}\ \emph {et~al.}(2014)\citenamefont
  {Ablowitz}, \citenamefont {Curtis},\ and\ \citenamefont
  {Ma}}]{Ablowitz:PRA2014}%
  \BibitemOpen
  \bibfield  {author} {\bibinfo {author} {\bibnamefont {Ablowitz},
  \bibfnamefont {Mark~J}}, \bibinfo {author} {\bibfnamefont {Christopher~W.}\
  \bibnamefont {Curtis}}, \ and\ \bibinfo {author} {\bibfnamefont {Yi-Ping}\
  \bibnamefont {Ma}}} (\bibinfo {year} {2014}),\ \bibfield  {title} {\enquote
  {\bibinfo {title} {Linear and nonlinear traveling edge waves in optical
  honeycomb lattices},}\ }\href
  {https://link.aps.org/doi/10.1103/PhysRevA.90.023813} {\bibfield  {journal}
  {\bibinfo  {journal} {Phys. Rev. A}\ }\textbf {\bibinfo {volume} {90}},\
  \bibinfo {pages} {023813}}\BibitemShut {NoStop}%
\bibitem [{\citenamefont {Ablowitz}\ \emph {et~al.}(2015)\citenamefont
  {Ablowitz}, \citenamefont {Curtis},\ and\ \citenamefont
  {Ma}}]{Ablowitz:20152DMat}%
  \BibitemOpen
  \bibfield  {author} {\bibinfo {author} {\bibnamefont {Ablowitz},
  \bibfnamefont {Mark~J}}, \bibinfo {author} {\bibfnamefont {Christopher~W}\
  \bibnamefont {Curtis}}, \ and\ \bibinfo {author} {\bibfnamefont {Yi-Ping}\
  \bibnamefont {Ma}}} (\bibinfo {year} {2015}),\ \bibfield  {title} {\enquote
  {\bibinfo {title} {Adiabatic dynamics of edge waves in photonic graphene},}\
  }\href {http://stacks.iop.org/2053-1583/2/i=2/a=024003} {\bibfield  {journal}
  {\bibinfo  {journal} {2D Materials}\ }\textbf {\bibinfo {volume}
  {2}}~(\bibinfo {number} {2}),\ \bibinfo {pages} {024003}}\BibitemShut
  {NoStop}%
\bibitem [{\citenamefont {Ablowitz}\ \emph {et~al.}(2013)\citenamefont
  {Ablowitz}, \citenamefont {Curtis},\ and\ \citenamefont
  {Zhu}}]{Ablowitz:2013PRA}%
  \BibitemOpen
  \bibfield  {author} {\bibinfo {author} {\bibnamefont {Ablowitz},
  \bibfnamefont {Mark~J}}, \bibinfo {author} {\bibfnamefont {Christopher~W.}\
  \bibnamefont {Curtis}}, \ and\ \bibinfo {author} {\bibfnamefont
  {Yi}~\bibnamefont {Zhu}}} (\bibinfo {year} {2013}),\ \bibfield  {title}
  {\enquote {\bibinfo {title} {Localized nonlinear edge states in honeycomb
  lattices},}\ }\href {https://link.aps.org/doi/10.1103/PhysRevA.88.013850}
  {\bibfield  {journal} {\bibinfo  {journal} {Phys. Rev. A}\ }\textbf {\bibinfo
  {volume} {88}},\ \bibinfo {pages} {013850}}\BibitemShut {NoStop}%
\bibitem [{\citenamefont {Ablowitz}\ and\ \citenamefont
  {Ma}(2015)}]{Ablowitz:2015OptLett}%
  \BibitemOpen
  \bibfield  {author} {\bibinfo {author} {\bibnamefont {Ablowitz},
  \bibfnamefont {Mark~J}}, \ and\ \bibinfo {author} {\bibfnamefont {Yi-Ping}\
  \bibnamefont {Ma}}} (\bibinfo {year} {2015}),\ \bibfield  {title} {\enquote
  {\bibinfo {title} {Strong transmission and reflection of edge modes in
  bounded photonic graphene},}\ }\href
  {http://ol.osa.org/abstract.cfm?URI=ol-40-20-4635} {\bibfield  {journal}
  {\bibinfo  {journal} {Opt. Lett.}\ }\textbf {\bibinfo {volume}
  {40}}~(\bibinfo {number} {20}),\ \bibinfo {pages} {4635--4638}}\BibitemShut
  {NoStop}%
\bibitem [{\citenamefont {Adams}\ and\ \citenamefont
  {Blount}(1959)}]{Adams:1959JPCS}%
  \BibitemOpen
  \bibfield  {author} {\bibinfo {author} {\bibnamefont {Adams}, \bibfnamefont
  {EN}}, \ and\ \bibinfo {author} {\bibfnamefont {E.I.}\ \bibnamefont
  {Blount}}} (\bibinfo {year} {1959}),\ \bibfield  {title} {\enquote {\bibinfo
  {title} {{Energy bands in the presence of an external force field-—II:
  Anomalous velocities}},}\ }\href
  {http://www.sciencedirect.com/science/article/pii/0022369759900046}
  {\bibfield  {journal} {\bibinfo  {journal} {J. Phys. Chem. Solids}\ }\textbf
  {\bibinfo {volume} {10}}~(\bibinfo {number} {4}),\ \bibinfo {pages} {286 --
  303}}\BibitemShut {NoStop}%
\bibitem [{\citenamefont {Aidelsburger}\ \emph {et~al.}(2013)\citenamefont
  {Aidelsburger}, \citenamefont {Atala}, \citenamefont {Lohse}, \citenamefont
  {Barreiro}, \citenamefont {Paredes},\ and\ \citenamefont
  {Bloch}}]{Aidelsburger:2013PRL}%
  \BibitemOpen
  \bibfield  {author} {\bibinfo {author} {\bibnamefont {Aidelsburger},
  \bibfnamefont {M}}, \bibinfo {author} {\bibfnamefont {M.}~\bibnamefont
  {Atala}}, \bibinfo {author} {\bibfnamefont {M.}~\bibnamefont {Lohse}},
  \bibinfo {author} {\bibfnamefont {J.~T.}\ \bibnamefont {Barreiro}}, \bibinfo
  {author} {\bibfnamefont {B.}~\bibnamefont {Paredes}}, \ and\ \bibinfo
  {author} {\bibfnamefont {I.}~\bibnamefont {Bloch}}} (\bibinfo {year}
  {2013}),\ \bibfield  {title} {\enquote {\bibinfo {title} {Realization of the
  {H}ofstadter {H}amiltonian with ultracold atoms in optical lattices},}\
  }\href {http://link.aps.org/doi/10.1103/PhysRevLett.111.185301} {\bibfield
  {journal} {\bibinfo  {journal} {Phys. Rev. Lett.}\ }\textbf {\bibinfo
  {volume} {111}},\ \bibinfo {pages} {185301}}\BibitemShut {NoStop}%
\bibitem [{\citenamefont {Aidelsburger}\ \emph {et~al.}(2015)\citenamefont
  {Aidelsburger}, \citenamefont {Lohse}, \citenamefont {Schweizer},
  \citenamefont {Atala}, \citenamefont {Barreiro}, \citenamefont {Nascimbene},
  \citenamefont {Cooper}, \citenamefont {Bloch},\ and\ \citenamefont
  {Goldman}}]{Aidelsburger:2014NatPhys}%
  \BibitemOpen
  \bibfield  {author} {\bibinfo {author} {\bibnamefont {Aidelsburger},
  \bibfnamefont {Monika}}, \bibinfo {author} {\bibfnamefont {Michael}\
  \bibnamefont {Lohse}}, \bibinfo {author} {\bibfnamefont {C}~\bibnamefont
  {Schweizer}}, \bibinfo {author} {\bibfnamefont {Marcos}\ \bibnamefont
  {Atala}}, \bibinfo {author} {\bibfnamefont {Julio~T}\ \bibnamefont
  {Barreiro}}, \bibinfo {author} {\bibfnamefont {S}~\bibnamefont {Nascimbene}},
  \bibinfo {author} {\bibfnamefont {NR}~\bibnamefont {Cooper}}, \bibinfo
  {author} {\bibfnamefont {Immanuel}\ \bibnamefont {Bloch}}, \ and\ \bibinfo
  {author} {\bibfnamefont {N}~\bibnamefont {Goldman}}} (\bibinfo {year}
  {2015}),\ \bibfield  {title} {\enquote {\bibinfo {title} {Measuring the
  {C}hern number of {H}ofstadter bands with ultracold bosonic atoms},}\ }\href
  {https://www.nature.com/nphys/journal/v11/n2/full/nphys3171.html} {\bibfield
  {journal} {\bibinfo  {journal} {Nat. Phys.}\ }\textbf {\bibinfo {volume}
  {11}}~(\bibinfo {number} {2}),\ \bibinfo {pages} {162--166}}\BibitemShut
  {NoStop}%
\bibitem [{\citenamefont {Albert}\ \emph {et~al.}(2015)\citenamefont {Albert},
  \citenamefont {Glazman},\ and\ \citenamefont
  {Jiang}}]{albert2015topological}%
  \BibitemOpen
  \bibfield  {author} {\bibinfo {author} {\bibnamefont {Albert}, \bibfnamefont
  {Victor~V}}, \bibinfo {author} {\bibfnamefont {Leonid~I}\ \bibnamefont
  {Glazman}}, \ and\ \bibinfo {author} {\bibfnamefont {Liang}\ \bibnamefont
  {Jiang}}} (\bibinfo {year} {2015}),\ \bibfield  {title} {\enquote {\bibinfo
  {title} {Topological properties of linear circuit lattices},}\ }\href
  {https://journals.aps.org/prl/abstract/10.1103/PhysRevLett.114.173902}
  {\bibfield  {journal} {\bibinfo  {journal} {Phys. Rev. Lett.}\ }\textbf
  {\bibinfo {volume} {114}}~(\bibinfo {number} {17}),\ \bibinfo {pages}
  {173902}}\BibitemShut {NoStop}%
\bibitem [{\citenamefont {Alexandradinata}\ \emph {et~al.}(2014)\citenamefont
  {Alexandradinata}, \citenamefont {Fang}, \citenamefont {Gilbert},\ and\
  \citenamefont {Bernevig}}]{Alexandradinata:2014PPL}%
  \BibitemOpen
  \bibfield  {author} {\bibinfo {author} {\bibnamefont {Alexandradinata},
  \bibfnamefont {A}}, \bibinfo {author} {\bibfnamefont {Chen}\ \bibnamefont
  {Fang}}, \bibinfo {author} {\bibfnamefont {Matthew~J}\ \bibnamefont
  {Gilbert}}, \ and\ \bibinfo {author} {\bibfnamefont {B~Andrei}\ \bibnamefont
  {Bernevig}}} (\bibinfo {year} {2014}),\ \bibfield  {title} {\enquote
  {\bibinfo {title} {Spin-orbit-free topological insulators without
  time-reversal symmetry},}\ }\href
  {https://journals.aps.org/prl/abstract/10.1103/PhysRevLett.113.116403}
  {\bibfield  {journal} {\bibinfo  {journal} {Phys. Rev. Lett.}\ }\textbf
  {\bibinfo {volume} {113}}~(\bibinfo {number} {11}),\ \bibinfo {pages}
  {116403}}\BibitemShut {NoStop}%
\bibitem [{\citenamefont {An}\ \emph {et~al.}(2017)\citenamefont {An},
  \citenamefont {Meier},\ and\ \citenamefont {Gadway}}]{An:2017SciAdv}%
  \BibitemOpen
  \bibfield  {author} {\bibinfo {author} {\bibnamefont {An}, \bibfnamefont
  {Fangzhao~Alex}}, \bibinfo {author} {\bibfnamefont {Eric~J.}\ \bibnamefont
  {Meier}}, \ and\ \bibinfo {author} {\bibfnamefont {Bryce}\ \bibnamefont
  {Gadway}}} (\bibinfo {year} {2017}),\ \bibfield  {title} {\enquote {\bibinfo
  {title} {Direct observation of chiral currents and magnetic reflection in
  atomic flux lattices},}\ }\href
  {http://advances.sciencemag.org/content/3/4/e1602685} {\bibfield  {journal}
  {\bibinfo  {journal} {Science Advances}\ }\textbf {\bibinfo {volume}
  {3}}~(\bibinfo {number} {4}),\ \bibinfo {pages} {e1602685}}\BibitemShut
  {NoStop}%
\bibitem [{\citenamefont {Anderson}\ \emph {et~al.}(2016)\citenamefont
  {Anderson}, \citenamefont {Ma}, \citenamefont {Owens}, \citenamefont
  {Schuster},\ and\ \citenamefont {Simon}}]{anderson2016engineering}%
  \BibitemOpen
  \bibfield  {author} {\bibinfo {author} {\bibnamefont {Anderson},
  \bibfnamefont {Brandon~M}}, \bibinfo {author} {\bibfnamefont {Ruichao}\
  \bibnamefont {Ma}}, \bibinfo {author} {\bibfnamefont {Clai}\ \bibnamefont
  {Owens}}, \bibinfo {author} {\bibfnamefont {David~I}\ \bibnamefont
  {Schuster}}, \ and\ \bibinfo {author} {\bibfnamefont {Jonathan}\ \bibnamefont
  {Simon}}} (\bibinfo {year} {2016}),\ \bibfield  {title} {\enquote {\bibinfo
  {title} {Engineering topological many-body materials in microwave cavity
  arrays},}\ }\href
  {https://journals.aps.org/prx/abstract/10.1103/PhysRevX.6.041043} {\bibfield
  {journal} {\bibinfo  {journal} {Phys. Rev. X}\ }\textbf {\bibinfo {volume}
  {6}}~(\bibinfo {number} {4}),\ \bibinfo {pages} {041043}}\BibitemShut
  {NoStop}%
\bibitem [{\citenamefont {Anderson}\ and\ \citenamefont
  {Subramania}(2017)}]{Anderson:OE2017}%
  \BibitemOpen
  \bibfield  {author} {\bibinfo {author} {\bibnamefont {Anderson},
  \bibfnamefont {P~Duke}}, \ and\ \bibinfo {author} {\bibfnamefont {Ganapathi}\
  \bibnamefont {Subramania}}} (\bibinfo {year} {2017}),\ \bibfield  {title}
  {\enquote {\bibinfo {title} {{Unidirectional edge states in topological
  honeycomb-lattice membrane photonic crystals}},}\ }\href
  {http://www.opticsexpress.org/abstract.cfm?URI=oe-25-19-23293} {\bibfield
  {journal} {\bibinfo  {journal} {Opt. Express}\ }\textbf {\bibinfo {volume}
  {25}}~(\bibinfo {number} {19}),\ \bibinfo {pages} {23293--23301}}\BibitemShut
  {NoStop}%
\bibitem [{\citenamefont {Angelakis}\ \emph {et~al.}(2014)\citenamefont
  {Angelakis}, \citenamefont {Das},\ and\ \citenamefont
  {Noh}}]{Angelakis:2014SciRep}%
  \BibitemOpen
  \bibfield  {author} {\bibinfo {author} {\bibnamefont {Angelakis},
  \bibfnamefont {D~G}}, \bibinfo {author} {\bibfnamefont {P.}~\bibnamefont
  {Das}}, \ and\ \bibinfo {author} {\bibfnamefont {C.}~\bibnamefont {Noh}}}
  (\bibinfo {year} {2014}),\ \bibfield  {title} {\enquote {\bibinfo {title}
  {{Probing the topological properties of the Jackiw-Rebbi model with
  light}},}\ }\href {https://www.nature.com/articles/srep06110} {\bibfield
  {journal} {\bibinfo  {journal} {Sci. Rep.}\ }\textbf {\bibinfo {volume}
  {4}},\ \bibinfo {pages} {6110}}\BibitemShut {NoStop}%
\bibitem [{\citenamefont {Angelakis}\ \emph {et~al.}(2007)\citenamefont
  {Angelakis}, \citenamefont {Santos},\ and\ \citenamefont
  {Bose}}]{Angelakis:PRA2007}%
  \BibitemOpen
  \bibfield  {author} {\bibinfo {author} {\bibnamefont {Angelakis},
  \bibfnamefont {Dimitris~G}}, \bibinfo {author} {\bibfnamefont
  {Marcelo~Franca}\ \bibnamefont {Santos}}, \ and\ \bibinfo {author}
  {\bibfnamefont {Sougato}\ \bibnamefont {Bose}}} (\bibinfo {year} {2007}),\
  \bibfield  {title} {\enquote {\bibinfo {title} {Photon-blockade-induced mott
  transitions and {XY} spin models in coupled cavity arrays},}\ }\href
  {https://journals.aps.org/pra/abstract/10.1103/PhysRevA.76.031805} {\bibfield
   {journal} {\bibinfo  {journal} {Phys. Rev. A}\ }\textbf {\bibinfo {volume}
  {76}}~(\bibinfo {number} {3})}\BibitemShut {NoStop}%
\bibitem [{\citenamefont {Anisimovas}\ \emph {et~al.}(2016)\citenamefont
  {Anisimovas}, \citenamefont {Ra\ifmmode \check{c}\else
  \v{c}\fi{}i\ifmmode~\bar{u}\else \={u}\fi{}nas}, \citenamefont {Str\"ater},
  \citenamefont {Eckardt}, \citenamefont {Spielman},\ and\ \citenamefont
  {Juzeli\ifmmode~\bar{u}\else \={u}\fi{}nas}}]{Anisimovas:2016PRA}%
  \BibitemOpen
  \bibfield  {author} {\bibinfo {author} {\bibnamefont {Anisimovas},
  \bibfnamefont {E}}, \bibinfo {author} {\bibfnamefont {M.}~\bibnamefont
  {Ra\ifmmode \check{c}\else \v{c}\fi{}i\ifmmode~\bar{u}\else \={u}\fi{}nas}},
  \bibinfo {author} {\bibfnamefont {C.}~\bibnamefont {Str\"ater}}, \bibinfo
  {author} {\bibfnamefont {A.}~\bibnamefont {Eckardt}}, \bibinfo {author}
  {\bibfnamefont {I.~B.}\ \bibnamefont {Spielman}}, \ and\ \bibinfo {author}
  {\bibfnamefont {G.}~\bibnamefont {Juzeli\ifmmode~\bar{u}\else
  \={u}\fi{}nas}}} (\bibinfo {year} {2016}),\ \bibfield  {title} {\enquote
  {\bibinfo {title} {Semisynthetic zigzag optical lattice for ultracold
  bosons},}\ }\href {https://link.aps.org/doi/10.1103/PhysRevA.94.063632}
  {\bibfield  {journal} {\bibinfo  {journal} {Phys. Rev. A}\ }\textbf {\bibinfo
  {volume} {94}},\ \bibinfo {pages} {063632}}\BibitemShut {NoStop}%
\bibitem [{\citenamefont {Ao}\ \emph {et~al.}(2009)\citenamefont {Ao},
  \citenamefont {Lin},\ and\ \citenamefont {Chan}}]{Ao:2009PRB}%
  \BibitemOpen
  \bibfield  {author} {\bibinfo {author} {\bibnamefont {Ao}, \bibfnamefont
  {Xianyu}}, \bibinfo {author} {\bibfnamefont {Zhifang}\ \bibnamefont {Lin}}, \
  and\ \bibinfo {author} {\bibfnamefont {CT}~\bibnamefont {Chan}}} (\bibinfo
  {year} {2009}),\ \bibfield  {title} {\enquote {\bibinfo {title} {One-way edge
  mode in a magneto-optical honeycomb photonic crystal},}\ }\href
  {https://journals.aps.org/prb/abstract/10.1103/PhysRevB.80.033105} {\bibfield
   {journal} {\bibinfo  {journal} {Phys. Rev. B}\ }\textbf {\bibinfo {volume}
  {80}}~(\bibinfo {number} {3}),\ \bibinfo {pages} {033105}}\BibitemShut
  {NoStop}%
\bibitem [{\citenamefont {Armitage}\ \emph {et~al.}(2018)\citenamefont
  {Armitage}, \citenamefont {Mele},\ and\ \citenamefont
  {Vishwanath}}]{Armitage:2017arXiv}%
  \BibitemOpen
  \bibfield  {author} {\bibinfo {author} {\bibnamefont {Armitage},
  \bibfnamefont {N~P}}, \bibinfo {author} {\bibfnamefont {E.~J.}\ \bibnamefont
  {Mele}}, \ and\ \bibinfo {author} {\bibfnamefont {Ashvin}\ \bibnamefont
  {Vishwanath}}} (\bibinfo {year} {2018}),\ \bibfield  {title} {\enquote
  {\bibinfo {title} {Weyl and {D}irac semimetals in three-dimensional
  solids},}\ }\href {https://link.aps.org/doi/10.1103/RevModPhys.90.015001}
  {\bibfield  {journal} {\bibinfo  {journal} {Rev. Mod. Phys.}\ }\textbf
  {\bibinfo {volume} {90}},\ \bibinfo {pages} {015001}}\BibitemShut {NoStop}%
\bibitem [{\citenamefont {Arovas}\ \emph {et~al.}(1984)\citenamefont {Arovas},
  \citenamefont {Schrieffer},\ and\ \citenamefont {Wilczek}}]{Arovas:1984PRL}%
  \BibitemOpen
  \bibfield  {author} {\bibinfo {author} {\bibnamefont {Arovas}, \bibfnamefont
  {Daniel}}, \bibinfo {author} {\bibfnamefont {J.~R.}\ \bibnamefont
  {Schrieffer}}, \ and\ \bibinfo {author} {\bibfnamefont {Frank}\ \bibnamefont
  {Wilczek}}} (\bibinfo {year} {1984}),\ \bibfield  {title} {\enquote {\bibinfo
  {title} {Fractional statistics and the quantum {H}all effect},}\ }\href
  {https://link.aps.org/doi/10.1103/PhysRevLett.53.722} {\bibfield  {journal}
  {\bibinfo  {journal} {Phys. Rev. Lett.}\ }\textbf {\bibinfo {volume} {53}},\
  \bibinfo {pages} {722--723}}\BibitemShut {NoStop}%
\bibitem [{\citenamefont {Array{\'a}s}\ \emph {et~al.}(2017)\citenamefont
  {Array{\'a}s}, \citenamefont {Bouwmeester},\ and\ \citenamefont
  {Trueba}}]{Arrayas:2017PhysRep}%
  \BibitemOpen
  \bibfield  {author} {\bibinfo {author} {\bibnamefont {Array{\'a}s},
  \bibfnamefont {M}}, \bibinfo {author} {\bibfnamefont {D}~\bibnamefont
  {Bouwmeester}}, \ and\ \bibinfo {author} {\bibfnamefont {JL}~\bibnamefont
  {Trueba}}} (\bibinfo {year} {2017}),\ \bibfield  {title} {\enquote {\bibinfo
  {title} {Knots in electromagnetism},}\ }\href
  {https://www.sciencedirect.com/science/article/pii/S0370157316303908}
  {\bibfield  {journal} {\bibinfo  {journal} {Physics Reports}\ }\textbf
  {\bibinfo {volume} {667}},\ \bibinfo {pages} {1 -- 61}}\BibitemShut {NoStop}%
\bibitem [{\citenamefont {Asatryan}\ \emph {et~al.}(2013)\citenamefont
  {Asatryan}, \citenamefont {Botten}, \citenamefont {Fang}, \citenamefont
  {Fan},\ and\ \citenamefont {McPhedran}}]{Asatryan:2013PRB}%
  \BibitemOpen
  \bibfield  {author} {\bibinfo {author} {\bibnamefont {Asatryan},
  \bibfnamefont {Ara~A}}, \bibinfo {author} {\bibfnamefont {Lindsay~C}\
  \bibnamefont {Botten}}, \bibinfo {author} {\bibfnamefont {Kejie}\
  \bibnamefont {Fang}}, \bibinfo {author} {\bibfnamefont {Shanhui}\
  \bibnamefont {Fan}}, \ and\ \bibinfo {author} {\bibfnamefont {Ross~C}\
  \bibnamefont {McPhedran}}} (\bibinfo {year} {2013}),\ \bibfield  {title}
  {\enquote {\bibinfo {title} {Local density of states of chiral {H}all edge
  states in gyrotropic photonic clusters},}\ }\href
  {https://journals.aps.org/prb/abstract/10.1103/PhysRevB.88.035127} {\bibfield
   {journal} {\bibinfo  {journal} {Phys. Rev. B}\ }\textbf {\bibinfo {volume}
  {88}}~(\bibinfo {number} {3}),\ \bibinfo {pages} {035127}}\BibitemShut
  {NoStop}%
\bibitem [{\citenamefont {Asatryan}\ \emph {et~al.}(2014)\citenamefont
  {Asatryan}, \citenamefont {Botten}, \citenamefont {Fang}, \citenamefont
  {Fan},\ and\ \citenamefont {McPhedran}}]{Asatryan:2014JOSAA}%
  \BibitemOpen
  \bibfield  {author} {\bibinfo {author} {\bibnamefont {Asatryan},
  \bibfnamefont {Ara~A}}, \bibinfo {author} {\bibfnamefont {Lindsay~C}\
  \bibnamefont {Botten}}, \bibinfo {author} {\bibfnamefont {Kejie}\
  \bibnamefont {Fang}}, \bibinfo {author} {\bibfnamefont {Shanhui}\
  \bibnamefont {Fan}}, \ and\ \bibinfo {author} {\bibfnamefont {Ross~C}\
  \bibnamefont {McPhedran}}} (\bibinfo {year} {2014}),\ \bibfield  {title}
  {\enquote {\bibinfo {title} {Two-dimensional {G}reen's tensor for gyrotropic
  clusters composed of circular cylinders},}\ }\href
  {https://www.osapublishing.org/josaa/abstract.cfm?uri=josaa-31-10-2294}
  {\bibfield  {journal} {\bibinfo  {journal} {J. Opt. Soc. Am. A}\ }\textbf
  {\bibinfo {volume} {31}}~(\bibinfo {number} {10}),\ \bibinfo {pages}
  {2294--2303}}\BibitemShut {NoStop}%
\bibitem [{\citenamefont {Asb\'oth}(2012)}]{Asboth:2012PRB}%
  \BibitemOpen
  \bibfield  {author} {\bibinfo {author} {\bibnamefont {Asb\'oth},
  \bibfnamefont {J~K}}} (\bibinfo {year} {2012}),\ \bibfield  {title} {\enquote
  {\bibinfo {title} {Symmetries, topological phases, and bound states in the
  one-dimensional quantum walk},}\ }\href
  {https://link.aps.org/doi/10.1103/PhysRevB.86.195414} {\bibfield  {journal}
  {\bibinfo  {journal} {Phys. Rev. B}\ }\textbf {\bibinfo {volume} {86}},\
  \bibinfo {pages} {195414}}\BibitemShut {NoStop}%
\bibitem [{\citenamefont {Ashcroft}\ and\ \citenamefont
  {Mermin}(1976)}]{AshcroftMermin}%
  \BibitemOpen
  \bibfield  {author} {\bibinfo {author} {\bibnamefont {Ashcroft},
  \bibfnamefont {Neil~W}}, \ and\ \bibinfo {author} {\bibfnamefont {N.~David}\
  \bibnamefont {Mermin}}} (\bibinfo {year} {1976}),\ \href@noop {} {\emph
  {\bibinfo {title} {Solid state physics}}}\ (\bibinfo  {publisher}
  {Brooks/Cole})\BibitemShut {NoStop}%
\bibitem [{\citenamefont {Aspelmeyer}\ \emph {et~al.}(2014)\citenamefont
  {Aspelmeyer}, \citenamefont {Kippenberg},\ and\ \citenamefont
  {Marquardt}}]{Aspelmeyer2014}%
  \BibitemOpen
  \bibfield  {author} {\bibinfo {author} {\bibnamefont {Aspelmeyer},
  \bibfnamefont {Markus}}, \bibinfo {author} {\bibfnamefont {Tobias~J}\
  \bibnamefont {Kippenberg}}, \ and\ \bibinfo {author} {\bibfnamefont
  {Florian}\ \bibnamefont {Marquardt}}} (\bibinfo {year} {2014}),\ \bibfield
  {title} {\enquote {\bibinfo {title} {Cavity optomechanics},}\ }\href
  {https://doi.org/10.1038/nature08524} {\bibfield  {journal} {\bibinfo
  {journal} {Rev. Mod. Phys.}\ }\textbf {\bibinfo {volume} {86}}~(\bibinfo
  {number} {4}),\ \bibinfo {pages} {1391}}\BibitemShut {NoStop}%
\bibitem [{\citenamefont {Avron}\ \emph {et~al.}(1983)\citenamefont {Avron},
  \citenamefont {Seiler},\ and\ \citenamefont {Simon}}]{Avron:1983PRL}%
  \BibitemOpen
  \bibfield  {author} {\bibinfo {author} {\bibnamefont {Avron}, \bibfnamefont
  {J~E}}, \bibinfo {author} {\bibfnamefont {R.}~\bibnamefont {Seiler}}, \ and\
  \bibinfo {author} {\bibfnamefont {B.}~\bibnamefont {Simon}}} (\bibinfo {year}
  {1983}),\ \bibfield  {title} {\enquote {\bibinfo {title} {Homotopy and
  quantization in condensed matter physics},}\ }\href
  {https://link.aps.org/doi/10.1103/PhysRevLett.51.51} {\bibfield  {journal}
  {\bibinfo  {journal} {Phys. Rev. Lett.}\ }\textbf {\bibinfo {volume} {51}},\
  \bibinfo {pages} {51--53}}\BibitemShut {NoStop}%
\bibitem [{\citenamefont {Azbel}(1964)}]{Azbel:1964JETP}%
  \BibitemOpen
  \bibfield  {author} {\bibinfo {author} {\bibnamefont {Azbel}, \bibfnamefont
  {Mark~Ya}}} (\bibinfo {year} {1964}),\ \bibfield  {title} {\enquote {\bibinfo
  {title} {Energy spectrum of a conduction electron in a magnetic field},}\
  }\href {http://www.jetp.ac.ru/cgi-bin/e/index/e/19/3/p634?a=list} {\bibfield
  {journal} {\bibinfo  {journal} {Sov. Phys. JETP}\ }\textbf {\bibinfo {volume}
  {19}}~(\bibinfo {number} {3}),\ \bibinfo {pages} {634--645}}\BibitemShut
  {NoStop}%
\bibitem [{\citenamefont {Baboux}\ \emph {et~al.}(2017)\citenamefont {Baboux},
  \citenamefont {Levy}, \citenamefont {Lema\^{\i}tre}, \citenamefont {G\'omez},
  \citenamefont {Galopin}, \citenamefont {Le~Gratiet}, \citenamefont {Sagnes},
  \citenamefont {Amo}, \citenamefont {Bloch},\ and\ \citenamefont
  {Akkermans}}]{Baboux:2017PRB}%
  \BibitemOpen
  \bibfield  {author} {\bibinfo {author} {\bibnamefont {Baboux}, \bibfnamefont
  {Florent}}, \bibinfo {author} {\bibfnamefont {Eli}\ \bibnamefont {Levy}},
  \bibinfo {author} {\bibfnamefont {Aristide}\ \bibnamefont {Lema\^{\i}tre}},
  \bibinfo {author} {\bibfnamefont {Carmen}\ \bibnamefont {G\'omez}}, \bibinfo
  {author} {\bibfnamefont {Elisabeth}\ \bibnamefont {Galopin}}, \bibinfo
  {author} {\bibfnamefont {Luc}\ \bibnamefont {Le~Gratiet}}, \bibinfo {author}
  {\bibfnamefont {Isabelle}\ \bibnamefont {Sagnes}}, \bibinfo {author}
  {\bibfnamefont {Alberto}\ \bibnamefont {Amo}}, \bibinfo {author}
  {\bibfnamefont {Jacqueline}\ \bibnamefont {Bloch}}, \ and\ \bibinfo {author}
  {\bibfnamefont {Eric}\ \bibnamefont {Akkermans}}} (\bibinfo {year} {2017}),\
  \bibfield  {title} {\enquote {\bibinfo {title} {Measuring topological
  invariants from generalized edge states in polaritonic quasicrystals},}\
  }\href {https://link.aps.org/doi/10.1103/PhysRevB.95.161114} {\bibfield
  {journal} {\bibinfo  {journal} {Phys. Rev. B}\ }\textbf {\bibinfo {volume}
  {95}},\ \bibinfo {pages} {161114}}\BibitemShut {NoStop}%
\bibitem [{\citenamefont {Bahari}\ \emph {et~al.}(2017)\citenamefont {Bahari},
  \citenamefont {Ndao}, \citenamefont {Vallini}, \citenamefont {El~Amili},
  \citenamefont {Fainman},\ and\ \citenamefont
  {Kant{\'e}}}]{Bahari:Science2017}%
  \BibitemOpen
  \bibfield  {author} {\bibinfo {author} {\bibnamefont {Bahari}, \bibfnamefont
  {Babak}}, \bibinfo {author} {\bibfnamefont {Abdoulaye}\ \bibnamefont {Ndao}},
  \bibinfo {author} {\bibfnamefont {Felipe}\ \bibnamefont {Vallini}}, \bibinfo
  {author} {\bibfnamefont {Abdelkrim}\ \bibnamefont {El~Amili}}, \bibinfo
  {author} {\bibfnamefont {Yeshaiahu}\ \bibnamefont {Fainman}}, \ and\ \bibinfo
  {author} {\bibfnamefont {Boubacar}\ \bibnamefont {Kant{\'e}}}} (\bibinfo
  {year} {2017}),\ \bibfield  {title} {\enquote {\bibinfo {title}
  {Nonreciprocal lasing in topological cavities of arbitrary geometries},}\
  }\href {http://science.sciencemag.org/content/358/6363/636} {\bibfield
  {journal} {\bibinfo  {journal} {Science}\ }\textbf {\bibinfo {volume}
  {358}}~(\bibinfo {number} {6363}),\ \bibinfo {pages} {636--640}}\BibitemShut
  {NoStop}%
\bibitem [{\citenamefont {Bahari}\ \emph {et~al.}(2016)\citenamefont {Bahari},
  \citenamefont {Tellez-Limon},\ and\ \citenamefont
  {Kant{\'e}}}]{Bahari:2016topological}%
  \BibitemOpen
  \bibfield  {author} {\bibinfo {author} {\bibnamefont {Bahari}, \bibfnamefont
  {Babak}}, \bibinfo {author} {\bibfnamefont {Ricardo}\ \bibnamefont
  {Tellez-Limon}}, \ and\ \bibinfo {author} {\bibfnamefont {Boubacar}\
  \bibnamefont {Kant{\'e}}}} (\bibinfo {year} {2016}),\ \bibfield  {title}
  {\enquote {\bibinfo {title} {Topological terahertz circuits using
  semiconductors},}\ }\href
  {https://aip.scitation.org/doi/full/10.1063/1.4963789} {\bibfield  {journal}
  {\bibinfo  {journal} {Appl. Phys. Lett.}\ }\textbf {\bibinfo {volume}
  {109}}~(\bibinfo {number} {14}),\ \bibinfo {pages} {143501}}\BibitemShut
  {NoStop}%
\bibitem [{\citenamefont {Balents}\ and\ \citenamefont
  {Fisher}(1996)}]{Balents:1996PRL}%
  \BibitemOpen
  \bibfield  {author} {\bibinfo {author} {\bibnamefont {Balents}, \bibfnamefont
  {Leon}}, \ and\ \bibinfo {author} {\bibfnamefont {Matthew P.~A.}\
  \bibnamefont {Fisher}}} (\bibinfo {year} {1996}),\ \bibfield  {title}
  {\enquote {\bibinfo {title} {Chiral surface states in the bulk quantum {H}all
  effect},}\ }\href {https://link.aps.org/doi/10.1103/PhysRevLett.76.2782}
  {\bibfield  {journal} {\bibinfo  {journal} {Phys. Rev. Lett.}\ }\textbf
  {\bibinfo {volume} {76}},\ \bibinfo {pages} {2782--2785}}\BibitemShut
  {NoStop}%
\bibitem [{\citenamefont {Bandres}\ \emph {et~al.}(2016)\citenamefont
  {Bandres}, \citenamefont {Rechtsman},\ and\ \citenamefont
  {Segev}}]{Bandres:2016PRX}%
  \BibitemOpen
  \bibfield  {author} {\bibinfo {author} {\bibnamefont {Bandres}, \bibfnamefont
  {Miguel~A}}, \bibinfo {author} {\bibfnamefont {Mikael~C.}\ \bibnamefont
  {Rechtsman}}, \ and\ \bibinfo {author} {\bibfnamefont {Mordechai}\
  \bibnamefont {Segev}}} (\bibinfo {year} {2016}),\ \bibfield  {title}
  {\enquote {\bibinfo {title} {Topological photonic quasicrystals: {F}ractal
  topological spectrum and protected transport},}\ }\href
  {http://link.aps.org/doi/10.1103/PhysRevX.6.011016} {\bibfield  {journal}
  {\bibinfo  {journal} {Phys. Rev. X}\ }\textbf {\bibinfo {volume} {6}},\
  \bibinfo {pages} {011016}}\BibitemShut {NoStop}%
\bibitem [{\citenamefont {Bandres}\ \emph {et~al.}(2018)\citenamefont
  {Bandres}, \citenamefont {Wittek}, \citenamefont {Harari}, \citenamefont
  {Parto}, \citenamefont {Ren}, \citenamefont {Segev}, \citenamefont
  {Christodoulides},\ and\ \citenamefont {Khajavikhan}}]{Bandres:Science2018}%
  \BibitemOpen
  \bibfield  {author} {\bibinfo {author} {\bibnamefont {Bandres}, \bibfnamefont
  {Miguel~A}}, \bibinfo {author} {\bibfnamefont {Steffen}\ \bibnamefont
  {Wittek}}, \bibinfo {author} {\bibfnamefont {Gal}\ \bibnamefont {Harari}},
  \bibinfo {author} {\bibfnamefont {Midya}\ \bibnamefont {Parto}}, \bibinfo
  {author} {\bibfnamefont {Jinhan}\ \bibnamefont {Ren}}, \bibinfo {author}
  {\bibfnamefont {Mordechai}\ \bibnamefont {Segev}}, \bibinfo {author}
  {\bibfnamefont {Demetrios~N.}\ \bibnamefont {Christodoulides}}, \ and\
  \bibinfo {author} {\bibfnamefont {Mercedeh}\ \bibnamefont {Khajavikhan}}}
  (\bibinfo {year} {2018}),\ \bibfield  {title} {\enquote {\bibinfo {title}
  {{Topological insulator laser: Experiments}},}\ }\href
  {http://science.sciencemag.org/content/early/2018/01/31/science.aar4005}
  {\bibfield  {journal} {\bibinfo  {journal} {Science}\ }\textbf {\bibinfo
  {volume} {359}},\ \bibinfo {pages} {eaar4005}}\BibitemShut {NoStop}%
\bibitem [{\citenamefont {Bansil}\ \emph {et~al.}(2016)\citenamefont {Bansil},
  \citenamefont {Lin},\ and\ \citenamefont {Das}}]{Bansil:2016RMP}%
  \BibitemOpen
  \bibfield  {author} {\bibinfo {author} {\bibnamefont {Bansil}, \bibfnamefont
  {A}}, \bibinfo {author} {\bibfnamefont {Hsin}\ \bibnamefont {Lin}}, \ and\
  \bibinfo {author} {\bibfnamefont {Tanmoy}\ \bibnamefont {Das}}} (\bibinfo
  {year} {2016}),\ \bibfield  {title} {\enquote {\bibinfo {title} {Colloquium:
  {T}opological band theory},}\ }\href
  {https://link.aps.org/doi/10.1103/RevModPhys.88.021004} {\bibfield  {journal}
  {\bibinfo  {journal} {Rev. Mod. Phys.}\ }\textbf {\bibinfo {volume} {88}},\
  \bibinfo {pages} {021004}}\BibitemShut {NoStop}%
\bibitem [{\citenamefont {Barbarino}\ \emph {et~al.}(2016)\citenamefont
  {Barbarino}, \citenamefont {Taddia}, \citenamefont {Rossini}, \citenamefont
  {Mazza},\ and\ \citenamefont {Fazio}}]{Barbarino:2016NJP}%
  \BibitemOpen
  \bibfield  {author} {\bibinfo {author} {\bibnamefont {Barbarino},
  \bibfnamefont {Simone}}, \bibinfo {author} {\bibfnamefont {Luca}\
  \bibnamefont {Taddia}}, \bibinfo {author} {\bibfnamefont {Davide}\
  \bibnamefont {Rossini}}, \bibinfo {author} {\bibfnamefont {Leonardo}\
  \bibnamefont {Mazza}}, \ and\ \bibinfo {author} {\bibfnamefont {Rosario}\
  \bibnamefont {Fazio}}} (\bibinfo {year} {2016}),\ \bibfield  {title}
  {\enquote {\bibinfo {title} {Synthetic gauge fields in synthetic dimensions:
  {I}nteractions and chiral edge modes},}\ }\href
  {http://iopscience.iop.org/article/10.1088/1367-2630/18/3/035010/meta}
  {\bibfield  {journal} {\bibinfo  {journal} {New J. Phys.}\ }\textbf {\bibinfo
  {volume} {18}}~(\bibinfo {number} {3}),\ \bibinfo {pages}
  {035010}}\BibitemShut {NoStop}%
\bibitem [{\citenamefont {Bardyn}\ \emph {et~al.}(2013)\citenamefont {Bardyn},
  \citenamefont {Baranov}, \citenamefont {Kraus}, \citenamefont {Rico},
  \citenamefont {İmamoğlu}, \citenamefont {Zoller},\ and\ \citenamefont
  {Diehl}}]{1367-2630-15-8-085001}%
  \BibitemOpen
  \bibfield  {author} {\bibinfo {author} {\bibnamefont {Bardyn}, \bibfnamefont
  {C-E}}, \bibinfo {author} {\bibfnamefont {M~A}\ \bibnamefont {Baranov}},
  \bibinfo {author} {\bibfnamefont {C~V}\ \bibnamefont {Kraus}}, \bibinfo
  {author} {\bibfnamefont {E}~\bibnamefont {Rico}}, \bibinfo {author}
  {\bibfnamefont {A}~\bibnamefont {İmamoğlu}}, \bibinfo {author}
  {\bibfnamefont {P}~\bibnamefont {Zoller}}, \ and\ \bibinfo {author}
  {\bibfnamefont {S}~\bibnamefont {Diehl}}} (\bibinfo {year} {2013}),\
  \bibfield  {title} {\enquote {\bibinfo {title} {Topology by dissipation},}\
  }\href {http://stacks.iop.org/1367-2630/15/i=8/a=085001} {\bibfield
  {journal} {\bibinfo  {journal} {New J. Phys.}\ }\textbf {\bibinfo {volume}
  {15}}~(\bibinfo {number} {8}),\ \bibinfo {pages} {085001}}\BibitemShut
  {NoStop}%
\bibitem [{\citenamefont {Bardyn}\ \emph {et~al.}(2012)\citenamefont {Bardyn},
  \citenamefont {Baranov}, \citenamefont {Rico}, \citenamefont {\ifmmode
  \dot{I}\else \.{I}\fi{}mamo\ifmmode~\breve{g}\else \u{g}\fi{}lu},
  \citenamefont {Zoller},\ and\ \citenamefont
  {Diehl}}]{PhysRevLett.109.130402}%
  \BibitemOpen
  \bibfield  {author} {\bibinfo {author} {\bibnamefont {Bardyn}, \bibfnamefont
  {C-E}}, \bibinfo {author} {\bibfnamefont {M.~A.}\ \bibnamefont {Baranov}},
  \bibinfo {author} {\bibfnamefont {E.}~\bibnamefont {Rico}}, \bibinfo {author}
  {\bibfnamefont {A.}~\bibnamefont {\ifmmode \dot{I}\else
  \.{I}\fi{}mamo\ifmmode~\breve{g}\else \u{g}\fi{}lu}}, \bibinfo {author}
  {\bibfnamefont {P.}~\bibnamefont {Zoller}}, \ and\ \bibinfo {author}
  {\bibfnamefont {S.}~\bibnamefont {Diehl}}} (\bibinfo {year} {2012}),\
  \bibfield  {title} {\enquote {\bibinfo {title} {Majorana modes in
  driven-dissipative atomic superfluids with a zero {C}hern number},}\ }\href
  {https://link.aps.org/doi/10.1103/PhysRevLett.109.130402} {\bibfield
  {journal} {\bibinfo  {journal} {Phys. Rev. Lett.}\ }\textbf {\bibinfo
  {volume} {109}},\ \bibinfo {pages} {130402}}\BibitemShut {NoStop}%
\bibitem [{\citenamefont {Bardyn}\ and\ \citenamefont {\ifmmode \dot{I}\else
  \.{I}\fi{}mamo\ifmmode~\check{g}\else \v{g}\fi{}lu}(2012)}]{Bardyn:2012PRL}%
  \BibitemOpen
  \bibfield  {author} {\bibinfo {author} {\bibnamefont {Bardyn}, \bibfnamefont
  {C-E}}, \ and\ \bibinfo {author} {\bibfnamefont {A.}~\bibnamefont {\ifmmode
  \dot{I}\else \.{I}\fi{}mamo\ifmmode~\check{g}\else \v{g}\fi{}lu}}} (\bibinfo
  {year} {2012}),\ \bibfield  {title} {\enquote {\bibinfo {title}
  {Majorana-like modes of light in a one-dimensional array of nonlinear
  cavities},}\ }\href {http://link.aps.org/doi/10.1103/PhysRevLett.109.253606}
  {\bibfield  {journal} {\bibinfo  {journal} {Phys. Rev. Lett.}\ }\textbf
  {\bibinfo {volume} {109}},\ \bibinfo {pages} {253606}}\BibitemShut {NoStop}%
\bibitem [{\citenamefont {Bardyn}\ \emph {et~al.}(2014)\citenamefont {Bardyn},
  \citenamefont {Huber},\ and\ \citenamefont {Zilberberg}}]{Bardyn:2014NJP}%
  \BibitemOpen
  \bibfield  {author} {\bibinfo {author} {\bibnamefont {Bardyn}, \bibfnamefont
  {Charles-Edouard}}, \bibinfo {author} {\bibfnamefont {Sebastian~D}\
  \bibnamefont {Huber}}, \ and\ \bibinfo {author} {\bibfnamefont {Oded}\
  \bibnamefont {Zilberberg}}} (\bibinfo {year} {2014}),\ \bibfield  {title}
  {\enquote {\bibinfo {title} {Measuring topological invariants in small
  photonic lattices},}\ }\href
  {http://iopscience.iop.org/article/10.1088/1367-2630/16/12/123013/meta}
  {\bibfield  {journal} {\bibinfo  {journal} {New J. Phys.}\ }\textbf {\bibinfo
  {volume} {16}}~(\bibinfo {number} {12}),\ \bibinfo {pages}
  {123013}}\BibitemShut {NoStop}%
\bibitem [{\citenamefont {Bardyn}\ \emph {et~al.}(2015)\citenamefont {Bardyn},
  \citenamefont {Karzig}, \citenamefont {Refael},\ and\ \citenamefont
  {Liew}}]{Bardyn:2015PRB}%
  \BibitemOpen
  \bibfield  {author} {\bibinfo {author} {\bibnamefont {Bardyn}, \bibfnamefont
  {Charles-Edouard}}, \bibinfo {author} {\bibfnamefont {Torsten}\ \bibnamefont
  {Karzig}}, \bibinfo {author} {\bibfnamefont {Gil}\ \bibnamefont {Refael}}, \
  and\ \bibinfo {author} {\bibfnamefont {Timothy C.~H.}\ \bibnamefont {Liew}}}
  (\bibinfo {year} {2015}),\ \bibfield  {title} {\enquote {\bibinfo {title}
  {Topological polaritons and excitons in garden-variety systems},}\ }\href
  {http://link.aps.org/doi/10.1103/PhysRevB.91.161413} {\bibfield  {journal}
  {\bibinfo  {journal} {Phys. Rev. B}\ }\textbf {\bibinfo {volume} {91}},\
  \bibinfo {pages} {161413}}\BibitemShut {NoStop}%
\bibitem [{\citenamefont {Bardyn}\ \emph {et~al.}(2016)\citenamefont {Bardyn},
  \citenamefont {Karzig}, \citenamefont {Refael},\ and\ \citenamefont
  {Liew}}]{Bardyn:2016PRB}%
  \BibitemOpen
  \bibfield  {author} {\bibinfo {author} {\bibnamefont {Bardyn}, \bibfnamefont
  {Charles-Edouard}}, \bibinfo {author} {\bibfnamefont {Torsten}\ \bibnamefont
  {Karzig}}, \bibinfo {author} {\bibfnamefont {Gil}\ \bibnamefont {Refael}}, \
  and\ \bibinfo {author} {\bibfnamefont {Timothy C.~H.}\ \bibnamefont {Liew}}}
  (\bibinfo {year} {2016}),\ \bibfield  {title} {\enquote {\bibinfo {title}
  {Chiral {B}ogoliubov excitations in nonlinear bosonic systems},}\ }\href
  {https://link.aps.org/doi/10.1103/PhysRevB.93.020502} {\bibfield  {journal}
  {\bibinfo  {journal} {Phys. Rev. B}\ }\textbf {\bibinfo {volume} {93}},\
  \bibinfo {pages} {020502}}\BibitemShut {NoStop}%
\bibitem [{\citenamefont {Barik}\ \emph {et~al.}(2018)\citenamefont {Barik},
  \citenamefont {Karasahin}, \citenamefont {Flower}, \citenamefont {Cai},
  \citenamefont {Miyake}, \citenamefont {DeGottardi}, \citenamefont {Hafezi},\
  and\ \citenamefont {Waks}}]{Barik:Science2018}%
  \BibitemOpen
  \bibfield  {author} {\bibinfo {author} {\bibnamefont {Barik}, \bibfnamefont
  {Sabyasachi}}, \bibinfo {author} {\bibfnamefont {Aziz}\ \bibnamefont
  {Karasahin}}, \bibinfo {author} {\bibfnamefont {Christopher}\ \bibnamefont
  {Flower}}, \bibinfo {author} {\bibfnamefont {Tao}\ \bibnamefont {Cai}},
  \bibinfo {author} {\bibfnamefont {Hirokazu}\ \bibnamefont {Miyake}}, \bibinfo
  {author} {\bibfnamefont {Wade}\ \bibnamefont {DeGottardi}}, \bibinfo {author}
  {\bibfnamefont {Mohammad}\ \bibnamefont {Hafezi}}, \ and\ \bibinfo {author}
  {\bibfnamefont {Edo}\ \bibnamefont {Waks}}} (\bibinfo {year} {2018}),\
  \bibfield  {title} {\enquote {\bibinfo {title} {A topological quantum optics
  interface},}\ }\href {http://science.sciencemag.org/content/359/6376/666}
  {\bibfield  {journal} {\bibinfo  {journal} {Science}\ }\textbf {\bibinfo
  {volume} {359}}~(\bibinfo {number} {6376}),\ \bibinfo {pages}
  {666--668}}\BibitemShut {NoStop}%
\bibitem [{\citenamefont {Barik}\ \emph {et~al.}(2016)\citenamefont {Barik},
  \citenamefont {Miyake}, \citenamefont {DeGottardi}, \citenamefont {Waks},\
  and\ \citenamefont {Hafezi}}]{Barik:NJP2016}%
  \BibitemOpen
  \bibfield  {author} {\bibinfo {author} {\bibnamefont {Barik}, \bibfnamefont
  {Sabyasachi}}, \bibinfo {author} {\bibfnamefont {Hirokazu}\ \bibnamefont
  {Miyake}}, \bibinfo {author} {\bibfnamefont {Wade}\ \bibnamefont
  {DeGottardi}}, \bibinfo {author} {\bibfnamefont {Edo}\ \bibnamefont {Waks}},
  \ and\ \bibinfo {author} {\bibfnamefont {Mohammad}\ \bibnamefont {Hafezi}}}
  (\bibinfo {year} {2016}),\ \bibfield  {title} {\enquote {\bibinfo {title}
  {{Two-dimensionally confined topological edge states in photonic
  crystals}},}\ }\href
  {https://iopscience.iop.org/article/10.1088/1367-2630/18/11/113013/meta}
  {\bibfield  {journal} {\bibinfo  {journal} {New J. Phys.}\ }\textbf {\bibinfo
  {volume} {18}}~(\bibinfo {number} {11}),\ \bibinfo {pages}
  {113013}}\BibitemShut {NoStop}%
\bibitem [{\citenamefont {Barkhofen}\ \emph {et~al.}(2017)\citenamefont
  {Barkhofen}, \citenamefont {Nitsche}, \citenamefont {Elster}, \citenamefont
  {Lorz}, \citenamefont {G\'abris}, \citenamefont {Jex},\ and\ \citenamefont
  {Silberhorn}}]{Barkhofen:2017PRA}%
  \BibitemOpen
  \bibfield  {author} {\bibinfo {author} {\bibnamefont {Barkhofen},
  \bibfnamefont {Sonja}}, \bibinfo {author} {\bibfnamefont {Thomas}\
  \bibnamefont {Nitsche}}, \bibinfo {author} {\bibfnamefont {Fabian}\
  \bibnamefont {Elster}}, \bibinfo {author} {\bibfnamefont {Lennart}\
  \bibnamefont {Lorz}}, \bibinfo {author} {\bibfnamefont {Aur\'el}\
  \bibnamefont {G\'abris}}, \bibinfo {author} {\bibfnamefont {Igor}\
  \bibnamefont {Jex}}, \ and\ \bibinfo {author} {\bibfnamefont {Christine}\
  \bibnamefont {Silberhorn}}} (\bibinfo {year} {2017}),\ \bibfield  {title}
  {\enquote {\bibinfo {title} {Measuring topological invariants in disordered
  discrete-time quantum walks},}\ }\href
  {https://link.aps.org/doi/10.1103/PhysRevA.96.033846} {\bibfield  {journal}
  {\bibinfo  {journal} {Phys. Rev. A}\ }\textbf {\bibinfo {volume} {96}},\
  \bibinfo {pages} {033846}}\BibitemShut {NoStop}%
\bibitem [{\citenamefont {Barnett}(2013)}]{Barnett:2013PRA}%
  \BibitemOpen
  \bibfield  {author} {\bibinfo {author} {\bibnamefont {Barnett}, \bibfnamefont
  {Ryan}}} (\bibinfo {year} {2013}),\ \bibfield  {title} {\enquote {\bibinfo
  {title} {Edge-state instabilities of bosons in a topological band},}\ }\href
  {http://link.aps.org/doi/10.1103/PhysRevA.88.063631} {\bibfield  {journal}
  {\bibinfo  {journal} {Phys. Rev. A}\ }\textbf {\bibinfo {volume} {88}},\
  \bibinfo {pages} {063631}}\BibitemShut {NoStop}%
\bibitem [{\citenamefont {Bayindir}\ \emph {et~al.}(2000)\citenamefont
  {Bayindir}, \citenamefont {Temelkuran},\ and\ \citenamefont
  {Ozbay}}]{Bayindir:PRL2000}%
  \BibitemOpen
  \bibfield  {author} {\bibinfo {author} {\bibnamefont {Bayindir},
  \bibfnamefont {Mehmet}}, \bibinfo {author} {\bibfnamefont {B}~\bibnamefont
  {Temelkuran}}, \ and\ \bibinfo {author} {\bibfnamefont {E}~\bibnamefont
  {Ozbay}}} (\bibinfo {year} {2000}),\ \bibfield  {title} {\enquote {\bibinfo
  {title} {Tight-binding description of the coupled defect modes in
  three-dimensional photonic crystals},}\ }\href
  {https://journals.aps.org/prl/abstract/10.1103/PhysRevLett.84.2140}
  {\bibfield  {journal} {\bibinfo  {journal} {Phys. Rev. Lett.}\ }\textbf
  {\bibinfo {volume} {84}}~(\bibinfo {number} {10}),\ \bibinfo {pages}
  {2140}}\BibitemShut {NoStop}%
\bibitem [{\citenamefont {Bellec}\ \emph {et~al.}(2017)\citenamefont {Bellec},
  \citenamefont {Michel}, \citenamefont {Zhang}, \citenamefont {Tzortzakis},\
  and\ \citenamefont {Delplace}}]{Bellec:2017EPL}%
  \BibitemOpen
  \bibfield  {author} {\bibinfo {author} {\bibnamefont {Bellec}, \bibfnamefont
  {M}}, \bibinfo {author} {\bibfnamefont {C.}~\bibnamefont {Michel}}, \bibinfo
  {author} {\bibfnamefont {H.}~\bibnamefont {Zhang}}, \bibinfo {author}
  {\bibfnamefont {S.}~\bibnamefont {Tzortzakis}}, \ and\ \bibinfo {author}
  {\bibfnamefont {P.}~\bibnamefont {Delplace}}} (\bibinfo {year} {2017}),\
  \bibfield  {title} {\enquote {\bibinfo {title} {Non-diffracting states in
  one-dimensional {F}loquet photonic topological insulators},}\ }\href
  {http://iopscience.iop.org/article/10.1209/0295-5075/119/14003} {\bibfield
  {journal} {\bibinfo  {journal} {EPL}\ }\textbf {\bibinfo {volume} {119}},\
  \bibinfo {pages} {14003}}\BibitemShut {NoStop}%
\bibitem [{\citenamefont {Bellec}\ \emph
  {et~al.}(2013{\natexlab{a}})\citenamefont {Bellec}, \citenamefont {Kuhl},
  \citenamefont {Montambaux},\ and\ \citenamefont
  {Mortessagne}}]{Bellec:2013PRB}%
  \BibitemOpen
  \bibfield  {author} {\bibinfo {author} {\bibnamefont {Bellec}, \bibfnamefont
  {Matthieu}}, \bibinfo {author} {\bibfnamefont {Ulrich}\ \bibnamefont {Kuhl}},
  \bibinfo {author} {\bibfnamefont {Gilles}\ \bibnamefont {Montambaux}}, \ and\
  \bibinfo {author} {\bibfnamefont {Fabrice}\ \bibnamefont {Mortessagne}}}
  (\bibinfo {year} {2013}{\natexlab{a}}),\ \bibfield  {title} {\enquote
  {\bibinfo {title} {Tight-binding couplings in microwave artificial
  graphene},}\ }\href {http://link.aps.org/doi/10.1103/PhysRevB.88.115437}
  {\bibfield  {journal} {\bibinfo  {journal} {Phys. Rev. B}\ }\textbf {\bibinfo
  {volume} {88}},\ \bibinfo {pages} {115437}}\BibitemShut {NoStop}%
\bibitem [{\citenamefont {Bellec}\ \emph
  {et~al.}(2013{\natexlab{b}})\citenamefont {Bellec}, \citenamefont {Kuhl},
  \citenamefont {Montambaux},\ and\ \citenamefont
  {Mortessagne}}]{Bellec:2013PRL}%
  \BibitemOpen
  \bibfield  {author} {\bibinfo {author} {\bibnamefont {Bellec}, \bibfnamefont
  {Matthieu}}, \bibinfo {author} {\bibfnamefont {Ulrich}\ \bibnamefont {Kuhl}},
  \bibinfo {author} {\bibfnamefont {Gilles}\ \bibnamefont {Montambaux}}, \ and\
  \bibinfo {author} {\bibfnamefont {Fabrice}\ \bibnamefont {Mortessagne}}}
  (\bibinfo {year} {2013}{\natexlab{b}}),\ \bibfield  {title} {\enquote
  {\bibinfo {title} {{Topological transition of Dirac Points in a microwave
  experiment}},}\ }\href
  {http://link.aps.org/doi/10.1103/PhysRevLett.110.033902} {\bibfield
  {journal} {\bibinfo  {journal} {Phys. Rev. Lett.}\ }\textbf {\bibinfo
  {volume} {110}}~(\bibinfo {number} {3}),\ \bibinfo {pages}
  {033902}}\BibitemShut {NoStop}%
\bibitem [{\citenamefont {Bellec}\ \emph {et~al.}(2014)\citenamefont {Bellec},
  \citenamefont {Kuhl}, \citenamefont {Montambaux},\ and\ \citenamefont
  {Mortessagne}}]{Bellec:2014NJP}%
  \BibitemOpen
  \bibfield  {author} {\bibinfo {author} {\bibnamefont {Bellec}, \bibfnamefont
  {Matthieu}}, \bibinfo {author} {\bibfnamefont {Ulrich}\ \bibnamefont {Kuhl}},
  \bibinfo {author} {\bibfnamefont {Gilles}\ \bibnamefont {Montambaux}}, \ and\
  \bibinfo {author} {\bibfnamefont {Fabrice}\ \bibnamefont {Mortessagne}}}
  (\bibinfo {year} {2014}),\ \bibfield  {title} {\enquote {\bibinfo {title}
  {Manipulation of edge states in microwave artificial graphene},}\ }\href
  {http://iopscience.iop.org/article/10.1088/1367-2630/16/11/113023} {\bibfield
   {journal} {\bibinfo  {journal} {New J. Phys.}\ }\textbf {\bibinfo {volume}
  {16}}~(\bibinfo {number} {11}),\ \bibinfo {pages} {113023}}\BibitemShut
  {NoStop}%
\bibitem [{\citenamefont {Bender}\ and\ \citenamefont
  {Boettcher}(1998)}]{bender1998real}%
  \BibitemOpen
  \bibfield  {author} {\bibinfo {author} {\bibnamefont {Bender}, \bibfnamefont
  {Carl~M}}, \ and\ \bibinfo {author} {\bibfnamefont {Stefan}\ \bibnamefont
  {Boettcher}}} (\bibinfo {year} {1998}),\ \bibfield  {title} {\enquote
  {\bibinfo {title} {Real spectra in non-{H}ermitian {H}amiltonians having
  $\mathcal{PT}$ symmetry},}\ }\href
  {https://journals.aps.org/prl/abstract/10.1103/PhysRevLett.80.5243}
  {\bibfield  {journal} {\bibinfo  {journal} {Phys. Rev. Lett.}\ }\textbf
  {\bibinfo {volume} {80}}~(\bibinfo {number} {24}),\ \bibinfo {pages}
  {5243}}\BibitemShut {NoStop}%
\bibitem [{\citenamefont {Berceanu}\ \emph {et~al.}(2016)\citenamefont
  {Berceanu}, \citenamefont {Price}, \citenamefont {Ozawa},\ and\ \citenamefont
  {Carusotto}}]{Berceanu:2016PRA}%
  \BibitemOpen
  \bibfield  {author} {\bibinfo {author} {\bibnamefont {Berceanu},
  \bibfnamefont {Andrei~C}}, \bibinfo {author} {\bibfnamefont {Hannah~M.}\
  \bibnamefont {Price}}, \bibinfo {author} {\bibfnamefont {Tomoki}\
  \bibnamefont {Ozawa}}, \ and\ \bibinfo {author} {\bibfnamefont {Iacopo}\
  \bibnamefont {Carusotto}}} (\bibinfo {year} {2016}),\ \bibfield  {title}
  {\enquote {\bibinfo {title} {Momentum-space {L}andau levels in
  driven-dissipative cavity arrays},}\ }\href
  {http://link.aps.org/doi/10.1103/PhysRevA.93.013827} {\bibfield  {journal}
  {\bibinfo  {journal} {Phys. Rev. A}\ }\textbf {\bibinfo {volume} {93}},\
  \bibinfo {pages} {013827}}\BibitemShut {NoStop}%
\bibitem [{\citenamefont {Bermudez}\ \emph {et~al.}(2011)\citenamefont
  {Bermudez}, \citenamefont {Schaetz},\ and\ \citenamefont
  {Porras}}]{Bermudez:2011PRL}%
  \BibitemOpen
  \bibfield  {author} {\bibinfo {author} {\bibnamefont {Bermudez},
  \bibfnamefont {Alejandro}}, \bibinfo {author} {\bibfnamefont {Tobias}\
  \bibnamefont {Schaetz}}, \ and\ \bibinfo {author} {\bibfnamefont {Diego}\
  \bibnamefont {Porras}}} (\bibinfo {year} {2011}),\ \bibfield  {title}
  {\enquote {\bibinfo {title} {Synthetic gauge fields for vibrational
  excitations of trapped ions},}\ }\href
  {https://link.aps.org/doi/10.1103/PhysRevLett.107.150501} {\bibfield
  {journal} {\bibinfo  {journal} {Phys. Rev. Lett.}\ }\textbf {\bibinfo
  {volume} {107}},\ \bibinfo {pages} {150501}}\BibitemShut {NoStop}%
\bibitem [{\citenamefont {Bernevig}\ and\ \citenamefont
  {Hughes}(2013)}]{BernevigBook}%
  \BibitemOpen
  \bibfield  {author} {\bibinfo {author} {\bibnamefont {Bernevig},
  \bibfnamefont {B~Andrei}}, \ and\ \bibinfo {author} {\bibfnamefont
  {Taylor~L.}\ \bibnamefont {Hughes}}} (\bibinfo {year} {2013}),\ \href@noop {}
  {\emph {\bibinfo {title} {Topological insulators and topological
  superconductors}}}\ (\bibinfo  {publisher} {Princeton University Press},\
  \bibinfo {address} {Princeton, NJ})\BibitemShut {NoStop}%
\bibitem [{\citenamefont {Bernevig}\ \emph {et~al.}(2007)\citenamefont
  {Bernevig}, \citenamefont {Hughes}, \citenamefont {Raghu},\ and\
  \citenamefont {Arovas}}]{bernevig:2018PRL}%
  \BibitemOpen
  \bibfield  {author} {\bibinfo {author} {\bibnamefont {Bernevig},
  \bibfnamefont {B~Andrei}}, \bibinfo {author} {\bibfnamefont {Taylor~L.}\
  \bibnamefont {Hughes}}, \bibinfo {author} {\bibfnamefont {Srinivas}\
  \bibnamefont {Raghu}}, \ and\ \bibinfo {author} {\bibfnamefont {Daniel~P.}\
  \bibnamefont {Arovas}}} (\bibinfo {year} {2007}),\ \bibfield  {title}
  {\enquote {\bibinfo {title} {Theory of the three-dimensional quantum {H}all
  effect in graphite},}\ }\href
  {https://link.aps.org/doi/10.1103/PhysRevLett.99.146804} {\bibfield
  {journal} {\bibinfo  {journal} {Phys. Rev. Lett.}\ }\textbf {\bibinfo
  {volume} {99}},\ \bibinfo {pages} {146804}}\BibitemShut {NoStop}%
\bibitem [{\citenamefont {Bernevig}\ \emph {et~al.}(2006)\citenamefont
  {Bernevig}, \citenamefont {Hughes},\ and\ \citenamefont
  {Zhang}}]{Bernevig:2006Science}%
  \BibitemOpen
  \bibfield  {author} {\bibinfo {author} {\bibnamefont {Bernevig},
  \bibfnamefont {B~Andrei}}, \bibinfo {author} {\bibfnamefont {Taylor~L.}\
  \bibnamefont {Hughes}}, \ and\ \bibinfo {author} {\bibfnamefont {Shou-Cheng}\
  \bibnamefont {Zhang}}} (\bibinfo {year} {2006}),\ \bibfield  {title}
  {\enquote {\bibinfo {title} {Quantum spin {H}all effect and topological phase
  transition in {HgTe} quantum wells},}\ }\href
  {http://science.sciencemag.org/content/314/5806/1757} {\bibfield  {journal}
  {\bibinfo  {journal} {Science}\ }\textbf {\bibinfo {volume} {314}}~(\bibinfo
  {number} {5806}),\ \bibinfo {pages} {1757--1761}}\BibitemShut {NoStop}%
\bibitem [{\citenamefont {Bernevig}\ and\ \citenamefont
  {Zhang}(2006)}]{Bernevig:2006PRL}%
  \BibitemOpen
  \bibfield  {author} {\bibinfo {author} {\bibnamefont {Bernevig},
  \bibfnamefont {B~Andrei}}, \ and\ \bibinfo {author} {\bibfnamefont
  {Shou-Cheng}\ \bibnamefont {Zhang}}} (\bibinfo {year} {2006}),\ \bibfield
  {title} {\enquote {\bibinfo {title} {Quantum spin {H}all effect},}\ }\href
  {https://link.aps.org/doi/10.1103/PhysRevLett.96.106802} {\bibfield
  {journal} {\bibinfo  {journal} {Phys. Rev. Lett.}\ }\textbf {\bibinfo
  {volume} {96}},\ \bibinfo {pages} {106802}}\BibitemShut {NoStop}%
\bibitem [{\citenamefont {Berry}(1984)}]{Berry:1984PRSLA}%
  \BibitemOpen
  \bibfield  {author} {\bibinfo {author} {\bibnamefont {Berry}, \bibfnamefont
  {Michael~V}}} (\bibinfo {year} {1984}),\ \bibfield  {title} {\enquote
  {\bibinfo {title} {Quantal phase factors accompanying adiabatic changes},}\
  }\href {http://rspa.royalsocietypublishing.org/content/392/1802/45}
  {\bibfield  {journal} {\bibinfo  {journal} {Proc. R. Soc. A}\ }\textbf
  {\bibinfo {volume} {392}}~(\bibinfo {number} {1802}),\ \bibinfo {pages}
  {45--57}}\BibitemShut {NoStop}%
\bibitem [{\citenamefont {Bi}\ and\ \citenamefont {Wang}(2015)}]{Bi:2015PRB}%
  \BibitemOpen
  \bibfield  {author} {\bibinfo {author} {\bibnamefont {Bi}, \bibfnamefont
  {Ren}}, \ and\ \bibinfo {author} {\bibfnamefont {Zhong}\ \bibnamefont
  {Wang}}} (\bibinfo {year} {2015}),\ \bibfield  {title} {\enquote {\bibinfo
  {title} {Unidirectional transport in electronic and photonic {W}eyl materials
  by {D}irac mass engineering},}\ }\href
  {https://journals.aps.org/prb/abstract/10.1103/PhysRevB.92.241109} {\bibfield
   {journal} {\bibinfo  {journal} {Phys. Rev. B}\ }\textbf {\bibinfo {volume}
  {92}}~(\bibinfo {number} {24}),\ \bibinfo {pages} {241109}}\BibitemShut
  {NoStop}%
\bibitem [{\citenamefont {Bi}\ \emph {et~al.}(2017{\natexlab{a}})\citenamefont
  {Bi}, \citenamefont {Yan}, \citenamefont {Lu},\ and\ \citenamefont
  {Wang}}]{Bi:2017arXiv}%
  \BibitemOpen
  \bibfield  {author} {\bibinfo {author} {\bibnamefont {Bi}, \bibfnamefont
  {Ren}}, \bibinfo {author} {\bibfnamefont {Zhongbo}\ \bibnamefont {Yan}},
  \bibinfo {author} {\bibfnamefont {Ling}\ \bibnamefont {Lu}}, \ and\ \bibinfo
  {author} {\bibfnamefont {Zhong}\ \bibnamefont {Wang}}} (\bibinfo {year}
  {2017}{\natexlab{a}}),\ \bibfield  {title} {\enquote {\bibinfo {title}
  {Nodal-knot semimetals},}\ }\href
  {https://link.aps.org/doi/10.1103/PhysRevB.96.201305} {\bibfield  {journal}
  {\bibinfo  {journal} {Phys. Rev. B}\ }\textbf {\bibinfo {volume} {96}},\
  \bibinfo {pages} {201305}}\BibitemShut {NoStop}%
\bibitem [{\citenamefont {Bi}\ \emph {et~al.}(2017{\natexlab{b}})\citenamefont
  {Bi}, \citenamefont {Yan}, \citenamefont {Lu},\ and\ \citenamefont
  {Wang}}]{Bi:2017PRB}%
  \BibitemOpen
  \bibfield  {author} {\bibinfo {author} {\bibnamefont {Bi}, \bibfnamefont
  {Ren}}, \bibinfo {author} {\bibfnamefont {Zhongbo}\ \bibnamefont {Yan}},
  \bibinfo {author} {\bibfnamefont {Ling}\ \bibnamefont {Lu}}, \ and\ \bibinfo
  {author} {\bibfnamefont {Zhong}\ \bibnamefont {Wang}}} (\bibinfo {year}
  {2017}{\natexlab{b}}),\ \bibfield  {title} {\enquote {\bibinfo {title}
  {Topological defects in {F}loquet systems: {A}nomalous chiral modes and
  topological invariant},}\ }\href
  {https://link.aps.org/doi/10.1103/PhysRevB.95.161115} {\bibfield  {journal}
  {\bibinfo  {journal} {Phys. Rev. B}\ }\textbf {\bibinfo {volume} {95}},\
  \bibinfo {pages} {161115}}\BibitemShut {NoStop}%
\bibitem [{\citenamefont {Biella}\ \emph {et~al.}(2017)\citenamefont {Biella},
  \citenamefont {Storme}, \citenamefont {Lebreuilly}, \citenamefont {Rossini},
  \citenamefont {Fazio}, \citenamefont {Carusotto},\ and\ \citenamefont
  {Ciuti}}]{Biella:PRA2017}%
  \BibitemOpen
  \bibfield  {author} {\bibinfo {author} {\bibnamefont {Biella}, \bibfnamefont
  {Alberto}}, \bibinfo {author} {\bibfnamefont {Florent}\ \bibnamefont
  {Storme}}, \bibinfo {author} {\bibfnamefont {Jos\'e}\ \bibnamefont
  {Lebreuilly}}, \bibinfo {author} {\bibfnamefont {Davide}\ \bibnamefont
  {Rossini}}, \bibinfo {author} {\bibfnamefont {Rosario}\ \bibnamefont
  {Fazio}}, \bibinfo {author} {\bibfnamefont {Iacopo}\ \bibnamefont
  {Carusotto}}, \ and\ \bibinfo {author} {\bibfnamefont {Cristiano}\
  \bibnamefont {Ciuti}}} (\bibinfo {year} {2017}),\ \bibfield  {title}
  {\enquote {\bibinfo {title} {Phase diagram of incoherently driven strongly
  correlated photonic lattices},}\ }\href
  {https://link.aps.org/doi/10.1103/PhysRevA.96.023839} {\bibfield  {journal}
  {\bibinfo  {journal} {Phys. Rev. A}\ }\textbf {\bibinfo {volume} {96}},\
  \bibinfo {pages} {023839}}\BibitemShut {NoStop}%
\bibitem [{\citenamefont {Bienias}\ \emph {et~al.}(2014)\citenamefont
  {Bienias}, \citenamefont {Choi}, \citenamefont {Firstenberg}, \citenamefont
  {Maghrebi}, \citenamefont {Gullans}, \citenamefont {Lukin}, \citenamefont
  {Gorshkov},\ and\ \citenamefont {B\"uchler}}]{Bienias:PRA2014}%
  \BibitemOpen
  \bibfield  {author} {\bibinfo {author} {\bibnamefont {Bienias}, \bibfnamefont
  {P}}, \bibinfo {author} {\bibfnamefont {S.}~\bibnamefont {Choi}}, \bibinfo
  {author} {\bibfnamefont {O.}~\bibnamefont {Firstenberg}}, \bibinfo {author}
  {\bibfnamefont {M.~F.}\ \bibnamefont {Maghrebi}}, \bibinfo {author}
  {\bibfnamefont {M.}~\bibnamefont {Gullans}}, \bibinfo {author} {\bibfnamefont
  {M.~D.}\ \bibnamefont {Lukin}}, \bibinfo {author} {\bibfnamefont {A.~V.}\
  \bibnamefont {Gorshkov}}, \ and\ \bibinfo {author} {\bibfnamefont {H.~P.}\
  \bibnamefont {B\"uchler}}} (\bibinfo {year} {2014}),\ \bibfield  {title}
  {\enquote {\bibinfo {title} {Scattering resonances and bound states for
  strongly interacting {R}ydberg polaritons},}\ }\href
  {https://link.aps.org/doi/10.1103/PhysRevA.90.053804} {\bibfield  {journal}
  {\bibinfo  {journal} {Phys. Rev. A}\ }\textbf {\bibinfo {volume} {90}},\
  \bibinfo {pages} {053804}}\BibitemShut {NoStop}%
\bibitem [{\citenamefont {Bilitewski}\ and\ \citenamefont
  {Cooper}(2016)}]{Bilitewski:2016PRA}%
  \BibitemOpen
  \bibfield  {author} {\bibinfo {author} {\bibnamefont {Bilitewski},
  \bibfnamefont {Thomas}}, \ and\ \bibinfo {author} {\bibfnamefont {Nigel~R.}\
  \bibnamefont {Cooper}}} (\bibinfo {year} {2016}),\ \bibfield  {title}
  {\enquote {\bibinfo {title} {{Synthetic dimensions in the strong-coupling
  limit: Supersolids and pair superfluids}},}\ }\href
  {https://link.aps.org/doi/10.1103/PhysRevA.94.023630} {\bibfield  {journal}
  {\bibinfo  {journal} {Phys. Rev. A}\ }\textbf {\bibinfo {volume} {94}},\
  \bibinfo {pages} {023630}}\BibitemShut {NoStop}%
\bibitem [{\citenamefont {Birnbaum}\ \emph {et~al.}(2005)\citenamefont
  {Birnbaum}, \citenamefont {Boca}, \citenamefont {Miller}, \citenamefont
  {Boozer}, \citenamefont {Northup},\ and\ \citenamefont
  {Kimble}}]{Birnbaum:Nature2005}%
  \BibitemOpen
  \bibfield  {author} {\bibinfo {author} {\bibnamefont {Birnbaum},
  \bibfnamefont {KM}}, \bibinfo {author} {\bibfnamefont {A}~\bibnamefont
  {Boca}}, \bibinfo {author} {\bibfnamefont {R}~\bibnamefont {Miller}},
  \bibinfo {author} {\bibfnamefont {AD}~\bibnamefont {Boozer}}, \bibinfo
  {author} {\bibfnamefont {TE}~\bibnamefont {Northup}}, \ and\ \bibinfo
  {author} {\bibfnamefont {HJ}~\bibnamefont {Kimble}}} (\bibinfo {year}
  {2005}),\ \bibfield  {title} {\enquote {\bibinfo {title} {Photon blockade in
  an optical cavity with one trapped atom},}\ }\href
  {https://www.nature.com/articles/nature03804} {\bibfield  {journal} {\bibinfo
   {journal} {Nature}\ }\textbf {\bibinfo {volume} {436}}~(\bibinfo {number}
  {7047}),\ \bibinfo {pages} {87--90}}\BibitemShut {NoStop}%
\bibitem [{\citenamefont {Biswas}\ and\ \citenamefont
  {Son}(2016)}]{biswas2016fractional}%
  \BibitemOpen
  \bibfield  {author} {\bibinfo {author} {\bibnamefont {Biswas}, \bibfnamefont
  {Rudro~R}}, \ and\ \bibinfo {author} {\bibfnamefont {Dam~Thanh}\ \bibnamefont
  {Son}}} (\bibinfo {year} {2016}),\ \bibfield  {title} {\enquote {\bibinfo
  {title} {Fractional charge and inter-{L}andau--level states at points of
  singular curvature},}\ }\href {https://www.pnas.org/content/113/31/8636}
  {\bibfield  {journal} {\bibinfo  {journal} {Proc. Natl. Acad. Scie. U.S.A.}\
  }\textbf {\bibinfo {volume} {113}}~(\bibinfo {number} {31}),\ \bibinfo
  {pages} {8636--8641}}\BibitemShut {NoStop}%
\bibitem [{\citenamefont {Bittner}\ \emph {et~al.}(2010)\citenamefont
  {Bittner}, \citenamefont {Dietz}, \citenamefont {Miski-Oglu}, \citenamefont
  {Oria~Iriarte}, \citenamefont {Richter},\ and\ \citenamefont
  {Sch\"afer}}]{Bittner:2010PRB}%
  \BibitemOpen
  \bibfield  {author} {\bibinfo {author} {\bibnamefont {Bittner}, \bibfnamefont
  {S}}, \bibinfo {author} {\bibfnamefont {B.}~\bibnamefont {Dietz}}, \bibinfo
  {author} {\bibfnamefont {M.}~\bibnamefont {Miski-Oglu}}, \bibinfo {author}
  {\bibfnamefont {P.}~\bibnamefont {Oria~Iriarte}}, \bibinfo {author}
  {\bibfnamefont {A.}~\bibnamefont {Richter}}, \ and\ \bibinfo {author}
  {\bibfnamefont {F.}~\bibnamefont {Sch\"afer}}} (\bibinfo {year} {2010}),\
  \bibfield  {title} {\enquote {\bibinfo {title} {Observation of a {D}irac
  point in microwave experiments with a photonic crystal modeling graphene},}\
  }\href {http://link.aps.org/doi/10.1103/PhysRevB.82.014301} {\bibfield
  {journal} {\bibinfo  {journal} {Phys. Rev. B}\ }\textbf {\bibinfo {volume}
  {82}},\ \bibinfo {pages} {014301}}\BibitemShut {NoStop}%
\bibitem [{\citenamefont {Bittner}\ \emph {et~al.}(2012)\citenamefont
  {Bittner}, \citenamefont {Dietz}, \citenamefont {Miski-Oglu},\ and\
  \citenamefont {Richter}}]{Bittner:2012PRB}%
  \BibitemOpen
  \bibfield  {author} {\bibinfo {author} {\bibnamefont {Bittner}, \bibfnamefont
  {S}}, \bibinfo {author} {\bibfnamefont {B.}~\bibnamefont {Dietz}}, \bibinfo
  {author} {\bibfnamefont {M.}~\bibnamefont {Miski-Oglu}}, \ and\ \bibinfo
  {author} {\bibfnamefont {A.}~\bibnamefont {Richter}}} (\bibinfo {year}
  {2012}),\ \bibfield  {title} {\enquote {\bibinfo {title} {{Extremal
  transmission through a microwave photonic crystal and the observation of edge
  states in a rectangular Dirac billiard}},}\ }\href
  {https://link.aps.org/doi/10.1103/PhysRevB.85.064301} {\bibfield  {journal}
  {\bibinfo  {journal} {Phys. Rev. B}\ }\textbf {\bibinfo {volume}
  {85}}~(\bibinfo {number} {6}),\ \bibinfo {pages} {064301}}\BibitemShut
  {NoStop}%
\bibitem [{\citenamefont {Bleckmann}\ \emph {et~al.}(2017)\citenamefont
  {Bleckmann}, \citenamefont {Cherpakova}, \citenamefont {Linden},\ and\
  \citenamefont {Alberti}}]{Bleckmann:2017PRB}%
  \BibitemOpen
  \bibfield  {author} {\bibinfo {author} {\bibnamefont {Bleckmann},
  \bibfnamefont {Felix}}, \bibinfo {author} {\bibfnamefont {Zlata}\
  \bibnamefont {Cherpakova}}, \bibinfo {author} {\bibfnamefont {Stefan}\
  \bibnamefont {Linden}}, \ and\ \bibinfo {author} {\bibfnamefont {Andrea}\
  \bibnamefont {Alberti}}} (\bibinfo {year} {2017}),\ \bibfield  {title}
  {\enquote {\bibinfo {title} {Spectral imaging of topological edge states in
  plasmonic waveguide arrays},}\ }\href
  {https://link.aps.org/doi/10.1103/PhysRevB.96.045417} {\bibfield  {journal}
  {\bibinfo  {journal} {Phys. Rev. B}\ }\textbf {\bibinfo {volume} {96}},\
  \bibinfo {pages} {045417}}\BibitemShut {NoStop}%
\bibitem [{\citenamefont {Bleu}\ \emph {et~al.}(2016)\citenamefont {Bleu},
  \citenamefont {Solnyshkov},\ and\ \citenamefont {Malpuech}}]{Bleu:2016PRB}%
  \BibitemOpen
  \bibfield  {author} {\bibinfo {author} {\bibnamefont {Bleu}, \bibfnamefont
  {O}}, \bibinfo {author} {\bibfnamefont {D.~D.}\ \bibnamefont {Solnyshkov}}, \
  and\ \bibinfo {author} {\bibfnamefont {G.}~\bibnamefont {Malpuech}}}
  (\bibinfo {year} {2016}),\ \bibfield  {title} {\enquote {\bibinfo {title}
  {Interacting quantum fluid in a polariton {C}hern insulator},}\ }\href
  {https://link.aps.org/doi/10.1103/PhysRevB.93.085438} {\bibfield  {journal}
  {\bibinfo  {journal} {Phys. Rev. B}\ }\textbf {\bibinfo {volume} {93}},\
  \bibinfo {pages} {085438}}\BibitemShut {NoStop}%
\bibitem [{\citenamefont {Bleu}\ \emph
  {et~al.}(2017{\natexlab{a}})\citenamefont {Bleu}, \citenamefont
  {Solnyshkov},\ and\ \citenamefont {Malpuech}}]{Bleu:2017PRB}%
  \BibitemOpen
  \bibfield  {author} {\bibinfo {author} {\bibnamefont {Bleu}, \bibfnamefont
  {O}}, \bibinfo {author} {\bibfnamefont {D.~D.}\ \bibnamefont {Solnyshkov}}, \
  and\ \bibinfo {author} {\bibfnamefont {G.}~\bibnamefont {Malpuech}}}
  (\bibinfo {year} {2017}{\natexlab{a}}),\ \bibfield  {title} {\enquote
  {\bibinfo {title} {Photonic versus electronic quantum anomalous {H}all
  effect},}\ }\href {https://link.aps.org/doi/10.1103/PhysRevB.95.115415}
  {\bibfield  {journal} {\bibinfo  {journal} {Phys. Rev. B}\ }\textbf {\bibinfo
  {volume} {95}},\ \bibinfo {pages} {115415}}\BibitemShut {NoStop}%
\bibitem [{\citenamefont {Bleu}\ \emph
  {et~al.}(2017{\natexlab{b}})\citenamefont {Bleu}, \citenamefont
  {Solnyshkov},\ and\ \citenamefont {Malpuech}}]{Bleu:PRB2017}%
  \BibitemOpen
  \bibfield  {author} {\bibinfo {author} {\bibnamefont {Bleu}, \bibfnamefont
  {O}}, \bibinfo {author} {\bibfnamefont {D~D}\ \bibnamefont {Solnyshkov}}, \
  and\ \bibinfo {author} {\bibfnamefont {G}~\bibnamefont {Malpuech}}} (\bibinfo
  {year} {2017}{\natexlab{b}}),\ \bibfield  {title} {\enquote {\bibinfo {title}
  {{Quantum valley Hall effect and perfect valley filter based on photonic
  analogs of transitional metal dichalcogenides}},}\ }\href
  {https://link.aps.org/doi/10.1103/PhysRevB.95.235431} {\bibfield  {journal}
  {\bibinfo  {journal} {Phys. Rev. B}\ }\textbf {\bibinfo {volume}
  {95}}~(\bibinfo {number} {23}),\ \bibinfo {pages} {235431}}\BibitemShut
  {NoStop}%
\bibitem [{\citenamefont {Bleu}\ \emph
  {et~al.}(2018{\natexlab{a}})\citenamefont {Bleu}, \citenamefont
  {Solnyshkov},\ and\ \citenamefont {Malpuech}}]{Bleu:2018PRB}%
  \BibitemOpen
  \bibfield  {author} {\bibinfo {author} {\bibnamefont {Bleu}, \bibfnamefont
  {O}}, \bibinfo {author} {\bibfnamefont {D.~D.}\ \bibnamefont {Solnyshkov}}, \
  and\ \bibinfo {author} {\bibfnamefont {G.}~\bibnamefont {Malpuech}}}
  (\bibinfo {year} {2018}{\natexlab{a}}),\ \bibfield  {title} {\enquote
  {\bibinfo {title} {Measuring the quantum geometric tensor in two-dimensional
  photonic and exciton-polariton systems},}\ }\href
  {https://link.aps.org/doi/10.1103/PhysRevB.97.195422} {\bibfield  {journal}
  {\bibinfo  {journal} {Phys. Rev. B}\ }\textbf {\bibinfo {volume} {97}},\
  \bibinfo {pages} {195422}}\BibitemShut {NoStop}%
\bibitem [{\citenamefont {Bleu}\ \emph
  {et~al.}(2018{\natexlab{b}})\citenamefont {Bleu}, \citenamefont {Malpuech},\
  and\ \citenamefont {Solnyshkov}}]{Bleu:arxiv2017}%
  \BibitemOpen
  \bibfield  {author} {\bibinfo {author} {\bibnamefont {Bleu}, \bibfnamefont
  {Olivier}}, \bibinfo {author} {\bibfnamefont {Guillaume}\ \bibnamefont
  {Malpuech}}, \ and\ \bibinfo {author} {\bibfnamefont {DD}~\bibnamefont
  {Solnyshkov}}} (\bibinfo {year} {2018}{\natexlab{b}}),\ \bibfield  {title}
  {\enquote {\bibinfo {title} {Robust quantum valley hall effect for vortices
  in an interacting bosonic quantum fluid},}\ }\href
  {https://www.nature.com/articles/s41467-018-06520-7} {\bibfield  {journal}
  {\bibinfo  {journal} {Nat. Commun.}\ }\textbf {\bibinfo {volume}
  {9}}~(\bibinfo {number} {1}),\ \bibinfo {pages} {3991}}\BibitemShut {NoStop}%
\bibitem [{\citenamefont {Bliokh}\ and\ \citenamefont
  {Bliokh}(2005)}]{Bliokh:AnnPhys2005}%
  \BibitemOpen
  \bibfield  {author} {\bibinfo {author} {\bibnamefont {Bliokh}, \bibfnamefont
  {K~Yu}}, \ and\ \bibinfo {author} {\bibfnamefont {Yu.~P.}\ \bibnamefont
  {Bliokh}}} (\bibinfo {year} {2005}),\ \bibfield  {title} {\enquote {\bibinfo
  {title} {{Spin gauge fields: From Berry phase to topological spin transport
  and Hall effects}},}\ }\href
  {http://www.sciencedirect.com/science/article/pii/S0003491605000394}
  {\bibfield  {journal} {\bibinfo  {journal} {Annals of Physics}\ }\textbf
  {\bibinfo {volume} {319}}~(\bibinfo {number} {1}),\ \bibinfo {pages} {13 --
  47}}\BibitemShut {NoStop}%
\bibitem [{\citenamefont {Bliokh}\ \emph {et~al.}(2014)\citenamefont {Bliokh},
  \citenamefont {Bekshaev},\ and\ \citenamefont
  {Nori}}]{Bliokh:2014Naturecommunications}%
  \BibitemOpen
  \bibfield  {author} {\bibinfo {author} {\bibnamefont {Bliokh}, \bibfnamefont
  {Konstantin~Y}}, \bibinfo {author} {\bibfnamefont {Aleksandr~Y}\ \bibnamefont
  {Bekshaev}}, \ and\ \bibinfo {author} {\bibfnamefont {Franco}\ \bibnamefont
  {Nori}}} (\bibinfo {year} {2014}),\ \bibfield  {title} {\enquote {\bibinfo
  {title} {Extraordinary momentum and spin in evanescent waves},}\ }\href
  {https://www.nature.com/articles/ncomms4300} {\bibfield  {journal} {\bibinfo
  {journal} {Nat. Commun.}\ }\textbf {\bibinfo {volume} {5}},\ \bibinfo {pages}
  {3300}}\BibitemShut {NoStop}%
\bibitem [{\citenamefont {Bliokh}\ and\ \citenamefont
  {Nori}(2012)}]{Bliokh:2012PRA}%
  \BibitemOpen
  \bibfield  {author} {\bibinfo {author} {\bibnamefont {Bliokh}, \bibfnamefont
  {Konstantin~Y}}, \ and\ \bibinfo {author} {\bibfnamefont {Franco}\
  \bibnamefont {Nori}}} (\bibinfo {year} {2012}),\ \bibfield  {title} {\enquote
  {\bibinfo {title} {Transverse spin of a surface polariton},}\ }\href
  {https://link.aps.org/doi/10.1103/PhysRevA.85.061801} {\bibfield  {journal}
  {\bibinfo  {journal} {Phys. Rev. A}\ }\textbf {\bibinfo {volume} {85}},\
  \bibinfo {pages} {061801}}\BibitemShut {NoStop}%
\bibitem [{\citenamefont {Bliokh}\ \emph
  {et~al.}(2015{\natexlab{a}})\citenamefont {Bliokh}, \citenamefont
  {Smirnova},\ and\ \citenamefont {Nori}}]{Bliokh:2015Science}%
  \BibitemOpen
  \bibfield  {author} {\bibinfo {author} {\bibnamefont {Bliokh}, \bibfnamefont
  {Konstantin~Y}}, \bibinfo {author} {\bibfnamefont {Daria}\ \bibnamefont
  {Smirnova}}, \ and\ \bibinfo {author} {\bibfnamefont {Franco}\ \bibnamefont
  {Nori}}} (\bibinfo {year} {2015}{\natexlab{a}}),\ \bibfield  {title}
  {\enquote {\bibinfo {title} {Quantum spin {H}all effect of light},}\ }\href
  {http://science.sciencemag.org/content/348/6242/1448} {\bibfield  {journal}
  {\bibinfo  {journal} {Science}\ }\textbf {\bibinfo {volume} {348}}~(\bibinfo
  {number} {6242}),\ \bibinfo {pages} {1448--1451}}\BibitemShut {NoStop}%
\bibitem [{\citenamefont {Bliokh}\ \emph
  {et~al.}(2015{\natexlab{b}})\citenamefont {Bliokh}, \citenamefont
  {Rodr{\'\i}guez-Fortu{\~n}o}, \citenamefont {Nori},\ and\ \citenamefont
  {Zayats}}]{bliokh:2015NatPhot}%
  \BibitemOpen
  \bibfield  {author} {\bibinfo {author} {\bibnamefont {Bliokh}, \bibfnamefont
  {KY}}, \bibinfo {author} {\bibfnamefont {FJ}~\bibnamefont
  {Rodr{\'\i}guez-Fortu{\~n}o}}, \bibinfo {author} {\bibfnamefont {Franco}\
  \bibnamefont {Nori}}, \ and\ \bibinfo {author} {\bibfnamefont {Anatoly~V}\
  \bibnamefont {Zayats}}} (\bibinfo {year} {2015}{\natexlab{b}}),\ \bibfield
  {title} {\enquote {\bibinfo {title} {Spin-orbit interactions of light},}\
  }\href
  {http://www.nature.com/nphoton/journal/v9/n12/abs/nphoton.2015.201.html}
  {\bibfield  {journal} {\bibinfo  {journal} {Nat. Photonics}\ }\textbf
  {\bibinfo {volume} {9}}~(\bibinfo {number} {12}),\ \bibinfo {pages}
  {796--808}}\BibitemShut {NoStop}%
\bibitem [{\citenamefont {Boada}\ \emph {et~al.}(2012)\citenamefont {Boada},
  \citenamefont {Celi}, \citenamefont {Latorre},\ and\ \citenamefont
  {Lewenstein}}]{Boada:2012PRL}%
  \BibitemOpen
  \bibfield  {author} {\bibinfo {author} {\bibnamefont {Boada}, \bibfnamefont
  {O}}, \bibinfo {author} {\bibfnamefont {A.}~\bibnamefont {Celi}}, \bibinfo
  {author} {\bibfnamefont {J.~I.}\ \bibnamefont {Latorre}}, \ and\ \bibinfo
  {author} {\bibfnamefont {M.}~\bibnamefont {Lewenstein}}} (\bibinfo {year}
  {2012}),\ \bibfield  {title} {\enquote {\bibinfo {title} {Quantum simulation
  of an extra dimension},}\ }\href
  {https://link.aps.org/doi/10.1103/PhysRevLett.108.133001} {\bibfield
  {journal} {\bibinfo  {journal} {Phys. Rev. Lett.}\ }\textbf {\bibinfo
  {volume} {108}},\ \bibinfo {pages} {133001}}\BibitemShut {NoStop}%
\bibitem [{\citenamefont {Boada}\ \emph {et~al.}(2015)\citenamefont {Boada},
  \citenamefont {Celi}, \citenamefont {Rodr{\'\i}guez-Laguna}, \citenamefont
  {Latorre},\ and\ \citenamefont {Lewenstein}}]{Boada:2015NJP}%
  \BibitemOpen
  \bibfield  {author} {\bibinfo {author} {\bibnamefont {Boada}, \bibfnamefont
  {Octavi}}, \bibinfo {author} {\bibfnamefont {Alessio}\ \bibnamefont {Celi}},
  \bibinfo {author} {\bibfnamefont {Javier}\ \bibnamefont
  {Rodr{\'\i}guez-Laguna}}, \bibinfo {author} {\bibfnamefont {Jos{\'e}~I}\
  \bibnamefont {Latorre}}, \ and\ \bibinfo {author} {\bibfnamefont {Maciej}\
  \bibnamefont {Lewenstein}}} (\bibinfo {year} {2015}),\ \bibfield  {title}
  {\enquote {\bibinfo {title} {Quantum simulation of non-trivial topology},}\
  }\href {http://iopscience.iop.org/article/10.1088/1367-2630/17/4/045007/meta}
  {\bibfield  {journal} {\bibinfo  {journal} {New J. Phys.}\ }\textbf {\bibinfo
  {volume} {17}}~(\bibinfo {number} {4}),\ \bibinfo {pages}
  {045007}}\BibitemShut {NoStop}%
\bibitem [{\citenamefont {Bourassa}\ \emph {et~al.}(2012)\citenamefont
  {Bourassa}, \citenamefont {Beaudoin}, \citenamefont {Gambetta},\ and\
  \citenamefont {Blais}}]{Bourassa:PRA2012}%
  \BibitemOpen
  \bibfield  {author} {\bibinfo {author} {\bibnamefont {Bourassa},
  \bibfnamefont {J}}, \bibinfo {author} {\bibfnamefont {F.}~\bibnamefont
  {Beaudoin}}, \bibinfo {author} {\bibfnamefont {Jay~M.}\ \bibnamefont
  {Gambetta}}, \ and\ \bibinfo {author} {\bibfnamefont {A.}~\bibnamefont
  {Blais}}} (\bibinfo {year} {2012}),\ \bibfield  {title} {\enquote {\bibinfo
  {title} {{Josephson-junction-embedded transmission-line resonators: From Kerr
  medium to in-line transmon}},}\ }\href
  {https://link.aps.org/doi/10.1103/PhysRevA.86.013814} {\bibfield  {journal}
  {\bibinfo  {journal} {Phys. Rev. A}\ }\textbf {\bibinfo {volume} {86}},\
  \bibinfo {pages} {013814}}\BibitemShut {NoStop}%
\bibitem [{\citenamefont {Boyd}(2008)}]{Boyd}%
  \BibitemOpen
  \bibfield  {author} {\bibinfo {author} {\bibnamefont {Boyd}, \bibfnamefont
  {R~W}}} (\bibinfo {year} {2008}),\ \href@noop {} {\emph {\bibinfo {title}
  {Nonlinear Optics}}}\ (\bibinfo  {publisher} {Academic Press},\ \bibinfo
  {address} {New York})\BibitemShut {NoStop}%
\bibitem [{\citenamefont {Bravo-Abad}\ \emph {et~al.}(2015)\citenamefont
  {Bravo-Abad}, \citenamefont {Lu}, \citenamefont {Fu}, \citenamefont
  {Buljan},\ and\ \citenamefont {Solja{\v{c}}i{\'c}}}]{Bravo:20152DM}%
  \BibitemOpen
  \bibfield  {author} {\bibinfo {author} {\bibnamefont {Bravo-Abad},
  \bibfnamefont {Jorge}}, \bibinfo {author} {\bibfnamefont {Ling}\ \bibnamefont
  {Lu}}, \bibinfo {author} {\bibfnamefont {Liang}\ \bibnamefont {Fu}}, \bibinfo
  {author} {\bibfnamefont {Hrvoje}\ \bibnamefont {Buljan}}, \ and\ \bibinfo
  {author} {\bibfnamefont {Marin}\ \bibnamefont {Solja{\v{c}}i{\'c}}}}
  (\bibinfo {year} {2015}),\ \bibfield  {title} {\enquote {\bibinfo {title}
  {Weyl points in photonic-crystal superlattices},}\ }\href
  {https://iopscience.iop.org/article/10.1088/2053-1583/2/3/034013} {\bibfield
  {journal} {\bibinfo  {journal} {2D Materials}\ }\textbf {\bibinfo {volume}
  {2}}~(\bibinfo {number} {3}),\ \bibinfo {pages} {034013}}\BibitemShut
  {NoStop}%
\bibitem [{\citenamefont {Brendel}\ \emph {et~al.}(2018)\citenamefont
  {Brendel}, \citenamefont {Peano}, \citenamefont {Painter},\ and\
  \citenamefont {Marquardt}}]{Brendel:arx2017}%
  \BibitemOpen
  \bibfield  {author} {\bibinfo {author} {\bibnamefont {Brendel}, \bibfnamefont
  {Christian}}, \bibinfo {author} {\bibfnamefont {Vittorio}\ \bibnamefont
  {Peano}}, \bibinfo {author} {\bibfnamefont {Oskar}\ \bibnamefont {Painter}},
  \ and\ \bibinfo {author} {\bibfnamefont {Florian}\ \bibnamefont {Marquardt}}}
  (\bibinfo {year} {2018}),\ \bibfield  {title} {\enquote {\bibinfo {title}
  {Snowflake phononic topological insulator at the nanoscale},}\ }\href
  {https://link.aps.org/doi/10.1103/PhysRevB.97.020102} {\bibfield  {journal}
  {\bibinfo  {journal} {Phys. Rev. B}\ }\textbf {\bibinfo {volume} {97}},\
  \bibinfo {pages} {020102}}\BibitemShut {NoStop}%
\bibitem [{\citenamefont {Broome}\ \emph {et~al.}(2010)\citenamefont {Broome},
  \citenamefont {Fedrizzi}, \citenamefont {Lanyon}, \citenamefont {Kassal},
  \citenamefont {Aspuru-Guzik},\ and\ \citenamefont {White}}]{Broome:2010PRL}%
  \BibitemOpen
  \bibfield  {author} {\bibinfo {author} {\bibnamefont {Broome}, \bibfnamefont
  {M~A}}, \bibinfo {author} {\bibfnamefont {A.}~\bibnamefont {Fedrizzi}},
  \bibinfo {author} {\bibfnamefont {B.~P.}\ \bibnamefont {Lanyon}}, \bibinfo
  {author} {\bibfnamefont {I.}~\bibnamefont {Kassal}}, \bibinfo {author}
  {\bibfnamefont {A.}~\bibnamefont {Aspuru-Guzik}}, \ and\ \bibinfo {author}
  {\bibfnamefont {A.~G.}\ \bibnamefont {White}}} (\bibinfo {year} {2010}),\
  \bibfield  {title} {\enquote {\bibinfo {title} {Discrete single-photon
  quantum walks with tunable decoherence},}\ }\href
  {https://link.aps.org/doi/10.1103/PhysRevLett.104.153602} {\bibfield
  {journal} {\bibinfo  {journal} {Phys. Rev. Lett.}\ }\textbf {\bibinfo
  {volume} {104}},\ \bibinfo {pages} {153602}}\BibitemShut {NoStop}%
\bibitem [{\citenamefont {Brouwer}(1998)}]{Brouwer:1998}%
  \BibitemOpen
  \bibfield  {author} {\bibinfo {author} {\bibnamefont {Brouwer}, \bibfnamefont
  {P~W}}} (\bibinfo {year} {1998}),\ \bibfield  {title} {\enquote {\bibinfo
  {title} {Scattering approach to parametric pumping},}\ }\href
  {https://link.aps.org/doi/10.1103/PhysRevB.58.R10135} {\bibfield  {journal}
  {\bibinfo  {journal} {Phys. Rev. B}\ }\textbf {\bibinfo {volume} {58}},\
  \bibinfo {pages} {R10135--R10138}}\BibitemShut {NoStop}%
\bibitem [{\citenamefont {Budich}\ and\ \citenamefont
  {Diehl}(2015)}]{PhysRevB.91.165140}%
  \BibitemOpen
  \bibfield  {author} {\bibinfo {author} {\bibnamefont {Budich}, \bibfnamefont
  {Jan~Carl}}, \ and\ \bibinfo {author} {\bibfnamefont {Sebastian}\
  \bibnamefont {Diehl}}} (\bibinfo {year} {2015}),\ \bibfield  {title}
  {\enquote {\bibinfo {title} {Topology of density matrices},}\ }\href
  {https://link.aps.org/doi/10.1103/PhysRevB.91.165140} {\bibfield  {journal}
  {\bibinfo  {journal} {Phys. Rev. B}\ }\textbf {\bibinfo {volume} {91}},\
  \bibinfo {pages} {165140}}\BibitemShut {NoStop}%
\bibitem [{\citenamefont {Budich}\ \emph {et~al.}(2015)\citenamefont {Budich},
  \citenamefont {Zoller},\ and\ \citenamefont {Diehl}}]{PhysRevA.91.042117}%
  \BibitemOpen
  \bibfield  {author} {\bibinfo {author} {\bibnamefont {Budich}, \bibfnamefont
  {Jan~Carl}}, \bibinfo {author} {\bibfnamefont {Peter}\ \bibnamefont
  {Zoller}}, \ and\ \bibinfo {author} {\bibfnamefont {Sebastian}\ \bibnamefont
  {Diehl}}} (\bibinfo {year} {2015}),\ \bibfield  {title} {\enquote {\bibinfo
  {title} {Dissipative preparation of {C}hern insulators},}\ }\href
  {https://link.aps.org/doi/10.1103/PhysRevA.91.042117} {\bibfield  {journal}
  {\bibinfo  {journal} {Phys. Rev. A}\ }\textbf {\bibinfo {volume} {91}},\
  \bibinfo {pages} {042117}}\BibitemShut {NoStop}%
\bibitem [{\citenamefont {Bukov}\ \emph {et~al.}(2015)\citenamefont {Bukov},
  \citenamefont {D'Alessio},\ and\ \citenamefont
  {Polkovnikov}}]{Bukov:2015AdPhys}%
  \BibitemOpen
  \bibfield  {author} {\bibinfo {author} {\bibnamefont {Bukov}, \bibfnamefont
  {Marin}}, \bibinfo {author} {\bibfnamefont {Luca}\ \bibnamefont {D'Alessio}},
  \ and\ \bibinfo {author} {\bibfnamefont {Anatoli}\ \bibnamefont
  {Polkovnikov}}} (\bibinfo {year} {2015}),\ \bibfield  {title} {\enquote
  {\bibinfo {title} {Universal high-frequency behavior of periodically driven
  systems: {F}rom dynamical stabilization to {F}loquet engineering},}\ }\href
  {http://www.tandfonline.com/doi/full/10.1080/00018732.2015.1055918}
  {\bibfield  {journal} {\bibinfo  {journal} {Adv. Phys.}\ }\textbf {\bibinfo
  {volume} {64}}~(\bibinfo {number} {2}),\ \bibinfo {pages}
  {139--226}}\BibitemShut {NoStop}%
\bibitem [{\citenamefont {Burkov}\ \emph {et~al.}(2011)\citenamefont {Burkov},
  \citenamefont {Hook},\ and\ \citenamefont {Balents}}]{Burkov:2011PRB}%
  \BibitemOpen
  \bibfield  {author} {\bibinfo {author} {\bibnamefont {Burkov}, \bibfnamefont
  {AA}}, \bibinfo {author} {\bibfnamefont {MD}~\bibnamefont {Hook}}, \ and\
  \bibinfo {author} {\bibfnamefont {Leon}\ \bibnamefont {Balents}}} (\bibinfo
  {year} {2011}),\ \bibfield  {title} {\enquote {\bibinfo {title} {Topological
  nodal semimetals},}\ }\href
  {https://journals.aps.org/prb/abstract/10.1103/PhysRevB.84.235126} {\bibfield
   {journal} {\bibinfo  {journal} {Phys. Rev. B}\ }\textbf {\bibinfo {volume}
  {84}}~(\bibinfo {number} {23}),\ \bibinfo {pages} {235126}}\BibitemShut
  {NoStop}%
\bibitem [{\citenamefont {Butcher}\ and\ \citenamefont
  {Cotter}(2008)}]{Butcher}%
  \BibitemOpen
  \bibfield  {author} {\bibinfo {author} {\bibnamefont {Butcher}, \bibfnamefont
  {P~N}}, \ and\ \bibinfo {author} {\bibfnamefont {D.}~\bibnamefont {Cotter}}}
  (\bibinfo {year} {2008}),\ \href@noop {} {\emph {\bibinfo {title} {The
  elements of nonlinear optics}}},\ Cambridge Studies in Modern Optics\
  (\bibinfo  {publisher} {Cambridge University Press},\ \bibinfo {address}
  {Cambridge, England})\BibitemShut {NoStop}%
\bibitem [{\citenamefont {B\"uttiker}(1988)}]{Buttiker:1988PRB}%
  \BibitemOpen
  \bibfield  {author} {\bibinfo {author} {\bibnamefont {B\"uttiker},
  \bibfnamefont {M}}} (\bibinfo {year} {1988}),\ \bibfield  {title} {\enquote
  {\bibinfo {title} {Absence of backscattering in the quantum {H}all effect in
  multiprobe conductors},}\ }\href
  {https://link.aps.org/doi/10.1103/PhysRevB.38.9375} {\bibfield  {journal}
  {\bibinfo  {journal} {Phys. Rev. B}\ }\textbf {\bibinfo {volume} {38}},\
  \bibinfo {pages} {9375--9389}}\BibitemShut {NoStop}%
\bibitem [{\citenamefont {Bzdusek}\ \emph {et~al.}(2016)\citenamefont
  {Bzdusek}, \citenamefont {Wu}, \citenamefont {R{\"u}egg}, \citenamefont
  {Sigrist},\ and\ \citenamefont {Soluyanov}}]{Bzdusek:2016Nature}%
  \BibitemOpen
  \bibfield  {author} {\bibinfo {author} {\bibnamefont {Bzdusek}, \bibfnamefont
  {Tom{\'a}s}}, \bibinfo {author} {\bibfnamefont {Quansheng}\ \bibnamefont
  {Wu}}, \bibinfo {author} {\bibfnamefont {Andreas}\ \bibnamefont {R{\"u}egg}},
  \bibinfo {author} {\bibfnamefont {Manfred}\ \bibnamefont {Sigrist}}, \ and\
  \bibinfo {author} {\bibfnamefont {Alexey~A}\ \bibnamefont {Soluyanov}}}
  (\bibinfo {year} {2016}),\ \bibfield  {title} {\enquote {\bibinfo {title}
  {Nodal-chain metals},}\ }\href {https://www.nature.com/articles/nature19099}
  {\bibfield  {journal} {\bibinfo  {journal} {Nature}\ }\textbf {\bibinfo
  {volume} {538}},\ \bibinfo {pages} {75--78}}\BibitemShut {NoStop}%
\bibitem [{\citenamefont {Cai}\ \emph {et~al.}(2017)\citenamefont {Cai},
  \citenamefont {Roslund}, \citenamefont {Ferrini}, \citenamefont {Arzani},
  \citenamefont {Xu}, \citenamefont {Fabre},\ and\ \citenamefont
  {Treps}}]{cai2017multimode}%
  \BibitemOpen
  \bibfield  {author} {\bibinfo {author} {\bibnamefont {Cai}, \bibfnamefont
  {Y}}, \bibinfo {author} {\bibfnamefont {J}~\bibnamefont {Roslund}}, \bibinfo
  {author} {\bibfnamefont {G}~\bibnamefont {Ferrini}}, \bibinfo {author}
  {\bibfnamefont {F}~\bibnamefont {Arzani}}, \bibinfo {author} {\bibfnamefont
  {X}~\bibnamefont {Xu}}, \bibinfo {author} {\bibfnamefont {C}~\bibnamefont
  {Fabre}}, \ and\ \bibinfo {author} {\bibfnamefont {N}~\bibnamefont {Treps}}}
  (\bibinfo {year} {2017}),\ \bibfield  {title} {\enquote {\bibinfo {title}
  {Multimode entanglement in reconfigurable graph states using optical
  frequency combs},}\ }\href {https://www.nature.com/articles/ncomms15645}
  {\bibfield  {journal} {\bibinfo  {journal} {Nat. Commun.}\ }\textbf {\bibinfo
  {volume} {8}},\ \bibinfo {pages} {15645}}\BibitemShut {NoStop}%
\bibitem [{\citenamefont {Callias}(1978)}]{Callias:1978CMP}%
  \BibitemOpen
  \bibfield  {author} {\bibinfo {author} {\bibnamefont {Callias}, \bibfnamefont
  {Constantine}}} (\bibinfo {year} {1978}),\ \bibfield  {title} {\enquote
  {\bibinfo {title} {Axial anomalies and index theorems on open spaces},}\
  }\href {https://link.springer.com/article/10.1007/BF01202525} {\bibfield
  {journal} {\bibinfo  {journal} {Commun. Math. Phys.}\ }\textbf {\bibinfo
  {volume} {62}}~(\bibinfo {number} {3}),\ \bibinfo {pages}
  {213--234}}\BibitemShut {NoStop}%
\bibitem [{\citenamefont {Calvanese~Strinati}\ \emph
  {et~al.}(2017)\citenamefont {Calvanese~Strinati}, \citenamefont {Cornfeld},
  \citenamefont {Rossini}, \citenamefont {Barbarino}, \citenamefont {Dalmonte},
  \citenamefont {Fazio}, \citenamefont {Sela},\ and\ \citenamefont
  {Mazza}}]{Strinati:PRX2017}%
  \BibitemOpen
  \bibfield  {author} {\bibinfo {author} {\bibnamefont {Calvanese~Strinati},
  \bibfnamefont {Marcello}}, \bibinfo {author} {\bibfnamefont {Eyal}\
  \bibnamefont {Cornfeld}}, \bibinfo {author} {\bibfnamefont {Davide}\
  \bibnamefont {Rossini}}, \bibinfo {author} {\bibfnamefont {Simone}\
  \bibnamefont {Barbarino}}, \bibinfo {author} {\bibfnamefont {Marcello}\
  \bibnamefont {Dalmonte}}, \bibinfo {author} {\bibfnamefont {Rosario}\
  \bibnamefont {Fazio}}, \bibinfo {author} {\bibfnamefont {Eran}\ \bibnamefont
  {Sela}}, \ and\ \bibinfo {author} {\bibfnamefont {Leonardo}\ \bibnamefont
  {Mazza}}} (\bibinfo {year} {2017}),\ \bibfield  {title} {\enquote {\bibinfo
  {title} {Laughlin-like states in bosonic and fermionic atomic synthetic
  ladders},}\ }\href {https://link.aps.org/doi/10.1103/PhysRevX.7.021033}
  {\bibfield  {journal} {\bibinfo  {journal} {Phys. Rev. X}\ }\textbf {\bibinfo
  {volume} {7}},\ \bibinfo {pages} {021033}}\BibitemShut {NoStop}%
\bibitem [{\citenamefont {Can}\ \emph {et~al.}(2016)\citenamefont {Can},
  \citenamefont {Chiu}, \citenamefont {Laskin},\ and\ \citenamefont
  {Wiegmann}}]{can2016emergent}%
  \BibitemOpen
  \bibfield  {author} {\bibinfo {author} {\bibnamefont {Can}, \bibfnamefont
  {T}}, \bibinfo {author} {\bibfnamefont {YH}~\bibnamefont {Chiu}}, \bibinfo
  {author} {\bibfnamefont {M}~\bibnamefont {Laskin}}, \ and\ \bibinfo {author}
  {\bibfnamefont {P}~\bibnamefont {Wiegmann}}} (\bibinfo {year} {2016}),\
  \bibfield  {title} {\enquote {\bibinfo {title} {Emergent conformal symmetry
  and geometric transport properties of quantum {H}all states on singular
  surfaces},}\ }\href
  {https://journals.aps.org/prl/abstract/10.1103/PhysRevLett.117.266803}
  {\bibfield  {journal} {\bibinfo  {journal} {Phys. Rev. Lett.}\ }\textbf
  {\bibinfo {volume} {117}}~(\bibinfo {number} {26}),\ \bibinfo {pages}
  {266803}}\BibitemShut {NoStop}%
\bibitem [{\citenamefont {Cardano}\ \emph {et~al.}(2017)\citenamefont
  {Cardano}, \citenamefont {D’Errico}, \citenamefont {Dauphin}, \citenamefont
  {Maffei}, \citenamefont {Piccirillo}, \citenamefont {de~Lisio}, \citenamefont
  {De~Filippis}, \citenamefont {Cataudella}, \citenamefont {Santamato},
  \citenamefont {Marrucci} \emph {et~al.}}]{Cardano:2017NatComm}%
  \BibitemOpen
  \bibfield  {author} {\bibinfo {author} {\bibnamefont {Cardano}, \bibfnamefont
  {Filippo}}, \bibinfo {author} {\bibfnamefont {Alessio}\ \bibnamefont
  {D’Errico}}, \bibinfo {author} {\bibfnamefont {Alexandre}\ \bibnamefont
  {Dauphin}}, \bibinfo {author} {\bibfnamefont {Maria}\ \bibnamefont {Maffei}},
  \bibinfo {author} {\bibfnamefont {Bruno}\ \bibnamefont {Piccirillo}},
  \bibinfo {author} {\bibfnamefont {Corrado}\ \bibnamefont {de~Lisio}},
  \bibinfo {author} {\bibfnamefont {Giulio}\ \bibnamefont {De~Filippis}},
  \bibinfo {author} {\bibfnamefont {Vittorio}\ \bibnamefont {Cataudella}},
  \bibinfo {author} {\bibfnamefont {Enrico}\ \bibnamefont {Santamato}},
  \bibinfo {author} {\bibfnamefont {Lorenzo}\ \bibnamefont {Marrucci}},  \emph
  {et~al.}} (\bibinfo {year} {2017}),\ \bibfield  {title} {\enquote {\bibinfo
  {title} {Detection of {Z}ak phases and topological invariants in a chiral
  quantum walk of twisted photons},}\ }\href
  {https://www.nature.com/articles/ncomms15516} {\bibfield  {journal} {\bibinfo
   {journal} {Nat. Commun.}\ }\textbf {\bibinfo {volume} {8}},\ \bibinfo
  {pages} {15516}}\BibitemShut {NoStop}%
\bibitem [{\citenamefont {Cardano}\ \emph {et~al.}(2016)\citenamefont
  {Cardano}, \citenamefont {Maffei}, \citenamefont {Massa}, \citenamefont
  {Piccirillo}, \citenamefont {De~Lisio}, \citenamefont {De~Filippis},
  \citenamefont {Cataudella}, \citenamefont {Santamato},\ and\ \citenamefont
  {Marrucci}}]{Cardano:2016NatComm}%
  \BibitemOpen
  \bibfield  {author} {\bibinfo {author} {\bibnamefont {Cardano}, \bibfnamefont
  {Filippo}}, \bibinfo {author} {\bibfnamefont {Maria}\ \bibnamefont {Maffei}},
  \bibinfo {author} {\bibfnamefont {Francesco}\ \bibnamefont {Massa}}, \bibinfo
  {author} {\bibfnamefont {Bruno}\ \bibnamefont {Piccirillo}}, \bibinfo
  {author} {\bibfnamefont {Corrado}\ \bibnamefont {De~Lisio}}, \bibinfo
  {author} {\bibfnamefont {Giulio}\ \bibnamefont {De~Filippis}}, \bibinfo
  {author} {\bibfnamefont {Vittorio}\ \bibnamefont {Cataudella}}, \bibinfo
  {author} {\bibfnamefont {Enrico}\ \bibnamefont {Santamato}}, \ and\ \bibinfo
  {author} {\bibfnamefont {Lorenzo}\ \bibnamefont {Marrucci}}} (\bibinfo {year}
  {2016}),\ \bibfield  {title} {\enquote {\bibinfo {title} {Statistical moments
  of quantum-walk dynamics reveal topological quantum transitions},}\ }\href
  {https://www.nature.com/articles/ncomms11439} {\bibfield  {journal} {\bibinfo
   {journal} {Nat. Commun.}\ }\textbf {\bibinfo {volume} {7}},\ \bibinfo
  {pages} {11439}}\BibitemShut {NoStop}%
\bibitem [{\citenamefont {Cardano}\ \emph {et~al.}(2015)\citenamefont
  {Cardano}, \citenamefont {Massa}, \citenamefont {Qassim}, \citenamefont
  {Karimi}, \citenamefont {Slussarenko}, \citenamefont {Paparo}, \citenamefont
  {de~Lisio}, \citenamefont {Sciarrino}, \citenamefont {Santamato},
  \citenamefont {Boyd} \emph {et~al.}}]{Cardano:2015SciAdv}%
  \BibitemOpen
  \bibfield  {author} {\bibinfo {author} {\bibnamefont {Cardano}, \bibfnamefont
  {Filippo}}, \bibinfo {author} {\bibfnamefont {Francesco}\ \bibnamefont
  {Massa}}, \bibinfo {author} {\bibfnamefont {Hammam}\ \bibnamefont {Qassim}},
  \bibinfo {author} {\bibfnamefont {Ebrahim}\ \bibnamefont {Karimi}}, \bibinfo
  {author} {\bibfnamefont {Sergei}\ \bibnamefont {Slussarenko}}, \bibinfo
  {author} {\bibfnamefont {Domenico}\ \bibnamefont {Paparo}}, \bibinfo {author}
  {\bibfnamefont {Corrado}\ \bibnamefont {de~Lisio}}, \bibinfo {author}
  {\bibfnamefont {Fabio}\ \bibnamefont {Sciarrino}}, \bibinfo {author}
  {\bibfnamefont {Enrico}\ \bibnamefont {Santamato}}, \bibinfo {author}
  {\bibfnamefont {Robert~W}\ \bibnamefont {Boyd}},  \emph {et~al.}} (\bibinfo
  {year} {2015}),\ \bibfield  {title} {\enquote {\bibinfo {title} {Quantum
  walks and wavepacket dynamics on a lattice with twisted photons},}\ }\href
  {http://advances.sciencemag.org/content/1/2/e1500087} {\bibfield  {journal}
  {\bibinfo  {journal} {Science advances}\ }\textbf {\bibinfo {volume}
  {1}}~(\bibinfo {number} {2}),\ \bibinfo {pages} {e1500087}}\BibitemShut
  {NoStop}%
\bibitem [{\citenamefont {Carlon~Zambon}\ \emph {et~al.}(2018)\citenamefont
  {Carlon~Zambon}, \citenamefont {St-Jean}, \citenamefont {Mili{\'c}evi{\'c}},
  \citenamefont {Lema{\^\i}tre}, \citenamefont {Harouri}, \citenamefont
  {LeGratiet}, \citenamefont {Bleu}, \citenamefont {Solnyshkov}, \citenamefont
  {Malpuech}, \citenamefont {Sagnes} \emph {et~al.}}]{Zambon:2018arXiv}%
  \BibitemOpen
  \bibfield  {author} {\bibinfo {author} {\bibnamefont {Carlon~Zambon},
  \bibfnamefont {Nicola}}, \bibinfo {author} {\bibfnamefont {Philippe}\
  \bibnamefont {St-Jean}}, \bibinfo {author} {\bibfnamefont {Marijana}\
  \bibnamefont {Mili{\'c}evi{\'c}}}, \bibinfo {author} {\bibfnamefont
  {Aristide}\ \bibnamefont {Lema{\^\i}tre}}, \bibinfo {author} {\bibfnamefont
  {Abdelmounaim}\ \bibnamefont {Harouri}}, \bibinfo {author} {\bibfnamefont
  {Luc}\ \bibnamefont {LeGratiet}}, \bibinfo {author} {\bibfnamefont {Olivier}\
  \bibnamefont {Bleu}}, \bibinfo {author} {\bibfnamefont {Dmitry}\ \bibnamefont
  {Solnyshkov}}, \bibinfo {author} {\bibfnamefont {Guillaume}\ \bibnamefont
  {Malpuech}}, \bibinfo {author} {\bibfnamefont {Isabelle}\ \bibnamefont
  {Sagnes}},  \emph {et~al.}} (\bibinfo {year} {2018}),\ \bibfield  {title}
  {\enquote {\bibinfo {title} {Optically controlling the emission chirality of
  microlasers},}\ }\href {https://arxiv.org/abs/1806.04590} {\bibinfo
  {journal} {arXiv:1806.04590}\ }\BibitemShut {NoStop}%
\bibitem [{\citenamefont {Carpentier}\ \emph {et~al.}(2015)\citenamefont
  {Carpentier}, \citenamefont {Delplace}, \citenamefont {Fruchart},\ and\
  \citenamefont {Gawedzki}}]{Carpentier:2015}%
  \BibitemOpen
\bibfield  {journal} {  }\bibfield  {author} {\bibinfo {author} {\bibnamefont
  {Carpentier}, \bibfnamefont {David}}, \bibinfo {author} {\bibfnamefont
  {Pierre}\ \bibnamefont {Delplace}}, \bibinfo {author} {\bibfnamefont
  {Michel}\ \bibnamefont {Fruchart}}, \ and\ \bibinfo {author} {\bibfnamefont
  {Krzysztof}\ \bibnamefont {Gawedzki}}} (\bibinfo {year} {2015}),\ \bibfield
  {title} {\enquote {\bibinfo {title} {Topological index for periodically
  driven time-reversal invariant 2{D} systems},}\ }\href
  {http://journals.aps.org/prl/abstract/10.1103/PhysRevLett.114.106806}
  {\bibfield  {journal} {\bibinfo  {journal} {Phys. Rev. Lett.}\ }\textbf
  {\bibinfo {volume} {114}}~(\bibinfo {number} {10}),\ \bibinfo {pages}
  {106806}}\BibitemShut {NoStop}%
\bibitem [{\citenamefont {Carusotto}\ \emph {et~al.}(2009)\citenamefont
  {Carusotto}, \citenamefont {Gerace}, \citenamefont {Tureci}, \citenamefont
  {De~Liberato}, \citenamefont {Ciuti},\ and\ \citenamefont
  {Imamo\ifmmode~\check{g}\else \v{g}\fi{}lu}}]{Carusotto:PRL2009}%
  \BibitemOpen
  \bibfield  {author} {\bibinfo {author} {\bibnamefont {Carusotto},
  \bibfnamefont {I}}, \bibinfo {author} {\bibfnamefont {D.}~\bibnamefont
  {Gerace}}, \bibinfo {author} {\bibfnamefont {H.~E.}\ \bibnamefont {Tureci}},
  \bibinfo {author} {\bibfnamefont {S.}~\bibnamefont {De~Liberato}}, \bibinfo
  {author} {\bibfnamefont {C.}~\bibnamefont {Ciuti}}, \ and\ \bibinfo {author}
  {\bibfnamefont {A.}~\bibnamefont {Imamo\ifmmode~\check{g}\else
  \v{g}\fi{}lu}}} (\bibinfo {year} {2009}),\ \bibfield  {title} {\enquote
  {\bibinfo {title} {Fermionized photons in an array of driven dissipative
  nonlinear cavities},}\ }\href
  {http://link.aps.org/doi/10.1103/PhysRevLett.103.033601} {\bibfield
  {journal} {\bibinfo  {journal} {Phys. Rev. Lett.}\ }\textbf {\bibinfo
  {volume} {103}},\ \bibinfo {pages} {033601}}\BibitemShut {NoStop}%
\bibitem [{\citenamefont {Carusotto}\ \emph {et~al.}(2010)\citenamefont
  {Carusotto}, \citenamefont {Volz},\ and\ \citenamefont
  {Imamoglu}}]{Carusotto:EPL2010}%
  \BibitemOpen
  \bibfield  {author} {\bibinfo {author} {\bibnamefont {Carusotto},
  \bibfnamefont {I}}, \bibinfo {author} {\bibfnamefont {T.}~\bibnamefont
  {Volz}}, \ and\ \bibinfo {author} {\bibfnamefont {A.}~\bibnamefont
  {Imamoglu}}} (\bibinfo {year} {2010}),\ \bibfield  {title} {\enquote
  {\bibinfo {title} {Feshbach blockade: {S}ingle-photon nonlinear optics using
  resonantly enhanced cavity polariton scattering from biexciton states},}\
  }\href {\doibase 10.1209/0295-5075/90/37001} {\bibfield  {journal} {\bibinfo
  {journal} {EPL}\ }\textbf {\bibinfo {volume} {90}}~(\bibinfo {number} {3}),\
  \bibinfo {pages} {37001}}\BibitemShut {NoStop}%
\bibitem [{\citenamefont {Carusotto}\ and\ \citenamefont
  {Ciuti}(2013)}]{carusotto:2013}%
  \BibitemOpen
  \bibfield  {author} {\bibinfo {author} {\bibnamefont {Carusotto},
  \bibfnamefont {Iacopo}}, \ and\ \bibinfo {author} {\bibfnamefont {Cristiano}\
  \bibnamefont {Ciuti}}} (\bibinfo {year} {2013}),\ \bibfield  {title}
  {\enquote {\bibinfo {title} {Quantum fluids of light},}\ }\href
  {http://link.aps.org/doi/10.1103/RevModPhys.85.299} {\bibfield  {journal}
  {\bibinfo  {journal} {Rev. Mod. Phys.}\ }\textbf {\bibinfo {volume} {85}},\
  \bibinfo {pages} {299--366}}\BibitemShut {NoStop}%
\bibitem [{\citenamefont {{Castro Neto}}\ \emph {et~al.}(2009)\citenamefont
  {{Castro Neto}}, \citenamefont {Guinea}, \citenamefont {Peres}, \citenamefont
  {Novoselov},\ and\ \citenamefont {Geim}}]{CastroNeto:2009RMP}%
  \BibitemOpen
  \bibfield  {author} {\bibinfo {author} {\bibnamefont {{Castro Neto}},
  \bibfnamefont {A~H}}, \bibinfo {author} {\bibfnamefont {F}~\bibnamefont
  {Guinea}}, \bibinfo {author} {\bibfnamefont {N~M~R}\ \bibnamefont {Peres}},
  \bibinfo {author} {\bibfnamefont {K~S}\ \bibnamefont {Novoselov}}, \ and\
  \bibinfo {author} {\bibfnamefont {A~K}\ \bibnamefont {Geim}}} (\bibinfo
  {year} {2009}),\ \bibfield  {title} {\enquote {\bibinfo {title} {{The
  electronic properties of graphene}},}\ }\href@noop {} {\bibfield  {journal}
  {\bibinfo  {journal} {Rev. Mod. Phys.}\ }\textbf {\bibinfo {volume}
  {81}}~(\bibinfo {number} {1}),\ \bibinfo {pages} {109--162}}\BibitemShut
  {NoStop}%
\bibitem [{\citenamefont {Cayssol}\ \emph {et~al.}(2013)\citenamefont
  {Cayssol}, \citenamefont {Dora}, \citenamefont {Simon},\ and\ \citenamefont
  {Moessner}}]{Cayssol:2013PS}%
  \BibitemOpen
  \bibfield  {author} {\bibinfo {author} {\bibnamefont {Cayssol}, \bibfnamefont
  {J}}, \bibinfo {author} {\bibfnamefont {B.}~\bibnamefont {Dora}}, \bibinfo
  {author} {\bibfnamefont {F.}~\bibnamefont {Simon}}, \ and\ \bibinfo {author}
  {\bibfnamefont {R.}~\bibnamefont {Moessner}}} (\bibinfo {year} {2013}),\
  \bibfield  {title} {\enquote {\bibinfo {title} {Floquet topological
  insulators},}\ }\href {https://doi.org/10.1002/pssr.201206451} {\bibfield
  {journal} {\bibinfo  {journal} {Phys. Status Solidi RRL}\ }\textbf {\bibinfo
  {volume} {7}},\ \bibinfo {pages} {101--108}}\BibitemShut {NoStop}%
\bibitem [{\citenamefont {Celi}\ \emph {et~al.}(2014)\citenamefont {Celi},
  \citenamefont {Massignan}, \citenamefont {Ruseckas}, \citenamefont {Goldman},
  \citenamefont {Spielman}, \citenamefont {Juzeli\ifmmode~\bar{u}\else
  \={u}\fi{}nas},\ and\ \citenamefont {Lewenstein}}]{Celi:2012PRL}%
  \BibitemOpen
  \bibfield  {author} {\bibinfo {author} {\bibnamefont {Celi}, \bibfnamefont
  {A}}, \bibinfo {author} {\bibfnamefont {P.}~\bibnamefont {Massignan}},
  \bibinfo {author} {\bibfnamefont {J.}~\bibnamefont {Ruseckas}}, \bibinfo
  {author} {\bibfnamefont {N.}~\bibnamefont {Goldman}}, \bibinfo {author}
  {\bibfnamefont {I.~B.}\ \bibnamefont {Spielman}}, \bibinfo {author}
  {\bibfnamefont {G.}~\bibnamefont {Juzeli\ifmmode~\bar{u}\else
  \={u}\fi{}nas}}, \ and\ \bibinfo {author} {\bibfnamefont {M.}~\bibnamefont
  {Lewenstein}}} (\bibinfo {year} {2014}),\ \bibfield  {title} {\enquote
  {\bibinfo {title} {Synthetic gauge fields in synthetic dimensions},}\ }\href
  {https://link.aps.org/doi/10.1103/PhysRevLett.112.043001} {\bibfield
  {journal} {\bibinfo  {journal} {Phys. Rev. Lett.}\ }\textbf {\bibinfo
  {volume} {112}},\ \bibinfo {pages} {043001}}\BibitemShut {NoStop}%
\bibitem [{\citenamefont {Juki\ifmmode~\acute{c}\else \'{c}\fi{}}\ and\
  \citenamefont {Buljan}(2013)}]{Jukic:2013PRA}%
  \BibitemOpen
  \bibfield  {author} {\bibinfo {author} {\bibnamefont
  {Juki\ifmmode~\acute{c}\else \'{c}\fi{}}, \bibfnamefont {D}}, \ and\ \bibinfo
  {author} {\bibfnamefont {H.}~\bibnamefont {Buljan}}} (\bibinfo {year}
  {2013}),\ \bibfield  {title} {\enquote {\bibinfo {title} {Four-dimensional
  photonic lattices and discrete tesseract solitons},}\ }\href
  {https://link.aps.org/doi/10.1103/PhysRevA.87.013814} {\bibfield  {journal}
  {\bibinfo  {journal} {Phys. Rev. A}\ }\textbf {\bibinfo {volume} {87}},\
  \bibinfo {pages} {013814}}\BibitemShut {NoStop}%
\bibitem [{\citenamefont {Chabanov}\ and\ \citenamefont
  {Genack}(2001)}]{Chabanov:2001}%
  \BibitemOpen
  \bibfield  {author} {\bibinfo {author} {\bibnamefont {Chabanov},
  \bibfnamefont {A~A}}, \ and\ \bibinfo {author} {\bibfnamefont
  {A.}~\bibnamefont {Genack}}} (\bibinfo {year} {2001}),\ \bibfield  {title}
  {\enquote {\bibinfo {title} {Statistics of dynamics of localized waves},}\
  }\href {https://journals.aps.org/prl/abstract/10.1103/PhysRevLett.87.233903}
  {\bibfield  {journal} {\bibinfo  {journal} {Phys. Rev. Lett.}\ }\textbf
  {\bibinfo {volume} {87}}~(\bibinfo {number} {23}),\ \bibinfo {pages}
  {233903}}\BibitemShut {NoStop}%
\bibitem [{\citenamefont {Chalker}\ and\ \citenamefont
  {Dohmen}(1995)}]{Chalker:1995PRL}%
  \BibitemOpen
  \bibfield  {author} {\bibinfo {author} {\bibnamefont {Chalker}, \bibfnamefont
  {J~T}}, \ and\ \bibinfo {author} {\bibfnamefont {A.}~\bibnamefont {Dohmen}}}
  (\bibinfo {year} {1995}),\ \bibfield  {title} {\enquote {\bibinfo {title}
  {Three-dimensional disordered conductors in a strong magnetic field:
  {S}urface states and quantum {H}all plateaus},}\ }\href
  {https://link.aps.org/doi/10.1103/PhysRevLett.75.4496} {\bibfield  {journal}
  {\bibinfo  {journal} {Phys. Rev. Lett.}\ }\textbf {\bibinfo {volume} {75}},\
  \bibinfo {pages} {4496--4499}}\BibitemShut {NoStop}%
\bibitem [{\citenamefont {Chang}\ \emph {et~al.}(2013)\citenamefont {Chang},
  \citenamefont {Zhang}, \citenamefont {Feng}, \citenamefont {Shen},
  \citenamefont {Zhang}, \citenamefont {Guo}, \citenamefont {Li}, \citenamefont
  {Ou}, \citenamefont {Wei}, \citenamefont {Wang} \emph
  {et~al.}}]{Chang:2013Science}%
  \BibitemOpen
  \bibfield  {author} {\bibinfo {author} {\bibnamefont {Chang}, \bibfnamefont
  {Cui-Zu}}, \bibinfo {author} {\bibfnamefont {Jinsong}\ \bibnamefont {Zhang}},
  \bibinfo {author} {\bibfnamefont {Xiao}\ \bibnamefont {Feng}}, \bibinfo
  {author} {\bibfnamefont {Jie}\ \bibnamefont {Shen}}, \bibinfo {author}
  {\bibfnamefont {Zuocheng}\ \bibnamefont {Zhang}}, \bibinfo {author}
  {\bibfnamefont {Minghua}\ \bibnamefont {Guo}}, \bibinfo {author}
  {\bibfnamefont {Kang}\ \bibnamefont {Li}}, \bibinfo {author} {\bibfnamefont
  {Yunbo}\ \bibnamefont {Ou}}, \bibinfo {author} {\bibfnamefont {Pang}\
  \bibnamefont {Wei}}, \bibinfo {author} {\bibfnamefont {Li-Li}\ \bibnamefont
  {Wang}},  \emph {et~al.}} (\bibinfo {year} {2013}),\ \bibfield  {title}
  {\enquote {\bibinfo {title} {Experimental observation of the quantum
  anomalous {H}all effect in a magnetic topological insulator},}\ }\href
  {http://science.sciencemag.org/content/340/6129/167} {\bibfield  {journal}
  {\bibinfo  {journal} {Science}\ }\textbf {\bibinfo {volume} {340}}~(\bibinfo
  {number} {6129}),\ \bibinfo {pages} {167--170}}\BibitemShut {NoStop}%
\bibitem [{\citenamefont {Chang}\ \emph {et~al.}(2014)\citenamefont {Chang},
  \citenamefont {Vuleti{\'c}},\ and\ \citenamefont
  {Lukin}}]{Chang:NatPhot2014}%
  \BibitemOpen
  \bibfield  {author} {\bibinfo {author} {\bibnamefont {Chang}, \bibfnamefont
  {Darrick~E}}, \bibinfo {author} {\bibfnamefont {Vladan}\ \bibnamefont
  {Vuleti{\'c}}}, \ and\ \bibinfo {author} {\bibfnamefont {Mikhail~D}\
  \bibnamefont {Lukin}}} (\bibinfo {year} {2014}),\ \bibfield  {title}
  {\enquote {\bibinfo {title} {Quantum nonlinear optics -- photon by photon},}\
  }\href {https://www.nature.com/articles/nphoton.2014.192} {\bibfield
  {journal} {\bibinfo  {journal} {Nat. Photonics}\ }\textbf {\bibinfo {volume}
  {8}}~(\bibinfo {number} {9}),\ \bibinfo {pages} {685--694}}\BibitemShut
  {NoStop}%
\bibitem [{\citenamefont {Chang}\ \emph {et~al.}(2017)\citenamefont {Chang},
  \citenamefont {Xiao}, \citenamefont {Chen},\ and\ \citenamefont
  {Chan}}]{Chang:2017PRB}%
  \BibitemOpen
  \bibfield  {author} {\bibinfo {author} {\bibnamefont {Chang}, \bibfnamefont
  {Ming-Li}}, \bibinfo {author} {\bibfnamefont {Meng}\ \bibnamefont {Xiao}},
  \bibinfo {author} {\bibfnamefont {Wen-Jie}\ \bibnamefont {Chen}}, \ and\
  \bibinfo {author} {\bibfnamefont {Che~Ting}\ \bibnamefont {Chan}}} (\bibinfo
  {year} {2017}),\ \bibfield  {title} {\enquote {\bibinfo {title} {Multiple
  {W}eyl points and the sign change of their topological charges in woodpile
  photonic crystals},}\ }\href
  {https://journals.aps.org/prb/abstract/10.1103/PhysRevB.95.125136} {\bibfield
   {journal} {\bibinfo  {journal} {Phys. Rev. B}\ }\textbf {\bibinfo {volume}
  {95}}~(\bibinfo {number} {12}),\ \bibinfo {pages} {125136}}\BibitemShut
  {NoStop}%
\bibitem [{\citenamefont {Cheianov}\ \emph {et~al.}(2007)\citenamefont
  {Cheianov}, \citenamefont {Fal'ko},\ and\ \citenamefont
  {Altshuler}}]{Cheianov:2007Science}%
  \BibitemOpen
  \bibfield  {author} {\bibinfo {author} {\bibnamefont {Cheianov},
  \bibfnamefont {Vadim~V}}, \bibinfo {author} {\bibfnamefont {Vladimir}\
  \bibnamefont {Fal'ko}}, \ and\ \bibinfo {author} {\bibfnamefont {B~L}\
  \bibnamefont {Altshuler}}} (\bibinfo {year} {2007}),\ \bibfield  {title}
  {\enquote {\bibinfo {title} {{The focusing of electron flow and a Veselago
  lens in graphene p-n junctions}},}\ }\href
  {http://www.sciencemag.org/content/315/5816/1252.short} {\bibfield  {journal}
  {\bibinfo  {journal} {Science}\ }\textbf {\bibinfo {volume} {315}}~(\bibinfo
  {number} {5816}),\ \bibinfo {pages} {1252--5}}\BibitemShut {NoStop}%
\bibitem [{\citenamefont {Chen}\ \emph {et~al.}(1990)\citenamefont {Chen},
  \citenamefont {Yin},\ and\ \citenamefont {Elliott}}]{Chen:PRL1990}%
  \BibitemOpen
  \bibfield  {author} {\bibinfo {author} {\bibnamefont {Chen}, \bibfnamefont
  {Ce}}, \bibinfo {author} {\bibfnamefont {Yi-Yian}\ \bibnamefont {Yin}}, \
  and\ \bibinfo {author} {\bibfnamefont {DS}~\bibnamefont {Elliott}}} (\bibinfo
  {year} {1990}),\ \bibfield  {title} {\enquote {\bibinfo {title} {Interference
  between optical transitions},}\ }\href
  {https://journals.aps.org/prl/abstract/10.1103/PhysRevLett.64.507} {\bibfield
   {journal} {\bibinfo  {journal} {Phys. Rev. Lett.}\ }\textbf {\bibinfo
  {volume} {64}}~(\bibinfo {number} {5}),\ \bibinfo {pages} {507}}\BibitemShut
  {NoStop}%
\bibitem [{\citenamefont {Chen}\ \emph {et~al.}(2014)\citenamefont {Chen},
  \citenamefont {Jiang}, \citenamefont {Chen}, \citenamefont {Zhu},
  \citenamefont {Zhou}, \citenamefont {Dong},\ and\ \citenamefont
  {Chan}}]{Chen:2014NatComm}%
  \BibitemOpen
  \bibfield  {author} {\bibinfo {author} {\bibnamefont {Chen}, \bibfnamefont
  {Wen-Jie}}, \bibinfo {author} {\bibfnamefont {Shao-Ji}\ \bibnamefont
  {Jiang}}, \bibinfo {author} {\bibfnamefont {Xiao-Dong}\ \bibnamefont {Chen}},
  \bibinfo {author} {\bibfnamefont {Baocheng}\ \bibnamefont {Zhu}}, \bibinfo
  {author} {\bibfnamefont {Lei}\ \bibnamefont {Zhou}}, \bibinfo {author}
  {\bibfnamefont {Jian-Wen}\ \bibnamefont {Dong}}, \ and\ \bibinfo {author}
  {\bibfnamefont {Che~Ting}\ \bibnamefont {Chan}}} (\bibinfo {year} {2014}),\
  \bibfield  {title} {\enquote {\bibinfo {title} {Experimental realization of
  photonic topological insulator in a uniaxial metacrystal waveguide},}\ }\href
  {http://www.nature.com/articles/ncomms6782} {\bibfield  {journal} {\bibinfo
  {journal} {Nat. Commun.}\ }\textbf {\bibinfo {volume} {5}},\ \bibinfo {pages}
  {5782}}\BibitemShut {NoStop}%
\bibitem [{\citenamefont {Chen}\ \emph {et~al.}(2016)\citenamefont {Chen},
  \citenamefont {Xiao},\ and\ \citenamefont {Chan}}]{Chen:2016NatComm}%
  \BibitemOpen
  \bibfield  {author} {\bibinfo {author} {\bibnamefont {Chen}, \bibfnamefont
  {Wen-Jie}}, \bibinfo {author} {\bibfnamefont {Meng}\ \bibnamefont {Xiao}}, \
  and\ \bibinfo {author} {\bibfnamefont {Che~Ting}\ \bibnamefont {Chan}}}
  (\bibinfo {year} {2016}),\ \bibfield  {title} {\enquote {\bibinfo {title}
  {Photonic crystals possessing multiple {W}eyl points and the experimental
  observation of robust surface states},}\ }\href
  {https://www.nature.com/articles/ncomms13038} {\bibfield  {journal} {\bibinfo
   {journal} {Nat. Commun.}\ }\textbf {\bibinfo {volume} {7}},\ \bibinfo
  {pages} {13038}}\BibitemShut {NoStop}%
\bibitem [{\citenamefont {Chen}\ and\ \citenamefont
  {Dong}(2016)}]{Chen:arx2016}%
  \BibitemOpen
  \bibfield  {author} {\bibinfo {author} {\bibnamefont {Chen}, \bibfnamefont
  {Xiao-Dong}}, \ and\ \bibinfo {author} {\bibfnamefont {Jian-Wen}\
  \bibnamefont {Dong}}} (\bibinfo {year} {2016}),\ \bibfield  {title} {\enquote
  {\bibinfo {title} {Valley-protected backscattering suppression in silicon
  photonic graphene},}\ }\href {http://arxiv.org/abs/1602.03352} {\bibinfo
  {journal} {arXiv:1602.03352}\ }\BibitemShut {NoStop}%
\bibitem [{\citenamefont {Cheng}\ \emph {et~al.}(2016)\citenamefont {Cheng},
  \citenamefont {Jouvaud}, \citenamefont {Ni}, \citenamefont {Mousavi},
  \citenamefont {Genack},\ and\ \citenamefont {Khanikaev}}]{cheng:2016NatMat}%
  \BibitemOpen
\bibfield  {journal} {  }\bibfield  {author} {\bibinfo {author} {\bibnamefont
  {Cheng}, \bibfnamefont {Xiaojun}}, \bibinfo {author} {\bibfnamefont
  {Camille}\ \bibnamefont {Jouvaud}}, \bibinfo {author} {\bibfnamefont {Xiang}\
  \bibnamefont {Ni}}, \bibinfo {author} {\bibfnamefont {S~Hossein}\
  \bibnamefont {Mousavi}}, \bibinfo {author} {\bibfnamefont {Azriel~Z}\
  \bibnamefont {Genack}}, \ and\ \bibinfo {author} {\bibfnamefont
  {Alexander~B}\ \bibnamefont {Khanikaev}}} (\bibinfo {year} {2016}),\
  \bibfield  {title} {\enquote {\bibinfo {title} {Robust reconfigurable
  electromagnetic pathways within a photonic topological insulator},}\ }\href
  {https://www.nature.com/articles/nmat4573} {\bibfield  {journal} {\bibinfo
  {journal} {Nat. Mater.}\ }\textbf {\bibinfo {volume} {15}}~(\bibinfo {number}
  {5}),\ \bibinfo {pages} {542--548}}\BibitemShut {NoStop}%
\bibitem [{\citenamefont {{Cherpakova}}\ \emph {et~al.}(2018)\citenamefont
  {{Cherpakova}}, \citenamefont {{J{\"o}rg}}, \citenamefont {{Dauer}},
  \citenamefont {{Letscher}}, \citenamefont {{Fleischhauer}}, \citenamefont
  {{Eggert}}, \citenamefont {{Linden}},\ and\ \citenamefont {{von
  Freymann}}}]{Cherpakova:2018arXiv}%
  \BibitemOpen
  \bibfield  {author} {\bibinfo {author} {\bibnamefont {{Cherpakova}},
  \bibfnamefont {Z}}, \bibinfo {author} {\bibfnamefont {C.}~\bibnamefont
  {{J{\"o}rg}}}, \bibinfo {author} {\bibfnamefont {C.}~\bibnamefont {{Dauer}}},
  \bibinfo {author} {\bibfnamefont {F.}~\bibnamefont {{Letscher}}}, \bibinfo
  {author} {\bibfnamefont {M.}~\bibnamefont {{Fleischhauer}}}, \bibinfo
  {author} {\bibfnamefont {S.}~\bibnamefont {{Eggert}}}, \bibinfo {author}
  {\bibfnamefont {S.}~\bibnamefont {{Linden}}}, \ and\ \bibinfo {author}
  {\bibfnamefont {G.}~\bibnamefont {{von Freymann}}}} (\bibinfo {year}
  {2018}),\ \bibfield  {title} {\enquote {\bibinfo {title} {{Depopulation of
  edge states under local periodic driving despite topological protection}},}\
  }\href {https://arxiv.org/abs/1807.02321} {\bibinfo  {journal}
  {arXiv:1807.02321}\ }\BibitemShut {NoStop}%
\bibitem [{\citenamefont {Chiao}\ \emph {et~al.}(1988)\citenamefont {Chiao},
  \citenamefont {Antaramian}, \citenamefont {Ganga}, \citenamefont {Jiao},
  \citenamefont {Wilkinson},\ and\ \citenamefont {Nathel}}]{Chiao:1988PRL}%
  \BibitemOpen
\bibfield  {journal} {  }\bibfield  {author} {\bibinfo {author} {\bibnamefont
  {Chiao}, \bibfnamefont {R~Y}}, \bibinfo {author} {\bibfnamefont
  {A.}~\bibnamefont {Antaramian}}, \bibinfo {author} {\bibfnamefont {K.~M.}\
  \bibnamefont {Ganga}}, \bibinfo {author} {\bibfnamefont {H.}~\bibnamefont
  {Jiao}}, \bibinfo {author} {\bibfnamefont {S.~R.}\ \bibnamefont {Wilkinson}},
  \ and\ \bibinfo {author} {\bibfnamefont {H.}~\bibnamefont {Nathel}}}
  (\bibinfo {year} {1988}),\ \bibfield  {title} {\enquote {\bibinfo {title}
  {Observation of a topological phase by means of a nonplanar {M}ach-{Z}ehnder
  interferometer},}\ }\href
  {https://link.aps.org/doi/10.1103/PhysRevLett.60.1214} {\bibfield  {journal}
  {\bibinfo  {journal} {Phys. Rev. Lett.}\ }\textbf {\bibinfo {volume} {60}},\
  \bibinfo {pages} {1214--1217}}\BibitemShut {NoStop}%
\bibitem [{\citenamefont {Chiu}\ \emph {et~al.}(2016)\citenamefont {Chiu},
  \citenamefont {Teo}, \citenamefont {Schnyder},\ and\ \citenamefont
  {Ryu}}]{Chiu:2016RMP}%
  \BibitemOpen
  \bibfield  {author} {\bibinfo {author} {\bibnamefont {Chiu}, \bibfnamefont
  {Ching-Kai}}, \bibinfo {author} {\bibfnamefont {Jeffrey C.~Y.}\ \bibnamefont
  {Teo}}, \bibinfo {author} {\bibfnamefont {Andreas~P.}\ \bibnamefont
  {Schnyder}}, \ and\ \bibinfo {author} {\bibfnamefont {Shinsei}\ \bibnamefont
  {Ryu}}} (\bibinfo {year} {2016}),\ \bibfield  {title} {\enquote {\bibinfo
  {title} {Classification of topological quantum matter with symmetries},}\
  }\href {https://link.aps.org/doi/10.1103/RevModPhys.88.035005} {\bibfield
  {journal} {\bibinfo  {journal} {Rev. Mod. Phys.}\ }\textbf {\bibinfo {volume}
  {88}},\ \bibinfo {pages} {035005}}\BibitemShut {NoStop}%
\bibitem [{\citenamefont {Cho}\ \emph {et~al.}(2008)\citenamefont {Cho},
  \citenamefont {Angelakis},\ and\ \citenamefont {Bose}}]{Cho:2008PRL}%
  \BibitemOpen
  \bibfield  {author} {\bibinfo {author} {\bibnamefont {Cho}, \bibfnamefont
  {Jaeyoon}}, \bibinfo {author} {\bibfnamefont {Dimitris~G.}\ \bibnamefont
  {Angelakis}}, \ and\ \bibinfo {author} {\bibfnamefont {Sougato}\ \bibnamefont
  {Bose}}} (\bibinfo {year} {2008}),\ \bibfield  {title} {\enquote {\bibinfo
  {title} {Fractional quantum {H}all state in coupled cavities},}\ }\href
  {https://link.aps.org/doi/10.1103/PhysRevLett.101.246809} {\bibfield
  {journal} {\bibinfo  {journal} {Phys. Rev. Lett.}\ }\textbf {\bibinfo
  {volume} {101}},\ \bibinfo {pages} {246809}}\BibitemShut {NoStop}%
\bibitem [{\citenamefont {Chong}\ \emph {et~al.}(2008)\citenamefont {Chong},
  \citenamefont {Wen},\ and\ \citenamefont {Solja\v{c}i\'{c}}}]{Chong:2008PRB}%
  \BibitemOpen
  \bibfield  {author} {\bibinfo {author} {\bibnamefont {Chong}, \bibfnamefont
  {Y~D}}, \bibinfo {author} {\bibfnamefont {Xiao-Gang}\ \bibnamefont {Wen}}, \
  and\ \bibinfo {author} {\bibfnamefont {Marin}\ \bibnamefont
  {Solja\v{c}i\'{c}}}} (\bibinfo {year} {2008}),\ \bibfield  {title} {\enquote
  {\bibinfo {title} {Effective theory of quadratic degeneracies},}\ }\href
  {https://journals.aps.org/prb/abstract/10.1103/PhysRevB.77.235125} {\bibfield
   {journal} {\bibinfo  {journal} {Phys. Rev. B}\ }\textbf {\bibinfo {volume}
  {77}},\ \bibinfo {pages} {235125}}\BibitemShut {NoStop}%
\bibitem [{\citenamefont {Christodoulides}\ \emph {et~al.}(2003)\citenamefont
  {Christodoulides}, \citenamefont {Lederer},\ and\ \citenamefont
  {Silberberg}}]{christodoulides2003discretizing}%
  \BibitemOpen
  \bibfield  {author} {\bibinfo {author} {\bibnamefont {Christodoulides},
  \bibfnamefont {Demetrios~N}}, \bibinfo {author} {\bibfnamefont {Falk}\
  \bibnamefont {Lederer}}, \ and\ \bibinfo {author} {\bibfnamefont {Yaron}\
  \bibnamefont {Silberberg}}} (\bibinfo {year} {2003}),\ \bibfield  {title}
  {\enquote {\bibinfo {title} {Discretizing light behaviour in linear and
  nonlinear waveguide lattices},}\ }\href
  {https://www.nature.com/articles/nature01936} {\bibfield  {journal} {\bibinfo
   {journal} {Nature}\ }\textbf {\bibinfo {volume} {424}}~(\bibinfo {number}
  {6950}),\ \bibinfo {pages} {817--823}}\BibitemShut {NoStop}%
\bibitem [{\citenamefont {Chu}\ \emph {et~al.}(2017)\citenamefont {Chu},
  \citenamefont {Kharel}, \citenamefont {Renninger}, \citenamefont {Burkhart},
  \citenamefont {Frunzio}, \citenamefont {Rakich},\ and\ \citenamefont
  {Schoelkopf}}]{Schoelkopf2017}%
  \BibitemOpen
  \bibfield  {author} {\bibinfo {author} {\bibnamefont {Chu}, \bibfnamefont
  {Yiwen}}, \bibinfo {author} {\bibfnamefont {Prashanta}\ \bibnamefont
  {Kharel}}, \bibinfo {author} {\bibfnamefont {William~H}\ \bibnamefont
  {Renninger}}, \bibinfo {author} {\bibfnamefont {Luke~D}\ \bibnamefont
  {Burkhart}}, \bibinfo {author} {\bibfnamefont {Luigi}\ \bibnamefont
  {Frunzio}}, \bibinfo {author} {\bibfnamefont {Peter~T}\ \bibnamefont
  {Rakich}}, \ and\ \bibinfo {author} {\bibfnamefont {Robert~J}\ \bibnamefont
  {Schoelkopf}}} (\bibinfo {year} {2017}),\ \bibfield  {title} {\enquote
  {\bibinfo {title} {Quantum acoustics with superconducting qubits},}\ }\href
  {http://science.sciencemag.org/content/early/2017/09/20/science.aao1511}
  {\bibfield  {journal} {\bibinfo  {journal} {Science}\ }\textbf {\bibinfo
  {volume} {358}},\ \bibinfo {pages} {199--202}}\BibitemShut {NoStop}%
\bibitem [{\citenamefont {Ciuti}\ \emph {et~al.}(1998)\citenamefont {Ciuti},
  \citenamefont {Savona}, \citenamefont {Piermarocchi}, \citenamefont
  {Quattropani},\ and\ \citenamefont {Schwendimann}}]{Ciuti:PRB1998}%
  \BibitemOpen
  \bibfield  {author} {\bibinfo {author} {\bibnamefont {Ciuti}, \bibfnamefont
  {C}}, \bibinfo {author} {\bibfnamefont {V}~\bibnamefont {Savona}}, \bibinfo
  {author} {\bibfnamefont {C}~\bibnamefont {Piermarocchi}}, \bibinfo {author}
  {\bibfnamefont {A}~\bibnamefont {Quattropani}}, \ and\ \bibinfo {author}
  {\bibfnamefont {P}~\bibnamefont {Schwendimann}}} (\bibinfo {year} {1998}),\
  \bibfield  {title} {\enquote {\bibinfo {title} {Role of the exchange of
  carriers in elastic exciton-exciton scattering in quantum wells},}\ }\href
  {https://journals.aps.org/prb/abstract/10.1103/PhysRevB.58.7926} {\bibfield
  {journal} {\bibinfo  {journal} {Phys. Rev. B}\ }\textbf {\bibinfo {volume}
  {58}}~(\bibinfo {number} {12}),\ \bibinfo {pages} {7926--7933}}\BibitemShut
  {NoStop}%
\bibitem [{\citenamefont {Claassen}\ \emph {et~al.}(2015)\citenamefont
  {Claassen}, \citenamefont {Lee}, \citenamefont {Thomale}, \citenamefont
  {Qi},\ and\ \citenamefont {Devereaux}}]{Claassen:PRL2015}%
  \BibitemOpen
  \bibfield  {author} {\bibinfo {author} {\bibnamefont {Claassen},
  \bibfnamefont {Martin}}, \bibinfo {author} {\bibfnamefont {Ching~Hua}\
  \bibnamefont {Lee}}, \bibinfo {author} {\bibfnamefont {Ronny}\ \bibnamefont
  {Thomale}}, \bibinfo {author} {\bibfnamefont {Xiao-Liang}\ \bibnamefont
  {Qi}}, \ and\ \bibinfo {author} {\bibfnamefont {Thomas~P}\ \bibnamefont
  {Devereaux}}} (\bibinfo {year} {2015}),\ \bibfield  {title} {\enquote
  {\bibinfo {title} {Position-momentum duality and fractional quantum {H}all
  effect in {C}hern insulators},}\ }\href
  {https://journals.aps.org/prl/abstract/10.1103/PhysRevLett.114.236802}
  {\bibfield  {journal} {\bibinfo  {journal} {Phys. Rev. Lett.}\ }\textbf
  {\bibinfo {volume} {114}}~(\bibinfo {number} {23}),\ \bibinfo {pages}
  {236802}}\BibitemShut {NoStop}%
\bibitem [{\citenamefont {Clark}\ \emph {et~al.}(2017)\citenamefont {Clark},
  \citenamefont {Lecocq}, \citenamefont {Simmonds}, \citenamefont {Aumentado},\
  and\ \citenamefont {Teufel}}]{Teufel2017}%
  \BibitemOpen
  \bibfield  {author} {\bibinfo {author} {\bibnamefont {Clark}, \bibfnamefont
  {Jeremy~B}}, \bibinfo {author} {\bibfnamefont {Florent}\ \bibnamefont
  {Lecocq}}, \bibinfo {author} {\bibfnamefont {Raymond~W}\ \bibnamefont
  {Simmonds}}, \bibinfo {author} {\bibfnamefont {Jos{\'e}}\ \bibnamefont
  {Aumentado}}, \ and\ \bibinfo {author} {\bibfnamefont {John~D}\ \bibnamefont
  {Teufel}}} (\bibinfo {year} {2017}),\ \bibfield  {title} {\enquote {\bibinfo
  {title} {Sideband cooling beyond the quantum backaction limit with squeezed
  light},}\ }\href {https://www.nature.com/articles/nature20604} {\bibfield
  {journal} {\bibinfo  {journal} {Nature}\ }\textbf {\bibinfo {volume}
  {541}}~(\bibinfo {number} {7636}),\ \bibinfo {pages} {191--195}}\BibitemShut
  {NoStop}%
\bibitem [{\citenamefont {Cohen-Tannoudji}\ \emph {et~al.}(2008)\citenamefont
  {Cohen-Tannoudji}, \citenamefont {Dupont-Roc},\ and\ \citenamefont
  {Grynberg}}]{CohenTannoudji4}%
  \BibitemOpen
  \bibfield  {author} {\bibinfo {author} {\bibnamefont {Cohen-Tannoudji},
  \bibfnamefont {Claude}}, \bibinfo {author} {\bibfnamefont {Jacques}\
  \bibnamefont {Dupont-Roc}}, \ and\ \bibinfo {author} {\bibfnamefont
  {Gilbert}\ \bibnamefont {Grynberg}}} (\bibinfo {year} {2008}),\ \href@noop {}
  {\emph {\bibinfo {title} {Atom—Photon Interactions}}}\ (\bibinfo
  {publisher} {Wiley-VCH Verlag GmbH},\ \bibinfo {address}
  {Weinheim})\BibitemShut {NoStop}%
\bibitem [{\citenamefont {Cominotti}\ and\ \citenamefont
  {Carusotto}(2013)}]{Cominotti:EPL2013}%
  \BibitemOpen
  \bibfield  {author} {\bibinfo {author} {\bibnamefont {Cominotti},
  \bibfnamefont {Marco}}, \ and\ \bibinfo {author} {\bibfnamefont {Iacopo}\
  \bibnamefont {Carusotto}}} (\bibinfo {year} {2013}),\ \bibfield  {title}
  {\enquote {\bibinfo {title} {Berry curvature effects in the {B}loch
  oscillations of a quantum particle under a strong (synthetic) magnetic
  field},}\ }\href
  {http://iopscience.iop.org/article/10.1209/0295-5075/103/10001/meta}
  {\bibfield  {journal} {\bibinfo  {journal} {EPL (Europhysics Letters)}\
  }\textbf {\bibinfo {volume} {103}}~(\bibinfo {number} {1}),\ \bibinfo {pages}
  {10001}}\BibitemShut {NoStop}%
\bibitem [{\citenamefont {Cooper}\ \emph {et~al.}(2010)\citenamefont {Cooper},
  \citenamefont {Gupta}, \citenamefont {Schneider}, \citenamefont {Green},
  \citenamefont {Assefa}, \citenamefont {Xia}, \citenamefont {Vlasov},\ and\
  \citenamefont {Mookherjea}}]{Cooper:2010}%
  \BibitemOpen
  \bibfield  {author} {\bibinfo {author} {\bibnamefont {Cooper}, \bibfnamefont
  {Michael~L}}, \bibinfo {author} {\bibfnamefont {Greeshma}\ \bibnamefont
  {Gupta}}, \bibinfo {author} {\bibfnamefont {Mark~A.}\ \bibnamefont
  {Schneider}}, \bibinfo {author} {\bibfnamefont {William M.~J.}\ \bibnamefont
  {Green}}, \bibinfo {author} {\bibfnamefont {Solomon}\ \bibnamefont {Assefa}},
  \bibinfo {author} {\bibfnamefont {Fengnian}\ \bibnamefont {Xia}}, \bibinfo
  {author} {\bibfnamefont {Yurii~A.}\ \bibnamefont {Vlasov}}, \ and\ \bibinfo
  {author} {\bibfnamefont {Shayan}\ \bibnamefont {Mookherjea}}} (\bibinfo
  {year} {2010}),\ \bibfield  {title} {\enquote {\bibinfo {title} {{Statistics
  of light transport in 235-ring silicon coupled-resonator optical
  waveguides}},}\ }\href
  {https://www.osapublishing.org/oe/abstract.cfm?uri=oe-18-25-26505} {\bibfield
   {journal} {\bibinfo  {journal} {Opt. Express}\ }\textbf {\bibinfo {volume}
  {18}}~(\bibinfo {number} {25}),\ \bibinfo {pages} {26505--26516}}\BibitemShut
  {NoStop}%
\bibitem [{\citenamefont {Cooper}(2008)}]{Cooper:AdvPhys2008}%
  \BibitemOpen
  \bibfield  {author} {\bibinfo {author} {\bibnamefont {Cooper}, \bibfnamefont
  {Nigel~R}}} (\bibinfo {year} {2008}),\ \bibfield  {title} {\enquote {\bibinfo
  {title} {Rapidly rotating atomic gases},}\ }\href
  {https://www.tandfonline.com/doi/abs/10.1080/00018730802564122} {\bibfield
  {journal} {\bibinfo  {journal} {Adv. Phys.}\ }\textbf {\bibinfo {volume}
  {57}}~(\bibinfo {number} {6}),\ \bibinfo {pages} {539--616}}\BibitemShut
  {NoStop}%
\bibitem [{\citenamefont {Creffield}\ \emph {et~al.}(2016)\citenamefont
  {Creffield}, \citenamefont {Pieplow}, \citenamefont {Sols},\ and\
  \citenamefont {Goldman}}]{Creffield:2016NJP}%
  \BibitemOpen
  \bibfield  {author} {\bibinfo {author} {\bibnamefont {Creffield},
  \bibfnamefont {C~E}}, \bibinfo {author} {\bibfnamefont {G}~\bibnamefont
  {Pieplow}}, \bibinfo {author} {\bibfnamefont {F}~\bibnamefont {Sols}}, \ and\
  \bibinfo {author} {\bibfnamefont {N}~\bibnamefont {Goldman}}} (\bibinfo
  {year} {2016}),\ \bibfield  {title} {\enquote {\bibinfo {title} {Realization
  of uniform synthetic magnetic fields by periodically shaking an optical
  square lattice},}\ }\href {http://stacks.iop.org/1367-2630/18/i=9/a=093013}
  {\bibfield  {journal} {\bibinfo  {journal} {New J. Phys.}\ }\textbf {\bibinfo
  {volume} {18}}~(\bibinfo {number} {9}),\ \bibinfo {pages}
  {093013}}\BibitemShut {NoStop}%
\bibitem [{\citenamefont {Crespi}\ \emph {et~al.}(2013)\citenamefont {Crespi},
  \citenamefont {Corrielli}, \citenamefont {Della~Valle}, \citenamefont
  {Osellame},\ and\ \citenamefont {Longhi}}]{crespi2013dynamic}%
  \BibitemOpen
  \bibfield  {author} {\bibinfo {author} {\bibnamefont {Crespi}, \bibfnamefont
  {Andrea}}, \bibinfo {author} {\bibfnamefont {Giacomo}\ \bibnamefont
  {Corrielli}}, \bibinfo {author} {\bibfnamefont {Giuseppe}\ \bibnamefont
  {Della~Valle}}, \bibinfo {author} {\bibfnamefont {Roberto}\ \bibnamefont
  {Osellame}}, \ and\ \bibinfo {author} {\bibfnamefont {Stefano}\ \bibnamefont
  {Longhi}}} (\bibinfo {year} {2013}),\ \bibfield  {title} {\enquote {\bibinfo
  {title} {Dynamic band collapse in photonic graphene},}\ }\href
  {https://iopscience.iop.org/article/10.1088/1367-2630/15/1/013012/meta}
  {\bibfield  {journal} {\bibinfo  {journal} {New J. Phys.}\ }\textbf {\bibinfo
  {volume} {15}}~(\bibinfo {number} {1}),\ \bibinfo {pages}
  {013012}}\BibitemShut {NoStop}%
\bibitem [{\citenamefont {Dalibard}\ \emph {et~al.}(2011)\citenamefont
  {Dalibard}, \citenamefont {Gerbier}, \citenamefont {Juzeli{\=u}nas},\ and\
  \citenamefont {{\"O}hberg}}]{Dalibard:RMP2011}%
  \BibitemOpen
  \bibfield  {author} {\bibinfo {author} {\bibnamefont {Dalibard},
  \bibfnamefont {Jean}}, \bibinfo {author} {\bibfnamefont {Fabrice}\
  \bibnamefont {Gerbier}}, \bibinfo {author} {\bibfnamefont {Gediminas}\
  \bibnamefont {Juzeli{\=u}nas}}, \ and\ \bibinfo {author} {\bibfnamefont
  {Patrik}\ \bibnamefont {{\"O}hberg}}} (\bibinfo {year} {2011}),\ \bibfield
  {title} {\enquote {\bibinfo {title} {Colloquium: {A}rtificial gauge
  potentials for neutral atoms},}\ }\href
  {https://journals.aps.org/rmp/abstract/10.1103/RevModPhys.83.1523} {\bibfield
   {journal} {\bibinfo  {journal} {Rev. Mod. Phys.}\ }\textbf {\bibinfo
  {volume} {83}}~(\bibinfo {number} {4}),\ \bibinfo {pages} {1523}}\BibitemShut
  {NoStop}%
\bibitem [{\citenamefont {Dana}\ \emph {et~al.}(1985)\citenamefont {Dana},
  \citenamefont {Avron},\ and\ \citenamefont {Zak}}]{Dana:1985JPC}%
  \BibitemOpen
  \bibfield  {author} {\bibinfo {author} {\bibnamefont {Dana}, \bibfnamefont
  {Itzhack}}, \bibinfo {author} {\bibfnamefont {Yosi}\ \bibnamefont {Avron}}, \
  and\ \bibinfo {author} {\bibfnamefont {J}~\bibnamefont {Zak}}} (\bibinfo
  {year} {1985}),\ \bibfield  {title} {\enquote {\bibinfo {title} {Quantised
  {H}all conductance in a perfect crystal},}\ }\href
  {http://iopscience.iop.org/article/10.1088/0022-3719/18/22/004/meta}
  {\bibfield  {journal} {\bibinfo  {journal} {J. Phys. C}\ }\textbf {\bibinfo
  {volume} {18}}~(\bibinfo {number} {22}),\ \bibinfo {pages}
  {L679}}\BibitemShut {NoStop}%
\bibitem [{\citenamefont {De~Nittis}\ and\ \citenamefont
  {Lein}(2017)}]{DeNittis:2017arXiv}%
  \BibitemOpen
  \bibfield  {author} {\bibinfo {author} {\bibnamefont {De~Nittis},
  \bibfnamefont {Giuseppe}}, \ and\ \bibinfo {author} {\bibfnamefont {Max}\
  \bibnamefont {Lein}}} (\bibinfo {year} {2017}),\ \bibfield  {title} {\enquote
  {\bibinfo {title} {Symmetry classification of topological photonic
  crystals},}\ }\href {https://arxiv.org/abs/1710.08104} {\bibinfo  {journal}
  {arXiv:1710.08104}\ }\BibitemShut {NoStop}%
\bibitem [{\citenamefont {Delplace}\ \emph {et~al.}(2011)\citenamefont
  {Delplace}, \citenamefont {Ullmo},\ and\ \citenamefont
  {Montambaux}}]{Delplace:2011PRB}%
  \BibitemOpen
\bibfield  {journal} {  }\bibfield  {author} {\bibinfo {author} {\bibnamefont
  {Delplace}, \bibfnamefont {P}}, \bibinfo {author} {\bibfnamefont
  {D.}~\bibnamefont {Ullmo}}, \ and\ \bibinfo {author} {\bibfnamefont
  {G.}~\bibnamefont {Montambaux}}} (\bibinfo {year} {2011}),\ \bibfield
  {title} {\enquote {\bibinfo {title} {{Zak phase and the existence of edge
  states in graphene}},}\ }\href
  {http://link.aps.org/doi/10.1103/PhysRevB.84.195452} {\bibfield  {journal}
  {\bibinfo  {journal} {Phys. Rev. B}\ }\textbf {\bibinfo {volume}
  {84}}~(\bibinfo {number} {19}),\ \bibinfo {pages} {195452}}\BibitemShut
  {NoStop}%
\bibitem [{\citenamefont {Deng}\ \emph {et~al.}(2015)\citenamefont {Deng},
  \citenamefont {Hong}, \citenamefont {Zheng},\ and\ \citenamefont
  {Shen}}]{Deng:2015AO}%
  \BibitemOpen
  \bibfield  {author} {\bibinfo {author} {\bibnamefont {Deng}, \bibfnamefont
  {Xiaohua}}, \bibinfo {author} {\bibfnamefont {Lujun}\ \bibnamefont {Hong}},
  \bibinfo {author} {\bibfnamefont {Xiaodong}\ \bibnamefont {Zheng}}, \ and\
  \bibinfo {author} {\bibfnamefont {Linfang}\ \bibnamefont {Shen}}} (\bibinfo
  {year} {2015}),\ \bibfield  {title} {\enquote {\bibinfo {title} {One-way
  regular electromagnetic mode immune to backscattering},}\ }\href
  {https://www.osapublishing.org/ao/abstract.cfm?uri=ao-54-14-4608} {\bibfield
  {journal} {\bibinfo  {journal} {Applied optics}\ }\textbf {\bibinfo {volume}
  {54}}~(\bibinfo {number} {14}),\ \bibinfo {pages} {4608--4612}}\BibitemShut
  {NoStop}%
\bibitem [{\citenamefont {Dennis}\ \emph {et~al.}(2009)\citenamefont {Dennis},
  \citenamefont {O'Holleran},\ and\ \citenamefont
  {Padgett}}]{Dennis:2009ProgOpt}%
  \BibitemOpen
  \bibfield  {author} {\bibinfo {author} {\bibnamefont {Dennis}, \bibfnamefont
  {Mark~R}}, \bibinfo {author} {\bibfnamefont {Kevin}\ \bibnamefont
  {O'Holleran}}, \ and\ \bibinfo {author} {\bibfnamefont {Miles~J}\
  \bibnamefont {Padgett}}} (\bibinfo {year} {2009}),\ \bibfield  {title}
  {\enquote {\bibinfo {title} {Singular optics: {O}ptical vortices and
  polarization singularities},}\ }\href
  {https://www.sciencedirect.com/science/article/pii/S0079663808002059}
  {\bibfield  {journal} {\bibinfo  {journal} {Progress in Optics}\ }\textbf
  {\bibinfo {volume} {53}},\ \bibinfo {pages} {293--363}}\BibitemShut {NoStop}%
\bibitem [{\citenamefont {Di~Liberto}\ \emph
  {et~al.}(2016{\natexlab{a}})\citenamefont {Di~Liberto}, \citenamefont
  {Hemmerich},\ and\ \citenamefont {Morais~Smith}}]{DiLiberto:2016PRL}%
  \BibitemOpen
  \bibfield  {author} {\bibinfo {author} {\bibnamefont {Di~Liberto},
  \bibfnamefont {M}}, \bibinfo {author} {\bibfnamefont {A.}~\bibnamefont
  {Hemmerich}}, \ and\ \bibinfo {author} {\bibfnamefont {C.}~\bibnamefont
  {Morais~Smith}}} (\bibinfo {year} {2016}{\natexlab{a}}),\ \bibfield  {title}
  {\enquote {\bibinfo {title} {Topological {V}arma superfluid in optical
  lattices},}\ }\href {https://link.aps.org/doi/10.1103/PhysRevLett.117.163001}
  {\bibfield  {journal} {\bibinfo  {journal} {Phys. Rev. Lett.}\ }\textbf
  {\bibinfo {volume} {117}},\ \bibinfo {pages} {163001}}\BibitemShut {NoStop}%
\bibitem [{\citenamefont {Di~Liberto}\ \emph
  {et~al.}(2016{\natexlab{b}})\citenamefont {Di~Liberto}, \citenamefont
  {Recati}, \citenamefont {Carusotto},\ and\ \citenamefont
  {Menotti}}]{DiLiberto:PRA2016}%
  \BibitemOpen
  \bibfield  {author} {\bibinfo {author} {\bibnamefont {Di~Liberto},
  \bibfnamefont {M}}, \bibinfo {author} {\bibfnamefont {A}~\bibnamefont
  {Recati}}, \bibinfo {author} {\bibfnamefont {I}~\bibnamefont {Carusotto}}, \
  and\ \bibinfo {author} {\bibfnamefont {C}~\bibnamefont {Menotti}}} (\bibinfo
  {year} {2016}{\natexlab{b}}),\ \bibfield  {title} {\enquote {\bibinfo {title}
  {{Two-body physics in the Su-Schrieffer-Heeger model}},}\ }\href
  {https://journals.aps.org/pra/abstract/10.1103/PhysRevA.94.062704} {\bibfield
   {journal} {\bibinfo  {journal} {Phys. Rev. A}\ }\textbf {\bibinfo {volume}
  {94}}~(\bibinfo {number} {6}),\ \bibinfo {pages} {062704}}\BibitemShut
  {NoStop}%
\bibitem [{\citenamefont {Diehl}\ \emph {et~al.}(2011)\citenamefont {Diehl},
  \citenamefont {Rico}, \citenamefont {Baranov},\ and\ \citenamefont
  {Zoller}}]{diehl2011topology}%
  \BibitemOpen
  \bibfield  {author} {\bibinfo {author} {\bibnamefont {Diehl}, \bibfnamefont
  {Sebastian}}, \bibinfo {author} {\bibfnamefont {Enrique}\ \bibnamefont
  {Rico}}, \bibinfo {author} {\bibfnamefont {Mikhail~A}\ \bibnamefont
  {Baranov}}, \ and\ \bibinfo {author} {\bibfnamefont {Peter}\ \bibnamefont
  {Zoller}}} (\bibinfo {year} {2011}),\ \bibfield  {title} {\enquote {\bibinfo
  {title} {Topology by dissipation in atomic quantum wires},}\ }\href
  {https://www.nature.com/articles/nphys2106} {\bibfield  {journal} {\bibinfo
  {journal} {Nat. Phys.}\ }\textbf {\bibinfo {volume} {7}}~(\bibinfo {number}
  {12}),\ \bibinfo {pages} {971--977}}\BibitemShut {NoStop}%
\bibitem [{\citenamefont {Downing}\ and\ \citenamefont
  {Weick}(2017)}]{Downing:PRB2017}%
  \BibitemOpen
  \bibfield  {author} {\bibinfo {author} {\bibnamefont {Downing}, \bibfnamefont
  {Charles~A}}, \ and\ \bibinfo {author} {\bibfnamefont {Guillaume}\
  \bibnamefont {Weick}}} (\bibinfo {year} {2017}),\ \bibfield  {title}
  {\enquote {\bibinfo {title} {Topological collective plasmons in bipartite
  chains of metallic nanoparticles},}\ }\href
  {https://journals.aps.org/prb/abstract/10.1103/PhysRevB.95.125426} {\bibfield
   {journal} {\bibinfo  {journal} {Phys. Rev. B}\ }\textbf {\bibinfo {volume}
  {95}}~(\bibinfo {number} {12}),\ \bibinfo {pages} {125426}}\BibitemShut
  {NoStop}%
\bibitem [{\citenamefont {Dreisow}\ \emph {et~al.}(2012)\citenamefont
  {Dreisow}, \citenamefont {Keil}, \citenamefont {T{\"{u}}nnermann},
  \citenamefont {Nolte}, \citenamefont {Longhi},\ and\ \citenamefont
  {Szameit}}]{Dreisow:2012EPL}%
  \BibitemOpen
  \bibfield  {author} {\bibinfo {author} {\bibnamefont {Dreisow}, \bibfnamefont
  {F}}, \bibinfo {author} {\bibfnamefont {R}~\bibnamefont {Keil}}, \bibinfo
  {author} {\bibfnamefont {A}~\bibnamefont {T{\"{u}}nnermann}}, \bibinfo
  {author} {\bibfnamefont {S}~\bibnamefont {Nolte}}, \bibinfo {author}
  {\bibfnamefont {S}~\bibnamefont {Longhi}}, \ and\ \bibinfo {author}
  {\bibfnamefont {A}~\bibnamefont {Szameit}}} (\bibinfo {year} {2012}),\
  \bibfield  {title} {\enquote {\bibinfo {title} {{Klein tunneling of light in
  waveguide superlattices}},}\ }\href
  {http://iopscience.iop.org/article/10.1209/0295-5075/97/10008} {\bibfield
  {journal} {\bibinfo  {journal} {EPL}\ }\textbf {\bibinfo {volume}
  {97}}~(\bibinfo {number} {1}),\ \bibinfo {pages} {10008}}\BibitemShut
  {NoStop}%
\bibitem [{\citenamefont {Dub{\v{c}}ek}\ \emph
  {et~al.}(2015{\natexlab{a}})\citenamefont {Dub{\v{c}}ek}, \citenamefont
  {Kennedy}, \citenamefont {Lu}, \citenamefont {Ketterle}, \citenamefont
  {Solja{\v{c}}i{\'c}},\ and\ \citenamefont {Buljan}}]{Dubvcek:2015PRL}%
  \BibitemOpen
  \bibfield  {author} {\bibinfo {author} {\bibnamefont {Dub{\v{c}}ek},
  \bibfnamefont {Tena}}, \bibinfo {author} {\bibfnamefont {Colin~J}\
  \bibnamefont {Kennedy}}, \bibinfo {author} {\bibfnamefont {Ling}\
  \bibnamefont {Lu}}, \bibinfo {author} {\bibfnamefont {Wolfgang}\ \bibnamefont
  {Ketterle}}, \bibinfo {author} {\bibfnamefont {Marin}\ \bibnamefont
  {Solja{\v{c}}i{\'c}}}, \ and\ \bibinfo {author} {\bibfnamefont {Hrvoje}\
  \bibnamefont {Buljan}}} (\bibinfo {year} {2015}{\natexlab{a}}),\ \bibfield
  {title} {\enquote {\bibinfo {title} {Weyl points in three-dimensional optical
  lattices: {S}ynthetic magnetic monopoles in momentum space},}\ }\href
  {https://journals.aps.org/prl/abstract/10.1103/PhysRevLett.114.225301}
  {\bibfield  {journal} {\bibinfo  {journal} {Phys. Rev. Lett.}\ }\textbf
  {\bibinfo {volume} {114}}~(\bibinfo {number} {22}),\ \bibinfo {pages}
  {225301}}\BibitemShut {NoStop}%
\bibitem [{\citenamefont {Dub{\v{c}}ek}\ \emph
  {et~al.}(2015{\natexlab{b}})\citenamefont {Dub{\v{c}}ek}, \citenamefont
  {Lelas}, \citenamefont {Juki{\'c}}, \citenamefont {Pezer}, \citenamefont
  {Solja{\v{c}}i{\'c}},\ and\ \citenamefont {Buljan}}]{Dubcek:2015NJP}%
  \BibitemOpen
  \bibfield  {author} {\bibinfo {author} {\bibnamefont {Dub{\v{c}}ek},
  \bibfnamefont {Tena}}, \bibinfo {author} {\bibfnamefont {Karlo}\ \bibnamefont
  {Lelas}}, \bibinfo {author} {\bibfnamefont {Dario}\ \bibnamefont
  {Juki{\'c}}}, \bibinfo {author} {\bibfnamefont {Robert}\ \bibnamefont
  {Pezer}}, \bibinfo {author} {\bibfnamefont {Marin}\ \bibnamefont
  {Solja{\v{c}}i{\'c}}}, \ and\ \bibinfo {author} {\bibfnamefont {Hrvoje}\
  \bibnamefont {Buljan}}} (\bibinfo {year} {2015}{\natexlab{b}}),\ \bibfield
  {title} {\enquote {\bibinfo {title} {The {H}arper--{H}ofstadter {H}amiltonian
  and conical diffraction in photonic lattices with grating assisted
  tunneling},}\ }\href
  {http://iopscience.iop.org/article/10.1088/1367-2630/17/12/125002/meta}
  {\bibfield  {journal} {\bibinfo  {journal} {New J. Phys.}\ }\textbf {\bibinfo
  {volume} {17}}~(\bibinfo {number} {12}),\ \bibinfo {pages}
  {125002}}\BibitemShut {NoStop}%
\bibitem [{\citenamefont {Dudarev}\ \emph {et~al.}(2004)\citenamefont
  {Dudarev}, \citenamefont {Diener}, \citenamefont {Carusotto},\ and\
  \citenamefont {Niu}}]{Dudarev:2004PRL}%
  \BibitemOpen
  \bibfield  {author} {\bibinfo {author} {\bibnamefont {Dudarev}, \bibfnamefont
  {Artem~M}}, \bibinfo {author} {\bibfnamefont {Roberto~B}\ \bibnamefont
  {Diener}}, \bibinfo {author} {\bibfnamefont {Iacopo}\ \bibnamefont
  {Carusotto}}, \ and\ \bibinfo {author} {\bibfnamefont {Qian}\ \bibnamefont
  {Niu}}} (\bibinfo {year} {2004}),\ \bibfield  {title} {\enquote {\bibinfo
  {title} {Spin-orbit coupling and {B}erry phase with ultracold atoms in 2{D}
  optical lattices},}\ }\href
  {https://journals.aps.org/prl/abstract/10.1103/PhysRevLett.92.153005}
  {\bibfield  {journal} {\bibinfo  {journal} {Phys. Rev. Lett.}\ }\textbf
  {\bibinfo {volume} {92}}~(\bibinfo {number} {15}),\ \bibinfo {pages}
  {153005}}\BibitemShut {NoStop}%
\bibitem [{\citenamefont {Dutta}\ and\ \citenamefont
  {Mueller}(2018)}]{Dutta:2017arXiv}%
  \BibitemOpen
  \bibfield  {author} {\bibinfo {author} {\bibnamefont {Dutta}, \bibfnamefont
  {Shovan}}, \ and\ \bibinfo {author} {\bibfnamefont {Erich~J.}\ \bibnamefont
  {Mueller}}} (\bibinfo {year} {2018}),\ \bibfield  {title} {\enquote {\bibinfo
  {title} {Coherent generation of photonic fractional quantum {H}all states in
  a cavity and the search for anyonic quasiparticles},}\ }\href
  {https://link.aps.org/doi/10.1103/PhysRevA.97.033825} {\bibfield  {journal}
  {\bibinfo  {journal} {Phys. Rev. A}\ }\textbf {\bibinfo {volume} {97}},\
  \bibinfo {pages} {033825}}\BibitemShut {NoStop}%
\bibitem [{\citenamefont {Eckardt}\ and\ \citenamefont
  {Anisimovas}(2015)}]{Eckardt:2015NJP}%
  \BibitemOpen
  \bibfield  {author} {\bibinfo {author} {\bibnamefont {Eckardt}, \bibfnamefont
  {A}}, \ and\ \bibinfo {author} {\bibfnamefont {E.}~\bibnamefont
  {Anisimovas}}} (\bibinfo {year} {2015}),\ \bibfield  {title} {\enquote
  {\bibinfo {title} {High-frequency approximation for periodically driven
  quantum systems from a {F}loquet-space perspective},}\ }\href
  {https://doi.org/10.1088/1367-2630/17/9/093039} {\bibfield  {journal}
  {\bibinfo  {journal} {New J. Phys.}\ }\textbf {\bibinfo {volume} {17}},\
  \bibinfo {pages} {093039}}\BibitemShut {NoStop}%
\bibitem [{\citenamefont {Eckardt}(2017)}]{Eckardt:2016Review}%
  \BibitemOpen
  \bibfield  {author} {\bibinfo {author} {\bibnamefont {Eckardt}, \bibfnamefont
  {Andr\'e}}} (\bibinfo {year} {2017}),\ \bibfield  {title} {\enquote {\bibinfo
  {title} {Colloquium: {A}tomic quantum gases in periodically driven optical
  lattices},}\ }\href {https://link.aps.org/doi/10.1103/RevModPhys.89.011004}
  {\bibfield  {journal} {\bibinfo  {journal} {Rev. Mod. Phys.}\ }\textbf
  {\bibinfo {volume} {89}},\ \bibinfo {pages} {011004}}\BibitemShut {NoStop}%
\bibitem [{\citenamefont {Efremidis}\ \emph {et~al.}(2002)\citenamefont
  {Efremidis}, \citenamefont {Sears}, \citenamefont {Christodoulides},
  \citenamefont {Fleischer},\ and\ \citenamefont
  {Segev}}]{efremidis2002discrete}%
  \BibitemOpen
  \bibfield  {author} {\bibinfo {author} {\bibnamefont {Efremidis},
  \bibfnamefont {Nikos~K}}, \bibinfo {author} {\bibfnamefont {Suzanne}\
  \bibnamefont {Sears}}, \bibinfo {author} {\bibfnamefont {Demetrios~N}\
  \bibnamefont {Christodoulides}}, \bibinfo {author} {\bibfnamefont {Jason~W}\
  \bibnamefont {Fleischer}}, \ and\ \bibinfo {author} {\bibfnamefont
  {Mordechai}\ \bibnamefont {Segev}}} (\bibinfo {year} {2002}),\ \bibfield
  {title} {\enquote {\bibinfo {title} {Discrete solitons in photorefractive
  optically induced photonic lattices},}\ }\href
  {https://journals.aps.org/pre/abstract/10.1103/PhysRevE.66.046602} {\bibfield
   {journal} {\bibinfo  {journal} {Phys. Rev. E}\ }\textbf {\bibinfo {volume}
  {66}}~(\bibinfo {number} {4}),\ \bibinfo {pages} {046602}}\BibitemShut
  {NoStop}%
\bibitem [{\citenamefont {Eisenberg}\ \emph {et~al.}(1998)\citenamefont
  {Eisenberg}, \citenamefont {Silberberg}, \citenamefont {Morandotti},
  \citenamefont {Boyd},\ and\ \citenamefont {Aitchison}}]{Eisenberg:PRL1998}%
  \BibitemOpen
  \bibfield  {author} {\bibinfo {author} {\bibnamefont {Eisenberg},
  \bibfnamefont {HS}}, \bibinfo {author} {\bibfnamefont {Ya}~\bibnamefont
  {Silberberg}}, \bibinfo {author} {\bibfnamefont {R}~\bibnamefont
  {Morandotti}}, \bibinfo {author} {\bibfnamefont {AR}~\bibnamefont {Boyd}}, \
  and\ \bibinfo {author} {\bibfnamefont {JS}~\bibnamefont {Aitchison}}}
  (\bibinfo {year} {1998}),\ \bibfield  {title} {\enquote {\bibinfo {title}
  {Discrete spatial optical solitons in waveguide arrays},}\ }\href
  {https://journals.aps.org/prl/abstract/10.1103/PhysRevLett.81.3383}
  {\bibfield  {journal} {\bibinfo  {journal} {Phys. Rev. Lett.}\ }\textbf
  {\bibinfo {volume} {81}}~(\bibinfo {number} {16}),\ \bibinfo {pages}
  {3383}}\BibitemShut {NoStop}%
\bibitem [{\citenamefont {Elliott}\ and\ \citenamefont
  {Franz}(2015)}]{Elliott:RMP2015}%
  \BibitemOpen
  \bibfield  {author} {\bibinfo {author} {\bibnamefont {Elliott}, \bibfnamefont
  {Steven~R}}, \ and\ \bibinfo {author} {\bibfnamefont {Marcel}\ \bibnamefont
  {Franz}}} (\bibinfo {year} {2015}),\ \bibfield  {title} {\enquote {\bibinfo
  {title} {{Colloquium: Majorana fermions in nuclear, particle, and solid-state
  physics}},}\ }\href {https://link.aps.org/doi/10.1103/RevModPhys.87.137}
  {\bibfield  {journal} {\bibinfo  {journal} {Rev. Mod. Phys.}\ }\textbf
  {\bibinfo {volume} {87}},\ \bibinfo {pages} {137--163}}\BibitemShut {NoStop}%
\bibitem [{\citenamefont {Engelhardt}\ \emph {et~al.}(2016)\citenamefont
  {Engelhardt}, \citenamefont {Benito}, \citenamefont {Platero},\ and\
  \citenamefont {Brandes}}]{Engelhardt:2016PRL}%
  \BibitemOpen
  \bibfield  {author} {\bibinfo {author} {\bibnamefont {Engelhardt},
  \bibfnamefont {G}}, \bibinfo {author} {\bibfnamefont {M.}~\bibnamefont
  {Benito}}, \bibinfo {author} {\bibfnamefont {G.}~\bibnamefont {Platero}}, \
  and\ \bibinfo {author} {\bibfnamefont {T.}~\bibnamefont {Brandes}}} (\bibinfo
  {year} {2016}),\ \bibfield  {title} {\enquote {\bibinfo {title} {Topological
  instabilities in ac-driven bosonic systems},}\ }\href
  {https://link.aps.org/doi/10.1103/PhysRevLett.117.045302} {\bibfield
  {journal} {\bibinfo  {journal} {Phys. Rev. Lett.}\ }\textbf {\bibinfo
  {volume} {117}},\ \bibinfo {pages} {045302}}\BibitemShut {NoStop}%
\bibitem [{\citenamefont {Engelhardt}\ and\ \citenamefont
  {Brandes}(2015)}]{Engelhardt:2015PRA}%
  \BibitemOpen
  \bibfield  {author} {\bibinfo {author} {\bibnamefont {Engelhardt},
  \bibfnamefont {G}}, \ and\ \bibinfo {author} {\bibfnamefont {T.}~\bibnamefont
  {Brandes}}} (\bibinfo {year} {2015}),\ \bibfield  {title} {\enquote {\bibinfo
  {title} {Topological {B}ogoliubov excitations in inversion-symmetric systems
  of interacting bosons},}\ }\href
  {https://link.aps.org/doi/10.1103/PhysRevA.91.053621} {\bibfield  {journal}
  {\bibinfo  {journal} {Phys. Rev. A}\ }\textbf {\bibinfo {volume} {91}},\
  \bibinfo {pages} {053621}}\BibitemShut {NoStop}%
\bibitem [{\citenamefont {Esaki}\ \emph {et~al.}(2011)\citenamefont {Esaki},
  \citenamefont {Sato}, \citenamefont {Hasebe},\ and\ \citenamefont
  {Kohmoto}}]{esaki2011edge}%
  \BibitemOpen
  \bibfield  {author} {\bibinfo {author} {\bibnamefont {Esaki}, \bibfnamefont
  {Kenta}}, \bibinfo {author} {\bibfnamefont {Masatoshi}\ \bibnamefont {Sato}},
  \bibinfo {author} {\bibfnamefont {Kazuki}\ \bibnamefont {Hasebe}}, \ and\
  \bibinfo {author} {\bibfnamefont {Mahito}\ \bibnamefont {Kohmoto}}} (\bibinfo
  {year} {2011}),\ \bibfield  {title} {\enquote {\bibinfo {title} {Edge states
  and topological phases in non-{H}ermitian systems},}\ }\href
  {https://arxiv.org/abs/1107.2079} {\bibfield  {journal} {\bibinfo  {journal}
  {Phys. Rev. B}\ }\textbf {\bibinfo {volume} {84}}~(\bibinfo {number} {20}),\
  \bibinfo {pages} {205128}}\BibitemShut {NoStop}%
\bibitem [{\citenamefont {Fang}\ \emph {et~al.}(2015)\citenamefont {Fang},
  \citenamefont {Chen}, \citenamefont {Kee},\ and\ \citenamefont
  {Fu}}]{Fang:2015PRB}%
  \BibitemOpen
  \bibfield  {author} {\bibinfo {author} {\bibnamefont {Fang}, \bibfnamefont
  {Chen}}, \bibinfo {author} {\bibfnamefont {Yige}\ \bibnamefont {Chen}},
  \bibinfo {author} {\bibfnamefont {Hae-Young}\ \bibnamefont {Kee}}, \ and\
  \bibinfo {author} {\bibfnamefont {Liang}\ \bibnamefont {Fu}}} (\bibinfo
  {year} {2015}),\ \bibfield  {title} {\enquote {\bibinfo {title} {Topological
  nodal line semimetals with and without spin-orbital coupling},}\ }\href
  {https://journals.aps.org/prb/abstract/10.1103/PhysRevB.92.081201} {\bibfield
   {journal} {\bibinfo  {journal} {Phys. Rev. B}\ }\textbf {\bibinfo {volume}
  {92}}~(\bibinfo {number} {8}),\ \bibinfo {pages} {081201}}\BibitemShut
  {NoStop}%
\bibitem [{\citenamefont {Fang}\ \emph
  {et~al.}(2016{\natexlab{a}})\citenamefont {Fang}, \citenamefont {Lu},
  \citenamefont {Liu},\ and\ \citenamefont {Fu}}]{Fang:2016NatPhys}%
  \BibitemOpen
  \bibfield  {author} {\bibinfo {author} {\bibnamefont {Fang}, \bibfnamefont
  {Chen}}, \bibinfo {author} {\bibfnamefont {Ling}\ \bibnamefont {Lu}},
  \bibinfo {author} {\bibfnamefont {Junwei}\ \bibnamefont {Liu}}, \ and\
  \bibinfo {author} {\bibfnamefont {Liang}\ \bibnamefont {Fu}}} (\bibinfo
  {year} {2016}{\natexlab{a}}),\ \bibfield  {title} {\enquote {\bibinfo {title}
  {Topological semimetals with helicoid surface states},}\ }\href
  {https://www.nature.com/articles/nphys3782} {\bibfield  {journal} {\bibinfo
  {journal} {Nat. Phys.}\ }\textbf {\bibinfo {volume} {12}},\ \bibinfo {pages}
  {936--941}}\BibitemShut {NoStop}%
\bibitem [{\citenamefont {Fang}\ \emph
  {et~al.}(2016{\natexlab{b}})\citenamefont {Fang}, \citenamefont {Weng},
  \citenamefont {Dai},\ and\ \citenamefont {Fang}}]{Fang:2016CPB}%
  \BibitemOpen
  \bibfield  {author} {\bibinfo {author} {\bibnamefont {Fang}, \bibfnamefont
  {Chen}}, \bibinfo {author} {\bibfnamefont {Hongming}\ \bibnamefont {Weng}},
  \bibinfo {author} {\bibfnamefont {Xi}~\bibnamefont {Dai}}, \ and\ \bibinfo
  {author} {\bibfnamefont {Zhong}\ \bibnamefont {Fang}}} (\bibinfo {year}
  {2016}{\natexlab{b}}),\ \bibfield  {title} {\enquote {\bibinfo {title}
  {Topological nodal line semimetals},}\ }\href
  {https://iopscience.iop.org/article/10.1088/1674-1056/25/11/117106/meta}
  {\bibfield  {journal} {\bibinfo  {journal} {Chinese Physics B}\ }\textbf
  {\bibinfo {volume} {25}}~(\bibinfo {number} {11}),\ \bibinfo {pages}
  {117106}}\BibitemShut {NoStop}%
\bibitem [{\citenamefont {Fang}\ and\ \citenamefont
  {Fan}(2013{\natexlab{a}})}]{Fang:2013PRL}%
  \BibitemOpen
  \bibfield  {author} {\bibinfo {author} {\bibnamefont {Fang}, \bibfnamefont
  {Kejie}}, \ and\ \bibinfo {author} {\bibfnamefont {Shanhui}\ \bibnamefont
  {Fan}}} (\bibinfo {year} {2013}{\natexlab{a}}),\ \bibfield  {title} {\enquote
  {\bibinfo {title} {Controlling the flow of light using the inhomogeneous
  effective gauge field that emerges from dynamic modulation},}\ }\href
  {https://link.aps.org/doi/10.1103/PhysRevLett.111.203901} {\bibfield
  {journal} {\bibinfo  {journal} {Phys. Rev. Lett.}\ }\textbf {\bibinfo
  {volume} {111}},\ \bibinfo {pages} {203901}}\BibitemShut {NoStop}%
\bibitem [{\citenamefont {Fang}\ and\ \citenamefont
  {Fan}(2013{\natexlab{b}})}]{Fang:2013PRA}%
  \BibitemOpen
  \bibfield  {author} {\bibinfo {author} {\bibnamefont {Fang}, \bibfnamefont
  {Kejie}}, \ and\ \bibinfo {author} {\bibfnamefont {Shanhui}\ \bibnamefont
  {Fan}}} (\bibinfo {year} {2013}{\natexlab{b}}),\ \bibfield  {title} {\enquote
  {\bibinfo {title} {Effective magnetic field for photons based on the
  magneto-optical effect},}\ }\href
  {https://link.aps.org/doi/10.1103/PhysRevA.88.043847} {\bibfield  {journal}
  {\bibinfo  {journal} {Phys. Rev. A}\ }\textbf {\bibinfo {volume} {88}},\
  \bibinfo {pages} {043847}}\BibitemShut {NoStop}%
\bibitem [{\citenamefont {Fang}\ \emph {et~al.}(2017)\citenamefont {Fang},
  \citenamefont {Luo}, \citenamefont {Metelmann}, \citenamefont {Matheny},
  \citenamefont {Marquardt}, \citenamefont {Clerk},\ and\ \citenamefont
  {Painter}}]{fang2016generalized}%
  \BibitemOpen
  \bibfield  {author} {\bibinfo {author} {\bibnamefont {Fang}, \bibfnamefont
  {Kejie}}, \bibinfo {author} {\bibfnamefont {Jie}\ \bibnamefont {Luo}},
  \bibinfo {author} {\bibfnamefont {Anja}\ \bibnamefont {Metelmann}}, \bibinfo
  {author} {\bibfnamefont {Matthew~H.}\ \bibnamefont {Matheny}}, \bibinfo
  {author} {\bibfnamefont {Florian}\ \bibnamefont {Marquardt}}, \bibinfo
  {author} {\bibfnamefont {Aashish~A.}\ \bibnamefont {Clerk}}, \ and\ \bibinfo
  {author} {\bibfnamefont {Oskar}\ \bibnamefont {Painter}}} (\bibinfo {year}
  {2017}),\ \bibfield  {title} {\enquote {\bibinfo {title} {Generalized
  non-reciprocity in an optomechanical circuit via synthetic magnetism and
  reservoir engineering},}\ }\href {https://doi.org/10.1038/nphys4009}
  {\bibfield  {journal} {\bibinfo  {journal} {Nat Phys}\ }\textbf {\bibinfo
  {volume} {13}}~(\bibinfo {number} {5}),\ \bibinfo {pages}
  {465--471}}\BibitemShut {NoStop}%
\bibitem [{\citenamefont {Fang}\ \emph {et~al.}(2011)\citenamefont {Fang},
  \citenamefont {Yu},\ and\ \citenamefont {Fan}}]{Fang:2011PRB}%
  \BibitemOpen
  \bibfield  {author} {\bibinfo {author} {\bibnamefont {Fang}, \bibfnamefont
  {Kejie}}, \bibinfo {author} {\bibfnamefont {Zongfu}\ \bibnamefont {Yu}}, \
  and\ \bibinfo {author} {\bibfnamefont {Shanhui}\ \bibnamefont {Fan}}}
  (\bibinfo {year} {2011}),\ \bibfield  {title} {\enquote {\bibinfo {title}
  {Microscopic theory of photonic one-way edge mode},}\ }\href
  {https://journals.aps.org/prb/abstract/10.1103/PhysRevB.84.075477} {\bibfield
   {journal} {\bibinfo  {journal} {Phys Rev B}\ }\textbf {\bibinfo {volume}
  {84}},\ \bibinfo {pages} {075477}}\BibitemShut {NoStop}%
\bibitem [{\citenamefont {Fang}\ \emph
  {et~al.}(2012{\natexlab{a}})\citenamefont {Fang}, \citenamefont {Yu},\ and\
  \citenamefont {Fan}}]{Fang:2012PRL}%
  \BibitemOpen
  \bibfield  {author} {\bibinfo {author} {\bibnamefont {Fang}, \bibfnamefont
  {Kejie}}, \bibinfo {author} {\bibfnamefont {Zongfu}\ \bibnamefont {Yu}}, \
  and\ \bibinfo {author} {\bibfnamefont {Shanhui}\ \bibnamefont {Fan}}}
  (\bibinfo {year} {2012}{\natexlab{a}}),\ \bibfield  {title} {\enquote
  {\bibinfo {title} {Photonic {A}haronov-{B}ohm effect based on dynamic
  modulation},}\ }\href
  {https://link.aps.org/doi/10.1103/PhysRevLett.108.153901} {\bibfield
  {journal} {\bibinfo  {journal} {Phys. Rev. Lett.}\ }\textbf {\bibinfo
  {volume} {108}},\ \bibinfo {pages} {153901}}\BibitemShut {NoStop}%
\bibitem [{\citenamefont {Fang}\ \emph
  {et~al.}(2012{\natexlab{b}})\citenamefont {Fang}, \citenamefont {Yu},\ and\
  \citenamefont {Fan}}]{Fang:2012NatPhot}%
  \BibitemOpen
  \bibfield  {author} {\bibinfo {author} {\bibnamefont {Fang}, \bibfnamefont
  {Kejie}}, \bibinfo {author} {\bibfnamefont {Zongfu}\ \bibnamefont {Yu}}, \
  and\ \bibinfo {author} {\bibfnamefont {Shanhui}\ \bibnamefont {Fan}}}
  (\bibinfo {year} {2012}{\natexlab{b}}),\ \bibfield  {title} {\enquote
  {\bibinfo {title} {Realizing effective magnetic field for photons by
  controlling the phase of dynamic modulation},}\ }\href
  {http://www.nature.com/nphoton/journal/v6/n11/full/nphoton.2012.236.html}
  {\bibfield  {journal} {\bibinfo  {journal} {Nat. Photonics}\ }\textbf
  {\bibinfo {volume} {6}}~(\bibinfo {number} {11}),\ \bibinfo {pages}
  {782--787}}\BibitemShut {NoStop}%
\bibitem [{\citenamefont {Faraon}\ \emph {et~al.}(2008)\citenamefont {Faraon},
  \citenamefont {Fushman}, \citenamefont {Englund}, \citenamefont {Stoltz},
  \citenamefont {Petroff},\ and\ \citenamefont
  {Vuckovic}}]{Faraon:NatPhys2008}%
  \BibitemOpen
  \bibfield  {author} {\bibinfo {author} {\bibnamefont {Faraon}, \bibfnamefont
  {Andrei}}, \bibinfo {author} {\bibfnamefont {Ilya}\ \bibnamefont {Fushman}},
  \bibinfo {author} {\bibfnamefont {Dirk}\ \bibnamefont {Englund}}, \bibinfo
  {author} {\bibfnamefont {Nick}\ \bibnamefont {Stoltz}}, \bibinfo {author}
  {\bibfnamefont {Pierre}\ \bibnamefont {Petroff}}, \ and\ \bibinfo {author}
  {\bibfnamefont {Jelena}\ \bibnamefont {Vuckovic}}} (\bibinfo {year} {2008}),\
  \bibfield  {title} {\enquote {\bibinfo {title} {Coherent generation of
  non-classical light on a chip via photon-induced tunnelling and blockade},}\
  }\href {https://www.nature.com/articles/nphys1078} {\bibfield  {journal}
  {\bibinfo  {journal} {Nat. Phys.}\ }\textbf {\bibinfo {volume} {4}}~(\bibinfo
  {number} {11}),\ \bibinfo {pages} {859--863}}\BibitemShut {NoStop}%
\bibitem [{\citenamefont {Fausti}\ \emph {et~al.}(2011)\citenamefont {Fausti},
  \citenamefont {Tobey}, \citenamefont {Dean}, \citenamefont {Kaiser},
  \citenamefont {Dienst}, \citenamefont {Hoffmann}, \citenamefont {Pyon},
  \citenamefont {Takayama}, \citenamefont {Takagi},\ and\ \citenamefont
  {Cavalleri}}]{Fausti:Science2011}%
  \BibitemOpen
  \bibfield  {author} {\bibinfo {author} {\bibnamefont {Fausti}, \bibfnamefont
  {Daniele}}, \bibinfo {author} {\bibfnamefont {RI}~\bibnamefont {Tobey}},
  \bibinfo {author} {\bibfnamefont {Nicky}\ \bibnamefont {Dean}}, \bibinfo
  {author} {\bibfnamefont {Stefan}\ \bibnamefont {Kaiser}}, \bibinfo {author}
  {\bibfnamefont {A}~\bibnamefont {Dienst}}, \bibinfo {author} {\bibfnamefont
  {Matthias~C}\ \bibnamefont {Hoffmann}}, \bibinfo {author} {\bibfnamefont
  {S}~\bibnamefont {Pyon}}, \bibinfo {author} {\bibfnamefont {T}~\bibnamefont
  {Takayama}}, \bibinfo {author} {\bibfnamefont {H}~\bibnamefont {Takagi}}, \
  and\ \bibinfo {author} {\bibfnamefont {Andrea}\ \bibnamefont {Cavalleri}}}
  (\bibinfo {year} {2011}),\ \bibfield  {title} {\enquote {\bibinfo {title}
  {Light-induced superconductivity in a stripe-ordered cuprate},}\ }\href
  {http://science.sciencemag.org/content/331/6014/189} {\bibfield  {journal}
  {\bibinfo  {journal} {Science}\ }\textbf {\bibinfo {volume} {331}}~(\bibinfo
  {number} {6014}),\ \bibinfo {pages} {189--191}}\BibitemShut {NoStop}%
\bibitem [{\citenamefont {Fefferman}\ and\ \citenamefont
  {Weinstein}(2012)}]{fefferman2012honeycomb}%
  \BibitemOpen
  \bibfield  {author} {\bibinfo {author} {\bibnamefont {Fefferman},
  \bibfnamefont {Charles}}, \ and\ \bibinfo {author} {\bibfnamefont {Michael}\
  \bibnamefont {Weinstein}}} (\bibinfo {year} {2012}),\ \bibfield  {title}
  {\enquote {\bibinfo {title} {Honeycomb lattice potentials and {D}irac
  points},}\ }\href
  {https://www.ams.org/journals/jams/2012-25-04/S0894-0347-2012-00745-0/}
  {\bibfield  {journal} {\bibinfo  {journal} {Journal of the American
  Mathematical Society}\ }\textbf {\bibinfo {volume} {25}}~(\bibinfo {number}
  {4}),\ \bibinfo {pages} {1169--1220}}\BibitemShut {NoStop}%
\bibitem [{\citenamefont {Fefferman}\ \emph {et~al.}(2014)\citenamefont
  {Fefferman}, \citenamefont {Lee-Thorp},\ and\ \citenamefont
  {Weinstein}}]{fefferman2014topologically}%
  \BibitemOpen
  \bibfield  {author} {\bibinfo {author} {\bibnamefont {Fefferman},
  \bibfnamefont {Charles~L}}, \bibinfo {author} {\bibfnamefont {James~P}\
  \bibnamefont {Lee-Thorp}}, \ and\ \bibinfo {author} {\bibfnamefont
  {Michael~I}\ \bibnamefont {Weinstein}}} (\bibinfo {year} {2014}),\ \bibfield
  {title} {\enquote {\bibinfo {title} {Topologically protected states in
  one-dimensional continuous systems and {D}irac points},}\ }\href
  {https://www.pnas.org/content/111/24/8759} {\bibfield  {journal} {\bibinfo
  {journal} {Proc. Natl. Acad. Scie. U.S.A.}\ }\textbf {\bibinfo {volume}
  {111}}~(\bibinfo {number} {24}),\ \bibinfo {pages} {8759--8763}}\BibitemShut
  {NoStop}%
\bibitem [{\citenamefont {Fefferman}\ and\ \citenamefont
  {Weinstein}(2014)}]{fefferman2014wave}%
  \BibitemOpen
  \bibfield  {author} {\bibinfo {author} {\bibnamefont {Fefferman},
  \bibfnamefont {Charles~L}}, \ and\ \bibinfo {author} {\bibfnamefont
  {Michael~I}\ \bibnamefont {Weinstein}}} (\bibinfo {year} {2014}),\ \bibfield
  {title} {\enquote {\bibinfo {title} {Wave packets in honeycomb structures and
  two-dimensional {D}irac equations},}\ }\href
  {https://link.springer.com/article/10.1007/s00220-013-1847-2} {\bibfield
  {journal} {\bibinfo  {journal} {Commun. Math. Phys.}\ }\textbf {\bibinfo
  {volume} {326}}~(\bibinfo {number} {1}),\ \bibinfo {pages}
  {251--286}}\BibitemShut {NoStop}%
\bibitem [{\citenamefont {Fefferman}\ \emph {et~al.}(2016)\citenamefont
  {Fefferman}, \citenamefont {Lee-Thorp},\ and\ \citenamefont
  {Weinstein}}]{Weinstein:2DMat2016}%
  \BibitemOpen
  \bibfield  {author} {\bibinfo {author} {\bibnamefont {Fefferman},
  \bibfnamefont {CL}}, \bibinfo {author} {\bibfnamefont {JP}~\bibnamefont
  {Lee-Thorp}}, \ and\ \bibinfo {author} {\bibfnamefont {MI}~\bibnamefont
  {Weinstein}}} (\bibinfo {year} {2016}),\ \bibfield  {title} {\enquote
  {\bibinfo {title} {Bifurcations of edge states—topologically protected and
  non-protected—in continuous 2{D} honeycomb structures},}\ }\href
  {http://iopscience.iop.org/article/10.1088/2053-1583/3/1/014008/meta}
  {\bibfield  {journal} {\bibinfo  {journal} {2D Materials}\ }\textbf {\bibinfo
  {volume} {3}}~(\bibinfo {number} {1}),\ \bibinfo {pages}
  {014008}}\BibitemShut {NoStop}%
\bibitem [{\citenamefont {Umucal\ifmmode \imath \else~\i \fi{}lar}\ and\
  \citenamefont {Carusotto}(2017)}]{Umucalilar:arXiv2017}%
  \BibitemOpen
  \bibfield  {author} {\bibinfo {author} {\bibnamefont {Umucal\ifmmode \imath
  \else~\i \fi{}lar}, \bibfnamefont {R~O}}, \ and\ \bibinfo {author}
  {\bibfnamefont {I.}~\bibnamefont {Carusotto}}} (\bibinfo {year} {2017}),\
  \bibfield  {title} {\enquote {\bibinfo {title} {Generation and spectroscopic
  signatures of a fractional quantum {H}all liquid of photons in an
  incoherently pumped optical cavity},}\ }\href
  {https://link.aps.org/doi/10.1103/PhysRevA.96.053808} {\bibfield  {journal}
  {\bibinfo  {journal} {Phys. Rev. A}\ }\textbf {\bibinfo {volume} {96}},\
  \bibinfo {pages} {053808}}\BibitemShut {NoStop}%
\bibitem [{\citenamefont {Firstenberg}\ \emph {et~al.}(2013)\citenamefont
  {Firstenberg}, \citenamefont {Peyronel}, \citenamefont {Liang}, \citenamefont
  {Gorshkov}, \citenamefont {Lukin},\ and\ \citenamefont
  {Vuleti{\'c}}}]{Firstenberg:Nature2013}%
  \BibitemOpen
  \bibfield  {author} {\bibinfo {author} {\bibnamefont {Firstenberg},
  \bibfnamefont {Ofer}}, \bibinfo {author} {\bibfnamefont {Thibault}\
  \bibnamefont {Peyronel}}, \bibinfo {author} {\bibfnamefont {Qi-Yu}\
  \bibnamefont {Liang}}, \bibinfo {author} {\bibfnamefont {Alexey~V}\
  \bibnamefont {Gorshkov}}, \bibinfo {author} {\bibfnamefont {Mikhail~D}\
  \bibnamefont {Lukin}}, \ and\ \bibinfo {author} {\bibfnamefont {Vladan}\
  \bibnamefont {Vuleti{\'c}}}} (\bibinfo {year} {2013}),\ \bibfield  {title}
  {\enquote {\bibinfo {title} {Attractive photons in a quantum nonlinear
  medium},}\ }\href {https://www.nature.com/articles/nature12512} {\bibfield
  {journal} {\bibinfo  {journal} {Nature}\ }\textbf {\bibinfo {volume}
  {502}}~(\bibinfo {number} {7469}),\ \bibinfo {pages} {71--75}}\BibitemShut
  {NoStop}%
\bibitem [{\citenamefont {Fitzpatrick}\ \emph {et~al.}(2017)\citenamefont
  {Fitzpatrick}, \citenamefont {Sundaresan}, \citenamefont {Li}, \citenamefont
  {Koch},\ and\ \citenamefont {Houck}}]{Fitzpatrick:PRX2017}%
  \BibitemOpen
  \bibfield  {author} {\bibinfo {author} {\bibnamefont {Fitzpatrick},
  \bibfnamefont {Mattias}}, \bibinfo {author} {\bibfnamefont {Neereja~M.}\
  \bibnamefont {Sundaresan}}, \bibinfo {author} {\bibfnamefont {Andy C.~Y.}\
  \bibnamefont {Li}}, \bibinfo {author} {\bibfnamefont {Jens}\ \bibnamefont
  {Koch}}, \ and\ \bibinfo {author} {\bibfnamefont {Andrew~A.}\ \bibnamefont
  {Houck}}} (\bibinfo {year} {2017}),\ \bibfield  {title} {\enquote {\bibinfo
  {title} {Observation of a dissipative phase transition in a one-dimensional
  circuit {QED} lattice},}\ }\href
  {https://link.aps.org/doi/10.1103/PhysRevX.7.011016} {\bibfield  {journal}
  {\bibinfo  {journal} {Phys. Rev. X}\ }\textbf {\bibinfo {volume} {7}},\
  \bibinfo {pages} {011016}}\BibitemShut {NoStop}%
\bibitem [{\citenamefont {Fl{\"a}schner}\ \emph {et~al.}(2018)\citenamefont
  {Fl{\"a}schner}, \citenamefont {Vogel}, \citenamefont {Tarnowski},
  \citenamefont {Rem}, \citenamefont {L{\"u}hmann}, \citenamefont {Heyl},
  \citenamefont {Budich}, \citenamefont {Mathey}, \citenamefont {Sengstock},\
  and\ \citenamefont {Weitenberg}}]{Flaschner:2016arxiv}%
  \BibitemOpen
  \bibfield  {author} {\bibinfo {author} {\bibnamefont {Fl{\"a}schner},
  \bibfnamefont {N}}, \bibinfo {author} {\bibfnamefont {D}~\bibnamefont
  {Vogel}}, \bibinfo {author} {\bibfnamefont {M}~\bibnamefont {Tarnowski}},
  \bibinfo {author} {\bibfnamefont {BS}~\bibnamefont {Rem}}, \bibinfo {author}
  {\bibfnamefont {D-S}\ \bibnamefont {L{\"u}hmann}}, \bibinfo {author}
  {\bibfnamefont {M}~\bibnamefont {Heyl}}, \bibinfo {author} {\bibfnamefont
  {JC}~\bibnamefont {Budich}}, \bibinfo {author} {\bibfnamefont
  {L}~\bibnamefont {Mathey}}, \bibinfo {author} {\bibfnamefont {K}~\bibnamefont
  {Sengstock}}, \ and\ \bibinfo {author} {\bibfnamefont {C}~\bibnamefont
  {Weitenberg}}} (\bibinfo {year} {2018}),\ \bibfield  {title} {\enquote
  {\bibinfo {title} {Observation of dynamical vortices after quenches in a
  system with topology},}\ }\href
  {https://www.nature.com/articles/s41567-017-0013-8} {\bibfield  {journal}
  {\bibinfo  {journal} {Nat. Phys.}\ }\textbf {\bibinfo {volume}
  {14}}~(\bibinfo {number} {3}),\ \bibinfo {pages} {265}}\BibitemShut {NoStop}%
\bibitem [{\citenamefont {Fleischer}\ \emph {et~al.}(2003)\citenamefont
  {Fleischer}, \citenamefont {Segev}, \citenamefont {Efremidis},\ and\
  \citenamefont {Christodoulides}}]{fleischer2003observation}%
  \BibitemOpen
  \bibfield  {author} {\bibinfo {author} {\bibnamefont {Fleischer},
  \bibfnamefont {Jason~W}}, \bibinfo {author} {\bibfnamefont {Mordechai}\
  \bibnamefont {Segev}}, \bibinfo {author} {\bibfnamefont {Nikolaos~K}\
  \bibnamefont {Efremidis}}, \ and\ \bibinfo {author} {\bibfnamefont
  {Demetrios~N}\ \bibnamefont {Christodoulides}}} (\bibinfo {year} {2003}),\
  \bibfield  {title} {\enquote {\bibinfo {title} {Observation of
  two-dimensional discrete solitons in optically induced nonlinear photonic
  lattices},}\ }\href {https://www.nature.com/articles/nature01452} {\bibfield
  {journal} {\bibinfo  {journal} {Nature}\ }\textbf {\bibinfo {volume}
  {422}}~(\bibinfo {number} {6928}),\ \bibinfo {pages} {147--150}}\BibitemShut
  {NoStop}%
\bibitem [{\citenamefont {Fleury}\ \emph {et~al.}(2015)\citenamefont {Fleury},
  \citenamefont {Sounas}, \citenamefont {Haberman},\ and\ \citenamefont
  {Al{\`u}}}]{Fleury:Acoustics2015}%
  \BibitemOpen
  \bibfield  {author} {\bibinfo {author} {\bibnamefont {Fleury}, \bibfnamefont
  {Romain}}, \bibinfo {author} {\bibfnamefont {Dimitrios}\ \bibnamefont
  {Sounas}}, \bibinfo {author} {\bibfnamefont {Michael~R}\ \bibnamefont
  {Haberman}}, \ and\ \bibinfo {author} {\bibfnamefont {Andrea}\ \bibnamefont
  {Al{\`u}}}} (\bibinfo {year} {2015}),\ \bibfield  {title} {\enquote {\bibinfo
  {title} {Nonreciprocal acoustics},}\ }\href
  {https://acousticstoday.org/nonreciprocal-acoustics-romain-fleury-dimitrios-sounas-michael-r-haberman-and-andrea-alu/}
  {\bibfield  {journal} {\bibinfo  {journal} {Acoustics Today}\ }\textbf
  {\bibinfo {volume} {11}},\ \bibinfo {pages} {14--21}}\BibitemShut {NoStop}%
\bibitem [{\citenamefont {Fleury}\ \emph {et~al.}(2014)\citenamefont {Fleury},
  \citenamefont {Sounas}, \citenamefont {Sieck}, \citenamefont {Haberman},\
  and\ \citenamefont {Al{\`u}}}]{Alu2014sound}%
  \BibitemOpen
  \bibfield  {author} {\bibinfo {author} {\bibnamefont {Fleury}, \bibfnamefont
  {Romain}}, \bibinfo {author} {\bibfnamefont {Dimitrios~L}\ \bibnamefont
  {Sounas}}, \bibinfo {author} {\bibfnamefont {Caleb~F}\ \bibnamefont {Sieck}},
  \bibinfo {author} {\bibfnamefont {Michael~R}\ \bibnamefont {Haberman}}, \
  and\ \bibinfo {author} {\bibfnamefont {Andrea}\ \bibnamefont {Al{\`u}}}}
  (\bibinfo {year} {2014}),\ \bibfield  {title} {\enquote {\bibinfo {title}
  {Sound isolation and giant linear nonreciprocity in a compact acoustic
  circulator},}\ }\href {https://doi.org/10.1126/science.1246957} {\bibfield
  {journal} {\bibinfo  {journal} {Science}\ }\textbf {\bibinfo {volume}
  {343}}~(\bibinfo {number} {6170}),\ \bibinfo {pages} {516--519}}\BibitemShut
  {NoStop}%
\bibitem [{\citenamefont {Franke-Arnold}\ \emph {et~al.}(2011)\citenamefont
  {Franke-Arnold}, \citenamefont {Gibson}, \citenamefont {Boyd},\ and\
  \citenamefont {Padgett}}]{franke2011rotary}%
  \BibitemOpen
  \bibfield  {author} {\bibinfo {author} {\bibnamefont {Franke-Arnold},
  \bibfnamefont {Sonja}}, \bibinfo {author} {\bibfnamefont {Graham}\
  \bibnamefont {Gibson}}, \bibinfo {author} {\bibfnamefont {Robert~W}\
  \bibnamefont {Boyd}}, \ and\ \bibinfo {author} {\bibfnamefont {Miles~J}\
  \bibnamefont {Padgett}}} (\bibinfo {year} {2011}),\ \bibfield  {title}
  {\enquote {\bibinfo {title} {Rotary photon drag enhanced by a slow-light
  medium},}\ }\href {http://science.sciencemag.org/content/333/6038/65}
  {\bibfield  {journal} {\bibinfo  {journal} {Science}\ }\textbf {\bibinfo
  {volume} {333}}~(\bibinfo {number} {6038}),\ \bibinfo {pages}
  {65--67}}\BibitemShut {NoStop}%
\bibitem [{\citenamefont {Freedman}\ \emph {et~al.}(2006)\citenamefont
  {Freedman}, \citenamefont {Bartal}, \citenamefont {Segev}, \citenamefont
  {Lifshitz}, \citenamefont {Christodoulides},\ and\ \citenamefont
  {Fleischer}}]{freedman2006wave}%
  \BibitemOpen
  \bibfield  {author} {\bibinfo {author} {\bibnamefont {Freedman},
  \bibfnamefont {Barak}}, \bibinfo {author} {\bibfnamefont {Guy}\ \bibnamefont
  {Bartal}}, \bibinfo {author} {\bibfnamefont {Mordechai}\ \bibnamefont
  {Segev}}, \bibinfo {author} {\bibfnamefont {Ron}\ \bibnamefont {Lifshitz}},
  \bibinfo {author} {\bibfnamefont {Demetrios~N}\ \bibnamefont
  {Christodoulides}}, \ and\ \bibinfo {author} {\bibfnamefont {Jason~W}\
  \bibnamefont {Fleischer}}} (\bibinfo {year} {2006}),\ \bibfield  {title}
  {\enquote {\bibinfo {title} {Wave and defect dynamics in nonlinear photonic
  quasicrystals},}\ }\href {https://www.nature.com/articles/nature04722}
  {\bibfield  {journal} {\bibinfo  {journal} {Nature}\ }\textbf {\bibinfo
  {volume} {440}}~(\bibinfo {number} {7088}),\ \bibinfo {pages}
  {1166--1169}}\BibitemShut {NoStop}%
\bibitem [{\citenamefont {Freedman}\ \emph {et~al.}(2007)\citenamefont
  {Freedman}, \citenamefont {Lifshitz}, \citenamefont {Fleischer},\ and\
  \citenamefont {Segev}}]{freedman2007phason}%
  \BibitemOpen
  \bibfield  {author} {\bibinfo {author} {\bibnamefont {Freedman},
  \bibfnamefont {Barak}}, \bibinfo {author} {\bibfnamefont {Ron}\ \bibnamefont
  {Lifshitz}}, \bibinfo {author} {\bibfnamefont {Jason~W}\ \bibnamefont
  {Fleischer}}, \ and\ \bibinfo {author} {\bibfnamefont {Mordechai}\
  \bibnamefont {Segev}}} (\bibinfo {year} {2007}),\ \bibfield  {title}
  {\enquote {\bibinfo {title} {Phason dynamics in nonlinear photonic
  quasicrystals},}\ }\href@noop {} {\bibfield  {journal} {\bibinfo  {journal}
  {Nat. Mater.}\ }\textbf {\bibinfo {volume} {6}}~(\bibinfo {number} {10}),\
  \bibinfo {pages} {776--781}}\BibitemShut {NoStop}%
\bibitem [{\citenamefont {Fr\"ohlich}\ and\ \citenamefont
  {Pedrini}(2000)}]{Frohlich:2000}%
  \BibitemOpen
  \bibfield  {author} {\bibinfo {author} {\bibnamefont {Fr\"ohlich},
  \bibfnamefont {J}}, \ and\ \bibinfo {author} {\bibfnamefont {B.}~\bibnamefont
  {Pedrini}}} (\bibinfo {year} {2000}),\ \bibfield  {title} {\enquote {\bibinfo
  {title} {New applications of the chiral anomaly},}\ }in\ \href
  {http://www.worldscientific.com/worldscibooks/10.1142/p195} {\emph {\bibinfo
  {booktitle} {Mathematical Physics 2000}}},\ \bibinfo {editor} {edited by\
  \bibinfo {editor} {\bibfnamefont {A.}~\bibnamefont {Fokas}}, \bibinfo
  {editor} {\bibfnamefont {A.}~\bibnamefont {Grigoryan}}, \bibinfo {editor}
  {\bibfnamefont {T.}~\bibnamefont {Kibble}}, \ and\ \bibinfo {editor}
  {\bibfnamefont {B.}~\bibnamefont {Zegarlinski}}}\ (\bibinfo  {publisher}
  {Imperial College Press, London, United Kingdom})\ pp.\ \bibinfo {pages}
  {9--47}\BibitemShut {NoStop}%
\bibitem [{\citenamefont {Fu}\ \emph {et~al.}(2011{\natexlab{a}})\citenamefont
  {Fu}, \citenamefont {Lian}, \citenamefont {Liu}, \citenamefont {Gan},\ and\
  \citenamefont {Li}}]{Fu:2011APL}%
  \BibitemOpen
  \bibfield  {author} {\bibinfo {author} {\bibnamefont {Fu}, \bibfnamefont
  {Jin-Xin}}, \bibinfo {author} {\bibfnamefont {Jin}\ \bibnamefont {Lian}},
  \bibinfo {author} {\bibfnamefont {Rong-Juan}\ \bibnamefont {Liu}}, \bibinfo
  {author} {\bibfnamefont {Lin}\ \bibnamefont {Gan}}, \ and\ \bibinfo {author}
  {\bibfnamefont {Zhi-Yuan}\ \bibnamefont {Li}}} (\bibinfo {year}
  {2011}{\natexlab{a}}),\ \bibfield  {title} {\enquote {\bibinfo {title}
  {Unidirectional channel-drop filter by one-way gyromagnetic photonic crystal
  waveguides},}\ }\href {https://aip.scitation.org/doi/full/10.1063/1.3593027}
  {\bibfield  {journal} {\bibinfo  {journal} {Appl. Phys. Lett.}\ }\textbf
  {\bibinfo {volume} {98}}~(\bibinfo {number} {21}),\ \bibinfo {pages}
  {211104--211104}}\BibitemShut {NoStop}%
\bibitem [{\citenamefont {Fu}\ \emph {et~al.}(2011{\natexlab{b}})\citenamefont
  {Fu}, \citenamefont {Liu}, \citenamefont {Gan},\ and\ \citenamefont
  {Li}}]{Fu:2011EPL}%
  \BibitemOpen
  \bibfield  {author} {\bibinfo {author} {\bibnamefont {Fu}, \bibfnamefont
  {Jin-Xin}}, \bibinfo {author} {\bibfnamefont {Rong-Juan}\ \bibnamefont
  {Liu}}, \bibinfo {author} {\bibfnamefont {Lin}\ \bibnamefont {Gan}}, \ and\
  \bibinfo {author} {\bibfnamefont {Zhi-Yuan}\ \bibnamefont {Li}}} (\bibinfo
  {year} {2011}{\natexlab{b}}),\ \bibfield  {title} {\enquote {\bibinfo {title}
  {Control and blockage of edge modes in magneto-optical photonic crystals},}\
  }\href {https://iopscience.iop.org/article/10.1209/0295-5075/93/24001}
  {\bibfield  {journal} {\bibinfo  {journal} {EPL (Europhysics Letters)}\
  }\textbf {\bibinfo {volume} {93}}~(\bibinfo {number} {2}),\ \bibinfo {pages}
  {24001}}\BibitemShut {NoStop}%
\bibitem [{\citenamefont {Fu}\ \emph {et~al.}(2010{\natexlab{a}})\citenamefont
  {Fu}, \citenamefont {Liu},\ and\ \citenamefont {Li}}]{Fu2010:2010EPL}%
  \BibitemOpen
  \bibfield  {author} {\bibinfo {author} {\bibnamefont {Fu}, \bibfnamefont
  {Jin-Xin}}, \bibinfo {author} {\bibfnamefont {Rong-Juan}\ \bibnamefont
  {Liu}}, \ and\ \bibinfo {author} {\bibfnamefont {Zhi-Yuan}\ \bibnamefont
  {Li}}} (\bibinfo {year} {2010}{\natexlab{a}}),\ \bibfield  {title} {\enquote
  {\bibinfo {title} {Experimental demonstration of tunable gyromagnetic
  photonic crystals controlled by dc magnetic fields},}\ }\href
  {https://iopscience.iop.org/article/10.1209/0295-5075/89/64003/meta}
  {\bibfield  {journal} {\bibinfo  {journal} {EPL (Europhysics Letters)}\
  }\textbf {\bibinfo {volume} {89}}~(\bibinfo {number} {6}),\ \bibinfo {pages}
  {64003}}\BibitemShut {NoStop}%
\bibitem [{\citenamefont {Fu}\ \emph {et~al.}(2010{\natexlab{b}})\citenamefont
  {Fu}, \citenamefont {Liu},\ and\ \citenamefont {Li}}]{Fu:2010APL}%
  \BibitemOpen
  \bibfield  {author} {\bibinfo {author} {\bibnamefont {Fu}, \bibfnamefont
  {Jin-Xin}}, \bibinfo {author} {\bibfnamefont {Rong-Juan}\ \bibnamefont
  {Liu}}, \ and\ \bibinfo {author} {\bibfnamefont {Zhi-Yuan}\ \bibnamefont
  {Li}}} (\bibinfo {year} {2010}{\natexlab{b}}),\ \bibfield  {title} {\enquote
  {\bibinfo {title} {Robust one-way modes in gyromagnetic photonic crystal
  waveguides with different interfaces},}\ }\href
  {https://aip.scitation.org/doi/full/10.1063/1.3470873} {\bibfield  {journal}
  {\bibinfo  {journal} {Appl. Phys. Lett.}\ }\textbf {\bibinfo {volume}
  {97}}~(\bibinfo {number} {4}),\ \bibinfo {pages}
  {041112--041112}}\BibitemShut {NoStop}%
\bibitem [{\citenamefont {Fu}(2011)}]{Fu:2011PRL}%
  \BibitemOpen
  \bibfield  {author} {\bibinfo {author} {\bibnamefont {Fu}, \bibfnamefont
  {Liang}}} (\bibinfo {year} {2011}),\ \bibfield  {title} {\enquote {\bibinfo
  {title} {Topological crystalline insulators},}\ }\href
  {https://link.aps.org/doi/10.1103/PhysRevLett.106.106802} {\bibfield
  {journal} {\bibinfo  {journal} {Phys. Rev. Lett.}\ }\textbf {\bibinfo
  {volume} {106}},\ \bibinfo {pages} {106802}}\BibitemShut {NoStop}%
\bibitem [{\citenamefont {Fu}\ and\ \citenamefont {Kane}(2006)}]{Fu:2006PRB}%
  \BibitemOpen
  \bibfield  {author} {\bibinfo {author} {\bibnamefont {Fu}, \bibfnamefont
  {Liang}}, \ and\ \bibinfo {author} {\bibfnamefont {C.~L.}\ \bibnamefont
  {Kane}}} (\bibinfo {year} {2006}),\ \bibfield  {title} {\enquote {\bibinfo
  {title} {Time reversal polarization and a ${Z}_{2}$ adiabatic spin pump},}\
  }\href {https://link.aps.org/doi/10.1103/PhysRevB.74.195312} {\bibfield
  {journal} {\bibinfo  {journal} {Phys. Rev. B}\ }\textbf {\bibinfo {volume}
  {74}},\ \bibinfo {pages} {195312}}\BibitemShut {NoStop}%
\bibitem [{\citenamefont {Fu}\ \emph {et~al.}(2007)\citenamefont {Fu},
  \citenamefont {Kane},\ and\ \citenamefont {Mele}}]{Fu:2007PRL}%
  \BibitemOpen
  \bibfield  {author} {\bibinfo {author} {\bibnamefont {Fu}, \bibfnamefont
  {Liang}}, \bibinfo {author} {\bibfnamefont {C.~L.}\ \bibnamefont {Kane}}, \
  and\ \bibinfo {author} {\bibfnamefont {E.~J.}\ \bibnamefont {Mele}}}
  (\bibinfo {year} {2007}),\ \bibfield  {title} {\enquote {\bibinfo {title}
  {Topological insulators in three dimensions},}\ }\href
  {https://link.aps.org/doi/10.1103/PhysRevLett.98.106803} {\bibfield
  {journal} {\bibinfo  {journal} {Phys. Rev. Lett.}\ }\textbf {\bibinfo
  {volume} {98}},\ \bibinfo {pages} {106803}}\BibitemShut {NoStop}%
\bibitem [{\citenamefont {Furukawa}\ and\ \citenamefont
  {Ueda}(2015)}]{Furukawa:2015NJP}%
  \BibitemOpen
  \bibfield  {author} {\bibinfo {author} {\bibnamefont {Furukawa},
  \bibfnamefont {Shunsuke}}, \ and\ \bibinfo {author} {\bibfnamefont
  {Masahito}\ \bibnamefont {Ueda}}} (\bibinfo {year} {2015}),\ \bibfield
  {title} {\enquote {\bibinfo {title} {{Excitation band topology and edge
  matter waves in Bose--Einstein condensates in optical lattices}},}\ }\href
  {http://iopscience.iop.org/article/10.1088/1367-2630/17/11/115014} {\bibfield
   {journal} {\bibinfo  {journal} {New J. Phys.}\ }\textbf {\bibinfo {volume}
  {17}}~(\bibinfo {number} {11}),\ \bibinfo {pages} {115014}}\BibitemShut
  {NoStop}%
\bibitem [{\citenamefont {Galilo}\ \emph {et~al.}(2015)\citenamefont {Galilo},
  \citenamefont {Lee},\ and\ \citenamefont {Barnett}}]{Galilo:2015PRL}%
  \BibitemOpen
  \bibfield  {author} {\bibinfo {author} {\bibnamefont {Galilo}, \bibfnamefont
  {Bogdan}}, \bibinfo {author} {\bibfnamefont {Derek K.~K.}\ \bibnamefont
  {Lee}}, \ and\ \bibinfo {author} {\bibfnamefont {Ryan}\ \bibnamefont
  {Barnett}}} (\bibinfo {year} {2015}),\ \bibfield  {title} {\enquote {\bibinfo
  {title} {Selective population of edge states in a {2D} topological band
  system},}\ }\href {https://link.aps.org/doi/10.1103/PhysRevLett.115.245302}
  {\bibfield  {journal} {\bibinfo  {journal} {Phys. Rev. Lett.}\ }\textbf
  {\bibinfo {volume} {115}},\ \bibinfo {pages} {245302}}\BibitemShut {NoStop}%
\bibitem [{\citenamefont {Gangaraj}\ and\ \citenamefont
  {Hanson}(2017)}]{Gangaraj:2017IEEE}%
  \BibitemOpen
  \bibfield  {author} {\bibinfo {author} {\bibnamefont {Gangaraj},
  \bibfnamefont {Seyyed Ali~Hassani}}, \ and\ \bibinfo {author} {\bibfnamefont
  {George~W}\ \bibnamefont {Hanson}}} (\bibinfo {year} {2017}),\ \bibfield
  {title} {\enquote {\bibinfo {title} {Topologically protected unidirectional
  surface states in biased ferrites: {D}uality and application to directional
  couplers},}\ }\href {https://ieeexplore.ieee.org/document/7496945} {\bibfield
   {journal} {\bibinfo  {journal} {IEEE Antennas and Wireless Propagation
  Letters}\ }\textbf {\bibinfo {volume} {16}},\ \bibinfo {pages}
  {449--452}}\BibitemShut {NoStop}%
\bibitem [{\citenamefont {Gao}\ \emph {et~al.}(2016{\natexlab{a}})\citenamefont
  {Gao}, \citenamefont {Gao}, \citenamefont {Shi}, \citenamefont {Yang},
  \citenamefont {Lin}, \citenamefont {Xu}, \citenamefont {Joannopoulos},
  \citenamefont {Solja{\v{c}}i{\'c}}, \citenamefont {Chen}, \citenamefont {Lu}
  \emph {et~al.}}]{Gao:2016NatCom}%
  \BibitemOpen
  \bibfield  {author} {\bibinfo {author} {\bibnamefont {Gao}, \bibfnamefont
  {Fei}}, \bibinfo {author} {\bibfnamefont {Zhen}\ \bibnamefont {Gao}},
  \bibinfo {author} {\bibfnamefont {Xihang}\ \bibnamefont {Shi}}, \bibinfo
  {author} {\bibfnamefont {Zhaoju}\ \bibnamefont {Yang}}, \bibinfo {author}
  {\bibfnamefont {Xiao}\ \bibnamefont {Lin}}, \bibinfo {author} {\bibfnamefont
  {Hongyi}\ \bibnamefont {Xu}}, \bibinfo {author} {\bibfnamefont {John~D}\
  \bibnamefont {Joannopoulos}}, \bibinfo {author} {\bibfnamefont {Marin}\
  \bibnamefont {Solja{\v{c}}i{\'c}}}, \bibinfo {author} {\bibfnamefont
  {Hongsheng}\ \bibnamefont {Chen}}, \bibinfo {author} {\bibfnamefont {Ling}\
  \bibnamefont {Lu}},  \emph {et~al.}} (\bibinfo {year} {2016}{\natexlab{a}}),\
  \bibfield  {title} {\enquote {\bibinfo {title} {Probing topological
  protection using a designer surface plasmon structure},}\ }\href
  {https://www.nature.com/articles/ncomms11619} {\bibfield  {journal} {\bibinfo
   {journal} {Nat. Commun.}\ }\textbf {\bibinfo {volume} {7}},\ \bibinfo
  {pages} {11619}}\BibitemShut {NoStop}%
\bibitem [{\citenamefont {Gao}\ \emph {et~al.}(2015)\citenamefont {Gao},
  \citenamefont {Lawrence}, \citenamefont {Yang}, \citenamefont {Liu},
  \citenamefont {Fang}, \citenamefont {B{\'e}ri}, \citenamefont {Li},\ and\
  \citenamefont {Zhang}}]{Gao:2015PRL}%
  \BibitemOpen
  \bibfield  {author} {\bibinfo {author} {\bibnamefont {Gao}, \bibfnamefont
  {Wenlong}}, \bibinfo {author} {\bibfnamefont {Mark}\ \bibnamefont
  {Lawrence}}, \bibinfo {author} {\bibfnamefont {Biao}\ \bibnamefont {Yang}},
  \bibinfo {author} {\bibfnamefont {Fu}~\bibnamefont {Liu}}, \bibinfo {author}
  {\bibfnamefont {Fengzhou}\ \bibnamefont {Fang}}, \bibinfo {author}
  {\bibfnamefont {Benjamin}\ \bibnamefont {B{\'e}ri}}, \bibinfo {author}
  {\bibfnamefont {Jensen}\ \bibnamefont {Li}}, \ and\ \bibinfo {author}
  {\bibfnamefont {Shuang}\ \bibnamefont {Zhang}}} (\bibinfo {year} {2015}),\
  \bibfield  {title} {\enquote {\bibinfo {title} {Topological photonic phase in
  chiral hyperbolic metamaterials},}\ }\href
  {https://journals.aps.org/prl/abstract/10.1103/PhysRevLett.114.037402}
  {\bibfield  {journal} {\bibinfo  {journal} {Phys. Rev. Lett.}\ }\textbf
  {\bibinfo {volume} {114}}~(\bibinfo {number} {3}),\ \bibinfo {pages}
  {037402}}\BibitemShut {NoStop}%
\bibitem [{\citenamefont {Gao}\ \emph {et~al.}(2016{\natexlab{b}})\citenamefont
  {Gao}, \citenamefont {Yang}, \citenamefont {Lawrence}, \citenamefont {Fang},
  \citenamefont {B{\'e}ri},\ and\ \citenamefont {Zhang}}]{Gao:2016NC}%
  \BibitemOpen
  \bibfield  {author} {\bibinfo {author} {\bibnamefont {Gao}, \bibfnamefont
  {Wenlong}}, \bibinfo {author} {\bibfnamefont {Biao}\ \bibnamefont {Yang}},
  \bibinfo {author} {\bibfnamefont {Mark}\ \bibnamefont {Lawrence}}, \bibinfo
  {author} {\bibfnamefont {Fengzhou}\ \bibnamefont {Fang}}, \bibinfo {author}
  {\bibfnamefont {Benjamin}\ \bibnamefont {B{\'e}ri}}, \ and\ \bibinfo {author}
  {\bibfnamefont {Shuang}\ \bibnamefont {Zhang}}} (\bibinfo {year}
  {2016}{\natexlab{b}}),\ \bibfield  {title} {\enquote {\bibinfo {title}
  {Photonic {W}eyl degeneracies in magnetized plasma},}\ }\href
  {https://www.nature.com/articles/ncomms12435} {\bibfield  {journal} {\bibinfo
   {journal} {Nat. Commun.}\ }\textbf {\bibinfo {volume} {7}},\ \bibinfo
  {pages} {12435}}\BibitemShut {NoStop}%
\bibitem [{\citenamefont {Gardiner}\ and\ \citenamefont
  {Collett}(1985)}]{Gardiner:PRA1985}%
  \BibitemOpen
  \bibfield  {author} {\bibinfo {author} {\bibnamefont {Gardiner},
  \bibfnamefont {C~W}}, \ and\ \bibinfo {author} {\bibfnamefont {M.~J.}\
  \bibnamefont {Collett}}} (\bibinfo {year} {1985}),\ \bibfield  {title}
  {\enquote {\bibinfo {title} {{Input and output in damped quantum systems:
  Quantum stochastic differential equations and the master equation}},}\ }\href
  {https://link.aps.org/doi/10.1103/PhysRevA.31.3761} {\bibfield  {journal}
  {\bibinfo  {journal} {Phys. Rev. A}\ }\textbf {\bibinfo {volume} {31}},\
  \bibinfo {pages} {3761--3774}}\BibitemShut {NoStop}%
\bibitem [{\citenamefont {Gardiner}\ and\ \citenamefont
  {Zoller}(2004)}]{QuantumNoise}%
  \BibitemOpen
  \bibfield  {author} {\bibinfo {author} {\bibnamefont {Gardiner},
  \bibfnamefont {C~W}}, \ and\ \bibinfo {author} {\bibfnamefont
  {P.}~\bibnamefont {Zoller}}} (\bibinfo {year} {2004}),\ \href@noop {} {\emph
  {\bibinfo {title} {Quantum Noise}}}\ (\bibinfo  {publisher} {Springer Verlag,
  Berlin})\BibitemShut {NoStop}%
\bibitem [{\citenamefont {Gbur}(2016)}]{Gbur:Book}%
  \BibitemOpen
  \bibfield  {author} {\bibinfo {author} {\bibnamefont {Gbur}, \bibfnamefont
  {Gregory~J}}} (\bibinfo {year} {2016}),\ \href@noop {} {\emph {\bibinfo
  {title} {Singular Optics}}}\ (\bibinfo  {publisher} {CRC Press},\ \bibinfo
  {address} {Boca Raton, FL})\BibitemShut {NoStop}%
\bibitem [{\citenamefont {Geerligs}\ \emph {et~al.}(1990)\citenamefont
  {Geerligs}, \citenamefont {Anderegg}, \citenamefont {Holweg}, \citenamefont
  {Mooij}, \citenamefont {Pothier}, \citenamefont {Esteve}, \citenamefont
  {Urbina},\ and\ \citenamefont {Devoret}}]{Geerligs:1990}%
  \BibitemOpen
  \bibfield  {author} {\bibinfo {author} {\bibnamefont {Geerligs},
  \bibfnamefont {L~J}}, \bibinfo {author} {\bibfnamefont {V.~F.}\ \bibnamefont
  {Anderegg}}, \bibinfo {author} {\bibfnamefont {P.~A.~M.}\ \bibnamefont
  {Holweg}}, \bibinfo {author} {\bibfnamefont {J.~E.}\ \bibnamefont {Mooij}},
  \bibinfo {author} {\bibfnamefont {H.}~\bibnamefont {Pothier}}, \bibinfo
  {author} {\bibfnamefont {D.}~\bibnamefont {Esteve}}, \bibinfo {author}
  {\bibfnamefont {C.}~\bibnamefont {Urbina}}, \ and\ \bibinfo {author}
  {\bibfnamefont {M.~H.}\ \bibnamefont {Devoret}}} (\bibinfo {year} {1990}),\
  \bibfield  {title} {\enquote {\bibinfo {title} {Frequency-locked turnstile
  device for single electrons},}\ }\href
  {https://link.aps.org/doi/10.1103/PhysRevLett.64.2691} {\bibfield  {journal}
  {\bibinfo  {journal} {Phys. Rev. Lett.}\ }\textbf {\bibinfo {volume} {64}},\
  \bibinfo {pages} {2691--2694}}\BibitemShut {NoStop}%
\bibitem [{\citenamefont {Gerace}\ \emph {et~al.}(2009)\citenamefont {Gerace},
  \citenamefont {T\"ureci}, \citenamefont {Imamoglu}, \citenamefont
  {Giovannetti},\ and\ \citenamefont {Fazio}}]{Gerace:NatPhys2009}%
  \BibitemOpen
  \bibfield  {author} {\bibinfo {author} {\bibnamefont {Gerace}, \bibfnamefont
  {Dario}}, \bibinfo {author} {\bibfnamefont {Hakan~E.}\ \bibnamefont
  {T\"ureci}}, \bibinfo {author} {\bibfnamefont {Atac}\ \bibnamefont
  {Imamoglu}}, \bibinfo {author} {\bibfnamefont {Vittorio}\ \bibnamefont
  {Giovannetti}}, \ and\ \bibinfo {author} {\bibfnamefont {Rosario}\
  \bibnamefont {Fazio}}} (\bibinfo {year} {2009}),\ \bibfield  {title}
  {\enquote {\bibinfo {title} {The quantum-optical {J}osephson
  interferometer},}\ }\href {https://www.nature.com/articles/nphys1223}
  {\bibfield  {journal} {\bibinfo  {journal} {Nat. Phys.}\ }\textbf {\bibinfo
  {volume} {5}}~(\bibinfo {number} {4}),\ \bibinfo {pages}
  {281--284}}\BibitemShut {NoStop}%
\bibitem [{\citenamefont {Gerry}\ and\ \citenamefont
  {Knight}(2005)}]{Berry:2005Book}%
  \BibitemOpen
  \bibfield  {author} {\bibinfo {author} {\bibnamefont {Gerry}, \bibfnamefont
  {Christopher}}, \ and\ \bibinfo {author} {\bibfnamefont {Peter}\ \bibnamefont
  {Knight}}} (\bibinfo {year} {2005}),\ \href@noop {} {\emph {\bibinfo {title}
  {Introductory quantum optics}}}\ (\bibinfo  {publisher} {Cambridge university
  press},\ \bibinfo {address} {Cambridge, England})\BibitemShut {NoStop}%
\bibitem [{\citenamefont {Ghulinyan}\ \emph {et~al.}(2014)\citenamefont
  {Ghulinyan}, \citenamefont {Manzano}, \citenamefont {Prtljaga}, \citenamefont
  {Bernard}, \citenamefont {Pavesi}, \citenamefont {Pucker},\ and\
  \citenamefont {Carusotto}}]{Ghulinyan:PRA2014}%
  \BibitemOpen
  \bibfield  {author} {\bibinfo {author} {\bibnamefont {Ghulinyan},
  \bibfnamefont {Mher}}, \bibinfo {author} {\bibfnamefont {Fernando~Ramiro}\
  \bibnamefont {Manzano}}, \bibinfo {author} {\bibfnamefont {Nikola}\
  \bibnamefont {Prtljaga}}, \bibinfo {author} {\bibfnamefont {Martino}\
  \bibnamefont {Bernard}}, \bibinfo {author} {\bibfnamefont {Lorenzo}\
  \bibnamefont {Pavesi}}, \bibinfo {author} {\bibfnamefont {Georg}\
  \bibnamefont {Pucker}}, \ and\ \bibinfo {author} {\bibfnamefont {Iacopo}\
  \bibnamefont {Carusotto}}} (\bibinfo {year} {2014}),\ \bibfield  {title}
  {\enquote {\bibinfo {title} {Intermode reactive coupling induced by
  waveguide-resonator interaction},}\ }\href
  {https://journals.aps.org/pra/abstract/10.1103/PhysRevA.90.053811} {\bibfield
   {journal} {\bibinfo  {journal} {Phys. Rev. A}\ }\textbf {\bibinfo {volume}
  {90}}~(\bibinfo {number} {5}),\ \bibinfo {pages} {053811}}\BibitemShut
  {NoStop}%
\bibitem [{\citenamefont {Goldman}\ \emph
  {et~al.}(2016{\natexlab{a}})\citenamefont {Goldman}, \citenamefont {Budich},\
  and\ \citenamefont {Zoller}}]{Goldman:2016NatPhys}%
  \BibitemOpen
  \bibfield  {author} {\bibinfo {author} {\bibnamefont {Goldman}, \bibfnamefont
  {N}}, \bibinfo {author} {\bibfnamefont {JC}~\bibnamefont {Budich}}, \ and\
  \bibinfo {author} {\bibfnamefont {P}~\bibnamefont {Zoller}}} (\bibinfo {year}
  {2016}{\natexlab{a}}),\ \bibfield  {title} {\enquote {\bibinfo {title}
  {Topological quantum matter with ultracold gases in optical lattices},}\
  }\href {https://www.nature.com/articles/nphys3803} {\bibfield  {journal}
  {\bibinfo  {journal} {Nat. Phys.}\ }\textbf {\bibinfo {volume}
  {12}}~(\bibinfo {number} {7}),\ \bibinfo {pages} {639--645}}\BibitemShut
  {NoStop}%
\bibitem [{\citenamefont {Goldman}\ and\ \citenamefont
  {Dalibard}(2014)}]{Goldman:PRX2014}%
  \BibitemOpen
  \bibfield  {author} {\bibinfo {author} {\bibnamefont {Goldman}, \bibfnamefont
  {N}}, \ and\ \bibinfo {author} {\bibfnamefont {J.}~\bibnamefont {Dalibard}}}
  (\bibinfo {year} {2014}),\ \bibfield  {title} {\enquote {\bibinfo {title}
  {Periodically driven quantum systems: {E}ffective {H}amiltonians and
  engineered gauge fields},}\ }\href
  {http://link.aps.org/doi/10.1103/PhysRevX.4.031027} {\bibfield  {journal}
  {\bibinfo  {journal} {Phys. Rev. X}\ }\textbf {\bibinfo {volume} {4}},\
  \bibinfo {pages} {031027}}\BibitemShut {NoStop}%
\bibitem [{\citenamefont {Goldman}\ \emph {et~al.}(2015)\citenamefont
  {Goldman}, \citenamefont {Dalibard}, \citenamefont {Aidelsburger},\ and\
  \citenamefont {Cooper}}]{Goldman:2015PRA}%
  \BibitemOpen
  \bibfield  {author} {\bibinfo {author} {\bibnamefont {Goldman}, \bibfnamefont
  {N}}, \bibinfo {author} {\bibfnamefont {J.}~\bibnamefont {Dalibard}},
  \bibinfo {author} {\bibfnamefont {M.}~\bibnamefont {Aidelsburger}}, \ and\
  \bibinfo {author} {\bibfnamefont {N.~R.}\ \bibnamefont {Cooper}}} (\bibinfo
  {year} {2015}),\ \bibfield  {title} {\enquote {\bibinfo {title} {Periodically
  driven quantum matter: {T}he case of resonant modulations},}\ }\href
  {http://link.aps.org/doi/10.1103/PhysRevA.91.033632} {\bibfield  {journal}
  {\bibinfo  {journal} {Phys. Rev. A}\ }\textbf {\bibinfo {volume} {91}},\
  \bibinfo {pages} {033632}}\BibitemShut {NoStop}%
\bibitem [{\citenamefont {Goldman}\ \emph
  {et~al.}(2016{\natexlab{b}})\citenamefont {Goldman}, \citenamefont {Jotzu},
  \citenamefont {Messer}, \citenamefont {G\"org}, \citenamefont {Desbuquois},\
  and\ \citenamefont {Esslinger}}]{Goldman:2016PRA}%
  \BibitemOpen
  \bibfield  {author} {\bibinfo {author} {\bibnamefont {Goldman}, \bibfnamefont
  {N}}, \bibinfo {author} {\bibfnamefont {G.}~\bibnamefont {Jotzu}}, \bibinfo
  {author} {\bibfnamefont {M.}~\bibnamefont {Messer}}, \bibinfo {author}
  {\bibfnamefont {F.}~\bibnamefont {G\"org}}, \bibinfo {author} {\bibfnamefont
  {R.}~\bibnamefont {Desbuquois}}, \ and\ \bibinfo {author} {\bibfnamefont
  {T.}~\bibnamefont {Esslinger}}} (\bibinfo {year} {2016}{\natexlab{b}}),\
  \bibfield  {title} {\enquote {\bibinfo {title} {Creating topological
  interfaces and detecting chiral edge modes in a two-dimensional optical
  lattice},}\ }\href {https://link.aps.org/doi/10.1103/PhysRevA.94.043611}
  {\bibfield  {journal} {\bibinfo  {journal} {Phys. Rev. A}\ }\textbf {\bibinfo
  {volume} {94}},\ \bibinfo {pages} {043611}}\BibitemShut {NoStop}%
\bibitem [{\citenamefont {Goldman}\ \emph {et~al.}(2010)\citenamefont
  {Goldman}, \citenamefont {Satija}, \citenamefont {Nikolic}, \citenamefont
  {Bermudez}, \citenamefont {Martin-Delgado}, \citenamefont {Lewenstein},\ and\
  \citenamefont {Spielman}}]{Goldman:2010PRL}%
  \BibitemOpen
  \bibfield  {author} {\bibinfo {author} {\bibnamefont {Goldman}, \bibfnamefont
  {N}}, \bibinfo {author} {\bibfnamefont {I.}~\bibnamefont {Satija}}, \bibinfo
  {author} {\bibfnamefont {P.}~\bibnamefont {Nikolic}}, \bibinfo {author}
  {\bibfnamefont {A.}~\bibnamefont {Bermudez}}, \bibinfo {author}
  {\bibfnamefont {M.~A.}\ \bibnamefont {Martin-Delgado}}, \bibinfo {author}
  {\bibfnamefont {M.}~\bibnamefont {Lewenstein}}, \ and\ \bibinfo {author}
  {\bibfnamefont {I.~B.}\ \bibnamefont {Spielman}}} (\bibinfo {year} {2010}),\
  \bibfield  {title} {\enquote {\bibinfo {title} {Realistic time-reversal
  invariant topological insulators with neutral atoms},}\ }\href
  {https://link.aps.org/doi/10.1103/PhysRevLett.105.255302} {\bibfield
  {journal} {\bibinfo  {journal} {Phys. Rev. Lett.}\ }\textbf {\bibinfo
  {volume} {105}},\ \bibinfo {pages} {255302}}\BibitemShut {NoStop}%
\bibitem [{\citenamefont {Goldman}\ \emph {et~al.}(2014)\citenamefont
  {Goldman}, \citenamefont {Juzeli{\=u}nas}, \citenamefont {{\"O}hberg},\ and\
  \citenamefont {Spielman}}]{Goldman:2014ROPP}%
  \BibitemOpen
  \bibfield  {author} {\bibinfo {author} {\bibnamefont {Goldman}, \bibfnamefont
  {Nathan}}, \bibinfo {author} {\bibfnamefont {G}~\bibnamefont
  {Juzeli{\=u}nas}}, \bibinfo {author} {\bibfnamefont {P}~\bibnamefont
  {{\"O}hberg}}, \ and\ \bibinfo {author} {\bibfnamefont {Ian~B}\ \bibnamefont
  {Spielman}}} (\bibinfo {year} {2014}),\ \bibfield  {title} {\enquote
  {\bibinfo {title} {Light-induced gauge fields for ultracold atoms},}\ }\href
  {http://iopscience.iop.org/article/10.1088/0034-4885/77/12/126401/meta}
  {\bibfield  {journal} {\bibinfo  {journal} {Rep. Prog. Phys.}\ }\textbf
  {\bibinfo {volume} {77}}~(\bibinfo {number} {12}),\ \bibinfo {pages}
  {126401}}\BibitemShut {NoStop}%
\bibitem [{\citenamefont {Gong}\ \emph {et~al.}(2018)\citenamefont {Gong},
  \citenamefont {Ashida}, \citenamefont {Kawabata}, \citenamefont {Takasan},
  \citenamefont {Higashikawa},\ and\ \citenamefont
  {Ueda}}]{gong2018topological}%
  \BibitemOpen
  \bibfield  {author} {\bibinfo {author} {\bibnamefont {Gong}, \bibfnamefont
  {Zongping}}, \bibinfo {author} {\bibfnamefont {Yuto}\ \bibnamefont {Ashida}},
  \bibinfo {author} {\bibfnamefont {Kohei}\ \bibnamefont {Kawabata}}, \bibinfo
  {author} {\bibfnamefont {Kazuaki}\ \bibnamefont {Takasan}}, \bibinfo {author}
  {\bibfnamefont {Sho}\ \bibnamefont {Higashikawa}}, \ and\ \bibinfo {author}
  {\bibfnamefont {Masahito}\ \bibnamefont {Ueda}}} (\bibinfo {year} {2018}),\
  \bibfield  {title} {\enquote {\bibinfo {title} {Topological phases of
  non-{H}ermitian systems},}\ }\href
  {https://link.aps.org/doi/10.1103/PhysRevX.8.031079} {\bibfield  {journal}
  {\bibinfo  {journal} {Phys. Rev. X}\ }\textbf {\bibinfo {volume} {8}},\
  \bibinfo {pages} {031079}}\BibitemShut {NoStop}%
\bibitem [{\citenamefont {Gopalakrishnan}\ \emph {et~al.}(2012)\citenamefont
  {Gopalakrishnan}, \citenamefont {Ghaemi},\ and\ \citenamefont
  {Ryu}}]{Gopalakrishnan:2012PRB}%
  \BibitemOpen
  \bibfield  {author} {\bibinfo {author} {\bibnamefont {Gopalakrishnan},
  \bibfnamefont {Sarang}}, \bibinfo {author} {\bibfnamefont {Pouyan}\
  \bibnamefont {Ghaemi}}, \ and\ \bibinfo {author} {\bibfnamefont {Shinsei}\
  \bibnamefont {Ryu}}} (\bibinfo {year} {2012}),\ \bibfield  {title} {\enquote
  {\bibinfo {title} {Non-{A}belian \textit{SU(2)} gauge fields through density
  wave order and strain in graphene},}\ }\href
  {https://link.aps.org/doi/10.1103/PhysRevB.86.081403} {\bibfield  {journal}
  {\bibinfo  {journal} {Phys. Rev. B}\ }\textbf {\bibinfo {volume}
  {86}}~(\bibinfo {number} {8}),\ \bibinfo {pages} {081403}}\BibitemShut
  {NoStop}%
\bibitem [{\citenamefont {Goren}\ \emph {et~al.}(2018)\citenamefont {Goren},
  \citenamefont {Plekhanov}, \citenamefont {Appas},\ and\ \citenamefont
  {Le~Hur}}]{Goren:2018PRB}%
  \BibitemOpen
  \bibfield  {author} {\bibinfo {author} {\bibnamefont {Goren}, \bibfnamefont
  {Tal}}, \bibinfo {author} {\bibfnamefont {Kirill}\ \bibnamefont {Plekhanov}},
  \bibinfo {author} {\bibfnamefont {F\'elicien}\ \bibnamefont {Appas}}, \ and\
  \bibinfo {author} {\bibfnamefont {Karyn}\ \bibnamefont {Le~Hur}}} (\bibinfo
  {year} {2018}),\ \bibfield  {title} {\enquote {\bibinfo {title} {Topological
  {Z}ak phase in strongly coupled {LC} circuits},}\ }\href
  {https://link.aps.org/doi/10.1103/PhysRevB.97.041106} {\bibfield  {journal}
  {\bibinfo  {journal} {Phys. Rev. B}\ }\textbf {\bibinfo {volume} {97}},\
  \bibinfo {pages} {041106}}\BibitemShut {NoStop}%
\bibitem [{\citenamefont {Gorlach}\ \emph
  {et~al.}(2018{\natexlab{a}})\citenamefont {Gorlach}, \citenamefont
  {Di~Liberto}, \citenamefont {Recati}, \citenamefont {Carusotto},
  \citenamefont {Poddubny},\ and\ \citenamefont {Menotti}}]{Gorlach:arxiv2018}%
  \BibitemOpen
  \bibfield  {author} {\bibinfo {author} {\bibnamefont {Gorlach}, \bibfnamefont
  {Maxim~A}}, \bibinfo {author} {\bibfnamefont {Marco}\ \bibnamefont
  {Di~Liberto}}, \bibinfo {author} {\bibfnamefont {Alessio}\ \bibnamefont
  {Recati}}, \bibinfo {author} {\bibfnamefont {Iacopo}\ \bibnamefont
  {Carusotto}}, \bibinfo {author} {\bibfnamefont {Alexander~N.}\ \bibnamefont
  {Poddubny}}, \ and\ \bibinfo {author} {\bibfnamefont {Chiara}\ \bibnamefont
  {Menotti}}} (\bibinfo {year} {2018}{\natexlab{a}}),\ \bibfield  {title}
  {\enquote {\bibinfo {title} {Simulation of two-boson bound states using
  arrays of driven-dissipative coupled linear optical resonators},}\ }\href
  {https://link.aps.org/doi/10.1103/PhysRevA.98.063625} {\bibfield  {journal}
  {\bibinfo  {journal} {Phys. Rev. A}\ }\textbf {\bibinfo {volume} {98}},\
  \bibinfo {pages} {063625}}\BibitemShut {NoStop}%
\bibitem [{\citenamefont {Gorlach}\ \emph
  {et~al.}(2018{\natexlab{b}})\citenamefont {Gorlach}, \citenamefont {Ni},
  \citenamefont {Smirnova}, \citenamefont {Korobkin}, \citenamefont {Zhirihin},
  \citenamefont {Slobozhanyuk}, \citenamefont {Belov}, \citenamefont
  {Al{\`{u}}},\ and\ \citenamefont {Khanikaev}}]{Gorlach:NatCom2018}%
  \BibitemOpen
  \bibfield  {author} {\bibinfo {author} {\bibnamefont {Gorlach}, \bibfnamefont
  {Maxim~A}}, \bibinfo {author} {\bibfnamefont {Xiang}\ \bibnamefont {Ni}},
  \bibinfo {author} {\bibfnamefont {Daria~A}\ \bibnamefont {Smirnova}},
  \bibinfo {author} {\bibfnamefont {Dmitry}\ \bibnamefont {Korobkin}}, \bibinfo
  {author} {\bibfnamefont {Dmitry}\ \bibnamefont {Zhirihin}}, \bibinfo {author}
  {\bibfnamefont {Alexey~P}\ \bibnamefont {Slobozhanyuk}}, \bibinfo {author}
  {\bibfnamefont {Pavel~A}\ \bibnamefont {Belov}}, \bibinfo {author}
  {\bibfnamefont {Andrea}\ \bibnamefont {Al{\`{u}}}}, \ and\ \bibinfo {author}
  {\bibfnamefont {Alexander~B}\ \bibnamefont {Khanikaev}}} (\bibinfo {year}
  {2018}{\natexlab{b}}),\ \bibfield  {title} {\enquote {\bibinfo {title}
  {{Far-field probing of leaky topological states in all-dielectric
  metasurfaces}},}\ }\href {https://doi.org/10.1038/s41467-018-03330-9}
  {\bibfield  {journal} {\bibinfo  {journal} {Nat. Commun.}\ }\textbf {\bibinfo
  {volume} {9}}~(\bibinfo {number} {1}),\ \bibinfo {pages} {909}}\BibitemShut
  {NoStop}%
\bibitem [{\citenamefont {Gorlach}\ and\ \citenamefont
  {Poddubny}(2017)}]{Gorlach:PRA2017}%
  \BibitemOpen
  \bibfield  {author} {\bibinfo {author} {\bibnamefont {Gorlach}, \bibfnamefont
  {Maxim~A}}, \ and\ \bibinfo {author} {\bibfnamefont {Alexander~N}\
  \bibnamefont {Poddubny}}} (\bibinfo {year} {2017}),\ \bibfield  {title}
  {\enquote {\bibinfo {title} {Interaction-induced two-photon edge states in an
  extended hubbard model realized in a cavity array},}\ }\href
  {https://journals.aps.org/pra/abstract/10.1103/PhysRevA.95.033831} {\bibfield
   {journal} {\bibinfo  {journal} {Phys. Rev. A}\ }\textbf {\bibinfo {volume}
  {95}}~(\bibinfo {number} {3}),\ \bibinfo {pages} {033831}}\BibitemShut
  {NoStop}%
\bibitem [{\citenamefont {Gorshkov}\ \emph {et~al.}(2011)\citenamefont
  {Gorshkov}, \citenamefont {Otterbach}, \citenamefont {Fleischhauer},
  \citenamefont {Pohl},\ and\ \citenamefont {Lukin}}]{Gorshkov:2011PRL}%
  \BibitemOpen
  \bibfield  {author} {\bibinfo {author} {\bibnamefont {Gorshkov},
  \bibfnamefont {Alexey~V}}, \bibinfo {author} {\bibfnamefont {Johannes}\
  \bibnamefont {Otterbach}}, \bibinfo {author} {\bibfnamefont {Michael}\
  \bibnamefont {Fleischhauer}}, \bibinfo {author} {\bibfnamefont {Thomas}\
  \bibnamefont {Pohl}}, \ and\ \bibinfo {author} {\bibfnamefont {Mikhail~D.}\
  \bibnamefont {Lukin}}} (\bibinfo {year} {2011}),\ \bibfield  {title}
  {\enquote {\bibinfo {title} {Photon-photon interactions via {R}ydberg
  blockade},}\ }\href {http://link.aps.org/doi/10.1103/PhysRevLett.107.133602}
  {\bibfield  {journal} {\bibinfo  {journal} {Phys. Rev. Lett.}\ }\textbf
  {\bibinfo {volume} {107}},\ \bibinfo {pages} {133602}}\BibitemShut {NoStop}%
\bibitem [{\citenamefont {Gra\ss}\ \emph {et~al.}(2014)\citenamefont {Gra\ss},
  \citenamefont {Celi},\ and\ \citenamefont {Lewenstein}}]{Grass:2014PRA}%
  \BibitemOpen
  \bibfield  {author} {\bibinfo {author} {\bibnamefont {Gra\ss}, \bibfnamefont
  {Tobias}}, \bibinfo {author} {\bibfnamefont {Alessio}\ \bibnamefont {Celi}},
  \ and\ \bibinfo {author} {\bibfnamefont {Maciej}\ \bibnamefont {Lewenstein}}}
  (\bibinfo {year} {2014}),\ \bibfield  {title} {\enquote {\bibinfo {title}
  {Quantum magnetism of ultracold atoms with a dynamical pseudospin degree of
  freedom},}\ }\href {https://link.aps.org/doi/10.1103/PhysRevA.90.043628}
  {\bibfield  {journal} {\bibinfo  {journal} {Phys. Rev. A}\ }\textbf {\bibinfo
  {volume} {90}},\ \bibinfo {pages} {043628}}\BibitemShut {NoStop}%
\bibitem [{\citenamefont {Gra\ss{}}\ \emph {et~al.}(2015)\citenamefont
  {Gra\ss{}}, \citenamefont {Muschik}, \citenamefont {Celi}, \citenamefont
  {Chhajlany},\ and\ \citenamefont {Lewenstein}}]{Grass:2015PRA}%
  \BibitemOpen
  \bibfield  {author} {\bibinfo {author} {\bibnamefont {Gra\ss{}},
  \bibfnamefont {Tobias}}, \bibinfo {author} {\bibfnamefont {Christine}\
  \bibnamefont {Muschik}}, \bibinfo {author} {\bibfnamefont {Alessio}\
  \bibnamefont {Celi}}, \bibinfo {author} {\bibfnamefont {Ravindra~W.}\
  \bibnamefont {Chhajlany}}, \ and\ \bibinfo {author} {\bibfnamefont {Maciej}\
  \bibnamefont {Lewenstein}}} (\bibinfo {year} {2015}),\ \bibfield  {title}
  {\enquote {\bibinfo {title} {Synthetic magnetic fluxes and topological order
  in one-dimensional spin systems},}\ }\href
  {https://link.aps.org/doi/10.1103/PhysRevA.91.063612} {\bibfield  {journal}
  {\bibinfo  {journal} {Phys. Rev. A}\ }\textbf {\bibinfo {volume} {91}},\
  \bibinfo {pages} {063612}}\BibitemShut {NoStop}%
\bibitem [{\citenamefont {Greentree}\ \emph {et~al.}(2006)\citenamefont
  {Greentree}, \citenamefont {Tahan}, \citenamefont {Cole},\ and\ \citenamefont
  {Hollenberg}}]{Greentree:2006}%
  \BibitemOpen
  \bibfield  {author} {\bibinfo {author} {\bibnamefont {Greentree},
  \bibfnamefont {Andrew~D}}, \bibinfo {author} {\bibfnamefont {Charles}\
  \bibnamefont {Tahan}}, \bibinfo {author} {\bibfnamefont {Jared~H.}\
  \bibnamefont {Cole}}, \ and\ \bibinfo {author} {\bibfnamefont {Lloyd C.~L.}\
  \bibnamefont {Hollenberg}}} (\bibinfo {year} {2006}),\ \bibfield  {title}
  {\enquote {\bibinfo {title} {Quantum phase transitions of light},}\ }\href
  {https://www.nature.com/articles/nphys466} {\bibfield  {journal} {\bibinfo
  {journal} {Nat. Phys.}\ }\textbf {\bibinfo {volume} {2}}~(\bibinfo {number}
  {12}),\ \bibinfo {pages} {856--861}}\BibitemShut {NoStop}%
\bibitem [{\citenamefont {Grosche}\ \emph {et~al.}(2016)\citenamefont
  {Grosche}, \citenamefont {Szameit},\ and\ \citenamefont
  {Ornigotti}}]{Grosche:2016PRA}%
  \BibitemOpen
  \bibfield  {author} {\bibinfo {author} {\bibnamefont {Grosche}, \bibfnamefont
  {Simon}}, \bibinfo {author} {\bibfnamefont {Alexander}\ \bibnamefont
  {Szameit}}, \ and\ \bibinfo {author} {\bibfnamefont {Marco}\ \bibnamefont
  {Ornigotti}}} (\bibinfo {year} {2016}),\ \bibfield  {title} {\enquote
  {\bibinfo {title} {{Spatial Goos-H{\"{a}}nchen shift in photonic
  graphene}},}\ }\href {http://link.aps.org/doi/10.1103/PhysRevA.94.063831}
  {\bibfield  {journal} {\bibinfo  {journal} {Phys. Rev. A}\ }\textbf {\bibinfo
  {volume} {94}}~(\bibinfo {number} {6}),\ \bibinfo {pages}
  {063831}}\BibitemShut {NoStop}%
\bibitem [{\citenamefont {Grusdt}\ and\ \citenamefont
  {Fleischhauer}(2013)}]{grusdt2013fractional}%
  \BibitemOpen
  \bibfield  {author} {\bibinfo {author} {\bibnamefont {Grusdt}, \bibfnamefont
  {Fabian}}, \ and\ \bibinfo {author} {\bibfnamefont {Michael}\ \bibnamefont
  {Fleischhauer}}} (\bibinfo {year} {2013}),\ \bibfield  {title} {\enquote
  {\bibinfo {title} {Fractional quantum {H}all physics with ultracold {R}ydberg
  gases in artificial gauge fields},}\ }\href
  {https://journals.aps.org/pra/abstract/10.1103/PhysRevA.87.043628} {\bibfield
   {journal} {\bibinfo  {journal} {Phys. Rev. A}\ }\textbf {\bibinfo {volume}
  {87}}~(\bibinfo {number} {4}),\ \bibinfo {pages} {043628}}\BibitemShut
  {NoStop}%
\bibitem [{\citenamefont {Grusdt}\ \emph {et~al.}(2014)\citenamefont {Grusdt},
  \citenamefont {Letscher}, \citenamefont {Hafezi},\ and\ \citenamefont
  {Fleischhauer}}]{Grusdt:PRL2014}%
  \BibitemOpen
  \bibfield  {author} {\bibinfo {author} {\bibnamefont {Grusdt}, \bibfnamefont
  {Fabian}}, \bibinfo {author} {\bibfnamefont {Fabian}\ \bibnamefont
  {Letscher}}, \bibinfo {author} {\bibfnamefont {Mohammad}\ \bibnamefont
  {Hafezi}}, \ and\ \bibinfo {author} {\bibfnamefont {Michael}\ \bibnamefont
  {Fleischhauer}}} (\bibinfo {year} {2014}),\ \bibfield  {title} {\enquote
  {\bibinfo {title} {Topological growing of {L}aughlin states in synthetic
  gauge fields},}\ }\href
  {https://journals.aps.org/prl/abstract/10.1103/PhysRevLett.113.155301}
  {\bibfield  {journal} {\bibinfo  {journal} {Phys. Rev. Lett.}\ }\textbf
  {\bibinfo {volume} {113}}~(\bibinfo {number} {15}),\ \bibinfo {pages}
  {155301}}\BibitemShut {NoStop}%
\bibitem [{\citenamefont {Grusdt}\ \emph {et~al.}(2013)\citenamefont {Grusdt},
  \citenamefont {Otterbach},\ and\ \citenamefont
  {Fleischhauer}}]{Fleischhauer:private}%
  \BibitemOpen
  \bibfield  {author} {\bibinfo {author} {\bibnamefont {Grusdt}, \bibfnamefont
  {Fabian}}, \bibinfo {author} {\bibfnamefont {Johannes}\ \bibnamefont
  {Otterbach}}, \ and\ \bibinfo {author} {\bibfnamefont {Michael}\ \bibnamefont
  {Fleischhauer}}} (\bibinfo {year} {2013}),\ \href@noop {} {}\bibinfo
  {howpublished} {personal communication}\BibitemShut {NoStop}%
\bibitem [{\citenamefont {Gu}\ \emph {et~al.}(2017)\citenamefont {Gu},
  \citenamefont {Kockum}, \citenamefont {Miranowicz}, \citenamefont {Liu},\
  and\ \citenamefont {Nori}}]{Gu:2017PhysRep}%
  \BibitemOpen
  \bibfield  {author} {\bibinfo {author} {\bibnamefont {Gu}, \bibfnamefont
  {Xiu}}, \bibinfo {author} {\bibfnamefont {Anton~Frisk}\ \bibnamefont
  {Kockum}}, \bibinfo {author} {\bibfnamefont {Adam}\ \bibnamefont
  {Miranowicz}}, \bibinfo {author} {\bibfnamefont {Yu-xi}\ \bibnamefont {Liu}},
  \ and\ \bibinfo {author} {\bibfnamefont {Franco}\ \bibnamefont {Nori}}}
  (\bibinfo {year} {2017}),\ \bibfield  {title} {\enquote {\bibinfo {title}
  {Microwave photonics with superconducting quantum circuits},}\ }\href
  {https://www.sciencedirect.com/science/article/pii/S0370157317303290}
  {\bibfield  {journal} {\bibinfo  {journal} {Physics Reports}\ }\textbf
  {\bibinfo {volume} {718-719}},\ \bibinfo {pages} {1--102}}\BibitemShut
  {NoStop}%
\bibitem [{\citenamefont {Guglielmon}\ \emph {et~al.}(2017)\citenamefont
  {Guglielmon}, \citenamefont {Huang}, \citenamefont {Chen},\ and\
  \citenamefont {Rechtsman}}]{guglielmon2017prediction}%
  \BibitemOpen
  \bibfield  {author} {\bibinfo {author} {\bibnamefont {Guglielmon},
  \bibfnamefont {J}}, \bibinfo {author} {\bibfnamefont {S.}~\bibnamefont
  {Huang}}, \bibinfo {author} {\bibfnamefont {K.}~\bibnamefont {Chen}}, \ and\
  \bibinfo {author} {\bibfnamefont {M.~C.}\ \bibnamefont {Rechtsman}}}
  (\bibinfo {year} {2017}),\ \bibfield  {title} {\enquote {\bibinfo {title}
  {Prediction and realization of a photonic topological phase transition},}\
  }in\ \href
  {https://www.osapublishing.org/abstract.cfm?uri=CLEO_QELS-2017-FM2G.6} {\emph
  {\bibinfo {booktitle} {2017 Conference on Lasers and Electro-Optics
  (CLEO)}}},\ pp.\ \bibinfo {pages} {FM2G--6}\BibitemShut {NoStop}%
\bibitem [{\citenamefont {Guinea}\ \emph {et~al.}(2010)\citenamefont {Guinea},
  \citenamefont {Katsnelson},\ and\ \citenamefont {Geim}}]{Guinea:2010NatPhys}%
  \BibitemOpen
  \bibfield  {author} {\bibinfo {author} {\bibnamefont {Guinea}, \bibfnamefont
  {F}}, \bibinfo {author} {\bibfnamefont {M.~I.}\ \bibnamefont {Katsnelson}}, \
  and\ \bibinfo {author} {\bibfnamefont {A.~K.}\ \bibnamefont {Geim}}}
  (\bibinfo {year} {2010}),\ \bibfield  {title} {\enquote {\bibinfo {title}
  {{Energy gaps and a zero-field quantum Hall effect in graphene by strain
  engineering}},}\ }\href {http://www.nature.com/doifinder/10.1038/nphys1420}
  {\bibfield  {journal} {\bibinfo  {journal} {Nat. Phys.}\ }\textbf {\bibinfo
  {volume} {6}}~(\bibinfo {number} {1}),\ \bibinfo {pages}
  {30--33}}\BibitemShut {NoStop}%
\bibitem [{\citenamefont {Gulevich}\ \emph {et~al.}(2017)\citenamefont
  {Gulevich}, \citenamefont {Yudin}, \citenamefont {Skryabin}, \citenamefont
  {Iorsh},\ and\ \citenamefont {Shelykh}}]{Gulevich:SciRep2017}%
  \BibitemOpen
  \bibfield  {author} {\bibinfo {author} {\bibnamefont {Gulevich},
  \bibfnamefont {Dmitry~R}}, \bibinfo {author} {\bibfnamefont {Dmitry}\
  \bibnamefont {Yudin}}, \bibinfo {author} {\bibfnamefont {Dmitry~V}\
  \bibnamefont {Skryabin}}, \bibinfo {author} {\bibfnamefont {Ivan~V}\
  \bibnamefont {Iorsh}}, \ and\ \bibinfo {author} {\bibfnamefont {Ivan~A}\
  \bibnamefont {Shelykh}}} (\bibinfo {year} {2017}),\ \bibfield  {title}
  {\enquote {\bibinfo {title} {{Exploring nonlinear topological states of
  matter with exciton-polaritons: Edge solitons in kagome lattice}},}\ }\href
  {https://www.nature.com/articles/s41598-017-01646-y} {\bibfield  {journal}
  {\bibinfo  {journal} {Sci. Rep.}\ }\textbf {\bibinfo {volume} {7}}~(\bibinfo
  {number} {1}),\ \bibinfo {pages} {1780}}\BibitemShut {NoStop}%
\bibitem [{\citenamefont {Gullans}\ \emph {et~al.}(2016)\citenamefont
  {Gullans}, \citenamefont {Thompson}, \citenamefont {Wang}, \citenamefont
  {Liang}, \citenamefont {Vuleti\ifmmode~\acute{c}\else \'{c}\fi{}},
  \citenamefont {Lukin},\ and\ \citenamefont {Gorshkov}}]{Gullans:PRL2016}%
  \BibitemOpen
  \bibfield  {author} {\bibinfo {author} {\bibnamefont {Gullans}, \bibfnamefont
  {M~J}}, \bibinfo {author} {\bibfnamefont {J.~D.}\ \bibnamefont {Thompson}},
  \bibinfo {author} {\bibfnamefont {Y.}~\bibnamefont {Wang}}, \bibinfo {author}
  {\bibfnamefont {Q.-Y.}\ \bibnamefont {Liang}}, \bibinfo {author}
  {\bibfnamefont {V.}~\bibnamefont {Vuleti\ifmmode~\acute{c}\else \'{c}\fi{}}},
  \bibinfo {author} {\bibfnamefont {M.~D.}\ \bibnamefont {Lukin}}, \ and\
  \bibinfo {author} {\bibfnamefont {A.~V.}\ \bibnamefont {Gorshkov}}} (\bibinfo
  {year} {2016}),\ \bibfield  {title} {\enquote {\bibinfo {title} {Effective
  field theory for {R}ydberg polaritons},}\ }\href
  {https://link.aps.org/doi/10.1103/PhysRevLett.117.113601} {\bibfield
  {journal} {\bibinfo  {journal} {Phys. Rev. Lett.}\ }\textbf {\bibinfo
  {volume} {117}},\ \bibinfo {pages} {113601}}\BibitemShut {NoStop}%
\bibitem [{\citenamefont {Guo}\ \emph {et~al.}(2017)\citenamefont {Guo},
  \citenamefont {Yang}, \citenamefont {Xia}, \citenamefont {Gao}, \citenamefont
  {Liu}, \citenamefont {Chen}, \citenamefont {Xiang},\ and\ \citenamefont
  {Zhang}}]{Guo:2017arXiv}%
  \BibitemOpen
  \bibfield  {author} {\bibinfo {author} {\bibnamefont {Guo}, \bibfnamefont
  {Qinghua}}, \bibinfo {author} {\bibfnamefont {Biao}\ \bibnamefont {Yang}},
  \bibinfo {author} {\bibfnamefont {Lingbo}\ \bibnamefont {Xia}}, \bibinfo
  {author} {\bibfnamefont {Wenlong}\ \bibnamefont {Gao}}, \bibinfo {author}
  {\bibfnamefont {Hongchao}\ \bibnamefont {Liu}}, \bibinfo {author}
  {\bibfnamefont {Jing}\ \bibnamefont {Chen}}, \bibinfo {author} {\bibfnamefont
  {Yuanjiang}\ \bibnamefont {Xiang}}, \ and\ \bibinfo {author} {\bibfnamefont
  {Shuang}\ \bibnamefont {Zhang}}} (\bibinfo {year} {2017}),\ \bibfield
  {title} {\enquote {\bibinfo {title} {Three dimensional photonic {D}irac
  points in metamaterials},}\ }\href
  {https://link.aps.org/doi/10.1103/PhysRevLett.119.213901} {\bibfield
  {journal} {\bibinfo  {journal} {Phys. Rev. Lett.}\ }\textbf {\bibinfo
  {volume} {119}},\ \bibinfo {pages} {213901}}\BibitemShut {NoStop}%
\bibitem [{\citenamefont {Guo}\ \emph {et~al.}(2018)\citenamefont {Guo},
  \citenamefont {You}, \citenamefont {Yang}, \citenamefont {Sellman},
  \citenamefont {Blythe}, \citenamefont {Liu}, \citenamefont {Xiang},
  \citenamefont {Li}, \citenamefont {Fan}, \citenamefont {Chen} \emph
  {et~al.}}]{guo2018observation}%
  \BibitemOpen
  \bibfield  {author} {\bibinfo {author} {\bibnamefont {Guo}, \bibfnamefont
  {Qinghua}}, \bibinfo {author} {\bibfnamefont {Oubo}\ \bibnamefont {You}},
  \bibinfo {author} {\bibfnamefont {Biao}\ \bibnamefont {Yang}}, \bibinfo
  {author} {\bibfnamefont {James~B}\ \bibnamefont {Sellman}}, \bibinfo {author}
  {\bibfnamefont {Edward}\ \bibnamefont {Blythe}}, \bibinfo {author}
  {\bibfnamefont {Hongchao}\ \bibnamefont {Liu}}, \bibinfo {author}
  {\bibfnamefont {Yuanjiang}\ \bibnamefont {Xiang}}, \bibinfo {author}
  {\bibfnamefont {Jensen}\ \bibnamefont {Li}}, \bibinfo {author} {\bibfnamefont
  {Dianyuan}\ \bibnamefont {Fan}}, \bibinfo {author} {\bibfnamefont {Jing}\
  \bibnamefont {Chen}},  \emph {et~al.}} (\bibinfo {year} {2018}),\ \bibfield
  {title} {\enquote {\bibinfo {title} {Observation of three-dimensional
  photonic {D}irac points and spin-polarized surface arcs},}\ }\href
  {https://arxiv.org/abs/1810.02354} {\bibinfo  {journal} {arXiv:1810.02354}\
  }\BibitemShut {NoStop}%
\bibitem [{\citenamefont {Gustafsson}\ \emph {et~al.}(2014)\citenamefont
  {Gustafsson}, \citenamefont {Aref}, \citenamefont {Kockum}, \citenamefont
  {Ekstr{\"o}m}, \citenamefont {Johansson},\ and\ \citenamefont
  {Delsing}}]{Delsing2014}%
  \BibitemOpen
\bibfield  {journal} {  }\bibfield  {author} {\bibinfo {author} {\bibnamefont
  {Gustafsson}, \bibfnamefont {Martin~V}}, \bibinfo {author} {\bibfnamefont
  {Thomas}\ \bibnamefont {Aref}}, \bibinfo {author} {\bibfnamefont
  {Anton~Frisk}\ \bibnamefont {Kockum}}, \bibinfo {author} {\bibfnamefont
  {Maria~K}\ \bibnamefont {Ekstr{\"o}m}}, \bibinfo {author} {\bibfnamefont
  {G{\"o}ran}\ \bibnamefont {Johansson}}, \ and\ \bibinfo {author}
  {\bibfnamefont {Per}\ \bibnamefont {Delsing}}} (\bibinfo {year} {2014}),\
  \bibfield  {title} {\enquote {\bibinfo {title} {Propagating phonons coupled
  to an artificial atom},}\ }\href
  {http://science.sciencemag.org/content/346/6206/207} {\bibfield  {journal}
  {\bibinfo  {journal} {Science}\ }\textbf {\bibinfo {volume} {346}}~(\bibinfo
  {number} {6206}),\ \bibinfo {pages} {207--211}}\BibitemShut {NoStop}%
\bibitem [{\citenamefont {Habraken}\ \emph {et~al.}(2012)\citenamefont
  {Habraken}, \citenamefont {Stannigel}, \citenamefont {Lukin}, \citenamefont
  {Zoller},\ and\ \citenamefont {Rabl}}]{habraken2012continuous}%
  \BibitemOpen
  \bibfield  {author} {\bibinfo {author} {\bibnamefont {Habraken},
  \bibfnamefont {SJM}}, \bibinfo {author} {\bibfnamefont {K}~\bibnamefont
  {Stannigel}}, \bibinfo {author} {\bibfnamefont {Mikhail~D}\ \bibnamefont
  {Lukin}}, \bibinfo {author} {\bibfnamefont {P}~\bibnamefont {Zoller}}, \ and\
  \bibinfo {author} {\bibfnamefont {P}~\bibnamefont {Rabl}}} (\bibinfo {year}
  {2012}),\ \bibfield  {title} {\enquote {\bibinfo {title} {Continuous mode
  cooling and phonon routers for phononic quantum networks},}\ }\href
  {https://doi.org/10.1088/1367-2630/14/11/115004} {\bibfield  {journal}
  {\bibinfo  {journal} {New J. Phys.}\ }\textbf {\bibinfo {volume}
  {14}}~(\bibinfo {number} {11}),\ \bibinfo {pages} {115004}}\BibitemShut
  {NoStop}%
\bibitem [{\citenamefont {Hadad}\ \emph {et~al.}(2016)\citenamefont {Hadad},
  \citenamefont {Khanikaev},\ and\ \citenamefont {Al\`u}}]{Hadad:PRB2016}%
  \BibitemOpen
  \bibfield  {author} {\bibinfo {author} {\bibnamefont {Hadad}, \bibfnamefont
  {Yakir}}, \bibinfo {author} {\bibfnamefont {Alexander~B.}\ \bibnamefont
  {Khanikaev}}, \ and\ \bibinfo {author} {\bibfnamefont {Andrea}\ \bibnamefont
  {Al\`u}}} (\bibinfo {year} {2016}),\ \bibfield  {title} {\enquote {\bibinfo
  {title} {Self-induced topological transitions and edge states supported by
  nonlinear staggered potentials},}\ }\href
  {https://link.aps.org/doi/10.1103/PhysRevB.93.155112} {\bibfield  {journal}
  {\bibinfo  {journal} {Phys. Rev. B}\ }\textbf {\bibinfo {volume} {93}},\
  \bibinfo {pages} {155112}}\BibitemShut {NoStop}%
\bibitem [{\citenamefont {Haddad}\ and\ \citenamefont
  {Carr}(2011)}]{Haddad:EPL2011}%
  \BibitemOpen
  \bibfield  {author} {\bibinfo {author} {\bibnamefont {Haddad}, \bibfnamefont
  {LH}}, \ and\ \bibinfo {author} {\bibfnamefont {LD}~\bibnamefont {Carr}}}
  (\bibinfo {year} {2011}),\ \bibfield  {title} {\enquote {\bibinfo {title}
  {{Relativistic linear stability equations for the nonlinear Dirac equation in
  Bose-Einstein condensates}},}\ }\href
  {https://iopscience.iop.org/article/10.1209/0295-5075/94/56002/meta}
  {\bibfield  {journal} {\bibinfo  {journal} {EPL (Europhysics Letters)}\
  }\textbf {\bibinfo {volume} {94}}~(\bibinfo {number} {5}),\ \bibinfo {pages}
  {56002}}\BibitemShut {NoStop}%
\bibitem [{\citenamefont {Haddad}\ and\ \citenamefont
  {Carr}(2015)}]{Haddad2:NJP2015}%
  \BibitemOpen
  \bibfield  {author} {\bibinfo {author} {\bibnamefont {Haddad}, \bibfnamefont
  {LH}}, \ and\ \bibinfo {author} {\bibfnamefont {Lincoln~D}\ \bibnamefont
  {Carr}}} (\bibinfo {year} {2015}),\ \bibfield  {title} {\enquote {\bibinfo
  {title} {{The nonlinear Dirac equation in Bose--Einstein condensates: vortex
  solutions and spectra in a weak harmonic trap}},}\ }\href
  {https://iopscience.iop.org/article/10.1088/1367-2630/17/11/113011}
  {\bibfield  {journal} {\bibinfo  {journal} {New J. Phys.}\ }\textbf {\bibinfo
  {volume} {17}}~(\bibinfo {number} {11}),\ \bibinfo {pages}
  {113011}}\BibitemShut {NoStop}%
\bibitem [{\citenamefont {Haddad}\ \emph {et~al.}(2015)\citenamefont {Haddad},
  \citenamefont {Weaver},\ and\ \citenamefont {Carr}}]{Haddad:NJP2015}%
  \BibitemOpen
  \bibfield  {author} {\bibinfo {author} {\bibnamefont {Haddad}, \bibfnamefont
  {LH}}, \bibinfo {author} {\bibfnamefont {CM}~\bibnamefont {Weaver}}, \ and\
  \bibinfo {author} {\bibfnamefont {Lincoln~D}\ \bibnamefont {Carr}}} (\bibinfo
  {year} {2015}),\ \bibfield  {title} {\enquote {\bibinfo {title} {The
  nonlinear {D}irac equation in {B}ose--{E}instein condensates: {I}.
  {R}elativistic solitons in armchair nanoribbon optical lattices},}\ }\href
  {https://iopscience.iop.org/article/10.1088/1367-2630/17/6/063033/meta}
  {\bibfield  {journal} {\bibinfo  {journal} {New J. Phys.}\ }\textbf {\bibinfo
  {volume} {17}}~(\bibinfo {number} {6}),\ \bibinfo {pages}
  {063033}}\BibitemShut {NoStop}%
\bibitem [{\citenamefont {Hafezi}\ \emph {et~al.}(2015)\citenamefont {Hafezi},
  \citenamefont {Adhikari},\ and\ \citenamefont {Taylor}}]{Hafezi:PRB2015}%
  \BibitemOpen
  \bibfield  {author} {\bibinfo {author} {\bibnamefont {Hafezi}, \bibfnamefont
  {M}}, \bibinfo {author} {\bibfnamefont {P}~\bibnamefont {Adhikari}}, \ and\
  \bibinfo {author} {\bibfnamefont {JM}~\bibnamefont {Taylor}}} (\bibinfo
  {year} {2015}),\ \bibfield  {title} {\enquote {\bibinfo {title} {Chemical
  potential for light by parametric coupling},}\ }\href
  {https://journals.aps.org/prb/abstract/10.1103/PhysRevB.92.174305} {\bibfield
   {journal} {\bibinfo  {journal} {Phys. Rev. B}\ }\textbf {\bibinfo {volume}
  {92}}~(\bibinfo {number} {17}),\ \bibinfo {pages} {174305}}\BibitemShut
  {NoStop}%
\bibitem [{\citenamefont {Hafezi}(2014)}]{Hafezi:2014PRL}%
  \BibitemOpen
  \bibfield  {author} {\bibinfo {author} {\bibnamefont {Hafezi}, \bibfnamefont
  {Mohammad}}} (\bibinfo {year} {2014}),\ \bibfield  {title} {\enquote
  {\bibinfo {title} {Measuring topological invariants in photonic systems},}\
  }\href {http://link.aps.org/doi/10.1103/PhysRevLett.112.210405} {\bibfield
  {journal} {\bibinfo  {journal} {Phys. Rev. Lett.}\ }\textbf {\bibinfo
  {volume} {112}},\ \bibinfo {pages} {210405}}\BibitemShut {NoStop}%
\bibitem [{\citenamefont {Hafezi}\ \emph {et~al.}(2014)\citenamefont {Hafezi},
  \citenamefont {Adhikari},\ and\ \citenamefont {Taylor}}]{Hafezi:PRB2014}%
  \BibitemOpen
  \bibfield  {author} {\bibinfo {author} {\bibnamefont {Hafezi}, \bibfnamefont
  {Mohammad}}, \bibinfo {author} {\bibfnamefont {Prabin}\ \bibnamefont
  {Adhikari}}, \ and\ \bibinfo {author} {\bibfnamefont {Jacob~M}\ \bibnamefont
  {Taylor}}} (\bibinfo {year} {2014}),\ \bibfield  {title} {\enquote {\bibinfo
  {title} {Engineering three-body interaction and {P}faffian states in circuit
  {QED} systems},}\ }\href
  {https://link.aps.org/doi/10.1103/PhysRevB.90.060503} {\bibfield  {journal}
  {\bibinfo  {journal} {Phys. Rev. B}\ }\textbf {\bibinfo {volume}
  {90}}~(\bibinfo {number} {6}),\ \bibinfo {pages} {060503}}\BibitemShut
  {NoStop}%
\bibitem [{\citenamefont {Hafezi}\ \emph {et~al.}(2011)\citenamefont {Hafezi},
  \citenamefont {Demler}, \citenamefont {Lukin},\ and\ \citenamefont
  {Taylor}}]{Hafezi:2011NatPhys}%
  \BibitemOpen
  \bibfield  {author} {\bibinfo {author} {\bibnamefont {Hafezi}, \bibfnamefont
  {Mohammad}}, \bibinfo {author} {\bibfnamefont {Eugene~A}\ \bibnamefont
  {Demler}}, \bibinfo {author} {\bibfnamefont {Mikhail~D}\ \bibnamefont
  {Lukin}}, \ and\ \bibinfo {author} {\bibfnamefont {Jacob~M}\ \bibnamefont
  {Taylor}}} (\bibinfo {year} {2011}),\ \bibfield  {title} {\enquote {\bibinfo
  {title} {Robust optical delay lines with topological protection},}\ }\href
  {http://www.nature.com/nphys/journal/v7/n11/full/nphys2063.html} {\bibfield
  {journal} {\bibinfo  {journal} {Nat. Phys.}\ }\textbf {\bibinfo {volume}
  {7}}~(\bibinfo {number} {11}),\ \bibinfo {pages} {907--912}}\BibitemShut
  {NoStop}%
\bibitem [{\citenamefont {Hafezi}\ \emph
  {et~al.}(2013{\natexlab{a}})\citenamefont {Hafezi}, \citenamefont {Lukin},\
  and\ \citenamefont {Taylor}}]{Hafezi:NJP2013}%
  \BibitemOpen
  \bibfield  {author} {\bibinfo {author} {\bibnamefont {Hafezi}, \bibfnamefont
  {Mohammad}}, \bibinfo {author} {\bibfnamefont {Mikhail~D}\ \bibnamefont
  {Lukin}}, \ and\ \bibinfo {author} {\bibfnamefont {Jacob~M}\ \bibnamefont
  {Taylor}}} (\bibinfo {year} {2013}{\natexlab{a}}),\ \bibfield  {title}
  {\enquote {\bibinfo {title} {Non-equilibrium fractional quantum {H}all state
  of light},}\ }\href {http://stacks.iop.org/1367-2630/15/i=6/a=063001}
  {\bibfield  {journal} {\bibinfo  {journal} {New J. Phys.}\ }\textbf {\bibinfo
  {volume} {15}}~(\bibinfo {number} {6}),\ \bibinfo {pages}
  {063001}}\BibitemShut {NoStop}%
\bibitem [{\citenamefont {Hafezi}\ \emph
  {et~al.}(2013{\natexlab{b}})\citenamefont {Hafezi}, \citenamefont {Mittal},
  \citenamefont {Fan}, \citenamefont {Migdall},\ and\ \citenamefont
  {Taylor}}]{Hafezi:2013NatPhot}%
  \BibitemOpen
  \bibfield  {author} {\bibinfo {author} {\bibnamefont {Hafezi}, \bibfnamefont
  {Mohammad}}, \bibinfo {author} {\bibfnamefont {S}~\bibnamefont {Mittal}},
  \bibinfo {author} {\bibfnamefont {J}~\bibnamefont {Fan}}, \bibinfo {author}
  {\bibfnamefont {A}~\bibnamefont {Migdall}}, \ and\ \bibinfo {author}
  {\bibfnamefont {JM}~\bibnamefont {Taylor}}} (\bibinfo {year}
  {2013}{\natexlab{b}}),\ \bibfield  {title} {\enquote {\bibinfo {title}
  {Imaging topological edge states in silicon photonics},}\ }\href
  {http://www.nature.com/nphoton/journal/v7/n12/full/nphoton.2013.274.html}
  {\bibfield  {journal} {\bibinfo  {journal} {Nat. Photonics}\ }\textbf
  {\bibinfo {volume} {7}}~(\bibinfo {number} {12}),\ \bibinfo {pages}
  {1001--1005}}\BibitemShut {NoStop}%
\bibitem [{\citenamefont {Hafezi}\ and\ \citenamefont
  {Rabl}(2012)}]{Hafezi:2012OPtExp}%
  \BibitemOpen
  \bibfield  {author} {\bibinfo {author} {\bibnamefont {Hafezi}, \bibfnamefont
  {Mohammad}}, \ and\ \bibinfo {author} {\bibfnamefont {Peter}\ \bibnamefont
  {Rabl}}} (\bibinfo {year} {2012}),\ \bibfield  {title} {\enquote {\bibinfo
  {title} {Optomechanically induced non-reciprocity in microring resonators},}\
  }\href {https://www.osapublishing.org/oe/abstract.cfm?uri=oe-20-7-7672}
  {\bibfield  {journal} {\bibinfo  {journal} {Opt. Express}\ }\textbf {\bibinfo
  {volume} {20}}~(\bibinfo {number} {7}),\ \bibinfo {pages}
  {7672--7684}}\BibitemShut {NoStop}%
\bibitem [{\citenamefont {Haldane}(1988)}]{Haldane:1988PRL}%
  \BibitemOpen
  \bibfield  {author} {\bibinfo {author} {\bibnamefont {Haldane}, \bibfnamefont
  {F~D~M}}} (\bibinfo {year} {1988}),\ \bibfield  {title} {\enquote {\bibinfo
  {title} {Model for a quantum {H}all effect without {L}andau levels:
  {C}ondensed-matter realization of the "parity anomaly"},}\ }\href
  {https://link.aps.org/doi/10.1103/PhysRevLett.61.2015} {\bibfield  {journal}
  {\bibinfo  {journal} {Phys. Rev. Lett.}\ }\textbf {\bibinfo {volume} {61}},\
  \bibinfo {pages} {2015--2018}}\BibitemShut {NoStop}%
\bibitem [{\citenamefont {Haldane}\ and\ \citenamefont
  {Raghu}(2008)}]{Haldane:2008PRL}%
  \BibitemOpen
  \bibfield  {author} {\bibinfo {author} {\bibnamefont {Haldane}, \bibfnamefont
  {F~D~M}}, \ and\ \bibinfo {author} {\bibfnamefont {S.}~\bibnamefont {Raghu}}}
  (\bibinfo {year} {2008}),\ \bibfield  {title} {\enquote {\bibinfo {title}
  {Possible realization of directional optical waveguides in photonic crystals
  with broken time-reversal symmetry},}\ }\href
  {http://link.aps.org/doi/10.1103/PhysRevLett.100.013904} {\bibfield
  {journal} {\bibinfo  {journal} {Phys. Rev. Lett.}\ }\textbf {\bibinfo
  {volume} {100}},\ \bibinfo {pages} {013904}}\BibitemShut {NoStop}%
\bibitem [{\citenamefont {Halperin}(1982)}]{Halperin1982}%
  \BibitemOpen
  \bibfield  {author} {\bibinfo {author} {\bibnamefont {Halperin},
  \bibfnamefont {B~I}}} (\bibinfo {year} {1982}),\ \bibfield  {title} {\enquote
  {\bibinfo {title} {Quantized hall conductance, current-carrying edge states,
  and the existence of extended states in a two-dimensional disordered
  potential},}\ }\href {\doibase 10.1103/PhysRevB.25.2185} {\bibfield
  {journal} {\bibinfo  {journal} {Phys. Rev. B}\ }\textbf {\bibinfo {volume}
  {25}},\ \bibinfo {pages} {2185--2190}}\BibitemShut {NoStop}%
\bibitem [{\citenamefont {Halperin}(1987)}]{Halperin:1987JJAP}%
  \BibitemOpen
  \bibfield  {author} {\bibinfo {author} {\bibnamefont {Halperin},
  \bibfnamefont {Bertrand~I}}} (\bibinfo {year} {1987}),\ \bibfield  {title}
  {\enquote {\bibinfo {title} {Possible states for a three-dimensional electron
  gas in a strong magnetic field},}\ }\href
  {https://iopscience.iop.org/article/10.7567/JJAPS.26S3.1913} {\bibfield
  {journal} {\bibinfo  {journal} {Japanese Journal of Applied Physics}\
  }\textbf {\bibinfo {volume} {26}}~(\bibinfo {number} {S3-3}),\ \bibinfo
  {pages} {1913}}\BibitemShut {NoStop}%
\bibitem [{\citenamefont {Hannay}(1985)}]{Hannay:1985JPA}%
  \BibitemOpen
  \bibfield  {author} {\bibinfo {author} {\bibnamefont {Hannay}, \bibfnamefont
  {J~H}}} (\bibinfo {year} {1985}),\ \bibfield  {title} {\enquote {\bibinfo
  {title} {Angle variable holonomy in adiabatic excursion of an integrable
  hamiltonian},}\ }\href {http://stacks.iop.org/0305-4470/18/i=2/a=011}
  {\bibfield  {journal} {\bibinfo  {journal} {J. Phys. A}\ }\textbf {\bibinfo
  {volume} {18}}~(\bibinfo {number} {2}),\ \bibinfo {pages} {221}}\BibitemShut
  {NoStop}%
\bibitem [{\citenamefont {Harari}\ \emph {et~al.}(2018)\citenamefont {Harari},
  \citenamefont {Bandres}, \citenamefont {Lumer}, \citenamefont {Rechtsman},
  \citenamefont {Chong}, \citenamefont {Khajavikhan}, \citenamefont
  {Christodoulides},\ and\ \citenamefont {Segev}}]{Harari:Science2018}%
  \BibitemOpen
  \bibfield  {author} {\bibinfo {author} {\bibnamefont {Harari}, \bibfnamefont
  {Gal}}, \bibinfo {author} {\bibfnamefont {Miguel~A.}\ \bibnamefont
  {Bandres}}, \bibinfo {author} {\bibfnamefont {Yaakov}\ \bibnamefont {Lumer}},
  \bibinfo {author} {\bibfnamefont {Mikael~C.}\ \bibnamefont {Rechtsman}},
  \bibinfo {author} {\bibfnamefont {Y.~D.}\ \bibnamefont {Chong}}, \bibinfo
  {author} {\bibfnamefont {Mercedeh}\ \bibnamefont {Khajavikhan}}, \bibinfo
  {author} {\bibfnamefont {Demetrios~N.}\ \bibnamefont {Christodoulides}}, \
  and\ \bibinfo {author} {\bibfnamefont {Mordechai}\ \bibnamefont {Segev}}}
  (\bibinfo {year} {2018}),\ \bibfield  {title} {\enquote {\bibinfo {title}
  {{Topological insulator laser: Theory}},}\ }\href
  {http://science.sciencemag.org/content/early/2018/01/31/science.aar4003}
  {\bibfield  {journal} {\bibinfo  {journal} {Science}\ }\textbf {\bibinfo
  {volume} {359}},\ \bibinfo {pages} {eaar4003}}\BibitemShut {NoStop}%
\bibitem [{\citenamefont {Harper}(1955)}]{Harper:1955PPSA}%
  \BibitemOpen
  \bibfield  {author} {\bibinfo {author} {\bibnamefont {Harper}, \bibfnamefont
  {P~G}}} (\bibinfo {year} {1955}),\ \bibfield  {title} {\enquote {\bibinfo
  {title} {Single band motion of conduction electrons in a uniform magnetic
  field},}\ }\href {http://stacks.iop.org/0370-1298/68/i=10/a=304} {\bibfield
  {journal} {\bibinfo  {journal} {Proceedings of the Physical Society. Section
  A}\ }\textbf {\bibinfo {volume} {68}}~(\bibinfo {number} {10}),\ \bibinfo
  {pages} {874}}\BibitemShut {NoStop}%
\bibitem [{\citenamefont {Harris}(1989)}]{Harris:PRL1989}%
  \BibitemOpen
  \bibfield  {author} {\bibinfo {author} {\bibnamefont {Harris}, \bibfnamefont
  {S~E}}} (\bibinfo {year} {1989}),\ \bibfield  {title} {\enquote {\bibinfo
  {title} {{Lasers without inversion: Interference of lifetime-broadened
  resonances}},}\ }\href {https://link.aps.org/doi/10.1103/PhysRevLett.62.1033}
  {\bibfield  {journal} {\bibinfo  {journal} {Phys. Rev. Lett.}\ }\textbf
  {\bibinfo {volume} {62}},\ \bibinfo {pages} {1033--1036}}\BibitemShut
  {NoStop}%
\bibitem [{\citenamefont {Hartmann}\ \emph {et~al.}(2006)\citenamefont
  {Hartmann}, \citenamefont {Brand\~ao},\ and\ \citenamefont
  {Plenio}}]{Hartmann:NatPhys2006}%
  \BibitemOpen
  \bibfield  {author} {\bibinfo {author} {\bibnamefont {Hartmann},
  \bibfnamefont {Michael~J}}, \bibinfo {author} {\bibfnamefont {Fernando G.
  S.~L.}\ \bibnamefont {Brand\~ao}}, \ and\ \bibinfo {author} {\bibfnamefont
  {Martin~B.}\ \bibnamefont {Plenio}}} (\bibinfo {year} {2006}),\ \bibfield
  {title} {\enquote {\bibinfo {title} {Strongly interacting polaritons in
  coupled arrays of cavities},}\ }\href
  {https://www.nature.com/articles/nphys462} {\bibfield  {journal} {\bibinfo
  {journal} {Nat. Phys.}\ }\textbf {\bibinfo {volume} {2}}~(\bibinfo {number}
  {12}),\ \bibinfo {pages} {849--855}}\BibitemShut {NoStop}%
\bibitem [{\citenamefont {Hartstein}\ \emph {et~al.}(1973)\citenamefont
  {Hartstein}, \citenamefont {Burstein}, \citenamefont {Maradudin},
  \citenamefont {Brewer},\ and\ \citenamefont {Wallis}}]{Hartstein:1973JOPC}%
  \BibitemOpen
  \bibfield  {author} {\bibinfo {author} {\bibnamefont {Hartstein},
  \bibfnamefont {A}}, \bibinfo {author} {\bibfnamefont {E}~\bibnamefont
  {Burstein}}, \bibinfo {author} {\bibfnamefont {AA}~\bibnamefont {Maradudin}},
  \bibinfo {author} {\bibfnamefont {R}~\bibnamefont {Brewer}}, \ and\ \bibinfo
  {author} {\bibfnamefont {RF}~\bibnamefont {Wallis}}} (\bibinfo {year}
  {1973}),\ \bibfield  {title} {\enquote {\bibinfo {title} {Surface polaritons
  on semi-infinite gyromagnetic media},}\ }\href
  {https://iopscience.iop.org/article/10.1088/0022-3719/6/7/016} {\bibfield
  {journal} {\bibinfo  {journal} {J. Phys. C}\ }\textbf {\bibinfo {volume}
  {6}}~(\bibinfo {number} {7}),\ \bibinfo {pages} {1266}}\BibitemShut {NoStop}%
\bibitem [{\citenamefont {Hasan}\ and\ \citenamefont
  {Kane}(2010)}]{Hasan:2010RMP}%
  \BibitemOpen
  \bibfield  {author} {\bibinfo {author} {\bibnamefont {Hasan}, \bibfnamefont
  {M~Z}}, \ and\ \bibinfo {author} {\bibfnamefont {C.~L.}\ \bibnamefont
  {Kane}}} (\bibinfo {year} {2010}),\ \bibfield  {title} {\enquote {\bibinfo
  {title} {{Colloquium: Topological insulators}},}\ }\href
  {https://link.aps.org/doi/10.1103/RevModPhys.82.3045} {\bibfield  {journal}
  {\bibinfo  {journal} {Rev. Mod. Phys.}\ }\textbf {\bibinfo {volume} {82}},\
  \bibinfo {pages} {3045--3067}}\BibitemShut {NoStop}%
\bibitem [{\citenamefont {Hatsugai}(1993{\natexlab{a}})}]{Hatsugai:1993PRL}%
  \BibitemOpen
  \bibfield  {author} {\bibinfo {author} {\bibnamefont {Hatsugai},
  \bibfnamefont {Yasuhiro}}} (\bibinfo {year} {1993}{\natexlab{a}}),\ \bibfield
   {title} {\enquote {\bibinfo {title} {Chern number and edge states in the
  integer quantum {H}all effect},}\ }\href
  {https://link.aps.org/doi/10.1103/PhysRevLett.71.3697} {\bibfield  {journal}
  {\bibinfo  {journal} {Phys. Rev. Lett.}\ }\textbf {\bibinfo {volume} {71}},\
  \bibinfo {pages} {3697--3700}}\BibitemShut {NoStop}%
\bibitem [{\citenamefont {Hatsugai}(1993{\natexlab{b}})}]{Hatsugai:1993PRB}%
  \BibitemOpen
  \bibfield  {author} {\bibinfo {author} {\bibnamefont {Hatsugai},
  \bibfnamefont {Yasuhiro}}} (\bibinfo {year} {1993}{\natexlab{b}}),\ \bibfield
   {title} {\enquote {\bibinfo {title} {Edge states in the integer quantum
  {H}all effect and the {R}iemann surface of the {B}loch function},}\ }\href
  {https://link.aps.org/doi/10.1103/PhysRevB.48.11851} {\bibfield  {journal}
  {\bibinfo  {journal} {Phys. Rev. B}\ }\textbf {\bibinfo {volume} {48}},\
  \bibinfo {pages} {11851--11862}}\BibitemShut {NoStop}%
\bibitem [{\citenamefont {Hatsugai}\ and\ \citenamefont
  {Kohmoto}(1990)}]{Hatsugai:1990}%
  \BibitemOpen
  \bibfield  {author} {\bibinfo {author} {\bibnamefont {Hatsugai},
  \bibfnamefont {Yasuhiro}}, \ and\ \bibinfo {author} {\bibfnamefont {Mahito}\
  \bibnamefont {Kohmoto}}} (\bibinfo {year} {1990}),\ \bibfield  {title}
  {\enquote {\bibinfo {title} {Energy spectrum and the quantum {H}all effect on
  the square lattice with next-nearest-neighbor hopping},}\ }\href
  {https://journals.aps.org/prb/abstract/10.1103/PhysRevB.42.8282} {\bibfield
  {journal} {\bibinfo  {journal} {Phys. Rev. B}\ }\textbf {\bibinfo {volume}
  {42}}~(\bibinfo {number} {13}),\ \bibinfo {pages} {8282}}\BibitemShut
  {NoStop}%
\bibitem [{\citenamefont {Haus}\ and\ \citenamefont
  {Huang}(1991)}]{Haus:1991IEEE}%
  \BibitemOpen
  \bibfield  {author} {\bibinfo {author} {\bibnamefont {Haus}, \bibfnamefont
  {Hermann~A}}, \ and\ \bibinfo {author} {\bibfnamefont {Weiping}\ \bibnamefont
  {Huang}}} (\bibinfo {year} {1991}),\ \bibfield  {title} {\enquote {\bibinfo
  {title} {Coupled-mode theory},}\ }\href
  {http://ieeexplore.ieee.org/document/104225/} {\bibfield  {journal} {\bibinfo
   {journal} {Proceedings of the IEEE}\ }\textbf {\bibinfo {volume}
  {79}}~(\bibinfo {number} {10}),\ \bibinfo {pages} {1505--1518}}\BibitemShut
  {NoStop}%
\bibitem [{\citenamefont {Hayward}\ \emph {et~al.}(2012)\citenamefont
  {Hayward}, \citenamefont {Martin},\ and\ \citenamefont
  {Greentree}}]{Hayward:2012PRL}%
  \BibitemOpen
  \bibfield  {author} {\bibinfo {author} {\bibnamefont {Hayward}, \bibfnamefont
  {Andrew L~C}}, \bibinfo {author} {\bibfnamefont {Andrew~M.}\ \bibnamefont
  {Martin}}, \ and\ \bibinfo {author} {\bibfnamefont {Andrew~D.}\ \bibnamefont
  {Greentree}}} (\bibinfo {year} {2012}),\ \bibfield  {title} {\enquote
  {\bibinfo {title} {Fractional quantum {H}all physics in
  {J}aynes-{C}ummings-{H}ubbard lattices},}\ }\href
  {https://link.aps.org/doi/10.1103/PhysRevLett.108.223602} {\bibfield
  {journal} {\bibinfo  {journal} {Phys. Rev. Lett.}\ }\textbf {\bibinfo
  {volume} {108}},\ \bibinfo {pages} {223602}}\BibitemShut {NoStop}%
\bibitem [{\citenamefont {He}\ \emph {et~al.}(2010{\natexlab{a}})\citenamefont
  {He}, \citenamefont {Chen}, \citenamefont {Lu}, \citenamefont {Li},
  \citenamefont {Wan}, \citenamefont {Qian}, \citenamefont {Yin},\ and\
  \citenamefont {Chen}}]{He:2010JAP}%
  \BibitemOpen
  \bibfield  {author} {\bibinfo {author} {\bibnamefont {He}, \bibfnamefont
  {Cheng}}, \bibinfo {author} {\bibfnamefont {Xiao-Lin}\ \bibnamefont {Chen}},
  \bibinfo {author} {\bibfnamefont {Ming-Hui}\ \bibnamefont {Lu}}, \bibinfo
  {author} {\bibfnamefont {Xue-Feng}\ \bibnamefont {Li}}, \bibinfo {author}
  {\bibfnamefont {Wei-Wei}\ \bibnamefont {Wan}}, \bibinfo {author}
  {\bibfnamefont {Xiao-Shi}\ \bibnamefont {Qian}}, \bibinfo {author}
  {\bibfnamefont {Ruo-Cheng}\ \bibnamefont {Yin}}, \ and\ \bibinfo {author}
  {\bibfnamefont {Yan-Feng}\ \bibnamefont {Chen}}} (\bibinfo {year}
  {2010}{\natexlab{a}}),\ \bibfield  {title} {\enquote {\bibinfo {title}
  {Left-handed and right-handed one-way edge modes in a gyromagnetic photonic
  crystal},}\ }\href {https://aip.scitation.org/doi/full/10.1063/1.3374470}
  {\bibfield  {journal} {\bibinfo  {journal} {Journal of Applied Physics}\
  }\textbf {\bibinfo {volume} {107}}~(\bibinfo {number} {12}),\ \bibinfo
  {pages} {123117}}\BibitemShut {NoStop}%
\bibitem [{\citenamefont {He}\ \emph {et~al.}(2010{\natexlab{b}})\citenamefont
  {He}, \citenamefont {Chen}, \citenamefont {Lu}, \citenamefont {Li},
  \citenamefont {Wan}, \citenamefont {Qian}, \citenamefont {Yin},\ and\
  \citenamefont {Chen}}]{He:2010APL}%
  \BibitemOpen
  \bibfield  {author} {\bibinfo {author} {\bibnamefont {He}, \bibfnamefont
  {Cheng}}, \bibinfo {author} {\bibfnamefont {Xiao-Lin}\ \bibnamefont {Chen}},
  \bibinfo {author} {\bibfnamefont {Ming-Hui}\ \bibnamefont {Lu}}, \bibinfo
  {author} {\bibfnamefont {Xue-Feng}\ \bibnamefont {Li}}, \bibinfo {author}
  {\bibfnamefont {Wei-Wei}\ \bibnamefont {Wan}}, \bibinfo {author}
  {\bibfnamefont {Xiao-Shi}\ \bibnamefont {Qian}}, \bibinfo {author}
  {\bibfnamefont {Ruo-Cheng}\ \bibnamefont {Yin}}, \ and\ \bibinfo {author}
  {\bibfnamefont {Yan-Feng}\ \bibnamefont {Chen}}} (\bibinfo {year}
  {2010}{\natexlab{b}}),\ \bibfield  {title} {\enquote {\bibinfo {title}
  {Tunable one-way cross-waveguide splitter based on gyromagnetic photonic
  crystal},}\ }\href {http://aip.scitation.org/doi/10.1063/1.3358386}
  {\bibfield  {journal} {\bibinfo  {journal} {Appl. Phys. Lett.}\ }\textbf
  {\bibinfo {volume} {96}}~(\bibinfo {number} {11}),\ \bibinfo {pages}
  {111111}}\BibitemShut {NoStop}%
\bibitem [{\citenamefont {He}\ \emph {et~al.}(2016{\natexlab{a}})\citenamefont
  {He}, \citenamefont {Ni}, \citenamefont {Ge}, \citenamefont {Sun},
  \citenamefont {Chen}, \citenamefont {Lu}, \citenamefont {Liu},\ and\
  \citenamefont {Chen}}]{He:NatPhys2016}%
  \BibitemOpen
  \bibfield  {author} {\bibinfo {author} {\bibnamefont {He}, \bibfnamefont
  {Cheng}}, \bibinfo {author} {\bibfnamefont {Xu}~\bibnamefont {Ni}}, \bibinfo
  {author} {\bibfnamefont {Hao}\ \bibnamefont {Ge}}, \bibinfo {author}
  {\bibfnamefont {Xiao-Chen}\ \bibnamefont {Sun}}, \bibinfo {author}
  {\bibfnamefont {Yan-Bin}\ \bibnamefont {Chen}}, \bibinfo {author}
  {\bibfnamefont {Ming-Hui}\ \bibnamefont {Lu}}, \bibinfo {author}
  {\bibfnamefont {Xiao-Ping}\ \bibnamefont {Liu}}, \ and\ \bibinfo {author}
  {\bibfnamefont {Yan-Feng}\ \bibnamefont {Chen}}} (\bibinfo {year}
  {2016}{\natexlab{a}}),\ \bibfield  {title} {\enquote {\bibinfo {title}
  {{Acoustic topological insulator and robust one-way sound transport}},}\
  }\href {https://www.nature.com/articles/nphys3867} {\bibfield  {journal}
  {\bibinfo  {journal} {Nat. Phys.}\ }\textbf {\bibinfo {volume}
  {12}}~(\bibinfo {number} {12}),\ \bibinfo {pages} {1124--1129}}\BibitemShut
  {NoStop}%
\bibitem [{\citenamefont {He}\ \emph {et~al.}(2016{\natexlab{b}})\citenamefont
  {He}, \citenamefont {Sun}, \citenamefont {Liu}, \citenamefont {Lu},
  \citenamefont {Chen}, \citenamefont {Feng},\ and\ \citenamefont
  {Chen}}]{He:2016PNAS}%
  \BibitemOpen
  \bibfield  {author} {\bibinfo {author} {\bibnamefont {He}, \bibfnamefont
  {Cheng}}, \bibinfo {author} {\bibfnamefont {Xiao-Chen}\ \bibnamefont {Sun}},
  \bibinfo {author} {\bibfnamefont {Xiao-Ping}\ \bibnamefont {Liu}}, \bibinfo
  {author} {\bibfnamefont {Ming-Hui}\ \bibnamefont {Lu}}, \bibinfo {author}
  {\bibfnamefont {Yulin}\ \bibnamefont {Chen}}, \bibinfo {author}
  {\bibfnamefont {Liang}\ \bibnamefont {Feng}}, \ and\ \bibinfo {author}
  {\bibfnamefont {Yan-Feng}\ \bibnamefont {Chen}}} (\bibinfo {year}
  {2016}{\natexlab{b}}),\ \bibfield  {title} {\enquote {\bibinfo {title}
  {Photonic topological insulator with broken time-reversal symmetry},}\ }\href
  {https://www.pnas.org/content/113/18/4924} {\bibfield  {journal} {\bibinfo
  {journal} {Proc. Natl. Acad. Scie. U.S.A.}\ }\textbf {\bibinfo {volume}
  {113}}~(\bibinfo {number} {18}),\ \bibinfo {pages} {4924--4928}}\BibitemShut
  {NoStop}%
\bibitem [{\citenamefont {Heisenberg}\ and\ \citenamefont
  {Euler}(1936)}]{Heisenberg:Zeit1936}%
  \BibitemOpen
  \bibfield  {author} {\bibinfo {author} {\bibnamefont {Heisenberg},
  \bibfnamefont {W}}, \ and\ \bibinfo {author} {\bibfnamefont {H}~\bibnamefont
  {Euler}}} (\bibinfo {year} {1936}),\ \bibfield  {title} {\enquote {\bibinfo
  {title} {Folgerungen aus der diracschen theorie des positrons},}\ }\href
  {https://link.springer.com/article/10.1007/BF01343663} {\bibfield  {journal}
  {\bibinfo  {journal} {Zeitschrift f{\"u}r Physik A Hadrons and Nuclei}\
  }\textbf {\bibinfo {volume} {98}}~(\bibinfo {number} {11}),\ \bibinfo {pages}
  {714--732}}\BibitemShut {NoStop}%
\bibitem [{\citenamefont {Hoffman}\ \emph {et~al.}(2011)\citenamefont
  {Hoffman}, \citenamefont {Srinivasan}, \citenamefont {Schmidt}, \citenamefont
  {Spietz}, \citenamefont {Aumentado}, \citenamefont {T\"ureci},\ and\
  \citenamefont {Houck}}]{Hoffman:PRL2011}%
  \BibitemOpen
  \bibfield  {author} {\bibinfo {author} {\bibnamefont {Hoffman}, \bibfnamefont
  {A~J}}, \bibinfo {author} {\bibfnamefont {S.~J.}\ \bibnamefont {Srinivasan}},
  \bibinfo {author} {\bibfnamefont {S.}~\bibnamefont {Schmidt}}, \bibinfo
  {author} {\bibfnamefont {L.}~\bibnamefont {Spietz}}, \bibinfo {author}
  {\bibfnamefont {J.}~\bibnamefont {Aumentado}}, \bibinfo {author}
  {\bibfnamefont {H.~E.}\ \bibnamefont {T\"ureci}}, \ and\ \bibinfo {author}
  {\bibfnamefont {A.~A.}\ \bibnamefont {Houck}}} (\bibinfo {year} {2011}),\
  \bibfield  {title} {\enquote {\bibinfo {title} {Dispersive photon blockade in
  a superconducting circuit},}\ }\href
  {https://link.aps.org/doi/10.1103/PhysRevLett.107.053602} {\bibfield
  {journal} {\bibinfo  {journal} {Phys. Rev. Lett.}\ }\textbf {\bibinfo
  {volume} {107}},\ \bibinfo {pages} {053602}}\BibitemShut {NoStop}%
\bibitem [{\citenamefont {Hofstadter}(1976)}]{Hofstadter:1976PRB}%
  \BibitemOpen
  \bibfield  {author} {\bibinfo {author} {\bibnamefont {Hofstadter},
  \bibfnamefont {Douglas~R}}} (\bibinfo {year} {1976}),\ \bibfield  {title}
  {\enquote {\bibinfo {title} {Energy levels and wave functions of {B}loch
  electrons in rational and irrational magnetic fields},}\ }\href
  {https://link.aps.org/doi/10.1103/PhysRevB.14.2239} {\bibfield  {journal}
  {\bibinfo  {journal} {Phys. Rev. B}\ }\textbf {\bibinfo {volume} {14}},\
  \bibinfo {pages} {2239--2249}}\BibitemShut {NoStop}%
\bibitem [{\citenamefont {Houck}\ \emph {et~al.}(2012)\citenamefont {Houck},
  \citenamefont {T{\"u}reci},\ and\ \citenamefont {Koch}}]{Houck:NatPHys2012}%
  \BibitemOpen
  \bibfield  {author} {\bibinfo {author} {\bibnamefont {Houck}, \bibfnamefont
  {Andrew~A}}, \bibinfo {author} {\bibfnamefont {Hakan~E}\ \bibnamefont
  {T{\"u}reci}}, \ and\ \bibinfo {author} {\bibfnamefont {Jens}\ \bibnamefont
  {Koch}}} (\bibinfo {year} {2012}),\ \bibfield  {title} {\enquote {\bibinfo
  {title} {On-chip quantum simulation with superconducting circuits},}\ }\href
  {https://www.nature.com/articles/nphys2251} {\bibfield  {journal} {\bibinfo
  {journal} {Nat. Phys.}\ }\textbf {\bibinfo {volume} {8}}~(\bibinfo {number}
  {4}),\ \bibinfo {pages} {292--299}}\BibitemShut {NoStop}%
\bibitem [{\citenamefont {Hsu}\ \emph {et~al.}(2016)\citenamefont {Hsu},
  \citenamefont {Zhen}, \citenamefont {Stone}, \citenamefont {Joannopoulos},\
  and\ \citenamefont {Solja{\v{c}}i{\'c}}}]{Hsu:2016NatRev}%
  \BibitemOpen
  \bibfield  {author} {\bibinfo {author} {\bibnamefont {Hsu}, \bibfnamefont
  {Chia~Wei}}, \bibinfo {author} {\bibfnamefont {Bo}~\bibnamefont {Zhen}},
  \bibinfo {author} {\bibfnamefont {A~Douglas}\ \bibnamefont {Stone}}, \bibinfo
  {author} {\bibfnamefont {John~D}\ \bibnamefont {Joannopoulos}}, \ and\
  \bibinfo {author} {\bibfnamefont {Marin}\ \bibnamefont {Solja{\v{c}}i{\'c}}}}
  (\bibinfo {year} {2016}),\ \bibfield  {title} {\enquote {\bibinfo {title}
  {Bound states in the continuum},}\ }\href
  {http://www.nature.com/articles/natrevmats201648} {\bibfield  {journal}
  {\bibinfo  {journal} {Nature Reviews Materials}\ }\textbf {\bibinfo {volume}
  {1}},\ \bibinfo {pages} {16048}}\BibitemShut {NoStop}%
\bibitem [{\citenamefont {Hu}\ \emph {et~al.}(2015)\citenamefont {Hu},
  \citenamefont {Pillay}, \citenamefont {Wu}, \citenamefont {Pasek},
  \citenamefont {Shum},\ and\ \citenamefont {Chong}}]{Hu:2015PRX}%
  \BibitemOpen
  \bibfield  {author} {\bibinfo {author} {\bibnamefont {Hu}, \bibfnamefont
  {Wenchao}}, \bibinfo {author} {\bibfnamefont {Jason~C.}\ \bibnamefont
  {Pillay}}, \bibinfo {author} {\bibfnamefont {Kan}\ \bibnamefont {Wu}},
  \bibinfo {author} {\bibfnamefont {Michael}\ \bibnamefont {Pasek}}, \bibinfo
  {author} {\bibfnamefont {Perry~Ping}\ \bibnamefont {Shum}}, \ and\ \bibinfo
  {author} {\bibfnamefont {Y.~D.}\ \bibnamefont {Chong}}} (\bibinfo {year}
  {2015}),\ \bibfield  {title} {\enquote {\bibinfo {title} {Measurement of a
  topological edge invariant in a microwave network},}\ }\href
  {http://link.aps.org/doi/10.1103/PhysRevX.5.011012} {\bibfield  {journal}
  {\bibinfo  {journal} {Phys. Rev. X}\ }\textbf {\bibinfo {volume} {5}},\
  \bibinfo {pages} {011012}}\BibitemShut {NoStop}%
\bibitem [{\citenamefont {Hu}\ and\ \citenamefont
  {Hughes}(2011)}]{hu2011absence}%
  \BibitemOpen
  \bibfield  {author} {\bibinfo {author} {\bibnamefont {Hu}, \bibfnamefont
  {Yi~Chen}}, \ and\ \bibinfo {author} {\bibfnamefont {Taylor~L}\ \bibnamefont
  {Hughes}}} (\bibinfo {year} {2011}),\ \bibfield  {title} {\enquote {\bibinfo
  {title} {Absence of topological insulator phases in non-{H}ermitian
  $\mathcal{PT}$-symmetric {H}amiltonians},}\ }\href
  {https://journals.aps.org/prb/abstract/10.1103/PhysRevB.84.153101} {\bibfield
   {journal} {\bibinfo  {journal} {Phys. Rev. B}\ }\textbf {\bibinfo {volume}
  {84}}~(\bibinfo {number} {15}),\ \bibinfo {pages} {153101}}\BibitemShut
  {NoStop}%
\bibitem [{\citenamefont {Hua}\ \emph {et~al.}(2016)\citenamefont {Hua},
  \citenamefont {Wen}, \citenamefont {Jiang}, \citenamefont {Hua},
  \citenamefont {Jiang},\ and\ \citenamefont {Xiao}}]{Hua:NatComm2015}%
  \BibitemOpen
  \bibfield  {author} {\bibinfo {author} {\bibnamefont {Hua}, \bibfnamefont
  {Shiyue}}, \bibinfo {author} {\bibfnamefont {Jianming}\ \bibnamefont {Wen}},
  \bibinfo {author} {\bibfnamefont {Xiaoshun}\ \bibnamefont {Jiang}}, \bibinfo
  {author} {\bibfnamefont {Qian}\ \bibnamefont {Hua}}, \bibinfo {author}
  {\bibfnamefont {Liang}\ \bibnamefont {Jiang}}, \ and\ \bibinfo {author}
  {\bibfnamefont {Min}\ \bibnamefont {Xiao}}} (\bibinfo {year} {2016}),\
  \bibfield  {title} {\enquote {\bibinfo {title} {Demonstration of a chip-based
  optical isolator with parametric amplification},}\ }\href
  {https://www.nature.com/articles/ncomms13657} {\bibfield  {journal} {\bibinfo
   {journal} {Nat. Commun.}\ }\textbf {\bibinfo {volume} {7}},\ \bibinfo
  {pages} {13657}}\BibitemShut {NoStop}%
\bibitem [{\citenamefont {Huang}(1987)}]{Huang}%
  \BibitemOpen
  \bibfield  {author} {\bibinfo {author} {\bibnamefont {Huang}, \bibfnamefont
  {K}}} (\bibinfo {year} {1987}),\ \href@noop {} {\emph {\bibinfo {title}
  {Statistical Mechanics}}}\ (\bibinfo  {publisher} {Wiley},\ \bibinfo
  {address} {New York})\BibitemShut {NoStop}%
\bibitem [{\citenamefont {Huang}\ \emph {et~al.}(2011)\citenamefont {Huang},
  \citenamefont {Lai}, \citenamefont {Hang}, \citenamefont {Zheng},\ and\
  \citenamefont {Chan}}]{Huang:2011NatMat}%
  \BibitemOpen
  \bibfield  {author} {\bibinfo {author} {\bibnamefont {Huang}, \bibfnamefont
  {Xueqin}}, \bibinfo {author} {\bibfnamefont {Yun}\ \bibnamefont {Lai}},
  \bibinfo {author} {\bibfnamefont {Zhi~Hong}\ \bibnamefont {Hang}}, \bibinfo
  {author} {\bibfnamefont {Huihuo}\ \bibnamefont {Zheng}}, \ and\ \bibinfo
  {author} {\bibfnamefont {CT}~\bibnamefont {Chan}}} (\bibinfo {year} {2011}),\
  \bibfield  {title} {\enquote {\bibinfo {title} {Dirac cones induced by
  accidental degeneracy in photonic crystals and zero-refractive-index
  materials},}\ }\href {https://www.nature.com/articles/nmat3030} {\bibfield
  {journal} {\bibinfo  {journal} {Nat. Mater.}\ }\textbf {\bibinfo {volume}
  {10}}~(\bibinfo {number} {8}),\ \bibinfo {pages} {582--586}}\BibitemShut
  {NoStop}%
\bibitem [{\citenamefont {Huber}(2016)}]{Huber:2016NatPhys}%
  \BibitemOpen
  \bibfield  {author} {\bibinfo {author} {\bibnamefont {Huber}, \bibfnamefont
  {Sebastian~D}}} (\bibinfo {year} {2016}),\ \bibfield  {title} {\enquote
  {\bibinfo {title} {Topological mechanics},}\ }\href
  {https://www.nature.com/articles/nphys3801} {\bibfield  {journal} {\bibinfo
  {journal} {Nat. Phys.}\ }\textbf {\bibinfo {volume} {12}}~(\bibinfo {number}
  {7}),\ \bibinfo {pages} {621--623}}\BibitemShut {NoStop}%
\bibitem [{\citenamefont {Iadecola}\ \emph {et~al.}(2016)\citenamefont
  {Iadecola}, \citenamefont {Schuster},\ and\ \citenamefont
  {Chamon}}]{Iadecola:2016PRL}%
  \BibitemOpen
  \bibfield  {author} {\bibinfo {author} {\bibnamefont {Iadecola},
  \bibfnamefont {Thomas}}, \bibinfo {author} {\bibfnamefont {Thomas}\
  \bibnamefont {Schuster}}, \ and\ \bibinfo {author} {\bibfnamefont {Claudio}\
  \bibnamefont {Chamon}}} (\bibinfo {year} {2016}),\ \bibfield  {title}
  {\enquote {\bibinfo {title} {Non-{A}belian braiding of light},}\ }\href
  {https://link.aps.org/doi/10.1103/PhysRevLett.117.073901} {\bibfield
  {journal} {\bibinfo  {journal} {Phys. Rev. Lett.}\ }\textbf {\bibinfo
  {volume} {117}},\ \bibinfo {pages} {073901}}\BibitemShut {NoStop}%
\bibitem [{\citenamefont {Imamoglu}\ \emph {et~al.}(1997)\citenamefont
  {Imamoglu}, \citenamefont {Schmidt}, \citenamefont {Woods},\ and\
  \citenamefont {Deutsch}}]{Imamoglu:PRL97}%
  \BibitemOpen
  \bibfield  {author} {\bibinfo {author} {\bibnamefont {Imamoglu},
  \bibfnamefont {A}}, \bibinfo {author} {\bibfnamefont {H}~\bibnamefont
  {Schmidt}}, \bibinfo {author} {\bibfnamefont {G}~\bibnamefont {Woods}}, \
  and\ \bibinfo {author} {\bibfnamefont {M}~\bibnamefont {Deutsch}}} (\bibinfo
  {year} {1997}),\ \bibfield  {title} {\enquote {\bibinfo {title} {Strongly
  interacting photons in a nonlinear cavity},}\ }\href
  {https://journals.aps.org/prl/abstract/10.1103/PhysRevLett.79.1467}
  {\bibfield  {journal} {\bibinfo  {journal} {Phys. Rev. Lett.}\ }\textbf
  {\bibinfo {volume} {79}}~(\bibinfo {number} {8}),\ \bibinfo {pages}
  {1467--1470}}\BibitemShut {NoStop}%
\bibitem [{\citenamefont {Imhof}\ \emph {et~al.}(2018)\citenamefont {Imhof},
  \citenamefont {Berger}, \citenamefont {Bayer}, \citenamefont {Brehm},
  \citenamefont {Molenkamp}, \citenamefont {Kiessling}, \citenamefont
  {Schindler}, \citenamefont {Lee}, \citenamefont {Greiter}, \citenamefont
  {Neupert} \emph {et~al.}}]{Imhof:2018NatPhys}%
  \BibitemOpen
  \bibfield  {author} {\bibinfo {author} {\bibnamefont {Imhof}, \bibfnamefont
  {Stefan}}, \bibinfo {author} {\bibfnamefont {Christian}\ \bibnamefont
  {Berger}}, \bibinfo {author} {\bibfnamefont {Florian}\ \bibnamefont {Bayer}},
  \bibinfo {author} {\bibfnamefont {Johannes}\ \bibnamefont {Brehm}}, \bibinfo
  {author} {\bibfnamefont {Laurens~W}\ \bibnamefont {Molenkamp}}, \bibinfo
  {author} {\bibfnamefont {Tobias}\ \bibnamefont {Kiessling}}, \bibinfo
  {author} {\bibfnamefont {Frank}\ \bibnamefont {Schindler}}, \bibinfo {author}
  {\bibfnamefont {Ching~Hua}\ \bibnamefont {Lee}}, \bibinfo {author}
  {\bibfnamefont {Martin}\ \bibnamefont {Greiter}}, \bibinfo {author}
  {\bibfnamefont {Titus}\ \bibnamefont {Neupert}},  \emph {et~al.}} (\bibinfo
  {year} {2018}),\ \bibfield  {title} {\enquote {\bibinfo {title}
  {Topolectrical-circuit realization of topological corner modes},}\ }\href
  {https://www.nature.com/articles/s41567-018-0246-1} {\bibfield  {journal}
  {\bibinfo  {journal} {Nat. Phys.}\ }\textbf {\bibinfo {volume}
  {14}}~(\bibinfo {number} {9}),\ \bibinfo {pages} {925}}\BibitemShut {NoStop}%
\bibitem [{\citenamefont {Inoue}\ and\ \citenamefont
  {Tanaka}(2010)}]{Inoue:PRL2010}%
  \BibitemOpen
  \bibfield  {author} {\bibinfo {author} {\bibnamefont {Inoue}, \bibfnamefont
  {Jun-ichi}}, \ and\ \bibinfo {author} {\bibfnamefont {Akihiro}\ \bibnamefont
  {Tanaka}}} (\bibinfo {year} {2010}),\ \bibfield  {title} {\enquote {\bibinfo
  {title} {Photoinduced transition between conventional and topological
  insulators in two-dimensional electronic systems},}\ }\href
  {https://link.aps.org/doi/10.1103/PhysRevLett.105.017401} {\bibfield
  {journal} {\bibinfo  {journal} {Phys. Rev. Lett.}\ }\textbf {\bibinfo
  {volume} {105}},\ \bibinfo {pages} {017401}}\BibitemShut {NoStop}%
\bibitem [{\citenamefont {Ivanov}\ \emph {et~al.}(2018)\citenamefont {Ivanov},
  \citenamefont {Letscher}, \citenamefont {Simon},\ and\ \citenamefont
  {Fleischhauer}}]{Ivanov:arxiv2018}%
  \BibitemOpen
  \bibfield  {author} {\bibinfo {author} {\bibnamefont {Ivanov}, \bibfnamefont
  {Peter~A}}, \bibinfo {author} {\bibfnamefont {Fabian}\ \bibnamefont
  {Letscher}}, \bibinfo {author} {\bibfnamefont {Jonathan}\ \bibnamefont
  {Simon}}, \ and\ \bibinfo {author} {\bibfnamefont {Michael}\ \bibnamefont
  {Fleischhauer}}} (\bibinfo {year} {2018}),\ \bibfield  {title} {\enquote
  {\bibinfo {title} {Adiabatic flux insertion and growing of {L}aughlin states
  of cavity {R}ydberg polaritons},}\ }\href
  {https://link.aps.org/doi/10.1103/PhysRevA.98.013847} {\bibfield  {journal}
  {\bibinfo  {journal} {Phys. Rev. A}\ }\textbf {\bibinfo {volume} {98}},\
  \bibinfo {pages} {013847}}\BibitemShut {NoStop}%
\bibitem [{\citenamefont {Jachymski}\ \emph {et~al.}(2016)\citenamefont
  {Jachymski}, \citenamefont {Bienias},\ and\ \citenamefont
  {B{\"u}chler}}]{Jachymski:PRL2016}%
  \BibitemOpen
  \bibfield  {author} {\bibinfo {author} {\bibnamefont {Jachymski},
  \bibfnamefont {Krzysztof}}, \bibinfo {author} {\bibfnamefont {Przemys{\l}aw}\
  \bibnamefont {Bienias}}, \ and\ \bibinfo {author} {\bibfnamefont
  {Hans~Peter}\ \bibnamefont {B{\"u}chler}}} (\bibinfo {year} {2016}),\
  \bibfield  {title} {\enquote {\bibinfo {title} {Three-body interaction of
  {R}ydberg slow-light polaritons},}\ }\href@noop {} {\bibfield  {journal}
  {\bibinfo  {journal} {Phys. Rev. Lett.}\ }\textbf {\bibinfo {volume}
  {117}}~(\bibinfo {number} {5}),\ \bibinfo {pages} {053601}}\BibitemShut
  {NoStop}%
\bibitem [{\citenamefont {Jackiw}\ and\ \citenamefont
  {Rebbi}(1976)}]{Jackiw:1976PRD}%
  \BibitemOpen
  \bibfield  {author} {\bibinfo {author} {\bibnamefont {Jackiw}, \bibfnamefont
  {R}}, \ and\ \bibinfo {author} {\bibfnamefont {C.}~\bibnamefont {Rebbi}}}
  (\bibinfo {year} {1976}),\ \bibfield  {title} {\enquote {\bibinfo {title}
  {Solitons with fermion number 1/2},}\ }\href
  {https://link.aps.org/doi/10.1103/PhysRevD.13.3398} {\bibfield  {journal}
  {\bibinfo  {journal} {Phys. Rev. D}\ }\textbf {\bibinfo {volume} {13}},\
  \bibinfo {pages} {3398--3409}}\BibitemShut {NoStop}%
\bibitem [{\citenamefont {Jacobs}\ \emph {et~al.}(2015)\citenamefont {Jacobs},
  \citenamefont {Miroshnichenko}, \citenamefont {Kivshar},\ and\ \citenamefont
  {Khanikaev}}]{Jacobs:2015NJP}%
  \BibitemOpen
  \bibfield  {author} {\bibinfo {author} {\bibnamefont {Jacobs}, \bibfnamefont
  {Daniel~A}}, \bibinfo {author} {\bibfnamefont {Andrey~E}\ \bibnamefont
  {Miroshnichenko}}, \bibinfo {author} {\bibfnamefont {Yuri~S}\ \bibnamefont
  {Kivshar}}, \ and\ \bibinfo {author} {\bibfnamefont {Alexander~B}\
  \bibnamefont {Khanikaev}}} (\bibinfo {year} {2015}),\ \bibfield  {title}
  {\enquote {\bibinfo {title} {Photonic topological {C}hern insulators based on
  {T}ellegen metacrystals},}\ }\href
  {https://iopscience.iop.org/article/10.1088/1367-2630/17/12/125015/meta}
  {\bibfield  {journal} {\bibinfo  {journal} {New J. Phys.}\ }\textbf {\bibinfo
  {volume} {17}}~(\bibinfo {number} {12}),\ \bibinfo {pages}
  {125015}}\BibitemShut {NoStop}%
\bibitem [{\citenamefont {Jacqmin}\ \emph {et~al.}(2014)\citenamefont
  {Jacqmin}, \citenamefont {Carusotto}, \citenamefont {Sagnes}, \citenamefont
  {Abbarchi}, \citenamefont {Solnyshkov}, \citenamefont {Malpuech},
  \citenamefont {Galopin}, \citenamefont {Lema\^{\i}tre}, \citenamefont
  {Bloch},\ and\ \citenamefont {Amo}}]{Jacqmin:2014PRL}%
  \BibitemOpen
  \bibfield  {author} {\bibinfo {author} {\bibnamefont {Jacqmin}, \bibfnamefont
  {T}}, \bibinfo {author} {\bibfnamefont {I.}~\bibnamefont {Carusotto}},
  \bibinfo {author} {\bibfnamefont {I.}~\bibnamefont {Sagnes}}, \bibinfo
  {author} {\bibfnamefont {M.}~\bibnamefont {Abbarchi}}, \bibinfo {author}
  {\bibfnamefont {D.~D.}\ \bibnamefont {Solnyshkov}}, \bibinfo {author}
  {\bibfnamefont {G.}~\bibnamefont {Malpuech}}, \bibinfo {author}
  {\bibfnamefont {E.}~\bibnamefont {Galopin}}, \bibinfo {author} {\bibfnamefont
  {A.}~\bibnamefont {Lema\^{\i}tre}}, \bibinfo {author} {\bibfnamefont
  {J.}~\bibnamefont {Bloch}}, \ and\ \bibinfo {author} {\bibfnamefont
  {A.}~\bibnamefont {Amo}}} (\bibinfo {year} {2014}),\ \bibfield  {title}
  {\enquote {\bibinfo {title} {Direct observation of {D}irac cones and a
  flatband in a honeycomb lattice for polaritons},}\ }\href
  {http://link.aps.org/doi/10.1103/PhysRevLett.112.116402} {\bibfield
  {journal} {\bibinfo  {journal} {Phys. Rev. Lett.}\ }\textbf {\bibinfo
  {volume} {112}},\ \bibinfo {pages} {116402}}\BibitemShut {NoStop}%
\bibitem [{\citenamefont {Jaksch}\ and\ \citenamefont
  {Zoller}(2003)}]{Jaksch:2003NJP}%
  \BibitemOpen
  \bibfield  {author} {\bibinfo {author} {\bibnamefont {Jaksch}, \bibfnamefont
  {D}}, \ and\ \bibinfo {author} {\bibfnamefont {P}~\bibnamefont {Zoller}}}
  (\bibinfo {year} {2003}),\ \bibfield  {title} {\enquote {\bibinfo {title}
  {Creation of effective magnetic fields in optical lattices: the {H}ofstadter
  butterfly for cold neutral atoms},}\ }\href
  {http://stacks.iop.org/1367-2630/5/i=1/a=356} {\bibfield  {journal} {\bibinfo
   {journal} {New J. Phys.}\ }\textbf {\bibinfo {volume} {5}}~(\bibinfo
  {number} {1}),\ \bibinfo {pages} {56}}\BibitemShut {NoStop}%
\bibitem [{\citenamefont {Jalas}\ \emph {et~al.}(2013)\citenamefont {Jalas},
  \citenamefont {Petrov}, \citenamefont {Eich}, \citenamefont {Freude},
  \citenamefont {Fan}, \citenamefont {Yu}, \citenamefont {Baets}, \citenamefont
  {Popovi{\'c}}, \citenamefont {Melloni}, \citenamefont {Joannopoulos},
  \citenamefont {Vanwolleghem}, \citenamefont {Doerr},\ and\ \citenamefont
  {Renner}}]{Jalas:NatPhot2013}%
  \BibitemOpen
  \bibfield  {author} {\bibinfo {author} {\bibnamefont {Jalas}, \bibfnamefont
  {Dirk}}, \bibinfo {author} {\bibfnamefont {Alexander}\ \bibnamefont
  {Petrov}}, \bibinfo {author} {\bibfnamefont {Manfred}\ \bibnamefont {Eich}},
  \bibinfo {author} {\bibfnamefont {Wolfgang}\ \bibnamefont {Freude}}, \bibinfo
  {author} {\bibfnamefont {Shanhui}\ \bibnamefont {Fan}}, \bibinfo {author}
  {\bibfnamefont {Zongfu}\ \bibnamefont {Yu}}, \bibinfo {author} {\bibfnamefont
  {Roel}\ \bibnamefont {Baets}}, \bibinfo {author} {\bibfnamefont
  {Milo{\v{s}}}\ \bibnamefont {Popovi{\'c}}}, \bibinfo {author} {\bibfnamefont
  {Andrea}\ \bibnamefont {Melloni}}, \bibinfo {author} {\bibfnamefont {John~D}\
  \bibnamefont {Joannopoulos}}, \bibinfo {author} {\bibfnamefont {Mathias}\
  \bibnamefont {Vanwolleghem}}, \bibinfo {author} {\bibfnamefont
  {Christopher~R.}\ \bibnamefont {Doerr}}, \ and\ \bibinfo {author}
  {\bibfnamefont {Hagen}\ \bibnamefont {Renner}}} (\bibinfo {year} {2013}),\
  \bibfield  {title} {\enquote {\bibinfo {title} {What is -- and what is not --
  an optical isolator},}\ }\href
  {https://www.nature.com/articles/nphoton.2013.185} {\bibfield  {journal}
  {\bibinfo  {journal} {Nat. Photonics}\ }\textbf {\bibinfo {volume}
  {7}}~(\bibinfo {number} {8}),\ \bibinfo {pages} {579--582}}\BibitemShut
  {NoStop}%
\bibitem [{\citenamefont {Jia}\ \emph {et~al.}(2018{\natexlab{a}})\citenamefont
  {Jia}, \citenamefont {Schine}, \citenamefont {Georgakopoulos}, \citenamefont
  {Ryou}, \citenamefont {Clark}, \citenamefont {Sommer},\ and\ \citenamefont
  {Simon}}]{Jia:arXiv2017}%
  \BibitemOpen
  \bibfield  {author} {\bibinfo {author} {\bibnamefont {Jia}, \bibfnamefont
  {Ningyuan}}, \bibinfo {author} {\bibfnamefont {Nathan}\ \bibnamefont
  {Schine}}, \bibinfo {author} {\bibfnamefont {Alexandros}\ \bibnamefont
  {Georgakopoulos}}, \bibinfo {author} {\bibfnamefont {Albert}\ \bibnamefont
  {Ryou}}, \bibinfo {author} {\bibfnamefont {Logan~W}\ \bibnamefont {Clark}},
  \bibinfo {author} {\bibfnamefont {Ariel}\ \bibnamefont {Sommer}}, \ and\
  \bibinfo {author} {\bibfnamefont {Jonathan}\ \bibnamefont {Simon}}} (\bibinfo
  {year} {2018}{\natexlab{a}}),\ \bibfield  {title} {\enquote {\bibinfo {title}
  {A strongly interacting polaritonic quantum dot},}\ }\href
  {https://www.nature.com/articles/s41567-018-0071-6} {\bibfield  {journal}
  {\bibinfo  {journal} {Nat. Phys.}\ }\textbf {\bibinfo {volume}
  {14}}~(\bibinfo {number} {6}),\ \bibinfo {pages} {550}}\BibitemShut {NoStop}%
\bibitem [{\citenamefont {Jia}\ \emph {et~al.}(2018{\natexlab{b}})\citenamefont
  {Jia}, \citenamefont {Schine}, \citenamefont {Georgakopoulos}, \citenamefont
  {Ryou}, \citenamefont {Sommer},\ and\ \citenamefont
  {Simon}}]{ningyuan2017photons}%
  \BibitemOpen
  \bibfield  {author} {\bibinfo {author} {\bibnamefont {Jia}, \bibfnamefont
  {Ningyuan}}, \bibinfo {author} {\bibfnamefont {Nathan}\ \bibnamefont
  {Schine}}, \bibinfo {author} {\bibfnamefont {Alexandros}\ \bibnamefont
  {Georgakopoulos}}, \bibinfo {author} {\bibfnamefont {Albert}\ \bibnamefont
  {Ryou}}, \bibinfo {author} {\bibfnamefont {Ariel}\ \bibnamefont {Sommer}}, \
  and\ \bibinfo {author} {\bibfnamefont {Jonathan}\ \bibnamefont {Simon}}}
  (\bibinfo {year} {2018}{\natexlab{b}}),\ \bibfield  {title} {\enquote
  {\bibinfo {title} {Photons and polaritons in a broken-time-reversal nonplanar
  resonator},}\ }\href {https://link.aps.org/doi/10.1103/PhysRevA.97.013802}
  {\bibfield  {journal} {\bibinfo  {journal} {Phys. Rev. A}\ }\textbf {\bibinfo
  {volume} {97}},\ \bibinfo {pages} {013802}}\BibitemShut {NoStop}%
\bibitem [{\citenamefont {Jia}\ and\ \citenamefont
  {Fleischer}(2009)}]{Jia:PRA2009}%
  \BibitemOpen
  \bibfield  {author} {\bibinfo {author} {\bibnamefont {Jia}, \bibfnamefont
  {Shu}}, \ and\ \bibinfo {author} {\bibfnamefont {Jason~W}\ \bibnamefont
  {Fleischer}}} (\bibinfo {year} {2009}),\ \bibfield  {title} {\enquote
  {\bibinfo {title} {Nonlinear light propagation in rotating waveguide
  arrays},}\ }\href
  {https://www.osapublishing.org/abstract.cfm?uri=FiO-2008-FThD6} {\bibfield
  {journal} {\bibinfo  {journal} {Phys. Rev. A}\ }\textbf {\bibinfo {volume}
  {79}}~(\bibinfo {number} {4}),\ \bibinfo {pages} {041804}}\BibitemShut
  {NoStop}%
\bibitem [{\citenamefont {Jin}\ \emph {et~al.}(2016)\citenamefont {Jin},
  \citenamefont {Lu}, \citenamefont {Wang}, \citenamefont {Fang}, \citenamefont
  {Joannopoulos}, \citenamefont {Solja{\v{c}}i{\'c}}, \citenamefont {Fu},\ and\
  \citenamefont {Fang}}]{Jin:2016NatComm}%
  \BibitemOpen
  \bibfield  {author} {\bibinfo {author} {\bibnamefont {Jin}, \bibfnamefont
  {Dafei}}, \bibinfo {author} {\bibfnamefont {Ling}\ \bibnamefont {Lu}},
  \bibinfo {author} {\bibfnamefont {Zhong}\ \bibnamefont {Wang}}, \bibinfo
  {author} {\bibfnamefont {Chen}\ \bibnamefont {Fang}}, \bibinfo {author}
  {\bibfnamefont {John~D}\ \bibnamefont {Joannopoulos}}, \bibinfo {author}
  {\bibfnamefont {Marin}\ \bibnamefont {Solja{\v{c}}i{\'c}}}, \bibinfo {author}
  {\bibfnamefont {Liang}\ \bibnamefont {Fu}}, \ and\ \bibinfo {author}
  {\bibfnamefont {Nicholas~X}\ \bibnamefont {Fang}}} (\bibinfo {year} {2016}),\
  \bibfield  {title} {\enquote {\bibinfo {title} {Topological
  magnetoplasmon},}\ }\href {https://www.nature.com/articles/ncomms13486}
  {\bibfield  {journal} {\bibinfo  {journal} {Nat. Commun.}\ }\textbf {\bibinfo
  {volume} {7}},\ \bibinfo {pages} {13486}}\BibitemShut {NoStop}%
\bibitem [{\citenamefont {Jitomirskaya}\ and\ \citenamefont
  {Marx}(2012)}]{Jitomirskaya:2012}%
  \BibitemOpen
  \bibfield  {author} {\bibinfo {author} {\bibnamefont {Jitomirskaya},
  \bibfnamefont {Svetlana}}, \ and\ \bibinfo {author} {\bibfnamefont
  {CA}~\bibnamefont {Marx}}} (\bibinfo {year} {2012}),\ \bibfield  {title}
  {\enquote {\bibinfo {title} {Analytic quasi-perodic cocycles with
  singularities and the {L}yapunov exponent of extended {H}arper’s model},}\
  }\href {https://link.springer.com/article/10.1007/s00220-012-1465-4}
  {\bibfield  {journal} {\bibinfo  {journal} {Commun. Math. Phys.}\ }\textbf
  {\bibinfo {volume} {316}}~(\bibinfo {number} {1}),\ \bibinfo {pages}
  {237--267}}\BibitemShut {NoStop}%
\bibitem [{\citenamefont {Joannopoulos}\ \emph {et~al.}(2011)\citenamefont
  {Joannopoulos}, \citenamefont {Johnson}, \citenamefont {Winn},\ and\
  \citenamefont {Meade}}]{Joannopoulos:2011book}%
  \BibitemOpen
  \bibfield  {author} {\bibinfo {author} {\bibnamefont {Joannopoulos},
  \bibfnamefont {John~D}}, \bibinfo {author} {\bibfnamefont {Steven~G}\
  \bibnamefont {Johnson}}, \bibinfo {author} {\bibfnamefont {Joshua~N}\
  \bibnamefont {Winn}}, \ and\ \bibinfo {author} {\bibfnamefont {Robert~D}\
  \bibnamefont {Meade}}} (\bibinfo {year} {2011}),\ \href@noop {} {\emph
  {\bibinfo {title} {Photonic crystals: {M}olding the flow of light}}}\
  (\bibinfo  {publisher} {Princeton university press},\ \bibinfo {address}
  {Princeton, NJ})\BibitemShut {NoStop}%
\bibitem [{\citenamefont {John}(1987)}]{John:1987PRL}%
  \BibitemOpen
  \bibfield  {author} {\bibinfo {author} {\bibnamefont {John}, \bibfnamefont
  {Sajeev}}} (\bibinfo {year} {1987}),\ \bibfield  {title} {\enquote {\bibinfo
  {title} {Strong localization of photons in certain disordered dielectric
  superlattices},}\ }\href
  {https://journals.aps.org/prl/abstract/10.1103/PhysRevLett.58.2486}
  {\bibfield  {journal} {\bibinfo  {journal} {Phys. Rev. Lett.}\ }\textbf
  {\bibinfo {volume} {58}},\ \bibinfo {pages} {2486--2489}}\BibitemShut
  {NoStop}%
\bibitem [{\citenamefont {J{\"o}rg}\ \emph {et~al.}(2017)\citenamefont
  {J{\"o}rg}, \citenamefont {Letscher}, \citenamefont {Fleischhauer},\ and\
  \citenamefont {von Freymann}}]{jorg2017dynamic}%
  \BibitemOpen
  \bibfield  {author} {\bibinfo {author} {\bibnamefont {J{\"o}rg},
  \bibfnamefont {Christina}}, \bibinfo {author} {\bibfnamefont {Fabian}\
  \bibnamefont {Letscher}}, \bibinfo {author} {\bibfnamefont {Michael}\
  \bibnamefont {Fleischhauer}}, \ and\ \bibinfo {author} {\bibfnamefont
  {Georg}\ \bibnamefont {von Freymann}}} (\bibinfo {year} {2017}),\ \bibfield
  {title} {\enquote {\bibinfo {title} {Dynamic defects in photonic {F}loquet
  topological insulators},}\ }\href
  {http://iopscience.iop.org/article/10.1088/1367-2630/aa7c82/meta} {\bibfield
  {journal} {\bibinfo  {journal} {New J. Phys.}\ }\textbf {\bibinfo {volume}
  {19}}~(\bibinfo {number} {8}),\ \bibinfo {pages} {083003}}\BibitemShut
  {NoStop}%
\bibitem [{\citenamefont {Jotzu}\ \emph {et~al.}(2014)\citenamefont {Jotzu},
  \citenamefont {Messer}, \citenamefont {Desbuquois}, \citenamefont {Lebrat},
  \citenamefont {Uehlinger}, \citenamefont {Greif},\ and\ \citenamefont
  {Esslinger}}]{Jotzu:2014Nat}%
  \BibitemOpen
  \bibfield  {author} {\bibinfo {author} {\bibnamefont {Jotzu}, \bibfnamefont
  {G}}, \bibinfo {author} {\bibfnamefont {M.}~\bibnamefont {Messer}}, \bibinfo
  {author} {\bibfnamefont {R.}~\bibnamefont {Desbuquois}}, \bibinfo {author}
  {\bibfnamefont {M.}~\bibnamefont {Lebrat}}, \bibinfo {author} {\bibfnamefont
  {T.}~\bibnamefont {Uehlinger}}, \bibinfo {author} {\bibfnamefont
  {D.}~\bibnamefont {Greif}}, \ and\ \bibinfo {author} {\bibfnamefont
  {T.}~\bibnamefont {Esslinger}}} (\bibinfo {year} {2014}),\ \bibfield  {title}
  {\enquote {\bibinfo {title} {Experimental realization of the topological
  {H}aldane model with ultracold fermions},}\ }\href
  {http://dx.doi.org/10.1038/nature13915} {\bibfield  {journal} {\bibinfo
  {journal} {Nature}\ }\textbf {\bibinfo {volume} {515}},\ \bibinfo {pages}
  {237}}\BibitemShut {NoStop}%
\bibitem [{\citenamefont {de~Juan}\ \emph {et~al.}(2012)\citenamefont
  {de~Juan}, \citenamefont {Sturla},\ and\ \citenamefont
  {Vozmediano}}]{DeJuan:2012PRL}%
  \BibitemOpen
  \bibfield  {author} {\bibinfo {author} {\bibnamefont {de~Juan}, \bibfnamefont
  {Fernando}}, \bibinfo {author} {\bibfnamefont {Mauricio}\ \bibnamefont
  {Sturla}}, \ and\ \bibinfo {author} {\bibfnamefont {Mar{\'{i}}a A.~H.}\
  \bibnamefont {Vozmediano}}} (\bibinfo {year} {2012}),\ \bibfield  {title}
  {\enquote {\bibinfo {title} {{Space dependent Fermi velocity in strained
  graphene}},}\ }\href
  {https://link.aps.org/doi/10.1103/PhysRevLett.108.227205} {\bibfield
  {journal} {\bibinfo  {journal} {Phys. Rev. Lett.}\ }\textbf {\bibinfo
  {volume} {108}}~(\bibinfo {number} {22}),\ \bibinfo {pages}
  {227205}}\BibitemShut {NoStop}%
\bibitem [{\citenamefont {J\"unemann}\ \emph {et~al.}(2017)\citenamefont
  {J\"unemann}, \citenamefont {Piga}, \citenamefont {Ran}, \citenamefont
  {Lewenstein}, \citenamefont {Rizzi},\ and\ \citenamefont
  {Bermudez}}]{Junemann:2017PRX}%
  \BibitemOpen
  \bibfield  {author} {\bibinfo {author} {\bibnamefont {J\"unemann},
  \bibfnamefont {J}}, \bibinfo {author} {\bibfnamefont {A.}~\bibnamefont
  {Piga}}, \bibinfo {author} {\bibfnamefont {S.-J.}\ \bibnamefont {Ran}},
  \bibinfo {author} {\bibfnamefont {M.}~\bibnamefont {Lewenstein}}, \bibinfo
  {author} {\bibfnamefont {M.}~\bibnamefont {Rizzi}}, \ and\ \bibinfo {author}
  {\bibfnamefont {A.}~\bibnamefont {Bermudez}}} (\bibinfo {year} {2017}),\
  \bibfield  {title} {\enquote {\bibinfo {title} {Exploring interacting
  topological insulators with ultracold atoms: {T}he synthetic {Creutz-Hubbard}
  model},}\ }\href {https://link.aps.org/doi/10.1103/PhysRevX.7.031057}
  {\bibfield  {journal} {\bibinfo  {journal} {Phys. Rev. X}\ }\textbf {\bibinfo
  {volume} {7}},\ \bibinfo {pages} {031057}}\BibitemShut {NoStop}%
\bibitem [{\citenamefont {Kane}\ and\ \citenamefont
  {Mele}(1997)}]{Kane:1997PRL}%
  \BibitemOpen
  \bibfield  {author} {\bibinfo {author} {\bibnamefont {Kane}, \bibfnamefont
  {C~L}}, \ and\ \bibinfo {author} {\bibfnamefont {E.~J.}\ \bibnamefont
  {Mele}}} (\bibinfo {year} {1997}),\ \bibfield  {title} {\enquote {\bibinfo
  {title} {Size, shape, and low energy electronic structure of carbon
  nanotubes},}\ }\href {https://link.aps.org/doi/10.1103/PhysRevLett.78.1932}
  {\bibfield  {journal} {\bibinfo  {journal} {Phys. Rev. Lett.}\ }\textbf
  {\bibinfo {volume} {78}}~(\bibinfo {number} {10}),\ \bibinfo {pages}
  {1932--1935}}\BibitemShut {NoStop}%
\bibitem [{\citenamefont {Kane}\ and\ \citenamefont
  {Mele}(2005{\natexlab{a}})}]{Kane:2005PRLb}%
  \BibitemOpen
  \bibfield  {author} {\bibinfo {author} {\bibnamefont {Kane}, \bibfnamefont
  {C~L}}, \ and\ \bibinfo {author} {\bibfnamefont {E.~J.}\ \bibnamefont
  {Mele}}} (\bibinfo {year} {2005}{\natexlab{a}}),\ \bibfield  {title}
  {\enquote {\bibinfo {title} {Quantum spin {H}all effect in graphene},}\
  }\href {https://link.aps.org/doi/10.1103/PhysRevLett.95.226801} {\bibfield
  {journal} {\bibinfo  {journal} {Phys. Rev. Lett.}\ }\textbf {\bibinfo
  {volume} {95}},\ \bibinfo {pages} {226801}}\BibitemShut {NoStop}%
\bibitem [{\citenamefont {Kane}\ and\ \citenamefont
  {Mele}(2005{\natexlab{b}})}]{Kane:2005PRLa}%
  \BibitemOpen
  \bibfield  {author} {\bibinfo {author} {\bibnamefont {Kane}, \bibfnamefont
  {C~L}}, \ and\ \bibinfo {author} {\bibfnamefont {E.~J.}\ \bibnamefont
  {Mele}}} (\bibinfo {year} {2005}{\natexlab{b}}),\ \bibfield  {title}
  {\enquote {\bibinfo {title} {${Z}_{2}$ topological order and the quantum spin
  {H}all effect},}\ }\href
  {https://link.aps.org/doi/10.1103/PhysRevLett.95.146802} {\bibfield
  {journal} {\bibinfo  {journal} {Phys. Rev. Lett.}\ }\textbf {\bibinfo
  {volume} {95}},\ \bibinfo {pages} {146802}}\BibitemShut {NoStop}%
\bibitem [{\citenamefont {Kane}\ and\ \citenamefont
  {Lubensky}(2014)}]{Kane:2014NatPhys}%
  \BibitemOpen
  \bibfield  {author} {\bibinfo {author} {\bibnamefont {Kane}, \bibfnamefont
  {CL}}, \ and\ \bibinfo {author} {\bibfnamefont {TC}~\bibnamefont {Lubensky}}}
  (\bibinfo {year} {2014}),\ \bibfield  {title} {\enquote {\bibinfo {title}
  {Topological boundary modes in isostatic lattices},}\ }\href
  {http://www.nature.com/nphys/journal/v10/n1/abs/nphys2835.html} {\bibfield
  {journal} {\bibinfo  {journal} {Nat. Phys.}\ }\textbf {\bibinfo {volume}
  {10}},\ \bibinfo {pages} {39--45}}\BibitemShut {NoStop}%
\bibitem [{\citenamefont {Kang}\ \emph {et~al.}(2018)\citenamefont {Kang},
  \citenamefont {Ni}, \citenamefont {Cheng}, \citenamefont {Khanikaev},\ and\
  \citenamefont {Genack}}]{Kang:2018NatComm}%
  \BibitemOpen
  \bibfield  {author} {\bibinfo {author} {\bibnamefont {Kang}, \bibfnamefont
  {Yuhao}}, \bibinfo {author} {\bibfnamefont {Xiang}\ \bibnamefont {Ni}},
  \bibinfo {author} {\bibfnamefont {Xiaojun}\ \bibnamefont {Cheng}}, \bibinfo
  {author} {\bibfnamefont {Alexander~B}\ \bibnamefont {Khanikaev}}, \ and\
  \bibinfo {author} {\bibfnamefont {Azriel~Z}\ \bibnamefont {Genack}}}
  (\bibinfo {year} {2018}),\ \bibfield  {title} {\enquote {\bibinfo {title}
  {Pseudo-spin--valley coupled edge states in a photonic topological
  insulator},}\ }\href {https://www.nature.com/articles/s41467-018-05408-w}
  {\bibfield  {journal} {\bibinfo  {journal} {Nat. Commun.}\ }\textbf {\bibinfo
  {volume} {9}}~(\bibinfo {number} {1}),\ \bibinfo {pages} {3029}}\BibitemShut
  {NoStop}%
\bibitem [{\citenamefont {Kapit}\ \emph {et~al.}(2014)\citenamefont {Kapit},
  \citenamefont {Hafezi},\ and\ \citenamefont {Simon}}]{Kapit:2014PRX}%
  \BibitemOpen
  \bibfield  {author} {\bibinfo {author} {\bibnamefont {Kapit}, \bibfnamefont
  {Eliot}}, \bibinfo {author} {\bibfnamefont {Mohammad}\ \bibnamefont
  {Hafezi}}, \ and\ \bibinfo {author} {\bibfnamefont {Steven~H.}\ \bibnamefont
  {Simon}}} (\bibinfo {year} {2014}),\ \bibfield  {title} {\enquote {\bibinfo
  {title} {Induced self-stabilization in fractional quantum {H}all states of
  light},}\ }\href {http://link.aps.org/doi/10.1103/PhysRevX.4.031039}
  {\bibfield  {journal} {\bibinfo  {journal} {Phys. Rev. X}\ }\textbf {\bibinfo
  {volume} {4}},\ \bibinfo {pages} {031039}}\BibitemShut {NoStop}%
\bibitem [{\citenamefont {Karplus}\ and\ \citenamefont
  {Luttinger}(1954)}]{Karplus:1954PR}%
  \BibitemOpen
  \bibfield  {author} {\bibinfo {author} {\bibnamefont {Karplus}, \bibfnamefont
  {Robert}}, \ and\ \bibinfo {author} {\bibfnamefont {J.~M.}\ \bibnamefont
  {Luttinger}}} (\bibinfo {year} {1954}),\ \bibfield  {title} {\enquote
  {\bibinfo {title} {Hall effect in ferromagnetics},}\ }\href
  {https://link.aps.org/doi/10.1103/PhysRev.95.1154} {\bibfield  {journal}
  {\bibinfo  {journal} {Phys. Rev.}\ }\textbf {\bibinfo {volume} {95}},\
  \bibinfo {pages} {1154--1160}}\BibitemShut {NoStop}%
\bibitem [{\citenamefont {Karplus}\ and\ \citenamefont
  {Neuman}(1951)}]{Karplus:PR1951}%
  \BibitemOpen
  \bibfield  {author} {\bibinfo {author} {\bibnamefont {Karplus}, \bibfnamefont
  {Robert}}, \ and\ \bibinfo {author} {\bibfnamefont {Maurice}\ \bibnamefont
  {Neuman}}} (\bibinfo {year} {1951}),\ \bibfield  {title} {\enquote {\bibinfo
  {title} {The scattering of light by light},}\ }\href
  {https://link.aps.org/doi/10.1103/PhysRev.83.776} {\bibfield  {journal}
  {\bibinfo  {journal} {Phys. Rev.}\ }\textbf {\bibinfo {volume} {83}},\
  \bibinfo {pages} {776--784}}\BibitemShut {NoStop}%
\bibitem [{\citenamefont {Kartashov}\ and\ \citenamefont
  {Skryabin}(2016)}]{Kartashov:Optica2016}%
  \BibitemOpen
  \bibfield  {author} {\bibinfo {author} {\bibnamefont {Kartashov},
  \bibfnamefont {Yaroslav~V}}, \ and\ \bibinfo {author} {\bibfnamefont
  {Dmitry~V}\ \bibnamefont {Skryabin}}} (\bibinfo {year} {2016}),\ \bibfield
  {title} {\enquote {\bibinfo {title} {Modulational instability and solitary
  waves in polariton topological insulators},}\ }\href@noop {} {\bibfield
  {journal} {\bibinfo  {journal} {Optica}\ }\textbf {\bibinfo {volume}
  {3}}~(\bibinfo {number} {11}),\ \bibinfo {pages} {1228--1236}}\BibitemShut
  {NoStop}%
\bibitem [{\citenamefont {Karzig}\ \emph {et~al.}(2015)\citenamefont {Karzig},
  \citenamefont {Bardyn}, \citenamefont {Lindner},\ and\ \citenamefont
  {Refael}}]{Karzig:2015PRX}%
  \BibitemOpen
  \bibfield  {author} {\bibinfo {author} {\bibnamefont {Karzig}, \bibfnamefont
  {Torsten}}, \bibinfo {author} {\bibfnamefont {Charles-Edouard}\ \bibnamefont
  {Bardyn}}, \bibinfo {author} {\bibfnamefont {Netanel~H.}\ \bibnamefont
  {Lindner}}, \ and\ \bibinfo {author} {\bibfnamefont {Gil}\ \bibnamefont
  {Refael}}} (\bibinfo {year} {2015}),\ \bibfield  {title} {\enquote {\bibinfo
  {title} {Topological polaritons},}\ }\href
  {http://link.aps.org/doi/10.1103/PhysRevX.5.031001} {\bibfield  {journal}
  {\bibinfo  {journal} {Phys. Rev. X}\ }\textbf {\bibinfo {volume} {5}},\
  \bibinfo {pages} {031001}}\BibitemShut {NoStop}%
\bibitem [{\citenamefont {Katan}\ \emph {et~al.}(2016)\citenamefont {Katan},
  \citenamefont {Bekenestein}, \citenamefont {Bandres}, \citenamefont {Lumer},
  \citenamefont {Yonatan},\ and\ \citenamefont {Segev}}]{Katan:CLEO2016}%
  \BibitemOpen
  \bibfield  {author} {\bibinfo {author} {\bibnamefont {Katan}, \bibfnamefont
  {Yaniv~Tenenbaum}}, \bibinfo {author} {\bibfnamefont {Rivka}\ \bibnamefont
  {Bekenestein}}, \bibinfo {author} {\bibfnamefont {Miguel~A}\ \bibnamefont
  {Bandres}}, \bibinfo {author} {\bibfnamefont {Yaakov}\ \bibnamefont {Lumer}},
  \bibinfo {author} {\bibfnamefont {Plotnik}\ \bibnamefont {Yonatan}}, \ and\
  \bibinfo {author} {\bibfnamefont {Mordechai}\ \bibnamefont {Segev}}}
  (\bibinfo {year} {2016}),\ \bibfield  {title} {\enquote {\bibinfo {title}
  {Induction of topological transport by long ranged nonlinearity},}\ }in\
  \href@noop {} {\emph {\bibinfo {booktitle} {CLEO: QELS\_Fundamental
  Science}}}\ (\bibinfo {organization} {Optical Society of America})\ pp.\
  \bibinfo {pages} {FM3A--6}\BibitemShut {NoStop}%
\bibitem [{\citenamefont {Kavokin}\ \emph {et~al.}(2005)\citenamefont
  {Kavokin}, \citenamefont {Malpuech},\ and\ \citenamefont
  {Glazov}}]{Kavokin:2005PRL}%
  \BibitemOpen
  \bibfield  {author} {\bibinfo {author} {\bibnamefont {Kavokin}, \bibfnamefont
  {Alexey}}, \bibinfo {author} {\bibfnamefont {Guillaume}\ \bibnamefont
  {Malpuech}}, \ and\ \bibinfo {author} {\bibfnamefont {Mikhail}\ \bibnamefont
  {Glazov}}} (\bibinfo {year} {2005}),\ \bibfield  {title} {\enquote {\bibinfo
  {title} {{Optical spin Hall effect}},}\ }\href
  {https://journals.aps.org/prl/abstract/10.1103/PhysRevLett.95.136601}
  {\bibfield  {journal} {\bibinfo  {journal} {Phys. Rev. Lett.}\ }\textbf
  {\bibinfo {volume} {95}}~(\bibinfo {number} {13}),\ \bibinfo {pages}
  {136601}}\BibitemShut {NoStop}%
\bibitem [{\citenamefont {Kawabata}\ \emph
  {et~al.}(2018{\natexlab{a}})\citenamefont {Kawabata}, \citenamefont {Ashida},
  \citenamefont {Katsura},\ and\ \citenamefont {Ueda}}]{kawabata2018parity}%
  \BibitemOpen
  \bibfield  {author} {\bibinfo {author} {\bibnamefont {Kawabata},
  \bibfnamefont {Kohei}}, \bibinfo {author} {\bibfnamefont {Yuto}\ \bibnamefont
  {Ashida}}, \bibinfo {author} {\bibfnamefont {Hosho}\ \bibnamefont {Katsura}},
  \ and\ \bibinfo {author} {\bibfnamefont {Masahito}\ \bibnamefont {Ueda}}}
  (\bibinfo {year} {2018}{\natexlab{a}}),\ \bibfield  {title} {\enquote
  {\bibinfo {title} {Parity-time-symmetric topological superconductor},}\
  }\href {https://link.aps.org/doi/10.1103/PhysRevB.98.085116} {\bibfield
  {journal} {\bibinfo  {journal} {Phys. Rev. B}\ }\textbf {\bibinfo {volume}
  {98}},\ \bibinfo {pages} {085116}}\BibitemShut {NoStop}%
\bibitem [{\citenamefont {Kawabata}\ \emph {et~al.}(2019)\citenamefont
  {Kawabata}, \citenamefont {Higashikawa}, \citenamefont {Gong}, \citenamefont
  {Ashida},\ and\ \citenamefont {Ueda}}]{kawabata2018topological}%
  \BibitemOpen
  \bibfield  {author} {\bibinfo {author} {\bibnamefont {Kawabata},
  \bibfnamefont {Kohei}}, \bibinfo {author} {\bibfnamefont {Sho}\ \bibnamefont
  {Higashikawa}}, \bibinfo {author} {\bibfnamefont {Zongping}\ \bibnamefont
  {Gong}}, \bibinfo {author} {\bibfnamefont {Yuto}\ \bibnamefont {Ashida}}, \
  and\ \bibinfo {author} {\bibfnamefont {Masahito}\ \bibnamefont {Ueda}}}
  (\bibinfo {year} {2019}),\ \bibfield  {title} {\enquote {\bibinfo {title}
  {Topological unification of time-reversal and particle-hole symmetries in
  non-{H}ermitian physics},}\ }\href
  {https://www.nature.com/articles/s41467-018-08254-y} {\bibfield  {journal}
  {\bibinfo  {journal} {Nat. Commun.}\ }\textbf {\bibinfo {volume}
  {10}}~(\bibinfo {number} {1}),\ \bibinfo {pages} {297}}\BibitemShut {NoStop}%
\bibitem [{\citenamefont {Kawabata}\ \emph
  {et~al.}(2018{\natexlab{b}})\citenamefont {Kawabata}, \citenamefont
  {Shiozaki},\ and\ \citenamefont {Ueda}}]{kawabata2018non}%
  \BibitemOpen
  \bibfield  {author} {\bibinfo {author} {\bibnamefont {Kawabata},
  \bibfnamefont {Kohei}}, \bibinfo {author} {\bibfnamefont {Ken}\ \bibnamefont
  {Shiozaki}}, \ and\ \bibinfo {author} {\bibfnamefont {Masahito}\ \bibnamefont
  {Ueda}}} (\bibinfo {year} {2018}{\natexlab{b}}),\ \bibfield  {title}
  {\enquote {\bibinfo {title} {Anomalous helical edge states in a
  non-{H}ermitian {C}hern insulator},}\ }\href
  {https://link.aps.org/doi/10.1103/PhysRevB.98.165148} {\bibfield  {journal}
  {\bibinfo  {journal} {Phys. Rev. B}\ }\textbf {\bibinfo {volume} {98}},\
  \bibinfo {pages} {165148}}\BibitemShut {NoStop}%
\bibitem [{\citenamefont {Kawakami}\ and\ \citenamefont
  {Hu}(2017)}]{Kawakami:2016arXiv}%
  \BibitemOpen
  \bibfield  {author} {\bibinfo {author} {\bibnamefont {Kawakami},
  \bibfnamefont {Takuto}}, \ and\ \bibinfo {author} {\bibfnamefont {Xiao}\
  \bibnamefont {Hu}}} (\bibinfo {year} {2017}),\ \bibfield  {title} {\enquote
  {\bibinfo {title} {Symmetry-guaranteed nodal-line semimetals in an fcc
  lattice},}\ }\href {https://link.aps.org/doi/10.1103/PhysRevB.96.235307}
  {\bibfield  {journal} {\bibinfo  {journal} {Phys. Rev. B}\ }\textbf {\bibinfo
  {volume} {96}},\ \bibinfo {pages} {235307}}\BibitemShut {NoStop}%
\bibitem [{\citenamefont {Kazimierczuk}\ \emph {et~al.}(2014)\citenamefont
  {Kazimierczuk}, \citenamefont {Fr{\"o}hlich}, \citenamefont {Scheel},
  \citenamefont {Stolz},\ and\ \citenamefont
  {Bayer}}]{Kazimierczuk:Nature2014}%
  \BibitemOpen
  \bibfield  {author} {\bibinfo {author} {\bibnamefont {Kazimierczuk},
  \bibfnamefont {Tomasz}}, \bibinfo {author} {\bibfnamefont {Dietmar}\
  \bibnamefont {Fr{\"o}hlich}}, \bibinfo {author} {\bibfnamefont {Stefan}\
  \bibnamefont {Scheel}}, \bibinfo {author} {\bibfnamefont {Heinrich}\
  \bibnamefont {Stolz}}, \ and\ \bibinfo {author} {\bibfnamefont {Manfred}\
  \bibnamefont {Bayer}}} (\bibinfo {year} {2014}),\ \bibfield  {title}
  {\enquote {\bibinfo {title} {Giant {R}ydberg excitons in the copper oxide
  $\mathrm{Cu_2 O}$},}\ }\href {https://www.nature.com/articles/nature13832}
  {\bibfield  {journal} {\bibinfo  {journal} {Nature}\ }\textbf {\bibinfo
  {volume} {514}}~(\bibinfo {number} {7522}),\ \bibinfo {pages}
  {343--347}}\BibitemShut {NoStop}%
\bibitem [{\citenamefont {Ke}\ \emph {et~al.}(2016)\citenamefont {Ke},
  \citenamefont {Qin}, \citenamefont {Mei}, \citenamefont {Zhong},
  \citenamefont {Kivshar},\ and\ \citenamefont {Lee}}]{Ke:2016}%
  \BibitemOpen
  \bibfield  {author} {\bibinfo {author} {\bibnamefont {Ke}, \bibfnamefont
  {Yongguan}}, \bibinfo {author} {\bibfnamefont {Xizhou}\ \bibnamefont {Qin}},
  \bibinfo {author} {\bibfnamefont {Feng}\ \bibnamefont {Mei}}, \bibinfo
  {author} {\bibfnamefont {Honghua}\ \bibnamefont {Zhong}}, \bibinfo {author}
  {\bibfnamefont {Yuri~S}\ \bibnamefont {Kivshar}}, \ and\ \bibinfo {author}
  {\bibfnamefont {Chaohong}\ \bibnamefont {Lee}}} (\bibinfo {year} {2016}),\
  \bibfield  {title} {\enquote {\bibinfo {title} {Topological phase transitions
  and {T}houless pumping of light in photonic waveguide arrays},}\ }\href
  {https://onlinelibrary.wiley.com/doi/abs/10.1002/lpor.201670069} {\bibfield
  {journal} {\bibinfo  {journal} {Laser Photonics Rev.}\ }\textbf {\bibinfo
  {volume} {10}}~(\bibinfo {number} {6}),\ \bibinfo {pages}
  {995--1001}}\BibitemShut {NoStop}%
\bibitem [{\citenamefont {Keil}\ \emph {et~al.}(2013)\citenamefont {Keil},
  \citenamefont {Zeuner}, \citenamefont {Dreisow}, \citenamefont {Heinrich},
  \citenamefont {T{\"u}nnermann}, \citenamefont {Nolte},\ and\ \citenamefont
  {Szameit}}]{Keil:2013NatComm}%
  \BibitemOpen
  \bibfield  {author} {\bibinfo {author} {\bibnamefont {Keil}, \bibfnamefont
  {Robert}}, \bibinfo {author} {\bibfnamefont {Julia~M}\ \bibnamefont
  {Zeuner}}, \bibinfo {author} {\bibfnamefont {Felix}\ \bibnamefont {Dreisow}},
  \bibinfo {author} {\bibfnamefont {Matthias}\ \bibnamefont {Heinrich}},
  \bibinfo {author} {\bibfnamefont {Andreas}\ \bibnamefont {T{\"u}nnermann}},
  \bibinfo {author} {\bibfnamefont {Stefan}\ \bibnamefont {Nolte}}, \ and\
  \bibinfo {author} {\bibfnamefont {Alexander}\ \bibnamefont {Szameit}}}
  (\bibinfo {year} {2013}),\ \bibfield  {title} {\enquote {\bibinfo {title}
  {The random mass {D}irac model and long-range correlations on an integrated
  optical platform},}\ }\href {https://www.nature.com/articles/ncomms2384}
  {\bibfield  {journal} {\bibinfo  {journal} {Nat. Commun.}\ }\textbf {\bibinfo
  {volume} {4}},\ \bibinfo {pages} {1368}}\BibitemShut {NoStop}%
\bibitem [{\citenamefont {Ketoja}\ and\ \citenamefont
  {Satija}(1997)}]{Ketoja:1997}%
  \BibitemOpen
  \bibfield  {author} {\bibinfo {author} {\bibnamefont {Ketoja}, \bibfnamefont
  {Jukka~A}}, \ and\ \bibinfo {author} {\bibfnamefont {Indubala~I}\
  \bibnamefont {Satija}}} (\bibinfo {year} {1997}),\ \bibfield  {title}
  {\enquote {\bibinfo {title} {The re-entrant phase diagram of the generalized
  {H}arper equation},}\ }\href
  {https://iopscience.iop.org/article/10.1088/0953-8984/9/5/016} {\bibfield
  {journal} {\bibinfo  {journal} {J. Phys. Condens. Matter}\ }\textbf {\bibinfo
  {volume} {9}}~(\bibinfo {number} {5}),\ \bibinfo {pages} {1123}}\BibitemShut
  {NoStop}%
\bibitem [{\citenamefont {Khanikaev}\ and\ \citenamefont
  {Al{\`u}}(2015)}]{Khanikaev:NatPhot2015}%
  \BibitemOpen
  \bibfield  {author} {\bibinfo {author} {\bibnamefont {Khanikaev},
  \bibfnamefont {Alexander~B}}, \ and\ \bibinfo {author} {\bibfnamefont
  {Andrea}\ \bibnamefont {Al{\`u}}}} (\bibinfo {year} {2015}),\ \bibfield
  {title} {\enquote {\bibinfo {title} {Optical isolators: {N}onlinear dynamic
  reciprocity},}\ }\href {https://www.nature.com/articles/nphoton.2015.86}
  {\bibfield  {journal} {\bibinfo  {journal} {Nat. Photonics}\ }\textbf
  {\bibinfo {volume} {9}}~(\bibinfo {number} {6}),\ \bibinfo {pages}
  {359--361}}\BibitemShut {NoStop}%
\bibitem [{\citenamefont {Khanikaev}\ \emph {et~al.}(2013)\citenamefont
  {Khanikaev}, \citenamefont {Mousavi}, \citenamefont {Tse}, \citenamefont
  {Kargarian}, \citenamefont {MacDonald},\ and\ \citenamefont
  {Shvets}}]{Khanikaev:2013NatMat}%
  \BibitemOpen
  \bibfield  {author} {\bibinfo {author} {\bibnamefont {Khanikaev},
  \bibfnamefont {Alexander~B}}, \bibinfo {author} {\bibfnamefont {S~Hossein}\
  \bibnamefont {Mousavi}}, \bibinfo {author} {\bibfnamefont {Wang-Kong}\
  \bibnamefont {Tse}}, \bibinfo {author} {\bibfnamefont {Mehdi}\ \bibnamefont
  {Kargarian}}, \bibinfo {author} {\bibfnamefont {Allan~H}\ \bibnamefont
  {MacDonald}}, \ and\ \bibinfo {author} {\bibfnamefont {Gennady}\ \bibnamefont
  {Shvets}}} (\bibinfo {year} {2013}),\ \bibfield  {title} {\enquote {\bibinfo
  {title} {Photonic topological insulators},}\ }\href
  {http://www.nature.com/nmat/journal/v12/n3/full/nmat3520.html} {\bibfield
  {journal} {\bibinfo  {journal} {Nat. Mater.}\ }\textbf {\bibinfo {volume}
  {12}}~(\bibinfo {number} {3}),\ \bibinfo {pages} {233--239}}\BibitemShut
  {NoStop}%
\bibitem [{\citenamefont {Khanikaev}\ and\ \citenamefont
  {Shvets}(2017)}]{khanikaev2017two}%
  \BibitemOpen
  \bibfield  {author} {\bibinfo {author} {\bibnamefont {Khanikaev},
  \bibfnamefont {Alexander~B}}, \ and\ \bibinfo {author} {\bibfnamefont
  {Gennady}\ \bibnamefont {Shvets}}} (\bibinfo {year} {2017}),\ \bibfield
  {title} {\enquote {\bibinfo {title} {Two-dimensional topological
  photonics},}\ }\href {https://www.nature.com/articles/s41566-017-0048-5}
  {\bibfield  {journal} {\bibinfo  {journal} {Nat. Photonics}\ }\textbf
  {\bibinfo {volume} {11}}~(\bibinfo {number} {12}),\ \bibinfo {pages}
  {763}}\BibitemShut {NoStop}%
\bibitem [{\citenamefont {Kim}\ \emph {et~al.}(2017)\citenamefont {Kim},
  \citenamefont {Xu}, \citenamefont {Taylor},\ and\ \citenamefont
  {Bahl}}]{bahl2016dynamically}%
  \BibitemOpen
  \bibfield  {author} {\bibinfo {author} {\bibnamefont {Kim}, \bibfnamefont
  {Seunghwi}}, \bibinfo {author} {\bibfnamefont {Xunnong}\ \bibnamefont {Xu}},
  \bibinfo {author} {\bibfnamefont {Jacob~M.}\ \bibnamefont {Taylor}}, \ and\
  \bibinfo {author} {\bibfnamefont {Gaurav}\ \bibnamefont {Bahl}}} (\bibinfo
  {year} {2017}),\ \bibfield  {title} {\enquote {\bibinfo {title} {Dynamically
  induced robust phonon transport and chiral cooling in an optomechanical
  system},}\ }\href {https://doi.org/10.1038/s41467-017-00247-7} {\bibfield
  {journal} {\bibinfo  {journal} {Nat. Commun.}\ }\textbf {\bibinfo {volume}
  {8}}~(\bibinfo {number} {1}),\ \bibinfo {pages} {205}}\BibitemShut {NoStop}%
\bibitem [{\citenamefont {Kiss}\ \emph {et~al.}(1994)\citenamefont {Kiss},
  \citenamefont {Janszky},\ and\ \citenamefont {Adam}}]{Kiss:1994PRA}%
  \BibitemOpen
  \bibfield  {author} {\bibinfo {author} {\bibnamefont {Kiss}, \bibfnamefont
  {T}}, \bibinfo {author} {\bibfnamefont {J.}~\bibnamefont {Janszky}}, \ and\
  \bibinfo {author} {\bibfnamefont {P.}~\bibnamefont {Adam}}} (\bibinfo {year}
  {1994}),\ \bibfield  {title} {\enquote {\bibinfo {title} {Time evolution of
  harmonic oscillators with time-dependent parameters: {A} step-function
  approximation},}\ }\href {https://link.aps.org/doi/10.1103/PhysRevA.49.4935}
  {\bibfield  {journal} {\bibinfo  {journal} {Phys. Rev. A}\ }\textbf {\bibinfo
  {volume} {49}},\ \bibinfo {pages} {4935--4942}}\BibitemShut {NoStop}%
\bibitem [{\citenamefont {Kitaev}\ \emph {et~al.}(2009)\citenamefont {Kitaev},
  \citenamefont {Lebedev},\ and\ \citenamefont
  {Feigel’man}}]{Kitaev:2009AIP}%
  \BibitemOpen
  \bibfield  {author} {\bibinfo {author} {\bibnamefont {Kitaev}, \bibfnamefont
  {Alexei}}, \bibinfo {author} {\bibfnamefont {Vladimir}\ \bibnamefont
  {Lebedev}}, \ and\ \bibinfo {author} {\bibfnamefont {Mikhail}\ \bibnamefont
  {Feigel’man}}} (\bibinfo {year} {2009}),\ \bibfield  {title} {\enquote
  {\bibinfo {title} {Periodic table for topological insulators and
  superconductors},}\ }\bibfield  {booktitle} {\emph {\bibinfo {booktitle} {AIP
  Conference Proceedings}},\ }\href
  {http://aip.scitation.org/doi/abs/10.1063/1.3149495} {\ \textbf {\bibinfo
  {volume} {1134}},\ \bibinfo {pages} {22--30}}\BibitemShut {NoStop}%
\bibitem [{\citenamefont {Kitagawa}\ \emph
  {et~al.}(2010{\natexlab{a}})\citenamefont {Kitagawa}, \citenamefont {Berg},
  \citenamefont {Rudner},\ and\ \citenamefont {Demler}}]{Kitagawa:2010PRB}%
  \BibitemOpen
  \bibfield  {author} {\bibinfo {author} {\bibnamefont {Kitagawa},
  \bibfnamefont {T}}, \bibinfo {author} {\bibfnamefont {E.}~\bibnamefont
  {Berg}}, \bibinfo {author} {\bibfnamefont {M.}~\bibnamefont {Rudner}}, \ and\
  \bibinfo {author} {\bibfnamefont {E.}~\bibnamefont {Demler}}} (\bibinfo
  {year} {2010}{\natexlab{a}}),\ \bibfield  {title} {\enquote {\bibinfo {title}
  {Topological characterization of periodically-driven quantum systems},}\
  }\href {https://doi.org/10.1103/PhysRevB.82.235114} {\bibfield  {journal}
  {\bibinfo  {journal} {Phys. Rev. B}\ }\textbf {\bibinfo {volume} {82}},\
  \bibinfo {pages} {235114}}\BibitemShut {NoStop}%
\bibitem [{\citenamefont {Kitagawa}(2012)}]{Kitagawa:2012QIP}%
  \BibitemOpen
  \bibfield  {author} {\bibinfo {author} {\bibnamefont {Kitagawa},
  \bibfnamefont {Takuya}}} (\bibinfo {year} {2012}),\ \bibfield  {title}
  {\enquote {\bibinfo {title} {Topological phenomena in quantum walks:
  {E}lementary introduction to the physics of topological phases},}\ }\href
  {https://link.springer.com/article/10.1007%2Fs11128-012-0425-4?LI=true}
  {\bibfield  {journal} {\bibinfo  {journal} {Quantum Information Processing}\
  }\textbf {\bibinfo {volume} {11}},\ \bibinfo {pages}
  {1107--1148}}\BibitemShut {NoStop}%
\bibitem [{\citenamefont {Kitagawa}\ \emph {et~al.}(2012)\citenamefont
  {Kitagawa}, \citenamefont {Broome}, \citenamefont {Fedrizzi}, \citenamefont
  {Rudner}, \citenamefont {Berg}, \citenamefont {Kassal}, \citenamefont
  {Aspuru-Guzik}, \citenamefont {Demler},\ and\ \citenamefont
  {White}}]{Kitagawa:2012NatComm}%
  \BibitemOpen
  \bibfield  {author} {\bibinfo {author} {\bibnamefont {Kitagawa},
  \bibfnamefont {Takuya}}, \bibinfo {author} {\bibfnamefont {Matthew~A}\
  \bibnamefont {Broome}}, \bibinfo {author} {\bibfnamefont {Alessandro}\
  \bibnamefont {Fedrizzi}}, \bibinfo {author} {\bibfnamefont {Mark~S}\
  \bibnamefont {Rudner}}, \bibinfo {author} {\bibfnamefont {Erez}\ \bibnamefont
  {Berg}}, \bibinfo {author} {\bibfnamefont {Ivan}\ \bibnamefont {Kassal}},
  \bibinfo {author} {\bibfnamefont {Al{\'a}n}\ \bibnamefont {Aspuru-Guzik}},
  \bibinfo {author} {\bibfnamefont {Eugene}\ \bibnamefont {Demler}}, \ and\
  \bibinfo {author} {\bibfnamefont {Andrew~G}\ \bibnamefont {White}}} (\bibinfo
  {year} {2012}),\ \bibfield  {title} {\enquote {\bibinfo {title} {Observation
  of topologically protected bound states in photonic quantum walks},}\ }\href
  {http://www.nature.com/articles/ncomms1872} {\bibfield  {journal} {\bibinfo
  {journal} {Nat. Commun.}\ }\textbf {\bibinfo {volume} {3}},\ \bibinfo {pages}
  {882}}\BibitemShut {NoStop}%
\bibitem [{\citenamefont {Kitagawa}\ \emph {et~al.}(2011)\citenamefont
  {Kitagawa}, \citenamefont {Oka}, \citenamefont {Brataas}, \citenamefont
  {Fu},\ and\ \citenamefont {Demler}}]{Kitagawa:2011PRB}%
  \BibitemOpen
  \bibfield  {author} {\bibinfo {author} {\bibnamefont {Kitagawa},
  \bibfnamefont {Takuya}}, \bibinfo {author} {\bibfnamefont {Takashi}\
  \bibnamefont {Oka}}, \bibinfo {author} {\bibfnamefont {Arne}\ \bibnamefont
  {Brataas}}, \bibinfo {author} {\bibfnamefont {Liang}\ \bibnamefont {Fu}}, \
  and\ \bibinfo {author} {\bibfnamefont {Eugene}\ \bibnamefont {Demler}}}
  (\bibinfo {year} {2011}),\ \bibfield  {title} {\enquote {\bibinfo {title}
  {Transport properties of nonequilibrium systems under the application of
  light: {P}hotoinduced quantum {H}all insulators without {L}andau levels},}\
  }\href {https://doi.org/10.1103/PhysRevB.84.235108} {\bibfield  {journal}
  {\bibinfo  {journal} {Phys. Rev. B}\ }\textbf {\bibinfo {volume} {84}},\
  \bibinfo {pages} {235108}}\BibitemShut {NoStop}%
\bibitem [{\citenamefont {Kitagawa}\ \emph
  {et~al.}(2010{\natexlab{b}})\citenamefont {Kitagawa}, \citenamefont {Rudner},
  \citenamefont {Berg},\ and\ \citenamefont {Demler}}]{Kitagawa:2010PRA}%
  \BibitemOpen
  \bibfield  {author} {\bibinfo {author} {\bibnamefont {Kitagawa},
  \bibfnamefont {Takuya}}, \bibinfo {author} {\bibfnamefont {Mark~S.}\
  \bibnamefont {Rudner}}, \bibinfo {author} {\bibfnamefont {Erez}\ \bibnamefont
  {Berg}}, \ and\ \bibinfo {author} {\bibfnamefont {Eugene}\ \bibnamefont
  {Demler}}} (\bibinfo {year} {2010}{\natexlab{b}}),\ \bibfield  {title}
  {\enquote {\bibinfo {title} {Exploring topological phases with quantum
  walks},}\ }\href {https://link.aps.org/doi/10.1103/PhysRevA.82.033429}
  {\bibfield  {journal} {\bibinfo  {journal} {Phys. Rev. A}\ }\textbf {\bibinfo
  {volume} {82}},\ \bibinfo {pages} {033429}}\BibitemShut {NoStop}%
\bibitem [{\citenamefont {Kivshar}\ and\ \citenamefont
  {Agrawal}(2003)}]{Kivshar:2003book}%
  \BibitemOpen
  \bibfield  {author} {\bibinfo {author} {\bibnamefont {Kivshar}, \bibfnamefont
  {Yuri~S}}, \ and\ \bibinfo {author} {\bibfnamefont {Govind}\ \bibnamefont
  {Agrawal}}} (\bibinfo {year} {2003}),\ \href@noop {} {\emph {\bibinfo {title}
  {Optical solitons: {F}rom fibers to photonic crystals}}}\ (\bibinfo
  {publisher} {Academic press},\ \bibinfo {address} {New York})\BibitemShut
  {NoStop}%
\bibitem [{\citenamefont {Klaers}\ \emph {et~al.}(2010)\citenamefont {Klaers},
  \citenamefont {Schmitt}, \citenamefont {Vewinger},\ and\ \citenamefont
  {Weitz}}]{Klaers:Nature2010}%
  \BibitemOpen
  \bibfield  {author} {\bibinfo {author} {\bibnamefont {Klaers}, \bibfnamefont
  {Jan}}, \bibinfo {author} {\bibfnamefont {Julian}\ \bibnamefont {Schmitt}},
  \bibinfo {author} {\bibfnamefont {Frank}\ \bibnamefont {Vewinger}}, \ and\
  \bibinfo {author} {\bibfnamefont {Martin}\ \bibnamefont {Weitz}}} (\bibinfo
  {year} {2010}),\ \bibfield  {title} {\enquote {\bibinfo {title}
  {{Bose-Einstein condensation of photons in an optical microcavity}},}\ }\href
  {https://www.nature.com/articles/nature09567} {\bibfield  {journal} {\bibinfo
   {journal} {Nature}\ }\textbf {\bibinfo {volume} {468}}~(\bibinfo {number}
  {7323}),\ \bibinfo {pages} {545--548}}\BibitemShut {NoStop}%
\bibitem [{\citenamefont {Klein}(1994)}]{Klein:1994ChemPhysLett}%
  \BibitemOpen
  \bibfield  {author} {\bibinfo {author} {\bibnamefont {Klein}, \bibfnamefont
  {DJ}}} (\bibinfo {year} {1994}),\ \bibfield  {title} {\enquote {\bibinfo
  {title} {{Graphitic polymer strips with edge states}},}\ }\href
  {http://www.sciencedirect.com/science/article/pii/0009261493E1378T}
  {\bibfield  {journal} {\bibinfo  {journal} {Chem. Phys. Lett.}\ }\textbf
  {\bibinfo {volume} {217}}~(\bibinfo {number} {3}),\ \bibinfo {pages}
  {261--265}}\BibitemShut {NoStop}%
\bibitem [{\citenamefont {Klembt}\ \emph {et~al.}(2018)\citenamefont {Klembt},
  \citenamefont {Harder}, \citenamefont {Egorov}, \citenamefont {Winkler},
  \citenamefont {Ge}, \citenamefont {Bandres}, \citenamefont {Emmerling},
  \citenamefont {Worschech}, \citenamefont {Liew}, \citenamefont {Segev} \emph
  {et~al.}}]{Klembt:arxiv2018}%
  \BibitemOpen
  \bibfield  {author} {\bibinfo {author} {\bibnamefont {Klembt}, \bibfnamefont
  {S}}, \bibinfo {author} {\bibfnamefont {TH}~\bibnamefont {Harder}}, \bibinfo
  {author} {\bibfnamefont {OA}~\bibnamefont {Egorov}}, \bibinfo {author}
  {\bibfnamefont {K}~\bibnamefont {Winkler}}, \bibinfo {author} {\bibfnamefont
  {R}~\bibnamefont {Ge}}, \bibinfo {author} {\bibfnamefont {MA}~\bibnamefont
  {Bandres}}, \bibinfo {author} {\bibfnamefont {M}~\bibnamefont {Emmerling}},
  \bibinfo {author} {\bibfnamefont {L}~\bibnamefont {Worschech}}, \bibinfo
  {author} {\bibfnamefont {TCH}\ \bibnamefont {Liew}}, \bibinfo {author}
  {\bibfnamefont {M}~\bibnamefont {Segev}},  \emph {et~al.}} (\bibinfo {year}
  {2018}),\ \bibfield  {title} {\enquote {\bibinfo {title} {Exciton-polariton
  topological insulator},}\ }\href
  {https://www.nature.com/articles/s41586-018-0601-5} {\bibfield  {journal}
  {\bibinfo  {journal} {Nature}\ }\textbf {\bibinfo {volume} {562}}~(\bibinfo
  {number} {7728}),\ \bibinfo {pages} {552}}\BibitemShut {NoStop}%
\bibitem [{\citenamefont {Klitzing}\ \emph {et~al.}(1980)\citenamefont
  {Klitzing}, \citenamefont {Dorda},\ and\ \citenamefont
  {Pepper}}]{Klitzing:1980PRL}%
  \BibitemOpen
  \bibfield  {author} {\bibinfo {author} {\bibnamefont {Klitzing},
  \bibfnamefont {K~v}}, \bibinfo {author} {\bibfnamefont {G.}~\bibnamefont
  {Dorda}}, \ and\ \bibinfo {author} {\bibfnamefont {M.}~\bibnamefont
  {Pepper}}} (\bibinfo {year} {1980}),\ \bibfield  {title} {\enquote {\bibinfo
  {title} {New method for high-accuracy determination of the fine-structure
  constant based on quantized {H}all resistance},}\ }\href
  {https://link.aps.org/doi/10.1103/PhysRevLett.45.494} {\bibfield  {journal}
  {\bibinfo  {journal} {Phys. Rev. Lett.}\ }\textbf {\bibinfo {volume} {45}},\
  \bibinfo {pages} {494--497}}\BibitemShut {NoStop}%
\bibitem [{\citenamefont {von Klitzing}(1986)}]{VonKlitzing:RMP1986}%
  \BibitemOpen
  \bibfield  {author} {\bibinfo {author} {\bibnamefont {von Klitzing},
  \bibfnamefont {Klaus}}} (\bibinfo {year} {1986}),\ \bibfield  {title}
  {\enquote {\bibinfo {title} {The quantized {H}all effect},}\ }\href
  {https://link.aps.org/doi/10.1103/RevModPhys.58.519} {\bibfield  {journal}
  {\bibinfo  {journal} {Rev. Mod. Phys.}\ }\textbf {\bibinfo {volume} {58}},\
  \bibinfo {pages} {519--531}}\BibitemShut {NoStop}%
\bibitem [{\citenamefont {Koch}\ \emph {et~al.}(2010)\citenamefont {Koch},
  \citenamefont {Houck}, \citenamefont {Hur},\ and\ \citenamefont
  {Girvin}}]{Koch:2010PRA}%
  \BibitemOpen
  \bibfield  {author} {\bibinfo {author} {\bibnamefont {Koch}, \bibfnamefont
  {Jens}}, \bibinfo {author} {\bibfnamefont {Andrew~A.}\ \bibnamefont {Houck}},
  \bibinfo {author} {\bibfnamefont {Karyn~Le}\ \bibnamefont {Hur}}, \ and\
  \bibinfo {author} {\bibfnamefont {S.~M.}\ \bibnamefont {Girvin}}} (\bibinfo
  {year} {2010}),\ \bibfield  {title} {\enquote {\bibinfo {title}
  {Time-reversal-symmetry breaking in circuit-{QED}-based photon lattices},}\
  }\href {http://link.aps.org/doi/10.1103/PhysRevA.82.043811} {\bibfield
  {journal} {\bibinfo  {journal} {Phys. Rev. A}\ }\textbf {\bibinfo {volume}
  {82}},\ \bibinfo {pages} {043811}}\BibitemShut {NoStop}%
\bibitem [{\citenamefont {Kohmoto}(1985)}]{Kohmoto:1985AnnPhys}%
  \BibitemOpen
  \bibfield  {author} {\bibinfo {author} {\bibnamefont {Kohmoto}, \bibfnamefont
  {Mahito}}} (\bibinfo {year} {1985}),\ \bibfield  {title} {\enquote {\bibinfo
  {title} {Topological invariant and the quantization of the {H}all
  conductance},}\ }\href
  {http://www.sciencedirect.com/science/article/pii/0003491685901484}
  {\bibfield  {journal} {\bibinfo  {journal} {Annals of Physics}\ }\textbf
  {\bibinfo {volume} {160}}~(\bibinfo {number} {2}),\ \bibinfo {pages} {343 --
  354}}\BibitemShut {NoStop}%
\bibitem [{\citenamefont {Kohmoto}\ \emph {et~al.}(1992)\citenamefont
  {Kohmoto}, \citenamefont {Halperin},\ and\ \citenamefont
  {Wu}}]{Kohmoto:1992PRB}%
  \BibitemOpen
  \bibfield  {author} {\bibinfo {author} {\bibnamefont {Kohmoto}, \bibfnamefont
  {Mahito}}, \bibinfo {author} {\bibfnamefont {Bertrand~I.}\ \bibnamefont
  {Halperin}}, \ and\ \bibinfo {author} {\bibfnamefont {Yong-Shi}\ \bibnamefont
  {Wu}}} (\bibinfo {year} {1992}),\ \bibfield  {title} {\enquote {\bibinfo
  {title} {Diophantine equation for the three-dimensional quantum {H}all
  effect},}\ }\href {https://link.aps.org/doi/10.1103/PhysRevB.45.13488}
  {\bibfield  {journal} {\bibinfo  {journal} {Phys. Rev. B}\ }\textbf {\bibinfo
  {volume} {45}},\ \bibinfo {pages} {13488--13493}}\BibitemShut {NoStop}%
\bibitem [{\citenamefont {Kolkowitz}\ \emph {et~al.}(2017)\citenamefont
  {Kolkowitz}, \citenamefont {Bromley}, \citenamefont {Bothwell}, \citenamefont
  {Wall}, \citenamefont {Marti}, \citenamefont {Koller}, \citenamefont {Zhang},
  \citenamefont {Rey},\ and\ \citenamefont {Ye}}]{Kolkowitz:2017Nature}%
  \BibitemOpen
  \bibfield  {author} {\bibinfo {author} {\bibnamefont {Kolkowitz},
  \bibfnamefont {S}}, \bibinfo {author} {\bibfnamefont {SL}~\bibnamefont
  {Bromley}}, \bibinfo {author} {\bibfnamefont {T}~\bibnamefont {Bothwell}},
  \bibinfo {author} {\bibfnamefont {ML}~\bibnamefont {Wall}}, \bibinfo {author}
  {\bibfnamefont {GE}~\bibnamefont {Marti}}, \bibinfo {author} {\bibfnamefont
  {AP}~\bibnamefont {Koller}}, \bibinfo {author} {\bibfnamefont
  {X}~\bibnamefont {Zhang}}, \bibinfo {author} {\bibfnamefont {AM}~\bibnamefont
  {Rey}}, \ and\ \bibinfo {author} {\bibfnamefont {J}~\bibnamefont {Ye}}}
  (\bibinfo {year} {2017}),\ \bibfield  {title} {\enquote {\bibinfo {title}
  {Spin--orbit-coupled fermions in an optical lattice clock},}\ }\href
  {https://www.nature.com/articles/nature20811} {\bibfield  {journal} {\bibinfo
   {journal} {Nature}\ }\textbf {\bibinfo {volume} {542}}~(\bibinfo {number}
  {7639}),\ \bibinfo {pages} {66--70}}\BibitemShut {NoStop}%
\bibitem [{\citenamefont {Kolovsky}(2011)}]{Kolovsky:2011EPL}%
  \BibitemOpen
  \bibfield  {author} {\bibinfo {author} {\bibnamefont {Kolovsky},
  \bibfnamefont {A~R}}} (\bibinfo {year} {2011}),\ \bibfield  {title} {\enquote
  {\bibinfo {title} {{Creating artificial magnetic fields for cold atoms by
  photon-assisted tunneling}},}\ }\href
  {https://doi.org/10.1209/0295-5075/93/20003} {\bibfield  {journal} {\bibinfo
  {journal} {EPL}\ }\textbf {\bibinfo {volume} {93}},\ \bibinfo {pages}
  {20003}}\BibitemShut {NoStop}%
\bibitem [{\citenamefont {K{\"o}nig}\ \emph {et~al.}(2007)\citenamefont
  {K{\"o}nig}, \citenamefont {Wiedmann}, \citenamefont {Br{\"u}ne},
  \citenamefont {Roth}, \citenamefont {Buhmann}, \citenamefont {Molenkamp},
  \citenamefont {Qi},\ and\ \citenamefont {Zhang}}]{Konig:2007Science}%
  \BibitemOpen
  \bibfield  {author} {\bibinfo {author} {\bibnamefont {K{\"o}nig},
  \bibfnamefont {Markus}}, \bibinfo {author} {\bibfnamefont {Steffen}\
  \bibnamefont {Wiedmann}}, \bibinfo {author} {\bibfnamefont {Christoph}\
  \bibnamefont {Br{\"u}ne}}, \bibinfo {author} {\bibfnamefont {Andreas}\
  \bibnamefont {Roth}}, \bibinfo {author} {\bibfnamefont {Hartmut}\
  \bibnamefont {Buhmann}}, \bibinfo {author} {\bibfnamefont {Laurens~W.}\
  \bibnamefont {Molenkamp}}, \bibinfo {author} {\bibfnamefont {Xiao-Liang}\
  \bibnamefont {Qi}}, \ and\ \bibinfo {author} {\bibfnamefont {Shou-Cheng}\
  \bibnamefont {Zhang}}} (\bibinfo {year} {2007}),\ \bibfield  {title}
  {\enquote {\bibinfo {title} {Quantum spin {H}all insulator state in {HgTe}
  quantum wells},}\ }\href {http://science.sciencemag.org/content/318/5851/766}
  {\bibfield  {journal} {\bibinfo  {journal} {Science}\ }\textbf {\bibinfo
  {volume} {318}}~(\bibinfo {number} {5851}),\ \bibinfo {pages}
  {766--770}}\BibitemShut {NoStop}%
\bibitem [{\citenamefont {Koshino}\ \emph {et~al.}(2002)\citenamefont
  {Koshino}, \citenamefont {Aoki},\ and\ \citenamefont
  {Halperin}}]{Koshino:2002PRB}%
  \BibitemOpen
  \bibfield  {author} {\bibinfo {author} {\bibnamefont {Koshino}, \bibfnamefont
  {Mikito}}, \bibinfo {author} {\bibfnamefont {Hideo}\ \bibnamefont {Aoki}}, \
  and\ \bibinfo {author} {\bibfnamefont {Bertrand~I.}\ \bibnamefont
  {Halperin}}} (\bibinfo {year} {2002}),\ \bibfield  {title} {\enquote
  {\bibinfo {title} {Wrapping current versus bulk integer quantum {H}all effect
  in three dimensions},}\ }\href
  {https://link.aps.org/doi/10.1103/PhysRevB.66.081301} {\bibfield  {journal}
  {\bibinfo  {journal} {Phys. Rev. B}\ }\textbf {\bibinfo {volume} {66}},\
  \bibinfo {pages} {081301}}\BibitemShut {NoStop}%
\bibitem [{\citenamefont {Kouwenhoven}\ \emph {et~al.}(1991)\citenamefont
  {Kouwenhoven}, \citenamefont {Johnson}, \citenamefont {van~der Vaart},
  \citenamefont {Harmans},\ and\ \citenamefont {Foxon}}]{Kouwenhoven:1991}%
  \BibitemOpen
  \bibfield  {author} {\bibinfo {author} {\bibnamefont {Kouwenhoven},
  \bibfnamefont {L~P}}, \bibinfo {author} {\bibfnamefont {A.~T.}\ \bibnamefont
  {Johnson}}, \bibinfo {author} {\bibfnamefont {N.~C.}\ \bibnamefont {van~der
  Vaart}}, \bibinfo {author} {\bibfnamefont {C.~J. P.~M.}\ \bibnamefont
  {Harmans}}, \ and\ \bibinfo {author} {\bibfnamefont {C.~T.}\ \bibnamefont
  {Foxon}}} (\bibinfo {year} {1991}),\ \bibfield  {title} {\enquote {\bibinfo
  {title} {Quantized current in a quantum-dot turnstile using oscillating
  tunnel barriers},}\ }\href
  {https://link.aps.org/doi/10.1103/PhysRevLett.67.1626} {\bibfield  {journal}
  {\bibinfo  {journal} {Phys. Rev. Lett.}\ }\textbf {\bibinfo {volume} {67}},\
  \bibinfo {pages} {1626--1629}}\BibitemShut {NoStop}%
\bibitem [{\citenamefont {Kozii}\ and\ \citenamefont
  {Fu}(2017)}]{kozii2017non}%
  \BibitemOpen
  \bibfield  {author} {\bibinfo {author} {\bibnamefont {Kozii}, \bibfnamefont
  {Vladyslav}}, \ and\ \bibinfo {author} {\bibfnamefont {Liang}\ \bibnamefont
  {Fu}}} (\bibinfo {year} {2017}),\ \bibfield  {title} {\enquote {\bibinfo
  {title} {Non-{H}ermitian topological theory of finite-lifetime
  quasiparticles: {P}rediction of bulk {F}ermi arc due to exceptional point},}\
  }\href {https://arxiv.org/abs/1708.05841} {\bibinfo  {journal}
  {arXiv:1708.05841}\ }\BibitemShut {NoStop}%
\bibitem [{\citenamefont {Kraus}\ \emph {et~al.}(2012)\citenamefont {Kraus},
  \citenamefont {Lahini}, \citenamefont {Ringel}, \citenamefont {Verbin},\ and\
  \citenamefont {Zilberberg}}]{Kraus:2012a}%
  \BibitemOpen
\bibfield  {journal} {  }\bibfield  {author} {\bibinfo {author} {\bibnamefont
  {Kraus}, \bibfnamefont {Yaacov~E}}, \bibinfo {author} {\bibfnamefont {Yoav}\
  \bibnamefont {Lahini}}, \bibinfo {author} {\bibfnamefont {Zohar}\
  \bibnamefont {Ringel}}, \bibinfo {author} {\bibfnamefont {Mor}\ \bibnamefont
  {Verbin}}, \ and\ \bibinfo {author} {\bibfnamefont {Oded}\ \bibnamefont
  {Zilberberg}}} (\bibinfo {year} {2012}),\ \bibfield  {title} {\enquote
  {\bibinfo {title} {Topological states and adiabatic pumping in
  quasicrystals},}\ }\href
  {https://link.aps.org/doi/10.1103/PhysRevLett.109.106402} {\bibfield
  {journal} {\bibinfo  {journal} {Phys. Rev. Lett.}\ }\textbf {\bibinfo
  {volume} {109}},\ \bibinfo {pages} {106402}}\BibitemShut {NoStop}%
\bibitem [{\citenamefont {Kraus}\ \emph {et~al.}(2013)\citenamefont {Kraus},
  \citenamefont {Ringel},\ and\ \citenamefont {Zilberberg}}]{Kraus:2013}%
  \BibitemOpen
  \bibfield  {author} {\bibinfo {author} {\bibnamefont {Kraus}, \bibfnamefont
  {Yaacov~E}}, \bibinfo {author} {\bibfnamefont {Zohar}\ \bibnamefont
  {Ringel}}, \ and\ \bibinfo {author} {\bibfnamefont {Oded}\ \bibnamefont
  {Zilberberg}}} (\bibinfo {year} {2013}),\ \bibfield  {title} {\enquote
  {\bibinfo {title} {Four-dimensional quantum {H}all effect in a
  two-dimensional quasicrystal},}\ }\href
  {https://link.aps.org/doi/10.1103/PhysRevLett.111.226401} {\bibfield
  {journal} {\bibinfo  {journal} {Phys. Rev. Lett.}\ }\textbf {\bibinfo
  {volume} {111}},\ \bibinfo {pages} {226401}}\BibitemShut {NoStop}%
\bibitem [{\citenamefont {Kraus}\ and\ \citenamefont
  {Zilberberg}(2012)}]{Kraus:2012b}%
  \BibitemOpen
  \bibfield  {author} {\bibinfo {author} {\bibnamefont {Kraus}, \bibfnamefont
  {Yaacov~E}}, \ and\ \bibinfo {author} {\bibfnamefont {Oded}\ \bibnamefont
  {Zilberberg}}} (\bibinfo {year} {2012}),\ \bibfield  {title} {\enquote
  {\bibinfo {title} {Topological equivalence between the {F}ibonacci
  quasicrystal and the {H}arper model},}\ }\href
  {https://link.aps.org/doi/10.1103/PhysRevLett.109.116404} {\bibfield
  {journal} {\bibinfo  {journal} {Phys. Rev. Lett.}\ }\textbf {\bibinfo
  {volume} {109}},\ \bibinfo {pages} {116404}}\BibitemShut {NoStop}%
\bibitem [{\citenamefont {Krimer}\ and\ \citenamefont
  {Khomeriki}(2011)}]{Krimer:PRA2011}%
  \BibitemOpen
  \bibfield  {author} {\bibinfo {author} {\bibnamefont {Krimer}, \bibfnamefont
  {Dmitry~O}}, \ and\ \bibinfo {author} {\bibfnamefont {Ramaz}\ \bibnamefont
  {Khomeriki}}} (\bibinfo {year} {2011}),\ \bibfield  {title} {\enquote
  {\bibinfo {title} {Realization of discrete quantum billiards in a
  two-dimensional optical lattice},}\ }\href
  {https://link.aps.org/doi/10.1103/PhysRevA.84.041807} {\bibfield  {journal}
  {\bibinfo  {journal} {Phys. Rev. A}\ }\textbf {\bibinfo {volume} {84}},\
  \bibinfo {pages} {041807}}\BibitemShut {NoStop}%
\bibitem [{\citenamefont {Kruk}\ \emph {et~al.}(2017)\citenamefont {Kruk},
  \citenamefont {Slobozhanyuk}, \citenamefont {Denkova}, \citenamefont
  {Poddubny}, \citenamefont {Kravchenko}, \citenamefont {Miroshnichenko},
  \citenamefont {Neshev},\ and\ \citenamefont {Kivshar}}]{Kruk:2017Small}%
  \BibitemOpen
  \bibfield  {author} {\bibinfo {author} {\bibnamefont {Kruk}, \bibfnamefont
  {Sergey}}, \bibinfo {author} {\bibfnamefont {Alexey}\ \bibnamefont
  {Slobozhanyuk}}, \bibinfo {author} {\bibfnamefont {Denitza}\ \bibnamefont
  {Denkova}}, \bibinfo {author} {\bibfnamefont {Alexander}\ \bibnamefont
  {Poddubny}}, \bibinfo {author} {\bibfnamefont {Ivan}\ \bibnamefont
  {Kravchenko}}, \bibinfo {author} {\bibfnamefont {Andrey}\ \bibnamefont
  {Miroshnichenko}}, \bibinfo {author} {\bibfnamefont {Dragomir}\ \bibnamefont
  {Neshev}}, \ and\ \bibinfo {author} {\bibfnamefont {Yuri}\ \bibnamefont
  {Kivshar}}} (\bibinfo {year} {2017}),\ \bibfield  {title} {\enquote {\bibinfo
  {title} {Edge states and topological phase transitions in chains of
  dielectric nanoparticles},}\ }\href
  {http://dx.doi.org/10.1002/smll.201603190} {\bibfield  {journal} {\bibinfo
  {journal} {Small}\ }\textbf {\bibinfo {volume} {13}}~(\bibinfo {number}
  {11}),\ \bibinfo {pages} {1603190}}\BibitemShut {NoStop}%
\bibitem [{\citenamefont {Kuhl}\ and\ \citenamefont
  {St\"ockmann}(1998)}]{Kuhl:1998PRL}%
  \BibitemOpen
  \bibfield  {author} {\bibinfo {author} {\bibnamefont {Kuhl}, \bibfnamefont
  {U}}, \ and\ \bibinfo {author} {\bibfnamefont {H.-J.}\ \bibnamefont
  {St\"ockmann}}} (\bibinfo {year} {1998}),\ \bibfield  {title} {\enquote
  {\bibinfo {title} {Microwave realization of the {H}ofstadter butterfly},}\
  }\href {https://link.aps.org/doi/10.1103/PhysRevLett.80.3232} {\bibfield
  {journal} {\bibinfo  {journal} {Phys. Rev. Lett.}\ }\textbf {\bibinfo
  {volume} {80}},\ \bibinfo {pages} {3232--3235}}\BibitemShut {NoStop}%
\bibitem [{\citenamefont {Kusudo}\ \emph {et~al.}(2013)\citenamefont {Kusudo},
  \citenamefont {Kim}, \citenamefont {L{\"{o}}ffler}, \citenamefont
  {H{\"{o}}fling}, \citenamefont {Forchel},\ and\ \citenamefont
  {Yamamoto}}]{Kusudo:2013PRB}%
  \BibitemOpen
  \bibfield  {author} {\bibinfo {author} {\bibnamefont {Kusudo}, \bibfnamefont
  {Kenichiro}}, \bibinfo {author} {\bibfnamefont {Na~Young}\ \bibnamefont
  {Kim}}, \bibinfo {author} {\bibfnamefont {Andreas}\ \bibnamefont
  {L{\"{o}}ffler}}, \bibinfo {author} {\bibfnamefont {Sven}\ \bibnamefont
  {H{\"{o}}fling}}, \bibinfo {author} {\bibfnamefont {Alfred}\ \bibnamefont
  {Forchel}}, \ and\ \bibinfo {author} {\bibfnamefont {Yoshihisa}\ \bibnamefont
  {Yamamoto}}} (\bibinfo {year} {2013}),\ \bibfield  {title} {\enquote
  {\bibinfo {title} {{Stochastic formation of polariton condensates in two
  degenerate orbital states}},}\ }\href
  {http://link.aps.org/doi/10.1103/PhysRevB.87.214503} {\bibfield  {journal}
  {\bibinfo  {journal} {Phys. Rev. B}\ }\textbf {\bibinfo {volume}
  {87}}~(\bibinfo {number} {21}),\ \bibinfo {pages} {214503}}\BibitemShut
  {NoStop}%
\bibitem [{\citenamefont {Lahini}\ \emph {et~al.}(2008)\citenamefont {Lahini},
  \citenamefont {Avidan}, \citenamefont {Pozzi}, \citenamefont {Sorel},
  \citenamefont {Morandotti}, \citenamefont {Christodoulides},\ and\
  \citenamefont {Silberberg}}]{lahini2008anderson}%
  \BibitemOpen
  \bibfield  {author} {\bibinfo {author} {\bibnamefont {Lahini}, \bibfnamefont
  {Yoav}}, \bibinfo {author} {\bibfnamefont {Assaf}\ \bibnamefont {Avidan}},
  \bibinfo {author} {\bibfnamefont {Francesca}\ \bibnamefont {Pozzi}}, \bibinfo
  {author} {\bibfnamefont {Marc}\ \bibnamefont {Sorel}}, \bibinfo {author}
  {\bibfnamefont {Roberto}\ \bibnamefont {Morandotti}}, \bibinfo {author}
  {\bibfnamefont {Demetrios~N}\ \bibnamefont {Christodoulides}}, \ and\
  \bibinfo {author} {\bibfnamefont {Yaron}\ \bibnamefont {Silberberg}}}
  (\bibinfo {year} {2008}),\ \bibfield  {title} {\enquote {\bibinfo {title}
  {Anderson localization and nonlinearity in one-dimensional disordered
  photonic lattices},}\ }\href
  {https://journals.aps.org/prl/abstract/10.1103/PhysRevLett.100.013906}
  {\bibfield  {journal} {\bibinfo  {journal} {Phys. Rev. Lett.}\ }\textbf
  {\bibinfo {volume} {100}}~(\bibinfo {number} {1}),\ \bibinfo {pages}
  {013906}}\BibitemShut {NoStop}%
\bibitem [{\citenamefont {Lai}\ \emph {et~al.}(2016)\citenamefont {Lai},
  \citenamefont {Ma}, \citenamefont {Bo}, \citenamefont {Anlage},\ and\
  \citenamefont {Shvets}}]{lai:2016SciRep}%
  \BibitemOpen
  \bibfield  {author} {\bibinfo {author} {\bibnamefont {Lai}, \bibfnamefont
  {Kueifu}}, \bibinfo {author} {\bibfnamefont {Tsuhsuang}\ \bibnamefont {Ma}},
  \bibinfo {author} {\bibfnamefont {Xiao}\ \bibnamefont {Bo}}, \bibinfo
  {author} {\bibfnamefont {Steven}\ \bibnamefont {Anlage}}, \ and\ \bibinfo
  {author} {\bibfnamefont {Gennady}\ \bibnamefont {Shvets}}} (\bibinfo {year}
  {2016}),\ \bibfield  {title} {\enquote {\bibinfo {title} {Experimental
  realization of a reflections-free compact delay line based on a photonic
  topological insulator},}\ }\href {https://www.nature.com/articles/srep28453}
  {\bibfield  {journal} {\bibinfo  {journal} {Sci. Rep.}\ }\textbf {\bibinfo
  {volume} {6}},\ \bibinfo {pages} {28453}}\BibitemShut {NoStop}%
\bibitem [{\citenamefont {Lai}\ and\ \citenamefont
  {Haus}(1989{\natexlab{a}})}]{Lai:PRA1989}%
  \BibitemOpen
  \bibfield  {author} {\bibinfo {author} {\bibnamefont {Lai}, \bibfnamefont
  {Y}}, \ and\ \bibinfo {author} {\bibfnamefont {H.~A.}\ \bibnamefont {Haus}}}
  (\bibinfo {year} {1989}{\natexlab{a}}),\ \bibfield  {title} {\enquote
  {\bibinfo {title} {{Quantum theory of solitons in optical fibers. I.
  Time-dependent Hartree approximation}},}\ }\href
  {http://link.aps.org/doi/10.1103/PhysRevA.40.844} {\bibfield  {journal}
  {\bibinfo  {journal} {Phys. Rev. A}\ }\textbf {\bibinfo {volume} {40}},\
  \bibinfo {pages} {844--853}}\BibitemShut {NoStop}%
\bibitem [{\citenamefont {Lai}\ and\ \citenamefont
  {Haus}(1989{\natexlab{b}})}]{Lai:PRA1989b}%
  \BibitemOpen
  \bibfield  {author} {\bibinfo {author} {\bibnamefont {Lai}, \bibfnamefont
  {Y}}, \ and\ \bibinfo {author} {\bibfnamefont {H.~A.}\ \bibnamefont {Haus}}}
  (\bibinfo {year} {1989}{\natexlab{b}}),\ \bibfield  {title} {\enquote
  {\bibinfo {title} {{Quantum theory of solitons in optical fibers. II. Exact
  solution}},}\ }\href {http://link.aps.org/doi/10.1103/PhysRevA.40.854}
  {\bibfield  {journal} {\bibinfo  {journal} {Phys. Rev. A}\ }\textbf {\bibinfo
  {volume} {40}},\ \bibinfo {pages} {854--866}}\BibitemShut {NoStop}%
\bibitem [{\citenamefont {Lang}\ \emph {et~al.}(2011)\citenamefont {Lang},
  \citenamefont {Bozyigit}, \citenamefont {Eichler}, \citenamefont {Steffen},
  \citenamefont {Fink}, \citenamefont {Abdumalikov}, \citenamefont {Baur},
  \citenamefont {Filipp}, \citenamefont {da~Silva}, \citenamefont {Blais},\
  and\ \citenamefont {Wallraff}}]{Lang:PRL2011}%
  \BibitemOpen
  \bibfield  {author} {\bibinfo {author} {\bibnamefont {Lang}, \bibfnamefont
  {C}}, \bibinfo {author} {\bibfnamefont {D.}~\bibnamefont {Bozyigit}},
  \bibinfo {author} {\bibfnamefont {C.}~\bibnamefont {Eichler}}, \bibinfo
  {author} {\bibfnamefont {L.}~\bibnamefont {Steffen}}, \bibinfo {author}
  {\bibfnamefont {J.~M.}\ \bibnamefont {Fink}}, \bibinfo {author}
  {\bibfnamefont {A.~A.}\ \bibnamefont {Abdumalikov}, \bibfnamefont {Jr.}},
  \bibinfo {author} {\bibfnamefont {M.}~\bibnamefont {Baur}}, \bibinfo {author}
  {\bibfnamefont {S.}~\bibnamefont {Filipp}}, \bibinfo {author} {\bibfnamefont
  {M.~P.}\ \bibnamefont {da~Silva}}, \bibinfo {author} {\bibfnamefont
  {A.}~\bibnamefont {Blais}}, \ and\ \bibinfo {author} {\bibfnamefont
  {A.}~\bibnamefont {Wallraff}}} (\bibinfo {year} {2011}),\ \bibfield  {title}
  {\enquote {\bibinfo {title} {Observation of resonant photon blockade at
  microwave frequencies using correlation function measurements},}\ }\href
  {https://journals.aps.org/prl/abstract/10.1103/PhysRevLett.106.243601}
  {\bibfield  {journal} {\bibinfo  {journal} {Phys. Rev. Lett.}\ }\textbf
  {\bibinfo {volume} {106}}~(\bibinfo {number} {24})}\BibitemShut {NoStop}%
\bibitem [{\citenamefont {Larr{\'e}}\ and\ \citenamefont
  {Carusotto}(2015)}]{Larre:PRA2015}%
  \BibitemOpen
  \bibfield  {author} {\bibinfo {author} {\bibnamefont {Larr{\'e}},
  \bibfnamefont {Pierre-{\'E}lie}}, \ and\ \bibinfo {author} {\bibfnamefont
  {Iacopo}\ \bibnamefont {Carusotto}}} (\bibinfo {year} {2015}),\ \bibfield
  {title} {\enquote {\bibinfo {title} {Propagation of a quantum fluid of light
  in a cavityless nonlinear optical medium: {G}eneral theory and response to
  quantum quenches},}\ }\href
  {https://journals.aps.org/pra/abstract/10.1103/PhysRevA.92.043802} {\bibfield
   {journal} {\bibinfo  {journal} {Phys. Rev. A}\ }\textbf {\bibinfo {volume}
  {92}}~(\bibinfo {number} {4}),\ \bibinfo {pages} {043802}}\BibitemShut
  {NoStop}%
\bibitem [{\citenamefont {Larr{\'e}}\ and\ \citenamefont
  {Carusotto}(2016)}]{Larre:EPJD2016}%
  \BibitemOpen
  \bibfield  {author} {\bibinfo {author} {\bibnamefont {Larr{\'e}},
  \bibfnamefont {Pierre-{\'E}lie}}, \ and\ \bibinfo {author} {\bibfnamefont
  {Iacopo}\ \bibnamefont {Carusotto}}} (\bibinfo {year} {2016}),\ \bibfield
  {title} {\enquote {\bibinfo {title} {Prethermalization in a quenched
  one-dimensional quantum fluid of light. intrinsic limits to the coherent
  propagation of a light beam in a nonlinear optical fiber},}\ }\href
  {https://link.springer.com/article/10.1140%2Fepjd%2Fe2016-60590-2} {\bibfield
   {journal} {\bibinfo  {journal} {Eur. Phys. J. D}\ }\textbf {\bibinfo
  {volume} {70}},\ \bibinfo {pages} {45}}\BibitemShut {NoStop}%
\bibitem [{\citenamefont {Laughlin}(1981)}]{Laughlin:1981}%
  \BibitemOpen
  \bibfield  {author} {\bibinfo {author} {\bibnamefont {Laughlin},
  \bibfnamefont {R~B}}} (\bibinfo {year} {1981}),\ \bibfield  {title} {\enquote
  {\bibinfo {title} {Quantized {H}all conductivity in two dimensions},}\ }\href
  {https://link.aps.org/doi/10.1103/PhysRevB.23.5632} {\bibfield  {journal}
  {\bibinfo  {journal} {Phys. Rev. B}\ }\textbf {\bibinfo {volume} {23}},\
  \bibinfo {pages} {5632--5633}}\BibitemShut {NoStop}%
\bibitem [{\citenamefont {Laughlin}(1983)}]{Laughlin:1983PRL}%
  \BibitemOpen
  \bibfield  {author} {\bibinfo {author} {\bibnamefont {Laughlin},
  \bibfnamefont {R~B}}} (\bibinfo {year} {1983}),\ \bibfield  {title} {\enquote
  {\bibinfo {title} {Anomalous quantum {H}all effect: {A}n incompressible
  quantum fluid with fractionally charged excitations},}\ }\href
  {https://link.aps.org/doi/10.1103/PhysRevLett.50.1395} {\bibfield  {journal}
  {\bibinfo  {journal} {Phys. Rev. Lett.}\ }\textbf {\bibinfo {volume} {50}},\
  \bibinfo {pages} {1395--1398}}\BibitemShut {NoStop}%
\bibitem [{\citenamefont {{\L}\c{a}cki}\ \emph {et~al.}(2016)\citenamefont
  {{\L}\c{a}cki}, \citenamefont {Pichler}, \citenamefont {Sterdyniak},
  \citenamefont {Lyras}, \citenamefont {Lembessis}, \citenamefont {Al-Dossary},
  \citenamefont {Budich},\ and\ \citenamefont {Zoller}}]{Lacki:2016PRA}%
  \BibitemOpen
  \bibfield  {author} {\bibinfo {author} {\bibnamefont {{\L}\c{a}cki},
  \bibfnamefont {Mateusz}}, \bibinfo {author} {\bibfnamefont {Hannes}\
  \bibnamefont {Pichler}}, \bibinfo {author} {\bibfnamefont {Antoine}\
  \bibnamefont {Sterdyniak}}, \bibinfo {author} {\bibfnamefont {Andreas}\
  \bibnamefont {Lyras}}, \bibinfo {author} {\bibfnamefont {Vassilis~E.}\
  \bibnamefont {Lembessis}}, \bibinfo {author} {\bibfnamefont {Omar}\
  \bibnamefont {Al-Dossary}}, \bibinfo {author} {\bibfnamefont {Jan~Carl}\
  \bibnamefont {Budich}}, \ and\ \bibinfo {author} {\bibfnamefont {Peter}\
  \bibnamefont {Zoller}}} (\bibinfo {year} {2016}),\ \bibfield  {title}
  {\enquote {\bibinfo {title} {Quantum {H}all physics with cold atoms in
  cylindrical optical lattices},}\ }\href
  {https://link.aps.org/doi/10.1103/PhysRevA.93.013604} {\bibfield  {journal}
  {\bibinfo  {journal} {Phys. Rev. A}\ }\textbf {\bibinfo {volume} {93}},\
  \bibinfo {pages} {013604}}\BibitemShut {NoStop}%
\bibitem [{\citenamefont {Le~Feber}\ \emph {et~al.}(2015)\citenamefont
  {Le~Feber}, \citenamefont {Rotenberg},\ and\ \citenamefont
  {Kuipers}}]{lefeber:2015naturecommunications}%
  \BibitemOpen
  \bibfield  {author} {\bibinfo {author} {\bibnamefont {Le~Feber},
  \bibfnamefont {B}}, \bibinfo {author} {\bibfnamefont {N}~\bibnamefont
  {Rotenberg}}, \ and\ \bibinfo {author} {\bibfnamefont {L}~\bibnamefont
  {Kuipers}}} (\bibinfo {year} {2015}),\ \bibfield  {title} {\enquote {\bibinfo
  {title} {Nanophotonic control of circular dipole emission},}\ }\href
  {https://www.nature.com/articles/ncomms7695} {\bibfield  {journal} {\bibinfo
  {journal} {Nat. Commun.}\ }\textbf {\bibinfo {volume} {6}},\ \bibinfo {pages}
  {6695}}\BibitemShut {NoStop}%
\bibitem [{\citenamefont {Lebreuilly}\ \emph {et~al.}(2017)\citenamefont
  {Lebreuilly}, \citenamefont {Biella}, \citenamefont {Storme}, \citenamefont
  {Rossini}, \citenamefont {Fazio}, \citenamefont {Ciuti},\ and\ \citenamefont
  {Carusotto}}]{Lebreuilly:PRA2017}%
  \BibitemOpen
  \bibfield  {author} {\bibinfo {author} {\bibnamefont {Lebreuilly},
  \bibfnamefont {Jos\'e}}, \bibinfo {author} {\bibfnamefont {Alberto}\
  \bibnamefont {Biella}}, \bibinfo {author} {\bibfnamefont {Florent}\
  \bibnamefont {Storme}}, \bibinfo {author} {\bibfnamefont {Davide}\
  \bibnamefont {Rossini}}, \bibinfo {author} {\bibfnamefont {Rosario}\
  \bibnamefont {Fazio}}, \bibinfo {author} {\bibfnamefont {Cristiano}\
  \bibnamefont {Ciuti}}, \ and\ \bibinfo {author} {\bibfnamefont {Iacopo}\
  \bibnamefont {Carusotto}}} (\bibinfo {year} {2017}),\ \bibfield  {title}
  {\enquote {\bibinfo {title} {Stabilizing strongly correlated photon fluids
  with non-{M}arkovian reservoirs},}\ }\href
  {https://link.aps.org/doi/10.1103/PhysRevA.96.033828} {\bibfield  {journal}
  {\bibinfo  {journal} {Phys. Rev. A}\ }\textbf {\bibinfo {volume} {96}},\
  \bibinfo {pages} {033828}}\BibitemShut {NoStop}%
\bibitem [{\citenamefont {Lebreuilly}\ \emph {et~al.}(2016)\citenamefont
  {Lebreuilly}, \citenamefont {Wouters},\ and\ \citenamefont
  {Carusotto}}]{Lebreuilly:CRAS2016}%
  \BibitemOpen
  \bibfield  {author} {\bibinfo {author} {\bibnamefont {Lebreuilly},
  \bibfnamefont {Jos{\'e}}}, \bibinfo {author} {\bibfnamefont {Michiel}\
  \bibnamefont {Wouters}}, \ and\ \bibinfo {author} {\bibfnamefont {Iacopo}\
  \bibnamefont {Carusotto}}} (\bibinfo {year} {2016}),\ \bibfield  {title}
  {\enquote {\bibinfo {title} {Towards strongly correlated photons in arrays of
  dissipative nonlinear cavities under a frequency-dependent incoherent
  pumping},}\ }\href
  {https://www.sciencedirect.com/science/article/pii/S1631070516300482}
  {\bibfield  {journal} {\bibinfo  {journal} {Comptes Rendus Physique}\
  }\textbf {\bibinfo {volume} {17}}~(\bibinfo {number} {8}),\ \bibinfo {pages}
  {836--860}}\BibitemShut {NoStop}%
\bibitem [{\citenamefont {Lee}\ \emph {et~al.}(2018)\citenamefont {Lee},
  \citenamefont {Imhof}, \citenamefont {Berger}, \citenamefont {Bayer},
  \citenamefont {Brehm}, \citenamefont {Molenkamp}, \citenamefont {Kiessling},\
  and\ \citenamefont {Thomale}}]{Lee:2018CommPhys}%
  \BibitemOpen
  \bibfield  {author} {\bibinfo {author} {\bibnamefont {Lee}, \bibfnamefont
  {Ching~Hua}}, \bibinfo {author} {\bibfnamefont {Stefan}\ \bibnamefont
  {Imhof}}, \bibinfo {author} {\bibfnamefont {Christian}\ \bibnamefont
  {Berger}}, \bibinfo {author} {\bibfnamefont {Florian}\ \bibnamefont {Bayer}},
  \bibinfo {author} {\bibfnamefont {Johannes}\ \bibnamefont {Brehm}}, \bibinfo
  {author} {\bibfnamefont {Laurens~W}\ \bibnamefont {Molenkamp}}, \bibinfo
  {author} {\bibfnamefont {Tobias}\ \bibnamefont {Kiessling}}, \ and\ \bibinfo
  {author} {\bibfnamefont {Ronny}\ \bibnamefont {Thomale}}} (\bibinfo {year}
  {2018}),\ \bibfield  {title} {\enquote {\bibinfo {title} {Topolectrical
  circuits},}\ }\href {https://www.nature.com/articles/s42005-018-0035-2}
  {\bibfield  {journal} {\bibinfo  {journal} {Commun. Phys.}\ }\textbf
  {\bibinfo {volume} {1}}~(\bibinfo {number} {1}),\ \bibinfo {pages}
  {39}}\BibitemShut {NoStop}%
\bibitem [{\citenamefont {Lee}\ \emph {et~al.}(2012)\citenamefont {Lee},
  \citenamefont {Lee}, \citenamefont {Park}, \citenamefont {Oh}, \citenamefont
  {Lee}, \citenamefont {Kim},\ and\ \citenamefont {Lee}}]{Lee:2012PRL}%
  \BibitemOpen
  \bibfield  {author} {\bibinfo {author} {\bibnamefont {Lee}, \bibfnamefont
  {Seung-Yeol}}, \bibinfo {author} {\bibfnamefont {Il-Min}\ \bibnamefont
  {Lee}}, \bibinfo {author} {\bibfnamefont {Junghyun}\ \bibnamefont {Park}},
  \bibinfo {author} {\bibfnamefont {Sewoong}\ \bibnamefont {Oh}}, \bibinfo
  {author} {\bibfnamefont {Wooyoung}\ \bibnamefont {Lee}}, \bibinfo {author}
  {\bibfnamefont {Kyoung-Youm}\ \bibnamefont {Kim}}, \ and\ \bibinfo {author}
  {\bibfnamefont {Byoungho}\ \bibnamefont {Lee}}} (\bibinfo {year} {2012}),\
  \bibfield  {title} {\enquote {\bibinfo {title} {Role of magnetic induction
  currents in nanoslit excitation of surface plasmon polaritons},}\ }\href
  {https://link.aps.org/doi/10.1103/PhysRevLett.108.213907} {\bibfield
  {journal} {\bibinfo  {journal} {Phys. Rev. Lett.}\ }\textbf {\bibinfo
  {volume} {108}},\ \bibinfo {pages} {213907}}\BibitemShut {NoStop}%
\bibitem [{\citenamefont {Lee}(2016)}]{lee2016anomalous}%
  \BibitemOpen
  \bibfield  {author} {\bibinfo {author} {\bibnamefont {Lee}, \bibfnamefont
  {Tony~E}}} (\bibinfo {year} {2016}),\ \bibfield  {title} {\enquote {\bibinfo
  {title} {Anomalous edge state in a non-{H}ermitian lattice},}\ }\href
  {https://journals.aps.org/prl/abstract/10.1103/PhysRevLett.116.133903}
  {\bibfield  {journal} {\bibinfo  {journal} {Phys. Rev. Lett.}\ }\textbf
  {\bibinfo {volume} {116}}~(\bibinfo {number} {13}),\ \bibinfo {pages}
  {133903}}\BibitemShut {NoStop}%
\bibitem [{\citenamefont {Lee-Thorp}\ \emph {et~al.}(2016)\citenamefont
  {Lee-Thorp}, \citenamefont {Vuki{\'c}evi{\'c}}, \citenamefont {Xu},
  \citenamefont {Yang}, \citenamefont {Fefferman}, \citenamefont {Wong},\ and\
  \citenamefont {Weinstein}}]{lee2016photonic}%
  \BibitemOpen
  \bibfield  {author} {\bibinfo {author} {\bibnamefont {Lee-Thorp},
  \bibfnamefont {James~P}}, \bibinfo {author} {\bibfnamefont {Iva}\
  \bibnamefont {Vuki{\'c}evi{\'c}}}, \bibinfo {author} {\bibfnamefont {Xinan}\
  \bibnamefont {Xu}}, \bibinfo {author} {\bibfnamefont {Jinghui}\ \bibnamefont
  {Yang}}, \bibinfo {author} {\bibfnamefont {Charles~L}\ \bibnamefont
  {Fefferman}}, \bibinfo {author} {\bibfnamefont {Chee~Wei}\ \bibnamefont
  {Wong}}, \ and\ \bibinfo {author} {\bibfnamefont {Michael~I}\ \bibnamefont
  {Weinstein}}} (\bibinfo {year} {2016}),\ \bibfield  {title} {\enquote
  {\bibinfo {title} {Photonic realization of topologically protected bound
  states in domain-wall waveguide arrays},}\ }\href
  {https://journals.aps.org/pra/abstract/10.1103/PhysRevA.93.033822} {\bibfield
   {journal} {\bibinfo  {journal} {Phys. Rev. A}\ }\textbf {\bibinfo {volume}
  {93}}~(\bibinfo {number} {3}),\ \bibinfo {pages} {033822}}\BibitemShut
  {NoStop}%
\bibitem [{\citenamefont {Letscher}\ \emph {et~al.}(2015)\citenamefont
  {Letscher}, \citenamefont {Grusdt},\ and\ \citenamefont
  {Fleischhauer}}]{Letscher:PRB2015}%
  \BibitemOpen
  \bibfield  {author} {\bibinfo {author} {\bibnamefont {Letscher},
  \bibfnamefont {Fabian}}, \bibinfo {author} {\bibfnamefont {Fabian}\
  \bibnamefont {Grusdt}}, \ and\ \bibinfo {author} {\bibfnamefont {Michael}\
  \bibnamefont {Fleischhauer}}} (\bibinfo {year} {2015}),\ \bibfield  {title}
  {\enquote {\bibinfo {title} {Growing quantum states with topological
  order},}\ }\href
  {https://journals.aps.org/prb/abstract/10.1103/PhysRevB.91.184302} {\bibfield
   {journal} {\bibinfo  {journal} {Phys. Rev. B}\ }\textbf {\bibinfo {volume}
  {91}}~(\bibinfo {number} {18}),\ \bibinfo {pages} {184302}}\BibitemShut
  {NoStop}%
\bibitem [{\citenamefont {Levi}\ \emph {et~al.}(2011)\citenamefont {Levi},
  \citenamefont {Rechtsman}, \citenamefont {Freedman}, \citenamefont
  {Schwartz}, \citenamefont {Manela},\ and\ \citenamefont
  {Segev}}]{levi2011disorder}%
  \BibitemOpen
  \bibfield  {author} {\bibinfo {author} {\bibnamefont {Levi}, \bibfnamefont
  {Liad}}, \bibinfo {author} {\bibfnamefont {Mikael}\ \bibnamefont
  {Rechtsman}}, \bibinfo {author} {\bibfnamefont {Barak}\ \bibnamefont
  {Freedman}}, \bibinfo {author} {\bibfnamefont {Tal}\ \bibnamefont
  {Schwartz}}, \bibinfo {author} {\bibfnamefont {Ofer}\ \bibnamefont {Manela}},
  \ and\ \bibinfo {author} {\bibfnamefont {Mordechai}\ \bibnamefont {Segev}}}
  (\bibinfo {year} {2011}),\ \bibfield  {title} {\enquote {\bibinfo {title}
  {Disorder-enhanced transport in photonic quasicrystals},}\ }\href
  {http://science.sciencemag.org/content/332/6037/1541} {\bibfield  {journal}
  {\bibinfo  {journal} {Science}\ }\textbf {\bibinfo {volume} {332}}~(\bibinfo
  {number} {6037}),\ \bibinfo {pages} {1541--1544}}\BibitemShut {NoStop}%
\bibitem [{\citenamefont {Leyder}\ \emph {et~al.}(2007)\citenamefont {Leyder},
  \citenamefont {Romanelli}, \citenamefont {Karr}, \citenamefont {Giacobino},
  \citenamefont {Liew}, \citenamefont {Glazov}, \citenamefont {Kavokin},
  \citenamefont {Malpuech},\ and\ \citenamefont
  {Bramati}}]{Leyder:NatPhys2007}%
  \BibitemOpen
  \bibfield  {author} {\bibinfo {author} {\bibnamefont {Leyder}, \bibfnamefont
  {C}}, \bibinfo {author} {\bibfnamefont {M.}~\bibnamefont {Romanelli}},
  \bibinfo {author} {\bibfnamefont {J.~Ph.}\ \bibnamefont {Karr}}, \bibinfo
  {author} {\bibfnamefont {E.}~\bibnamefont {Giacobino}}, \bibinfo {author}
  {\bibfnamefont {T.~C.~H.}\ \bibnamefont {Liew}}, \bibinfo {author}
  {\bibfnamefont {M.~M.}\ \bibnamefont {Glazov}}, \bibinfo {author}
  {\bibfnamefont {A.~V.}\ \bibnamefont {Kavokin}}, \bibinfo {author}
  {\bibfnamefont {G.}~\bibnamefont {Malpuech}}, \ and\ \bibinfo {author}
  {\bibfnamefont {A.}~\bibnamefont {Bramati}}} (\bibinfo {year} {2007}),\
  \bibfield  {title} {\enquote {\bibinfo {title} {{Observation of the optical
  spin Hall effect}},}\ }\href {https://www.nature.com/articles/nphys676}
  {\bibfield  {journal} {\bibinfo  {journal} {Nat. Phys.}\ }\textbf {\bibinfo
  {volume} {3}}~(\bibinfo {number} {9}),\ \bibinfo {pages}
  {628--631}}\BibitemShut {NoStop}%
\bibitem [{\citenamefont {Leykam}\ \emph
  {et~al.}(2017{\natexlab{a}})\citenamefont {Leykam}, \citenamefont {Bliokh},
  \citenamefont {Huang}, \citenamefont {Chong},\ and\ \citenamefont
  {Nori}}]{leykam2017edge}%
  \BibitemOpen
  \bibfield  {author} {\bibinfo {author} {\bibnamefont {Leykam}, \bibfnamefont
  {Daniel}}, \bibinfo {author} {\bibfnamefont {Konstantin~Y}\ \bibnamefont
  {Bliokh}}, \bibinfo {author} {\bibfnamefont {Chunli}\ \bibnamefont {Huang}},
  \bibinfo {author} {\bibfnamefont {YD}~\bibnamefont {Chong}}, \ and\ \bibinfo
  {author} {\bibfnamefont {Franco}\ \bibnamefont {Nori}}} (\bibinfo {year}
  {2017}{\natexlab{a}}),\ \bibfield  {title} {\enquote {\bibinfo {title} {Edge
  modes, degeneracies, and topological numbers in non-{H}ermitian systems},}\
  }\href {https://journals.aps.org/prl/abstract/10.1103/PhysRevLett.118.040401}
  {\bibfield  {journal} {\bibinfo  {journal} {Phys. Rev. Lett.}\ }\textbf
  {\bibinfo {volume} {118}}~(\bibinfo {number} {4}),\ \bibinfo {pages}
  {040401}}\BibitemShut {NoStop}%
\bibitem [{\citenamefont {Leykam}\ and\ \citenamefont
  {Chong}(2016)}]{Leykam:PRL2016}%
  \BibitemOpen
  \bibfield  {author} {\bibinfo {author} {\bibnamefont {Leykam}, \bibfnamefont
  {Daniel}}, \ and\ \bibinfo {author} {\bibfnamefont {Y.~D.}\ \bibnamefont
  {Chong}}} (\bibinfo {year} {2016}),\ \bibfield  {title} {\enquote {\bibinfo
  {title} {Edge solitons in nonlinear-photonic topological insulators},}\
  }\href {https://link.aps.org/doi/10.1103/PhysRevLett.117.143901} {\bibfield
  {journal} {\bibinfo  {journal} {Phys. Rev. Lett.}\ }\textbf {\bibinfo
  {volume} {117}},\ \bibinfo {pages} {143901}}\BibitemShut {NoStop}%
\bibitem [{\citenamefont {Leykam}\ \emph
  {et~al.}(2017{\natexlab{b}})\citenamefont {Leykam}, \citenamefont {Flach},\
  and\ \citenamefont {Chong}}]{leykam2017flat}%
  \BibitemOpen
  \bibfield  {author} {\bibinfo {author} {\bibnamefont {Leykam}, \bibfnamefont
  {Daniel}}, \bibinfo {author} {\bibfnamefont {Sergej}\ \bibnamefont {Flach}},
  \ and\ \bibinfo {author} {\bibfnamefont {Y.~D.}\ \bibnamefont {Chong}}}
  (\bibinfo {year} {2017}{\natexlab{b}}),\ \bibfield  {title} {\enquote
  {\bibinfo {title} {Flat bands in lattices with non-{H}ermitian coupling},}\
  }\href {https://link.aps.org/doi/10.1103/PhysRevB.96.064305} {\bibfield
  {journal} {\bibinfo  {journal} {Phys. Rev. B}\ }\textbf {\bibinfo {volume}
  {96}},\ \bibinfo {pages} {064305}}\BibitemShut {NoStop}%
\bibitem [{\citenamefont {Leykam}\ \emph {et~al.}(2016)\citenamefont {Leykam},
  \citenamefont {Rechtsman},\ and\ \citenamefont
  {Chong}}]{Leykam:2016PRLanomalous}%
  \BibitemOpen
  \bibfield  {author} {\bibinfo {author} {\bibnamefont {Leykam}, \bibfnamefont
  {Daniel}}, \bibinfo {author} {\bibfnamefont {M.~C.}\ \bibnamefont
  {Rechtsman}}, \ and\ \bibinfo {author} {\bibfnamefont {Y.~D.}\ \bibnamefont
  {Chong}}} (\bibinfo {year} {2016}),\ \bibfield  {title} {\enquote {\bibinfo
  {title} {Anomalous topological phases and unpaired {D}irac cones in photonic
  {F}loquet topological insulators},}\ }\href
  {http://link.aps.org/doi/10.1103/PhysRevLett.117.013902} {\bibfield
  {journal} {\bibinfo  {journal} {Phys. Rev. Lett.}\ }\textbf {\bibinfo
  {volume} {117}},\ \bibinfo {pages} {013902}}\BibitemShut {NoStop}%
\bibitem [{\citenamefont {Li}\ \emph {et~al.}(2018)\citenamefont {Li},
  \citenamefont {Wang}, \citenamefont {Xiong}, \citenamefont {Lou},
  \citenamefont {Chen}, \citenamefont {Wu}, \citenamefont {Poo}, \citenamefont
  {Jiang},\ and\ \citenamefont {John}}]{Li:2018NatComm}%
  \BibitemOpen
  \bibfield  {author} {\bibinfo {author} {\bibnamefont {Li}, \bibfnamefont
  {Fei-Fei}}, \bibinfo {author} {\bibfnamefont {Hai-Xiao}\ \bibnamefont
  {Wang}}, \bibinfo {author} {\bibfnamefont {Zhan}\ \bibnamefont {Xiong}},
  \bibinfo {author} {\bibfnamefont {Qun}\ \bibnamefont {Lou}}, \bibinfo
  {author} {\bibfnamefont {Ping}\ \bibnamefont {Chen}}, \bibinfo {author}
  {\bibfnamefont {Rui-Xin}\ \bibnamefont {Wu}}, \bibinfo {author}
  {\bibfnamefont {Yin}\ \bibnamefont {Poo}}, \bibinfo {author} {\bibfnamefont
  {Jian-Hua}\ \bibnamefont {Jiang}}, \ and\ \bibinfo {author} {\bibfnamefont
  {Sajeev}\ \bibnamefont {John}}} (\bibinfo {year} {2018}),\ \bibfield  {title}
  {\enquote {\bibinfo {title} {{Topological light-trapping on a
  dislocation}},}\ }\href {https://doi.org/10.1038/s41467-018-04861-x}
  {\bibfield  {journal} {\bibinfo  {journal} {Nat. Commun.}\ }\textbf {\bibinfo
  {volume} {9}}~(\bibinfo {number} {1}),\ \bibinfo {pages} {2462}}\BibitemShut
  {NoStop}%
\bibitem [{\citenamefont {Li}\ \emph {et~al.}(2015{\natexlab{a}})\citenamefont
  {Li}, \citenamefont {Li},\ and\ \citenamefont {Wu}}]{Li:2015APL}%
  \BibitemOpen
  \bibfield  {author} {\bibinfo {author} {\bibnamefont {Li}, \bibfnamefont
  {Qing-Bo}}, \bibinfo {author} {\bibfnamefont {Zhen}\ \bibnamefont {Li}}, \
  and\ \bibinfo {author} {\bibfnamefont {Rui-xin}\ \bibnamefont {Wu}}}
  (\bibinfo {year} {2015}{\natexlab{a}}),\ \bibfield  {title} {\enquote
  {\bibinfo {title} {Bending self-collimated one-way light by using
  gyromagnetic photonic crystals},}\ }\href
  {https://aip.scitation.org/doi/full/10.1063/1.4938008} {\bibfield  {journal}
  {\bibinfo  {journal} {Appl. Phys. Lett.}\ }\textbf {\bibinfo {volume}
  {107}}~(\bibinfo {number} {24}),\ \bibinfo {pages} {241907}}\BibitemShut
  {NoStop}%
\bibitem [{\citenamefont {Li}\ \emph {et~al.}(2013)\citenamefont {Li},
  \citenamefont {Wu}, \citenamefont {Yang},\ and\ \citenamefont
  {Sun}}]{Li:2013CPL}%
  \BibitemOpen
  \bibfield  {author} {\bibinfo {author} {\bibnamefont {Li}, \bibfnamefont
  {Qing-Bo}}, \bibinfo {author} {\bibfnamefont {Rui-Xin}\ \bibnamefont {Wu}},
  \bibinfo {author} {\bibfnamefont {Yan}\ \bibnamefont {Yang}}, \ and\ \bibinfo
  {author} {\bibfnamefont {Hui-Ling}\ \bibnamefont {Sun}}} (\bibinfo {year}
  {2013}),\ \bibfield  {title} {\enquote {\bibinfo {title} {Modeling 2{D}
  gyromagnetic photonic crystals by modified {FDTD} method},}\ }\href
  {https://iopscience.iop.org/article/10.1088/0256-307X/30/7/074208} {\bibfield
   {journal} {\bibinfo  {journal} {Chinese Physics Letters}\ }\textbf {\bibinfo
  {volume} {30}}~(\bibinfo {number} {7}),\ \bibinfo {pages}
  {074208}}\BibitemShut {NoStop}%
\bibitem [{\citenamefont {Li}\ \emph {et~al.}(2015{\natexlab{b}})\citenamefont
  {Li}, \citenamefont {Sengupta}, \citenamefont {Batrouni}, \citenamefont
  {Miniatura},\ and\ \citenamefont {Gr\'emaud}}]{Li:2015PRA}%
  \BibitemOpen
  \bibfield  {author} {\bibinfo {author} {\bibnamefont {Li}, \bibfnamefont
  {Yun}}, \bibinfo {author} {\bibfnamefont {Pinaki}\ \bibnamefont {Sengupta}},
  \bibinfo {author} {\bibfnamefont {George~G.}\ \bibnamefont {Batrouni}},
  \bibinfo {author} {\bibfnamefont {Christian}\ \bibnamefont {Miniatura}}, \
  and\ \bibinfo {author} {\bibfnamefont {Beno\^{\i}t}\ \bibnamefont
  {Gr\'emaud}}} (\bibinfo {year} {2015}{\natexlab{b}}),\ \bibfield  {title}
  {\enquote {\bibinfo {title} {Berry curvature of interacting bosons in a
  honeycomb lattice},}\ }\href
  {https://link.aps.org/doi/10.1103/PhysRevA.92.043605} {\bibfield  {journal}
  {\bibinfo  {journal} {Phys. Rev. A}\ }\textbf {\bibinfo {volume} {92}},\
  \bibinfo {pages} {043605}}\BibitemShut {NoStop}%
\bibitem [{\citenamefont {Li}\ \emph {et~al.}(2009)\citenamefont {Li},
  \citenamefont {Aydin},\ and\ \citenamefont {Ozbay}}]{Li:2009PRE}%
  \BibitemOpen
  \bibfield  {author} {\bibinfo {author} {\bibnamefont {Li}, \bibfnamefont
  {Zhaofeng}}, \bibinfo {author} {\bibfnamefont {Koray}\ \bibnamefont {Aydin}},
  \ and\ \bibinfo {author} {\bibfnamefont {Ekmel}\ \bibnamefont {Ozbay}}}
  (\bibinfo {year} {2009}),\ \bibfield  {title} {\enquote {\bibinfo {title}
  {Determination of the effective constitutive parameters of bianisotropic
  metamaterials from reflection and transmission coefficients},}\ }\href
  {https://link.aps.org/doi/10.1103/PhysRevE.79.026610} {\bibfield  {journal}
  {\bibinfo  {journal} {Phys. Rev. E}\ }\textbf {\bibinfo {volume} {79}},\
  \bibinfo {pages} {026610}}\BibitemShut {NoStop}%
\bibitem [{\citenamefont {Li}\ \emph {et~al.}(2015{\natexlab{c}})\citenamefont
  {Li}, \citenamefont {Wu}, \citenamefont {Li},\ and\ \citenamefont
  {Poo}}]{Li:2015AO}%
  \BibitemOpen
  \bibfield  {author} {\bibinfo {author} {\bibnamefont {Li}, \bibfnamefont
  {Zhen}}, \bibinfo {author} {\bibfnamefont {Rui-xin}\ \bibnamefont {Wu}},
  \bibinfo {author} {\bibfnamefont {Qing-bo}\ \bibnamefont {Li}}, \ and\
  \bibinfo {author} {\bibfnamefont {Yin}\ \bibnamefont {Poo}}} (\bibinfo {year}
  {2015}{\natexlab{c}}),\ \bibfield  {title} {\enquote {\bibinfo {title}
  {Realization of self-guided unidirectional waveguides by a chain of
  gyromagnetic rods},}\ }\href
  {https://www.osapublishing.org/ao/abstract.cfm?uri=ao-54-6-1267} {\bibfield
  {journal} {\bibinfo  {journal} {Applied optics}\ }\textbf {\bibinfo {volume}
  {54}}~(\bibinfo {number} {6}),\ \bibinfo {pages} {1267--1272}}\BibitemShut
  {NoStop}%
\bibitem [{\citenamefont {Li}\ \emph {et~al.}(2014{\natexlab{a}})\citenamefont
  {Li}, \citenamefont {Wu}, \citenamefont {Poo}, \citenamefont {Li},
  \citenamefont {Liu},\ and\ \citenamefont {Li}}]{Li:2014APA}%
  \BibitemOpen
  \bibfield  {author} {\bibinfo {author} {\bibnamefont {Li}, \bibfnamefont
  {Zhen}}, \bibinfo {author} {\bibfnamefont {Rui-xin}\ \bibnamefont {Wu}},
  \bibinfo {author} {\bibfnamefont {Yin}\ \bibnamefont {Poo}}, \bibinfo
  {author} {\bibfnamefont {Qing-bo}\ \bibnamefont {Li}}, \bibinfo {author}
  {\bibfnamefont {Rong-juan}\ \bibnamefont {Liu}}, \ and\ \bibinfo {author}
  {\bibfnamefont {Zhi-yuan}\ \bibnamefont {Li}}} (\bibinfo {year}
  {2014}{\natexlab{a}}),\ \bibfield  {title} {\enquote {\bibinfo {title} {An
  experimental study on the bandwidth and tunability of {MSP}-based one-way
  transmission},}\ }\href
  {https://link.springer.com/article/10.1007/s00339-014-8680-0} {\bibfield
  {journal} {\bibinfo  {journal} {Applied Physics A}\ }\textbf {\bibinfo
  {volume} {117}}~(\bibinfo {number} {2}),\ \bibinfo {pages}
  {451--454}}\BibitemShut {NoStop}%
\bibitem [{\citenamefont {Li}\ \emph {et~al.}(2014{\natexlab{b}})\citenamefont
  {Li}, \citenamefont {Wu}, \citenamefont {Poo}, \citenamefont {Lin},\ and\
  \citenamefont {Li}}]{Li:2014JO}%
  \BibitemOpen
  \bibfield  {author} {\bibinfo {author} {\bibnamefont {Li}, \bibfnamefont
  {Zhen}}, \bibinfo {author} {\bibfnamefont {Rui-xin}\ \bibnamefont {Wu}},
  \bibinfo {author} {\bibfnamefont {Yin}\ \bibnamefont {Poo}}, \bibinfo
  {author} {\bibfnamefont {Zhi-fang}\ \bibnamefont {Lin}}, \ and\ \bibinfo
  {author} {\bibfnamefont {Qing-Bo}\ \bibnamefont {Li}}} (\bibinfo {year}
  {2014}{\natexlab{b}}),\ \bibfield  {title} {\enquote {\bibinfo {title}
  {Fusing electromagnetic one-way edge states to achieve broadband
  unidirectional wave transmission},}\ }\href
  {https://iopscience.iop.org/article/10.1088/2040-8978/16/12/125004}
  {\bibfield  {journal} {\bibinfo  {journal} {Journal of Optics}\ }\textbf
  {\bibinfo {volume} {16}}~(\bibinfo {number} {12}),\ \bibinfo {pages}
  {125004}}\BibitemShut {NoStop}%
\bibitem [{\citenamefont {Li}\ \emph {et~al.}(2014{\natexlab{c}})\citenamefont
  {Li}, \citenamefont {Liu}, \citenamefont {Gan}, \citenamefont {Fu},\ and\
  \citenamefont {Lian}}]{Li:2014IJMPB}%
  \BibitemOpen
  \bibfield  {author} {\bibinfo {author} {\bibnamefont {Li}, \bibfnamefont
  {Zhi-Yuan}}, \bibinfo {author} {\bibfnamefont {Rong-Juan}\ \bibnamefont
  {Liu}}, \bibinfo {author} {\bibfnamefont {Lin}\ \bibnamefont {Gan}}, \bibinfo
  {author} {\bibfnamefont {Jin-Xin}\ \bibnamefont {Fu}}, \ and\ \bibinfo
  {author} {\bibfnamefont {Jin}\ \bibnamefont {Lian}}} (\bibinfo {year}
  {2014}{\natexlab{c}}),\ \bibfield  {title} {\enquote {\bibinfo {title}
  {Nonreciprocal electromagnetic devices in gyromagnetic photonic crystals},}\
  }\href {https://www.worldscientific.com/doi/abs/10.1142/S0217979214410100}
  {\bibfield  {journal} {\bibinfo  {journal} {Int. J. Mod. Phys. B}\ }\textbf
  {\bibinfo {volume} {28}}~(\bibinfo {number} {02}),\ \bibinfo {pages}
  {1441010}}\BibitemShut {NoStop}%
\bibitem [{\citenamefont {Lian}\ \emph
  {et~al.}(2012{\natexlab{a}})\citenamefont {Lian}, \citenamefont {Fu},
  \citenamefont {Gan},\ and\ \citenamefont {Li}}]{Lian:2012CPL}%
  \BibitemOpen
  \bibfield  {author} {\bibinfo {author} {\bibnamefont {Lian}, \bibfnamefont
  {Jin}}, \bibinfo {author} {\bibfnamefont {Jin-Xin}\ \bibnamefont {Fu}},
  \bibinfo {author} {\bibfnamefont {Lin}\ \bibnamefont {Gan}}, \ and\ \bibinfo
  {author} {\bibfnamefont {Zhi-Yuan}\ \bibnamefont {Li}}} (\bibinfo {year}
  {2012}{\natexlab{a}}),\ \bibfield  {title} {\enquote {\bibinfo {title}
  {Experimental realization of a magnetically tunable cavity in a gyromagnetic
  photonic crystal},}\ }\href
  {http://cpl.iphy.ac.cn/10.1088/0256-307X/29/7/074208} {\bibfield  {journal}
  {\bibinfo  {journal} {Chinese Physics Letters}\ }\textbf {\bibinfo {volume}
  {29}}~(\bibinfo {number} {7}),\ \bibinfo {pages} {074208}}\BibitemShut
  {NoStop}%
\bibitem [{\citenamefont {Lian}\ \emph
  {et~al.}(2012{\natexlab{b}})\citenamefont {Lian}, \citenamefont {Fu},
  \citenamefont {Gan},\ and\ \citenamefont {Li}}]{Lian:2012PRB}%
  \BibitemOpen
  \bibfield  {author} {\bibinfo {author} {\bibnamefont {Lian}, \bibfnamefont
  {Jin}}, \bibinfo {author} {\bibfnamefont {Jin-Xin}\ \bibnamefont {Fu}},
  \bibinfo {author} {\bibfnamefont {Lin}\ \bibnamefont {Gan}}, \ and\ \bibinfo
  {author} {\bibfnamefont {Zhi-Yuan}\ \bibnamefont {Li}}} (\bibinfo {year}
  {2012}{\natexlab{b}}),\ \bibfield  {title} {\enquote {\bibinfo {title}
  {Robust and disorder-immune magnetically tunable one-way waveguides in a
  gyromagnetic photonic crystal},}\ }\href
  {https://journals.aps.org/prb/abstract/10.1103/PhysRevB.85.125108} {\bibfield
   {journal} {\bibinfo  {journal} {Phys. Rev. B}\ }\textbf {\bibinfo {volume}
  {85}}~(\bibinfo {number} {12}),\ \bibinfo {pages} {125108}}\BibitemShut
  {NoStop}%
\bibitem [{\citenamefont {Lin}\ \emph {et~al.}(2017)\citenamefont {Lin},
  \citenamefont {Hu}, \citenamefont {Chen}, \citenamefont {Lee},\ and\
  \citenamefont {Zhang}}]{Lin:2017PRB}%
  \BibitemOpen
  \bibfield  {author} {\bibinfo {author} {\bibnamefont {Lin}, \bibfnamefont
  {Jun~Yu}}, \bibinfo {author} {\bibfnamefont {Nai~Chao}\ \bibnamefont {Hu}},
  \bibinfo {author} {\bibfnamefont {You~Jian}\ \bibnamefont {Chen}}, \bibinfo
  {author} {\bibfnamefont {Ching~Hua}\ \bibnamefont {Lee}}, \ and\ \bibinfo
  {author} {\bibfnamefont {Xiao}\ \bibnamefont {Zhang}}} (\bibinfo {year}
  {2017}),\ \bibfield  {title} {\enquote {\bibinfo {title} {Line nodes, {D}irac
  points, and lifshitz transition in two-dimensional nonsymmorphic photonic
  crystals},}\ }\href {https://link.aps.org/doi/10.1103/PhysRevB.96.075438}
  {\bibfield  {journal} {\bibinfo  {journal} {Phys. Rev. B}\ }\textbf {\bibinfo
  {volume} {96}},\ \bibinfo {pages} {075438}}\BibitemShut {NoStop}%
\bibitem [{\citenamefont {Lin}\ and\ \citenamefont {Fan}(2014)}]{Lin:2014PRX}%
  \BibitemOpen
  \bibfield  {author} {\bibinfo {author} {\bibnamefont {Lin}, \bibfnamefont
  {Qian}}, \ and\ \bibinfo {author} {\bibfnamefont {Shanhui}\ \bibnamefont
  {Fan}}} (\bibinfo {year} {2014}),\ \bibfield  {title} {\enquote {\bibinfo
  {title} {Light guiding by effective gauge field for photons},}\ }\href
  {https://link.aps.org/doi/10.1103/PhysRevX.4.031031} {\bibfield  {journal}
  {\bibinfo  {journal} {Phys. Rev. X}\ }\textbf {\bibinfo {volume} {4}},\
  \bibinfo {pages} {031031}}\BibitemShut {NoStop}%
\bibitem [{\citenamefont {Lin}\ and\ \citenamefont {Fan}(2015)}]{Lin:2015NJP}%
  \BibitemOpen
  \bibfield  {author} {\bibinfo {author} {\bibnamefont {Lin}, \bibfnamefont
  {Qian}}, \ and\ \bibinfo {author} {\bibfnamefont {Shanhui}\ \bibnamefont
  {Fan}}} (\bibinfo {year} {2015}),\ \bibfield  {title} {\enquote {\bibinfo
  {title} {Resonator-free realization of effective magnetic field for
  photons},}\ }\href {http://stacks.iop.org/1367-2630/17/i=7/a=075008}
  {\bibfield  {journal} {\bibinfo  {journal} {New J. Phys.}\ }\textbf {\bibinfo
  {volume} {17}}~(\bibinfo {number} {7}),\ \bibinfo {pages}
  {075008}}\BibitemShut {NoStop}%
\bibitem [{\citenamefont {Lin}\ \emph {et~al.}(2018)\citenamefont {Lin},
  \citenamefont {Sun}, \citenamefont {Xiao}, \citenamefont {Zhang},\ and\
  \citenamefont {Fan}}]{Lin:2018arxiv}%
  \BibitemOpen
  \bibfield  {author} {\bibinfo {author} {\bibnamefont {Lin}, \bibfnamefont
  {Qian}}, \bibinfo {author} {\bibfnamefont {Xiao-Qi}\ \bibnamefont {Sun}},
  \bibinfo {author} {\bibfnamefont {Meng}\ \bibnamefont {Xiao}}, \bibinfo
  {author} {\bibfnamefont {Shou-Cheng}\ \bibnamefont {Zhang}}, \ and\ \bibinfo
  {author} {\bibfnamefont {Shanhui}\ \bibnamefont {Fan}}} (\bibinfo {year}
  {2018}),\ \bibfield  {title} {\enquote {\bibinfo {title} {A three-dimensional
  photonic topological insulator using a two-dimensional ring resonator lattice
  with a synthetic frequency dimension},}\ }\href
  {http://advances.sciencemag.org/content/4/10/eaat2774} {\bibfield  {journal}
  {\bibinfo  {journal} {Science Advances}\ }\textbf {\bibinfo {volume}
  {4}}~(\bibinfo {number} {10}),\ \bibinfo {pages} {eaat2774}}\BibitemShut
  {NoStop}%
\bibitem [{\citenamefont {Lin}\ \emph {et~al.}(2016)\citenamefont {Lin},
  \citenamefont {Xiao}, \citenamefont {Yuan},\ and\ \citenamefont
  {Fan}}]{Lin:2016NatComm}%
  \BibitemOpen
  \bibfield  {author} {\bibinfo {author} {\bibnamefont {Lin}, \bibfnamefont
  {Qian}}, \bibinfo {author} {\bibfnamefont {Meng}\ \bibnamefont {Xiao}},
  \bibinfo {author} {\bibfnamefont {Luqi}\ \bibnamefont {Yuan}}, \ and\
  \bibinfo {author} {\bibfnamefont {Shanhui}\ \bibnamefont {Fan}}} (\bibinfo
  {year} {2016}),\ \bibfield  {title} {\enquote {\bibinfo {title} {Photonic
  {W}eyl point in a two-dimensional resonator lattice with a synthetic
  frequency dimension},}\ }\href {https://www.nature.com/articles/ncomms13731}
  {\bibfield  {journal} {\bibinfo  {journal} {Nat. Commun.}\ }\textbf {\bibinfo
  {volume} {7}},\ \bibinfo {pages} {13731}}\BibitemShut {NoStop}%
\bibitem [{\citenamefont {Lin}\ \emph {et~al.}(1998)\citenamefont {Lin},
  \citenamefont {Chow}, \citenamefont {Hietala}, \citenamefont {Villeneuve},\
  and\ \citenamefont {Joannopoulos}}]{Lin:Science1998}%
  \BibitemOpen
  \bibfield  {author} {\bibinfo {author} {\bibnamefont {Lin}, \bibfnamefont
  {Shawn-Yu}}, \bibinfo {author} {\bibfnamefont {Edmund}\ \bibnamefont {Chow}},
  \bibinfo {author} {\bibfnamefont {Vince}\ \bibnamefont {Hietala}}, \bibinfo
  {author} {\bibfnamefont {Pierre~R}\ \bibnamefont {Villeneuve}}, \ and\
  \bibinfo {author} {\bibfnamefont {JD}~\bibnamefont {Joannopoulos}}} (\bibinfo
  {year} {1998}),\ \bibfield  {title} {\enquote {\bibinfo {title} {Experimental
  demonstration of guiding and bending of electromagnetic waves in a photonic
  crystal},}\ }\href {http://science.sciencemag.org/content/282/5387/274}
  {\bibfield  {journal} {\bibinfo  {journal} {Science}\ }\textbf {\bibinfo
  {volume} {282}}~(\bibinfo {number} {5387}),\ \bibinfo {pages}
  {274--276}}\BibitemShut {NoStop}%
\bibitem [{\citenamefont {Lindner}\ \emph {et~al.}(2011)\citenamefont
  {Lindner}, \citenamefont {Refael},\ and\ \citenamefont
  {Galitski}}]{Lindner:2011NatPhys}%
  \BibitemOpen
  \bibfield  {author} {\bibinfo {author} {\bibnamefont {Lindner}, \bibfnamefont
  {N~H}}, \bibinfo {author} {\bibfnamefont {G.}~\bibnamefont {Refael}}, \ and\
  \bibinfo {author} {\bibfnamefont {V.}~\bibnamefont {Galitski}}} (\bibinfo
  {year} {2011}),\ \bibfield  {title} {\enquote {\bibinfo {title} {Floquet
  topological insulator in semiconductor quantum wells},}\ }\href
  {https://doi.org/10.1038/nphys1926} {\bibfield  {journal} {\bibinfo
  {journal} {Nat. Phys.}\ }\textbf {\bibinfo {volume} {7}},\ \bibinfo {pages}
  {490--495}}\BibitemShut {NoStop}%
\bibitem [{\citenamefont {Ling}\ \emph {et~al.}(2015)\citenamefont {Ling},
  \citenamefont {Xiao}, \citenamefont {Chan}, \citenamefont {Yu},\ and\
  \citenamefont {Fung}}]{Ling:2015OptExp}%
  \BibitemOpen
  \bibfield  {author} {\bibinfo {author} {\bibnamefont {Ling}, \bibfnamefont
  {CW}}, \bibinfo {author} {\bibfnamefont {Meng}\ \bibnamefont {Xiao}},
  \bibinfo {author} {\bibfnamefont {CT}~\bibnamefont {Chan}}, \bibinfo {author}
  {\bibfnamefont {SF}~\bibnamefont {Yu}}, \ and\ \bibinfo {author}
  {\bibfnamefont {Kin~Hung}\ \bibnamefont {Fung}}} (\bibinfo {year} {2015}),\
  \bibfield  {title} {\enquote {\bibinfo {title} {Topological edge plasmon
  modes between diatomic chains of plasmonic nanoparticles},}\ }\href
  {https://www.osapublishing.org/oe/abstract.cfm?uri=oe-23-3-2021} {\bibfield
  {journal} {\bibinfo  {journal} {Opt. Express}\ }\textbf {\bibinfo {volume}
  {23}}~(\bibinfo {number} {3}),\ \bibinfo {pages} {2021--2031}}\BibitemShut
  {NoStop}%
\bibitem [{\citenamefont {Lira}\ \emph {et~al.}(2012)\citenamefont {Lira},
  \citenamefont {Yu}, \citenamefont {Fan},\ and\ \citenamefont
  {Lipson}}]{Lira:2012PRL}%
  \BibitemOpen
  \bibfield  {author} {\bibinfo {author} {\bibnamefont {Lira}, \bibfnamefont
  {Hugo}}, \bibinfo {author} {\bibfnamefont {Zongfu}\ \bibnamefont {Yu}},
  \bibinfo {author} {\bibfnamefont {Shanhui}\ \bibnamefont {Fan}}, \ and\
  \bibinfo {author} {\bibfnamefont {Michal}\ \bibnamefont {Lipson}}} (\bibinfo
  {year} {2012}),\ \bibfield  {title} {\enquote {\bibinfo {title} {Electrically
  driven nonreciprocity induced by interband photonic transition on a silicon
  chip},}\ }\href {https://link.aps.org/doi/10.1103/PhysRevLett.109.033901}
  {\bibfield  {journal} {\bibinfo  {journal} {Phys. Rev. Lett.}\ }\textbf
  {\bibinfo {volume} {109}},\ \bibinfo {pages} {033901}}\BibitemShut {NoStop}%
\bibitem [{\citenamefont {Liu}\ \emph {et~al.}(2017)\citenamefont {Liu},
  \citenamefont {Gao}, \citenamefont {Yang},\ and\ \citenamefont
  {Zhang}}]{Liu:2017PRL}%
  \BibitemOpen
  \bibfield  {author} {\bibinfo {author} {\bibnamefont {Liu}, \bibfnamefont
  {Changxu}}, \bibinfo {author} {\bibfnamefont {Wenlong}\ \bibnamefont {Gao}},
  \bibinfo {author} {\bibfnamefont {Biao}\ \bibnamefont {Yang}}, \ and\
  \bibinfo {author} {\bibfnamefont {Shuang}\ \bibnamefont {Zhang}}} (\bibinfo
  {year} {2017}),\ \bibfield  {title} {\enquote {\bibinfo {title}
  {Disorder-induced topological state transition in photonic metamaterials},}\
  }\href {https://link.aps.org/doi/10.1103/PhysRevLett.119.183901} {\bibfield
  {journal} {\bibinfo  {journal} {Phys. Rev. Lett.}\ }\textbf {\bibinfo
  {volume} {119}},\ \bibinfo {pages} {183901}}\BibitemShut {NoStop}%
\bibitem [{\citenamefont {Liu}\ \emph {et~al.}(2012)\citenamefont {Liu},
  \citenamefont {Shen},\ and\ \citenamefont {He}}]{Liu:2012OL}%
  \BibitemOpen
  \bibfield  {author} {\bibinfo {author} {\bibnamefont {Liu}, \bibfnamefont
  {Kexin}}, \bibinfo {author} {\bibfnamefont {Linfang}\ \bibnamefont {Shen}}, \
  and\ \bibinfo {author} {\bibfnamefont {Sailing}\ \bibnamefont {He}}}
  (\bibinfo {year} {2012}),\ \bibfield  {title} {\enquote {\bibinfo {title}
  {One-way edge mode in a gyromagnetic photonic crystal slab},}\ }\href
  {https://www.osapublishing.org/ol/abstract.cfm?uri=ol-37-19-4110} {\bibfield
  {journal} {\bibinfo  {journal} {Opt. Lett.}\ }\textbf {\bibinfo {volume}
  {37}}~(\bibinfo {number} {19}),\ \bibinfo {pages} {4110--4112}}\BibitemShut
  {NoStop}%
\bibitem [{\citenamefont {Liu}\ \emph {et~al.}(2010)\citenamefont {Liu},
  \citenamefont {Lu}, \citenamefont {Lin},\ and\ \citenamefont
  {Chui}}]{Liu:2010APL}%
  \BibitemOpen
  \bibfield  {author} {\bibinfo {author} {\bibnamefont {Liu}, \bibfnamefont
  {Shiyang}}, \bibinfo {author} {\bibfnamefont {Wanli}\ \bibnamefont {Lu}},
  \bibinfo {author} {\bibfnamefont {Zhifang}\ \bibnamefont {Lin}}, \ and\
  \bibinfo {author} {\bibfnamefont {ST}~\bibnamefont {Chui}}} (\bibinfo {year}
  {2010}),\ \bibfield  {title} {\enquote {\bibinfo {title} {Magnetically
  controllable unidirectional electromagnetic waveguiding devices designed with
  metamaterials},}\ }\href {https://aip.scitation.org/doi/10.1063/1.3520141}
  {\bibfield  {journal} {\bibinfo  {journal} {Appl. Phys. Lett.}\ }\textbf
  {\bibinfo {volume} {97}}~(\bibinfo {number} {20}),\ \bibinfo {pages}
  {201113}}\BibitemShut {NoStop}%
\bibitem [{\citenamefont {Liu}\ and\ \citenamefont
  {Zhang}(2011)}]{liu2011metamaterials}%
  \BibitemOpen
  \bibfield  {author} {\bibinfo {author} {\bibnamefont {Liu}, \bibfnamefont
  {Yongmin}}, \ and\ \bibinfo {author} {\bibfnamefont {Xiang}\ \bibnamefont
  {Zhang}}} (\bibinfo {year} {2011}),\ \bibfield  {title} {\enquote {\bibinfo
  {title} {Metamaterials: a new frontier of science and technology},}\
  }\href@noop {} {\bibfield  {journal} {\bibinfo  {journal} {Chemical Society
  Reviews}\ }\textbf {\bibinfo {volume} {40}}~(\bibinfo {number} {5}),\
  \bibinfo {pages} {2494--2507}}\BibitemShut {NoStop}%
\bibitem [{\citenamefont {Livi}\ \emph {et~al.}(2016)\citenamefont {Livi},
  \citenamefont {Cappellini}, \citenamefont {Diem}, \citenamefont {Franchi},
  \citenamefont {Clivati}, \citenamefont {Frittelli}, \citenamefont {Levi},
  \citenamefont {Calonico}, \citenamefont {Catani}, \citenamefont {Inguscio},\
  and\ \citenamefont {Fallani}}]{Livi:2016PRL}%
  \BibitemOpen
  \bibfield  {author} {\bibinfo {author} {\bibnamefont {Livi}, \bibfnamefont
  {L~F}}, \bibinfo {author} {\bibfnamefont {G.}~\bibnamefont {Cappellini}},
  \bibinfo {author} {\bibfnamefont {M.}~\bibnamefont {Diem}}, \bibinfo {author}
  {\bibfnamefont {L.}~\bibnamefont {Franchi}}, \bibinfo {author} {\bibfnamefont
  {C.}~\bibnamefont {Clivati}}, \bibinfo {author} {\bibfnamefont
  {M.}~\bibnamefont {Frittelli}}, \bibinfo {author} {\bibfnamefont
  {F.}~\bibnamefont {Levi}}, \bibinfo {author} {\bibfnamefont {D.}~\bibnamefont
  {Calonico}}, \bibinfo {author} {\bibfnamefont {J.}~\bibnamefont {Catani}},
  \bibinfo {author} {\bibfnamefont {M.}~\bibnamefont {Inguscio}}, \ and\
  \bibinfo {author} {\bibfnamefont {L.}~\bibnamefont {Fallani}}} (\bibinfo
  {year} {2016}),\ \bibfield  {title} {\enquote {\bibinfo {title} {Synthetic
  dimensions and spin-orbit coupling with an optical clock transition},}\
  }\href {https://link.aps.org/doi/10.1103/PhysRevLett.117.220401} {\bibfield
  {journal} {\bibinfo  {journal} {Phys. Rev. Lett.}\ }\textbf {\bibinfo
  {volume} {117}},\ \bibinfo {pages} {220401}}\BibitemShut {NoStop}%
\bibitem [{\citenamefont {Lodahl}\ \emph {et~al.}(2017)\citenamefont {Lodahl},
  \citenamefont {Mahmoodian}, \citenamefont {Stobbe}, \citenamefont
  {Rauschenbeutel}, \citenamefont {Schneeweiss}, \citenamefont {Volz},
  \citenamefont {Pichler},\ and\ \citenamefont {Zoller}}]{Lodahl:Nat2017}%
  \BibitemOpen
  \bibfield  {author} {\bibinfo {author} {\bibnamefont {Lodahl}, \bibfnamefont
  {Peter}}, \bibinfo {author} {\bibfnamefont {Sahand}\ \bibnamefont
  {Mahmoodian}}, \bibinfo {author} {\bibfnamefont {S{\o}ren}\ \bibnamefont
  {Stobbe}}, \bibinfo {author} {\bibfnamefont {Arno}\ \bibnamefont
  {Rauschenbeutel}}, \bibinfo {author} {\bibfnamefont {Philipp}\ \bibnamefont
  {Schneeweiss}}, \bibinfo {author} {\bibfnamefont {J{\"{u}}rgen}\ \bibnamefont
  {Volz}}, \bibinfo {author} {\bibfnamefont {Hannes}\ \bibnamefont {Pichler}},
  \ and\ \bibinfo {author} {\bibfnamefont {Peter}\ \bibnamefont {Zoller}}}
  (\bibinfo {year} {2017}),\ \bibfield  {title} {\enquote {\bibinfo {title}
  {{Chiral quantum optics}},}\ }\href
  {https://www.nature.com/articles/nature21037} {\bibfield  {journal} {\bibinfo
   {journal} {Nature}\ }\textbf {\bibinfo {volume} {541}}~(\bibinfo {number}
  {7638}),\ \bibinfo {pages} {473--480}}\BibitemShut {NoStop}%
\bibitem [{\citenamefont {Lohse}\ \emph {et~al.}(2018)\citenamefont {Lohse},
  \citenamefont {Schweizer}, \citenamefont {Price}, \citenamefont
  {Zilberberg},\ and\ \citenamefont {Bloch}}]{Lohse:2018}%
  \BibitemOpen
  \bibfield  {author} {\bibinfo {author} {\bibnamefont {Lohse}, \bibfnamefont
  {Michael}}, \bibinfo {author} {\bibfnamefont {Christian}\ \bibnamefont
  {Schweizer}}, \bibinfo {author} {\bibfnamefont {Hannah~M.}\ \bibnamefont
  {Price}}, \bibinfo {author} {\bibfnamefont {Oded}\ \bibnamefont
  {Zilberberg}}, \ and\ \bibinfo {author} {\bibfnamefont {Immanuel}\
  \bibnamefont {Bloch}}} (\bibinfo {year} {2018}),\ \bibfield  {title}
  {\enquote {\bibinfo {title} {Exploring 4{D} quantum {H}all physics with a
  2{D} topological charge pump},}\ }\href
  {https://www.nature.com/articles/nature25000} {\bibfield  {journal} {\bibinfo
   {journal} {Nature}\ }\textbf {\bibinfo {volume} {553}},\ \bibinfo {pages}
  {55}}\BibitemShut {NoStop}%
\bibitem [{\citenamefont {Lohse}\ \emph {et~al.}(2016)\citenamefont {Lohse},
  \citenamefont {Schweizer}, \citenamefont {Zilberberg}, \citenamefont
  {Aidelsburger},\ and\ \citenamefont {Bloch}}]{Lohse:2016NatPhys}%
  \BibitemOpen
  \bibfield  {author} {\bibinfo {author} {\bibnamefont {Lohse}, \bibfnamefont
  {Michael}}, \bibinfo {author} {\bibfnamefont {Christian}\ \bibnamefont
  {Schweizer}}, \bibinfo {author} {\bibfnamefont {Oded}\ \bibnamefont
  {Zilberberg}}, \bibinfo {author} {\bibfnamefont {Monika}\ \bibnamefont
  {Aidelsburger}}, \ and\ \bibinfo {author} {\bibfnamefont {Immanuel}\
  \bibnamefont {Bloch}}} (\bibinfo {year} {2016}),\ \bibfield  {title}
  {\enquote {\bibinfo {title} {A {T}houless quantum pump with ultracold bosonic
  atoms in an optical superlattice},}\ }\href
  {https://www.nature.com/articles/nphys3584} {\bibfield  {journal} {\bibinfo
  {journal} {Nat. Phys.}\ }\textbf {\bibinfo {volume} {12}}~(\bibinfo {number}
  {4}),\ \bibinfo {pages} {350}}\BibitemShut {NoStop}%
\bibitem [{\citenamefont {Longhi}\ \emph {et~al.}(2018)\citenamefont {Longhi},
  \citenamefont {Kominis},\ and\ \citenamefont {Kovanis}}]{Longhi:2018EPL}%
  \BibitemOpen
  \bibfield  {author} {\bibinfo {author} {\bibnamefont {Longhi}, \bibfnamefont
  {S}}, \bibinfo {author} {\bibfnamefont {Y}~\bibnamefont {Kominis}}, \ and\
  \bibinfo {author} {\bibfnamefont {V}~\bibnamefont {Kovanis}}} (\bibinfo
  {year} {2018}),\ \bibfield  {title} {\enquote {\bibinfo {title} {Presence of
  temporal dynamical instabilities in topological insulator lasers},}\ }\href
  {http://iopscience.iop.org/article/10.1209/0295-5075/122/14004} {\bibfield
  {journal} {\bibinfo  {journal} {EPL (Europhysics Letters)}\ }\textbf
  {\bibinfo {volume} {122}}~(\bibinfo {number} {1}),\ \bibinfo {pages}
  {14004}}\BibitemShut {NoStop}%
\bibitem [{\citenamefont {Longhi}(2011)}]{Longhi:OL2011}%
  \BibitemOpen
  \bibfield  {author} {\bibinfo {author} {\bibnamefont {Longhi}, \bibfnamefont
  {Stefano}}} (\bibinfo {year} {2011}),\ \bibfield  {title} {\enquote {\bibinfo
  {title} {Photonic {B}loch oscillations of correlated particles},}\ }\href
  {http://ol.osa.org/abstract.cfm?URI=ol-36-16-3248} {\bibfield  {journal}
  {\bibinfo  {journal} {Opt. Lett.}\ }\textbf {\bibinfo {volume}
  {36}}~(\bibinfo {number} {16}),\ \bibinfo {pages} {3248--3250}}\BibitemShut
  {NoStop}%
\bibitem [{\citenamefont {Longhi}(2013)}]{Longhi:2013OptLett}%
  \BibitemOpen
  \bibfield  {author} {\bibinfo {author} {\bibnamefont {Longhi}, \bibfnamefont
  {Stefano}}} (\bibinfo {year} {2013}),\ \bibfield  {title} {\enquote {\bibinfo
  {title} {Effective magnetic fields for photons in waveguide and coupled
  resonator lattices},}\ }\href
  {http://ol.osa.org/abstract.cfm?URI=ol-38-18-3570} {\bibfield  {journal}
  {\bibinfo  {journal} {Opt. Lett.}\ }\textbf {\bibinfo {volume}
  {38}}~(\bibinfo {number} {18}),\ \bibinfo {pages} {3570--3573}}\BibitemShut
  {NoStop}%
\bibitem [{\citenamefont {Longhi}(2015)}]{longhi2015synthetic}%
  \BibitemOpen
  \bibfield  {author} {\bibinfo {author} {\bibnamefont {Longhi}, \bibfnamefont
  {Stefano}}} (\bibinfo {year} {2015}),\ \bibfield  {title} {\enquote {\bibinfo
  {title} {Synthetic gauge fields for light beams in optical resonators},}\
  }\href {https://www.osapublishing.org/ol/abstract.cfm?uri=ol-40-13-2941}
  {\bibfield  {journal} {\bibinfo  {journal} {Opt. Lett.}\ }\textbf {\bibinfo
  {volume} {40}}~(\bibinfo {number} {13}),\ \bibinfo {pages}
  {2941--2944}}\BibitemShut {NoStop}%
\bibitem [{\citenamefont {Lu}\ \emph {et~al.}(2016{\natexlab{a}})\citenamefont
  {Lu}, \citenamefont {Schemmer}, \citenamefont {Aycock}, \citenamefont
  {Genkina}, \citenamefont {Sugawa},\ and\ \citenamefont
  {Spielman}}]{LuSpielman:2016}%
  \BibitemOpen
  \bibfield  {author} {\bibinfo {author} {\bibnamefont {Lu}, \bibfnamefont
  {H-I}}, \bibinfo {author} {\bibfnamefont {M.}~\bibnamefont {Schemmer}},
  \bibinfo {author} {\bibfnamefont {L.~M.}\ \bibnamefont {Aycock}}, \bibinfo
  {author} {\bibfnamefont {D.}~\bibnamefont {Genkina}}, \bibinfo {author}
  {\bibfnamefont {S.}~\bibnamefont {Sugawa}}, \ and\ \bibinfo {author}
  {\bibfnamefont {I.~B.}\ \bibnamefont {Spielman}}} (\bibinfo {year}
  {2016}{\natexlab{a}}),\ \bibfield  {title} {\enquote {\bibinfo {title}
  {Geometrical pumping with a {Bose-Einstein} condensate},}\ }\href
  {https://link.aps.org/doi/10.1103/PhysRevLett.116.200402} {\bibfield
  {journal} {\bibinfo  {journal} {Phys. Rev. Lett.}\ }\textbf {\bibinfo
  {volume} {116}},\ \bibinfo {pages} {200402}}\BibitemShut {NoStop}%
\bibitem [{\citenamefont {Lu}\ \emph {et~al.}(2013{\natexlab{a}})\citenamefont
  {Lu}, \citenamefont {Shen}, \citenamefont {Deng}, \citenamefont {Li},\ and\
  \citenamefont {Zheng}}]{Lu:2013AO}%
  \BibitemOpen
  \bibfield  {author} {\bibinfo {author} {\bibnamefont {Lu}, \bibfnamefont
  {Jie}}, \bibinfo {author} {\bibfnamefont {Linfang}\ \bibnamefont {Shen}},
  \bibinfo {author} {\bibfnamefont {Xiaohua}\ \bibnamefont {Deng}}, \bibinfo
  {author} {\bibfnamefont {Xiaoer}\ \bibnamefont {Li}}, \ and\ \bibinfo
  {author} {\bibfnamefont {Xiaodong}\ \bibnamefont {Zheng}}} (\bibinfo {year}
  {2013}{\natexlab{a}}),\ \bibfield  {title} {\enquote {\bibinfo {title}
  {Impact of photonic crystal boundary shape on the existence of one-way edge
  mode},}\ }\href
  {https://www.osapublishing.org/ao/abstract.cfm?uri=ao-52-21-5216} {\bibfield
  {journal} {\bibinfo  {journal} {Applied optics}\ }\textbf {\bibinfo {volume}
  {52}}~(\bibinfo {number} {21}),\ \bibinfo {pages} {5216--5220}}\BibitemShut
  {NoStop}%
\bibitem [{\citenamefont {Lu}\ \emph {et~al.}(2016{\natexlab{b}})\citenamefont
  {Lu}, \citenamefont {Fang}, \citenamefont {Fu}, \citenamefont {Johnson},
  \citenamefont {Joannopoulos},\ and\ \citenamefont
  {Solja{\v{c}}i{\'c}}}]{Lu:2016NatPhys}%
  \BibitemOpen
  \bibfield  {author} {\bibinfo {author} {\bibnamefont {Lu}, \bibfnamefont
  {Ling}}, \bibinfo {author} {\bibfnamefont {Chen}\ \bibnamefont {Fang}},
  \bibinfo {author} {\bibfnamefont {Liang}\ \bibnamefont {Fu}}, \bibinfo
  {author} {\bibfnamefont {Steven~G}\ \bibnamefont {Johnson}}, \bibinfo
  {author} {\bibfnamefont {John~D}\ \bibnamefont {Joannopoulos}}, \ and\
  \bibinfo {author} {\bibfnamefont {Marin}\ \bibnamefont {Solja{\v{c}}i{\'c}}}}
  (\bibinfo {year} {2016}{\natexlab{b}}),\ \bibfield  {title} {\enquote
  {\bibinfo {title} {Symmetry-protected topological photonic crystal in three
  dimensions},}\ }\href
  {http://www.nature.com/nphys/journal/v12/n4/full/nphys3611.html} {\bibfield
  {journal} {\bibinfo  {journal} {Nat. Phys.}\ }\textbf {\bibinfo {volume}
  {12}}~(\bibinfo {number} {4}),\ \bibinfo {pages} {337--340}}\BibitemShut
  {NoStop}%
\bibitem [{\citenamefont {Lu}\ \emph {et~al.}(2013{\natexlab{b}})\citenamefont
  {Lu}, \citenamefont {Fu}, \citenamefont {Joannopoulos},\ and\ \citenamefont
  {Solja{\v{c}}i{\'c}}}]{Lu:2013NatPhot}%
  \BibitemOpen
  \bibfield  {author} {\bibinfo {author} {\bibnamefont {Lu}, \bibfnamefont
  {Ling}}, \bibinfo {author} {\bibfnamefont {Liang}\ \bibnamefont {Fu}},
  \bibinfo {author} {\bibfnamefont {John~D}\ \bibnamefont {Joannopoulos}}, \
  and\ \bibinfo {author} {\bibfnamefont {Marin}\ \bibnamefont
  {Solja{\v{c}}i{\'c}}}} (\bibinfo {year} {2013}{\natexlab{b}}),\ \bibfield
  {title} {\enquote {\bibinfo {title} {Weyl points and line nodes in gyroid
  photonic crystals},}\ }\href
  {http://www.nature.com/nphoton/journal/v7/n4/full/nphoton.2013.42.html}
  {\bibfield  {journal} {\bibinfo  {journal} {Nat. Photonics}\ }\textbf
  {\bibinfo {volume} {7}}~(\bibinfo {number} {4}),\ \bibinfo {pages}
  {294--299}}\BibitemShut {NoStop}%
\bibitem [{\citenamefont {Lu}\ \emph {et~al.}(2018)\citenamefont {Lu},
  \citenamefont {Gao},\ and\ \citenamefont {Wang}}]{Lu:2016arXiv}%
  \BibitemOpen
  \bibfield  {author} {\bibinfo {author} {\bibnamefont {Lu}, \bibfnamefont
  {Ling}}, \bibinfo {author} {\bibfnamefont {Haozhe}\ \bibnamefont {Gao}}, \
  and\ \bibinfo {author} {\bibfnamefont {Zhong}\ \bibnamefont {Wang}}}
  (\bibinfo {year} {2018}),\ \bibfield  {title} {\enquote {\bibinfo {title}
  {Topological one-way fiber of second {C}hern number},}\ }\href
  {https://www.nature.com/articles/s41467-018-07817-3} {\bibfield  {journal}
  {\bibinfo  {journal} {Nat. Commun.}\ }\textbf {\bibinfo {volume} {9}},\
  \bibinfo {pages} {5384}}\BibitemShut {NoStop}%
\bibitem [{\citenamefont {Lu}\ \emph {et~al.}(2014)\citenamefont {Lu},
  \citenamefont {Joannopoulos},\ and\ \citenamefont
  {Solja{\v{c}}i{\'c}}}]{lu2014topological}%
  \BibitemOpen
  \bibfield  {author} {\bibinfo {author} {\bibnamefont {Lu}, \bibfnamefont
  {Ling}}, \bibinfo {author} {\bibfnamefont {John~D}\ \bibnamefont
  {Joannopoulos}}, \ and\ \bibinfo {author} {\bibfnamefont {Marin}\
  \bibnamefont {Solja{\v{c}}i{\'c}}}} (\bibinfo {year} {2014}),\ \bibfield
  {title} {\enquote {\bibinfo {title} {Topological photonics},}\ }\href
  {https://www.nature.com/articles/nphoton.2014.248} {\bibfield  {journal}
  {\bibinfo  {journal} {Nat. Photonics}\ }\textbf {\bibinfo {volume}
  {8}}~(\bibinfo {number} {11}),\ \bibinfo {pages} {821--829}}\BibitemShut
  {NoStop}%
\bibitem [{\citenamefont {Lu}\ \emph {et~al.}(2016{\natexlab{c}})\citenamefont
  {Lu}, \citenamefont {Joannopoulos},\ and\ \citenamefont
  {Solja{\v{c}}i{\'c}}}]{Lu:2016NatPhys2}%
  \BibitemOpen
  \bibfield  {author} {\bibinfo {author} {\bibnamefont {Lu}, \bibfnamefont
  {Ling}}, \bibinfo {author} {\bibfnamefont {John~D}\ \bibnamefont
  {Joannopoulos}}, \ and\ \bibinfo {author} {\bibfnamefont {Marin}\
  \bibnamefont {Solja{\v{c}}i{\'c}}}} (\bibinfo {year} {2016}{\natexlab{c}}),\
  \bibfield  {title} {\enquote {\bibinfo {title} {Topological states in
  photonic systems},}\ }\href {https://www.nature.com/articles/nphys3796}
  {\bibfield  {journal} {\bibinfo  {journal} {Nat. Phys.}\ }\textbf {\bibinfo
  {volume} {12}}~(\bibinfo {number} {7}),\ \bibinfo {pages}
  {626--629}}\BibitemShut {NoStop}%
\bibitem [{\citenamefont {Lu}\ \emph {et~al.}(2015)\citenamefont {Lu},
  \citenamefont {Wang}, \citenamefont {Ye}, \citenamefont {Ran}, \citenamefont
  {Fu}, \citenamefont {Joannopoulos},\ and\ \citenamefont
  {Solja{\v{c}}i{\'c}}}]{Lu:2015Science}%
  \BibitemOpen
  \bibfield  {author} {\bibinfo {author} {\bibnamefont {Lu}, \bibfnamefont
  {Ling}}, \bibinfo {author} {\bibfnamefont {Zhiyu}\ \bibnamefont {Wang}},
  \bibinfo {author} {\bibfnamefont {Dexin}\ \bibnamefont {Ye}}, \bibinfo
  {author} {\bibfnamefont {Lixin}\ \bibnamefont {Ran}}, \bibinfo {author}
  {\bibfnamefont {Liang}\ \bibnamefont {Fu}}, \bibinfo {author} {\bibfnamefont
  {John~D}\ \bibnamefont {Joannopoulos}}, \ and\ \bibinfo {author}
  {\bibfnamefont {Marin}\ \bibnamefont {Solja{\v{c}}i{\'c}}}} (\bibinfo {year}
  {2015}),\ \bibfield  {title} {\enquote {\bibinfo {title} {Experimental
  observation of {W}eyl points},}\ }\href
  {http://science.sciencemag.org/content/349/6248/622} {\bibfield  {journal}
  {\bibinfo  {journal} {Science}\ }\textbf {\bibinfo {volume} {349}}~(\bibinfo
  {number} {6248}),\ \bibinfo {pages} {622--624}}\BibitemShut {NoStop}%
\bibitem [{\citenamefont {Lu}\ \emph {et~al.}(2019)\citenamefont {Lu},
  \citenamefont {Jia}, \citenamefont {Su}, \citenamefont {Owens}, \citenamefont
  {Juzeli\ifmmode~\bar{u}\else \={u}\fi{}nas}, \citenamefont {Schuster},\ and\
  \citenamefont {Simon}}]{Lu:arxiv2018}%
  \BibitemOpen
  \bibfield  {author} {\bibinfo {author} {\bibnamefont {Lu}, \bibfnamefont
  {Yuehui}}, \bibinfo {author} {\bibfnamefont {Ningyuan}\ \bibnamefont {Jia}},
  \bibinfo {author} {\bibfnamefont {Lin}\ \bibnamefont {Su}}, \bibinfo {author}
  {\bibfnamefont {Clai}\ \bibnamefont {Owens}}, \bibinfo {author}
  {\bibfnamefont {Gediminas}\ \bibnamefont {Juzeli\ifmmode~\bar{u}\else
  \={u}\fi{}nas}}, \bibinfo {author} {\bibfnamefont {David~I.}\ \bibnamefont
  {Schuster}}, \ and\ \bibinfo {author} {\bibfnamefont {Jonathan}\ \bibnamefont
  {Simon}}} (\bibinfo {year} {2019}),\ \bibfield  {title} {\enquote {\bibinfo
  {title} {Probing the {B}erry curvature and {F}ermi arcs of a {W}eyl
  circuit},}\ }\href {https://link.aps.org/doi/10.1103/PhysRevB.99.020302}
  {\bibfield  {journal} {\bibinfo  {journal} {Phys. Rev. B}\ }\textbf {\bibinfo
  {volume} {99}},\ \bibinfo {pages} {020302}}\BibitemShut {NoStop}%
\bibitem [{\citenamefont {Lumer}\ \emph {et~al.}(2018)\citenamefont {Lumer},
  \citenamefont {Bandres}, \citenamefont {Heinrich}, \citenamefont {Maczewsky},
  \citenamefont {Herzig-Sheinfux}, \citenamefont {Szameit},\ and\ \citenamefont
  {Segev}}]{Lumer:2018arXiv}%
  \BibitemOpen
  \bibfield  {author} {\bibinfo {author} {\bibnamefont {Lumer}, \bibfnamefont
  {Yaakov}}, \bibinfo {author} {\bibfnamefont {Miguel~A}\ \bibnamefont
  {Bandres}}, \bibinfo {author} {\bibfnamefont {Matthias}\ \bibnamefont
  {Heinrich}}, \bibinfo {author} {\bibfnamefont {Lukas}\ \bibnamefont
  {Maczewsky}}, \bibinfo {author} {\bibfnamefont {Hanan}\ \bibnamefont
  {Herzig-Sheinfux}}, \bibinfo {author} {\bibfnamefont {Alexander}\
  \bibnamefont {Szameit}}, \ and\ \bibinfo {author} {\bibfnamefont {Mordechai}\
  \bibnamefont {Segev}}} (\bibinfo {year} {2018}),\ \bibfield  {title}
  {\enquote {\bibinfo {title} {Observation of light guiding by artificial gauge
  fields},}\ }\href {https://arxiv.org/abs/1808.10207} {\bibinfo  {journal}
  {arXiv:1808.10207}\ }\BibitemShut {NoStop}%
\bibitem [{\citenamefont {Lumer}\ \emph
  {et~al.}(2013{\natexlab{a}})\citenamefont {Lumer}, \citenamefont {Plotnik},
  \citenamefont {Rechtsman},\ and\ \citenamefont {Segev}}]{Lumer:PRL2013b}%
  \BibitemOpen
\bibfield  {journal} {  }\bibfield  {author} {\bibinfo {author} {\bibnamefont
  {Lumer}, \bibfnamefont {Yaakov}}, \bibinfo {author} {\bibfnamefont {Yonatan}\
  \bibnamefont {Plotnik}}, \bibinfo {author} {\bibfnamefont {Mikael~C}\
  \bibnamefont {Rechtsman}}, \ and\ \bibinfo {author} {\bibfnamefont
  {Mordechai}\ \bibnamefont {Segev}}} (\bibinfo {year} {2013}{\natexlab{a}}),\
  \bibfield  {title} {\enquote {\bibinfo {title} {Nonlinearly induced
  $\mathcal{PT}$ transition in photonic systems},}\ }\href
  {https://journals.aps.org/prl/abstract/10.1103/PhysRevLett.111.263901}
  {\bibfield  {journal} {\bibinfo  {journal} {Phys. Rev. Lett.}\ }\textbf
  {\bibinfo {volume} {111}}~(\bibinfo {number} {26}),\ \bibinfo {pages}
  {263901}}\BibitemShut {NoStop}%
\bibitem [{\citenamefont {Lumer}\ \emph
  {et~al.}(2013{\natexlab{b}})\citenamefont {Lumer}, \citenamefont {Plotnik},
  \citenamefont {Rechtsman},\ and\ \citenamefont {Segev}}]{Lumer:PRL2013}%
  \BibitemOpen
  \bibfield  {author} {\bibinfo {author} {\bibnamefont {Lumer}, \bibfnamefont
  {Yaakov}}, \bibinfo {author} {\bibfnamefont {Yonatan}\ \bibnamefont
  {Plotnik}}, \bibinfo {author} {\bibfnamefont {Mikael~C}\ \bibnamefont
  {Rechtsman}}, \ and\ \bibinfo {author} {\bibfnamefont {Mordechai}\
  \bibnamefont {Segev}}} (\bibinfo {year} {2013}{\natexlab{b}}),\ \bibfield
  {title} {\enquote {\bibinfo {title} {Self-localized states in photonic
  topological insulators},}\ }\href
  {https://journals.aps.org/prl/abstract/10.1103/PhysRevLett.111.243905}
  {\bibfield  {journal} {\bibinfo  {journal} {Phys. Rev. Lett.}\ }\textbf
  {\bibinfo {volume} {111}}~(\bibinfo {number} {24}),\ \bibinfo {pages}
  {243905}}\BibitemShut {NoStop}%
\bibitem [{\citenamefont {Lumer}\ \emph {et~al.}(2016)\citenamefont {Lumer},
  \citenamefont {Rechtsman}, \citenamefont {Plotnik},\ and\ \citenamefont
  {Segev}}]{Lumer:PRA2016}%
  \BibitemOpen
  \bibfield  {author} {\bibinfo {author} {\bibnamefont {Lumer}, \bibfnamefont
  {Yaakov}}, \bibinfo {author} {\bibfnamefont {Mikael~C.}\ \bibnamefont
  {Rechtsman}}, \bibinfo {author} {\bibfnamefont {Yonatan}\ \bibnamefont
  {Plotnik}}, \ and\ \bibinfo {author} {\bibfnamefont {Mordechai}\ \bibnamefont
  {Segev}}} (\bibinfo {year} {2016}),\ \bibfield  {title} {\enquote {\bibinfo
  {title} {Instability of bosonic topological edge states in the presence of
  interactions},}\ }\href {https://link.aps.org/doi/10.1103/PhysRevA.94.021801}
  {\bibfield  {journal} {\bibinfo  {journal} {Phys. Rev. A}\ }\textbf {\bibinfo
  {volume} {94}},\ \bibinfo {pages} {021801}}\BibitemShut {NoStop}%
\bibitem [{\citenamefont {Luo}\ \emph {et~al.}(2018)\citenamefont {Luo},
  \citenamefont {Yu},\ and\ \citenamefont {Weng}}]{Luo:2018arXiv}%
  \BibitemOpen
  \bibfield  {author} {\bibinfo {author} {\bibnamefont {Luo}, \bibfnamefont
  {Kaifa}}, \bibinfo {author} {\bibfnamefont {Rui}\ \bibnamefont {Yu}}, \ and\
  \bibinfo {author} {\bibfnamefont {Hongming}\ \bibnamefont {Weng}}} (\bibinfo
  {year} {2018}),\ \bibfield  {title} {\enquote {\bibinfo {title} {Topological
  nodal states in circuit lattice},}\ }\href
  {https://spj.sciencemag.org/research/2018/6793752/} {\bibfield  {journal}
  {\bibinfo  {journal} {Research}\ }\textbf {\bibinfo {volume} {2018}},\
  \bibinfo {pages} {6793752}}\BibitemShut {NoStop}%
\bibitem [{\citenamefont {Luo}\ \emph {et~al.}(2015)\citenamefont {Luo},
  \citenamefont {Zhou}, \citenamefont {Li}, \citenamefont {Xu}, \citenamefont
  {Guo},\ and\ \citenamefont {Zhou}}]{Luo:2015NatComm}%
  \BibitemOpen
  \bibfield  {author} {\bibinfo {author} {\bibnamefont {Luo}, \bibfnamefont
  {Xi-Wang}}, \bibinfo {author} {\bibfnamefont {Xingxiang}\ \bibnamefont
  {Zhou}}, \bibinfo {author} {\bibfnamefont {Chuan-Feng}\ \bibnamefont {Li}},
  \bibinfo {author} {\bibfnamefont {Jin-Shi}\ \bibnamefont {Xu}}, \bibinfo
  {author} {\bibfnamefont {Guang-Can}\ \bibnamefont {Guo}}, \ and\ \bibinfo
  {author} {\bibfnamefont {Zheng-Wei}\ \bibnamefont {Zhou}}} (\bibinfo {year}
  {2015}),\ \bibfield  {title} {\enquote {\bibinfo {title} {Quantum simulation
  of 2{D} topological physics in a 1{D} array of optical cavities},}\ }\href
  {http://www.nature.com/articles/ncomms8704} {\bibfield  {journal} {\bibinfo
  {journal} {Nat. Commun.}\ }\textbf {\bibinfo {volume} {6}},\ \bibinfo {pages}
  {7704}}\BibitemShut {NoStop}%
\bibitem [{\citenamefont {Luo}\ \emph {et~al.}(2017)\citenamefont {Luo},
  \citenamefont {Zhou}, \citenamefont {Xu}, \citenamefont {Li}, \citenamefont
  {Guo}, \citenamefont {Zhang},\ and\ \citenamefont {Zhou}}]{Luo:2017NatComm}%
  \BibitemOpen
  \bibfield  {author} {\bibinfo {author} {\bibnamefont {Luo}, \bibfnamefont
  {Xi-Wang}}, \bibinfo {author} {\bibfnamefont {Xingxiang}\ \bibnamefont
  {Zhou}}, \bibinfo {author} {\bibfnamefont {Jin-Shi}\ \bibnamefont {Xu}},
  \bibinfo {author} {\bibfnamefont {Chuan-Feng}\ \bibnamefont {Li}}, \bibinfo
  {author} {\bibfnamefont {Guang-Can}\ \bibnamefont {Guo}}, \bibinfo {author}
  {\bibfnamefont {Chuanwei}\ \bibnamefont {Zhang}}, \ and\ \bibinfo {author}
  {\bibfnamefont {Zheng-Wei}\ \bibnamefont {Zhou}}} (\bibinfo {year} {2017}),\
  \bibfield  {title} {\enquote {\bibinfo {title} {Synthetic-lattice enabled
  all-optical devices based on orbital angular momentum of light},}\ }\href
  {https://www.nature.com/articles/ncomms16097} {\bibfield  {journal} {\bibinfo
   {journal} {Nat. Commun.}\ }\textbf {\bibinfo {volume} {8}},\ \bibinfo
  {pages} {ncomms16097}}\BibitemShut {NoStop}%
\bibitem [{\citenamefont {Luo}\ \emph {et~al.}(2016)\citenamefont {Luo},
  \citenamefont {Zhou}, \citenamefont {Liu}, \citenamefont {Qiu},\ and\
  \citenamefont {Yu}}]{Luo:2016APL}%
  \BibitemOpen
  \bibfield  {author} {\bibinfo {author} {\bibnamefont {Luo}, \bibfnamefont
  {Xiaoguang}}, \bibinfo {author} {\bibfnamefont {Ming}\ \bibnamefont {Zhou}},
  \bibinfo {author} {\bibfnamefont {Jingfeng}\ \bibnamefont {Liu}}, \bibinfo
  {author} {\bibfnamefont {Teng}\ \bibnamefont {Qiu}}, \ and\ \bibinfo {author}
  {\bibfnamefont {Zongfu}\ \bibnamefont {Yu}}} (\bibinfo {year} {2016}),\
  \bibfield  {title} {\enquote {\bibinfo {title} {Magneto-optical metamaterials
  with extraordinarily strong magneto-optical effect},}\ }\href
  {https://aip.scitation.org/doi/full/10.1063/1.4945051} {\bibfield  {journal}
  {\bibinfo  {journal} {Appl. Phys. Lett.}\ }\textbf {\bibinfo {volume}
  {108}}~(\bibinfo {number} {13}),\ \bibinfo {pages} {131104}}\BibitemShut
  {NoStop}%
\bibitem [{\citenamefont {Lustig}\ \emph {et~al.}(2018)\citenamefont {Lustig},
  \citenamefont {Weimann}, \citenamefont {Plotnik}, \citenamefont {Lumer},
  \citenamefont {Bandres}, \citenamefont {Szameit},\ and\ \citenamefont
  {Segev}}]{Lustig:arxiv2018}%
  \BibitemOpen
  \bibfield  {author} {\bibinfo {author} {\bibnamefont {Lustig}, \bibfnamefont
  {Eran}}, \bibinfo {author} {\bibfnamefont {Steffen}\ \bibnamefont {Weimann}},
  \bibinfo {author} {\bibfnamefont {Yonatan}\ \bibnamefont {Plotnik}}, \bibinfo
  {author} {\bibfnamefont {Yaakov}\ \bibnamefont {Lumer}}, \bibinfo {author}
  {\bibfnamefont {Miguel~A}\ \bibnamefont {Bandres}}, \bibinfo {author}
  {\bibfnamefont {Alexander}\ \bibnamefont {Szameit}}, \ and\ \bibinfo {author}
  {\bibfnamefont {Mordechai}\ \bibnamefont {Segev}}} (\bibinfo {year} {2018}),\
  \bibfield  {title} {\enquote {\bibinfo {title} {Photonic topological
  insulator in synthetic dimensions},}\ }\href
  {https://arxiv.org/abs/1807.01983} {\bibinfo  {journal} {arXiv:1807.01983}\
  }\BibitemShut {NoStop}%
\bibitem [{\citenamefont {Luttinger}(1951)}]{Luttinger:1951PR}%
  \BibitemOpen
\bibfield  {journal} {  }\bibfield  {author} {\bibinfo {author} {\bibnamefont
  {Luttinger}, \bibfnamefont {J~M}}} (\bibinfo {year} {1951}),\ \bibfield
  {title} {\enquote {\bibinfo {title} {The effect of a magnetic field on
  electrons in a periodic potential},}\ }\href
  {https://link.aps.org/doi/10.1103/PhysRev.84.814} {\bibfield  {journal}
  {\bibinfo  {journal} {Phys. Rev.}\ }\textbf {\bibinfo {volume} {84}},\
  \bibinfo {pages} {814--817}}\BibitemShut {NoStop}%
\bibitem [{\citenamefont {Ma}\ \emph {et~al.}(2017)\citenamefont {Ma},
  \citenamefont {Owens}, \citenamefont {LaChapelle}, \citenamefont {Schuster},\
  and\ \citenamefont {Simon}}]{ma2016hamiltonian}%
  \BibitemOpen
  \bibfield  {author} {\bibinfo {author} {\bibnamefont {Ma}, \bibfnamefont
  {Ruichao}}, \bibinfo {author} {\bibfnamefont {Clai}\ \bibnamefont {Owens}},
  \bibinfo {author} {\bibfnamefont {Aman}\ \bibnamefont {LaChapelle}}, \bibinfo
  {author} {\bibfnamefont {David~I.}\ \bibnamefont {Schuster}}, \ and\ \bibinfo
  {author} {\bibfnamefont {Jonathan}\ \bibnamefont {Simon}}} (\bibinfo {year}
  {2017}),\ \bibfield  {title} {\enquote {\bibinfo {title} {Hamiltonian
  tomography of photonic lattices},}\ }\href
  {https://link.aps.org/doi/10.1103/PhysRevA.95.062120} {\bibfield  {journal}
  {\bibinfo  {journal} {Phys. Rev. A}\ }\textbf {\bibinfo {volume} {95}},\
  \bibinfo {pages} {062120}}\BibitemShut {NoStop}%
\bibitem [{\citenamefont {Ma}\ \emph {et~al.}(2019)\citenamefont {Ma},
  \citenamefont {Saxberg}, \citenamefont {Owens}, \citenamefont {Leung},
  \citenamefont {Lu}, \citenamefont {Simon},\ and\ \citenamefont
  {Schuster}}]{Ma:2019Nature}%
  \BibitemOpen
  \bibfield  {author} {\bibinfo {author} {\bibnamefont {Ma}, \bibfnamefont
  {Ruichao}}, \bibinfo {author} {\bibfnamefont {Brendan}\ \bibnamefont
  {Saxberg}}, \bibinfo {author} {\bibfnamefont {Clai}\ \bibnamefont {Owens}},
  \bibinfo {author} {\bibfnamefont {Nelson}\ \bibnamefont {Leung}}, \bibinfo
  {author} {\bibfnamefont {Yao}\ \bibnamefont {Lu}}, \bibinfo {author}
  {\bibfnamefont {Jonathan}\ \bibnamefont {Simon}}, \ and\ \bibinfo {author}
  {\bibfnamefont {David~I}\ \bibnamefont {Schuster}}} (\bibinfo {year}
  {2019}),\ \bibfield  {title} {\enquote {\bibinfo {title} {A dissipatively
  stabilized {M}ott insulator of photons},}\ }\href
  {https://www.nature.com/articles/s41586-019-0897-9} {\bibfield  {journal}
  {\bibinfo  {journal} {Nature}\ }\textbf {\bibinfo {volume} {566}}~(\bibinfo
  {number} {7742}),\ \bibinfo {pages} {51}}\BibitemShut {NoStop}%
\bibitem [{\citenamefont {Ma}\ \emph {et~al.}(2015)\citenamefont {Ma},
  \citenamefont {Khanikaev}, \citenamefont {Mousavi},\ and\ \citenamefont
  {Shvets}}]{Ma:2015PRL}%
  \BibitemOpen
  \bibfield  {author} {\bibinfo {author} {\bibnamefont {Ma}, \bibfnamefont
  {Tzuhsuan}}, \bibinfo {author} {\bibfnamefont {Alexander~B.}\ \bibnamefont
  {Khanikaev}}, \bibinfo {author} {\bibfnamefont {S.~Hossein}\ \bibnamefont
  {Mousavi}}, \ and\ \bibinfo {author} {\bibfnamefont {Gennady}\ \bibnamefont
  {Shvets}}} (\bibinfo {year} {2015}),\ \bibfield  {title} {\enquote {\bibinfo
  {title} {Guiding electromagnetic waves around sharp corners: {T}opologically
  protected photonic transport in metawaveguides},}\ }\href
  {https://link.aps.org/doi/10.1103/PhysRevLett.114.127401} {\bibfield
  {journal} {\bibinfo  {journal} {Phys. Rev. Lett.}\ }\textbf {\bibinfo
  {volume} {114}},\ \bibinfo {pages} {127401}}\BibitemShut {NoStop}%
\bibitem [{\citenamefont {Ma}\ and\ \citenamefont {Shvets}(2016)}]{Ma:NJP2016}%
  \BibitemOpen
  \bibfield  {author} {\bibinfo {author} {\bibnamefont {Ma}, \bibfnamefont
  {Tzuhsuan}}, \ and\ \bibinfo {author} {\bibfnamefont {Gennady}\ \bibnamefont
  {Shvets}}} (\bibinfo {year} {2016}),\ \bibfield  {title} {\enquote {\bibinfo
  {title} {{All-Si valley-Hall photonic topological insulator}},}\ }\href
  {http://stacks.iop.org/1367-2630/18/i=2/a=025012?key=crossref.6843ccb7b3111b680b4eb3904f525ddd}
  {\bibfield  {journal} {\bibinfo  {journal} {New J. Phys.}\ }\textbf {\bibinfo
  {volume} {18}}~(\bibinfo {number} {2}),\ \bibinfo {pages}
  {025012}}\BibitemShut {NoStop}%
\bibitem [{\citenamefont {MacDonald}\ and\ \citenamefont
  {St\ifmmode~\check{r}\else \v{r}\fi{}eda}(1984)}]{MacDonald:1984PRB}%
  \BibitemOpen
  \bibfield  {author} {\bibinfo {author} {\bibnamefont {MacDonald},
  \bibfnamefont {A~H}}, \ and\ \bibinfo {author} {\bibfnamefont
  {P.}~\bibnamefont {St\ifmmode~\check{r}\else \v{r}\fi{}eda}}} (\bibinfo
  {year} {1984}),\ \bibfield  {title} {\enquote {\bibinfo {title} {Quantized
  {H}all effect and edge currents},}\ }\href
  {https://link.aps.org/doi/10.1103/PhysRevB.29.1616} {\bibfield  {journal}
  {\bibinfo  {journal} {Phys. Rev. B}\ }\textbf {\bibinfo {volume} {29}},\
  \bibinfo {pages} {1616--1619}}\BibitemShut {NoStop}%
\bibitem [{\citenamefont {Maczewsky}\ \emph {et~al.}(2017)\citenamefont
  {Maczewsky}, \citenamefont {Zeuner}, \citenamefont {Nolte},\ and\
  \citenamefont {Szameit}}]{Maczewsky:2017NatCom}%
  \BibitemOpen
  \bibfield  {author} {\bibinfo {author} {\bibnamefont {Maczewsky},
  \bibfnamefont {Lukas~J}}, \bibinfo {author} {\bibfnamefont {Julia~M}\
  \bibnamefont {Zeuner}}, \bibinfo {author} {\bibfnamefont {Stefan}\
  \bibnamefont {Nolte}}, \ and\ \bibinfo {author} {\bibfnamefont {Alexander}\
  \bibnamefont {Szameit}}} (\bibinfo {year} {2017}),\ \bibfield  {title}
  {\enquote {\bibinfo {title} {Observation of photonic anomalous {F}loquet
  topological insulators},}\ }\href
  {https://www.nature.com/articles/ncomms13756} {\bibfield  {journal} {\bibinfo
   {journal} {Nat. Commun.}\ }\textbf {\bibinfo {volume} {8}},\ \bibinfo
  {pages} {13756}}\BibitemShut {NoStop}%
\bibitem [{\citenamefont {Maffei}\ \emph {et~al.}(2018)\citenamefont {Maffei},
  \citenamefont {Dauphin}, \citenamefont {Cardano}, \citenamefont
  {Lewenstein},\ and\ \citenamefont {Massignan}}]{Maffei:2017NJP}%
  \BibitemOpen
  \bibfield  {author} {\bibinfo {author} {\bibnamefont {Maffei}, \bibfnamefont
  {Maria}}, \bibinfo {author} {\bibfnamefont {Alexandre}\ \bibnamefont
  {Dauphin}}, \bibinfo {author} {\bibfnamefont {Filippo}\ \bibnamefont
  {Cardano}}, \bibinfo {author} {\bibfnamefont {Maciej}\ \bibnamefont
  {Lewenstein}}, \ and\ \bibinfo {author} {\bibfnamefont {Pietro}\ \bibnamefont
  {Massignan}}} (\bibinfo {year} {2018}),\ \bibfield  {title} {\enquote
  {\bibinfo {title} {Topological characterization of chiral models through
  their long time dynamics},}\ }\href
  {http://iopscience.iop.org/article/10.1088/1367-2630/aa9d4c} {\bibfield
  {journal} {\bibinfo  {journal} {New J. Phys.}\ }\textbf {\bibinfo {volume}
  {20}}~(\bibinfo {number} {1}),\ \bibinfo {pages} {013023}}\BibitemShut
  {NoStop}%
\bibitem [{\citenamefont {Maghrebi}\ \emph
  {et~al.}(2015{\natexlab{a}})\citenamefont {Maghrebi}, \citenamefont
  {Gullans}, \citenamefont {Bienias}, \citenamefont {Choi}, \citenamefont
  {Martin}, \citenamefont {Firstenberg}, \citenamefont {Lukin}, \citenamefont
  {B\"uchler},\ and\ \citenamefont {Gorshkov}}]{Maghrebi:PRL2015}%
  \BibitemOpen
  \bibfield  {author} {\bibinfo {author} {\bibnamefont {Maghrebi},
  \bibfnamefont {M~F}}, \bibinfo {author} {\bibfnamefont {M.~J.}\ \bibnamefont
  {Gullans}}, \bibinfo {author} {\bibfnamefont {P.}~\bibnamefont {Bienias}},
  \bibinfo {author} {\bibfnamefont {S.}~\bibnamefont {Choi}}, \bibinfo {author}
  {\bibfnamefont {I.}~\bibnamefont {Martin}}, \bibinfo {author} {\bibfnamefont
  {O.}~\bibnamefont {Firstenberg}}, \bibinfo {author} {\bibfnamefont {M.~D.}\
  \bibnamefont {Lukin}}, \bibinfo {author} {\bibfnamefont {H.~P.}\ \bibnamefont
  {B\"uchler}}, \ and\ \bibinfo {author} {\bibfnamefont {A.~V.}\ \bibnamefont
  {Gorshkov}}} (\bibinfo {year} {2015}{\natexlab{a}}),\ \bibfield  {title}
  {\enquote {\bibinfo {title} {Coulomb bound states of strongly interacting
  photons},}\ }\href {https://link.aps.org/doi/10.1103/PhysRevLett.115.123601}
  {\bibfield  {journal} {\bibinfo  {journal} {Phys. Rev. Lett.}\ }\textbf
  {\bibinfo {volume} {115}},\ \bibinfo {pages} {123601}}\BibitemShut {NoStop}%
\bibitem [{\citenamefont {Maghrebi}\ \emph
  {et~al.}(2015{\natexlab{b}})\citenamefont {Maghrebi}, \citenamefont {Yao},
  \citenamefont {Hafezi}, \citenamefont {Pohl}, \citenamefont {Firstenberg},\
  and\ \citenamefont {Gorshkov}}]{Maghrebi:2015PRA}%
  \BibitemOpen
  \bibfield  {author} {\bibinfo {author} {\bibnamefont {Maghrebi},
  \bibfnamefont {Mohammad~F}}, \bibinfo {author} {\bibfnamefont {Norman~Y.}\
  \bibnamefont {Yao}}, \bibinfo {author} {\bibfnamefont {Mohammad}\
  \bibnamefont {Hafezi}}, \bibinfo {author} {\bibfnamefont {Thomas}\
  \bibnamefont {Pohl}}, \bibinfo {author} {\bibfnamefont {Ofer}\ \bibnamefont
  {Firstenberg}}, \ and\ \bibinfo {author} {\bibfnamefont {Alexey~V.}\
  \bibnamefont {Gorshkov}}} (\bibinfo {year} {2015}{\natexlab{b}}),\ \bibfield
  {title} {\enquote {\bibinfo {title} {Fractional quantum {H}all states of
  {R}ydberg polaritons},}\ }\href
  {http://link.aps.org/doi/10.1103/PhysRevA.91.033838} {\bibfield  {journal}
  {\bibinfo  {journal} {Phys. Rev. A}\ }\textbf {\bibinfo {volume} {91}},\
  \bibinfo {pages} {033838}}\BibitemShut {NoStop}%
\bibitem [{\citenamefont {Makris}\ \emph {et~al.}(2008)\citenamefont {Makris},
  \citenamefont {El-Ganainy}, \citenamefont {Christodoulides},\ and\
  \citenamefont {Musslimani}}]{makris2008beam}%
  \BibitemOpen
  \bibfield  {author} {\bibinfo {author} {\bibnamefont {Makris}, \bibfnamefont
  {Konstantinos~G}}, \bibinfo {author} {\bibfnamefont {Ramy}\ \bibnamefont
  {El-Ganainy}}, \bibinfo {author} {\bibfnamefont {Demetrios~N}\ \bibnamefont
  {Christodoulides}}, \ and\ \bibinfo {author} {\bibfnamefont {Ziad~H}\
  \bibnamefont {Musslimani}}} (\bibinfo {year} {2008}),\ \bibfield  {title}
  {\enquote {\bibinfo {title} {Beam dynamics in $\mathcal{PT}$ symmetric
  optical lattices},}\ }\href
  {https://journals.aps.org/prl/abstract/10.1103/PhysRevLett.100.103904}
  {\bibfield  {journal} {\bibinfo  {journal} {Phys. Rev. Lett.}\ }\textbf
  {\bibinfo {volume} {100}}~(\bibinfo {number} {10}),\ \bibinfo {pages}
  {103904}}\BibitemShut {NoStop}%
\bibitem [{\citenamefont {Malkova}\ \emph {et~al.}(2009)\citenamefont
  {Malkova}, \citenamefont {Hromada}, \citenamefont {Wang}, \citenamefont
  {Bryant},\ and\ \citenamefont {Chen}}]{Malkova:2009OptLett}%
  \BibitemOpen
  \bibfield  {author} {\bibinfo {author} {\bibnamefont {Malkova}, \bibfnamefont
  {Natalia}}, \bibinfo {author} {\bibfnamefont {Ivan}\ \bibnamefont {Hromada}},
  \bibinfo {author} {\bibfnamefont {Xiaosheng}\ \bibnamefont {Wang}}, \bibinfo
  {author} {\bibfnamefont {Garnett}\ \bibnamefont {Bryant}}, \ and\ \bibinfo
  {author} {\bibfnamefont {Zhigang}\ \bibnamefont {Chen}}} (\bibinfo {year}
  {2009}),\ \bibfield  {title} {\enquote {\bibinfo {title} {Observation of
  optical {S}hockley-like surface states in photonic superlattices},}\ }\href
  {https://www.osapublishing.org/ol/abstract.cfm?uri=ol-34-11-1633} {\bibfield
  {journal} {\bibinfo  {journal} {Opt. Lett.}\ }\textbf {\bibinfo {volume}
  {34}}~(\bibinfo {number} {11}),\ \bibinfo {pages} {1633--1635}}\BibitemShut
  {NoStop}%
\bibitem [{\citenamefont {Malzard}\ \emph {et~al.}(2015)\citenamefont
  {Malzard}, \citenamefont {Poli},\ and\ \citenamefont
  {Schomerus}}]{malzard2015topologically}%
  \BibitemOpen
  \bibfield  {author} {\bibinfo {author} {\bibnamefont {Malzard}, \bibfnamefont
  {Simon}}, \bibinfo {author} {\bibfnamefont {Charles}\ \bibnamefont {Poli}}, \
  and\ \bibinfo {author} {\bibfnamefont {Henning}\ \bibnamefont {Schomerus}}}
  (\bibinfo {year} {2015}),\ \bibfield  {title} {\enquote {\bibinfo {title}
  {Topologically protected defect states in open photonic systems with
  non-{H}ermitian charge-conjugation and parity-time symmetry},}\ }\href
  {https://journals.aps.org/prl/abstract/10.1103/PhysRevLett.115.200402}
  {\bibfield  {journal} {\bibinfo  {journal} {Phys. Rev. Lett.}\ }\textbf
  {\bibinfo {volume} {115}}~(\bibinfo {number} {20}),\ \bibinfo {pages}
  {200402}}\BibitemShut {NoStop}%
\bibitem [{\citenamefont {Malzard}\ and\ \citenamefont
  {Schomerus}(2018{\natexlab{a}})}]{malzard2018bulk}%
  \BibitemOpen
  \bibfield  {author} {\bibinfo {author} {\bibnamefont {Malzard}, \bibfnamefont
  {Simon}}, \ and\ \bibinfo {author} {\bibfnamefont {Henning}\ \bibnamefont
  {Schomerus}}} (\bibinfo {year} {2018}{\natexlab{a}}),\ \bibfield  {title}
  {\enquote {\bibinfo {title} {Bulk and edge-state arcs in non-{H}ermitian
  coupled-resonator arrays},}\ }\href
  {https://link.aps.org/doi/10.1103/PhysRevA.98.033807} {\bibfield  {journal}
  {\bibinfo  {journal} {Phys. Rev. A}\ }\textbf {\bibinfo {volume} {98}},\
  \bibinfo {pages} {033807}}\BibitemShut {NoStop}%
\bibitem [{\citenamefont {Malzard}\ and\ \citenamefont
  {Schomerus}(2018{\natexlab{b}})}]{malzard:2018NJP}%
  \BibitemOpen
  \bibfield  {author} {\bibinfo {author} {\bibnamefont {Malzard}, \bibfnamefont
  {Simon}}, \ and\ \bibinfo {author} {\bibfnamefont {Henning}\ \bibnamefont
  {Schomerus}}} (\bibinfo {year} {2018}{\natexlab{b}}),\ \bibfield  {title}
  {\enquote {\bibinfo {title} {Nonlinear mode competition and
  symmetry-protected power oscillations in topological lasers},}\ }\href
  {http://iopscience.iop.org/article/10.1088/1367-2630/aac9e0/meta} {\bibfield
  {journal} {\bibinfo  {journal} {New J. Phys.}\ }\textbf {\bibinfo {volume}
  {20}},\ \bibinfo {pages} {063044}}\BibitemShut {NoStop}%
\bibitem [{\citenamefont {Mancini}\ \emph {et~al.}(2015)\citenamefont
  {Mancini}, \citenamefont {Pagano}, \citenamefont {Cappellini}, \citenamefont
  {Livi}, \citenamefont {Rider}, \citenamefont {Catani}, \citenamefont {Sias},
  \citenamefont {Zoller}, \citenamefont {Inguscio}, \citenamefont {Dalmonte},\
  and\ \citenamefont {Fallani}}]{Mancini:2015Science}%
  \BibitemOpen
  \bibfield  {author} {\bibinfo {author} {\bibnamefont {Mancini}, \bibfnamefont
  {M}}, \bibinfo {author} {\bibfnamefont {G.}~\bibnamefont {Pagano}}, \bibinfo
  {author} {\bibfnamefont {G.}~\bibnamefont {Cappellini}}, \bibinfo {author}
  {\bibfnamefont {L.}~\bibnamefont {Livi}}, \bibinfo {author} {\bibfnamefont
  {M.}~\bibnamefont {Rider}}, \bibinfo {author} {\bibfnamefont
  {J.}~\bibnamefont {Catani}}, \bibinfo {author} {\bibfnamefont
  {C.}~\bibnamefont {Sias}}, \bibinfo {author} {\bibfnamefont {P.}~\bibnamefont
  {Zoller}}, \bibinfo {author} {\bibfnamefont {M.}~\bibnamefont {Inguscio}},
  \bibinfo {author} {\bibfnamefont {M.}~\bibnamefont {Dalmonte}}, \ and\
  \bibinfo {author} {\bibfnamefont {L.}~\bibnamefont {Fallani}}} (\bibinfo
  {year} {2015}),\ \bibfield  {title} {\enquote {\bibinfo {title} {Observation
  of chiral edge states with neutral fermions in synthetic {H}all ribbons},}\
  }\href {http://science.sciencemag.org/content/349/6255/1510} {\bibfield
  {journal} {\bibinfo  {journal} {Science}\ }\textbf {\bibinfo {volume}
  {349}}~(\bibinfo {number} {6255}),\ \bibinfo {pages}
  {1510--1513}}\BibitemShut {NoStop}%
\bibitem [{\citenamefont {Mansha}\ and\ \citenamefont
  {Chong}(2017)}]{Mansha:2017arXiv}%
  \BibitemOpen
  \bibfield  {author} {\bibinfo {author} {\bibnamefont {Mansha}, \bibfnamefont
  {Shampy}}, \ and\ \bibinfo {author} {\bibfnamefont {Y.~D.}\ \bibnamefont
  {Chong}}} (\bibinfo {year} {2017}),\ \bibfield  {title} {\enquote {\bibinfo
  {title} {Robust edge states in amorphous gyromagnetic photonic lattices},}\
  }\href {https://link.aps.org/doi/10.1103/PhysRevB.96.121405} {\bibfield
  {journal} {\bibinfo  {journal} {Phys. Rev. B}\ }\textbf {\bibinfo {volume}
  {96}},\ \bibinfo {pages} {121405}}\BibitemShut {NoStop}%
\bibitem [{\citenamefont {Marqu\'es}\ \emph {et~al.}(2002)\citenamefont
  {Marqu\'es}, \citenamefont {Medina},\ and\ \citenamefont
  {Rafii-El-Idrissi}}]{Marques:2002PRB}%
  \BibitemOpen
  \bibfield  {author} {\bibinfo {author} {\bibnamefont {Marqu\'es},
  \bibfnamefont {Ricardo}}, \bibinfo {author} {\bibfnamefont {Francisco}\
  \bibnamefont {Medina}}, \ and\ \bibinfo {author} {\bibfnamefont {Rachid}\
  \bibnamefont {Rafii-El-Idrissi}}} (\bibinfo {year} {2002}),\ \bibfield
  {title} {\enquote {\bibinfo {title} {Role of bianisotropy in negative
  permeability and left-handed metamaterials},}\ }\href
  {https://link.aps.org/doi/10.1103/PhysRevB.65.144440} {\bibfield  {journal}
  {\bibinfo  {journal} {Phys. Rev. B}\ }\textbf {\bibinfo {volume} {65}},\
  \bibinfo {pages} {144440}}\BibitemShut {NoStop}%
\bibitem [{\citenamefont {Martinez~Alvarez}\ \emph {et~al.}(2018)\citenamefont
  {Martinez~Alvarez}, \citenamefont {Vargas}, \citenamefont {Berdakin},\ and\
  \citenamefont {Foa~Torres}}]{alvarez2018topological}%
  \BibitemOpen
  \bibfield  {author} {\bibinfo {author} {\bibnamefont {Martinez~Alvarez},
  \bibfnamefont {V~M}}, \bibinfo {author} {\bibfnamefont {J.~E.~Barrios}\
  \bibnamefont {Vargas}}, \bibinfo {author} {\bibfnamefont {M.}~\bibnamefont
  {Berdakin}}, \ and\ \bibinfo {author} {\bibfnamefont {L.~E.~F.}\ \bibnamefont
  {Foa~Torres}}} (\bibinfo {year} {2018}),\ \bibfield  {title} {\enquote
  {\bibinfo {title} {Topological states of non-hermitian systems},}\ }\href
  {https://link.springer.com/article/10.1140%2Fepjst%2Fe2018-800091-5}
  {\bibfield  {journal} {\bibinfo  {journal} {Euro. Phys. J. Spec. Top.}\
  }\textbf {\bibinfo {volume} {227}}~(\bibinfo {number} {12}),\ \bibinfo
  {pages} {1295--1308}}\BibitemShut {NoStop}%
\bibitem [{\citenamefont {Mei}\ \emph {et~al.}(2016)\citenamefont {Mei},
  \citenamefont {Xue}, \citenamefont {Zhang}, \citenamefont {Tian},
  \citenamefont {Lee},\ and\ \citenamefont {Zhu}}]{Mei:2016QST}%
  \BibitemOpen
  \bibfield  {author} {\bibinfo {author} {\bibnamefont {Mei}, \bibfnamefont
  {Feng}}, \bibinfo {author} {\bibfnamefont {Zheng-Yuan}\ \bibnamefont {Xue}},
  \bibinfo {author} {\bibfnamefont {Dan-Wei}\ \bibnamefont {Zhang}}, \bibinfo
  {author} {\bibfnamefont {Lin}\ \bibnamefont {Tian}}, \bibinfo {author}
  {\bibfnamefont {Chaohong}\ \bibnamefont {Lee}}, \ and\ \bibinfo {author}
  {\bibfnamefont {Shi-Liang}\ \bibnamefont {Zhu}}} (\bibinfo {year} {2016}),\
  \bibfield  {title} {\enquote {\bibinfo {title} {Witnessing topological {W}eyl
  semimetal phase in a minimal circuit-{QED} lattice},}\ }\href
  {http://iopscience.iop.org/article/10.1088/2058-9565/1/1/015006/meta}
  {\bibfield  {journal} {\bibinfo  {journal} {Quantum Science and Technology}\
  }\textbf {\bibinfo {volume} {1}}~(\bibinfo {number} {1}),\ \bibinfo {pages}
  {015006}}\BibitemShut {NoStop}%
\bibitem [{\citenamefont {Mei}\ \emph {et~al.}(2015)\citenamefont {Mei},
  \citenamefont {You}, \citenamefont {Nie}, \citenamefont {Fazio},
  \citenamefont {Zhu},\ and\ \citenamefont {Kwek}}]{Mei:2015}%
  \BibitemOpen
  \bibfield  {author} {\bibinfo {author} {\bibnamefont {Mei}, \bibfnamefont
  {Feng}}, \bibinfo {author} {\bibfnamefont {Jia-Bin}\ \bibnamefont {You}},
  \bibinfo {author} {\bibfnamefont {Wei}\ \bibnamefont {Nie}}, \bibinfo
  {author} {\bibfnamefont {Rosario}\ \bibnamefont {Fazio}}, \bibinfo {author}
  {\bibfnamefont {Shi-Liang}\ \bibnamefont {Zhu}}, \ and\ \bibinfo {author}
  {\bibfnamefont {L.~C.}\ \bibnamefont {Kwek}}} (\bibinfo {year} {2015}),\
  \bibfield  {title} {\enquote {\bibinfo {title} {Simulation and detection of
  photonic {C}hern insulators in a one-dimensional circuit-{QED} lattice},}\
  }\href {https://link.aps.org/doi/10.1103/PhysRevA.92.041805} {\bibfield
  {journal} {\bibinfo  {journal} {Phys. Rev. A}\ }\textbf {\bibinfo {volume}
  {92}},\ \bibinfo {pages} {041805}}\BibitemShut {NoStop}%
\bibitem [{\citenamefont {Meier}\ \emph {et~al.}(2018)\citenamefont {Meier},
  \citenamefont {An}, \citenamefont {Dauphin}, \citenamefont {Maffei},
  \citenamefont {Massignan}, \citenamefont {Hughes},\ and\ \citenamefont
  {Gadway}}]{Meier:2018arXiv}%
  \BibitemOpen
  \bibfield  {author} {\bibinfo {author} {\bibnamefont {Meier}, \bibfnamefont
  {Eric~J}}, \bibinfo {author} {\bibfnamefont {Fangzhao~Alex}\ \bibnamefont
  {An}}, \bibinfo {author} {\bibfnamefont {Alexandre}\ \bibnamefont {Dauphin}},
  \bibinfo {author} {\bibfnamefont {Maria}\ \bibnamefont {Maffei}}, \bibinfo
  {author} {\bibfnamefont {Pietro}\ \bibnamefont {Massignan}}, \bibinfo
  {author} {\bibfnamefont {Taylor~L}\ \bibnamefont {Hughes}}, \ and\ \bibinfo
  {author} {\bibfnamefont {Bryce}\ \bibnamefont {Gadway}}} (\bibinfo {year}
  {2018}),\ \bibfield  {title} {\enquote {\bibinfo {title} {Observation of the
  topological {A}nderson insulator in disordered atomic wires},}\ }\href
  {http://science.sciencemag.org/content/362/6417/929} {\bibfield  {journal}
  {\bibinfo  {journal} {Science}\ }\textbf {\bibinfo {volume} {362}}~(\bibinfo
  {number} {6417}),\ \bibinfo {pages} {929--933}}\BibitemShut {NoStop}%
\bibitem [{\citenamefont {Mekis}\ \emph {et~al.}(1996)\citenamefont {Mekis},
  \citenamefont {Chen}, \citenamefont {Kurland}, \citenamefont {Fan},
  \citenamefont {Villeneuve},\ and\ \citenamefont
  {Joannopoulos}}]{Mekis:PRL1996}%
  \BibitemOpen
  \bibfield  {author} {\bibinfo {author} {\bibnamefont {Mekis}, \bibfnamefont
  {Attila}}, \bibinfo {author} {\bibfnamefont {JC}~\bibnamefont {Chen}},
  \bibinfo {author} {\bibfnamefont {I}~\bibnamefont {Kurland}}, \bibinfo
  {author} {\bibfnamefont {Shanhui}\ \bibnamefont {Fan}}, \bibinfo {author}
  {\bibfnamefont {Pierre~R}\ \bibnamefont {Villeneuve}}, \ and\ \bibinfo
  {author} {\bibfnamefont {JD}~\bibnamefont {Joannopoulos}}} (\bibinfo {year}
  {1996}),\ \bibfield  {title} {\enquote {\bibinfo {title} {High transmission
  through sharp bends in photonic crystal waveguides},}\ }\href
  {https://journals.aps.org/prl/abstract/10.1103/PhysRevLett.77.3787}
  {\bibfield  {journal} {\bibinfo  {journal} {Phys. Rev. Lett.}\ }\textbf
  {\bibinfo {volume} {77}}~(\bibinfo {number} {18}),\ \bibinfo {pages}
  {3787}}\BibitemShut {NoStop}%
\bibitem [{\citenamefont {Meng}(2003)}]{Meng:2003JPA}%
  \BibitemOpen
  \bibfield  {author} {\bibinfo {author} {\bibnamefont {Meng}, \bibfnamefont
  {Guowu}}} (\bibinfo {year} {2003}),\ \bibfield  {title} {\enquote {\bibinfo
  {title} {Geometric construction of the quantum {H}all effect in all even
  dimensions},}\ }\href
  {http://iopscience.iop.org/article/10.1088/0305-4470/36/36/301/meta}
  {\bibfield  {journal} {\bibinfo  {journal} {J. Phys. A}\ }\textbf {\bibinfo
  {volume} {36}}~(\bibinfo {number} {36}),\ \bibinfo {pages}
  {9415}}\BibitemShut {NoStop}%
\bibitem [{\citenamefont {Metelmann}\ and\ \citenamefont
  {Clerk}(2015)}]{Clerk2015}%
  \BibitemOpen
  \bibfield  {author} {\bibinfo {author} {\bibnamefont {Metelmann},
  \bibfnamefont {A}}, \ and\ \bibinfo {author} {\bibfnamefont {A.~A.}\
  \bibnamefont {Clerk}}} (\bibinfo {year} {2015}),\ \bibfield  {title}
  {\enquote {\bibinfo {title} {Nonreciprocal photon transmission and
  amplification via reservoir engineering},}\ }\href
  {https://link.aps.org/doi/10.1103/PhysRevX.5.021025} {\bibfield  {journal}
  {\bibinfo  {journal} {Phys. Rev. X}\ }\textbf {\bibinfo {volume} {5}},\
  \bibinfo {pages} {021025}}\BibitemShut {NoStop}%
\bibitem [{\citenamefont {Mikami}\ \emph {et~al.}(2016)\citenamefont {Mikami},
  \citenamefont {Kitamura}, \citenamefont {Yasuda}, \citenamefont {Tsuji},
  \citenamefont {Oka},\ and\ \citenamefont {Aoki}}]{Mikami:2015PRB}%
  \BibitemOpen
  \bibfield  {author} {\bibinfo {author} {\bibnamefont {Mikami}, \bibfnamefont
  {T}}, \bibinfo {author} {\bibfnamefont {S.}~\bibnamefont {Kitamura}},
  \bibinfo {author} {\bibfnamefont {K.}~\bibnamefont {Yasuda}}, \bibinfo
  {author} {\bibfnamefont {N.}~\bibnamefont {Tsuji}}, \bibinfo {author}
  {\bibfnamefont {T.}~\bibnamefont {Oka}}, \ and\ \bibinfo {author}
  {\bibfnamefont {H.}~\bibnamefont {Aoki}}} (\bibinfo {year} {2016}),\
  \bibfield  {title} {\enquote {\bibinfo {title} {Brillouin-{W}igner theory for
  high-frequency expansion in periodically driven systems: {A}pplication to
  {F}loquet topological insulators},}\ }\href
  {https://doi.org/10.1103/PhysRevB.93.144307} {\bibfield  {journal} {\bibinfo
  {journal} {Phys. Rev. B}\ }\textbf {\bibinfo {volume} {93}},\ \bibinfo
  {pages} {144307}}\BibitemShut {NoStop}%
\bibitem [{\citenamefont {Mili{\'c}evi{\'c}}\ \emph {et~al.}(2018)\citenamefont
  {Mili{\'c}evi{\'c}}, \citenamefont {Montambaux}, \citenamefont {Ozawa},
  \citenamefont {Sagnes}, \citenamefont {Lema{\^\i}tre}, \citenamefont
  {Gratiet}, \citenamefont {Harouri}, \citenamefont {Bloch},\ and\
  \citenamefont {Amo}}]{Milicevic:2018arXiv}%
  \BibitemOpen
  \bibfield  {author} {\bibinfo {author} {\bibnamefont {Mili{\'c}evi{\'c}},
  \bibfnamefont {M}}, \bibinfo {author} {\bibfnamefont {G}~\bibnamefont
  {Montambaux}}, \bibinfo {author} {\bibfnamefont {T}~\bibnamefont {Ozawa}},
  \bibinfo {author} {\bibfnamefont {I}~\bibnamefont {Sagnes}}, \bibinfo
  {author} {\bibfnamefont {A}~\bibnamefont {Lema{\^\i}tre}}, \bibinfo {author}
  {\bibfnamefont {L~Le}\ \bibnamefont {Gratiet}}, \bibinfo {author}
  {\bibfnamefont {A}~\bibnamefont {Harouri}}, \bibinfo {author} {\bibfnamefont
  {J}~\bibnamefont {Bloch}}, \ and\ \bibinfo {author} {\bibfnamefont
  {A}~\bibnamefont {Amo}}} (\bibinfo {year} {2018}),\ \bibfield  {title}
  {\enquote {\bibinfo {title} {{Tilted and type-III Dirac cones emerging from
  flat bands in photonic orbital graphene}},}\ }\href
  {https://arxiv.org/abs/1807.08650} {\bibinfo  {journal} {arXiv:1807.08650}\
  }\BibitemShut {NoStop}%
\bibitem [{\citenamefont {Mili{\'c}evi{\'c}}\ \emph {et~al.}(2015)\citenamefont
  {Mili{\'c}evi{\'c}}, \citenamefont {Ozawa}, \citenamefont {Andreakou},
  \citenamefont {Carusotto}, \citenamefont {Jacqmin}, \citenamefont {Galopin},
  \citenamefont {Lema{\^\i}tre}, \citenamefont {Le~Gratiet}, \citenamefont
  {Sagnes}, \citenamefont {Bloch} \emph {et~al.}}]{Milicevic:20152DMat}%
  \BibitemOpen
\bibfield  {journal} {  }\bibfield  {author} {\bibinfo {author} {\bibnamefont
  {Mili{\'c}evi{\'c}}, \bibfnamefont {M}}, \bibinfo {author} {\bibfnamefont
  {T}~\bibnamefont {Ozawa}}, \bibinfo {author} {\bibfnamefont {P}~\bibnamefont
  {Andreakou}}, \bibinfo {author} {\bibfnamefont {I}~\bibnamefont {Carusotto}},
  \bibinfo {author} {\bibfnamefont {T}~\bibnamefont {Jacqmin}}, \bibinfo
  {author} {\bibfnamefont {E}~\bibnamefont {Galopin}}, \bibinfo {author}
  {\bibfnamefont {A}~\bibnamefont {Lema{\^\i}tre}}, \bibinfo {author}
  {\bibfnamefont {L}~\bibnamefont {Le~Gratiet}}, \bibinfo {author}
  {\bibfnamefont {I}~\bibnamefont {Sagnes}}, \bibinfo {author} {\bibfnamefont
  {J}~\bibnamefont {Bloch}},  \emph {et~al.}} (\bibinfo {year} {2015}),\
  \bibfield  {title} {\enquote {\bibinfo {title} {Edge states in polariton
  honeycomb lattices},}\ }\href
  {http://iopscience.iop.org/article/10.1088/2053-1583/2/3/034012} {\bibfield
  {journal} {\bibinfo  {journal} {2D Materials}\ }\textbf {\bibinfo {volume}
  {2}}~(\bibinfo {number} {3}),\ \bibinfo {pages} {034012}}\BibitemShut
  {NoStop}%
\bibitem [{\citenamefont {Mili{\'{c}}evi{\'{c}}}\ \emph
  {et~al.}(2017)\citenamefont {Mili{\'{c}}evi{\'{c}}}, \citenamefont {Ozawa},
  \citenamefont {Montambaux}, \citenamefont {Carusotto}, \citenamefont
  {Galopin}, \citenamefont {Lema{\^{i}}tre}, \citenamefont {{Le Gratiet}},
  \citenamefont {Sagnes}, \citenamefont {Bloch},\ and\ \citenamefont
  {Amo}}]{Milicevic:2017PRL}%
  \BibitemOpen
  \bibfield  {author} {\bibinfo {author} {\bibnamefont {Mili{\'{c}}evi{\'{c}}},
  \bibfnamefont {M}}, \bibinfo {author} {\bibfnamefont {T.}~\bibnamefont
  {Ozawa}}, \bibinfo {author} {\bibfnamefont {G.}~\bibnamefont {Montambaux}},
  \bibinfo {author} {\bibfnamefont {I.}~\bibnamefont {Carusotto}}, \bibinfo
  {author} {\bibfnamefont {E.}~\bibnamefont {Galopin}}, \bibinfo {author}
  {\bibfnamefont {A.}~\bibnamefont {Lema{\^{i}}tre}}, \bibinfo {author}
  {\bibfnamefont {L.}~\bibnamefont {{Le Gratiet}}}, \bibinfo {author}
  {\bibfnamefont {I.}~\bibnamefont {Sagnes}}, \bibinfo {author} {\bibfnamefont
  {J.}~\bibnamefont {Bloch}}, \ and\ \bibinfo {author} {\bibfnamefont
  {A.}~\bibnamefont {Amo}}} (\bibinfo {year} {2017}),\ \bibfield  {title}
  {\enquote {\bibinfo {title} {Orbital edge states in a photonic honeycomb
  lattice},}\ }\href {http://link.aps.org/doi/10.1103/PhysRevLett.118.107403}
  {\bibfield  {journal} {\bibinfo  {journal} {Phys. Rev. Lett.}\ }\textbf
  {\bibinfo {volume} {118}}~(\bibinfo {number} {10}),\ \bibinfo {pages}
  {107403}}\BibitemShut {NoStop}%
\bibitem [{\citenamefont {Milman}\ \emph {et~al.}(2000)\citenamefont {Milman},
  \citenamefont {Castin},\ and\ \citenamefont {Davidovich}}]{Milman:PRA2000}%
  \BibitemOpen
  \bibfield  {author} {\bibinfo {author} {\bibnamefont {Milman}, \bibfnamefont
  {P\'erola}}, \bibinfo {author} {\bibfnamefont {Yvan}\ \bibnamefont {Castin}},
  \ and\ \bibinfo {author} {\bibfnamefont {Luiz}\ \bibnamefont {Davidovich}}}
  (\bibinfo {year} {2000}),\ \bibfield  {title} {\enquote {\bibinfo {title}
  {Decoherence as phase diffusion},}\ }\href
  {https://link.aps.org/doi/10.1103/PhysRevA.61.063803} {\bibfield  {journal}
  {\bibinfo  {journal} {Phys. Rev. A}\ }\textbf {\bibinfo {volume} {61}},\
  \bibinfo {pages} {063803}}\BibitemShut {NoStop}%
\bibitem [{\citenamefont {Minkov}\ and\ \citenamefont
  {Savona}(2016)}]{Minkov:2016Optica}%
  \BibitemOpen
  \bibfield  {author} {\bibinfo {author} {\bibnamefont {Minkov}, \bibfnamefont
  {Momchil}}, \ and\ \bibinfo {author} {\bibfnamefont {Vincenzo}\ \bibnamefont
  {Savona}}} (\bibinfo {year} {2016}),\ \bibfield  {title} {\enquote {\bibinfo
  {title} {Haldane quantum {H}all effect for light in a dynamically modulated
  array of resonators},}\ }\href
  {http://www.osapublishing.org/optica/abstract.cfm?URI=optica-3-2-200}
  {\bibfield  {journal} {\bibinfo  {journal} {Optica}\ }\textbf {\bibinfo
  {volume} {3}}~(\bibinfo {number} {2}),\ \bibinfo {pages}
  {200--206}}\BibitemShut {NoStop}%
\bibitem [{\citenamefont {Mirek}\ \emph {et~al.}(2017)\citenamefont {Mirek},
  \citenamefont {Kr{\'{o}}l}, \citenamefont {Lekenta}, \citenamefont {Rousset},
  \citenamefont {Nawrocki}, \citenamefont {Kulczykowski}, \citenamefont
  {Matuszewski}, \citenamefont {Szczytko}, \citenamefont {Pacuski},\ and\
  \citenamefont {Pi{\c{e}}tka}}]{Mirek:2017PRB}%
  \BibitemOpen
  \bibfield  {author} {\bibinfo {author} {\bibnamefont {Mirek}, \bibfnamefont
  {R}}, \bibinfo {author} {\bibfnamefont {M.}~\bibnamefont {Kr{\'{o}}l}},
  \bibinfo {author} {\bibfnamefont {K.}~\bibnamefont {Lekenta}}, \bibinfo
  {author} {\bibfnamefont {J.-G.}\ \bibnamefont {Rousset}}, \bibinfo {author}
  {\bibfnamefont {M.}~\bibnamefont {Nawrocki}}, \bibinfo {author}
  {\bibfnamefont {M.}~\bibnamefont {Kulczykowski}}, \bibinfo {author}
  {\bibfnamefont {M.}~\bibnamefont {Matuszewski}}, \bibinfo {author}
  {\bibfnamefont {J.}~\bibnamefont {Szczytko}}, \bibinfo {author}
  {\bibfnamefont {W.}~\bibnamefont {Pacuski}}, \ and\ \bibinfo {author}
  {\bibfnamefont {B.}~\bibnamefont {Pi{\c{e}}tka}}} (\bibinfo {year} {2017}),\
  \bibfield  {title} {\enquote {\bibinfo {title} {{Angular dependence of giant
  Zeeman effect for semimagnetic cavity polaritons}},}\ }\href
  {https://link.aps.org/doi/10.1103/PhysRevB.95.085429} {\bibfield  {journal}
  {\bibinfo  {journal} {Phys. Rev. B}\ }\textbf {\bibinfo {volume}
  {95}}~(\bibinfo {number} {8}),\ \bibinfo {pages} {085429}}\BibitemShut
  {NoStop}%
\bibitem [{\citenamefont {Mitsch}\ \emph {et~al.}(2014)\citenamefont {Mitsch},
  \citenamefont {Sayrin}, \citenamefont {Albrecht}, \citenamefont
  {Schneeweiss},\ and\ \citenamefont
  {Rauschenbeutel}}]{Mitsch:2014naturecommunications}%
  \BibitemOpen
  \bibfield  {author} {\bibinfo {author} {\bibnamefont {Mitsch}, \bibfnamefont
  {R}}, \bibinfo {author} {\bibfnamefont {C}~\bibnamefont {Sayrin}}, \bibinfo
  {author} {\bibfnamefont {B}~\bibnamefont {Albrecht}}, \bibinfo {author}
  {\bibfnamefont {P}~\bibnamefont {Schneeweiss}}, \ and\ \bibinfo {author}
  {\bibfnamefont {A}~\bibnamefont {Rauschenbeutel}}} (\bibinfo {year} {2014}),\
  \bibfield  {title} {\enquote {\bibinfo {title} {Quantum state-controlled
  directional spontaneous emission of photons into a nanophotonic waveguide},}\
  }\href {https://www.nature.com/articles/ncomms6713} {\bibfield  {journal}
  {\bibinfo  {journal} {Nat. Commun.}\ }\textbf {\bibinfo {volume} {5}},\
  \bibinfo {pages} {5713}}\BibitemShut {NoStop}%
\bibitem [{\citenamefont {Mittal}\ \emph {et~al.}(2014)\citenamefont {Mittal},
  \citenamefont {Fan}, \citenamefont {Faez}, \citenamefont {Migdall},
  \citenamefont {Taylor},\ and\ \citenamefont {Hafezi}}]{Mittal:2014PRL}%
  \BibitemOpen
  \bibfield  {author} {\bibinfo {author} {\bibnamefont {Mittal}, \bibfnamefont
  {S}}, \bibinfo {author} {\bibfnamefont {J.}~\bibnamefont {Fan}}, \bibinfo
  {author} {\bibfnamefont {S.}~\bibnamefont {Faez}}, \bibinfo {author}
  {\bibfnamefont {A.}~\bibnamefont {Migdall}}, \bibinfo {author} {\bibfnamefont
  {J.~M.}\ \bibnamefont {Taylor}}, \ and\ \bibinfo {author} {\bibfnamefont
  {M.}~\bibnamefont {Hafezi}}} (\bibinfo {year} {2014}),\ \bibfield  {title}
  {\enquote {\bibinfo {title} {Topologically robust transport of photons in a
  synthetic gauge field},}\ }\href
  {https://link.aps.org/doi/10.1103/PhysRevLett.113.087403} {\bibfield
  {journal} {\bibinfo  {journal} {Phys. Rev. Lett.}\ }\textbf {\bibinfo
  {volume} {113}},\ \bibinfo {pages} {087403}}\BibitemShut {NoStop}%
\bibitem [{\citenamefont {Mittal}\ \emph
  {et~al.}(2016{\natexlab{a}})\citenamefont {Mittal}, \citenamefont {Ganeshan},
  \citenamefont {Fan}, \citenamefont {Vaezi},\ and\ \citenamefont
  {Hafezi}}]{Mittal:2016NatPhot}%
  \BibitemOpen
  \bibfield  {author} {\bibinfo {author} {\bibnamefont {Mittal}, \bibfnamefont
  {Sunil}}, \bibinfo {author} {\bibfnamefont {Sriram}\ \bibnamefont
  {Ganeshan}}, \bibinfo {author} {\bibfnamefont {Jingyun}\ \bibnamefont {Fan}},
  \bibinfo {author} {\bibfnamefont {Abolhassan}\ \bibnamefont {Vaezi}}, \ and\
  \bibinfo {author} {\bibfnamefont {Mohammad}\ \bibnamefont {Hafezi}}}
  (\bibinfo {year} {2016}{\natexlab{a}}),\ \bibfield  {title} {\enquote
  {\bibinfo {title} {Measurement of topological invariants in a 2{D} photonic
  system},}\ }\href
  {http://www.nature.com/nphoton/journal/v10/n3/full/nphoton.2016.10.html}
  {\bibfield  {journal} {\bibinfo  {journal} {Nat. Photonics}\ }\textbf
  {\bibinfo {volume} {10}}~(\bibinfo {number} {3}),\ \bibinfo {pages}
  {180}}\BibitemShut {NoStop}%
\bibitem [{\citenamefont {Mittal}\ and\ \citenamefont
  {Hafezi}(2017)}]{Mittal2017}%
  \BibitemOpen
  \bibfield  {author} {\bibinfo {author} {\bibnamefont {Mittal}, \bibfnamefont
  {Sunil}}, \ and\ \bibinfo {author} {\bibfnamefont {Mohammad}\ \bibnamefont
  {Hafezi}}} (\bibinfo {year} {2017}),\ \bibfield  {title} {\enquote {\bibinfo
  {title} {Topologically robust generation of correlated photon pairs},}\
  }\href {https://arxiv.org/abs/1709.09984} {\bibinfo  {journal}
  {arXiv:1709.09984}\ }\BibitemShut {NoStop}%
\bibitem [{\citenamefont {Mittal}\ \emph
  {et~al.}(2016{\natexlab{b}})\citenamefont {Mittal}, \citenamefont {Orre},\
  and\ \citenamefont {Hafezi}}]{Mittal:2016OptEx}%
  \BibitemOpen
\bibfield  {journal} {  }\bibfield  {author} {\bibinfo {author} {\bibnamefont
  {Mittal}, \bibfnamefont {Sunil}}, \bibinfo {author} {\bibfnamefont
  {Venkata~Vikram}\ \bibnamefont {Orre}}, \ and\ \bibinfo {author}
  {\bibfnamefont {Mohammad}\ \bibnamefont {Hafezi}}} (\bibinfo {year}
  {2016}{\natexlab{b}}),\ \bibfield  {title} {\enquote {\bibinfo {title}
  {Topologically robust transport of entangled photons in a 2{D} photonic
  system},}\ }\href
  {http://www.opticsexpress.org/abstract.cfm?URI=oe-24-14-15631} {\bibfield
  {journal} {\bibinfo  {journal} {Opt. Express}\ }\textbf {\bibinfo {volume}
  {24}}~(\bibinfo {number} {14}),\ \bibinfo {pages} {15631--15641}}\BibitemShut
  {NoStop}%
\bibitem [{\citenamefont {Miyake}\ \emph {et~al.}(2013)\citenamefont {Miyake},
  \citenamefont {Siviloglou}, \citenamefont {Kennedy}, \citenamefont {Burton},\
  and\ \citenamefont {Ketterle}}]{Miyake:2013PRL}%
  \BibitemOpen
  \bibfield  {author} {\bibinfo {author} {\bibnamefont {Miyake}, \bibfnamefont
  {H}}, \bibinfo {author} {\bibfnamefont {G.~A.}\ \bibnamefont {Siviloglou}},
  \bibinfo {author} {\bibfnamefont {C.~J.}\ \bibnamefont {Kennedy}}, \bibinfo
  {author} {\bibfnamefont {W.~C.}\ \bibnamefont {Burton}}, \ and\ \bibinfo
  {author} {\bibfnamefont {W.}~\bibnamefont {Ketterle}}} (\bibinfo {year}
  {2013}),\ \bibfield  {title} {\enquote {\bibinfo {title} {Realizing the
  {H}arper {H}amiltonian with laser-assisted tunneling in optical lattices},}\
  }\href {http://dx.doi.org/10.1103/physrevlett.111.185302} {\bibfield
  {journal} {\bibinfo  {journal} {Phys. Rev. Lett.}\ }\textbf {\bibinfo
  {volume} {111}},\ \bibinfo {pages} {185302}}\BibitemShut {NoStop}%
\bibitem [{\citenamefont {Moitra}\ \emph {et~al.}(2013)\citenamefont {Moitra},
  \citenamefont {Yang}, \citenamefont {Anderson}, \citenamefont {Kravchenko},
  \citenamefont {Briggs},\ and\ \citenamefont
  {Valentine}}]{Moitra:2013NatPhot}%
  \BibitemOpen
  \bibfield  {author} {\bibinfo {author} {\bibnamefont {Moitra}, \bibfnamefont
  {Parikshit}}, \bibinfo {author} {\bibfnamefont {Yuanmu}\ \bibnamefont
  {Yang}}, \bibinfo {author} {\bibfnamefont {Zachary}\ \bibnamefont
  {Anderson}}, \bibinfo {author} {\bibfnamefont {Ivan~I}\ \bibnamefont
  {Kravchenko}}, \bibinfo {author} {\bibfnamefont {Dayrl~P}\ \bibnamefont
  {Briggs}}, \ and\ \bibinfo {author} {\bibfnamefont {Jason}\ \bibnamefont
  {Valentine}}} (\bibinfo {year} {2013}),\ \bibfield  {title} {\enquote
  {\bibinfo {title} {Realization of an all-dielectric zero-index optical
  metamaterial},}\ }\href {https://www.nature.com/articles/nphoton.2013.214}
  {\bibfield  {journal} {\bibinfo  {journal} {Nat. Photonics}\ }\textbf
  {\bibinfo {volume} {7}}~(\bibinfo {number} {10}),\ \bibinfo {pages}
  {791--795}}\BibitemShut {NoStop}%
\bibitem [{\citenamefont {Mondragon-Shem}\ \emph {et~al.}(2014)\citenamefont
  {Mondragon-Shem}, \citenamefont {Hughes}, \citenamefont {Song},\ and\
  \citenamefont {Prodan}}]{Mondragon:PRL2014}%
  \BibitemOpen
  \bibfield  {author} {\bibinfo {author} {\bibnamefont {Mondragon-Shem},
  \bibfnamefont {Ian}}, \bibinfo {author} {\bibfnamefont {Taylor~L}\
  \bibnamefont {Hughes}}, \bibinfo {author} {\bibfnamefont {Juntao}\
  \bibnamefont {Song}}, \ and\ \bibinfo {author} {\bibfnamefont {Emil}\
  \bibnamefont {Prodan}}} (\bibinfo {year} {2014}),\ \bibfield  {title}
  {\enquote {\bibinfo {title} {{Topological criticality in the chiral-symmetric
  AIII class at strong disorder}},}\ }\href
  {https://journals.aps.org/prl/abstract/10.1103/PhysRevLett.113.046802}
  {\bibfield  {journal} {\bibinfo  {journal} {Phys. Rev. Lett.}\ }\textbf
  {\bibinfo {volume} {113}}~(\bibinfo {number} {4}),\ \bibinfo {pages}
  {46802}}\BibitemShut {NoStop}%
\bibitem [{\citenamefont {Mong}\ and\ \citenamefont
  {Shivamoggi}(2011)}]{Mong:2011PRB}%
  \BibitemOpen
  \bibfield  {author} {\bibinfo {author} {\bibnamefont {Mong}, \bibfnamefont
  {Roger S~K}}, \ and\ \bibinfo {author} {\bibfnamefont {Vasudha}\ \bibnamefont
  {Shivamoggi}}} (\bibinfo {year} {2011}),\ \bibfield  {title} {\enquote
  {\bibinfo {title} {{Edge states and the bulk-boundary correspondence in Dirac
  Hamiltonians}},}\ }\href
  {http://journals.aps.org/prb/abstract/10.1103/PhysRevB.83.125109} {\bibfield
  {journal} {\bibinfo  {journal} {Phys. Rev. B}\ }\textbf {\bibinfo {volume}
  {83}}~(\bibinfo {number} {12}),\ \bibinfo {pages} {125109}}\BibitemShut
  {NoStop}%
\bibitem [{\citenamefont {Montambaux}\ and\ \citenamefont
  {Kohmoto}(1990)}]{montambaux:1990PRB}%
  \BibitemOpen
  \bibfield  {author} {\bibinfo {author} {\bibnamefont {Montambaux},
  \bibfnamefont {G}}, \ and\ \bibinfo {author} {\bibfnamefont {M.}~\bibnamefont
  {Kohmoto}}} (\bibinfo {year} {1990}),\ \bibfield  {title} {\enquote {\bibinfo
  {title} {Quantized {H}all effect in three dimensions},}\ }\href
  {https://link.aps.org/doi/10.1103/PhysRevB.41.11417} {\bibfield  {journal}
  {\bibinfo  {journal} {Phys. Rev. B}\ }\textbf {\bibinfo {volume} {41}},\
  \bibinfo {pages} {11417--11421}}\BibitemShut {NoStop}%
\bibitem [{\citenamefont {Montambaux}\ \emph {et~al.}(2009)\citenamefont
  {Montambaux}, \citenamefont {Pi{\'{e}}chon}, \citenamefont {Fuchs},\ and\
  \citenamefont {Goerbig}}]{Montambaux:2009PRB}%
  \BibitemOpen
  \bibfield  {author} {\bibinfo {author} {\bibnamefont {Montambaux},
  \bibfnamefont {G}}, \bibinfo {author} {\bibfnamefont {F.}~\bibnamefont
  {Pi{\'{e}}chon}}, \bibinfo {author} {\bibfnamefont {J.-N.}\ \bibnamefont
  {Fuchs}}, \ and\ \bibinfo {author} {\bibfnamefont {M.~O.}\ \bibnamefont
  {Goerbig}}} (\bibinfo {year} {2009}),\ \bibfield  {title} {\enquote {\bibinfo
  {title} {{Merging of Dirac points in a two-dimensional crystal}},}\ }\href
  {http://link.aps.org/doi/10.1103/PhysRevB.80.153412} {\bibfield  {journal}
  {\bibinfo  {journal} {Phys. Rev. B}\ }\textbf {\bibinfo {volume}
  {80}}~(\bibinfo {number} {15}),\ \bibinfo {pages} {153412}}\BibitemShut
  {NoStop}%
\bibitem [{\citenamefont {Moore}\ and\ \citenamefont
  {Read}(1991)}]{Moore:NPB1991}%
  \BibitemOpen
  \bibfield  {author} {\bibinfo {author} {\bibnamefont {Moore}, \bibfnamefont
  {Gregory}}, \ and\ \bibinfo {author} {\bibfnamefont {Nicholas}\ \bibnamefont
  {Read}}} (\bibinfo {year} {1991}),\ \bibfield  {title} {\enquote {\bibinfo
  {title} {Nonabelions in the fractional quantum {H}all effect},}\ }\href
  {http://www.sciencedirect.com/science/article/pii/055032139190407O}
  {\bibfield  {journal} {\bibinfo  {journal} {Nuclear Physics B}\ }\textbf
  {\bibinfo {volume} {360}}~(\bibinfo {number} {2}),\ \bibinfo {pages} {362 --
  396}}\BibitemShut {NoStop}%
\bibitem [{\citenamefont {Moore}\ and\ \citenamefont
  {Balents}(2007)}]{Moore:2007PRB}%
  \BibitemOpen
  \bibfield  {author} {\bibinfo {author} {\bibnamefont {Moore}, \bibfnamefont
  {J~E}}, \ and\ \bibinfo {author} {\bibfnamefont {L.}~\bibnamefont {Balents}}}
  (\bibinfo {year} {2007}),\ \bibfield  {title} {\enquote {\bibinfo {title}
  {Topological invariants of time-reversal-invariant band structures},}\ }\href
  {https://link.aps.org/doi/10.1103/PhysRevB.75.121306} {\bibfield  {journal}
  {\bibinfo  {journal} {Phys. Rev. B}\ }\textbf {\bibinfo {volume} {75}},\
  \bibinfo {pages} {121306}}\BibitemShut {NoStop}%
\bibitem [{\citenamefont {Moos}\ \emph {et~al.}(2015)\citenamefont {Moos},
  \citenamefont {H\"oning}, \citenamefont {Unanyan},\ and\ \citenamefont
  {Fleischhauer}}]{Moos:PRA2015}%
  \BibitemOpen
  \bibfield  {author} {\bibinfo {author} {\bibnamefont {Moos}, \bibfnamefont
  {Matthias}}, \bibinfo {author} {\bibfnamefont {Michael}\ \bibnamefont
  {H\"oning}}, \bibinfo {author} {\bibfnamefont {Razmik}\ \bibnamefont
  {Unanyan}}, \ and\ \bibinfo {author} {\bibfnamefont {Michael}\ \bibnamefont
  {Fleischhauer}}} (\bibinfo {year} {2015}),\ \bibfield  {title} {\enquote
  {\bibinfo {title} {Many-body physics of {R}ydberg dark-state polaritons in
  the strongly interacting regime},}\ }\href
  {https://link.aps.org/doi/10.1103/PhysRevA.92.053846} {\bibfield  {journal}
  {\bibinfo  {journal} {Phys. Rev. A}\ }\textbf {\bibinfo {volume} {92}},\
  \bibinfo {pages} {053846}}\BibitemShut {NoStop}%
\bibitem [{\citenamefont {M\"ott\"onen}\ \emph {et~al.}(2008)\citenamefont
  {M\"ott\"onen}, \citenamefont {Vartiainen},\ and\ \citenamefont
  {Pekola}}]{Pekola:2008}%
  \BibitemOpen
  \bibfield  {author} {\bibinfo {author} {\bibnamefont {M\"ott\"onen},
  \bibfnamefont {Mikko}}, \bibinfo {author} {\bibfnamefont {Juha~J.}\
  \bibnamefont {Vartiainen}}, \ and\ \bibinfo {author} {\bibfnamefont
  {Jukka~P.}\ \bibnamefont {Pekola}}} (\bibinfo {year} {2008}),\ \bibfield
  {title} {\enquote {\bibinfo {title} {Experimental determination of the
  {B}erry phase in a superconducting charge pump},}\ }\href
  {https://link.aps.org/doi/10.1103/PhysRevLett.100.177201} {\bibfield
  {journal} {\bibinfo  {journal} {Phys. Rev. Lett.}\ }\textbf {\bibinfo
  {volume} {100}},\ \bibinfo {pages} {177201}}\BibitemShut {NoStop}%
\bibitem [{\citenamefont {Mukherjee}\ \emph {et~al.}(2018)\citenamefont
  {Mukherjee}, \citenamefont {Chandrasekharan}, \citenamefont {{\"O}hberg},
  \citenamefont {Goldman},\ and\ \citenamefont {Thomson}}]{mukherjee2017state}%
  \BibitemOpen
  \bibfield  {author} {\bibinfo {author} {\bibnamefont {Mukherjee},
  \bibfnamefont {Sebabrata}}, \bibinfo {author} {\bibfnamefont {Harikumar~K}\
  \bibnamefont {Chandrasekharan}}, \bibinfo {author} {\bibfnamefont {Patrik}\
  \bibnamefont {{\"O}hberg}}, \bibinfo {author} {\bibfnamefont {Nathan}\
  \bibnamefont {Goldman}}, \ and\ \bibinfo {author} {\bibfnamefont {Robert~R}\
  \bibnamefont {Thomson}}} (\bibinfo {year} {2018}),\ \bibfield  {title}
  {\enquote {\bibinfo {title} {State-recycling and time-resolved imaging in
  topological photonic lattices},}\ }\href
  {https://www.nature.com/articles/s41467-018-06723-y} {\bibfield  {journal}
  {\bibinfo  {journal} {Nat. Commun.}\ }\textbf {\bibinfo {volume}
  {9}}~(\bibinfo {number} {1}),\ \bibinfo {pages} {4209}}\BibitemShut {NoStop}%
\bibitem [{\citenamefont {Mukherjee}\ \emph {et~al.}(2015)\citenamefont
  {Mukherjee}, \citenamefont {Spracklen}, \citenamefont {Choudhury},
  \citenamefont {Goldman}, \citenamefont {{\"O}hberg}, \citenamefont
  {Andersson},\ and\ \citenamefont {Thomson}}]{Mukherjee:2015NJP}%
  \BibitemOpen
  \bibfield  {author} {\bibinfo {author} {\bibnamefont {Mukherjee},
  \bibfnamefont {Sebabrata}}, \bibinfo {author} {\bibfnamefont {Alexander}\
  \bibnamefont {Spracklen}}, \bibinfo {author} {\bibfnamefont {Debaditya}\
  \bibnamefont {Choudhury}}, \bibinfo {author} {\bibfnamefont {Nathan}\
  \bibnamefont {Goldman}}, \bibinfo {author} {\bibfnamefont {Patrik}\
  \bibnamefont {{\"O}hberg}}, \bibinfo {author} {\bibfnamefont {Erika}\
  \bibnamefont {Andersson}}, \ and\ \bibinfo {author} {\bibfnamefont
  {Robert~R}\ \bibnamefont {Thomson}}} (\bibinfo {year} {2015}),\ \bibfield
  {title} {\enquote {\bibinfo {title} {Modulation-assisted tunneling in
  laser-fabricated photonic {W}annier-{S}tark ladders},}\ }\href
  {http://iopscience.iop.org/article/10.1088/1367-2630/17/11/115002/meta}
  {\bibfield  {journal} {\bibinfo  {journal} {New J. Phys.}\ }\textbf {\bibinfo
  {volume} {17}}~(\bibinfo {number} {11}),\ \bibinfo {pages}
  {115002}}\BibitemShut {NoStop}%
\bibitem [{\citenamefont {Mukherjee}\ \emph {et~al.}(2017)\citenamefont
  {Mukherjee}, \citenamefont {Spracklen}, \citenamefont {Valiente},
  \citenamefont {Andersson}, \citenamefont {{\"O}hberg}, \citenamefont
  {Goldman},\ and\ \citenamefont {Thomson}}]{Mukherjee:2017NatCom}%
  \BibitemOpen
  \bibfield  {author} {\bibinfo {author} {\bibnamefont {Mukherjee},
  \bibfnamefont {Sebabrata}}, \bibinfo {author} {\bibfnamefont {Alexander}\
  \bibnamefont {Spracklen}}, \bibinfo {author} {\bibfnamefont {Manuel}\
  \bibnamefont {Valiente}}, \bibinfo {author} {\bibfnamefont {Erika}\
  \bibnamefont {Andersson}}, \bibinfo {author} {\bibfnamefont {Patrik}\
  \bibnamefont {{\"O}hberg}}, \bibinfo {author} {\bibfnamefont {Nathan}\
  \bibnamefont {Goldman}}, \ and\ \bibinfo {author} {\bibfnamefont {Robert~R}\
  \bibnamefont {Thomson}}} (\bibinfo {year} {2017}),\ \bibfield  {title}
  {\enquote {\bibinfo {title} {Experimental observation of anomalous
  topological edge modes in a slowly driven photonic lattice},}\ }\href
  {https://www.nature.com/articles/ncomms13918} {\bibfield  {journal} {\bibinfo
   {journal} {Nat. Commun.}\ }\textbf {\bibinfo {volume} {8}},\ \bibinfo
  {pages} {13918}}\BibitemShut {NoStop}%
\bibitem [{\citenamefont {Mukherjee}\ \emph {et~al.}(2016)\citenamefont
  {Mukherjee}, \citenamefont {Valiente}, \citenamefont {Goldman}, \citenamefont
  {Spracklen}, \citenamefont {Andersson}, \citenamefont {\"Ohberg},\ and\
  \citenamefont {Thomson}}]{Mukherjee:PRA2016}%
  \BibitemOpen
  \bibfield  {author} {\bibinfo {author} {\bibnamefont {Mukherjee},
  \bibfnamefont {Sebabrata}}, \bibinfo {author} {\bibfnamefont {Manuel}\
  \bibnamefont {Valiente}}, \bibinfo {author} {\bibfnamefont {Nathan}\
  \bibnamefont {Goldman}}, \bibinfo {author} {\bibfnamefont {Alexander}\
  \bibnamefont {Spracklen}}, \bibinfo {author} {\bibfnamefont {Erika}\
  \bibnamefont {Andersson}}, \bibinfo {author} {\bibfnamefont {Patrik}\
  \bibnamefont {\"Ohberg}}, \ and\ \bibinfo {author} {\bibfnamefont
  {Robert~R.}\ \bibnamefont {Thomson}}} (\bibinfo {year} {2016}),\ \bibfield
  {title} {\enquote {\bibinfo {title} {Observation of pair tunneling and
  coherent destruction of tunneling in arrays of optical waveguides},}\ }\href
  {https://link.aps.org/doi/10.1103/PhysRevA.94.053853} {\bibfield  {journal}
  {\bibinfo  {journal} {Phys. Rev. A}\ }\textbf {\bibinfo {volume} {94}},\
  \bibinfo {pages} {053853}}\BibitemShut {NoStop}%
\bibitem [{\citenamefont {Nagaosa}\ \emph {et~al.}(2010)\citenamefont
  {Nagaosa}, \citenamefont {Sinova}, \citenamefont {Onoda}, \citenamefont
  {MacDonald},\ and\ \citenamefont {Ong}}]{Nagaosa:2010RMP}%
  \BibitemOpen
  \bibfield  {author} {\bibinfo {author} {\bibnamefont {Nagaosa}, \bibfnamefont
  {Naoto}}, \bibinfo {author} {\bibfnamefont {Jairo}\ \bibnamefont {Sinova}},
  \bibinfo {author} {\bibfnamefont {Shigeki}\ \bibnamefont {Onoda}}, \bibinfo
  {author} {\bibfnamefont {A.~H.}\ \bibnamefont {MacDonald}}, \ and\ \bibinfo
  {author} {\bibfnamefont {N.~P.}\ \bibnamefont {Ong}}} (\bibinfo {year}
  {2010}),\ \bibfield  {title} {\enquote {\bibinfo {title} {Anomalous {H}all
  effect},}\ }\href {https://link.aps.org/doi/10.1103/RevModPhys.82.1539}
  {\bibfield  {journal} {\bibinfo  {journal} {Rev. Mod. Phys.}\ }\textbf
  {\bibinfo {volume} {82}},\ \bibinfo {pages} {1539--1592}}\BibitemShut
  {NoStop}%
\bibitem [{\citenamefont {Nakada}\ \emph {et~al.}(1996)\citenamefont {Nakada},
  \citenamefont {Fujita}, \citenamefont {Dresselhaus},\ and\ \citenamefont
  {Dresselhaus}}]{Nakada:1996PRB}%
  \BibitemOpen
  \bibfield  {author} {\bibinfo {author} {\bibnamefont {Nakada}, \bibfnamefont
  {Kyoko}}, \bibinfo {author} {\bibfnamefont {Mitsutaka}\ \bibnamefont
  {Fujita}}, \bibinfo {author} {\bibfnamefont {Gene}\ \bibnamefont
  {Dresselhaus}}, \ and\ \bibinfo {author} {\bibfnamefont {Mildred}\
  \bibnamefont {Dresselhaus}}} (\bibinfo {year} {1996}),\ \bibfield  {title}
  {\enquote {\bibinfo {title} {{Edge state in graphene ribbons: Nanometer size
  effect and edge shape dependence}},}\ }\href
  {https://journals.aps.org/prb/abstract/10.1103/PhysRevB.54.17954} {\bibfield
  {journal} {\bibinfo  {journal} {Phys. Rev. B}\ }\textbf {\bibinfo {volume}
  {54}}~(\bibinfo {number} {24}),\ \bibinfo {pages} {17954--17961}}\BibitemShut
  {NoStop}%
\bibitem [{\citenamefont {Nakahara}(2003)}]{NakaharaBook}%
  \BibitemOpen
  \bibfield  {author} {\bibinfo {author} {\bibnamefont {Nakahara},
  \bibfnamefont {Mikio}}} (\bibinfo {year} {2003}),\ \href@noop {} {\emph
  {\bibinfo {title} {Geometry, Topology and Physics, Second Edition}}}\
  (\bibinfo  {publisher} {CRC Press},\ \bibinfo {address} {Boca Raton,
  FL})\BibitemShut {NoStop}%
\bibitem [{\citenamefont {Nakajima}\ \emph {et~al.}(2016)\citenamefont
  {Nakajima}, \citenamefont {Tomita}, \citenamefont {Taie}, \citenamefont
  {Ichinose}, \citenamefont {Ozawa}, \citenamefont {Wang}, \citenamefont
  {Troyer},\ and\ \citenamefont {Takahashi}}]{Nakajima:2016NatPhys}%
  \BibitemOpen
  \bibfield  {author} {\bibinfo {author} {\bibnamefont {Nakajima},
  \bibfnamefont {Shuta}}, \bibinfo {author} {\bibfnamefont {Takafumi}\
  \bibnamefont {Tomita}}, \bibinfo {author} {\bibfnamefont {Shintaro}\
  \bibnamefont {Taie}}, \bibinfo {author} {\bibfnamefont {Tomohiro}\
  \bibnamefont {Ichinose}}, \bibinfo {author} {\bibfnamefont {Hideki}\
  \bibnamefont {Ozawa}}, \bibinfo {author} {\bibfnamefont {Lei}\ \bibnamefont
  {Wang}}, \bibinfo {author} {\bibfnamefont {Matthias}\ \bibnamefont {Troyer}},
  \ and\ \bibinfo {author} {\bibfnamefont {Yoshiro}\ \bibnamefont {Takahashi}}}
  (\bibinfo {year} {2016}),\ \bibfield  {title} {\enquote {\bibinfo {title}
  {Topological {T}houless pumping of ultracold fermions},}\ }\href
  {https://www.nature.com/articles/nphys3622} {\bibfield  {journal} {\bibinfo
  {journal} {Nat. Phys.}\ }\textbf {\bibinfo {volume} {12}}~(\bibinfo {number}
  {4}),\ \bibinfo {pages} {296}}\BibitemShut {NoStop}%
\bibitem [{\citenamefont {Nalitov}\ \emph {et~al.}(2015)\citenamefont
  {Nalitov}, \citenamefont {Solnyshkov},\ and\ \citenamefont
  {Malpuech}}]{Nalitov:2015PRL}%
  \BibitemOpen
  \bibfield  {author} {\bibinfo {author} {\bibnamefont {Nalitov}, \bibfnamefont
  {A~V}}, \bibinfo {author} {\bibfnamefont {D.~D.}\ \bibnamefont {Solnyshkov}},
  \ and\ \bibinfo {author} {\bibfnamefont {G.}~\bibnamefont {Malpuech}}}
  (\bibinfo {year} {2015}),\ \bibfield  {title} {\enquote {\bibinfo {title}
  {Polariton $\mathbb{Z}$ topological insulator},}\ }\href
  {http://link.aps.org/doi/10.1103/PhysRevLett.114.116401} {\bibfield
  {journal} {\bibinfo  {journal} {Phys. Rev. Lett.}\ }\textbf {\bibinfo
  {volume} {114}},\ \bibinfo {pages} {116401}}\BibitemShut {NoStop}%
\bibitem [{\citenamefont {Nathan}\ and\ \citenamefont
  {Rudner}(2015)}]{Nathan:2015}%
  \BibitemOpen
  \bibfield  {author} {\bibinfo {author} {\bibnamefont {Nathan}, \bibfnamefont
  {Frederik}}, \ and\ \bibinfo {author} {\bibfnamefont {Mark~S}\ \bibnamefont
  {Rudner}}} (\bibinfo {year} {2015}),\ \bibfield  {title} {\enquote {\bibinfo
  {title} {Topological singularities and the general classification of
  {F}loquet{\textendash}{B}loch systems},}\ }\href
  {https://doi.org/10.1088%2F1367-2630%2F17%2F12%2F125014} {\bibfield
  {journal} {\bibinfo  {journal} {New J. Phys.}\ }\textbf {\bibinfo {volume}
  {17}}~(\bibinfo {number} {12}),\ \bibinfo {pages} {125014}}\BibitemShut
  {NoStop}%
\bibitem [{\citenamefont {Nayak}\ \emph {et~al.}(2008)\citenamefont {Nayak},
  \citenamefont {Simon}, \citenamefont {Stern}, \citenamefont {Freedman},\ and\
  \citenamefont {Das~Sarma}}]{Nayak:RMP2008}%
  \BibitemOpen
  \bibfield  {author} {\bibinfo {author} {\bibnamefont {Nayak}, \bibfnamefont
  {Chetan}}, \bibinfo {author} {\bibfnamefont {Steven~H.}\ \bibnamefont
  {Simon}}, \bibinfo {author} {\bibfnamefont {Ady}\ \bibnamefont {Stern}},
  \bibinfo {author} {\bibfnamefont {Michael}\ \bibnamefont {Freedman}}, \ and\
  \bibinfo {author} {\bibfnamefont {Sankar}\ \bibnamefont {Das~Sarma}}}
  (\bibinfo {year} {2008}),\ \bibfield  {title} {\enquote {\bibinfo {title}
  {Non-{A}belian anyons and topological quantum computation},}\ }\href
  {https://link.aps.org/doi/10.1103/RevModPhys.80.1083} {\bibfield  {journal}
  {\bibinfo  {journal} {Rev. Mod. Phys.}\ }\textbf {\bibinfo {volume} {80}},\
  \bibinfo {pages} {1083--1159}}\BibitemShut {NoStop}%
\bibitem [{\citenamefont {Neto}\ \emph {et~al.}(2009)\citenamefont {Neto},
  \citenamefont {Guinea}, \citenamefont {Peres}, \citenamefont {Novoselov},\
  and\ \citenamefont {Geim}}]{neto2009electronic}%
  \BibitemOpen
  \bibfield  {author} {\bibinfo {author} {\bibnamefont {Neto}, \bibfnamefont
  {AH~Castro}}, \bibinfo {author} {\bibfnamefont {F}~\bibnamefont {Guinea}},
  \bibinfo {author} {\bibfnamefont {Nuno~MR}\ \bibnamefont {Peres}}, \bibinfo
  {author} {\bibfnamefont {Kostya~S}\ \bibnamefont {Novoselov}}, \ and\
  \bibinfo {author} {\bibfnamefont {Andre~K}\ \bibnamefont {Geim}}} (\bibinfo
  {year} {2009}),\ \bibfield  {title} {\enquote {\bibinfo {title} {The
  electronic properties of graphene},}\ }\href
  {https://journals.aps.org/rmp/abstract/10.1103/RevModPhys.81.109} {\bibfield
  {journal} {\bibinfo  {journal} {Rev. Mod. Phys.}\ }\textbf {\bibinfo {volume}
  {81}}~(\bibinfo {number} {1}),\ \bibinfo {pages} {109}}\BibitemShut {NoStop}%
\bibitem [{\citenamefont {Ningyuan}\ \emph {et~al.}(2016)\citenamefont
  {Ningyuan}, \citenamefont {Georgakopoulos}, \citenamefont {Ryou},
  \citenamefont {Schine}, \citenamefont {Sommer},\ and\ \citenamefont
  {Simon}}]{Ningyuan:PRA2016}%
  \BibitemOpen
  \bibfield  {author} {\bibinfo {author} {\bibnamefont {Ningyuan},
  \bibfnamefont {Jia}}, \bibinfo {author} {\bibfnamefont {Alexandros}\
  \bibnamefont {Georgakopoulos}}, \bibinfo {author} {\bibfnamefont {Albert}\
  \bibnamefont {Ryou}}, \bibinfo {author} {\bibfnamefont {Nathan}\ \bibnamefont
  {Schine}}, \bibinfo {author} {\bibfnamefont {Ariel}\ \bibnamefont {Sommer}},
  \ and\ \bibinfo {author} {\bibfnamefont {Jonathan}\ \bibnamefont {Simon}}}
  (\bibinfo {year} {2016}),\ \bibfield  {title} {\enquote {\bibinfo {title}
  {Observation and characterization of cavity {R}ydberg polaritons},}\ }\href
  {https://journals.aps.org/pra/abstract/10.1103/PhysRevA.93.041802} {\bibfield
   {journal} {\bibinfo  {journal} {Phys. Rev. A}\ }\textbf {\bibinfo {volume}
  {93}}~(\bibinfo {number} {4}),\ \bibinfo {pages} {041802}}\BibitemShut
  {NoStop}%
\bibitem [{\citenamefont {Ningyuan}\ \emph {et~al.}(2015)\citenamefont
  {Ningyuan}, \citenamefont {Owens}, \citenamefont {Sommer}, \citenamefont
  {Schuster},\ and\ \citenamefont {Simon}}]{ningyuan2015time}%
  \BibitemOpen
  \bibfield  {author} {\bibinfo {author} {\bibnamefont {Ningyuan},
  \bibfnamefont {Jia}}, \bibinfo {author} {\bibfnamefont {Clai}\ \bibnamefont
  {Owens}}, \bibinfo {author} {\bibfnamefont {Ariel}\ \bibnamefont {Sommer}},
  \bibinfo {author} {\bibfnamefont {David}\ \bibnamefont {Schuster}}, \ and\
  \bibinfo {author} {\bibfnamefont {Jonathan}\ \bibnamefont {Simon}}} (\bibinfo
  {year} {2015}),\ \bibfield  {title} {\enquote {\bibinfo {title} {Time-and
  site-resolved dynamics in a topological circuit},}\ }\href
  {https://journals.aps.org/prx/abstract/10.1103/PhysRevX.5.021031} {\bibfield
  {journal} {\bibinfo  {journal} {Phys. Rev. X}\ }\textbf {\bibinfo {volume}
  {5}}~(\bibinfo {number} {2}),\ \bibinfo {pages} {021031}}\BibitemShut
  {NoStop}%
\bibitem [{\citenamefont {Niu}\ \emph {et~al.}(1985)\citenamefont {Niu},
  \citenamefont {Thouless},\ and\ \citenamefont {Wu}}]{Niu:1985PRB}%
  \BibitemOpen
  \bibfield  {author} {\bibinfo {author} {\bibnamefont {Niu}, \bibfnamefont
  {Qian}}, \bibinfo {author} {\bibfnamefont {D.~J.}\ \bibnamefont {Thouless}},
  \ and\ \bibinfo {author} {\bibfnamefont {Yong-Shi}\ \bibnamefont {Wu}}}
  (\bibinfo {year} {1985}),\ \bibfield  {title} {\enquote {\bibinfo {title}
  {Quantized {H}all conductance as a topological invariant},}\ }\href
  {https://link.aps.org/doi/10.1103/PhysRevB.31.3372} {\bibfield  {journal}
  {\bibinfo  {journal} {Phys. Rev. B}\ }\textbf {\bibinfo {volume} {31}},\
  \bibinfo {pages} {3372--3377}}\BibitemShut {NoStop}%
\bibitem [{\citenamefont {Niu}\ and\ \citenamefont
  {Thouless}(1984)}]{Niu:1984}%
  \BibitemOpen
  \bibfield  {author} {\bibinfo {author} {\bibnamefont {Niu}, \bibfnamefont
  {Qian}}, \ and\ \bibinfo {author} {\bibfnamefont {DJ}~\bibnamefont
  {Thouless}}} (\bibinfo {year} {1984}),\ \bibfield  {title} {\enquote
  {\bibinfo {title} {Quantised adiabatic charge transport in the presence of
  substrate disorder and many-body interaction},}\ }\href
  {https://iopscience.iop.org/article/10.1088/0305-4470/17/12/016} {\bibfield
  {journal} {\bibinfo  {journal} {J. Phys. A}\ }\textbf {\bibinfo {volume}
  {17}}~(\bibinfo {number} {12}),\ \bibinfo {pages} {2453}}\BibitemShut
  {NoStop}%
\bibitem [{\citenamefont {Nixon}\ \emph {et~al.}(2013)\citenamefont {Nixon},
  \citenamefont {Ronen}, \citenamefont {Friesem},\ and\ \citenamefont
  {Davidson}}]{Nixon:2013PRL}%
  \BibitemOpen
  \bibfield  {author} {\bibinfo {author} {\bibnamefont {Nixon}, \bibfnamefont
  {Micha}}, \bibinfo {author} {\bibfnamefont {Eitan}\ \bibnamefont {Ronen}},
  \bibinfo {author} {\bibfnamefont {Asher~A.}\ \bibnamefont {Friesem}}, \ and\
  \bibinfo {author} {\bibfnamefont {Nir}\ \bibnamefont {Davidson}}} (\bibinfo
  {year} {2013}),\ \bibfield  {title} {\enquote {\bibinfo {title} {Observing
  geometric frustration with thousands of coupled lasers},}\ }\href
  {http://link.aps.org/doi/10.1103/PhysRevLett.110.184102} {\bibfield
  {journal} {\bibinfo  {journal} {Phys. Rev. Lett.}\ }\textbf {\bibinfo
  {volume} {110}}~(\bibinfo {number} {18}),\ \bibinfo {pages}
  {184102}}\BibitemShut {NoStop}%
\bibitem [{\citenamefont {Noda}(2016)}]{Noda:OFCC16}%
  \BibitemOpen
  \bibfield  {author} {\bibinfo {author} {\bibnamefont {Noda}, \bibfnamefont
  {Susumu}}} (\bibinfo {year} {2016}),\ \bibfield  {title} {\enquote {\bibinfo
  {title} {Photonic-crystal cavities},}\ }in\ \href
  {http://www.osapublishing.org/abstract.cfm?URI=OFC-2016-Th1K.1} {\emph
  {\bibinfo {booktitle} {Optical Fiber Communication Conference}}}\ (\bibinfo
  {publisher} {Optical Society of America})\ p.\ \bibinfo {pages}
  {Th1K.1}\BibitemShut {NoStop}%
\bibitem [{\citenamefont {Noh}\ and\ \citenamefont
  {Angelakis}(2017)}]{Noh:ROPP2017}%
  \BibitemOpen
  \bibfield  {author} {\bibinfo {author} {\bibnamefont {Noh}, \bibfnamefont
  {Changsuk}}, \ and\ \bibinfo {author} {\bibfnamefont {Dimitris~G}\
  \bibnamefont {Angelakis}}} (\bibinfo {year} {2017}),\ \bibfield  {title}
  {\enquote {\bibinfo {title} {Quantum simulations and many-body physics with
  light},}\ }\href {http://stacks.iop.org/0034-4885/80/i=1/a=016401} {\bibfield
   {journal} {\bibinfo  {journal} {Rep. Prog. Phys.}\ }\textbf {\bibinfo
  {volume} {80}}~(\bibinfo {number} {1}),\ \bibinfo {pages}
  {016401}}\BibitemShut {NoStop}%
\bibitem [{\citenamefont {Noh}\ \emph {et~al.}(2018{\natexlab{a}})\citenamefont
  {Noh}, \citenamefont {Benalcazar}, \citenamefont {Huang}, \citenamefont
  {Collins}, \citenamefont {Chen}, \citenamefont {Hughes},\ and\ \citenamefont
  {Rechtsman}}]{Noh:2018NatPhot}%
  \BibitemOpen
  \bibfield  {author} {\bibinfo {author} {\bibnamefont {Noh}, \bibfnamefont
  {Jiho}}, \bibinfo {author} {\bibfnamefont {Wladimir~A}\ \bibnamefont
  {Benalcazar}}, \bibinfo {author} {\bibfnamefont {Sheng}\ \bibnamefont
  {Huang}}, \bibinfo {author} {\bibfnamefont {Matthew~J}\ \bibnamefont
  {Collins}}, \bibinfo {author} {\bibfnamefont {Kevin~P}\ \bibnamefont {Chen}},
  \bibinfo {author} {\bibfnamefont {Taylor~L}\ \bibnamefont {Hughes}}, \ and\
  \bibinfo {author} {\bibfnamefont {Mikael~C}\ \bibnamefont {Rechtsman}}}
  (\bibinfo {year} {2018}{\natexlab{a}}),\ \bibfield  {title} {\enquote
  {\bibinfo {title} {Topological protection of photonic mid-gap defect
  modes},}\ }\href {https://www.nature.com/articles/s41566-018-0179-3}
  {\bibfield  {journal} {\bibinfo  {journal} {Nat. Photonics}\ }\textbf
  {\bibinfo {volume} {12}},\ \bibinfo {pages} {408–415}}\BibitemShut
  {NoStop}%
\bibitem [{\citenamefont {Noh}\ \emph {et~al.}(2018{\natexlab{b}})\citenamefont
  {Noh}, \citenamefont {Huang}, \citenamefont {Chen},\ and\ \citenamefont
  {Rechtsman}}]{Noh:arx2017}%
  \BibitemOpen
  \bibfield  {author} {\bibinfo {author} {\bibnamefont {Noh}, \bibfnamefont
  {Jiho}}, \bibinfo {author} {\bibfnamefont {Sheng}\ \bibnamefont {Huang}},
  \bibinfo {author} {\bibfnamefont {Kevin~P.}\ \bibnamefont {Chen}}, \ and\
  \bibinfo {author} {\bibfnamefont {Mikael~C.}\ \bibnamefont {Rechtsman}}}
  (\bibinfo {year} {2018}{\natexlab{b}}),\ \bibfield  {title} {\enquote
  {\bibinfo {title} {Observation of photonic topological valley {H}all edge
  states},}\ }\href {https://link.aps.org/doi/10.1103/PhysRevLett.120.063902}
  {\bibfield  {journal} {\bibinfo  {journal} {Phys. Rev. Lett.}\ }\textbf
  {\bibinfo {volume} {120}},\ \bibinfo {pages} {063902}}\BibitemShut {NoStop}%
\bibitem [{\citenamefont {Noh}\ \emph {et~al.}(2017)\citenamefont {Noh},
  \citenamefont {Huang}, \citenamefont {Leykam}, \citenamefont {Chong},
  \citenamefont {Chen},\ and\ \citenamefont {Rechtsman}}]{Noh:2017NatPhys}%
  \BibitemOpen
  \bibfield  {author} {\bibinfo {author} {\bibnamefont {Noh}, \bibfnamefont
  {Jiho}}, \bibinfo {author} {\bibfnamefont {Sheng}\ \bibnamefont {Huang}},
  \bibinfo {author} {\bibfnamefont {Daniel}\ \bibnamefont {Leykam}}, \bibinfo
  {author} {\bibfnamefont {YD}~\bibnamefont {Chong}}, \bibinfo {author}
  {\bibfnamefont {Kevin~P}\ \bibnamefont {Chen}}, \ and\ \bibinfo {author}
  {\bibfnamefont {Mikael~C}\ \bibnamefont {Rechtsman}}} (\bibinfo {year}
  {2017}),\ \bibfield  {title} {\enquote {\bibinfo {title} {Experimental
  observation of optical {W}eyl points and {F}ermi arc-like surface states},}\
  }\href {https://www.nature.com/articles/nphys4072} {\bibfield  {journal}
  {\bibinfo  {journal} {Nat. Phys.}\ }\textbf {\bibinfo {volume}
  {13}}~(\bibinfo {number} {6}),\ \bibinfo {pages} {611--617}}\BibitemShut
  {NoStop}%
\bibitem [{\citenamefont {Nunnenkamp}\ \emph {et~al.}(2011)\citenamefont
  {Nunnenkamp}, \citenamefont {Koch},\ and\ \citenamefont
  {Girvin}}]{Nunnenkamp:NJP2011}%
  \BibitemOpen
  \bibfield  {author} {\bibinfo {author} {\bibnamefont {Nunnenkamp},
  \bibfnamefont {A}}, \bibinfo {author} {\bibfnamefont {Jens}\ \bibnamefont
  {Koch}}, \ and\ \bibinfo {author} {\bibfnamefont {S~M}\ \bibnamefont
  {Girvin}}} (\bibinfo {year} {2011}),\ \bibfield  {title} {\enquote {\bibinfo
  {title} {{Synthetic gauge fields and homodyne transmission in
  Jaynes-–Cummings lattices}},}\ }\href
  {http://stacks.iop.org/1367-2630/13/i=9/a=095008} {\bibfield  {journal}
  {\bibinfo  {journal} {New J. Phys.}\ }\textbf {\bibinfo {volume}
  {13}}~(\bibinfo {number} {9}),\ \bibinfo {pages} {095008}}\BibitemShut
  {NoStop}%
\bibitem [{\citenamefont {Ochiai}(2015{\natexlab{a}})}]{Ochiai:2015STAM}%
  \BibitemOpen
  \bibfield  {author} {\bibinfo {author} {\bibnamefont {Ochiai}, \bibfnamefont
  {Tetsuyuki}}} (\bibinfo {year} {2015}{\natexlab{a}}),\ \bibfield  {title}
  {\enquote {\bibinfo {title} {Non-reciprocity and topology in optics:
  {O}ne-way road for light via surface magnon polariton},}\ }\href
  {https://www.tandfonline.com/doi/full/10.1088/1468-6996/16/1/014401}
  {\bibfield  {journal} {\bibinfo  {journal} {Sci. Technol. Adv. Mater.}\
  }\textbf {\bibinfo {volume} {16}}~(\bibinfo {number} {1}),\ \bibinfo {pages}
  {014401}}\BibitemShut {NoStop}%
\bibitem [{\citenamefont {Ochiai}(2015{\natexlab{b}})}]{Ochiai:2015JPSJ}%
  \BibitemOpen
  \bibfield  {author} {\bibinfo {author} {\bibnamefont {Ochiai}, \bibfnamefont
  {Tetsuyuki}}} (\bibinfo {year} {2015}{\natexlab{b}}),\ \bibfield  {title}
  {\enquote {\bibinfo {title} {Time-reversal-violating photonic topological
  insulators with helical edge states},}\ }\href
  {https://journals.jps.jp/doi/10.7566/JPSJ.84.054401} {\bibfield  {journal}
  {\bibinfo  {journal} {J. Phys. Soc. Jpn.}\ }\textbf {\bibinfo {volume}
  {84}}~(\bibinfo {number} {5}),\ \bibinfo {pages} {054401}}\BibitemShut
  {NoStop}%
\bibitem [{\citenamefont {Ochiai}(2016)}]{Ochiai:2016JOP}%
  \BibitemOpen
  \bibfield  {author} {\bibinfo {author} {\bibnamefont {Ochiai}, \bibfnamefont
  {Tetsuyuki}}} (\bibinfo {year} {2016}),\ \bibfield  {title} {\enquote
  {\bibinfo {title} {Floquet--{W}eyl and {F}loquet-topological-insulator phases
  in a stacked two-dimensional ring-network lattice},}\ }\href
  {https://iopscience.iop.org/article/10.1088/0953-8984/28/42/425501}
  {\bibfield  {journal} {\bibinfo  {journal} {J. Phys. Condens. Matter}\
  }\textbf {\bibinfo {volume} {28}}~(\bibinfo {number} {42}),\ \bibinfo {pages}
  {425501}}\BibitemShut {NoStop}%
\bibitem [{\citenamefont {Ochiai}(2017)}]{Ochiai:2017arXiv}%
  \BibitemOpen
  \bibfield  {author} {\bibinfo {author} {\bibnamefont {Ochiai}, \bibfnamefont
  {Tetsuyuki}}} (\bibinfo {year} {2017}),\ \bibfield  {title} {\enquote
  {\bibinfo {title} {Gapless surface states originating from accidentally
  degenerate quadratic band touching in a three-dimensional tetragonal photonic
  crystal},}\ }\href {https://link.aps.org/doi/10.1103/PhysRevA.96.043842}
  {\bibfield  {journal} {\bibinfo  {journal} {Phys. Rev. A}\ }\textbf {\bibinfo
  {volume} {96}},\ \bibinfo {pages} {043842}}\BibitemShut {NoStop}%
\bibitem [{\citenamefont {O'Connell}\ \emph {et~al.}(2010)\citenamefont
  {O'Connell}, \citenamefont {Hofheinz}, \citenamefont {Ansmann}, \citenamefont
  {Bialczak}, \citenamefont {Lenander}, \citenamefont {Lucero}, \citenamefont
  {Neeley}, \citenamefont {Sank}, \citenamefont {Wang}, \citenamefont {Weides}
  \emph {et~al.}}]{Cleland2010}%
  \BibitemOpen
  \bibfield  {author} {\bibinfo {author} {\bibnamefont {O'Connell},
  \bibfnamefont {Aaron~D}}, \bibinfo {author} {\bibfnamefont {Max}\
  \bibnamefont {Hofheinz}}, \bibinfo {author} {\bibfnamefont {Markus}\
  \bibnamefont {Ansmann}}, \bibinfo {author} {\bibfnamefont {Radoslaw~C}\
  \bibnamefont {Bialczak}}, \bibinfo {author} {\bibfnamefont {Mike}\
  \bibnamefont {Lenander}}, \bibinfo {author} {\bibfnamefont {Erik}\
  \bibnamefont {Lucero}}, \bibinfo {author} {\bibfnamefont {Matthew}\
  \bibnamefont {Neeley}}, \bibinfo {author} {\bibfnamefont {Daniel}\
  \bibnamefont {Sank}}, \bibinfo {author} {\bibfnamefont {H}~\bibnamefont
  {Wang}}, \bibinfo {author} {\bibfnamefont {M}~\bibnamefont {Weides}},  \emph
  {et~al.}} (\bibinfo {year} {2010}),\ \bibfield  {title} {\enquote {\bibinfo
  {title} {Quantum ground state and single-phonon control of a mechanical
  resonator},}\ }\href {https://www.nature.com/articles/nature08967} {\bibfield
   {journal} {\bibinfo  {journal} {Nature}\ }\textbf {\bibinfo {volume}
  {464}}~(\bibinfo {number} {7289}),\ \bibinfo {pages} {697--703}}\BibitemShut
  {NoStop}%
\bibitem [{\citenamefont {O'Connor}\ \emph {et~al.}(2014)\citenamefont
  {O'Connor}, \citenamefont {Ginzburg}, \citenamefont
  {Rodr{\'\i}guez-Fortu{\~n}o}, \citenamefont {Wurtz},\ and\ \citenamefont
  {Zayats}}]{oconnor:2014naturecommunications}%
  \BibitemOpen
  \bibfield  {author} {\bibinfo {author} {\bibnamefont {O'Connor},
  \bibfnamefont {D}}, \bibinfo {author} {\bibfnamefont {P}~\bibnamefont
  {Ginzburg}}, \bibinfo {author} {\bibfnamefont {FJ}~\bibnamefont
  {Rodr{\'\i}guez-Fortu{\~n}o}}, \bibinfo {author} {\bibfnamefont
  {GA}~\bibnamefont {Wurtz}}, \ and\ \bibinfo {author} {\bibfnamefont
  {AV}~\bibnamefont {Zayats}}} (\bibinfo {year} {2014}),\ \bibfield  {title}
  {\enquote {\bibinfo {title} {Spin--orbit coupling in surface plasmon
  scattering by nanostructures},}\ }\href
  {https://www.nature.com/articles/ncomms6327} {\bibfield  {journal} {\bibinfo
  {journal} {Nat. Commun.}\ }\textbf {\bibinfo {volume} {5}},\ \bibinfo {pages}
  {5327}}\BibitemShut {NoStop}%
\bibitem [{\citenamefont {Oka}\ and\ \citenamefont {Aoki}(2009)}]{Oka:2009PRB}%
  \BibitemOpen
  \bibfield  {author} {\bibinfo {author} {\bibnamefont {Oka}, \bibfnamefont
  {Takashi}}, \ and\ \bibinfo {author} {\bibfnamefont {Hideo}\ \bibnamefont
  {Aoki}}} (\bibinfo {year} {2009}),\ \bibfield  {title} {\enquote {\bibinfo
  {title} {Photovoltaic {H}all effect in graphene},}\ }\href
  {https://doi.org/10.1103/PhysRevB.79.081406} {\bibfield  {journal} {\bibinfo
  {journal} {Phys. Rev. B}\ }\textbf {\bibinfo {volume} {79}},\ \bibinfo
  {pages} {081406}}\BibitemShut {NoStop}%
\bibitem [{\citenamefont {Onbasli}\ \emph {et~al.}(2016)\citenamefont
  {Onbasli}, \citenamefont {Beran}, \citenamefont {Zahradn{\'\i}k},
  \citenamefont {Ku{\v{c}}era}, \citenamefont {Anto{\v{s}}}, \citenamefont
  {Mistr{\'\i}k}, \citenamefont {Dionne}, \citenamefont {Veis},\ and\
  \citenamefont {Ross}}]{Onbasli:2016SR}%
  \BibitemOpen
  \bibfield  {author} {\bibinfo {author} {\bibnamefont {Onbasli}, \bibfnamefont
  {Mehmet~C}}, \bibinfo {author} {\bibfnamefont {Luk{\'a}{\v{s}}}\ \bibnamefont
  {Beran}}, \bibinfo {author} {\bibfnamefont {Martin}\ \bibnamefont
  {Zahradn{\'\i}k}}, \bibinfo {author} {\bibfnamefont {Miroslav}\ \bibnamefont
  {Ku{\v{c}}era}}, \bibinfo {author} {\bibfnamefont {Roman}\ \bibnamefont
  {Anto{\v{s}}}}, \bibinfo {author} {\bibfnamefont {Jan}\ \bibnamefont
  {Mistr{\'\i}k}}, \bibinfo {author} {\bibfnamefont {Gerald~F}\ \bibnamefont
  {Dionne}}, \bibinfo {author} {\bibfnamefont {Martin}\ \bibnamefont {Veis}}, \
  and\ \bibinfo {author} {\bibfnamefont {Caroline~A}\ \bibnamefont {Ross}}}
  (\bibinfo {year} {2016}),\ \bibfield  {title} {\enquote {\bibinfo {title}
  {Optical and magneto-optical behavior of {C}erium {Y}ttrium {I}ron {G}arnet
  thin films at wavelengths of 200--1770 nm},}\ }\href
  {https://www.nature.com/articles/srep23640} {\bibfield  {journal} {\bibinfo
  {journal} {Sci. Rep.}\ }\textbf {\bibinfo {volume} {6}},\ \bibinfo {pages}
  {23640}}\BibitemShut {NoStop}%
\bibitem [{\citenamefont {Ota}\ \emph {et~al.}(2018)\citenamefont {Ota},
  \citenamefont {Katsumi}, \citenamefont {Watanabe}, \citenamefont {Iwamoto},\
  and\ \citenamefont {Arakawa}}]{Ota:2018NatComm}%
  \BibitemOpen
  \bibfield  {author} {\bibinfo {author} {\bibnamefont {Ota}, \bibfnamefont
  {Yasutomo}}, \bibinfo {author} {\bibfnamefont {Ryota}\ \bibnamefont
  {Katsumi}}, \bibinfo {author} {\bibfnamefont {Katsuyuki}\ \bibnamefont
  {Watanabe}}, \bibinfo {author} {\bibfnamefont {Satoshi}\ \bibnamefont
  {Iwamoto}}, \ and\ \bibinfo {author} {\bibfnamefont {Yasuhiko}\ \bibnamefont
  {Arakawa}}} (\bibinfo {year} {2018}),\ \bibfield  {title} {\enquote {\bibinfo
  {title} {Topological photonic crystal nanocavity laser},}\ }\href
  {https://www.nature.com/articles/s42005-018-0083-7} {\bibfield  {journal}
  {\bibinfo  {journal} {Commun. Phys.}\ }\textbf {\bibinfo {volume}
  {1}}~(\bibinfo {number} {1}),\ \bibinfo {pages} {86}}\BibitemShut {NoStop}%
\bibitem [{\citenamefont {Otterbach}\ \emph {et~al.}(2010)\citenamefont
  {Otterbach}, \citenamefont {Ruseckas}, \citenamefont {Unanyan}, \citenamefont
  {Juzeli{\=u}nas},\ and\ \citenamefont
  {Fleischhauer}}]{otterbach2010effective}%
  \BibitemOpen
  \bibfield  {author} {\bibinfo {author} {\bibnamefont {Otterbach},
  \bibfnamefont {J}}, \bibinfo {author} {\bibfnamefont {J}~\bibnamefont
  {Ruseckas}}, \bibinfo {author} {\bibfnamefont {RG}~\bibnamefont {Unanyan}},
  \bibinfo {author} {\bibfnamefont {G}~\bibnamefont {Juzeli{\=u}nas}}, \ and\
  \bibinfo {author} {\bibfnamefont {M}~\bibnamefont {Fleischhauer}}} (\bibinfo
  {year} {2010}),\ \bibfield  {title} {\enquote {\bibinfo {title} {Effective
  magnetic fields for stationary light},}\ }\href
  {https://journals.aps.org/prl/abstract/10.1103/PhysRevLett.104.033903}
  {\bibfield  {journal} {\bibinfo  {journal} {Phys. Rev. Lett.}\ }\textbf
  {\bibinfo {volume} {104}}~(\bibinfo {number} {3}),\ \bibinfo {pages}
  {033903}}\BibitemShut {NoStop}%
\bibitem [{\citenamefont {Owens}\ \emph {et~al.}(2018)\citenamefont {Owens},
  \citenamefont {LaChapelle}, \citenamefont {Saxberg}, \citenamefont
  {Anderson}, \citenamefont {Ma}, \citenamefont {Simon},\ and\ \citenamefont
  {Schuster}}]{Owens:2018PRA}%
  \BibitemOpen
  \bibfield  {author} {\bibinfo {author} {\bibnamefont {Owens}, \bibfnamefont
  {Clai}}, \bibinfo {author} {\bibfnamefont {Aman}\ \bibnamefont {LaChapelle}},
  \bibinfo {author} {\bibfnamefont {Brendan}\ \bibnamefont {Saxberg}}, \bibinfo
  {author} {\bibfnamefont {Brandon~M.}\ \bibnamefont {Anderson}}, \bibinfo
  {author} {\bibfnamefont {Ruichao}\ \bibnamefont {Ma}}, \bibinfo {author}
  {\bibfnamefont {Jonathan}\ \bibnamefont {Simon}}, \ and\ \bibinfo {author}
  {\bibfnamefont {David~I.}\ \bibnamefont {Schuster}}} (\bibinfo {year}
  {2018}),\ \bibfield  {title} {\enquote {\bibinfo {title} {Quarter-flux
  {H}ofstadter lattice in a qubit-compatible microwave cavity array},}\ }\href
  {https://link.aps.org/doi/10.1103/PhysRevA.97.013818} {\bibfield  {journal}
  {\bibinfo  {journal} {Phys. Rev. A}\ }\textbf {\bibinfo {volume} {97}},\
  \bibinfo {pages} {013818}}\BibitemShut {NoStop}%
\bibitem [{\citenamefont {Ozawa}(2018)}]{Ozawa:2018PRB}%
  \BibitemOpen
  \bibfield  {author} {\bibinfo {author} {\bibnamefont {Ozawa}, \bibfnamefont
  {Tomoki}}} (\bibinfo {year} {2018}),\ \bibfield  {title} {\enquote {\bibinfo
  {title} {Steady-state {H}all response and quantum geometry of
  driven-dissipative lattices},}\ }\href
  {https://link.aps.org/doi/10.1103/PhysRevB.97.041108} {\bibfield  {journal}
  {\bibinfo  {journal} {Phys. Rev. B}\ }\textbf {\bibinfo {volume} {97}},\
  \bibinfo {pages} {041108(R)}}\BibitemShut {NoStop}%
\bibitem [{\citenamefont {Ozawa}\ \emph {et~al.}(2017)\citenamefont {Ozawa},
  \citenamefont {Amo}, \citenamefont {Bloch},\ and\ \citenamefont
  {Carusotto}}]{Ozawa:2017PRA}%
  \BibitemOpen
  \bibfield  {author} {\bibinfo {author} {\bibnamefont {Ozawa}, \bibfnamefont
  {Tomoki}}, \bibinfo {author} {\bibfnamefont {Alberto}\ \bibnamefont {Amo}},
  \bibinfo {author} {\bibfnamefont {Jacqueline}\ \bibnamefont {Bloch}}, \ and\
  \bibinfo {author} {\bibfnamefont {Iacopo}\ \bibnamefont {Carusotto}}}
  (\bibinfo {year} {2017}),\ \bibfield  {title} {\enquote {\bibinfo {title}
  {Klein tunneling in driven-dissipative photonic graphene},}\ }\href
  {https://link.aps.org/doi/10.1103/PhysRevA.96.013813} {\bibfield  {journal}
  {\bibinfo  {journal} {Phys. Rev. A}\ }\textbf {\bibinfo {volume} {96}},\
  \bibinfo {pages} {013813}}\BibitemShut {NoStop}%
\bibitem [{\citenamefont {Ozawa}\ and\ \citenamefont
  {Carusotto}(2014)}]{Ozawa:2014PRL}%
  \BibitemOpen
  \bibfield  {author} {\bibinfo {author} {\bibnamefont {Ozawa}, \bibfnamefont
  {Tomoki}}, \ and\ \bibinfo {author} {\bibfnamefont {Iacopo}\ \bibnamefont
  {Carusotto}}} (\bibinfo {year} {2014}),\ \bibfield  {title} {\enquote
  {\bibinfo {title} {Anomalous and quantum {H}all effects in lossy photonic
  lattices},}\ }\href {http://link.aps.org/doi/10.1103/PhysRevLett.112.133902}
  {\bibfield  {journal} {\bibinfo  {journal} {Phys. Rev. Lett.}\ }\textbf
  {\bibinfo {volume} {112}},\ \bibinfo {pages} {133902}}\BibitemShut {NoStop}%
\bibitem [{\citenamefont {Ozawa}\ and\ \citenamefont
  {Carusotto}(2017)}]{Ozawa:2017PRL}%
  \BibitemOpen
  \bibfield  {author} {\bibinfo {author} {\bibnamefont {Ozawa}, \bibfnamefont
  {Tomoki}}, \ and\ \bibinfo {author} {\bibfnamefont {Iacopo}\ \bibnamefont
  {Carusotto}}} (\bibinfo {year} {2017}),\ \bibfield  {title} {\enquote
  {\bibinfo {title} {Synthetic dimensions with magnetic fields and local
  interactions in photonic lattices},}\ }\href
  {https://link.aps.org/doi/10.1103/PhysRevLett.118.013601} {\bibfield
  {journal} {\bibinfo  {journal} {Phys. Rev. Lett.}\ }\textbf {\bibinfo
  {volume} {118}},\ \bibinfo {pages} {013601}}\BibitemShut {NoStop}%
\bibitem [{\citenamefont {Ozawa}\ \emph {et~al.}(2015)\citenamefont {Ozawa},
  \citenamefont {Price},\ and\ \citenamefont {Carusotto}}]{Ozawa:2015PRA}%
  \BibitemOpen
  \bibfield  {author} {\bibinfo {author} {\bibnamefont {Ozawa}, \bibfnamefont
  {Tomoki}}, \bibinfo {author} {\bibfnamefont {Hannah~M}\ \bibnamefont
  {Price}}, \ and\ \bibinfo {author} {\bibfnamefont {Iacopo}\ \bibnamefont
  {Carusotto}}} (\bibinfo {year} {2015}),\ \bibfield  {title} {\enquote
  {\bibinfo {title} {Momentum-space {H}arper-{H}ofstadter model},}\ }\href
  {https://journals.aps.org/pra/abstract/10.1103/PhysRevA.92.023609} {\bibfield
   {journal} {\bibinfo  {journal} {Phys. Rev. A}\ }\textbf {\bibinfo {volume}
  {92}}~(\bibinfo {number} {2}),\ \bibinfo {pages} {023609}}\BibitemShut
  {NoStop}%
\bibitem [{\citenamefont {Ozawa}\ \emph
  {et~al.}(2016{\natexlab{a}})\citenamefont {Ozawa}, \citenamefont {Price},\
  and\ \citenamefont {Carusotto}}]{Ozawa:2016PRB}%
  \BibitemOpen
  \bibfield  {author} {\bibinfo {author} {\bibnamefont {Ozawa}, \bibfnamefont
  {Tomoki}}, \bibinfo {author} {\bibfnamefont {Hannah~M}\ \bibnamefont
  {Price}}, \ and\ \bibinfo {author} {\bibfnamefont {Iacopo}\ \bibnamefont
  {Carusotto}}} (\bibinfo {year} {2016}{\natexlab{a}}),\ \bibfield  {title}
  {\enquote {\bibinfo {title} {Quantum {H}all effect in momentum space},}\
  }\href {https://journals.aps.org/prb/abstract/10.1103/PhysRevB.93.195201}
  {\bibfield  {journal} {\bibinfo  {journal} {Phys. Rev. B}\ }\textbf {\bibinfo
  {volume} {93}}~(\bibinfo {number} {19}),\ \bibinfo {pages}
  {195201}}\BibitemShut {NoStop}%
\bibitem [{\citenamefont {Ozawa}\ \emph
  {et~al.}(2016{\natexlab{b}})\citenamefont {Ozawa}, \citenamefont {Price},
  \citenamefont {Goldman}, \citenamefont {Zilberberg},\ and\ \citenamefont
  {Carusotto}}]{Ozawa:2016PRA}%
  \BibitemOpen
  \bibfield  {author} {\bibinfo {author} {\bibnamefont {Ozawa}, \bibfnamefont
  {Tomoki}}, \bibinfo {author} {\bibfnamefont {Hannah~M.}\ \bibnamefont
  {Price}}, \bibinfo {author} {\bibfnamefont {Nathan}\ \bibnamefont {Goldman}},
  \bibinfo {author} {\bibfnamefont {Oded}\ \bibnamefont {Zilberberg}}, \ and\
  \bibinfo {author} {\bibfnamefont {Iacopo}\ \bibnamefont {Carusotto}}}
  (\bibinfo {year} {2016}{\natexlab{b}}),\ \bibfield  {title} {\enquote
  {\bibinfo {title} {Synthetic dimensions in integrated photonics: {F}rom
  optical isolation to four-dimensional quantum {H}all physics},}\ }\href
  {http://link.aps.org/doi/10.1103/PhysRevA.93.043827} {\bibfield  {journal}
  {\bibinfo  {journal} {Phys. Rev. A}\ }\textbf {\bibinfo {volume} {93}},\
  \bibinfo {pages} {043827}}\BibitemShut {NoStop}%
\bibitem [{\citenamefont {Painter}\ \emph {et~al.}(1999)\citenamefont
  {Painter}, \citenamefont {Lee}, \citenamefont {Scherer}, \citenamefont
  {Yariv}, \citenamefont {O'brien}, \citenamefont {Dapkus},\ and\ \citenamefont
  {Kim}}]{Painter:Science1999}%
  \BibitemOpen
  \bibfield  {author} {\bibinfo {author} {\bibnamefont {Painter}, \bibfnamefont
  {Oskar}}, \bibinfo {author} {\bibfnamefont {RK}~\bibnamefont {Lee}}, \bibinfo
  {author} {\bibfnamefont {Axel}\ \bibnamefont {Scherer}}, \bibinfo {author}
  {\bibfnamefont {A}~\bibnamefont {Yariv}}, \bibinfo {author} {\bibfnamefont
  {JD}~\bibnamefont {O'brien}}, \bibinfo {author} {\bibfnamefont
  {PD}~\bibnamefont {Dapkus}}, \ and\ \bibinfo {author} {\bibfnamefont
  {I}~\bibnamefont {Kim}}} (\bibinfo {year} {1999}),\ \bibfield  {title}
  {\enquote {\bibinfo {title} {Two-dimensional photonic band-gap defect mode
  laser},}\ }\href {http://science.sciencemag.org/content/284/5421/1819}
  {\bibfield  {journal} {\bibinfo  {journal} {Science}\ }\textbf {\bibinfo
  {volume} {284}}~(\bibinfo {number} {5421}),\ \bibinfo {pages}
  {1819--1821}}\BibitemShut {NoStop}%
\bibitem [{\citenamefont {Pan}\ \emph {et~al.}(2018)\citenamefont {Pan},
  \citenamefont {Zhao}, \citenamefont {Miao}, \citenamefont {Longhi},\ and\
  \citenamefont {Feng}}]{pan2018photonic}%
  \BibitemOpen
  \bibfield  {author} {\bibinfo {author} {\bibnamefont {Pan}, \bibfnamefont
  {Mingsen}}, \bibinfo {author} {\bibfnamefont {Han}\ \bibnamefont {Zhao}},
  \bibinfo {author} {\bibfnamefont {Pei}\ \bibnamefont {Miao}}, \bibinfo
  {author} {\bibfnamefont {Stefano}\ \bibnamefont {Longhi}}, \ and\ \bibinfo
  {author} {\bibfnamefont {Liang}\ \bibnamefont {Feng}}} (\bibinfo {year}
  {2018}),\ \bibfield  {title} {\enquote {\bibinfo {title} {Photonic zero mode
  in a non-{H}ermitian photonic lattice},}\ }\href
  {https://www.nature.com/articles/s41467-018-03822-8} {\bibfield  {journal}
  {\bibinfo  {journal} {Nat. Commun.}\ }\textbf {\bibinfo {volume}
  {9}}~(\bibinfo {number} {1}),\ \bibinfo {pages} {1308}}\BibitemShut {NoStop}%
\bibitem [{\citenamefont {Pancharatnam}(1956)}]{Pancharatnam:1956PIAS}%
  \BibitemOpen
  \bibfield  {author} {\bibinfo {author} {\bibnamefont {Pancharatnam},
  \bibfnamefont {S}}} (\bibinfo {year} {1956}),\ \bibfield  {title} {\enquote
  {\bibinfo {title} {Generalized theory of interference, and its
  applications},}\ }\href {\doibase 10.1007/BF03046050} {\bibfield  {journal}
  {\bibinfo  {journal} {Proc. Indian Acad. Sci. A}\ }\textbf {\bibinfo {volume}
  {44}}~(\bibinfo {number} {5}),\ \bibinfo {pages} {247--262}}\BibitemShut
  {NoStop}%
\bibitem [{\citenamefont {Parto}\ \emph {et~al.}(2018)\citenamefont {Parto},
  \citenamefont {Wittek}, \citenamefont {Hodaei}, \citenamefont {Harari},
  \citenamefont {Bandres}, \citenamefont {Ren}, \citenamefont {Rechtsman},
  \citenamefont {Segev}, \citenamefont {Christodoulides},\ and\ \citenamefont
  {Khajavikhan}}]{Parto:PRL2018}%
  \BibitemOpen
  \bibfield  {author} {\bibinfo {author} {\bibnamefont {Parto}, \bibfnamefont
  {Midya}}, \bibinfo {author} {\bibfnamefont {Steffen}\ \bibnamefont {Wittek}},
  \bibinfo {author} {\bibfnamefont {Hossein}\ \bibnamefont {Hodaei}}, \bibinfo
  {author} {\bibfnamefont {Gal}\ \bibnamefont {Harari}}, \bibinfo {author}
  {\bibfnamefont {Miguel~A}\ \bibnamefont {Bandres}}, \bibinfo {author}
  {\bibfnamefont {Jinhan}\ \bibnamefont {Ren}}, \bibinfo {author}
  {\bibfnamefont {Mikael~C}\ \bibnamefont {Rechtsman}}, \bibinfo {author}
  {\bibfnamefont {Mordechai}\ \bibnamefont {Segev}}, \bibinfo {author}
  {\bibfnamefont {Demetrios~N}\ \bibnamefont {Christodoulides}}, \ and\
  \bibinfo {author} {\bibfnamefont {Mercedeh}\ \bibnamefont {Khajavikhan}}}
  (\bibinfo {year} {2018}),\ \bibfield  {title} {\enquote {\bibinfo {title}
  {Edge-mode lasing in 1{D} topological active arrays},}\ }\href
  {https://link.aps.org/doi/10.1103/PhysRevLett.120.113901} {\bibfield
  {journal} {\bibinfo  {journal} {Phys. Rev. Lett.}\ }\textbf {\bibinfo
  {volume} {120}}~(\bibinfo {number} {11}),\ \bibinfo {pages}
  {113901}}\BibitemShut {NoStop}%
\bibitem [{\citenamefont {Peano}\ \emph
  {et~al.}(2016{\natexlab{a}})\citenamefont {Peano}, \citenamefont {Houde},
  \citenamefont {Brendel}, \citenamefont {Marquardt},\ and\ \citenamefont
  {Clerk}}]{Peano:2016NatComm}%
  \BibitemOpen
  \bibfield  {author} {\bibinfo {author} {\bibnamefont {Peano}, \bibfnamefont
  {Vittorio}}, \bibinfo {author} {\bibfnamefont {Martin}\ \bibnamefont
  {Houde}}, \bibinfo {author} {\bibfnamefont {Christian}\ \bibnamefont
  {Brendel}}, \bibinfo {author} {\bibfnamefont {Florian}\ \bibnamefont
  {Marquardt}}, \ and\ \bibinfo {author} {\bibfnamefont {Aashish~A}\
  \bibnamefont {Clerk}}} (\bibinfo {year} {2016}{\natexlab{a}}),\ \bibfield
  {title} {\enquote {\bibinfo {title} {Topological phase transitions and chiral
  inelastic transport induced by the squeezing of light},}\ }\href
  {http://www.nature.com/articles/ncomms10779} {\bibfield  {journal} {\bibinfo
  {journal} {Nat. Commun.}\ }\textbf {\bibinfo {volume} {7}},\ \bibinfo {pages}
  {10779}}\BibitemShut {NoStop}%
\bibitem [{\citenamefont {Peano}\ \emph
  {et~al.}(2016{\natexlab{b}})\citenamefont {Peano}, \citenamefont {Houde},
  \citenamefont {Marquardt},\ and\ \citenamefont {Clerk}}]{Peano:2016PRX}%
  \BibitemOpen
  \bibfield  {author} {\bibinfo {author} {\bibnamefont {Peano}, \bibfnamefont
  {Vittorio}}, \bibinfo {author} {\bibfnamefont {Martin}\ \bibnamefont
  {Houde}}, \bibinfo {author} {\bibfnamefont {Florian}\ \bibnamefont
  {Marquardt}}, \ and\ \bibinfo {author} {\bibfnamefont {Aashish~A.}\
  \bibnamefont {Clerk}}} (\bibinfo {year} {2016}{\natexlab{b}}),\ \bibfield
  {title} {\enquote {\bibinfo {title} {Topological quantum fluctuations and
  traveling wave amplifiers},}\ }\href
  {http://link.aps.org/doi/10.1103/PhysRevX.6.041026} {\bibfield  {journal}
  {\bibinfo  {journal} {Phys. Rev. X}\ }\textbf {\bibinfo {volume} {6}},\
  \bibinfo {pages} {041026}}\BibitemShut {NoStop}%
\bibitem [{\citenamefont {Peierls}(1933)}]{Peierls:1933ZPhys}%
  \BibitemOpen
  \bibfield  {author} {\bibinfo {author} {\bibnamefont {Peierls}, \bibfnamefont
  {R}}} (\bibinfo {year} {1933}),\ \bibfield  {title} {\enquote {\bibinfo
  {title} {Zur theorie des diamagnetismus von leitungselektronen},}\ }\href
  {https://link.springer.com/article/10.1007/BF01342591} {\bibfield  {journal}
  {\bibinfo  {journal} {Z. Phys.}\ }\textbf {\bibinfo {volume} {80}},\ \bibinfo
  {pages} {763--791}}\BibitemShut {NoStop}%
\bibitem [{\citenamefont {Peleg}\ \emph {et~al.}(2007)\citenamefont {Peleg},
  \citenamefont {Bartal}, \citenamefont {Freedman}, \citenamefont {Manela},
  \citenamefont {Segev},\ and\ \citenamefont
  {Christodoulides}}]{Peleg:2007PRL}%
  \BibitemOpen
  \bibfield  {author} {\bibinfo {author} {\bibnamefont {Peleg}, \bibfnamefont
  {Or}}, \bibinfo {author} {\bibfnamefont {Guy}\ \bibnamefont {Bartal}},
  \bibinfo {author} {\bibfnamefont {Barak}\ \bibnamefont {Freedman}}, \bibinfo
  {author} {\bibfnamefont {Ofer}\ \bibnamefont {Manela}}, \bibinfo {author}
  {\bibfnamefont {Mordechai}\ \bibnamefont {Segev}}, \ and\ \bibinfo {author}
  {\bibfnamefont {Demetrios~N}\ \bibnamefont {Christodoulides}}} (\bibinfo
  {year} {2007}),\ \bibfield  {title} {\enquote {\bibinfo {title} {{Conical
  diffraction and gap solitons in honeycomb photonic lattices}},}\ }\href
  {https://doi.org/10.1103/PhysRevLett.98.103901} {\bibfield  {journal}
  {\bibinfo  {journal} {Phys. Rev. Lett.}\ }\textbf {\bibinfo {volume}
  {98}}~(\bibinfo {number} {10}),\ \bibinfo {pages} {103901}}\BibitemShut
  {NoStop}%
\bibitem [{\citenamefont {Pelegr\'{\i}}\ \emph {et~al.}(2017)\citenamefont
  {Pelegr\'{\i}}, \citenamefont {Polo}, \citenamefont {Turpin}, \citenamefont
  {Lewenstein}, \citenamefont {Mompart},\ and\ \citenamefont
  {Ahufinger}}]{Pelegri:2017PRA}%
  \BibitemOpen
  \bibfield  {author} {\bibinfo {author} {\bibnamefont {Pelegr\'{\i}},
  \bibfnamefont {G}}, \bibinfo {author} {\bibfnamefont {J.}~\bibnamefont
  {Polo}}, \bibinfo {author} {\bibfnamefont {A.}~\bibnamefont {Turpin}},
  \bibinfo {author} {\bibfnamefont {M.}~\bibnamefont {Lewenstein}}, \bibinfo
  {author} {\bibfnamefont {J.}~\bibnamefont {Mompart}}, \ and\ \bibinfo
  {author} {\bibfnamefont {V.}~\bibnamefont {Ahufinger}}} (\bibinfo {year}
  {2017}),\ \bibfield  {title} {\enquote {\bibinfo {title} {Single-atom
  edgelike states via quantum interference},}\ }\href
  {https://link.aps.org/doi/10.1103/PhysRevA.95.013614} {\bibfield  {journal}
  {\bibinfo  {journal} {Phys. Rev. A}\ }\textbf {\bibinfo {volume} {95}},\
  \bibinfo {pages} {013614}}\BibitemShut {NoStop}%
\bibitem [{\citenamefont {Pendry}\ \emph {et~al.}(1999)\citenamefont {Pendry},
  \citenamefont {Holden}, \citenamefont {Robbins},\ and\ \citenamefont
  {Stewart}}]{pendry:1999IEEE}%
  \BibitemOpen
  \bibfield  {author} {\bibinfo {author} {\bibnamefont {Pendry}, \bibfnamefont
  {John~B}}, \bibinfo {author} {\bibfnamefont {A~J\_}\ \bibnamefont {Holden}},
  \bibinfo {author} {\bibfnamefont {DJ}~\bibnamefont {Robbins}}, \ and\
  \bibinfo {author} {\bibfnamefont {WJ}~\bibnamefont {Stewart}}} (\bibinfo
  {year} {1999}),\ \bibfield  {title} {\enquote {\bibinfo {title} {Magnetism
  from conductors and enhanced nonlinear phenomena},}\ }\href
  {https://ieeexplore.ieee.org/document/798002} {\bibfield  {journal} {\bibinfo
   {journal} {IEEE Trans. Microwave Theory Tech.}\ }\textbf {\bibinfo {volume}
  {47}}~(\bibinfo {number} {11}),\ \bibinfo {pages} {2075--2084}}\BibitemShut
  {NoStop}%
\bibitem [{\citenamefont {Peropadre}\ \emph {et~al.}(2013)\citenamefont
  {Peropadre}, \citenamefont {Zueco}, \citenamefont {Wulschner}, \citenamefont
  {Deppe}, \citenamefont {Marx}, \citenamefont {Gross},\ and\ \citenamefont
  {Garc\'{\i}a-Ripoll}}]{Peropadre:2013PRB}%
  \BibitemOpen
  \bibfield  {author} {\bibinfo {author} {\bibnamefont {Peropadre},
  \bibfnamefont {Borja}}, \bibinfo {author} {\bibfnamefont {David}\
  \bibnamefont {Zueco}}, \bibinfo {author} {\bibfnamefont {Friedrich}\
  \bibnamefont {Wulschner}}, \bibinfo {author} {\bibfnamefont {Frank}\
  \bibnamefont {Deppe}}, \bibinfo {author} {\bibfnamefont {Achim}\ \bibnamefont
  {Marx}}, \bibinfo {author} {\bibfnamefont {Rudolf}\ \bibnamefont {Gross}}, \
  and\ \bibinfo {author} {\bibfnamefont {Juan~Jos\'e}\ \bibnamefont
  {Garc\'{\i}a-Ripoll}}} (\bibinfo {year} {2013}),\ \bibfield  {title}
  {\enquote {\bibinfo {title} {Tunable coupling engineering between
  superconducting resonators: {F}rom sidebands to effective gauge fields},}\
  }\href {https://link.aps.org/doi/10.1103/PhysRevB.87.134504} {\bibfield
  {journal} {\bibinfo  {journal} {Phys. Rev. B}\ }\textbf {\bibinfo {volume}
  {87}},\ \bibinfo {pages} {134504}}\BibitemShut {NoStop}%
\bibitem [{\citenamefont {Petersen}\ \emph {et~al.}(2014)\citenamefont
  {Petersen}, \citenamefont {Volz},\ and\ \citenamefont
  {Rauschenbeutel}}]{Petersen:2014science}%
  \BibitemOpen
  \bibfield  {author} {\bibinfo {author} {\bibnamefont {Petersen},
  \bibfnamefont {Jan}}, \bibinfo {author} {\bibfnamefont {J{\"u}rgen}\
  \bibnamefont {Volz}}, \ and\ \bibinfo {author} {\bibfnamefont {Arno}\
  \bibnamefont {Rauschenbeutel}}} (\bibinfo {year} {2014}),\ \bibfield  {title}
  {\enquote {\bibinfo {title} {Chiral nanophotonic waveguide interface based on
  spin-orbit interaction of light},}\ }\href
  {http://science.sciencemag.org/content/346/6205/67} {\bibfield  {journal}
  {\bibinfo  {journal} {Science}\ }\textbf {\bibinfo {volume} {346}}~(\bibinfo
  {number} {6205}),\ \bibinfo {pages} {67--71}}\BibitemShut {NoStop}%
\bibitem [{\citenamefont {Petrescu}\ \emph {et~al.}(2012)\citenamefont
  {Petrescu}, \citenamefont {Houck},\ and\ \citenamefont
  {Le~Hur}}]{Petrescu:2012PRA}%
  \BibitemOpen
  \bibfield  {author} {\bibinfo {author} {\bibnamefont {Petrescu},
  \bibfnamefont {Alexandru}}, \bibinfo {author} {\bibfnamefont {Andrew~A.}\
  \bibnamefont {Houck}}, \ and\ \bibinfo {author} {\bibfnamefont {Karyn}\
  \bibnamefont {Le~Hur}}} (\bibinfo {year} {2012}),\ \bibfield  {title}
  {\enquote {\bibinfo {title} {Anomalous {H}all effects of light and chiral
  edge modes on the {K}agom\'e lattice},}\ }\href
  {https://link.aps.org/doi/10.1103/PhysRevA.86.053804} {\bibfield  {journal}
  {\bibinfo  {journal} {Phys. Rev. A}\ }\textbf {\bibinfo {volume} {86}},\
  \bibinfo {pages} {053804}}\BibitemShut {NoStop}%
\bibitem [{\citenamefont {Petrosyan}\ \emph {et~al.}(2011)\citenamefont
  {Petrosyan}, \citenamefont {Otterbach},\ and\ \citenamefont
  {Fleischhauer}}]{Petrosyan:PRL2011}%
  \BibitemOpen
  \bibfield  {author} {\bibinfo {author} {\bibnamefont {Petrosyan},
  \bibfnamefont {David}}, \bibinfo {author} {\bibfnamefont {Johannes}\
  \bibnamefont {Otterbach}}, \ and\ \bibinfo {author} {\bibfnamefont {Michael}\
  \bibnamefont {Fleischhauer}}} (\bibinfo {year} {2011}),\ \bibfield  {title}
  {\enquote {\bibinfo {title} {Electromagnetically induced transparency with
  {R}ydberg atoms},}\ }\href
  {https://link.aps.org/doi/10.1103/PhysRevLett.107.213601} {\bibfield
  {journal} {\bibinfo  {journal} {Phys. Rev. Lett.}\ }\textbf {\bibinfo
  {volume} {107}},\ \bibinfo {pages} {213601}}\BibitemShut {NoStop}%
\bibitem [{\citenamefont {Peyronel}\ \emph {et~al.}(2012)\citenamefont
  {Peyronel}, \citenamefont {Firstenberg}, \citenamefont {Liang}, \citenamefont
  {Hofferberth}, \citenamefont {Gorshkov}, \citenamefont {Pohl}, \citenamefont
  {Lukin},\ and\ \citenamefont {Vuleti{\'c}}}]{Peyronel:Nature2012}%
  \BibitemOpen
  \bibfield  {author} {\bibinfo {author} {\bibnamefont {Peyronel},
  \bibfnamefont {Thibault}}, \bibinfo {author} {\bibfnamefont {Ofer}\
  \bibnamefont {Firstenberg}}, \bibinfo {author} {\bibfnamefont {Qi-Yu}\
  \bibnamefont {Liang}}, \bibinfo {author} {\bibfnamefont {Sebastian}\
  \bibnamefont {Hofferberth}}, \bibinfo {author} {\bibfnamefont {Alexey~V}\
  \bibnamefont {Gorshkov}}, \bibinfo {author} {\bibfnamefont {Thomas}\
  \bibnamefont {Pohl}}, \bibinfo {author} {\bibfnamefont {Mikhail~D}\
  \bibnamefont {Lukin}}, \ and\ \bibinfo {author} {\bibfnamefont {Vladan}\
  \bibnamefont {Vuleti{\'c}}}} (\bibinfo {year} {2012}),\ \bibfield  {title}
  {\enquote {\bibinfo {title} {Quantum nonlinear optics with single photons
  enabled by strongly interacting atoms},}\ }\href
  {https://www.nature.com/articles/nature11361} {\bibfield  {journal} {\bibinfo
   {journal} {Nature}\ }\textbf {\bibinfo {volume} {488}}~(\bibinfo {number}
  {7409}),\ \bibinfo {pages} {57--60}}\BibitemShut {NoStop}%
\bibitem [{\citenamefont {Phillips}(1996)}]{phillips1996self}%
  \BibitemOpen
  \bibfield  {author} {\bibinfo {author} {\bibnamefont {Phillips},
  \bibfnamefont {John}}} (\bibinfo {year} {1996}),\ \bibfield  {title}
  {\enquote {\bibinfo {title} {Self-adjoint {F}redholm operators and spectral
  flow},}\ }\href {http://dx.doi.org/10.4153/CMB-1996-054-4} {\bibfield
  {journal} {\bibinfo  {journal} {Can. Math. Bull.}\ }\textbf {\bibinfo
  {volume} {39}}~(\bibinfo {number} {4}),\ \bibinfo {pages}
  {460--467}}\BibitemShut {NoStop}%
\bibitem [{\citenamefont {Pilozzi}\ and\ \citenamefont
  {Conti}(2016)}]{Pilozzi:2016PRB}%
  \BibitemOpen
  \bibfield  {author} {\bibinfo {author} {\bibnamefont {Pilozzi}, \bibfnamefont
  {Laura}}, \ and\ \bibinfo {author} {\bibfnamefont {Claudio}\ \bibnamefont
  {Conti}}} (\bibinfo {year} {2016}),\ \bibfield  {title} {\enquote {\bibinfo
  {title} {Topological lasing in resonant photonic structures},}\ }\href
  {https://link.aps.org/doi/10.1103/PhysRevB.93.195317} {\bibfield  {journal}
  {\bibinfo  {journal} {Phys. Rev. B}\ }\textbf {\bibinfo {volume} {93}},\
  \bibinfo {pages} {195317}}\BibitemShut {NoStop}%
\bibitem [{\citenamefont {Pitaevskii}\ and\ \citenamefont
  {Stringari}(2016)}]{BECbook}%
  \BibitemOpen
  \bibfield  {author} {\bibinfo {author} {\bibnamefont {Pitaevskii},
  \bibfnamefont {L~P}}, \ and\ \bibinfo {author} {\bibfnamefont
  {S.}~\bibnamefont {Stringari}}} (\bibinfo {year} {2016}),\ \href@noop {}
  {\emph {\bibinfo {title} {{Bose Einstein condensation and superfluidity}}}}\
  (\bibinfo  {publisher} {Clarendon Press, Oxford})\BibitemShut {NoStop}%
\bibitem [{\citenamefont {Plekhanov}\ \emph {et~al.}(2017)\citenamefont
  {Plekhanov}, \citenamefont {Roux},\ and\ \citenamefont
  {Le~Hur}}]{plekhanov2017floquet}%
  \BibitemOpen
  \bibfield  {author} {\bibinfo {author} {\bibnamefont {Plekhanov},
  \bibfnamefont {Kirill}}, \bibinfo {author} {\bibfnamefont {Guillaume}\
  \bibnamefont {Roux}}, \ and\ \bibinfo {author} {\bibfnamefont {Karyn}\
  \bibnamefont {Le~Hur}}} (\bibinfo {year} {2017}),\ \bibfield  {title}
  {\enquote {\bibinfo {title} {{Floquet engineering of Haldane Chern insulators
  and chiral bosonic phase transitions}},}\ }\href
  {https://journals.aps.org/prb/abstract/10.1103/PhysRevB.95.045102} {\bibfield
   {journal} {\bibinfo  {journal} {Phys. Rev. B}\ }\textbf {\bibinfo {volume}
  {95}}~(\bibinfo {number} {4}),\ \bibinfo {pages} {045102}}\BibitemShut
  {NoStop}%
\bibitem [{\citenamefont {Plotnik}\ \emph {et~al.}(2016)\citenamefont
  {Plotnik}, \citenamefont {Bandres}, \citenamefont {St{\"u}tzer},
  \citenamefont {Lumer}, \citenamefont {Rechtsman}, \citenamefont {Szameit},\
  and\ \citenamefont {Segev}}]{plotnik2016analogue}%
  \BibitemOpen
  \bibfield  {author} {\bibinfo {author} {\bibnamefont {Plotnik}, \bibfnamefont
  {Y}}, \bibinfo {author} {\bibfnamefont {MA}~\bibnamefont {Bandres}}, \bibinfo
  {author} {\bibfnamefont {S}~\bibnamefont {St{\"u}tzer}}, \bibinfo {author}
  {\bibfnamefont {Y}~\bibnamefont {Lumer}}, \bibinfo {author} {\bibfnamefont
  {MC}~\bibnamefont {Rechtsman}}, \bibinfo {author} {\bibfnamefont
  {A}~\bibnamefont {Szameit}}, \ and\ \bibinfo {author} {\bibfnamefont
  {M}~\bibnamefont {Segev}}} (\bibinfo {year} {2016}),\ \bibfield  {title}
  {\enquote {\bibinfo {title} {Analogue of {R}ashba pseudo-spin-orbit coupling
  in photonic lattices by gauge field engineering},}\ }\href
  {https://journals.aps.org/prb/abstract/10.1103/PhysRevB.94.020301} {\bibfield
   {journal} {\bibinfo  {journal} {Phys. Rev. B}\ }\textbf {\bibinfo {volume}
  {94}}~(\bibinfo {number} {2}),\ \bibinfo {pages} {020301}}\BibitemShut
  {NoStop}%
\bibitem [{\citenamefont {Plotnik}\ \emph {et~al.}(2014)\citenamefont
  {Plotnik}, \citenamefont {Rechtsman}, \citenamefont {Song}, \citenamefont
  {Heinrich}, \citenamefont {Zeuner}, \citenamefont {Nolte}, \citenamefont
  {Lumer}, \citenamefont {Malkova}, \citenamefont {Xu}, \citenamefont {Szameit}
  \emph {et~al.}}]{Plotnik:2014NatMat}%
  \BibitemOpen
  \bibfield  {author} {\bibinfo {author} {\bibnamefont {Plotnik}, \bibfnamefont
  {Yonatan}}, \bibinfo {author} {\bibfnamefont {Mikael~C}\ \bibnamefont
  {Rechtsman}}, \bibinfo {author} {\bibfnamefont {Daohong}\ \bibnamefont
  {Song}}, \bibinfo {author} {\bibfnamefont {Matthias}\ \bibnamefont
  {Heinrich}}, \bibinfo {author} {\bibfnamefont {Julia~M}\ \bibnamefont
  {Zeuner}}, \bibinfo {author} {\bibfnamefont {Stefan}\ \bibnamefont {Nolte}},
  \bibinfo {author} {\bibfnamefont {Yaakov}\ \bibnamefont {Lumer}}, \bibinfo
  {author} {\bibfnamefont {Natalia}\ \bibnamefont {Malkova}}, \bibinfo {author}
  {\bibfnamefont {Jingjun}\ \bibnamefont {Xu}}, \bibinfo {author}
  {\bibfnamefont {Alexander}\ \bibnamefont {Szameit}},  \emph {et~al.}}
  (\bibinfo {year} {2014}),\ \bibfield  {title} {\enquote {\bibinfo {title}
  {Observation of unconventional edge states in ‘photonic graphene’},}\
  }\href
  {http://www.nature.com/nphoton/journal/v7/n2/full/nphoton.2012.302.html}
  {\bibfield  {journal} {\bibinfo  {journal} {Nat. Mater.}\ }\textbf {\bibinfo
  {volume} {13}}~(\bibinfo {number} {1}),\ \bibinfo {pages}
  {57--62}}\BibitemShut {NoStop}%
\bibitem [{\citenamefont {Poddubny}\ \emph {et~al.}(2014)\citenamefont
  {Poddubny}, \citenamefont {Miroshnichenko}, \citenamefont {Slobozhanyuk},\
  and\ \citenamefont {Kivshar}}]{Poddubny:2014ACS}%
  \BibitemOpen
  \bibfield  {author} {\bibinfo {author} {\bibnamefont {Poddubny},
  \bibfnamefont {Alexander}}, \bibinfo {author} {\bibfnamefont {Andrey}\
  \bibnamefont {Miroshnichenko}}, \bibinfo {author} {\bibfnamefont {Alexey}\
  \bibnamefont {Slobozhanyuk}}, \ and\ \bibinfo {author} {\bibfnamefont {Yuri}\
  \bibnamefont {Kivshar}}} (\bibinfo {year} {2014}),\ \bibfield  {title}
  {\enquote {\bibinfo {title} {Topological {M}ajorana states in zigzag chains
  of plasmonic nanoparticles},}\ }\href
  {http://pubs.acs.org/doi/abs/10.1021/ph4000949} {\bibfield  {journal}
  {\bibinfo  {journal} {ACS Photonics}\ }\textbf {\bibinfo {volume}
  {1}}~(\bibinfo {number} {2}),\ \bibinfo {pages} {101--105}}\BibitemShut
  {NoStop}%
\bibitem [{\citenamefont {Poli}\ \emph {et~al.}(2015)\citenamefont {Poli},
  \citenamefont {Bellec}, \citenamefont {Kuhl}, \citenamefont {Mortessagne},\
  and\ \citenamefont {Schomerus}}]{Poli:2015NatComm}%
  \BibitemOpen
  \bibfield  {author} {\bibinfo {author} {\bibnamefont {Poli}, \bibfnamefont
  {Charles}}, \bibinfo {author} {\bibfnamefont {Matthieu}\ \bibnamefont
  {Bellec}}, \bibinfo {author} {\bibfnamefont {Ulrich}\ \bibnamefont {Kuhl}},
  \bibinfo {author} {\bibfnamefont {Fabrice}\ \bibnamefont {Mortessagne}}, \
  and\ \bibinfo {author} {\bibfnamefont {Henning}\ \bibnamefont {Schomerus}}}
  (\bibinfo {year} {2015}),\ \bibfield  {title} {\enquote {\bibinfo {title}
  {Selective enhancement of topologically induced interface states in a
  dielectric resonator chain},}\ }\href
  {http://www.nature.com/articles/ncomms7710} {\bibfield  {journal} {\bibinfo
  {journal} {Nat. Commun.}\ }\textbf {\bibinfo {volume} {6}},\ \bibinfo {pages}
  {6710}}\BibitemShut {NoStop}%
\bibitem [{\citenamefont {Polkovnikov}\ \emph {et~al.}(2011)\citenamefont
  {Polkovnikov}, \citenamefont {Sengupta}, \citenamefont {Silva},\ and\
  \citenamefont {Vengalattore}}]{Polkovnikov:RMP2011}%
  \BibitemOpen
  \bibfield  {author} {\bibinfo {author} {\bibnamefont {Polkovnikov},
  \bibfnamefont {Anatoli}}, \bibinfo {author} {\bibfnamefont {Krishnendu}\
  \bibnamefont {Sengupta}}, \bibinfo {author} {\bibfnamefont {Alessandro}\
  \bibnamefont {Silva}}, \ and\ \bibinfo {author} {\bibfnamefont {Mukund}\
  \bibnamefont {Vengalattore}}} (\bibinfo {year} {2011}),\ \bibfield  {title}
  {\enquote {\bibinfo {title} {{Colloquium: Nonequilibrium dynamics of closed
  interacting quantum systems}},}\ }\href
  {https://link.aps.org/doi/10.1103/RevModPhys.83.863} {\bibfield  {journal}
  {\bibinfo  {journal} {Rev. Mod. Phys.}\ }\textbf {\bibinfo {volume} {83}},\
  \bibinfo {pages} {863--883}}\BibitemShut {NoStop}%
\bibitem [{\citenamefont {Poo}\ \emph {et~al.}(2016)\citenamefont {Poo},
  \citenamefont {He}, \citenamefont {Xiao}, \citenamefont {Lu}, \citenamefont
  {Wu},\ and\ \citenamefont {Chen}}]{Poo:2016SR}%
  \BibitemOpen
  \bibfield  {author} {\bibinfo {author} {\bibnamefont {Poo}, \bibfnamefont
  {Yin}}, \bibinfo {author} {\bibfnamefont {Cheng}\ \bibnamefont {He}},
  \bibinfo {author} {\bibfnamefont {Chao}\ \bibnamefont {Xiao}}, \bibinfo
  {author} {\bibfnamefont {Ming-Hui}\ \bibnamefont {Lu}}, \bibinfo {author}
  {\bibfnamefont {Rui-Xin}\ \bibnamefont {Wu}}, \ and\ \bibinfo {author}
  {\bibfnamefont {Yan-Feng}\ \bibnamefont {Chen}}} (\bibinfo {year} {2016}),\
  \bibfield  {title} {\enquote {\bibinfo {title} {Manipulating one-way space
  wave and its refraction by time-reversal and parity symmetry breaking},}\
  }\href {https://www.nature.com/articles/srep29380} {\bibfield  {journal}
  {\bibinfo  {journal} {Sci. Rep.}\ }\textbf {\bibinfo {volume} {6}},\ \bibinfo
  {pages} {29380}}\BibitemShut {NoStop}%
\bibitem [{\citenamefont {Poo}\ \emph {et~al.}(2011)\citenamefont {Poo},
  \citenamefont {Wu}, \citenamefont {Lin}, \citenamefont {Yang},\ and\
  \citenamefont {Chan}}]{Poo:2011PRL}%
  \BibitemOpen
  \bibfield  {author} {\bibinfo {author} {\bibnamefont {Poo}, \bibfnamefont
  {Yin}}, \bibinfo {author} {\bibfnamefont {Rui-xin}\ \bibnamefont {Wu}},
  \bibinfo {author} {\bibfnamefont {Zhifang}\ \bibnamefont {Lin}}, \bibinfo
  {author} {\bibfnamefont {Yan}\ \bibnamefont {Yang}}, \ and\ \bibinfo {author}
  {\bibfnamefont {C.~T.}\ \bibnamefont {Chan}}} (\bibinfo {year} {2011}),\
  \bibfield  {title} {\enquote {\bibinfo {title} {Experimental realization of
  self-guiding unidirectional electromagnetic edge states},}\ }\href
  {https://link.aps.org/doi/10.1103/PhysRevLett.106.093903} {\bibfield
  {journal} {\bibinfo  {journal} {Phys. Rev. Lett.}\ }\textbf {\bibinfo
  {volume} {106}},\ \bibinfo {pages} {093903}}\BibitemShut {NoStop}%
\bibitem [{\citenamefont {Poo}\ \emph {et~al.}(2012)\citenamefont {Poo},
  \citenamefont {Wu}, \citenamefont {Liu}, \citenamefont {Yang}, \citenamefont
  {Lin},\ and\ \citenamefont {Chui}}]{Poo:2012APL}%
  \BibitemOpen
  \bibfield  {author} {\bibinfo {author} {\bibnamefont {Poo}, \bibfnamefont
  {Yin}}, \bibinfo {author} {\bibfnamefont {Rui-xin}\ \bibnamefont {Wu}},
  \bibinfo {author} {\bibfnamefont {Shiyang}\ \bibnamefont {Liu}}, \bibinfo
  {author} {\bibfnamefont {Yan}\ \bibnamefont {Yang}}, \bibinfo {author}
  {\bibfnamefont {Zhifang}\ \bibnamefont {Lin}}, \ and\ \bibinfo {author}
  {\bibfnamefont {ST}~\bibnamefont {Chui}}} (\bibinfo {year} {2012}),\
  \bibfield  {title} {\enquote {\bibinfo {title} {Experimental demonstration of
  surface morphology independent electromagnetic chiral edge states originated
  from magnetic plasmon resonance},}\ }\href
  {https://aip.scitation.org/doi/10.1063/1.4747810} {\bibfield  {journal}
  {\bibinfo  {journal} {Appl. Phys. Lett.}\ }\textbf {\bibinfo {volume}
  {101}}~(\bibinfo {number} {8}),\ \bibinfo {pages} {081912}}\BibitemShut
  {NoStop}%
\bibitem [{\citenamefont {Poshakinskiy}\ and\ \citenamefont
  {Poddubny}(2017)}]{Sasha2017}%
  \BibitemOpen
  \bibfield  {author} {\bibinfo {author} {\bibnamefont {Poshakinskiy},
  \bibfnamefont {A~V}}, \ and\ \bibinfo {author} {\bibfnamefont {A.~N.}\
  \bibnamefont {Poddubny}}} (\bibinfo {year} {2017}),\ \bibfield  {title}
  {\enquote {\bibinfo {title} {Phonoritonic crystals with a synthetic magnetic
  field for an acoustic diode},}\ }\href
  {https://link.aps.org/doi/10.1103/PhysRevLett.118.156801} {\bibfield
  {journal} {\bibinfo  {journal} {Phys. Rev. Lett.}\ }\textbf {\bibinfo
  {volume} {118}},\ \bibinfo {pages} {156801}}\BibitemShut {NoStop}%
\bibitem [{\citenamefont {Poshakinskiy}\ \emph {et~al.}(2014)\citenamefont
  {Poshakinskiy}, \citenamefont {Poddubny}, \citenamefont {Pilozzi},\ and\
  \citenamefont {Ivchenko}}]{Poshakinskiy:2014PRL}%
  \BibitemOpen
  \bibfield  {author} {\bibinfo {author} {\bibnamefont {Poshakinskiy},
  \bibfnamefont {A~V}}, \bibinfo {author} {\bibfnamefont {A.~N.}\ \bibnamefont
  {Poddubny}}, \bibinfo {author} {\bibfnamefont {L.}~\bibnamefont {Pilozzi}}, \
  and\ \bibinfo {author} {\bibfnamefont {E.~L.}\ \bibnamefont {Ivchenko}}}
  (\bibinfo {year} {2014}),\ \bibfield  {title} {\enquote {\bibinfo {title}
  {Radiative topological states in resonant photonic crystals},}\ }\href
  {https://link.aps.org/doi/10.1103/PhysRevLett.112.107403} {\bibfield
  {journal} {\bibinfo  {journal} {Phys. Rev. Lett.}\ }\textbf {\bibinfo
  {volume} {112}},\ \bibinfo {pages} {107403}}\BibitemShut {NoStop}%
\bibitem [{\citenamefont {Pothier}\ \emph {et~al.}(1991)\citenamefont
  {Pothier}, \citenamefont {Lafarge}, \citenamefont {Orfila}, \citenamefont
  {Urbina}, \citenamefont {Esteve},\ and\ \citenamefont
  {Devoret}}]{Pothier:1991}%
  \BibitemOpen
  \bibfield  {author} {\bibinfo {author} {\bibnamefont {Pothier}, \bibfnamefont
  {H}}, \bibinfo {author} {\bibfnamefont {P.}~\bibnamefont {Lafarge}}, \bibinfo
  {author} {\bibfnamefont {P.F.}\ \bibnamefont {Orfila}}, \bibinfo {author}
  {\bibfnamefont {C.}~\bibnamefont {Urbina}}, \bibinfo {author} {\bibfnamefont
  {D.}~\bibnamefont {Esteve}}, \ and\ \bibinfo {author} {\bibfnamefont {M.H.}\
  \bibnamefont {Devoret}}} (\bibinfo {year} {1991}),\ \bibfield  {title}
  {\enquote {\bibinfo {title} {Single electron pump fabricated with ultrasmall
  normal tunnel junctions},}\ }\href
  {http://www.sciencedirect.com/science/article/pii/0921452691903329}
  {\bibfield  {journal} {\bibinfo  {journal} {Physica B: Condensed Matter}\
  }\textbf {\bibinfo {volume} {169}}~(\bibinfo {number} {1}),\ \bibinfo {pages}
  {573 -- 574}}\BibitemShut {NoStop}%
\bibitem [{\citenamefont {Pothier}\ \emph {et~al.}(1992)\citenamefont
  {Pothier}, \citenamefont {Lafarge}, \citenamefont {Urbina}, \citenamefont
  {Esteve},\ and\ \citenamefont {Devoret}}]{Pothier:1992}%
  \BibitemOpen
  \bibfield  {author} {\bibinfo {author} {\bibnamefont {Pothier}, \bibfnamefont
  {H}}, \bibinfo {author} {\bibfnamefont {P.}~\bibnamefont {Lafarge}}, \bibinfo
  {author} {\bibfnamefont {C.}~\bibnamefont {Urbina}}, \bibinfo {author}
  {\bibfnamefont {D.}~\bibnamefont {Esteve}}, \ and\ \bibinfo {author}
  {\bibfnamefont {M.~H.}\ \bibnamefont {Devoret}}} (\bibinfo {year} {1992}),\
  \bibfield  {title} {\enquote {\bibinfo {title} {Single-electron pump based on
  charging effects},}\ }\href {http://stacks.iop.org/0295-5075/17/i=3/a=011}
  {\bibfield  {journal} {\bibinfo  {journal} {EPL (Europhysics Letters)}\
  }\textbf {\bibinfo {volume} {17}}~(\bibinfo {number} {3}),\ \bibinfo {pages}
  {249}}\BibitemShut {NoStop}%
\bibitem [{\citenamefont {Prange}\ \emph {et~al.}(1989)\citenamefont {Prange},
  \citenamefont {Girvin}, \citenamefont {Cage}, \citenamefont {Klitzing},
  \citenamefont {Chang}, \citenamefont {Duncan}, \citenamefont {Haldane},
  \citenamefont {Laughlin}, \citenamefont {Pruisken},\ and\ \citenamefont
  {Thouless}}]{PrangeBook}%
  \BibitemOpen
  \bibfield  {author} {\bibinfo {author} {\bibnamefont {Prange}, \bibfnamefont
  {Richard~E}}, \bibinfo {author} {\bibfnamefont {Steven~M.}\ \bibnamefont
  {Girvin}}, \bibinfo {author} {\bibfnamefont {Marvin~E}\ \bibnamefont {Cage}},
  \bibinfo {author} {\bibfnamefont {Kv}~\bibnamefont {Klitzing}}, \bibinfo
  {author} {\bibfnamefont {AM}~\bibnamefont {Chang}}, \bibinfo {author}
  {\bibfnamefont {F}~\bibnamefont {Duncan}}, \bibinfo {author} {\bibfnamefont
  {M}~\bibnamefont {Haldane}}, \bibinfo {author} {\bibfnamefont
  {RB}~\bibnamefont {Laughlin}}, \bibinfo {author} {\bibfnamefont {AMM}\
  \bibnamefont {Pruisken}}, \ and\ \bibinfo {author} {\bibfnamefont
  {DJ}~\bibnamefont {Thouless}}} (\bibinfo {year} {1989}),\ \href@noop {}
  {\emph {\bibinfo {title} {The quantum {H}all effect}}}\ (\bibinfo
  {publisher} {Springer},\ \bibinfo {address} {New York})\BibitemShut {NoStop}%
\bibitem [{\citenamefont {{Price}}(2018)}]{Price:2018arxiv}%
  \BibitemOpen
  \bibfield  {author} {\bibinfo {author} {\bibnamefont {{Price}}, \bibfnamefont
  {H~M}}} (\bibinfo {year} {2018}),\ \bibfield  {title} {\enquote {\bibinfo
  {title} {{Four-dimensional topological lattices without gauge fields}},}\
  }\href {https://arxiv.org/abs/1806.05263} {\bibinfo  {journal}
  {arXiv:1806.05263}\ }\BibitemShut {NoStop}%
\bibitem [{\citenamefont {Price}\ \emph {et~al.}(2015)\citenamefont {Price},
  \citenamefont {Zilberberg}, \citenamefont {Ozawa}, \citenamefont
  {Carusotto},\ and\ \citenamefont {Goldman}}]{Price:2015PRL}%
  \BibitemOpen
\bibfield  {journal} {  }\bibfield  {author} {\bibinfo {author} {\bibnamefont
  {Price}, \bibfnamefont {H~M}}, \bibinfo {author} {\bibfnamefont
  {O.}~\bibnamefont {Zilberberg}}, \bibinfo {author} {\bibfnamefont
  {T.}~\bibnamefont {Ozawa}}, \bibinfo {author} {\bibfnamefont
  {I.}~\bibnamefont {Carusotto}}, \ and\ \bibinfo {author} {\bibfnamefont
  {N.}~\bibnamefont {Goldman}}} (\bibinfo {year} {2015}),\ \bibfield  {title}
  {\enquote {\bibinfo {title} {Four-dimensional quantum {H}all effect with
  ultracold atoms},}\ }\href
  {https://link.aps.org/doi/10.1103/PhysRevLett.115.195303} {\bibfield
  {journal} {\bibinfo  {journal} {Phys. Rev. Lett.}\ }\textbf {\bibinfo
  {volume} {115}},\ \bibinfo {pages} {195303}}\BibitemShut {NoStop}%
\bibitem [{\citenamefont {Price}\ \emph {et~al.}(2014)\citenamefont {Price},
  \citenamefont {Ozawa},\ and\ \citenamefont {Carusotto}}]{Price:2014PRL}%
  \BibitemOpen
  \bibfield  {author} {\bibinfo {author} {\bibnamefont {Price}, \bibfnamefont
  {Hannah~M}}, \bibinfo {author} {\bibfnamefont {Tomoki}\ \bibnamefont
  {Ozawa}}, \ and\ \bibinfo {author} {\bibfnamefont {Iacopo}\ \bibnamefont
  {Carusotto}}} (\bibinfo {year} {2014}),\ \bibfield  {title} {\enquote
  {\bibinfo {title} {Quantum mechanics with a momentum-space artificial
  magnetic field},}\ }\href
  {http://link.aps.org/doi/10.1103/PhysRevLett.113.190403} {\bibfield
  {journal} {\bibinfo  {journal} {Phys. Rev. Lett.}\ }\textbf {\bibinfo
  {volume} {113}},\ \bibinfo {pages} {190403}}\BibitemShut {NoStop}%
\bibitem [{\citenamefont {Price}\ \emph {et~al.}(2017)\citenamefont {Price},
  \citenamefont {Ozawa},\ and\ \citenamefont {Goldman}}]{Price:2017PRA}%
  \BibitemOpen
  \bibfield  {author} {\bibinfo {author} {\bibnamefont {Price}, \bibfnamefont
  {Hannah~M}}, \bibinfo {author} {\bibfnamefont {Tomoki}\ \bibnamefont
  {Ozawa}}, \ and\ \bibinfo {author} {\bibfnamefont {Nathan}\ \bibnamefont
  {Goldman}}} (\bibinfo {year} {2017}),\ \bibfield  {title} {\enquote {\bibinfo
  {title} {Synthetic dimensions for cold atoms from shaking a harmonic trap},}\
  }\href {https://link.aps.org/doi/10.1103/PhysRevA.95.023607} {\bibfield
  {journal} {\bibinfo  {journal} {Phys. Rev. A}\ }\textbf {\bibinfo {volume}
  {95}},\ \bibinfo {pages} {023607}}\BibitemShut {NoStop}%
\bibitem [{\citenamefont {Price}\ \emph {et~al.}(2016)\citenamefont {Price},
  \citenamefont {Zilberberg}, \citenamefont {Ozawa}, \citenamefont
  {Carusotto},\ and\ \citenamefont {Goldman}}]{Price:2016PRA}%
  \BibitemOpen
  \bibfield  {author} {\bibinfo {author} {\bibnamefont {Price}, \bibfnamefont
  {Hannah~M}}, \bibinfo {author} {\bibfnamefont {Oded}\ \bibnamefont
  {Zilberberg}}, \bibinfo {author} {\bibfnamefont {Tomoki}\ \bibnamefont
  {Ozawa}}, \bibinfo {author} {\bibfnamefont {Iacopo}\ \bibnamefont
  {Carusotto}}, \ and\ \bibinfo {author} {\bibfnamefont {Nathan}\ \bibnamefont
  {Goldman}}} (\bibinfo {year} {2016}),\ \bibfield  {title} {\enquote {\bibinfo
  {title} {Measurement of {C}hern numbers through center-of-mass responses},}\
  }\href {https://journals.aps.org/prb/abstract/10.1103/PhysRevB.93.245113}
  {\bibfield  {journal} {\bibinfo  {journal} {Phys. Rev. B}\ }\textbf {\bibinfo
  {volume} {93}}~(\bibinfo {number} {24}),\ \bibinfo {pages}
  {245113}}\BibitemShut {NoStop}%
\bibitem [{\citenamefont {Price}\ and\ \citenamefont
  {Cooper}(2012)}]{Price:2012PRA}%
  \BibitemOpen
  \bibfield  {author} {\bibinfo {author} {\bibnamefont {Price}, \bibfnamefont
  {HM}}, \ and\ \bibinfo {author} {\bibfnamefont {NR}~\bibnamefont {Cooper}}}
  (\bibinfo {year} {2012}),\ \bibfield  {title} {\enquote {\bibinfo {title}
  {Mapping the {B}erry curvature from semiclassical dynamics in optical
  lattices},}\ }\href
  {https://journals.aps.org/pra/abstract/10.1103/PhysRevA.85.033620} {\bibfield
   {journal} {\bibinfo  {journal} {Phys. Rev. A}\ }\textbf {\bibinfo {volume}
  {85}}~(\bibinfo {number} {3}),\ \bibinfo {pages} {033620}}\BibitemShut
  {NoStop}%
\bibitem [{\citenamefont {Prodan}\ and\ \citenamefont
  {Schulz-Baldes}(2016)}]{prodan2016bulk}%
  \BibitemOpen
  \bibfield  {author} {\bibinfo {author} {\bibnamefont {Prodan}, \bibfnamefont
  {Emil}}, \ and\ \bibinfo {author} {\bibfnamefont {Hermann}\ \bibnamefont
  {Schulz-Baldes}}} (\bibinfo {year} {2016}),\ \href@noop {} {\emph {\bibinfo
  {title} {Bulk and Boundary Invariants for Complex Topological Insulators:
  {F}rom {K}-Theory to Physics}}}\ (\bibinfo  {publisher}
  {Springer})\BibitemShut {NoStop}%
\bibitem [{\citenamefont {Qi}\ \emph {et~al.}(2008)\citenamefont {Qi},
  \citenamefont {Hughes},\ and\ \citenamefont {Zhang}}]{Qi:2008PRB}%
  \BibitemOpen
  \bibfield  {author} {\bibinfo {author} {\bibnamefont {Qi}, \bibfnamefont
  {Xiao-Liang}}, \bibinfo {author} {\bibfnamefont {Taylor~L.}\ \bibnamefont
  {Hughes}}, \ and\ \bibinfo {author} {\bibfnamefont {Shou-Cheng}\ \bibnamefont
  {Zhang}}} (\bibinfo {year} {2008}),\ \bibfield  {title} {\enquote {\bibinfo
  {title} {Topological field theory of time-reversal invariant insulators},}\
  }\href {https://link.aps.org/doi/10.1103/PhysRevB.78.195424} {\bibfield
  {journal} {\bibinfo  {journal} {Phys. Rev. B}\ }\textbf {\bibinfo {volume}
  {78}},\ \bibinfo {pages} {195424}}\BibitemShut {NoStop}%
\bibitem [{\citenamefont {Qi}\ \emph {et~al.}(2006)\citenamefont {Qi},
  \citenamefont {Wu},\ and\ \citenamefont {Zhang}}]{Qi:2006PRB}%
  \BibitemOpen
  \bibfield  {author} {\bibinfo {author} {\bibnamefont {Qi}, \bibfnamefont
  {Xiao-Liang}}, \bibinfo {author} {\bibfnamefont {Yong-Shi}\ \bibnamefont
  {Wu}}, \ and\ \bibinfo {author} {\bibfnamefont {Shou-Cheng}\ \bibnamefont
  {Zhang}}} (\bibinfo {year} {2006}),\ \bibfield  {title} {\enquote {\bibinfo
  {title} {General theorem relating the bulk topological number to edge states
  in two-dimensional insulators},}\ }\href
  {https://link.aps.org/doi/10.1103/PhysRevB.74.045125} {\bibfield  {journal}
  {\bibinfo  {journal} {Phys. Rev. B}\ }\textbf {\bibinfo {volume} {74}},\
  \bibinfo {pages} {045125}}\BibitemShut {NoStop}%
\bibitem [{\citenamefont {Qi}\ and\ \citenamefont {Zhang}(2011)}]{Qi:2011RMP}%
  \BibitemOpen
  \bibfield  {author} {\bibinfo {author} {\bibnamefont {Qi}, \bibfnamefont
  {Xiao-Liang}}, \ and\ \bibinfo {author} {\bibfnamefont {Shou-Cheng}\
  \bibnamefont {Zhang}}} (\bibinfo {year} {2011}),\ \bibfield  {title}
  {\enquote {\bibinfo {title} {Topological insulators and superconductors},}\
  }\href {https://link.aps.org/doi/10.1103/RevModPhys.83.1057} {\bibfield
  {journal} {\bibinfo  {journal} {Rev. Mod. Phys.}\ }\textbf {\bibinfo {volume}
  {83}},\ \bibinfo {pages} {1057--1110}}\BibitemShut {NoStop}%
\bibitem [{\citenamefont {Qiu}\ \emph {et~al.}(2017)\citenamefont {Qiu},
  \citenamefont {Liang}, \citenamefont {Qiu}, \citenamefont {Chen},
  \citenamefont {Ren}, \citenamefont {Lin}, \citenamefont {Wang}, \citenamefont
  {Kan},\ and\ \citenamefont {Pan}}]{Qiu:OE2017}%
  \BibitemOpen
  \bibfield  {author} {\bibinfo {author} {\bibnamefont {Qiu}, \bibfnamefont
  {Pingping}}, \bibinfo {author} {\bibfnamefont {Rui}\ \bibnamefont {Liang}},
  \bibinfo {author} {\bibfnamefont {Weibin}\ \bibnamefont {Qiu}}, \bibinfo
  {author} {\bibfnamefont {Houbo}\ \bibnamefont {Chen}}, \bibinfo {author}
  {\bibfnamefont {Junbo}\ \bibnamefont {Ren}}, \bibinfo {author} {\bibfnamefont
  {Zhili}\ \bibnamefont {Lin}}, \bibinfo {author} {\bibfnamefont {Jia-Xian}\
  \bibnamefont {Wang}}, \bibinfo {author} {\bibfnamefont {Qiang}\ \bibnamefont
  {Kan}}, \ and\ \bibinfo {author} {\bibfnamefont {Jiao-Qing}\ \bibnamefont
  {Pan}}} (\bibinfo {year} {2017}),\ \bibfield  {title} {\enquote {\bibinfo
  {title} {{Topologically protected edge states in graphene plasmonic
  crystals}},}\ }\href
  {http://www.opticsexpress.org/abstract.cfm?URI=oe-25-19-22587} {\bibfield
  {journal} {\bibinfo  {journal} {Opt. Express}\ }\textbf {\bibinfo {volume}
  {25}}~(\bibinfo {number} {19}),\ \bibinfo {pages} {22587--22594}}\BibitemShut
  {NoStop}%
\bibitem [{\citenamefont {Qiu}\ \emph {et~al.}(2011)\citenamefont {Qiu},
  \citenamefont {Wang},\ and\ \citenamefont {Solja{\v{c}}i{\'c}}}]{Qiu:2011OE}%
  \BibitemOpen
  \bibfield  {author} {\bibinfo {author} {\bibnamefont {Qiu}, \bibfnamefont
  {W}}, \bibinfo {author} {\bibfnamefont {Z}~\bibnamefont {Wang}}, \ and\
  \bibinfo {author} {\bibfnamefont {M}~\bibnamefont {Solja{\v{c}}i{\'c}}}}
  (\bibinfo {year} {2011}),\ \bibfield  {title} {\enquote {\bibinfo {title}
  {Broadband circulators based on directional coupling of one-way
  waveguides},}\ }\href
  {https://www.osapublishing.org/oe/abstract.cfm?uri=oe-19-22-22248} {\bibfield
   {journal} {\bibinfo  {journal} {Opt. Express}\ }\textbf {\bibinfo {volume}
  {19}}~(\bibinfo {number} {22}),\ \bibinfo {pages} {22248}}\BibitemShut
  {NoStop}%
\bibitem [{\citenamefont {Quelle}\ \emph {et~al.}(2017)\citenamefont {Quelle},
  \citenamefont {Weitenberg}, \citenamefont {Sengstock},\ and\ \citenamefont
  {Smith}}]{Quelle:2017arxiv}%
  \BibitemOpen
  \bibfield  {author} {\bibinfo {author} {\bibnamefont {Quelle}, \bibfnamefont
  {A}}, \bibinfo {author} {\bibfnamefont {C}~\bibnamefont {Weitenberg}},
  \bibinfo {author} {\bibfnamefont {K}~\bibnamefont {Sengstock}}, \ and\
  \bibinfo {author} {\bibfnamefont {C~Morais}\ \bibnamefont {Smith}}} (\bibinfo
  {year} {2017}),\ \bibfield  {title} {\enquote {\bibinfo {title} {Driving
  protocol for a {F}loquet topological phase without static counterpart},}\
  }\href {http://iopscience.iop.org/article/10.1088/1367-2630/aa8646/meta}
  {\bibfield  {journal} {\bibinfo  {journal} {New J. Phys.}\ }\textbf {\bibinfo
  {volume} {19}}~(\bibinfo {number} {11}),\ \bibinfo {pages}
  {113010}}\BibitemShut {NoStop}%
\bibitem [{\citenamefont {Raghu}\ and\ \citenamefont
  {Haldane}(2008)}]{Raghu:2008PRA}%
  \BibitemOpen
  \bibfield  {author} {\bibinfo {author} {\bibnamefont {Raghu}, \bibfnamefont
  {S}}, \ and\ \bibinfo {author} {\bibfnamefont {F.~D.~M.}\ \bibnamefont
  {Haldane}}} (\bibinfo {year} {2008}),\ \bibfield  {title} {\enquote {\bibinfo
  {title} {Analogs of quantum-{H}all-effect edge states in photonic
  crystals},}\ }\href {http://link.aps.org/doi/10.1103/PhysRevA.78.033834}
  {\bibfield  {journal} {\bibinfo  {journal} {Phys. Rev. A}\ }\textbf {\bibinfo
  {volume} {78}},\ \bibinfo {pages} {033834}}\BibitemShut {NoStop}%
\bibitem [{\citenamefont {Rahav}\ \emph {et~al.}(2003)\citenamefont {Rahav},
  \citenamefont {Gilary},\ and\ \citenamefont {Fishman}}]{Rahav:2003PRA}%
  \BibitemOpen
  \bibfield  {author} {\bibinfo {author} {\bibnamefont {Rahav}, \bibfnamefont
  {Saar}}, \bibinfo {author} {\bibfnamefont {Ido}\ \bibnamefont {Gilary}}, \
  and\ \bibinfo {author} {\bibfnamefont {Shmuel}\ \bibnamefont {Fishman}}}
  (\bibinfo {year} {2003}),\ \bibfield  {title} {\enquote {\bibinfo {title}
  {Effective {H}amiltonians for periodically driven systems},}\ }\href
  {https://doi.org/10.1103/PhysRevA.68.013820} {\bibfield  {journal} {\bibinfo
  {journal} {Phys. Rev. A}\ }\textbf {\bibinfo {volume} {68}},\ \bibinfo
  {pages} {013820}}\BibitemShut {NoStop}%
\bibitem [{\citenamefont {Rasmussen}\ \emph {et~al.}(2000)\citenamefont
  {Rasmussen}, \citenamefont {Cretegny}, \citenamefont {Kevrekidis},\ and\
  \citenamefont {Gr{\o}nbech-Jensen}}]{Rasmussen:PRL2000}%
  \BibitemOpen
  \bibfield  {author} {\bibinfo {author} {\bibnamefont {Rasmussen},
  \bibfnamefont {K{\O}}}, \bibinfo {author} {\bibfnamefont {T}~\bibnamefont
  {Cretegny}}, \bibinfo {author} {\bibfnamefont {PG}~\bibnamefont
  {Kevrekidis}}, \ and\ \bibinfo {author} {\bibfnamefont {Niels}\ \bibnamefont
  {Gr{\o}nbech-Jensen}}} (\bibinfo {year} {2000}),\ \bibfield  {title}
  {\enquote {\bibinfo {title} {Statistical mechanics of a discrete nonlinear
  system},}\ }\href
  {https://journals.aps.org/prl/abstract/10.1103/PhysRevLett.84.3740}
  {\bibfield  {journal} {\bibinfo  {journal} {Phys. Rev. Lett.}\ }\textbf
  {\bibinfo {volume} {84}}~(\bibinfo {number} {17}),\ \bibinfo {pages}
  {3740}}\BibitemShut {NoStop}%
\bibitem [{\citenamefont {Rechtsman}\ \emph {et~al.}(2016)\citenamefont
  {Rechtsman}, \citenamefont {Lumer}, \citenamefont {Plotnik}, \citenamefont
  {Perez-Leija}, \citenamefont {Szameit},\ and\ \citenamefont
  {Segev}}]{Rechtsman:2016Optica}%
  \BibitemOpen
  \bibfield  {author} {\bibinfo {author} {\bibnamefont {Rechtsman},
  \bibfnamefont {Mikael~C}}, \bibinfo {author} {\bibfnamefont {Yaakov}\
  \bibnamefont {Lumer}}, \bibinfo {author} {\bibfnamefont {Yonatan}\
  \bibnamefont {Plotnik}}, \bibinfo {author} {\bibfnamefont {Armando}\
  \bibnamefont {Perez-Leija}}, \bibinfo {author} {\bibfnamefont {Alexander}\
  \bibnamefont {Szameit}}, \ and\ \bibinfo {author} {\bibfnamefont {Mordechai}\
  \bibnamefont {Segev}}} (\bibinfo {year} {2016}),\ \bibfield  {title}
  {\enquote {\bibinfo {title} {Topological protection of photonic path
  entanglement},}\ }\href
  {http://www.osapublishing.org/optica/abstract.cfm?URI=optica-3-9-925}
  {\bibfield  {journal} {\bibinfo  {journal} {Optica}\ }\textbf {\bibinfo
  {volume} {3}}~(\bibinfo {number} {9}),\ \bibinfo {pages}
  {925--930}}\BibitemShut {NoStop}%
\bibitem [{\citenamefont {Rechtsman}\ \emph
  {et~al.}(2013{\natexlab{a}})\citenamefont {Rechtsman}, \citenamefont
  {Plotnik}, \citenamefont {Zeuner}, \citenamefont {Song}, \citenamefont
  {Chen}, \citenamefont {Szameit},\ and\ \citenamefont
  {Segev}}]{Rechtsman:2013PRL}%
  \BibitemOpen
  \bibfield  {author} {\bibinfo {author} {\bibnamefont {Rechtsman},
  \bibfnamefont {Mikael~C}}, \bibinfo {author} {\bibfnamefont {Yonatan}\
  \bibnamefont {Plotnik}}, \bibinfo {author} {\bibfnamefont {Julia~M.}\
  \bibnamefont {Zeuner}}, \bibinfo {author} {\bibfnamefont {Daohong}\
  \bibnamefont {Song}}, \bibinfo {author} {\bibfnamefont {Zhigang}\
  \bibnamefont {Chen}}, \bibinfo {author} {\bibfnamefont {Alexander}\
  \bibnamefont {Szameit}}, \ and\ \bibinfo {author} {\bibfnamefont {Mordechai}\
  \bibnamefont {Segev}}} (\bibinfo {year} {2013}{\natexlab{a}}),\ \bibfield
  {title} {\enquote {\bibinfo {title} {Topological creation and destruction of
  edge states in photonic graphene},}\ }\href
  {http://link.aps.org/doi/10.1103/PhysRevLett.111.103901} {\bibfield
  {journal} {\bibinfo  {journal} {Phys. Rev. Lett.}\ }\textbf {\bibinfo
  {volume} {111}},\ \bibinfo {pages} {103901}}\BibitemShut {NoStop}%
\bibitem [{\citenamefont {Rechtsman}\ \emph
  {et~al.}(2013{\natexlab{b}})\citenamefont {Rechtsman}, \citenamefont
  {Zeuner}, \citenamefont {Plotnik}, \citenamefont {Lumer}, \citenamefont
  {Podolsky}, \citenamefont {Dreisow}, \citenamefont {Nolte}, \citenamefont
  {Segev},\ and\ \citenamefont {Szameit}}]{Rechtsman:2013Nature}%
  \BibitemOpen
  \bibfield  {author} {\bibinfo {author} {\bibnamefont {Rechtsman},
  \bibfnamefont {Mikael~C}}, \bibinfo {author} {\bibfnamefont {Julia~M}\
  \bibnamefont {Zeuner}}, \bibinfo {author} {\bibfnamefont {Yonatan}\
  \bibnamefont {Plotnik}}, \bibinfo {author} {\bibfnamefont {Yaakov}\
  \bibnamefont {Lumer}}, \bibinfo {author} {\bibfnamefont {Daniel}\
  \bibnamefont {Podolsky}}, \bibinfo {author} {\bibfnamefont {Felix}\
  \bibnamefont {Dreisow}}, \bibinfo {author} {\bibfnamefont {Stefan}\
  \bibnamefont {Nolte}}, \bibinfo {author} {\bibfnamefont {Mordechai}\
  \bibnamefont {Segev}}, \ and\ \bibinfo {author} {\bibfnamefont {Alexander}\
  \bibnamefont {Szameit}}} (\bibinfo {year} {2013}{\natexlab{b}}),\ \bibfield
  {title} {\enquote {\bibinfo {title} {Photonic {F}loquet topological
  insulators},}\ }\href
  {http://www.nature.com/nature/journal/v496/n7444/full/nature12066.html}
  {\bibfield  {journal} {\bibinfo  {journal} {Nature}\ }\textbf {\bibinfo
  {volume} {496}}~(\bibinfo {number} {7444}),\ \bibinfo {pages}
  {196--200}}\BibitemShut {NoStop}%
\bibitem [{\citenamefont {Rechtsman}\ \emph
  {et~al.}(2013{\natexlab{c}})\citenamefont {Rechtsman}, \citenamefont
  {Zeuner}, \citenamefont {T{\"u}nnermann}, \citenamefont {Nolte},
  \citenamefont {Segev},\ and\ \citenamefont
  {Szameit}}]{Rechtsman:2013NatPhot}%
  \BibitemOpen
  \bibfield  {author} {\bibinfo {author} {\bibnamefont {Rechtsman},
  \bibfnamefont {Mikael~C}}, \bibinfo {author} {\bibfnamefont {Julia~M}\
  \bibnamefont {Zeuner}}, \bibinfo {author} {\bibfnamefont {Andreas}\
  \bibnamefont {T{\"u}nnermann}}, \bibinfo {author} {\bibfnamefont {Stefan}\
  \bibnamefont {Nolte}}, \bibinfo {author} {\bibfnamefont {Mordechai}\
  \bibnamefont {Segev}}, \ and\ \bibinfo {author} {\bibfnamefont {Alexander}\
  \bibnamefont {Szameit}}} (\bibinfo {year} {2013}{\natexlab{c}}),\ \bibfield
  {title} {\enquote {\bibinfo {title} {Strain-induced pseudomagnetic field and
  photonic {L}andau levels in dielectric structures},}\ }\href
  {http://www.nature.com/nphoton/journal/v7/n2/full/nphoton.2012.302.html}
  {\bibfield  {journal} {\bibinfo  {journal} {Nat. Photonics}\ }\textbf
  {\bibinfo {volume} {7}}~(\bibinfo {number} {2}),\ \bibinfo {pages}
  {153--158}}\BibitemShut {NoStop}%
\bibitem [{\citenamefont {Reichl}\ and\ \citenamefont
  {Mueller}(2014)}]{Reichl:2014PRA}%
  \BibitemOpen
  \bibfield  {author} {\bibinfo {author} {\bibnamefont {Reichl}, \bibfnamefont
  {Matthew~D}}, \ and\ \bibinfo {author} {\bibfnamefont {Erich~J.}\
  \bibnamefont {Mueller}}} (\bibinfo {year} {2014}),\ \bibfield  {title}
  {\enquote {\bibinfo {title} {Floquet edge states with ultracold atoms},}\
  }\href {https://link.aps.org/doi/10.1103/PhysRevA.89.063628} {\bibfield
  {journal} {\bibinfo  {journal} {Phys. Rev. A}\ }\textbf {\bibinfo {volume}
  {89}},\ \bibinfo {pages} {063628}}\BibitemShut {NoStop}%
\bibitem [{\citenamefont {Reinhard}\ \emph {et~al.}(2012)\citenamefont
  {Reinhard}, \citenamefont {Volz}, \citenamefont {Winger}, \citenamefont
  {Badolato}, \citenamefont {Hennessy}, \citenamefont {Hu},\ and\ \citenamefont
  {Imamo{\u{g}}lu}}]{Reinhard:NatPhot2011}%
  \BibitemOpen
  \bibfield  {author} {\bibinfo {author} {\bibnamefont {Reinhard},
  \bibfnamefont {Andreas}}, \bibinfo {author} {\bibfnamefont {Thomas}\
  \bibnamefont {Volz}}, \bibinfo {author} {\bibfnamefont {Martin}\ \bibnamefont
  {Winger}}, \bibinfo {author} {\bibfnamefont {Antonio}\ \bibnamefont
  {Badolato}}, \bibinfo {author} {\bibfnamefont {Kevin~J}\ \bibnamefont
  {Hennessy}}, \bibinfo {author} {\bibfnamefont {Evelyn~L}\ \bibnamefont {Hu}},
  \ and\ \bibinfo {author} {\bibfnamefont {Ata{\c{c}}}\ \bibnamefont
  {Imamo{\u{g}}lu}}} (\bibinfo {year} {2012}),\ \bibfield  {title} {\enquote
  {\bibinfo {title} {Strongly correlated photons on a chip},}\ }\href
  {https://www.nature.com/articles/nphoton.2011.321} {\bibfield  {journal}
  {\bibinfo  {journal} {Nat. Photonics}\ }\textbf {\bibinfo {volume}
  {6}}~(\bibinfo {number} {2}),\ \bibinfo {pages} {93--96}}\BibitemShut
  {NoStop}%
\bibitem [{\citenamefont {Resta}(1994)}]{Resta:1994RMP}%
  \BibitemOpen
  \bibfield  {author} {\bibinfo {author} {\bibnamefont {Resta}, \bibfnamefont
  {Raffaele}}} (\bibinfo {year} {1994}),\ \bibfield  {title} {\enquote
  {\bibinfo {title} {Macroscopic polarization in crystalline dielectrics: {T}he
  geometric phase approach},}\ }\href
  {https://link.aps.org/doi/10.1103/RevModPhys.66.899} {\bibfield  {journal}
  {\bibinfo  {journal} {Rev. Mod. Phys.}\ }\textbf {\bibinfo {volume} {66}},\
  \bibinfo {pages} {899--915}}\BibitemShut {NoStop}%
\bibitem [{\citenamefont {Resta}(2011)}]{Resta:2011EPJB}%
  \BibitemOpen
  \bibfield  {author} {\bibinfo {author} {\bibnamefont {Resta}, \bibfnamefont
  {Raffaele}}} (\bibinfo {year} {2011}),\ \bibfield  {title} {\enquote
  {\bibinfo {title} {The insulating state of matter: {A} geometrical theory},}\
  }\href {https://link.springer.com/article/10.1140%2Fepjb%2Fe2010-10874-4}
  {\bibfield  {journal} {\bibinfo  {journal} {Eur. Phys. J. B}\ }\textbf
  {\bibinfo {volume} {79}}~(\bibinfo {number} {2}),\ \bibinfo {pages}
  {121--137}}\BibitemShut {NoStop}%
\bibitem [{\citenamefont {Richard}\ \emph {et~al.}(2005)\citenamefont
  {Richard}, \citenamefont {Kasprzak}, \citenamefont {Romestain}, \citenamefont
  {Andre},\ and\ \citenamefont {Dang}}]{Richard:PRL2005}%
  \BibitemOpen
  \bibfield  {author} {\bibinfo {author} {\bibnamefont {Richard}, \bibfnamefont
  {M}}, \bibinfo {author} {\bibfnamefont {J}~\bibnamefont {Kasprzak}}, \bibinfo
  {author} {\bibfnamefont {R}~\bibnamefont {Romestain}}, \bibinfo {author}
  {\bibfnamefont {R}~\bibnamefont {Andre}}, \ and\ \bibinfo {author}
  {\bibfnamefont {LS}~\bibnamefont {Dang}}} (\bibinfo {year} {2005}),\
  \bibfield  {title} {\enquote {\bibinfo {title} {Spontaneous coherent phase
  transition of polaritons in {CdTe} microcavities},}\ }\href
  {https://journals.aps.org/prl/abstract/10.1103/PhysRevLett.94.187401}
  {\bibfield  {journal} {\bibinfo  {journal} {Phys. Rev. Lett.}\ }\textbf
  {\bibinfo {volume} {94}}~(\bibinfo {number} {18})}\BibitemShut {NoStop}%
\bibitem [{\citenamefont {Ringel}\ \emph {et~al.}(2014)\citenamefont {Ringel},
  \citenamefont {Pletyukhov},\ and\ \citenamefont {Gritsev}}]{Ringel:NJP2014}%
  \BibitemOpen
  \bibfield  {author} {\bibinfo {author} {\bibnamefont {Ringel}, \bibfnamefont
  {Matous}}, \bibinfo {author} {\bibfnamefont {Mikhail}\ \bibnamefont
  {Pletyukhov}}, \ and\ \bibinfo {author} {\bibfnamefont {Vladimir}\
  \bibnamefont {Gritsev}}} (\bibinfo {year} {2014}),\ \bibfield  {title}
  {\enquote {\bibinfo {title} {Topologically protected strongly correlated
  states of photons},}\ }\href
  {http://stacks.iop.org/1367-2630/16/i=11/a=113030} {\bibfield  {journal}
  {\bibinfo  {journal} {New J. Phys.}\ }\textbf {\bibinfo {volume}
  {16}}~(\bibinfo {number} {11}),\ \bibinfo {pages} {113030}}\BibitemShut
  {NoStop}%
\bibitem [{\citenamefont {Rodr{\'\i}guez-Fortu{\~n}o}\ \emph
  {et~al.}(2013)\citenamefont {Rodr{\'\i}guez-Fortu{\~n}o}, \citenamefont
  {Marino}, \citenamefont {Ginzburg}, \citenamefont {O{\textquoteright}Connor},
  \citenamefont {Mart{\'\i}nez}, \citenamefont {Wurtz},\ and\ \citenamefont
  {Zayats}}]{Rodriguez:2013science}%
  \BibitemOpen
  \bibfield  {author} {\bibinfo {author} {\bibnamefont
  {Rodr{\'\i}guez-Fortu{\~n}o}, \bibfnamefont {Francisco~J}}, \bibinfo {author}
  {\bibfnamefont {Giuseppe}\ \bibnamefont {Marino}}, \bibinfo {author}
  {\bibfnamefont {Pavel}\ \bibnamefont {Ginzburg}}, \bibinfo {author}
  {\bibfnamefont {Daniel}\ \bibnamefont {O{\textquoteright}Connor}}, \bibinfo
  {author} {\bibfnamefont {Alejandro}\ \bibnamefont {Mart{\'\i}nez}}, \bibinfo
  {author} {\bibfnamefont {Gregory~A.}\ \bibnamefont {Wurtz}}, \ and\ \bibinfo
  {author} {\bibfnamefont {Anatoly~V.}\ \bibnamefont {Zayats}}} (\bibinfo
  {year} {2013}),\ \bibfield  {title} {\enquote {\bibinfo {title} {Near-field
  interference for the unidirectional excitation of electromagnetic guided
  modes},}\ }\href {http://science.sciencemag.org/content/340/6130/328}
  {\bibfield  {journal} {\bibinfo  {journal} {Science}\ }\textbf {\bibinfo
  {volume} {340}}~(\bibinfo {number} {6130}),\ \bibinfo {pages}
  {328--330}}\BibitemShut {NoStop}%
\bibitem [{\citenamefont {Rosanov}(2002)}]{Rosanov}%
  \BibitemOpen
  \bibfield  {author} {\bibinfo {author} {\bibnamefont {Rosanov}, \bibfnamefont
  {N~N}}} (\bibinfo {year} {2002}),\ \href@noop {} {\emph {\bibinfo {title}
  {Spatial hysteresis and optical patterns}}}\ (\bibinfo  {publisher}
  {Springer},\ \bibinfo {address} {New York})\BibitemShut {NoStop}%
\bibitem [{\citenamefont {Rosenthal}\ \emph {et~al.}(2018)\citenamefont
  {Rosenthal}, \citenamefont {Ehrlich}, \citenamefont {Rudner}, \citenamefont
  {Higginbotham},\ and\ \citenamefont {Lehnert}}]{Rosenthal:2018PRB}%
  \BibitemOpen
  \bibfield  {author} {\bibinfo {author} {\bibnamefont {Rosenthal},
  \bibfnamefont {Eric~I}}, \bibinfo {author} {\bibfnamefont {Nicole~K.}\
  \bibnamefont {Ehrlich}}, \bibinfo {author} {\bibfnamefont {Mark~S.}\
  \bibnamefont {Rudner}}, \bibinfo {author} {\bibfnamefont {Andrew~P.}\
  \bibnamefont {Higginbotham}}, \ and\ \bibinfo {author} {\bibfnamefont
  {K.~W.}\ \bibnamefont {Lehnert}}} (\bibinfo {year} {2018}),\ \bibfield
  {title} {\enquote {\bibinfo {title} {Topological phase transition measured in
  a dissipative metamaterial},}\ }\href
  {https://link.aps.org/doi/10.1103/PhysRevB.97.220301} {\bibfield  {journal}
  {\bibinfo  {journal} {Phys. Rev. B}\ }\textbf {\bibinfo {volume} {97}},\
  \bibinfo {pages} {220301}}\BibitemShut {NoStop}%
\bibitem [{\citenamefont {Roushan}\ \emph
  {et~al.}(2017{\natexlab{a}})\citenamefont {Roushan}, \citenamefont {Neill},
  \citenamefont {Tangpanitanon}, \citenamefont {Bastidas}, \citenamefont
  {Megrant}, \citenamefont {Barends}, \citenamefont {Chen}, \citenamefont
  {Chen}, \citenamefont {Chiaro}, \citenamefont {Dunsworth} \emph
  {et~al.}}]{Roushan:Science2017}%
  \BibitemOpen
  \bibfield  {author} {\bibinfo {author} {\bibnamefont {Roushan}, \bibfnamefont
  {P}}, \bibinfo {author} {\bibfnamefont {C}~\bibnamefont {Neill}}, \bibinfo
  {author} {\bibfnamefont {J}~\bibnamefont {Tangpanitanon}}, \bibinfo {author}
  {\bibfnamefont {VM}~\bibnamefont {Bastidas}}, \bibinfo {author}
  {\bibfnamefont {A}~\bibnamefont {Megrant}}, \bibinfo {author} {\bibfnamefont
  {R}~\bibnamefont {Barends}}, \bibinfo {author} {\bibfnamefont
  {Y}~\bibnamefont {Chen}}, \bibinfo {author} {\bibfnamefont {Z}~\bibnamefont
  {Chen}}, \bibinfo {author} {\bibfnamefont {B}~\bibnamefont {Chiaro}},
  \bibinfo {author} {\bibfnamefont {A}~\bibnamefont {Dunsworth}},  \emph
  {et~al.}} (\bibinfo {year} {2017}{\natexlab{a}}),\ \bibfield  {title}
  {\enquote {\bibinfo {title} {Spectroscopic signatures of localization with
  interacting photons in superconducting qubits},}\ }\href
  {http://science.sciencemag.org/content/358/6367/1175} {\bibfield  {journal}
  {\bibinfo  {journal} {Science}\ }\textbf {\bibinfo {volume} {358}}~(\bibinfo
  {number} {6367}),\ \bibinfo {pages} {1175--1179}}\BibitemShut {NoStop}%
\bibitem [{\citenamefont {Roushan}\ \emph
  {et~al.}(2017{\natexlab{b}})\citenamefont {Roushan}, \citenamefont {Neill},
  \citenamefont {Megrant}, \citenamefont {Chen}, \citenamefont {Babbush},
  \citenamefont {Barends}, \citenamefont {Campbell}, \citenamefont {Chen},
  \citenamefont {Chiaro}, \citenamefont {Dunsworth} \emph
  {et~al.}}]{Roushan:2016NatPhys}%
  \BibitemOpen
  \bibfield  {author} {\bibinfo {author} {\bibnamefont {Roushan}, \bibfnamefont
  {Pedram}}, \bibinfo {author} {\bibfnamefont {Charles}\ \bibnamefont {Neill}},
  \bibinfo {author} {\bibfnamefont {Anthony}\ \bibnamefont {Megrant}}, \bibinfo
  {author} {\bibfnamefont {Yu}~\bibnamefont {Chen}}, \bibinfo {author}
  {\bibfnamefont {Ryan}\ \bibnamefont {Babbush}}, \bibinfo {author}
  {\bibfnamefont {Rami}\ \bibnamefont {Barends}}, \bibinfo {author}
  {\bibfnamefont {Brooks}\ \bibnamefont {Campbell}}, \bibinfo {author}
  {\bibfnamefont {Zijun}\ \bibnamefont {Chen}}, \bibinfo {author}
  {\bibfnamefont {Ben}\ \bibnamefont {Chiaro}}, \bibinfo {author}
  {\bibfnamefont {Andrew}\ \bibnamefont {Dunsworth}},  \emph {et~al.}}
  (\bibinfo {year} {2017}{\natexlab{b}}),\ \bibfield  {title} {\enquote
  {\bibinfo {title} {Chiral ground-state currents of interacting photons in a
  synthetic magnetic field},}\ }\href
  {https://www.nature.com/articles/nphys3930} {\bibfield  {journal} {\bibinfo
  {journal} {Nat. Phys.}\ }\textbf {\bibinfo {volume} {13}}~(\bibinfo {number}
  {2}),\ \bibinfo {pages} {146}}\BibitemShut {NoStop}%
\bibitem [{\citenamefont {Roy}\ \emph {et~al.}(2017)\citenamefont {Roy},
  \citenamefont {Wilson},\ and\ \citenamefont {Firstenberg}}]{Roy:RMP2017}%
  \BibitemOpen
  \bibfield  {author} {\bibinfo {author} {\bibnamefont {Roy}, \bibfnamefont
  {Dibyendu}}, \bibinfo {author} {\bibfnamefont {C.~M.}\ \bibnamefont
  {Wilson}}, \ and\ \bibinfo {author} {\bibfnamefont {Ofer}\ \bibnamefont
  {Firstenberg}}} (\bibinfo {year} {2017}),\ \bibfield  {title} {\enquote
  {\bibinfo {title} {{Colloquium: Strongly interacting photons in
  one-dimensional continuum}},}\ }\href
  {https://link.aps.org/doi/10.1103/RevModPhys.89.021001} {\bibfield  {journal}
  {\bibinfo  {journal} {Rev. Mod. Phys.}\ }\textbf {\bibinfo {volume} {89}},\
  \bibinfo {pages} {021001}}\BibitemShut {NoStop}%
\bibitem [{\citenamefont {Roy}(2009)}]{Roy:2009PRB}%
  \BibitemOpen
  \bibfield  {author} {\bibinfo {author} {\bibnamefont {Roy}, \bibfnamefont
  {Rahul}}} (\bibinfo {year} {2009}),\ \bibfield  {title} {\enquote {\bibinfo
  {title} {Topological phases and the quantum spin {H}all effect in three
  dimensions},}\ }\href {https://link.aps.org/doi/10.1103/PhysRevB.79.195322}
  {\bibfield  {journal} {\bibinfo  {journal} {Phys. Rev. B}\ }\textbf {\bibinfo
  {volume} {79}},\ \bibinfo {pages} {195322}}\BibitemShut {NoStop}%
\bibitem [{\citenamefont {Roy}\ \emph {et~al.}(2018)\citenamefont {Roy},
  \citenamefont {Kolodrubetz}, \citenamefont {Goldman},\ and\ \citenamefont
  {Grushin}}]{Roy:2017arXiv}%
  \BibitemOpen
  \bibfield  {author} {\bibinfo {author} {\bibnamefont {Roy}, \bibfnamefont
  {Sthitadhi}}, \bibinfo {author} {\bibfnamefont {Michael}\ \bibnamefont
  {Kolodrubetz}}, \bibinfo {author} {\bibfnamefont {Nathan}\ \bibnamefont
  {Goldman}}, \ and\ \bibinfo {author} {\bibfnamefont {Adolfo~G}\ \bibnamefont
  {Grushin}}} (\bibinfo {year} {2018}),\ \bibfield  {title} {\enquote {\bibinfo
  {title} {Tunable axial gauge fields in engineered {W}eyl semimetals:
  {S}emiclassical analysis and optical lattice implementations},}\ }\href
  {http://iopscience.iop.org/article/10.1088/2053-1583/aaa577/meta} {\bibfield
  {journal} {\bibinfo  {journal} {2D Materials}\ }\textbf {\bibinfo {volume}
  {5}}~(\bibinfo {number} {2}),\ \bibinfo {pages} {024001}}\BibitemShut
  {NoStop}%
\bibitem [{\citenamefont {Rudner}\ \emph {et~al.}(2013)\citenamefont {Rudner},
  \citenamefont {Lindner}, \citenamefont {Berg},\ and\ \citenamefont
  {Levin}}]{Rudner:2013PRX}%
  \BibitemOpen
  \bibfield  {author} {\bibinfo {author} {\bibnamefont {Rudner}, \bibfnamefont
  {Mark~S}}, \bibinfo {author} {\bibfnamefont {Netanel~H}\ \bibnamefont
  {Lindner}}, \bibinfo {author} {\bibfnamefont {Erez}\ \bibnamefont {Berg}}, \
  and\ \bibinfo {author} {\bibfnamefont {Michael}\ \bibnamefont {Levin}}}
  (\bibinfo {year} {2013}),\ \bibfield  {title} {\enquote {\bibinfo {title}
  {Anomalous edge states and the bulk-edge correspondence for periodically
  driven two-dimensional systems},}\ }\href
  {https://doi.org/10.1103/physrevx.3.031005} {\bibfield  {journal} {\bibinfo
  {journal} {Phys. Rev. X}\ }\textbf {\bibinfo {volume} {3}}~(\bibinfo {number}
  {3}),\ \bibinfo {pages} {031005}}\BibitemShut {NoStop}%
\bibitem [{\citenamefont {Rudner}\ and\ \citenamefont
  {Levitov}(2009)}]{rudner2009topological}%
  \BibitemOpen
  \bibfield  {author} {\bibinfo {author} {\bibnamefont {Rudner}, \bibfnamefont
  {MS}}, \ and\ \bibinfo {author} {\bibfnamefont {LS}~\bibnamefont {Levitov}}}
  (\bibinfo {year} {2009}),\ \bibfield  {title} {\enquote {\bibinfo {title}
  {Topological transition in a non-{H}ermitian quantum walk},}\ }\href
  {http://link.aps.org/doi/10.1103/PhysRevLett.102.065703} {\bibfield
  {journal} {\bibinfo  {journal} {Phys. Rev. Lett.}\ }\textbf {\bibinfo
  {volume} {102}}~(\bibinfo {number} {6}),\ \bibinfo {pages}
  {065703}}\BibitemShut {NoStop}%
\bibitem [{\citenamefont {Ruesink}\ \emph {et~al.}(2016)\citenamefont
  {Ruesink}, \citenamefont {Miri}, \citenamefont {Al{\`u}},\ and\ \citenamefont
  {Verhagen}}]{alu2016nonreciprocity}%
  \BibitemOpen
  \bibfield  {author} {\bibinfo {author} {\bibnamefont {Ruesink}, \bibfnamefont
  {Freek}}, \bibinfo {author} {\bibfnamefont {Mohammad-Ali}\ \bibnamefont
  {Miri}}, \bibinfo {author} {\bibfnamefont {Andrea}\ \bibnamefont {Al{\`u}}},
  \ and\ \bibinfo {author} {\bibfnamefont {Ewold}\ \bibnamefont {Verhagen}}}
  (\bibinfo {year} {2016}),\ \bibfield  {title} {\enquote {\bibinfo {title}
  {Nonreciprocity and magnetic-free isolation based on optomechanical
  interactions},}\ }\href {https://doi.org/10.1038/ncomms13662} {\bibfield
  {journal} {\bibinfo  {journal} {Nat. Commun.}\ }\textbf {\bibinfo {volume}
  {7}},\ \bibinfo {pages} {13662}}\BibitemShut {NoStop}%
\bibitem [{\citenamefont {R{\"u}ter}\ \emph {et~al.}(2010)\citenamefont
  {R{\"u}ter}, \citenamefont {Makris}, \citenamefont {El-Ganainy},
  \citenamefont {Christodoulides}, \citenamefont {Segev},\ and\ \citenamefont
  {Kip}}]{ruter2010observation}%
  \BibitemOpen
  \bibfield  {author} {\bibinfo {author} {\bibnamefont {R{\"u}ter},
  \bibfnamefont {Christian~E}}, \bibinfo {author} {\bibfnamefont
  {Konstantinos~G}\ \bibnamefont {Makris}}, \bibinfo {author} {\bibfnamefont
  {Ramy}\ \bibnamefont {El-Ganainy}}, \bibinfo {author} {\bibfnamefont
  {Demetrios~N}\ \bibnamefont {Christodoulides}}, \bibinfo {author}
  {\bibfnamefont {Mordechai}\ \bibnamefont {Segev}}, \ and\ \bibinfo {author}
  {\bibfnamefont {Detlef}\ \bibnamefont {Kip}}} (\bibinfo {year} {2010}),\
  \bibfield  {title} {\enquote {\bibinfo {title} {Observation of parity--time
  symmetry in optics},}\ }\href {https://www.nature.com/articles/nphys1515}
  {\bibfield  {journal} {\bibinfo  {journal} {Nat. Phys.}\ }\textbf {\bibinfo
  {volume} {6}}~(\bibinfo {number} {3}),\ \bibinfo {pages}
  {192--195}}\BibitemShut {NoStop}%
\bibitem [{\citenamefont {Ryu}\ and\ \citenamefont
  {Hatsugai}(2002)}]{Ryu:2002PRL}%
  \BibitemOpen
  \bibfield  {author} {\bibinfo {author} {\bibnamefont {Ryu}, \bibfnamefont
  {Shinsei}}, \ and\ \bibinfo {author} {\bibfnamefont {Yasuhiro}\ \bibnamefont
  {Hatsugai}}} (\bibinfo {year} {2002}),\ \bibfield  {title} {\enquote
  {\bibinfo {title} {Topological origin of zero-energy edge states in
  particle-hole symmetric systems},}\ }\href
  {https://link.aps.org/doi/10.1103/PhysRevLett.89.077002} {\bibfield
  {journal} {\bibinfo  {journal} {Phys. Rev. Lett.}\ }\textbf {\bibinfo
  {volume} {89}},\ \bibinfo {pages} {077002}}\BibitemShut {NoStop}%
\bibitem [{\citenamefont {Ryu}\ \emph {et~al.}(2010)\citenamefont {Ryu},
  \citenamefont {Schnyder}, \citenamefont {Furusaki},\ and\ \citenamefont
  {Ludwig}}]{Ryu:2010NJP}%
  \BibitemOpen
  \bibfield  {author} {\bibinfo {author} {\bibnamefont {Ryu}, \bibfnamefont
  {Shinsei}}, \bibinfo {author} {\bibfnamefont {Andreas~P}\ \bibnamefont
  {Schnyder}}, \bibinfo {author} {\bibfnamefont {Akira}\ \bibnamefont
  {Furusaki}}, \ and\ \bibinfo {author} {\bibfnamefont {Andreas~WW}\
  \bibnamefont {Ludwig}}} (\bibinfo {year} {2010}),\ \bibfield  {title}
  {\enquote {\bibinfo {title} {Topological insulators and superconductors:
  {T}enfold way and dimensional hierarchy},}\ }\href
  {http://iopscience.iop.org/article/10.1088/1367-2630/12/6/065010/meta}
  {\bibfield  {journal} {\bibinfo  {journal} {New J. Phys.}\ }\textbf {\bibinfo
  {volume} {12}},\ \bibinfo {pages} {065010}}\BibitemShut {NoStop}%
\bibitem [{\citenamefont {Saei Ghareh~Naz}\ \emph {et~al.}(2018)\citenamefont
  {Saei Ghareh~Naz}, \citenamefont {Fulga}, \citenamefont {Ma}, \citenamefont
  {Schmidt},\ and\ \citenamefont {van~den Brink}}]{Naz:2017arXiv}%
  \BibitemOpen
  \bibfield  {author} {\bibinfo {author} {\bibnamefont {Saei Ghareh~Naz},
  \bibfnamefont {Ehsan}}, \bibinfo {author} {\bibfnamefont {Ion~Cosma}\
  \bibnamefont {Fulga}}, \bibinfo {author} {\bibfnamefont {Libo}\ \bibnamefont
  {Ma}}, \bibinfo {author} {\bibfnamefont {Oliver~G.}\ \bibnamefont {Schmidt}},
  \ and\ \bibinfo {author} {\bibfnamefont {Jeroen}\ \bibnamefont {van~den
  Brink}}} (\bibinfo {year} {2018}),\ \bibfield  {title} {\enquote {\bibinfo
  {title} {Topological phase transition in a stretchable photonic crystal},}\
  }\href {https://link.aps.org/doi/10.1103/PhysRevA.98.033830} {\bibfield
  {journal} {\bibinfo  {journal} {Phys. Rev. A}\ }\textbf {\bibinfo {volume}
  {98}},\ \bibinfo {pages} {033830}}\BibitemShut {NoStop}%
\bibitem [{\citenamefont {Saffman}\ \emph {et~al.}(2010)\citenamefont
  {Saffman}, \citenamefont {Walker},\ and\ \citenamefont
  {M{\o}lmer}}]{Saffman:RMP2010}%
  \BibitemOpen
  \bibfield  {author} {\bibinfo {author} {\bibnamefont {Saffman}, \bibfnamefont
  {Mark}}, \bibinfo {author} {\bibfnamefont {Thad~G}\ \bibnamefont {Walker}}, \
  and\ \bibinfo {author} {\bibfnamefont {Klaus}\ \bibnamefont {M{\o}lmer}}}
  (\bibinfo {year} {2010}),\ \bibfield  {title} {\enquote {\bibinfo {title}
  {Quantum information with {R}ydberg atoms},}\ }\href
  {https://journals.aps.org/rmp/abstract/10.1103/RevModPhys.82.2313} {\bibfield
   {journal} {\bibinfo  {journal} {Rev. Mod. Phys.}\ }\textbf {\bibinfo
  {volume} {82}}~(\bibinfo {number} {3}),\ \bibinfo {pages} {2313}}\BibitemShut
  {NoStop}%
\bibitem [{\citenamefont {Sala}\ \emph {et~al.}(2015)\citenamefont {Sala},
  \citenamefont {Solnyshkov}, \citenamefont {Carusotto}, \citenamefont
  {Jacqmin}, \citenamefont {Lema{\^\i}tre}, \citenamefont {Ter{\c{c}}as},
  \citenamefont {Nalitov}, \citenamefont {Abbarchi}, \citenamefont {Galopin},
  \citenamefont {Sagnes} \emph {et~al.}}]{Sala:PRX2015}%
  \BibitemOpen
  \bibfield  {author} {\bibinfo {author} {\bibnamefont {Sala}, \bibfnamefont
  {VG}}, \bibinfo {author} {\bibfnamefont {DD}~\bibnamefont {Solnyshkov}},
  \bibinfo {author} {\bibfnamefont {I}~\bibnamefont {Carusotto}}, \bibinfo
  {author} {\bibfnamefont {T}~\bibnamefont {Jacqmin}}, \bibinfo {author}
  {\bibfnamefont {A}~\bibnamefont {Lema{\^\i}tre}}, \bibinfo {author}
  {\bibfnamefont {H}~\bibnamefont {Ter{\c{c}}as}}, \bibinfo {author}
  {\bibfnamefont {A}~\bibnamefont {Nalitov}}, \bibinfo {author} {\bibfnamefont
  {M}~\bibnamefont {Abbarchi}}, \bibinfo {author} {\bibfnamefont
  {E}~\bibnamefont {Galopin}}, \bibinfo {author} {\bibfnamefont
  {I}~\bibnamefont {Sagnes}},  \emph {et~al.}} (\bibinfo {year} {2015}),\
  \bibfield  {title} {\enquote {\bibinfo {title} {Spin-orbit coupling for
  photons and polaritons in microstructures},}\ }\href
  {https://journals.aps.org/prx/abstract/10.1103/PhysRevX.5.011034} {\bibfield
  {journal} {\bibinfo  {journal} {Phys. Rev. X}\ }\textbf {\bibinfo {volume}
  {5}}~(\bibinfo {number} {1}),\ \bibinfo {pages} {011034}}\BibitemShut
  {NoStop}%
\bibitem [{\citenamefont {Saleh}\ and\ \citenamefont
  {Teich}(2007)}]{saleh_book}%
  \BibitemOpen
  \bibfield  {author} {\bibinfo {author} {\bibnamefont {Saleh}, \bibfnamefont
  {BEA}}, \ and\ \bibinfo {author} {\bibfnamefont {M.C.}\ \bibnamefont
  {Teich}}} (\bibinfo {year} {2007}),\ \href
  {https://books.google.it/books?id=Ve8eAQAAIAAJ} {\emph {\bibinfo {title}
  {Fundamentals of Photonics}}},\ Wiley Series in Pure and Applied Optics\
  (\bibinfo  {publisher} {Wiley},\ \bibinfo {address} {New York})\BibitemShut
  {NoStop}%
\bibitem [{\citenamefont {Salerno}\ \emph {et~al.}(2015)\citenamefont
  {Salerno}, \citenamefont {Ozawa}, \citenamefont {Price},\ and\ \citenamefont
  {Carusotto}}]{Salerno:20152DMat}%
  \BibitemOpen
  \bibfield  {author} {\bibinfo {author} {\bibnamefont {Salerno}, \bibfnamefont
  {Grazia}}, \bibinfo {author} {\bibfnamefont {Tomoki}\ \bibnamefont {Ozawa}},
  \bibinfo {author} {\bibfnamefont {Hannah~M}\ \bibnamefont {Price}}, \ and\
  \bibinfo {author} {\bibfnamefont {Iacopo}\ \bibnamefont {Carusotto}}}
  (\bibinfo {year} {2015}),\ \bibfield  {title} {\enquote {\bibinfo {title}
  {How to directly observe {L}andau levels in driven-dissipative strained
  honeycomb lattices},}\ }\href
  {http://iopscience.iop.org/article/10.1088/2053-1583/2/3/034015} {\bibfield
  {journal} {\bibinfo  {journal} {2D Materials}\ }\textbf {\bibinfo {volume}
  {2}}~(\bibinfo {number} {3}),\ \bibinfo {pages} {034015}}\BibitemShut
  {NoStop}%
\bibitem [{\citenamefont {Salerno}\ \emph {et~al.}(2016)\citenamefont
  {Salerno}, \citenamefont {Ozawa}, \citenamefont {Price},\ and\ \citenamefont
  {Carusotto}}]{Salerno:2016PRB}%
  \BibitemOpen
  \bibfield  {author} {\bibinfo {author} {\bibnamefont {Salerno}, \bibfnamefont
  {Grazia}}, \bibinfo {author} {\bibfnamefont {Tomoki}\ \bibnamefont {Ozawa}},
  \bibinfo {author} {\bibfnamefont {Hannah~M.}\ \bibnamefont {Price}}, \ and\
  \bibinfo {author} {\bibfnamefont {Iacopo}\ \bibnamefont {Carusotto}}}
  (\bibinfo {year} {2016}),\ \bibfield  {title} {\enquote {\bibinfo {title}
  {Floquet topological system based on frequency-modulated classical coupled
  harmonic oscillators},}\ }\href
  {https://link.aps.org/doi/10.1103/PhysRevB.93.085105} {\bibfield  {journal}
  {\bibinfo  {journal} {Phys. Rev. B}\ }\textbf {\bibinfo {volume} {93}},\
  \bibinfo {pages} {085105}}\BibitemShut {NoStop}%
\bibitem [{\citenamefont {Salerno}\ \emph {et~al.}(2017)\citenamefont
  {Salerno}, \citenamefont {Ozawa}, \citenamefont {Price},\ and\ \citenamefont
  {Carusotto}}]{Salerno:2017PRB}%
  \BibitemOpen
  \bibfield  {author} {\bibinfo {author} {\bibnamefont {Salerno}, \bibfnamefont
  {Grazia}}, \bibinfo {author} {\bibfnamefont {Tomoki}\ \bibnamefont {Ozawa}},
  \bibinfo {author} {\bibfnamefont {Hannah~M.}\ \bibnamefont {Price}}, \ and\
  \bibinfo {author} {\bibfnamefont {Iacopo}\ \bibnamefont {Carusotto}}}
  (\bibinfo {year} {2017}),\ \bibfield  {title} {\enquote {\bibinfo {title}
  {Propagating edge states in strained honeycomb lattices},}\ }\href
  {https://link.aps.org/doi/10.1103/PhysRevB.95.245418} {\bibfield  {journal}
  {\bibinfo  {journal} {Phys. Rev. B}\ }\textbf {\bibinfo {volume} {95}},\
  \bibinfo {pages} {245418}}\BibitemShut {NoStop}%
\bibitem [{\citenamefont {Sayrin}\ \emph {et~al.}(2015)\citenamefont {Sayrin},
  \citenamefont {Junge}, \citenamefont {Mitsch}, \citenamefont {Albrecht},
  \citenamefont {O'Shea}, \citenamefont {Schneeweiss}, \citenamefont {Volz},\
  and\ \citenamefont {Rauschenbeutel}}]{Sayrin:2015PRX}%
  \BibitemOpen
  \bibfield  {author} {\bibinfo {author} {\bibnamefont {Sayrin}, \bibfnamefont
  {Cl\'ement}}, \bibinfo {author} {\bibfnamefont {Christian}\ \bibnamefont
  {Junge}}, \bibinfo {author} {\bibfnamefont {Rudolf}\ \bibnamefont {Mitsch}},
  \bibinfo {author} {\bibfnamefont {Bernhard}\ \bibnamefont {Albrecht}},
  \bibinfo {author} {\bibfnamefont {Danny}\ \bibnamefont {O'Shea}}, \bibinfo
  {author} {\bibfnamefont {Philipp}\ \bibnamefont {Schneeweiss}}, \bibinfo
  {author} {\bibfnamefont {J\"urgen}\ \bibnamefont {Volz}}, \ and\ \bibinfo
  {author} {\bibfnamefont {Arno}\ \bibnamefont {Rauschenbeutel}}} (\bibinfo
  {year} {2015}),\ \bibfield  {title} {\enquote {\bibinfo {title} {Nanophotonic
  optical isolator controlled by the internal state of cold atoms},}\ }\href
  {https://link.aps.org/doi/10.1103/PhysRevX.5.041036} {\bibfield  {journal}
  {\bibinfo  {journal} {Phys. Rev. X}\ }\textbf {\bibinfo {volume} {5}},\
  \bibinfo {pages} {041036}}\BibitemShut {NoStop}%
\bibitem [{\citenamefont {Schine}\ \emph {et~al.}(2016)\citenamefont {Schine},
  \citenamefont {Ryou}, \citenamefont {Gromov}, \citenamefont {Sommer},\ and\
  \citenamefont {Simon}}]{Schine:2016Nature}%
  \BibitemOpen
  \bibfield  {author} {\bibinfo {author} {\bibnamefont {Schine}, \bibfnamefont
  {Nathan}}, \bibinfo {author} {\bibfnamefont {Albert}\ \bibnamefont {Ryou}},
  \bibinfo {author} {\bibfnamefont {Andrey}\ \bibnamefont {Gromov}}, \bibinfo
  {author} {\bibfnamefont {Ariel}\ \bibnamefont {Sommer}}, \ and\ \bibinfo
  {author} {\bibfnamefont {Jonathan}\ \bibnamefont {Simon}}} (\bibinfo {year}
  {2016}),\ \bibfield  {title} {\enquote {\bibinfo {title} {Synthetic {L}andau
  levels for photons},}\ }\href
  {http://www.nature.com/nature/journal/v534/n7609/full/nature17943.html}
  {\bibfield  {journal} {\bibinfo  {journal} {Nature}\ }\textbf {\bibinfo
  {volume} {534}},\ \bibinfo {pages} {671--675}}\BibitemShut {NoStop}%
\bibitem [{\citenamefont {Schmidt}\ \emph {et~al.}(2015)\citenamefont
  {Schmidt}, \citenamefont {Kessler}, \citenamefont {Peano}, \citenamefont
  {Painter},\ and\ \citenamefont {Marquardt}}]{Schmidt:2015Opt}%
  \BibitemOpen
  \bibfield  {author} {\bibinfo {author} {\bibnamefont {Schmidt}, \bibfnamefont
  {M}}, \bibinfo {author} {\bibfnamefont {S.}~\bibnamefont {Kessler}}, \bibinfo
  {author} {\bibfnamefont {V.}~\bibnamefont {Peano}}, \bibinfo {author}
  {\bibfnamefont {O.}~\bibnamefont {Painter}}, \ and\ \bibinfo {author}
  {\bibfnamefont {F.}~\bibnamefont {Marquardt}}} (\bibinfo {year} {2015}),\
  \bibfield  {title} {\enquote {\bibinfo {title} {Optomechanical creation of
  magnetic fields for photons on a lattice},}\ }\href
  {http://www.osapublishing.org/optica/abstract.cfm?URI=optica-2-7-635}
  {\bibfield  {journal} {\bibinfo  {journal} {Optica}\ }\textbf {\bibinfo
  {volume} {2}}~(\bibinfo {number} {7}),\ \bibinfo {pages}
  {635--641}}\BibitemShut {NoStop}%
\bibitem [{\citenamefont {Schnyder}\ \emph {et~al.}(2009)\citenamefont
  {Schnyder}, \citenamefont {Ryu}, \citenamefont {Furusaki}, \citenamefont
  {Ludwig}, \citenamefont {Lebedev},\ and\ \citenamefont
  {Feigel’man}}]{Schnyder:2009AIP}%
  \BibitemOpen
  \bibfield  {author} {\bibinfo {author} {\bibnamefont {Schnyder},
  \bibfnamefont {Andreas~P}}, \bibinfo {author} {\bibfnamefont {Shinsei}\
  \bibnamefont {Ryu}}, \bibinfo {author} {\bibfnamefont {Akira}\ \bibnamefont
  {Furusaki}}, \bibinfo {author} {\bibfnamefont {Andreas~WW}\ \bibnamefont
  {Ludwig}}, \bibinfo {author} {\bibfnamefont {Vladimir}\ \bibnamefont
  {Lebedev}}, \ and\ \bibinfo {author} {\bibfnamefont {Mikhail}\ \bibnamefont
  {Feigel’man}}} (\bibinfo {year} {2009}),\ \bibfield  {title} {\enquote
  {\bibinfo {title} {Classification of topological insulators and
  superconductors},}\ }\bibfield  {booktitle} {\emph {\bibinfo {booktitle} {AIP
  Conference Proceedings}},\ }\href
  {http://aip.scitation.org/doi/abs/10.1063/1.3149481} {\ \textbf {\bibinfo
  {volume} {1134}}~(\bibinfo {number} {1}),\ \bibinfo {pages}
  {10--21}}\BibitemShut {NoStop}%
\bibitem [{\citenamefont {Schoelkopf}\ and\ \citenamefont
  {Girvin}(2008)}]{Schoelkopf:2008Nature}%
  \BibitemOpen
  \bibfield  {author} {\bibinfo {author} {\bibnamefont {Schoelkopf},
  \bibfnamefont {RJ}}, \ and\ \bibinfo {author} {\bibfnamefont
  {SM}~\bibnamefont {Girvin}}} (\bibinfo {year} {2008}),\ \bibfield  {title}
  {\enquote {\bibinfo {title} {Wiring up quantum systems},}\ }\href
  {http://www.nature.com/nature/journal/v451/n7179/full/451664a.html}
  {\bibfield  {journal} {\bibinfo  {journal} {Nature}\ }\textbf {\bibinfo
  {volume} {451}}~(\bibinfo {number} {7179}),\ \bibinfo {pages}
  {664--669}}\BibitemShut {NoStop}%
\bibitem [{\citenamefont {Schomerus}(2013)}]{Schomerus:2013OptLett}%
  \BibitemOpen
  \bibfield  {author} {\bibinfo {author} {\bibnamefont {Schomerus},
  \bibfnamefont {Henning}}} (\bibinfo {year} {2013}),\ \bibfield  {title}
  {\enquote {\bibinfo {title} {Topologically protected midgap states in complex
  photonic lattices},}\ }\href
  {https://www.osapublishing.org/ol/abstract.cfm?uri=ol-38-11-1912} {\bibfield
  {journal} {\bibinfo  {journal} {Opt. Lett.}\ }\textbf {\bibinfo {volume}
  {38}}~(\bibinfo {number} {11}),\ \bibinfo {pages} {1912--1914}}\BibitemShut
  {NoStop}%
\bibitem [{\citenamefont {Schomerus}\ and\ \citenamefont
  {Halpern}(2013)}]{Schomerus:2013PRL}%
  \BibitemOpen
  \bibfield  {author} {\bibinfo {author} {\bibnamefont {Schomerus},
  \bibfnamefont {Henning}}, \ and\ \bibinfo {author} {\bibfnamefont
  {Nicole~Yunger}\ \bibnamefont {Halpern}}} (\bibinfo {year} {2013}),\
  \bibfield  {title} {\enquote {\bibinfo {title} {{Parity anomaly and
  Landau-level lasing in strained photonic honeycomb lattices}},}\ }\href
  {https://link.aps.org/doi/10.1103/PhysRevLett.110.013903} {\bibfield
  {journal} {\bibinfo  {journal} {Phys. Rev. Lett.}\ }\textbf {\bibinfo
  {volume} {110}}~(\bibinfo {number} {1}),\ \bibinfo {pages}
  {013903}}\BibitemShut {NoStop}%
\bibitem [{\citenamefont {Schreiber}\ \emph {et~al.}(2012)\citenamefont
  {Schreiber}, \citenamefont {G{\'a}bris}, \citenamefont {Rohde}, \citenamefont
  {Laiho}, \citenamefont {{\v S}tefa{\v n}{\'a}k}, \citenamefont {Poto{\v
  c}ek}, \citenamefont {Hamilton}, \citenamefont {Jex},\ and\ \citenamefont
  {Silberhorn}}]{Schreiber:Science2012}%
  \BibitemOpen
  \bibfield  {author} {\bibinfo {author} {\bibnamefont {Schreiber},
  \bibfnamefont {Andreas}}, \bibinfo {author} {\bibfnamefont {Aur{\'e}l}\
  \bibnamefont {G{\'a}bris}}, \bibinfo {author} {\bibfnamefont {Peter~P.}\
  \bibnamefont {Rohde}}, \bibinfo {author} {\bibfnamefont {Kaisa}\ \bibnamefont
  {Laiho}}, \bibinfo {author} {\bibfnamefont {Martin}\ \bibnamefont {{\v
  S}tefa{\v n}{\'a}k}}, \bibinfo {author} {\bibfnamefont {V{\'a}clav}\
  \bibnamefont {Poto{\v c}ek}}, \bibinfo {author} {\bibfnamefont {Craig}\
  \bibnamefont {Hamilton}}, \bibinfo {author} {\bibfnamefont {Igor}\
  \bibnamefont {Jex}}, \ and\ \bibinfo {author} {\bibfnamefont {Christine}\
  \bibnamefont {Silberhorn}}} (\bibinfo {year} {2012}),\ \bibfield  {title}
  {\enquote {\bibinfo {title} {A 2{D} quantum walk simulation of two-particle
  dynamics},}\ }\href {http://science.sciencemag.org/content/336/6077/55}
  {\bibfield  {journal} {\bibinfo  {journal} {Science}\ }\textbf {\bibinfo
  {volume} {336}}~(\bibinfo {number} {6077}),\ \bibinfo {pages}
  {55--58}}\BibitemShut {NoStop}%
\bibitem [{\citenamefont {Schwartz}\ and\ \citenamefont
  {Fischer}(2013)}]{Schwartz:2013OptExp}%
  \BibitemOpen
  \bibfield  {author} {\bibinfo {author} {\bibnamefont {Schwartz},
  \bibfnamefont {Alon}}, \ and\ \bibinfo {author} {\bibfnamefont {Baruch}\
  \bibnamefont {Fischer}}} (\bibinfo {year} {2013}),\ \bibfield  {title}
  {\enquote {\bibinfo {title} {Laser mode hyper-combs},}\ }\href
  {https://www.osapublishing.org/oe/abstract.cfm?uri=oe-21-5-6196} {\bibfield
  {journal} {\bibinfo  {journal} {Opt. Express}\ }\textbf {\bibinfo {volume}
  {21}}~(\bibinfo {number} {5}),\ \bibinfo {pages} {6196--6204}}\BibitemShut
  {NoStop}%
\bibitem [{\citenamefont {Schwartz}\ \emph {et~al.}(2007)\citenamefont
  {Schwartz}, \citenamefont {Bartal}, \citenamefont {Fishman},\ and\
  \citenamefont {Segev}}]{schwartz2007transport}%
  \BibitemOpen
  \bibfield  {author} {\bibinfo {author} {\bibnamefont {Schwartz},
  \bibfnamefont {Tal}}, \bibinfo {author} {\bibfnamefont {Guy}\ \bibnamefont
  {Bartal}}, \bibinfo {author} {\bibfnamefont {Shmuel}\ \bibnamefont
  {Fishman}}, \ and\ \bibinfo {author} {\bibfnamefont {Mordechai}\ \bibnamefont
  {Segev}}} (\bibinfo {year} {2007}),\ \bibfield  {title} {\enquote {\bibinfo
  {title} {Transport and {A}nderson localization in disordered two-dimensional
  photonic lattices},}\ }\href {https://www.nature.com/articles/nature05623}
  {\bibfield  {journal} {\bibinfo  {journal} {Nature}\ }\textbf {\bibinfo
  {volume} {446}}~(\bibinfo {number} {7131}),\ \bibinfo {pages}
  {52--55}}\BibitemShut {NoStop}%
\bibitem [{\citenamefont {Secl\'i}(2017)}]{Secli:MSc}%
  \BibitemOpen
  \bibfield  {author} {\bibinfo {author} {\bibnamefont {Secl\'i}, \bibfnamefont
  {Matteo}}} (\bibinfo {year} {2017}),\ \emph {\bibinfo {title} {Edge State
  Lasing in a 2{D} Topological Photonic System}},\ \href@noop {} {Master's
  thesis}\ (\bibinfo  {school} {University of Trento})\BibitemShut {NoStop}%
\bibitem [{\citenamefont {Segev}\ \emph {et~al.}(1992)\citenamefont {Segev},
  \citenamefont {Crosignani}, \citenamefont {Yariv},\ and\ \citenamefont
  {Fischer}}]{Segev:PRL1992}%
  \BibitemOpen
  \bibfield  {author} {\bibinfo {author} {\bibnamefont {Segev}, \bibfnamefont
  {Mordechai}}, \bibinfo {author} {\bibfnamefont {Bruno}\ \bibnamefont
  {Crosignani}}, \bibinfo {author} {\bibfnamefont {Amnon}\ \bibnamefont
  {Yariv}}, \ and\ \bibinfo {author} {\bibfnamefont {Baruch}\ \bibnamefont
  {Fischer}}} (\bibinfo {year} {1992}),\ \bibfield  {title} {\enquote {\bibinfo
  {title} {Spatial solitons in photorefractive media},}\ }\href
  {https://journals.aps.org/prl/abstract/10.1103/PhysRevLett.68.923} {\bibfield
   {journal} {\bibinfo  {journal} {Phys. Rev. Lett.}\ }\textbf {\bibinfo
  {volume} {68}}~(\bibinfo {number} {7}),\ \bibinfo {pages} {923}}\BibitemShut
  {NoStop}%
\bibitem [{\citenamefont {Serdyukov}\ \emph {et~al.}(2001)\citenamefont
  {Serdyukov}, \citenamefont {Semchenko}, \citenamefont {Tertyakov},\ and\
  \citenamefont {Sihvola}}]{serdyukov2001electromagnetics}%
  \BibitemOpen
  \bibfield  {author} {\bibinfo {author} {\bibnamefont {Serdyukov},
  \bibfnamefont {A}}, \bibinfo {author} {\bibfnamefont {Igor}\ \bibnamefont
  {Semchenko}}, \bibinfo {author} {\bibfnamefont {S}~\bibnamefont {Tertyakov}},
  \ and\ \bibinfo {author} {\bibfnamefont {A}~\bibnamefont {Sihvola}}}
  (\bibinfo {year} {2001}),\ \href@noop {} {\emph {\bibinfo {title}
  {Electromagnetics of bi-anisotropic materials-Theory and Application}}},\
  Vol.~\bibinfo {volume} {11}\ (\bibinfo  {publisher} {Gordon and Breach
  Science Publishers},\ \bibinfo {address} {New York})\BibitemShut {NoStop}%
\bibitem [{\citenamefont {Shelby}\ \emph {et~al.}(2001)\citenamefont {Shelby},
  \citenamefont {Smith},\ and\ \citenamefont {Schultz}}]{Shelby:2001Sci}%
  \BibitemOpen
  \bibfield  {author} {\bibinfo {author} {\bibnamefont {Shelby}, \bibfnamefont
  {R~A}}, \bibinfo {author} {\bibfnamefont {D.~R.}\ \bibnamefont {Smith}}, \
  and\ \bibinfo {author} {\bibfnamefont {S.}~\bibnamefont {Schultz}}} (\bibinfo
  {year} {2001}),\ \bibfield  {title} {\enquote {\bibinfo {title} {Experimental
  verification of a negative index of refraction},}\ }\href
  {http://science.sciencemag.org/content/292/5514/77} {\bibfield  {journal}
  {\bibinfo  {journal} {Science}\ }\textbf {\bibinfo {volume} {292}}~(\bibinfo
  {number} {5514}),\ \bibinfo {pages} {77--79}}\BibitemShut {NoStop}%
\bibitem [{\citenamefont {Shen}\ \emph {et~al.}(2018)\citenamefont {Shen},
  \citenamefont {Zhen},\ and\ \citenamefont {Fu}}]{shen2018topological}%
  \BibitemOpen
  \bibfield  {author} {\bibinfo {author} {\bibnamefont {Shen}, \bibfnamefont
  {Huitao}}, \bibinfo {author} {\bibfnamefont {Bo}~\bibnamefont {Zhen}}, \ and\
  \bibinfo {author} {\bibfnamefont {Liang}\ \bibnamefont {Fu}}} (\bibinfo
  {year} {2018}),\ \bibfield  {title} {\enquote {\bibinfo {title} {Topological
  band theory for non-{H}ermitian {H}amiltonians},}\ }\href
  {https://journals.aps.org/prl/abstract/10.1103/PhysRevLett.120.146402}
  {\bibfield  {journal} {\bibinfo  {journal} {Phys. Rev. Lett.}\ }\textbf
  {\bibinfo {volume} {120}}~(\bibinfo {number} {14}),\ \bibinfo {pages}
  {146402}}\BibitemShut {NoStop}%
\bibitem [{\citenamefont {Shen}\ \emph {et~al.}(2015)\citenamefont {Shen},
  \citenamefont {You}, \citenamefont {Wang},\ and\ \citenamefont
  {Deng}}]{Shen:2015OE}%
  \BibitemOpen
  \bibfield  {author} {\bibinfo {author} {\bibnamefont {Shen}, \bibfnamefont
  {Linfang}}, \bibinfo {author} {\bibfnamefont {Yun}\ \bibnamefont {You}},
  \bibinfo {author} {\bibfnamefont {Zhuoyuan}\ \bibnamefont {Wang}}, \ and\
  \bibinfo {author} {\bibfnamefont {Xiaohua}\ \bibnamefont {Deng}}} (\bibinfo
  {year} {2015}),\ \bibfield  {title} {\enquote {\bibinfo {title}
  {Backscattering-immune one-way surface magnetoplasmons at terahertz
  frequencies},}\ }\href
  {https://www.osapublishing.org/oe/abstract.cfm?uri=oe-23-2-950} {\bibfield
  {journal} {\bibinfo  {journal} {Opt. Express}\ }\textbf {\bibinfo {volume}
  {23}}~(\bibinfo {number} {2}),\ \bibinfo {pages} {950--962}}\BibitemShut
  {NoStop}%
\bibitem [{\citenamefont {Shen}\ \emph {et~al.}(2016)\citenamefont {Shen},
  \citenamefont {Zhang}, \citenamefont {Chen}, \citenamefont {Zou},
  \citenamefont {Xiao}, \citenamefont {Zou}, \citenamefont {Sun}, \citenamefont
  {Guo},\ and\ \citenamefont {Dong}}]{shen2016experimental}%
  \BibitemOpen
  \bibfield  {author} {\bibinfo {author} {\bibnamefont {Shen}, \bibfnamefont
  {Zhen}}, \bibinfo {author} {\bibfnamefont {Yan-Lei}\ \bibnamefont {Zhang}},
  \bibinfo {author} {\bibfnamefont {Yuan}\ \bibnamefont {Chen}}, \bibinfo
  {author} {\bibfnamefont {Chang-Ling}\ \bibnamefont {Zou}}, \bibinfo {author}
  {\bibfnamefont {Yun-Feng}\ \bibnamefont {Xiao}}, \bibinfo {author}
  {\bibfnamefont {Xu-Bo}\ \bibnamefont {Zou}}, \bibinfo {author} {\bibfnamefont
  {Fang-Wen}\ \bibnamefont {Sun}}, \bibinfo {author} {\bibfnamefont
  {Guang-Can}\ \bibnamefont {Guo}}, \ and\ \bibinfo {author} {\bibfnamefont
  {Chun-Hua}\ \bibnamefont {Dong}}} (\bibinfo {year} {2016}),\ \bibfield
  {title} {\enquote {\bibinfo {title} {Experimental realization of
  optomechanically induced non-reciprocity},}\ }\href
  {https://doi.org/10.1038/nphoton.2016.161} {\bibfield  {journal} {\bibinfo
  {journal} {Nat. Photonics}\ }\textbf {\bibinfo {volume} {10}},\ \bibinfo
  {pages} {657--661}}\BibitemShut {NoStop}%
\bibitem [{\citenamefont {Sheng}\ \emph {et~al.}(2006)\citenamefont {Sheng},
  \citenamefont {Weng}, \citenamefont {Sheng},\ and\ \citenamefont
  {Haldane}}]{Sheng:2006PRL}%
  \BibitemOpen
  \bibfield  {author} {\bibinfo {author} {\bibnamefont {Sheng}, \bibfnamefont
  {D~N}}, \bibinfo {author} {\bibfnamefont {Z.~Y.}\ \bibnamefont {Weng}},
  \bibinfo {author} {\bibfnamefont {L.}~\bibnamefont {Sheng}}, \ and\ \bibinfo
  {author} {\bibfnamefont {F.~D.~M.}\ \bibnamefont {Haldane}}} (\bibinfo {year}
  {2006}),\ \bibfield  {title} {\enquote {\bibinfo {title} {Quantum spin-{H}all
  effect and topologically invariant {C}hern numbers},}\ }\href
  {https://link.aps.org/doi/10.1103/PhysRevLett.97.036808} {\bibfield
  {journal} {\bibinfo  {journal} {Phys. Rev. Lett.}\ }\textbf {\bibinfo
  {volume} {97}},\ \bibinfo {pages} {036808}}\BibitemShut {NoStop}%
\bibitem [{\citenamefont {Shi}\ \emph {et~al.}(2017)\citenamefont {Shi},
  \citenamefont {Kimble},\ and\ \citenamefont {Cirac}}]{Shi:2017PNAS}%
  \BibitemOpen
  \bibfield  {author} {\bibinfo {author} {\bibnamefont {Shi}, \bibfnamefont
  {T}}, \bibinfo {author} {\bibfnamefont {HJ}~\bibnamefont {Kimble}}, \ and\
  \bibinfo {author} {\bibfnamefont {JI}~\bibnamefont {Cirac}}} (\bibinfo {year}
  {2017}),\ \bibfield  {title} {\enquote {\bibinfo {title} {Topological
  phenomena in classical optical networks},}\ }\href
  {http://www.pnas.org/content/114/43/E8967} {\bibfield  {journal} {\bibinfo
  {journal} {Proc. Natl. Acad. Scie. U.S.A.}\ }\textbf {\bibinfo {volume}
  {114}},\ \bibinfo {pages} {E8967--E8976}}\BibitemShut {NoStop}%
\bibitem [{\citenamefont {Shi}\ \emph {et~al.}(2015)\citenamefont {Shi},
  \citenamefont {Yu},\ and\ \citenamefont {Fan}}]{Shi:NatPhot2015}%
  \BibitemOpen
  \bibfield  {author} {\bibinfo {author} {\bibnamefont {Shi}, \bibfnamefont
  {Yu}}, \bibinfo {author} {\bibfnamefont {Zongfu}\ \bibnamefont {Yu}}, \ and\
  \bibinfo {author} {\bibfnamefont {Shanhui}\ \bibnamefont {Fan}}} (\bibinfo
  {year} {2015}),\ \bibfield  {title} {\enquote {\bibinfo {title} {Limitations
  of nonlinear optical isolators due to dynamic reciprocity},}\ }\href
  {https://www.nature.com/articles/nphoton.2015.79} {\bibfield  {journal}
  {\bibinfo  {journal} {Nat. Photonics}\ }\textbf {\bibinfo {volume}
  {9}}~(\bibinfo {number} {6}),\ \bibinfo {pages} {388--392}}\BibitemShut
  {NoStop}%
\bibitem [{\citenamefont {Shindou}\ \emph
  {et~al.}(2013{\natexlab{a}})\citenamefont {Shindou}, \citenamefont
  {Matsumoto}, \citenamefont {Murakami},\ and\ \citenamefont
  {Ohe}}]{Shindou:2013PRBb}%
  \BibitemOpen
  \bibfield  {author} {\bibinfo {author} {\bibnamefont {Shindou}, \bibfnamefont
  {Ryuichi}}, \bibinfo {author} {\bibfnamefont {Ryo}\ \bibnamefont
  {Matsumoto}}, \bibinfo {author} {\bibfnamefont {Shuichi}\ \bibnamefont
  {Murakami}}, \ and\ \bibinfo {author} {\bibfnamefont {Jun-ichiro}\
  \bibnamefont {Ohe}}} (\bibinfo {year} {2013}{\natexlab{a}}),\ \bibfield
  {title} {\enquote {\bibinfo {title} {Topological chiral magnonic edge mode in
  a magnonic crystal},}\ }\href
  {https://link.aps.org/doi/10.1103/PhysRevB.87.174427} {\bibfield  {journal}
  {\bibinfo  {journal} {Phys. Rev. B}\ }\textbf {\bibinfo {volume} {87}},\
  \bibinfo {pages} {174427}}\BibitemShut {NoStop}%
\bibitem [{\citenamefont {Shindou}\ and\ \citenamefont
  {Ohe}(2014)}]{Shindou:2014PRB}%
  \BibitemOpen
  \bibfield  {author} {\bibinfo {author} {\bibnamefont {Shindou}, \bibfnamefont
  {Ryuichi}}, \ and\ \bibinfo {author} {\bibfnamefont {Jun-ichiro}\
  \bibnamefont {Ohe}}} (\bibinfo {year} {2014}),\ \bibfield  {title} {\enquote
  {\bibinfo {title} {Magnetostatic wave analog of integer quantum {H}all state
  in patterned magnetic films},}\ }\href
  {https://link.aps.org/doi/10.1103/PhysRevB.89.054412} {\bibfield  {journal}
  {\bibinfo  {journal} {Phys. Rev. B}\ }\textbf {\bibinfo {volume} {89}},\
  \bibinfo {pages} {054412}}\BibitemShut {NoStop}%
\bibitem [{\citenamefont {Shindou}\ \emph
  {et~al.}(2013{\natexlab{b}})\citenamefont {Shindou}, \citenamefont {Ohe},
  \citenamefont {Matsumoto}, \citenamefont {Murakami},\ and\ \citenamefont
  {Saitoh}}]{Shindou:2013PRBa}%
  \BibitemOpen
  \bibfield  {author} {\bibinfo {author} {\bibnamefont {Shindou}, \bibfnamefont
  {Ryuichi}}, \bibinfo {author} {\bibfnamefont {Jun-ichiro}\ \bibnamefont
  {Ohe}}, \bibinfo {author} {\bibfnamefont {Ryo}\ \bibnamefont {Matsumoto}},
  \bibinfo {author} {\bibfnamefont {Shuichi}\ \bibnamefont {Murakami}}, \ and\
  \bibinfo {author} {\bibfnamefont {Eiji}\ \bibnamefont {Saitoh}}} (\bibinfo
  {year} {2013}{\natexlab{b}}),\ \bibfield  {title} {\enquote {\bibinfo {title}
  {Chiral spin-wave edge modes in dipolar magnetic thin films},}\ }\href
  {https://link.aps.org/doi/10.1103/PhysRevB.87.174402} {\bibfield  {journal}
  {\bibinfo  {journal} {Phys. Rev. B}\ }\textbf {\bibinfo {volume} {87}},\
  \bibinfo {pages} {174402}}\BibitemShut {NoStop}%
\bibitem [{\citenamefont {Siegman}(1986)}]{siegman1986lasers}%
  \BibitemOpen
  \bibfield  {author} {\bibinfo {author} {\bibnamefont {Siegman}, \bibfnamefont
  {Anthony~E}}} (\bibinfo {year} {1986}),\ \href@noop {} {\emph {\bibinfo
  {title} {Lasers}}}\ (\bibinfo  {publisher} {University Science Books},\
  \bibinfo {address} {Sausalito, CA})\BibitemShut {NoStop}%
\bibitem [{\citenamefont {Silberberg}\ \emph {et~al.}(2009)\citenamefont
  {Silberberg}, \citenamefont {Lahini}, \citenamefont {Bromberg}, \citenamefont
  {Small},\ and\ \citenamefont {Morandotti}}]{Silberberg:PRL2009}%
  \BibitemOpen
  \bibfield  {author} {\bibinfo {author} {\bibnamefont {Silberberg},
  \bibfnamefont {Yaron}}, \bibinfo {author} {\bibfnamefont {Yoav}\ \bibnamefont
  {Lahini}}, \bibinfo {author} {\bibfnamefont {Yaron}\ \bibnamefont
  {Bromberg}}, \bibinfo {author} {\bibfnamefont {Eran}\ \bibnamefont {Small}},
  \ and\ \bibinfo {author} {\bibfnamefont {Roberto}\ \bibnamefont
  {Morandotti}}} (\bibinfo {year} {2009}),\ \bibfield  {title} {\enquote
  {\bibinfo {title} {Universal correlations in a nonlinear periodic 1{D}
  system},}\ }\href
  {https://journals.aps.org/prl/abstract/10.1103/PhysRevLett.102.233904}
  {\bibfield  {journal} {\bibinfo  {journal} {Phys. Rev. Lett.}\ }\textbf
  {\bibinfo {volume} {102}}~(\bibinfo {number} {23}),\ \bibinfo {pages}
  {233904}}\BibitemShut {NoStop}%
\bibitem [{\citenamefont {Silveirinha}(2015)}]{Silveirinha:2015PRB}%
  \BibitemOpen
  \bibfield  {author} {\bibinfo {author} {\bibnamefont {Silveirinha},
  \bibfnamefont {M{\'a}rio~G}}} (\bibinfo {year} {2015}),\ \bibfield  {title}
  {\enquote {\bibinfo {title} {Chern invariants for continuous media},}\ }\href
  {https://journals.aps.org/prb/abstract/10.1103/PhysRevB.92.125153} {\bibfield
   {journal} {\bibinfo  {journal} {Phys. Rev. B}\ }\textbf {\bibinfo {volume}
  {92}}~(\bibinfo {number} {12}),\ \bibinfo {pages} {125153}}\BibitemShut
  {NoStop}%
\bibitem [{\citenamefont
  {Silveirinha}(2016{\natexlab{a}})}]{Silveirinha:2016PRB}%
  \BibitemOpen
  \bibfield  {author} {\bibinfo {author} {\bibnamefont {Silveirinha},
  \bibfnamefont {M{\'a}rio~G}}} (\bibinfo {year} {2016}{\natexlab{a}}),\
  \bibfield  {title} {\enquote {\bibinfo {title} {Bulk-edge correspondence for
  topological photonic continua},}\ }\href
  {https://journals.aps.org/prb/abstract/10.1103/PhysRevB.94.205105} {\bibfield
   {journal} {\bibinfo  {journal} {Phys. Rev. B}\ }\textbf {\bibinfo {volume}
  {94}}~(\bibinfo {number} {20}),\ \bibinfo {pages} {205105}}\BibitemShut
  {NoStop}%
\bibitem [{\citenamefont
  {Silveirinha}(2016{\natexlab{b}})}]{Silveirinha:2016PRB-Z2}%
  \BibitemOpen
  \bibfield  {author} {\bibinfo {author} {\bibnamefont {Silveirinha},
  \bibfnamefont {M{\'a}rio~G}}} (\bibinfo {year} {2016}{\natexlab{b}}),\
  \bibfield  {title} {\enquote {\bibinfo {title} {${Z}_{2}$ topological index
  for continuous photonic materials},}\ }\href
  {https://journals.aps.org/prb/abstract/10.1103/PhysRevB.93.075110} {\bibfield
   {journal} {\bibinfo  {journal} {Phys. Rev. B}\ }\textbf {\bibinfo {volume}
  {93}}~(\bibinfo {number} {7}),\ \bibinfo {pages} {075110}}\BibitemShut
  {NoStop}%
\bibitem [{\citenamefont {Simon}(1983)}]{Simon:1983PRL}%
  \BibitemOpen
  \bibfield  {author} {\bibinfo {author} {\bibnamefont {Simon}, \bibfnamefont
  {Barry}}} (\bibinfo {year} {1983}),\ \bibfield  {title} {\enquote {\bibinfo
  {title} {Holonomy, the quantum adiabatic theorem, and {B}erry's phase},}\
  }\href {https://link.aps.org/doi/10.1103/PhysRevLett.51.2167} {\bibfield
  {journal} {\bibinfo  {journal} {Phys. Rev. Lett.}\ }\textbf {\bibinfo
  {volume} {51}},\ \bibinfo {pages} {2167--2170}}\BibitemShut {NoStop}%
\bibitem [{\citenamefont {Sinev}\ \emph {et~al.}(2015)\citenamefont {Sinev},
  \citenamefont {Mukhin}, \citenamefont {Slobozhanyuk}, \citenamefont
  {Poddubny}, \citenamefont {Miroshnichenko}, \citenamefont {Samusev},\ and\
  \citenamefont {Kivshar}}]{Sinev:2015Nanoscale}%
  \BibitemOpen
  \bibfield  {author} {\bibinfo {author} {\bibnamefont {Sinev}, \bibfnamefont
  {Ivan~S}}, \bibinfo {author} {\bibfnamefont {Ivan~S}\ \bibnamefont {Mukhin}},
  \bibinfo {author} {\bibfnamefont {Alexey~P}\ \bibnamefont {Slobozhanyuk}},
  \bibinfo {author} {\bibfnamefont {Alexander~N}\ \bibnamefont {Poddubny}},
  \bibinfo {author} {\bibfnamefont {Andrey~E}\ \bibnamefont {Miroshnichenko}},
  \bibinfo {author} {\bibfnamefont {Anton~K}\ \bibnamefont {Samusev}}, \ and\
  \bibinfo {author} {\bibfnamefont {Yuri~S}\ \bibnamefont {Kivshar}}} (\bibinfo
  {year} {2015}),\ \bibfield  {title} {\enquote {\bibinfo {title} {Mapping
  plasmonic topological states at the nanoscale},}\ }\href
  {http://pubs.rsc.org/en/Content/ArticleLanding/2015/NR/c5nr00231a#!divAbstract}
  {\bibfield  {journal} {\bibinfo  {journal} {Nanoscale}\ }\textbf {\bibinfo
  {volume} {7}}~(\bibinfo {number} {28}),\ \bibinfo {pages}
  {11904--11908}}\BibitemShut {NoStop}%
\bibitem [{\citenamefont {Siroki}\ \emph {et~al.}(2017)\citenamefont {Siroki},
  \citenamefont {Huidobro},\ and\ \citenamefont {Giannini}}]{Siroki:PRB2017}%
  \BibitemOpen
  \bibfield  {author} {\bibinfo {author} {\bibnamefont {Siroki}, \bibfnamefont
  {Gleb}}, \bibinfo {author} {\bibfnamefont {Paloma~A.}\ \bibnamefont
  {Huidobro}}, \ and\ \bibinfo {author} {\bibfnamefont {Vincenzo}\ \bibnamefont
  {Giannini}}} (\bibinfo {year} {2017}),\ \bibfield  {title} {\enquote
  {\bibinfo {title} {{Topological photonics: From crystals to particles}},}\
  }\href {http://link.aps.org/doi/10.1103/PhysRevB.96.041408} {\bibfield
  {journal} {\bibinfo  {journal} {Phys. Rev. B}\ }\textbf {\bibinfo {volume}
  {96}}~(\bibinfo {number} {4}),\ \bibinfo {pages} {041408}}\BibitemShut
  {NoStop}%
\bibitem [{\citenamefont {Skirlo}\ \emph {et~al.}(2015)\citenamefont {Skirlo},
  \citenamefont {Lu}, \citenamefont {Igarashi}, \citenamefont {Yan},
  \citenamefont {Joannopoulos},\ and\ \citenamefont {Solja\ifmmode
  \check{c}\else \v{c}\fi{}i\ifmmode~\acute{c}\else
  \'{c}\fi{}}}]{Skirlo:2015PRL}%
  \BibitemOpen
  \bibfield  {author} {\bibinfo {author} {\bibnamefont {Skirlo}, \bibfnamefont
  {Scott~A}}, \bibinfo {author} {\bibfnamefont {Ling}\ \bibnamefont {Lu}},
  \bibinfo {author} {\bibfnamefont {Yuichi}\ \bibnamefont {Igarashi}}, \bibinfo
  {author} {\bibfnamefont {Qinghui}\ \bibnamefont {Yan}}, \bibinfo {author}
  {\bibfnamefont {John}\ \bibnamefont {Joannopoulos}}, \ and\ \bibinfo {author}
  {\bibfnamefont {Marin}\ \bibnamefont {Solja\ifmmode \check{c}\else
  \v{c}\fi{}i\ifmmode~\acute{c}\else \'{c}\fi{}}}} (\bibinfo {year} {2015}),\
  \bibfield  {title} {\enquote {\bibinfo {title} {Experimental observation of
  large {C}hern numbers in photonic crystals},}\ }\href
  {http://link.aps.org/doi/10.1103/PhysRevLett.115.253901} {\bibfield
  {journal} {\bibinfo  {journal} {Phys. Rev. Lett.}\ }\textbf {\bibinfo
  {volume} {115}},\ \bibinfo {pages} {253901}}\BibitemShut {NoStop}%
\bibitem [{\citenamefont {Skirlo}\ \emph {et~al.}(2014)\citenamefont {Skirlo},
  \citenamefont {Lu},\ and\ \citenamefont
  {Solja{\v{c}}i{\'c}}}]{Skirlo:2014PRL}%
  \BibitemOpen
  \bibfield  {author} {\bibinfo {author} {\bibnamefont {Skirlo}, \bibfnamefont
  {Scott~A}}, \bibinfo {author} {\bibfnamefont {Ling}\ \bibnamefont {Lu}}, \
  and\ \bibinfo {author} {\bibfnamefont {Marin}\ \bibnamefont
  {Solja{\v{c}}i{\'c}}}} (\bibinfo {year} {2014}),\ \bibfield  {title}
  {\enquote {\bibinfo {title} {Multimode one-way waveguides of large {C}hern
  numbers},}\ }\href
  {https://journals.aps.org/prl/abstract/10.1103/PhysRevLett.113.113904}
  {\bibfield  {journal} {\bibinfo  {journal} {Phys. Rev. Lett.}\ }\textbf
  {\bibinfo {volume} {113}}~(\bibinfo {number} {11}),\ \bibinfo {pages}
  {113904}}\BibitemShut {NoStop}%
\bibitem [{\citenamefont {Slobozhanyuk}\ \emph {et~al.}(2017)\citenamefont
  {Slobozhanyuk}, \citenamefont {Mousavi}, \citenamefont {Ni}, \citenamefont
  {Smirnova}, \citenamefont {Kivshar},\ and\ \citenamefont
  {Khanikaev}}]{Slobozhanyuk:2016NatPhot}%
  \BibitemOpen
  \bibfield  {author} {\bibinfo {author} {\bibnamefont {Slobozhanyuk},
  \bibfnamefont {Alexey}}, \bibinfo {author} {\bibfnamefont {S~Hossein}\
  \bibnamefont {Mousavi}}, \bibinfo {author} {\bibfnamefont {Xiang}\
  \bibnamefont {Ni}}, \bibinfo {author} {\bibfnamefont {Daria}\ \bibnamefont
  {Smirnova}}, \bibinfo {author} {\bibfnamefont {Yuri~S}\ \bibnamefont
  {Kivshar}}, \ and\ \bibinfo {author} {\bibfnamefont {Alexander~B}\
  \bibnamefont {Khanikaev}}} (\bibinfo {year} {2017}),\ \bibfield  {title}
  {\enquote {\bibinfo {title} {Three-dimensional all-dielectric photonic
  topological insulator},}\ }\href
  {https://www.nature.com/articles/nphoton.2016.253} {\bibfield  {journal}
  {\bibinfo  {journal} {Nat. Photonics}\ }\textbf {\bibinfo {volume} {11}},\
  \bibinfo {pages} {130--136}}\BibitemShut {NoStop}%
\bibitem [{\citenamefont {Slobozhanyuk}\ \emph {et~al.}(2019)\citenamefont
  {Slobozhanyuk}, \citenamefont {Shchelokova}, \citenamefont {Ni},
  \citenamefont {Hossein~Mousavi}, \citenamefont {Smirnova}, \citenamefont
  {Belov}, \citenamefont {Al{\`u}}, \citenamefont {Kivshar},\ and\
  \citenamefont {Khanikaev}}]{slobozhanyuk:2017arxiv}%
  \BibitemOpen
  \bibfield  {author} {\bibinfo {author} {\bibnamefont {Slobozhanyuk},
  \bibfnamefont {Alexey}}, \bibinfo {author} {\bibfnamefont {Alena~V}\
  \bibnamefont {Shchelokova}}, \bibinfo {author} {\bibfnamefont {Xiang}\
  \bibnamefont {Ni}}, \bibinfo {author} {\bibfnamefont {S}~\bibnamefont
  {Hossein~Mousavi}}, \bibinfo {author} {\bibfnamefont {Daria~A}\ \bibnamefont
  {Smirnova}}, \bibinfo {author} {\bibfnamefont {Pavel~A}\ \bibnamefont
  {Belov}}, \bibinfo {author} {\bibfnamefont {Andrea}\ \bibnamefont {Al{\`u}}},
  \bibinfo {author} {\bibfnamefont {Yuri~S}\ \bibnamefont {Kivshar}}, \ and\
  \bibinfo {author} {\bibfnamefont {Alexander~B}\ \bibnamefont {Khanikaev}}}
  (\bibinfo {year} {2019}),\ \bibfield  {title} {\enquote {\bibinfo {title}
  {Near-field imaging of spin-locked edge states in all-dielectric topological
  metasurfaces},}\ }\href {https://aip.scitation.org/doi/10.1063/1.5055601}
  {\bibfield  {journal} {\bibinfo  {journal} {Appl. Phys. Lett.}\ }\textbf
  {\bibinfo {volume} {114}}~(\bibinfo {number} {3}),\ \bibinfo {pages}
  {031103}}\BibitemShut {NoStop}%
\bibitem [{\citenamefont {Slobozhanyuk}\ \emph
  {et~al.}(2016{\natexlab{a}})\citenamefont {Slobozhanyuk}, \citenamefont
  {Khanikaev}, \citenamefont {Filonov}, \citenamefont {Smirnova}, \citenamefont
  {Miroshnichenko},\ and\ \citenamefont {Kivshar}}]{slobozhanyuk:2016SciRep}%
  \BibitemOpen
  \bibfield  {author} {\bibinfo {author} {\bibnamefont {Slobozhanyuk},
  \bibfnamefont {Alexey~P}}, \bibinfo {author} {\bibfnamefont {Alexander~B}\
  \bibnamefont {Khanikaev}}, \bibinfo {author} {\bibfnamefont {Dmitry~S}\
  \bibnamefont {Filonov}}, \bibinfo {author} {\bibfnamefont {Daria~A}\
  \bibnamefont {Smirnova}}, \bibinfo {author} {\bibfnamefont {Andrey~E}\
  \bibnamefont {Miroshnichenko}}, \ and\ \bibinfo {author} {\bibfnamefont
  {Yuri~S}\ \bibnamefont {Kivshar}}} (\bibinfo {year} {2016}{\natexlab{a}}),\
  \bibfield  {title} {\enquote {\bibinfo {title} {Experimental demonstration of
  topological effects in bianisotropic metamaterials},}\ }\href
  {https://www.nature.com/articles/srep22270} {\bibfield  {journal} {\bibinfo
  {journal} {Sci. Rep.}\ }\textbf {\bibinfo {volume} {6}},\ \bibinfo {pages}
  {22270}}\BibitemShut {NoStop}%
\bibitem [{\citenamefont {Slobozhanyuk}\ \emph {et~al.}(2015)\citenamefont
  {Slobozhanyuk}, \citenamefont {Poddubny}, \citenamefont {Miroshnichenko},
  \citenamefont {Belov},\ and\ \citenamefont {Kivshar}}]{Slobozhanyuk:2015PRL}%
  \BibitemOpen
  \bibfield  {author} {\bibinfo {author} {\bibnamefont {Slobozhanyuk},
  \bibfnamefont {Alexey~P}}, \bibinfo {author} {\bibfnamefont {Alexander~N.}\
  \bibnamefont {Poddubny}}, \bibinfo {author} {\bibfnamefont {Andrey~E.}\
  \bibnamefont {Miroshnichenko}}, \bibinfo {author} {\bibfnamefont {Pavel~A.}\
  \bibnamefont {Belov}}, \ and\ \bibinfo {author} {\bibfnamefont {Yuri~S.}\
  \bibnamefont {Kivshar}}} (\bibinfo {year} {2015}),\ \bibfield  {title}
  {\enquote {\bibinfo {title} {Subwavelength topological edge states in
  optically resonant dielectric structures},}\ }\href
  {https://link.aps.org/doi/10.1103/PhysRevLett.114.123901} {\bibfield
  {journal} {\bibinfo  {journal} {Phys. Rev. Lett.}\ }\textbf {\bibinfo
  {volume} {114}},\ \bibinfo {pages} {123901}}\BibitemShut {NoStop}%
\bibitem [{\citenamefont {Slobozhanyuk}\ \emph
  {et~al.}(2016{\natexlab{b}})\citenamefont {Slobozhanyuk}, \citenamefont
  {Poddubny}, \citenamefont {Sinev}, \citenamefont {Samusev}, \citenamefont
  {Yu}, \citenamefont {Kuznetsov}, \citenamefont {Miroshnichenko},\ and\
  \citenamefont {Kivshar}}]{Slobozhanyuk:2016LaserPhotRev}%
  \BibitemOpen
  \bibfield  {author} {\bibinfo {author} {\bibnamefont {Slobozhanyuk},
  \bibfnamefont {Alexey~P}}, \bibinfo {author} {\bibfnamefont {Alexander~N}\
  \bibnamefont {Poddubny}}, \bibinfo {author} {\bibfnamefont {Ivan~S}\
  \bibnamefont {Sinev}}, \bibinfo {author} {\bibfnamefont {Anton~K}\
  \bibnamefont {Samusev}}, \bibinfo {author} {\bibfnamefont {Ye~Feng}\
  \bibnamefont {Yu}}, \bibinfo {author} {\bibfnamefont {Arseniy~I}\
  \bibnamefont {Kuznetsov}}, \bibinfo {author} {\bibfnamefont {Andrey~E}\
  \bibnamefont {Miroshnichenko}}, \ and\ \bibinfo {author} {\bibfnamefont
  {Yuri~S}\ \bibnamefont {Kivshar}}} (\bibinfo {year} {2016}{\natexlab{b}}),\
  \bibfield  {title} {\enquote {\bibinfo {title} {Enhanced photonic spin {H}all
  effect with subwavelength topological edge states},}\ }\href
  {http://onlinelibrary.wiley.com/doi/10.1002/lpor.201600042/full} {\bibfield
  {journal} {\bibinfo  {journal} {Laser Photonics Rev.}\ }\textbf {\bibinfo
  {volume} {10}}~(\bibinfo {number} {4}),\ \bibinfo {pages}
  {656--664}}\BibitemShut {NoStop}%
\bibitem [{\citenamefont {S{\"o}llner}\ \emph {et~al.}(2015)\citenamefont
  {S{\"o}llner}, \citenamefont {Mahmoodian}, \citenamefont {Hansen},
  \citenamefont {Midolo}, \citenamefont {Javadi}, \citenamefont
  {Kir{\v{s}}ansk{\.e}}, \citenamefont {Pregnolato}, \citenamefont {El-Ella},
  \citenamefont {Lee}, \citenamefont {Song} \emph
  {et~al.}}]{Sollner:2015naturenanotech}%
  \BibitemOpen
  \bibfield  {author} {\bibinfo {author} {\bibnamefont {S{\"o}llner},
  \bibfnamefont {Immo}}, \bibinfo {author} {\bibfnamefont {Sahand}\
  \bibnamefont {Mahmoodian}}, \bibinfo {author} {\bibfnamefont
  {Sofie~Lindskov}\ \bibnamefont {Hansen}}, \bibinfo {author} {\bibfnamefont
  {Leonardo}\ \bibnamefont {Midolo}}, \bibinfo {author} {\bibfnamefont {Alisa}\
  \bibnamefont {Javadi}}, \bibinfo {author} {\bibfnamefont {Gabija}\
  \bibnamefont {Kir{\v{s}}ansk{\.e}}}, \bibinfo {author} {\bibfnamefont
  {Tommaso}\ \bibnamefont {Pregnolato}}, \bibinfo {author} {\bibfnamefont
  {Haitham}\ \bibnamefont {El-Ella}}, \bibinfo {author} {\bibfnamefont
  {Eun~Hye}\ \bibnamefont {Lee}}, \bibinfo {author} {\bibfnamefont {Jin~Dong}\
  \bibnamefont {Song}},  \emph {et~al.}} (\bibinfo {year} {2015}),\ \bibfield
  {title} {\enquote {\bibinfo {title} {Deterministic photon--emitter coupling
  in chiral photonic circuits},}\ }\href
  {http://www.nature.com/nnano/journal/v10/n9/full/nnano.2015.159.html}
  {\bibfield  {journal} {\bibinfo  {journal} {Nat. Nanotechnol.}\ }\textbf
  {\bibinfo {volume} {10}}~(\bibinfo {number} {9}),\ \bibinfo {pages}
  {775--778}}\BibitemShut {NoStop}%
\bibitem [{\citenamefont {Solnyshkov}\ \emph {et~al.}(2018)\citenamefont
  {Solnyshkov}, \citenamefont {Bleu},\ and\ \citenamefont
  {Malpuech}}]{Solnyshkov:APL2018}%
  \BibitemOpen
  \bibfield  {author} {\bibinfo {author} {\bibnamefont {Solnyshkov},
  \bibfnamefont {D~D}}, \bibinfo {author} {\bibfnamefont {O}~\bibnamefont
  {Bleu}}, \ and\ \bibinfo {author} {\bibfnamefont {G}~\bibnamefont
  {Malpuech}}} (\bibinfo {year} {2018}),\ \bibfield  {title} {\enquote
  {\bibinfo {title} {{Topological optical isolator based on polariton
  graphene}},}\ }\href {https://doi.org/10.1063/1.5018902} {\bibfield
  {journal} {\bibinfo  {journal} {Appl. Phys. Lett.}\ }\textbf {\bibinfo
  {volume} {112}}~(\bibinfo {number} {3}),\ \bibinfo {pages}
  {031106}}\BibitemShut {NoStop}%
\bibitem [{\citenamefont {Solnyshkov}\ \emph
  {et~al.}(2016{\natexlab{a}})\citenamefont {Solnyshkov}, \citenamefont
  {Nalitov},\ and\ \citenamefont {Malpuech}}]{Solnyshkov:2016PRL}%
  \BibitemOpen
  \bibfield  {author} {\bibinfo {author} {\bibnamefont {Solnyshkov},
  \bibfnamefont {D~D}}, \bibinfo {author} {\bibfnamefont {A.~V.}\ \bibnamefont
  {Nalitov}}, \ and\ \bibinfo {author} {\bibfnamefont {G.}~\bibnamefont
  {Malpuech}}} (\bibinfo {year} {2016}{\natexlab{a}}),\ \bibfield  {title}
  {\enquote {\bibinfo {title} {Kibble-{Z}urek mechanism in topologically
  nontrivial zigzag chains of polariton micropillars},}\ }\href
  {https://link.aps.org/doi/10.1103/PhysRevLett.116.046402} {\bibfield
  {journal} {\bibinfo  {journal} {Phys. Rev. Lett.}\ }\textbf {\bibinfo
  {volume} {116}},\ \bibinfo {pages} {046402}}\BibitemShut {NoStop}%
\bibitem [{\citenamefont {Solnyshkov}\ \emph
  {et~al.}(2016{\natexlab{b}})\citenamefont {Solnyshkov}, \citenamefont
  {Nalitov}, \citenamefont {Teklu}, \citenamefont {Franck},\ and\ \citenamefont
  {Malpuech}}]{Solnyshkov:2016prb}%
  \BibitemOpen
  \bibfield  {author} {\bibinfo {author} {\bibnamefont {Solnyshkov},
  \bibfnamefont {Dmitry}}, \bibinfo {author} {\bibfnamefont {Anton}\
  \bibnamefont {Nalitov}}, \bibinfo {author} {\bibfnamefont {Berihu}\
  \bibnamefont {Teklu}}, \bibinfo {author} {\bibfnamefont {Louis}\ \bibnamefont
  {Franck}}, \ and\ \bibinfo {author} {\bibfnamefont {Guillaume}\ \bibnamefont
  {Malpuech}}} (\bibinfo {year} {2016}{\natexlab{b}}),\ \bibfield  {title}
  {\enquote {\bibinfo {title} {{Spin-dependent Klein tunneling in polariton
  graphene with photonic spin-orbit interaction}},}\ }\href
  {http://link.aps.org/doi/10.1103/PhysRevB.93.085404} {\bibfield  {journal}
  {\bibinfo  {journal} {Phys. Rev. B}\ }\textbf {\bibinfo {volume}
  {93}}~(\bibinfo {number} {8}),\ \bibinfo {pages} {085404}}\BibitemShut
  {NoStop}%
\bibitem [{\citenamefont {Soluyanov}\ \emph {et~al.}(2015)\citenamefont
  {Soluyanov}, \citenamefont {Gresch}, \citenamefont {Wang}, \citenamefont
  {Wu}, \citenamefont {Troyer}, \citenamefont {Dai},\ and\ \citenamefont
  {Bernevig}}]{Soluyanov:2015Nature}%
  \BibitemOpen
  \bibfield  {author} {\bibinfo {author} {\bibnamefont {Soluyanov},
  \bibfnamefont {Alexey~A}}, \bibinfo {author} {\bibfnamefont {Dominik}\
  \bibnamefont {Gresch}}, \bibinfo {author} {\bibfnamefont {Zhijun}\
  \bibnamefont {Wang}}, \bibinfo {author} {\bibfnamefont {QuanSheng}\
  \bibnamefont {Wu}}, \bibinfo {author} {\bibfnamefont {Matthias}\ \bibnamefont
  {Troyer}}, \bibinfo {author} {\bibfnamefont {Xi}~\bibnamefont {Dai}}, \ and\
  \bibinfo {author} {\bibfnamefont {B~Andrei}\ \bibnamefont {Bernevig}}}
  (\bibinfo {year} {2015}),\ \bibfield  {title} {\enquote {\bibinfo {title}
  {Type-{II} {W}eyl semimetals},}\ }\href
  {https://www.nature.com/articles/nature15768} {\bibfield  {journal} {\bibinfo
   {journal} {Nature}\ }\textbf {\bibinfo {volume} {527}}~(\bibinfo {number}
  {7579}),\ \bibinfo {pages} {495--498}}\BibitemShut {NoStop}%
\bibitem [{\citenamefont {{Sommer}}\ \emph {et~al.}(2015)\citenamefont
  {{Sommer}}, \citenamefont {{B{\"u}chler}},\ and\ \citenamefont
  {{Simon}}}]{Sommer:arxiv2015}%
  \BibitemOpen
  \bibfield  {author} {\bibinfo {author} {\bibnamefont {{Sommer}},
  \bibfnamefont {A}}, \bibinfo {author} {\bibfnamefont {H.~P.}\ \bibnamefont
  {{B{\"u}chler}}}, \ and\ \bibinfo {author} {\bibfnamefont {J.}~\bibnamefont
  {{Simon}}}} (\bibinfo {year} {2015}),\ \bibfield  {title} {\enquote {\bibinfo
  {title} {{Quantum Crystals and {L}aughlin Droplets of Cavity {R}ydberg
  Polaritons}},}\ }\href@noop {} {\bibinfo  {journal} {arXiv:1506.00341}\
  }\BibitemShut {NoStop}%
\bibitem [{\citenamefont {Sommer}\ and\ \citenamefont
  {Simon}(2016)}]{sommer2016engineering}%
  \BibitemOpen
\bibfield  {journal} {  }\bibfield  {author} {\bibinfo {author} {\bibnamefont
  {Sommer}, \bibfnamefont {Ariel}}, \ and\ \bibinfo {author} {\bibfnamefont
  {Jonathan}\ \bibnamefont {Simon}}} (\bibinfo {year} {2016}),\ \bibfield
  {title} {\enquote {\bibinfo {title} {Engineering photonic {F}loquet
  {H}amiltonians through {F}abry--{P}{\'e}rot resonators},}\ }\href
  {https://iopscience.iop.org/article/10.1088/1367-2630/18/3/035008/meta}
  {\bibfield  {journal} {\bibinfo  {journal} {New J. Phys.}\ }\textbf {\bibinfo
  {volume} {18}}~(\bibinfo {number} {3}),\ \bibinfo {pages}
  {035008}}\BibitemShut {NoStop}%
\bibitem [{\citenamefont {Song}\ \emph {et~al.}(2015)\citenamefont {Song},
  \citenamefont {Paltoglou}, \citenamefont {Liu}, \citenamefont {Zhu},
  \citenamefont {Gallardo}, \citenamefont {Tang}, \citenamefont {Xu},
  \citenamefont {Ablowitz}, \citenamefont {Efremidis},\ and\ \citenamefont
  {Chen}}]{Song:2015NatComm}%
  \BibitemOpen
  \bibfield  {author} {\bibinfo {author} {\bibnamefont {Song}, \bibfnamefont
  {Daohong}}, \bibinfo {author} {\bibfnamefont {Vassilis}\ \bibnamefont
  {Paltoglou}}, \bibinfo {author} {\bibfnamefont {Sheng}\ \bibnamefont {Liu}},
  \bibinfo {author} {\bibfnamefont {Yi}~\bibnamefont {Zhu}}, \bibinfo {author}
  {\bibfnamefont {Daniel}\ \bibnamefont {Gallardo}}, \bibinfo {author}
  {\bibfnamefont {Liqin}\ \bibnamefont {Tang}}, \bibinfo {author}
  {\bibfnamefont {Jingjun}\ \bibnamefont {Xu}}, \bibinfo {author}
  {\bibfnamefont {Mark}\ \bibnamefont {Ablowitz}}, \bibinfo {author}
  {\bibfnamefont {Nikolaos~K}\ \bibnamefont {Efremidis}}, \ and\ \bibinfo
  {author} {\bibfnamefont {Zhigang}\ \bibnamefont {Chen}}} (\bibinfo {year}
  {2015}),\ \bibfield  {title} {\enquote {\bibinfo {title} {{Unveiling
  pseudospin and angular momentum in photonic graphene}},}\ }\href
  {http://www.nature.com/ncomms/2015/150217/ncomms7272/full/ncomms7272.html}
  {\bibfield  {journal} {\bibinfo  {journal} {Nat. Commun.}\ }\textbf {\bibinfo
  {volume} {6}},\ \bibinfo {pages} {6272}}\BibitemShut {NoStop}%
\bibitem [{\citenamefont {Spivak}\ \emph {et~al.}(1995)\citenamefont {Spivak},
  \citenamefont {Zhou},\ and\ \citenamefont {Beal~Monod}}]{Spivak:1995}%
  \BibitemOpen
  \bibfield  {author} {\bibinfo {author} {\bibnamefont {Spivak}, \bibfnamefont
  {B}}, \bibinfo {author} {\bibfnamefont {F.}~\bibnamefont {Zhou}}, \ and\
  \bibinfo {author} {\bibfnamefont {M.~T.}\ \bibnamefont {Beal~Monod}}}
  (\bibinfo {year} {1995}),\ \bibfield  {title} {\enquote {\bibinfo {title}
  {Mesoscopic mechanisms of the photovoltaic effect and microwave absorption in
  granular metals},}\ }\href
  {https://link.aps.org/doi/10.1103/PhysRevB.51.13226} {\bibfield  {journal}
  {\bibinfo  {journal} {Phys. Rev. B}\ }\textbf {\bibinfo {volume} {51}},\
  \bibinfo {pages} {13226--13230}}\BibitemShut {NoStop}%
\bibitem [{\citenamefont {St-Jean}\ \emph {et~al.}(2017)\citenamefont
  {St-Jean}, \citenamefont {Goblot}, \citenamefont {Galopin}, \citenamefont
  {Lema{\^\i}tre}, \citenamefont {Ozawa}, \citenamefont {Le~Gratiet},
  \citenamefont {Sagnes}, \citenamefont {Bloch},\ and\ \citenamefont
  {Amo}}]{St-Jean:2017NatPhot}%
  \BibitemOpen
  \bibfield  {author} {\bibinfo {author} {\bibnamefont {St-Jean}, \bibfnamefont
  {P}}, \bibinfo {author} {\bibfnamefont {V}~\bibnamefont {Goblot}}, \bibinfo
  {author} {\bibfnamefont {E}~\bibnamefont {Galopin}}, \bibinfo {author}
  {\bibfnamefont {A}~\bibnamefont {Lema{\^\i}tre}}, \bibinfo {author}
  {\bibfnamefont {T}~\bibnamefont {Ozawa}}, \bibinfo {author} {\bibfnamefont
  {L}~\bibnamefont {Le~Gratiet}}, \bibinfo {author} {\bibfnamefont
  {I}~\bibnamefont {Sagnes}}, \bibinfo {author} {\bibfnamefont {J}~\bibnamefont
  {Bloch}}, \ and\ \bibinfo {author} {\bibfnamefont {A}~\bibnamefont {Amo}}}
  (\bibinfo {year} {2017}),\ \bibfield  {title} {\enquote {\bibinfo {title}
  {Lasing in topological edge states of a one-dimensional lattice},}\ }\href
  {https://www.nature.com/articles/s41566-017-0006-2} {\bibfield  {journal}
  {\bibinfo  {journal} {Nat. Photonics}\ }\textbf {\bibinfo {volume}
  {11}}~(\bibinfo {number} {10}),\ \bibinfo {pages} {651}}\BibitemShut
  {NoStop}%
\bibitem [{\citenamefont {Stannigel}\ \emph {et~al.}(2012)\citenamefont
  {Stannigel}, \citenamefont {Komar}, \citenamefont {Habraken}, \citenamefont
  {Bennett}, \citenamefont {Lukin}, \citenamefont {Zoller},\ and\ \citenamefont
  {Rabl}}]{stannigel2012optomechanical}%
  \BibitemOpen
  \bibfield  {author} {\bibinfo {author} {\bibnamefont {Stannigel},
  \bibfnamefont {K}}, \bibinfo {author} {\bibfnamefont {Peter}\ \bibnamefont
  {Komar}}, \bibinfo {author} {\bibfnamefont {SJM}\ \bibnamefont {Habraken}},
  \bibinfo {author} {\bibfnamefont {SD}~\bibnamefont {Bennett}}, \bibinfo
  {author} {\bibfnamefont {Mikhail~D}\ \bibnamefont {Lukin}}, \bibinfo {author}
  {\bibfnamefont {P}~\bibnamefont {Zoller}}, \ and\ \bibinfo {author}
  {\bibfnamefont {P}~\bibnamefont {Rabl}}} (\bibinfo {year} {2012}),\ \bibfield
   {title} {\enquote {\bibinfo {title} {Optomechanical quantum information
  processing with photons and phonons},}\ }\href
  {https://doi.org/10.1103/PhysRevLett.109.013603} {\bibfield  {journal}
  {\bibinfo  {journal} {Phys. Rev. Lett.}\ }\textbf {\bibinfo {volume}
  {109}}~(\bibinfo {number} {1}),\ \bibinfo {pages} {013603}}\BibitemShut
  {NoStop}%
\bibitem [{\citenamefont {St{\"o}rmer}\ \emph {et~al.}(1986)\citenamefont
  {St{\"o}rmer}, \citenamefont {Eisenstein}, \citenamefont {Gossard},
  \citenamefont {Wiegmann},\ and\ \citenamefont {Baldwin}}]{Stormer:1986PRL}%
  \BibitemOpen
  \bibfield  {author} {\bibinfo {author} {\bibnamefont {St{\"o}rmer},
  \bibfnamefont {HL}}, \bibinfo {author} {\bibfnamefont {JP}~\bibnamefont
  {Eisenstein}}, \bibinfo {author} {\bibfnamefont {AC}~\bibnamefont {Gossard}},
  \bibinfo {author} {\bibfnamefont {W}~\bibnamefont {Wiegmann}}, \ and\
  \bibinfo {author} {\bibfnamefont {K}~\bibnamefont {Baldwin}}} (\bibinfo
  {year} {1986}),\ \bibfield  {title} {\enquote {\bibinfo {title} {Quantization
  of the {H}all effect in an anisotropic three-dimensional electronic
  system},}\ }\href
  {https://journals.aps.org/prl/abstract/10.1103/PhysRevLett.56.85} {\bibfield
  {journal} {\bibinfo  {journal} {Phys. Rev. Lett.}\ }\textbf {\bibinfo
  {volume} {56}}~(\bibinfo {number} {1}),\ \bibinfo {pages} {85}}\BibitemShut
  {NoStop}%
\bibitem [{\citenamefont {Stuhl}\ \emph {et~al.}(2015)\citenamefont {Stuhl},
  \citenamefont {Lu}, \citenamefont {Aycock}, \citenamefont {Genkina},\ and\
  \citenamefont {Spielman}}]{Stuhl:2015Science}%
  \BibitemOpen
  \bibfield  {author} {\bibinfo {author} {\bibnamefont {Stuhl}, \bibfnamefont
  {B~K}}, \bibinfo {author} {\bibfnamefont {H.-I.}\ \bibnamefont {Lu}},
  \bibinfo {author} {\bibfnamefont {L.~M.}\ \bibnamefont {Aycock}}, \bibinfo
  {author} {\bibfnamefont {D.}~\bibnamefont {Genkina}}, \ and\ \bibinfo
  {author} {\bibfnamefont {I.~B.}\ \bibnamefont {Spielman}}} (\bibinfo {year}
  {2015}),\ \bibfield  {title} {\enquote {\bibinfo {title} {Visualizing edge
  states with an atomic {B}ose gas in the quantum {H}all regime},}\ }\href
  {http://science.sciencemag.org/content/349/6255/1514} {\bibfield  {journal}
  {\bibinfo  {journal} {Science}\ }\textbf {\bibinfo {volume} {349}}~(\bibinfo
  {number} {6255}),\ \bibinfo {pages} {1514--1518}}\BibitemShut {NoStop}%
\bibitem [{\citenamefont {St{\"u}tzer}\ \emph {et~al.}(2018)\citenamefont
  {St{\"u}tzer}, \citenamefont {Plotnik}, \citenamefont {Lumer}, \citenamefont
  {Titum}, \citenamefont {Lindner}, \citenamefont {Segev}, \citenamefont
  {Rechtsman},\ and\ \citenamefont {Szameit}}]{Stutzer:2018Naure}%
  \BibitemOpen
  \bibfield  {author} {\bibinfo {author} {\bibnamefont {St{\"u}tzer},
  \bibfnamefont {Simon}}, \bibinfo {author} {\bibfnamefont {Yonatan}\
  \bibnamefont {Plotnik}}, \bibinfo {author} {\bibfnamefont {Yaakov}\
  \bibnamefont {Lumer}}, \bibinfo {author} {\bibfnamefont {Paraj}\ \bibnamefont
  {Titum}}, \bibinfo {author} {\bibfnamefont {Netanel~H}\ \bibnamefont
  {Lindner}}, \bibinfo {author} {\bibfnamefont {Mordechai}\ \bibnamefont
  {Segev}}, \bibinfo {author} {\bibfnamefont {Mikael~C}\ \bibnamefont
  {Rechtsman}}, \ and\ \bibinfo {author} {\bibfnamefont {Alexander}\
  \bibnamefont {Szameit}}} (\bibinfo {year} {2018}),\ \bibfield  {title}
  {\enquote {\bibinfo {title} {Photonic topological {A}nderson insulators},}\
  }\href {https://www.nature.com/articles/s41586-018-0418-2} {\bibfield
  {journal} {\bibinfo  {journal} {Nature}\ }\textbf {\bibinfo {volume}
  {560}}~(\bibinfo {number} {7719}),\ \bibinfo {pages} {461}}\BibitemShut
  {NoStop}%
\bibitem [{\citenamefont {Su}\ \emph {et~al.}(1979)\citenamefont {Su},
  \citenamefont {Schrieffer},\ and\ \citenamefont {Heeger}}]{Su:1979}%
  \BibitemOpen
  \bibfield  {author} {\bibinfo {author} {\bibnamefont {Su}, \bibfnamefont
  {W~P}}, \bibinfo {author} {\bibfnamefont {J.~R.}\ \bibnamefont {Schrieffer}},
  \ and\ \bibinfo {author} {\bibfnamefont {A.~J.}\ \bibnamefont {Heeger}}}
  (\bibinfo {year} {1979}),\ \bibfield  {title} {\enquote {\bibinfo {title}
  {Solitons in polyacetylene},}\ }\href
  {http://link.aps.org/doi/10.1103/PhysRevLett.42.1698} {\bibfield  {journal}
  {\bibinfo  {journal} {Phys. Rev. Lett.}\ }\textbf {\bibinfo {volume} {42}},\
  \bibinfo {pages} {1698--1701}}\BibitemShut {NoStop}%
\bibitem [{\citenamefont {Sugawa}\ \emph {et~al.}(2018)\citenamefont {Sugawa},
  \citenamefont {Salces-Carcoba}, \citenamefont {Perry}, \citenamefont {Yue},\
  and\ \citenamefont {Spielman}}]{Sugawa:2016arXiv}%
  \BibitemOpen
  \bibfield  {author} {\bibinfo {author} {\bibnamefont {Sugawa}, \bibfnamefont
  {Seiji}}, \bibinfo {author} {\bibfnamefont {Francisco}\ \bibnamefont
  {Salces-Carcoba}}, \bibinfo {author} {\bibfnamefont {Abigail~R}\ \bibnamefont
  {Perry}}, \bibinfo {author} {\bibfnamefont {Yuchen}\ \bibnamefont {Yue}}, \
  and\ \bibinfo {author} {\bibfnamefont {IB}~\bibnamefont {Spielman}}}
  (\bibinfo {year} {2018}),\ \bibfield  {title} {\enquote {\bibinfo {title}
  {Second {C}hern number of a quantum-simulated non-{A}belian {Y}ang
  monopole},}\ }\href {http://science.sciencemag.org/content/360/6396/1429}
  {\bibfield  {journal} {\bibinfo  {journal} {Science}\ }\textbf {\bibinfo
  {volume} {360}}~(\bibinfo {number} {6396}),\ \bibinfo {pages}
  {1429--1434}}\BibitemShut {NoStop}%
\bibitem [{\citenamefont {Sun}\ \emph {et~al.}(2017{\natexlab{a}})\citenamefont
  {Sun}, \citenamefont {Luo}, \citenamefont {Gong}, \citenamefont {Guo},\ and\
  \citenamefont {Zhou}}]{Sun:2017PRA}%
  \BibitemOpen
  \bibfield  {author} {\bibinfo {author} {\bibnamefont {Sun}, \bibfnamefont
  {Bo~Ye}}, \bibinfo {author} {\bibfnamefont {Xi~Wang}\ \bibnamefont {Luo}},
  \bibinfo {author} {\bibfnamefont {Ming}\ \bibnamefont {Gong}}, \bibinfo
  {author} {\bibfnamefont {Guang~Can}\ \bibnamefont {Guo}}, \ and\ \bibinfo
  {author} {\bibfnamefont {Zheng~Wei}\ \bibnamefont {Zhou}}} (\bibinfo {year}
  {2017}{\natexlab{a}}),\ \bibfield  {title} {\enquote {\bibinfo {title} {Weyl
  semimetal phases and implementation in degenerate optical cavities},}\ }\href
  {https://link.aps.org/doi/10.1103/PhysRevA.96.013857} {\bibfield  {journal}
  {\bibinfo  {journal} {Phys. Rev. A}\ }\textbf {\bibinfo {volume} {96}},\
  \bibinfo {pages} {013857}}\BibitemShut {NoStop}%
\bibitem [{\citenamefont {Sun}\ \emph {et~al.}(2017{\natexlab{b}})\citenamefont
  {Sun}, \citenamefont {He}, \citenamefont {Liu}, \citenamefont {Lu},
  \citenamefont {Zhu},\ and\ \citenamefont {Chen}}]{Sun:2017PQE}%
  \BibitemOpen
  \bibfield  {author} {\bibinfo {author} {\bibnamefont {Sun}, \bibfnamefont
  {Xiao-Chen}}, \bibinfo {author} {\bibfnamefont {Cheng}\ \bibnamefont {He}},
  \bibinfo {author} {\bibfnamefont {Xiao-Ping}\ \bibnamefont {Liu}}, \bibinfo
  {author} {\bibfnamefont {Ming-Hui}\ \bibnamefont {Lu}}, \bibinfo {author}
  {\bibfnamefont {Shi-Ning}\ \bibnamefont {Zhu}}, \ and\ \bibinfo {author}
  {\bibfnamefont {Yan-Feng}\ \bibnamefont {Chen}}} (\bibinfo {year}
  {2017}{\natexlab{b}}),\ \bibfield  {title} {\enquote {\bibinfo {title}
  {Two-dimensional topological photonic systems},}\ }\href
  {https://www.sciencedirect.com/science/article/pii/S0079672717300290}
  {\bibfield  {journal} {\bibinfo  {journal} {Prog. Quantum Electron.}\
  }\textbf {\bibinfo {volume} {55}},\ \bibinfo {pages} {52--73}}\BibitemShut
  {NoStop}%
\bibitem [{\citenamefont {Sun}\ \emph {et~al.}(2018)\citenamefont {Sun},
  \citenamefont {Leykam}, \citenamefont {Nenni}, \citenamefont {Song},
  \citenamefont {Chen}, \citenamefont {Chong},\ and\ \citenamefont
  {Chen}}]{Sun:2018PRL}%
  \BibitemOpen
  \bibfield  {author} {\bibinfo {author} {\bibnamefont {Sun}, \bibfnamefont
  {Yong}}, \bibinfo {author} {\bibfnamefont {Daniel}\ \bibnamefont {Leykam}},
  \bibinfo {author} {\bibfnamefont {Stephen}\ \bibnamefont {Nenni}}, \bibinfo
  {author} {\bibfnamefont {Daohong}\ \bibnamefont {Song}}, \bibinfo {author}
  {\bibfnamefont {Hong}\ \bibnamefont {Chen}}, \bibinfo {author} {\bibfnamefont
  {Y.~D.}\ \bibnamefont {Chong}}, \ and\ \bibinfo {author} {\bibfnamefont
  {Zhigang}\ \bibnamefont {Chen}}} (\bibinfo {year} {2018}),\ \bibfield
  {title} {\enquote {\bibinfo {title} {Observation of valley
  {L}andau-{Z}ener-{B}loch oscillations and pseudospin imbalance in photonic
  graphene},}\ }\href {https://link.aps.org/doi/10.1103/PhysRevLett.121.033904}
  {\bibfield  {journal} {\bibinfo  {journal} {Phys. Rev. Lett.}\ }\textbf
  {\bibinfo {volume} {121}},\ \bibinfo {pages} {033904}}\BibitemShut {NoStop}%
\bibitem [{\citenamefont {Suszalski}\ and\ \citenamefont
  {Zakrzewski}(2016)}]{Suszalski:2016PRA}%
  \BibitemOpen
  \bibfield  {author} {\bibinfo {author} {\bibnamefont {Suszalski},
  \bibfnamefont {Dominik}}, \ and\ \bibinfo {author} {\bibfnamefont {Jakub}\
  \bibnamefont {Zakrzewski}}} (\bibinfo {year} {2016}),\ \bibfield  {title}
  {\enquote {\bibinfo {title} {Different lattice geometries with a synthetic
  dimension},}\ }\href {https://link.aps.org/doi/10.1103/PhysRevA.94.033602}
  {\bibfield  {journal} {\bibinfo  {journal} {Phys. Rev. A}\ }\textbf {\bibinfo
  {volume} {94}},\ \bibinfo {pages} {033602}}\BibitemShut {NoStop}%
\bibitem [{\citenamefont {Switkes}\ \emph {et~al.}(1999)\citenamefont
  {Switkes}, \citenamefont {Marcus}, \citenamefont {Campman},\ and\
  \citenamefont {Gossard}}]{Switkes:1999}%
  \BibitemOpen
  \bibfield  {author} {\bibinfo {author} {\bibnamefont {Switkes}, \bibfnamefont
  {M}}, \bibinfo {author} {\bibfnamefont {C.~M.}\ \bibnamefont {Marcus}},
  \bibinfo {author} {\bibfnamefont {K.}~\bibnamefont {Campman}}, \ and\
  \bibinfo {author} {\bibfnamefont {A.~C.}\ \bibnamefont {Gossard}}} (\bibinfo
  {year} {1999}),\ \bibfield  {title} {\enquote {\bibinfo {title} {An adiabatic
  quantum electron pump},}\ }\href
  {http://science.sciencemag.org/content/283/5409/1905} {\bibfield  {journal}
  {\bibinfo  {journal} {Science}\ }\textbf {\bibinfo {volume} {283}}~(\bibinfo
  {number} {5409}),\ \bibinfo {pages} {1905--1908}}\BibitemShut {NoStop}%
\bibitem [{\citenamefont {Szameit}\ \emph {et~al.}(2007)\citenamefont
  {Szameit}, \citenamefont {Dreisow}, \citenamefont {Pertsch}, \citenamefont
  {Nolte},\ and\ \citenamefont {T{\"u}nnermann}}]{szameit2007control}%
  \BibitemOpen
  \bibfield  {author} {\bibinfo {author} {\bibnamefont {Szameit}, \bibfnamefont
  {Alexander}}, \bibinfo {author} {\bibfnamefont {Felix}\ \bibnamefont
  {Dreisow}}, \bibinfo {author} {\bibfnamefont {Thomas}\ \bibnamefont
  {Pertsch}}, \bibinfo {author} {\bibfnamefont {Stefan}\ \bibnamefont {Nolte}},
  \ and\ \bibinfo {author} {\bibfnamefont {Andreas}\ \bibnamefont
  {T{\"u}nnermann}}} (\bibinfo {year} {2007}),\ \bibfield  {title} {\enquote
  {\bibinfo {title} {Control of directional evanescent coupling in fs laser
  written waveguides},}\ }\href
  {https://www.osapublishing.org/oe/abstract.cfm?uri=oe-15-4-1579} {\bibfield
  {journal} {\bibinfo  {journal} {Opt. Express}\ }\textbf {\bibinfo {volume}
  {15}}~(\bibinfo {number} {4}),\ \bibinfo {pages} {1579--1587}}\BibitemShut
  {NoStop}%
\bibitem [{\citenamefont {Szameit}\ and\ \citenamefont
  {Nolte}(2010)}]{Szameit:2010JPB}%
  \BibitemOpen
  \bibfield  {author} {\bibinfo {author} {\bibnamefont {Szameit}, \bibfnamefont
  {Alexander}}, \ and\ \bibinfo {author} {\bibfnamefont {Stefan}\ \bibnamefont
  {Nolte}}} (\bibinfo {year} {2010}),\ \bibfield  {title} {\enquote {\bibinfo
  {title} {Discrete optics in femtosecond-laser-written photonic structures},}\
  }\href {https://iopscience.iop.org/article/10.1088/0953-4075/43/16/163001}
  {\bibfield  {journal} {\bibinfo  {journal} {J. Phys. B}\ }\textbf {\bibinfo
  {volume} {43}}~(\bibinfo {number} {16}),\ \bibinfo {pages}
  {163001}}\BibitemShut {NoStop}%
\bibitem [{\citenamefont {Szameit}\ \emph {et~al.}(2011)\citenamefont
  {Szameit}, \citenamefont {Rechtsman}, \citenamefont {Bahat-Treidel},\ and\
  \citenamefont {Segev}}]{szameit2011p}%
  \BibitemOpen
  \bibfield  {author} {\bibinfo {author} {\bibnamefont {Szameit}, \bibfnamefont
  {Alexander}}, \bibinfo {author} {\bibfnamefont {Mikael~C}\ \bibnamefont
  {Rechtsman}}, \bibinfo {author} {\bibfnamefont {Omri}\ \bibnamefont
  {Bahat-Treidel}}, \ and\ \bibinfo {author} {\bibfnamefont {Mordechai}\
  \bibnamefont {Segev}}} (\bibinfo {year} {2011}),\ \bibfield  {title}
  {\enquote {\bibinfo {title} {$\mathcal{PT}$-symmetry in honeycomb photonic
  lattices},}\ }\href
  {https://journals.aps.org/pra/abstract/10.1103/PhysRevA.84.021806} {\bibfield
   {journal} {\bibinfo  {journal} {Phys. Rev. A}\ }\textbf {\bibinfo {volume}
  {84}}~(\bibinfo {number} {2}),\ \bibinfo {pages} {021806}}\BibitemShut
  {NoStop}%
\bibitem [{\citenamefont {Taddia}\ \emph {et~al.}(2017)\citenamefont {Taddia},
  \citenamefont {Cornfeld}, \citenamefont {Rossini}, \citenamefont {Mazza},
  \citenamefont {Sela},\ and\ \citenamefont {Fazio}}]{Taddia:2017PRL}%
  \BibitemOpen
  \bibfield  {author} {\bibinfo {author} {\bibnamefont {Taddia}, \bibfnamefont
  {Luca}}, \bibinfo {author} {\bibfnamefont {Eyal}\ \bibnamefont {Cornfeld}},
  \bibinfo {author} {\bibfnamefont {Davide}\ \bibnamefont {Rossini}}, \bibinfo
  {author} {\bibfnamefont {Leonardo}\ \bibnamefont {Mazza}}, \bibinfo {author}
  {\bibfnamefont {Eran}\ \bibnamefont {Sela}}, \ and\ \bibinfo {author}
  {\bibfnamefont {Rosario}\ \bibnamefont {Fazio}}} (\bibinfo {year} {2017}),\
  \bibfield  {title} {\enquote {\bibinfo {title} {Topological fractional
  pumping with alkaline-earth-like atoms in synthetic lattices},}\ }\href
  {https://link.aps.org/doi/10.1103/PhysRevLett.118.230402} {\bibfield
  {journal} {\bibinfo  {journal} {Phys. Rev. Lett.}\ }\textbf {\bibinfo
  {volume} {118}},\ \bibinfo {pages} {230402}}\BibitemShut {NoStop}%
\bibitem [{\citenamefont {Tai}\ \emph {et~al.}(2017{\natexlab{a}})\citenamefont
  {Tai}, \citenamefont {Lukin}, \citenamefont {Rispoli}, \citenamefont
  {Schittko}, \citenamefont {Menke}, \citenamefont {Borgnia}, \citenamefont
  {Preiss}, \citenamefont {Grusdt}, \citenamefont {Kaufman},\ and\
  \citenamefont {Greiner}}]{tai2017microscopy}%
  \BibitemOpen
  \bibfield  {author} {\bibinfo {author} {\bibnamefont {Tai}, \bibfnamefont
  {M~Eric}}, \bibinfo {author} {\bibfnamefont {Alexander}\ \bibnamefont
  {Lukin}}, \bibinfo {author} {\bibfnamefont {Matthew}\ \bibnamefont
  {Rispoli}}, \bibinfo {author} {\bibfnamefont {Robert}\ \bibnamefont
  {Schittko}}, \bibinfo {author} {\bibfnamefont {Tim}\ \bibnamefont {Menke}},
  \bibinfo {author} {\bibfnamefont {Dan}\ \bibnamefont {Borgnia}}, \bibinfo
  {author} {\bibfnamefont {Philipp~M}\ \bibnamefont {Preiss}}, \bibinfo
  {author} {\bibfnamefont {Fabian}\ \bibnamefont {Grusdt}}, \bibinfo {author}
  {\bibfnamefont {Adam~M}\ \bibnamefont {Kaufman}}, \ and\ \bibinfo {author}
  {\bibfnamefont {Markus}\ \bibnamefont {Greiner}}} (\bibinfo {year}
  {2017}{\natexlab{a}}),\ \bibfield  {title} {\enquote {\bibinfo {title}
  {Microscopy of the interacting {H}arper--{H}ofstadter model in the two-body
  limit},}\ }\href {https://www.nature.com/articles/nature22811} {\bibfield
  {journal} {\bibinfo  {journal} {Nature}\ }\textbf {\bibinfo {volume}
  {546}}~(\bibinfo {number} {7659}),\ \bibinfo {pages} {519}}\BibitemShut
  {NoStop}%
\bibitem [{\citenamefont {Tai}\ \emph {et~al.}(2017{\natexlab{b}})\citenamefont
  {Tai}, \citenamefont {Lukin}, \citenamefont {Rispoli}, \citenamefont
  {Schittko}, \citenamefont {Menke}, \citenamefont {Borgnia}, \citenamefont
  {Preiss}, \citenamefont {Grusdt}, \citenamefont {Kaufman},\ and\
  \citenamefont {Greiner}}]{Tai:arxiv2016}%
  \BibitemOpen
  \bibfield  {author} {\bibinfo {author} {\bibnamefont {Tai}, \bibfnamefont
  {M~Eric}}, \bibinfo {author} {\bibfnamefont {Alexander}\ \bibnamefont
  {Lukin}}, \bibinfo {author} {\bibfnamefont {Matthew}\ \bibnamefont
  {Rispoli}}, \bibinfo {author} {\bibfnamefont {Robert}\ \bibnamefont
  {Schittko}}, \bibinfo {author} {\bibfnamefont {Tim}\ \bibnamefont {Menke}},
  \bibinfo {author} {\bibfnamefont {Dan}\ \bibnamefont {Borgnia}}, \bibinfo
  {author} {\bibfnamefont {Philipp~M}\ \bibnamefont {Preiss}}, \bibinfo
  {author} {\bibfnamefont {Fabian}\ \bibnamefont {Grusdt}}, \bibinfo {author}
  {\bibfnamefont {Adam~M}\ \bibnamefont {Kaufman}}, \ and\ \bibinfo {author}
  {\bibfnamefont {Markus}\ \bibnamefont {Greiner}}} (\bibinfo {year}
  {2017}{\natexlab{b}}),\ \bibfield  {title} {\enquote {\bibinfo {title}
  {Microscopy of the interacting {H}arper--{H}ofstadter model in the two-body
  limit},}\ }\href {https://www.nature.com/articles/nature22811} {\bibfield
  {journal} {\bibinfo  {journal} {Nature}\ }\textbf {\bibinfo {volume}
  {546}}~(\bibinfo {number} {7659}),\ \bibinfo {pages} {519}}\BibitemShut
  {NoStop}%
\bibitem [{\citenamefont {Takemura}\ \emph {et~al.}(2014)\citenamefont
  {Takemura}, \citenamefont {Trebaol}, \citenamefont {Wouters}, \citenamefont
  {Portella-Oberli},\ and\ \citenamefont {Deveaud}}]{Takemura:NatPhys2014}%
  \BibitemOpen
  \bibfield  {author} {\bibinfo {author} {\bibnamefont {Takemura},
  \bibfnamefont {Naotomo}}, \bibinfo {author} {\bibfnamefont {St{\'e}phane}\
  \bibnamefont {Trebaol}}, \bibinfo {author} {\bibfnamefont {Michiel}\
  \bibnamefont {Wouters}}, \bibinfo {author} {\bibfnamefont {Marcia~T}\
  \bibnamefont {Portella-Oberli}}, \ and\ \bibinfo {author} {\bibfnamefont
  {Beno{\^\i}t}\ \bibnamefont {Deveaud}}} (\bibinfo {year} {2014}),\ \bibfield
  {title} {\enquote {\bibinfo {title} {Polaritonic {F}eshbach resonance},}\
  }\href {https://www.nature.com/articles/nphys2999} {\bibfield  {journal}
  {\bibinfo  {journal} {Nat. Phys.}\ }\textbf {\bibinfo {volume}
  {10}}~(\bibinfo {number} {7}),\ \bibinfo {pages} {500--504}}\BibitemShut
  {NoStop}%
\bibitem [{\citenamefont {Tan}\ \emph {et~al.}(2014)\citenamefont {Tan},
  \citenamefont {Sun}, \citenamefont {Chen},\ and\ \citenamefont
  {Shen}}]{Tan:2014SciRep}%
  \BibitemOpen
  \bibfield  {author} {\bibinfo {author} {\bibnamefont {Tan}, \bibfnamefont
  {Wei}}, \bibinfo {author} {\bibfnamefont {Yong}\ \bibnamefont {Sun}},
  \bibinfo {author} {\bibfnamefont {Hong}\ \bibnamefont {Chen}}, \ and\
  \bibinfo {author} {\bibfnamefont {Shun-Qing}\ \bibnamefont {Shen}}} (\bibinfo
  {year} {2014}),\ \bibfield  {title} {\enquote {\bibinfo {title} {Photonic
  simulation of topological excitations in metamaterials},}\ }\href
  {https://www.nature.com/articles/srep03842} {\bibfield  {journal} {\bibinfo
  {journal} {Sci. Rep.}\ }\textbf {\bibinfo {volume} {4}},\ \bibinfo {pages}
  {3842}}\BibitemShut {NoStop}%
\bibitem [{\citenamefont {Tanese}\ \emph {et~al.}(2013)\citenamefont {Tanese},
  \citenamefont {Flayac}, \citenamefont {Solnyshkov}, \citenamefont {Amo},
  \citenamefont {Lema{\^\i}tre}, \citenamefont {Galopin}, \citenamefont
  {Braive}, \citenamefont {Senellart}, \citenamefont {Sagnes}, \citenamefont
  {Malpuech} \emph {et~al.}}]{Tanese:NatComm2013}%
  \BibitemOpen
  \bibfield  {author} {\bibinfo {author} {\bibnamefont {Tanese}, \bibfnamefont
  {D}}, \bibinfo {author} {\bibfnamefont {H}~\bibnamefont {Flayac}}, \bibinfo
  {author} {\bibfnamefont {D}~\bibnamefont {Solnyshkov}}, \bibinfo {author}
  {\bibfnamefont {A}~\bibnamefont {Amo}}, \bibinfo {author} {\bibfnamefont
  {A}~\bibnamefont {Lema{\^\i}tre}}, \bibinfo {author} {\bibfnamefont
  {E}~\bibnamefont {Galopin}}, \bibinfo {author} {\bibfnamefont
  {R}~\bibnamefont {Braive}}, \bibinfo {author} {\bibfnamefont {P}~\bibnamefont
  {Senellart}}, \bibinfo {author} {\bibfnamefont {I}~\bibnamefont {Sagnes}},
  \bibinfo {author} {\bibfnamefont {G}~\bibnamefont {Malpuech}},  \emph
  {et~al.}} (\bibinfo {year} {2013}),\ \bibfield  {title} {\enquote {\bibinfo
  {title} {Polariton condensation in solitonic gap states in a one-dimensional
  periodic potential},}\ }\href {https://www.nature.com/articles/ncomms2760}
  {\bibfield  {journal} {\bibinfo  {journal} {Nat. Commun.}\ }\textbf {\bibinfo
  {volume} {4}},\ \bibinfo {pages} {1749}}\BibitemShut {NoStop}%
\bibitem [{\citenamefont {Tangpanitanon}\ \emph {et~al.}(2016)\citenamefont
  {Tangpanitanon}, \citenamefont {Bastidas}, \citenamefont {Al-Assam},
  \citenamefont {Roushan}, \citenamefont {Jaksch},\ and\ \citenamefont
  {Angelakis}}]{Tangpanitanon:PRL2016}%
  \BibitemOpen
  \bibfield  {author} {\bibinfo {author} {\bibnamefont {Tangpanitanon},
  \bibfnamefont {Jirawat}}, \bibinfo {author} {\bibfnamefont {Victor~M.}\
  \bibnamefont {Bastidas}}, \bibinfo {author} {\bibfnamefont {Sarah}\
  \bibnamefont {Al-Assam}}, \bibinfo {author} {\bibfnamefont {Pedram}\
  \bibnamefont {Roushan}}, \bibinfo {author} {\bibfnamefont {Dieter}\
  \bibnamefont {Jaksch}}, \ and\ \bibinfo {author} {\bibfnamefont
  {Dimitris~G.}\ \bibnamefont {Angelakis}}} (\bibinfo {year} {2016}),\
  \bibfield  {title} {\enquote {\bibinfo {title} {Topological pumping of
  photons in nonlinear resonator arrays},}\ }\href
  {https://link.aps.org/doi/10.1103/PhysRevLett.117.213603} {\bibfield
  {journal} {\bibinfo  {journal} {Phys. Rev. Lett.}\ }\textbf {\bibinfo
  {volume} {117}},\ \bibinfo {pages} {213603}}\BibitemShut {NoStop}%
\bibitem [{\citenamefont {Tarasinski}\ \emph {et~al.}(2014)\citenamefont
  {Tarasinski}, \citenamefont {Asb\'oth},\ and\ \citenamefont
  {Dahlhaus}}]{Tarasinski:2014PRA}%
  \BibitemOpen
  \bibfield  {author} {\bibinfo {author} {\bibnamefont {Tarasinski},
  \bibfnamefont {B}}, \bibinfo {author} {\bibfnamefont {J.~K.}\ \bibnamefont
  {Asb\'oth}}, \ and\ \bibinfo {author} {\bibfnamefont {J.~P.}\ \bibnamefont
  {Dahlhaus}}} (\bibinfo {year} {2014}),\ \bibfield  {title} {\enquote
  {\bibinfo {title} {Scattering theory of topological phases in discrete-time
  quantum walks},}\ }\href
  {https://link.aps.org/doi/10.1103/PhysRevA.89.042327} {\bibfield  {journal}
  {\bibinfo  {journal} {Phys. Rev. A}\ }\textbf {\bibinfo {volume} {89}},\
  \bibinfo {pages} {042327}}\BibitemShut {NoStop}%
\bibitem [{\citenamefont {Tarruell}\ \emph {et~al.}(2012)\citenamefont
  {Tarruell}, \citenamefont {Greif}, \citenamefont {Uehlinger}, \citenamefont
  {Jotzu},\ and\ \citenamefont {Esslinger}}]{Tarruell:2012Nature}%
  \BibitemOpen
  \bibfield  {author} {\bibinfo {author} {\bibnamefont {Tarruell},
  \bibfnamefont {Leticia}}, \bibinfo {author} {\bibfnamefont {Daniel}\
  \bibnamefont {Greif}}, \bibinfo {author} {\bibfnamefont {Thomas}\
  \bibnamefont {Uehlinger}}, \bibinfo {author} {\bibfnamefont {Gregor}\
  \bibnamefont {Jotzu}}, \ and\ \bibinfo {author} {\bibfnamefont {Tilman}\
  \bibnamefont {Esslinger}}} (\bibinfo {year} {2012}),\ \bibfield  {title}
  {\enquote {\bibinfo {title} {{Creating, moving and merging {D}irac points
  with a Fermi gas in a tunable honeycomb lattice}},}\ }\href
  {https://www.nature.com/articles/nature10871} {\bibfield  {journal} {\bibinfo
   {journal} {Nature}\ }\textbf {\bibinfo {volume} {483}}~(\bibinfo {number}
  {7389}),\ \bibinfo {pages} {302--305}}\BibitemShut {NoStop}%
\bibitem [{\citenamefont {Teo}\ and\ \citenamefont {Kane}(2010)}]{Teo:2010PRB}%
  \BibitemOpen
  \bibfield  {author} {\bibinfo {author} {\bibnamefont {Teo}, \bibfnamefont
  {Jeffrey C~Y}}, \ and\ \bibinfo {author} {\bibfnamefont {C.~L.}\ \bibnamefont
  {Kane}}} (\bibinfo {year} {2010}),\ \bibfield  {title} {\enquote {\bibinfo
  {title} {Topological defects and gapless modes in insulators and
  superconductors},}\ }\href
  {https://link.aps.org/doi/10.1103/PhysRevB.82.115120} {\bibfield  {journal}
  {\bibinfo  {journal} {Phys. Rev. B}\ }\textbf {\bibinfo {volume} {82}},\
  \bibinfo {pages} {115120}}\BibitemShut {NoStop}%
\bibitem [{\citenamefont {Thouless}(1983)}]{Thouless:1983PRB}%
  \BibitemOpen
  \bibfield  {author} {\bibinfo {author} {\bibnamefont {Thouless},
  \bibfnamefont {D~J}}} (\bibinfo {year} {1983}),\ \bibfield  {title} {\enquote
  {\bibinfo {title} {Quantization of particle transport},}\ }\href
  {https://link.aps.org/doi/10.1103/PhysRevB.27.6083} {\bibfield  {journal}
  {\bibinfo  {journal} {Phys. Rev. B}\ }\textbf {\bibinfo {volume} {27}},\
  \bibinfo {pages} {6083--6087}}\BibitemShut {NoStop}%
\bibitem [{\citenamefont {Thouless}\ \emph {et~al.}(1982)\citenamefont
  {Thouless}, \citenamefont {Kohmoto}, \citenamefont {Nightingale},\ and\
  \citenamefont {den Nijs}}]{Thouless:1982PRL}%
  \BibitemOpen
  \bibfield  {author} {\bibinfo {author} {\bibnamefont {Thouless},
  \bibfnamefont {D~J}}, \bibinfo {author} {\bibfnamefont {M.}~\bibnamefont
  {Kohmoto}}, \bibinfo {author} {\bibfnamefont {M.~P.}\ \bibnamefont
  {Nightingale}}, \ and\ \bibinfo {author} {\bibfnamefont {M.}~\bibnamefont
  {den Nijs}}} (\bibinfo {year} {1982}),\ \bibfield  {title} {\enquote
  {\bibinfo {title} {Quantized {H}all conductance in a two-dimensional periodic
  potential},}\ }\href {https://link.aps.org/doi/10.1103/PhysRevLett.49.405}
  {\bibfield  {journal} {\bibinfo  {journal} {Phys. Rev. Lett.}\ }\textbf
  {\bibinfo {volume} {49}},\ \bibinfo {pages} {405--408}}\BibitemShut {NoStop}%
\bibitem [{\citenamefont {Tomita}\ and\ \citenamefont
  {Chiao}(1986)}]{Tomita:1986PRL}%
  \BibitemOpen
  \bibfield  {author} {\bibinfo {author} {\bibnamefont {Tomita}, \bibfnamefont
  {Akira}}, \ and\ \bibinfo {author} {\bibfnamefont {Raymond~Y.}\ \bibnamefont
  {Chiao}}} (\bibinfo {year} {1986}),\ \bibfield  {title} {\enquote {\bibinfo
  {title} {Observation of {B}erry's topological phase by use of an optical
  fiber},}\ }\href {https://link.aps.org/doi/10.1103/PhysRevLett.57.937}
  {\bibfield  {journal} {\bibinfo  {journal} {Phys. Rev. Lett.}\ }\textbf
  {\bibinfo {volume} {57}},\ \bibinfo {pages} {937--940}}\BibitemShut {NoStop}%
\bibitem [{\citenamefont {Tomka}\ \emph {et~al.}(2015)\citenamefont {Tomka},
  \citenamefont {Pletyukhov},\ and\ \citenamefont
  {Gritsev}}]{Tomka:SciRep2015}%
  \BibitemOpen
  \bibfield  {author} {\bibinfo {author} {\bibnamefont {Tomka}, \bibfnamefont
  {Michael}}, \bibinfo {author} {\bibfnamefont {Mikhail}\ \bibnamefont
  {Pletyukhov}}, \ and\ \bibinfo {author} {\bibfnamefont {Vladimir}\
  \bibnamefont {Gritsev}}} (\bibinfo {year} {2015}),\ \bibfield  {title}
  {\enquote {\bibinfo {title} {Supersymmetry in quantum optics and in
  spin-orbit coupled systems},}\ }\href
  {https://www.nature.com/articles/srep13097} {\bibfield  {journal} {\bibinfo
  {journal} {Sci. Rep.}\ }\textbf {\bibinfo {volume} {5}},\ \bibinfo {pages}
  {13097}}\BibitemShut {NoStop}%
\bibitem [{\citenamefont {{Tong}}(2016)}]{Tong:QHbook}%
  \BibitemOpen
  \bibfield  {author} {\bibinfo {author} {\bibnamefont {{Tong}}, \bibfnamefont
  {D}}} (\bibinfo {year} {2016}),\ \bibfield  {title} {\enquote {\bibinfo
  {title} {Lectures on the quantum {H}all effect},}\ }\href
  {https://arxiv.org/abs/1606.06687} {\bibinfo  {journal} {arXiv:1606.06687}\
  }\BibitemShut {NoStop}%
\bibitem [{\citenamefont {Tsomokos}\ \emph {et~al.}(2010)\citenamefont
  {Tsomokos}, \citenamefont {Ashhab},\ and\ \citenamefont
  {Nori}}]{Tsomokos:2010PRA}%
  \BibitemOpen
\bibfield  {journal} {  }\bibfield  {author} {\bibinfo {author} {\bibnamefont
  {Tsomokos}, \bibfnamefont {Dimitris~I}}, \bibinfo {author} {\bibfnamefont
  {Sahel}\ \bibnamefont {Ashhab}}, \ and\ \bibinfo {author} {\bibfnamefont
  {Franco}\ \bibnamefont {Nori}}} (\bibinfo {year} {2010}),\ \bibfield  {title}
  {\enquote {\bibinfo {title} {Using superconducting qubit circuits to engineer
  exotic lattice systems},}\ }\href
  {https://link.aps.org/doi/10.1103/PhysRevA.82.052311} {\bibfield  {journal}
  {\bibinfo  {journal} {Phys. Rev. A}\ }\textbf {\bibinfo {volume} {82}},\
  \bibinfo {pages} {052311}}\BibitemShut {NoStop}%
\bibitem [{\citenamefont {Tsui}\ \emph {et~al.}(1982)\citenamefont {Tsui},
  \citenamefont {Stormer},\ and\ \citenamefont {Gossard}}]{Tsui:1982PRL}%
  \BibitemOpen
  \bibfield  {author} {\bibinfo {author} {\bibnamefont {Tsui}, \bibfnamefont
  {D~C}}, \bibinfo {author} {\bibfnamefont {H.~L.}\ \bibnamefont {Stormer}}, \
  and\ \bibinfo {author} {\bibfnamefont {A.~C.}\ \bibnamefont {Gossard}}}
  (\bibinfo {year} {1982}),\ \bibfield  {title} {\enquote {\bibinfo {title}
  {Two-dimensional magnetotransport in the extreme quantum limit},}\ }\href
  {https://link.aps.org/doi/10.1103/PhysRevLett.48.1559} {\bibfield  {journal}
  {\bibinfo  {journal} {Phys. Rev. Lett.}\ }\textbf {\bibinfo {volume} {48}},\
  \bibinfo {pages} {1559--1562}}\BibitemShut {NoStop}%
\bibitem [{\citenamefont {T\"ureci}\ \emph {et~al.}(2007)\citenamefont
  {T\"ureci}, \citenamefont {Stone},\ and\ \citenamefont
  {Ge}}]{Tureci:PRA2007}%
  \BibitemOpen
  \bibfield  {author} {\bibinfo {author} {\bibnamefont {T\"ureci},
  \bibfnamefont {Hakan~E}}, \bibinfo {author} {\bibfnamefont {A.~Douglas}\
  \bibnamefont {Stone}}, \ and\ \bibinfo {author} {\bibfnamefont
  {Li}~\bibnamefont {Ge}}} (\bibinfo {year} {2007}),\ \bibfield  {title}
  {\enquote {\bibinfo {title} {Theory of the spatial structure of nonlinear
  lasing modes},}\ }\href {https://link.aps.org/doi/10.1103/PhysRevA.76.013813}
  {\bibfield  {journal} {\bibinfo  {journal} {Phys. Rev. A}\ }\textbf {\bibinfo
  {volume} {76}},\ \bibinfo {pages} {013813}}\BibitemShut {NoStop}%
\bibitem [{\citenamefont {Umucal{\i}lar}\ and\ \citenamefont
  {Carusotto}(2011)}]{Umucalilar:2011PRA}%
  \BibitemOpen
  \bibfield  {author} {\bibinfo {author} {\bibnamefont {Umucal{\i}lar},
  \bibfnamefont {R~O}}, \ and\ \bibinfo {author} {\bibfnamefont
  {I.}~\bibnamefont {Carusotto}}} (\bibinfo {year} {2011}),\ \bibfield  {title}
  {\enquote {\bibinfo {title} {Artificial gauge field for photons in coupled
  cavity arrays},}\ }\href {http://link.aps.org/doi/10.1103/PhysRevA.84.043804}
  {\bibfield  {journal} {\bibinfo  {journal} {Phys. Rev. A}\ }\textbf {\bibinfo
  {volume} {84}},\ \bibinfo {pages} {043804}}\BibitemShut {NoStop}%
\bibitem [{\citenamefont {Umucal{\i}lar}\ and\ \citenamefont
  {Carusotto}(2012)}]{Umucalilar:2012PRL}%
  \BibitemOpen
  \bibfield  {author} {\bibinfo {author} {\bibnamefont {Umucal{\i}lar},
  \bibfnamefont {R~O}}, \ and\ \bibinfo {author} {\bibfnamefont
  {I.}~\bibnamefont {Carusotto}}} (\bibinfo {year} {2012}),\ \bibfield  {title}
  {\enquote {\bibinfo {title} {Fractional quantum {H}all states of photons in
  an array of dissipative coupled cavities},}\ }\href
  {http://link.aps.org/doi/10.1103/PhysRevLett.108.206809} {\bibfield
  {journal} {\bibinfo  {journal} {Phys. Rev. Lett.}\ }\textbf {\bibinfo
  {volume} {108}},\ \bibinfo {pages} {206809}}\BibitemShut {NoStop}%
\bibitem [{\citenamefont {Umucal{\i}lar}\ and\ \citenamefont
  {Carusotto}(2013)}]{Umucalilar:PLA2013}%
  \BibitemOpen
  \bibfield  {author} {\bibinfo {author} {\bibnamefont {Umucal{\i}lar},
  \bibfnamefont {RO}}, \ and\ \bibinfo {author} {\bibfnamefont {I}~\bibnamefont
  {Carusotto}}} (\bibinfo {year} {2013}),\ \bibfield  {title} {\enquote
  {\bibinfo {title} {Many-body braiding phases in a rotating strongly
  correlated photon gas},}\ }\href
  {http://www.sciencedirect.com/science/article/pii/S0375960113005860}
  {\bibfield  {journal} {\bibinfo  {journal} {Phys. Lett. A}\ }\textbf
  {\bibinfo {volume} {377}}~(\bibinfo {number} {34}),\ \bibinfo {pages}
  {2074--2078}}\BibitemShut {NoStop}%
\bibitem [{\citenamefont {Van~Mechelen}\ and\ \citenamefont
  {Jacob}(2016)}]{VanMechelen:2016Optica}%
  \BibitemOpen
  \bibfield  {author} {\bibinfo {author} {\bibnamefont {Van~Mechelen},
  \bibfnamefont {Todd}}, \ and\ \bibinfo {author} {\bibfnamefont {Zubin}\
  \bibnamefont {Jacob}}} (\bibinfo {year} {2016}),\ \bibfield  {title}
  {\enquote {\bibinfo {title} {Universal spin-momentum locking of evanescent
  waves},}\ }\href
  {http://www.osapublishing.org/optica/abstract.cfm?URI=optica-3-2-118}
  {\bibfield  {journal} {\bibinfo  {journal} {Optica}\ }\textbf {\bibinfo
  {volume} {3}}~(\bibinfo {number} {2}),\ \bibinfo {pages}
  {118--126}}\BibitemShut {NoStop}%
\bibitem [{\citenamefont {Verbin}\ \emph {et~al.}(2013)\citenamefont {Verbin},
  \citenamefont {Zilberberg}, \citenamefont {Kraus}, \citenamefont {Lahini},\
  and\ \citenamefont {Silberberg}}]{Verbin:2013}%
  \BibitemOpen
  \bibfield  {author} {\bibinfo {author} {\bibnamefont {Verbin}, \bibfnamefont
  {Mor}}, \bibinfo {author} {\bibfnamefont {Oded}\ \bibnamefont {Zilberberg}},
  \bibinfo {author} {\bibfnamefont {Yaacov~E.}\ \bibnamefont {Kraus}}, \bibinfo
  {author} {\bibfnamefont {Yoav}\ \bibnamefont {Lahini}}, \ and\ \bibinfo
  {author} {\bibfnamefont {Yaron}\ \bibnamefont {Silberberg}}} (\bibinfo {year}
  {2013}),\ \bibfield  {title} {\enquote {\bibinfo {title} {Observation of
  topological phase transitions in photonic quasicrystals},}\ }\href
  {https://link.aps.org/doi/10.1103/PhysRevLett.110.076403} {\bibfield
  {journal} {\bibinfo  {journal} {Phys. Rev. Lett.}\ }\textbf {\bibinfo
  {volume} {110}},\ \bibinfo {pages} {076403}}\BibitemShut {NoStop}%
\bibitem [{\citenamefont {Verbin}\ \emph {et~al.}(2015)\citenamefont {Verbin},
  \citenamefont {Zilberberg}, \citenamefont {Lahini}, \citenamefont {Kraus},\
  and\ \citenamefont {Silberberg}}]{Verbin:2015}%
  \BibitemOpen
  \bibfield  {author} {\bibinfo {author} {\bibnamefont {Verbin}, \bibfnamefont
  {Mor}}, \bibinfo {author} {\bibfnamefont {Oded}\ \bibnamefont {Zilberberg}},
  \bibinfo {author} {\bibfnamefont {Yoav}\ \bibnamefont {Lahini}}, \bibinfo
  {author} {\bibfnamefont {Yaacov~E.}\ \bibnamefont {Kraus}}, \ and\ \bibinfo
  {author} {\bibfnamefont {Yaron}\ \bibnamefont {Silberberg}}} (\bibinfo {year}
  {2015}),\ \bibfield  {title} {\enquote {\bibinfo {title} {Topological pumping
  over a photonic {F}ibonacci quasicrystal},}\ }\href
  {https://link.aps.org/doi/10.1103/PhysRevB.91.064201} {\bibfield  {journal}
  {\bibinfo  {journal} {Phys. Rev. B}\ }\textbf {\bibinfo {volume} {91}},\
  \bibinfo {pages} {064201}}\BibitemShut {NoStop}%
\bibitem [{\citenamefont {Vignolo}\ \emph {et~al.}(2016)\citenamefont
  {Vignolo}, \citenamefont {Bellec}, \citenamefont {B{\"o}hm}, \citenamefont
  {Camara}, \citenamefont {Gambaudo}, \citenamefont {Kuhl},\ and\ \citenamefont
  {Mortessagne}}]{Vignolo:PRB2016}%
  \BibitemOpen
  \bibfield  {author} {\bibinfo {author} {\bibnamefont {Vignolo}, \bibfnamefont
  {Patrizia}}, \bibinfo {author} {\bibfnamefont {Matthieu}\ \bibnamefont
  {Bellec}}, \bibinfo {author} {\bibfnamefont {Julian}\ \bibnamefont
  {B{\"o}hm}}, \bibinfo {author} {\bibfnamefont {Abdoulaye}\ \bibnamefont
  {Camara}}, \bibinfo {author} {\bibfnamefont {Jean-Marc}\ \bibnamefont
  {Gambaudo}}, \bibinfo {author} {\bibfnamefont {Ulrich}\ \bibnamefont {Kuhl}},
  \ and\ \bibinfo {author} {\bibfnamefont {Fabrice}\ \bibnamefont
  {Mortessagne}}} (\bibinfo {year} {2016}),\ \bibfield  {title} {\enquote
  {\bibinfo {title} {Energy landscape in a {P}enrose tiling},}\ }\href
  {https://journals.aps.org/prb/abstract/10.1103/PhysRevB.93.075141} {\bibfield
   {journal} {\bibinfo  {journal} {Phys. Rev. B}\ }\textbf {\bibinfo {volume}
  {93}}~(\bibinfo {number} {7}),\ \bibinfo {pages} {075141}}\BibitemShut
  {NoStop}%
\bibitem [{\citenamefont {Vocke}\ \emph {et~al.}(2015)\citenamefont {Vocke},
  \citenamefont {Roger}, \citenamefont {Marino}, \citenamefont {Wright},
  \citenamefont {Carusotto}, \citenamefont {Clerici},\ and\ \citenamefont
  {Faccio}}]{Vocke:Optica2015}%
  \BibitemOpen
  \bibfield  {author} {\bibinfo {author} {\bibnamefont {Vocke}, \bibfnamefont
  {David}}, \bibinfo {author} {\bibfnamefont {Thomas}\ \bibnamefont {Roger}},
  \bibinfo {author} {\bibfnamefont {Francesco}\ \bibnamefont {Marino}},
  \bibinfo {author} {\bibfnamefont {Ewan~M}\ \bibnamefont {Wright}}, \bibinfo
  {author} {\bibfnamefont {Iacopo}\ \bibnamefont {Carusotto}}, \bibinfo
  {author} {\bibfnamefont {Matteo}\ \bibnamefont {Clerici}}, \ and\ \bibinfo
  {author} {\bibfnamefont {Daniele}\ \bibnamefont {Faccio}}} (\bibinfo {year}
  {2015}),\ \bibfield  {title} {\enquote {\bibinfo {title} {Experimental
  characterization of nonlocal photon fluids},}\ }\href
  {https://www.osapublishing.org/optica/abstract.cfm?uri=optica-2-5-484}
  {\bibfield  {journal} {\bibinfo  {journal} {Optica}\ }\textbf {\bibinfo
  {volume} {2}}~(\bibinfo {number} {5}),\ \bibinfo {pages}
  {484--490}}\BibitemShut {NoStop}%
\bibitem [{\citenamefont {Volovik}(2009)}]{Volovik:2009Book}%
  \BibitemOpen
  \bibfield  {author} {\bibinfo {author} {\bibnamefont {Volovik}, \bibfnamefont
  {G~E}}} (\bibinfo {year} {2009}),\ \href@noop {} {\emph {\bibinfo {title}
  {The Universe in a Helium Droplet}}}\ (\bibinfo  {publisher} {Oxford
  university press},\ \bibinfo {address} {New York})\BibitemShut {NoStop}%
\bibitem [{\citenamefont {Wallace}(1947)}]{Wallace:1947PR}%
  \BibitemOpen
  \bibfield  {author} {\bibinfo {author} {\bibnamefont {Wallace}, \bibfnamefont
  {P~R}}} (\bibinfo {year} {1947}),\ \bibfield  {title} {\enquote {\bibinfo
  {title} {The band theory of graphite},}\ }\href
  {https://link.aps.org/doi/10.1103/PhysRev.71.622} {\bibfield  {journal}
  {\bibinfo  {journal} {Phys. Rev.}\ }\textbf {\bibinfo {volume} {71}},\
  \bibinfo {pages} {622--634}}\BibitemShut {NoStop}%
\bibitem [{\citenamefont {Walls}\ and\ \citenamefont
  {Milburn}(2006)}]{QuantumOptics}%
  \BibitemOpen
  \bibfield  {author} {\bibinfo {author} {\bibnamefont {Walls}, \bibfnamefont
  {D~F}}, \ and\ \bibinfo {author} {\bibfnamefont {G.}~\bibnamefont {Milburn}}}
  (\bibinfo {year} {2006}),\ \href@noop {} {\emph {\bibinfo {title} {Quantum
  Optics}}}\ (\bibinfo  {publisher} {Springer Verlag, Berlin})\BibitemShut
  {NoStop}%
\bibitem [{\citenamefont {Wan}\ \emph {et~al.}(2011)\citenamefont {Wan},
  \citenamefont {Turner}, \citenamefont {Vishwanath},\ and\ \citenamefont
  {Savrasov}}]{Wan:2011PRB}%
  \BibitemOpen
  \bibfield  {author} {\bibinfo {author} {\bibnamefont {Wan}, \bibfnamefont
  {Xiangang}}, \bibinfo {author} {\bibfnamefont {Ari~M}\ \bibnamefont
  {Turner}}, \bibinfo {author} {\bibfnamefont {Ashvin}\ \bibnamefont
  {Vishwanath}}, \ and\ \bibinfo {author} {\bibfnamefont {Sergey~Y}\
  \bibnamefont {Savrasov}}} (\bibinfo {year} {2011}),\ \bibfield  {title}
  {\enquote {\bibinfo {title} {Topological semimetal and {F}ermi-arc surface
  states in the electronic structure of pyrochlore iridates},}\ }\href
  {https://journals.aps.org/prb/abstract/10.1103/PhysRevB.83.205101} {\bibfield
   {journal} {\bibinfo  {journal} {Phys. Rev. B}\ }\textbf {\bibinfo {volume}
  {83}}~(\bibinfo {number} {20}),\ \bibinfo {pages} {205101}}\BibitemShut
  {NoStop}%
\bibitem [{\citenamefont {Wang}\ \emph
  {et~al.}(2015{\natexlab{a}})\citenamefont {Wang}, \citenamefont {Qiu},
  \citenamefont {Rakich},\ and\ \citenamefont {Wang}}]{Wang:2015CLEO}%
  \BibitemOpen
  \bibfield  {author} {\bibinfo {author} {\bibnamefont {Wang}, \bibfnamefont
  {Danlu}}, \bibinfo {author} {\bibfnamefont {Chengwei}\ \bibnamefont {Qiu}},
  \bibinfo {author} {\bibfnamefont {Peter~T}\ \bibnamefont {Rakich}}, \ and\
  \bibinfo {author} {\bibfnamefont {Zheng}\ \bibnamefont {Wang}}} (\bibinfo
  {year} {2015}{\natexlab{a}}),\ \bibfield  {title} {\enquote {\bibinfo {title}
  {Guide-wave photonic pulling force using one-way photonic chiral edge
  states},}\ }in\ \href
  {https://www.osapublishing.org/abstract.cfm?uri=CLEO_QELS-2015-FM2D.7} {\emph
  {\bibinfo {booktitle} {CLEO: QELS\_Fundamental Science}}}\ (\bibinfo
  {organization} {Optical Society of America})\ pp.\ \bibinfo {pages}
  {FM2D--7}\BibitemShut {NoStop}%
\bibitem [{\citenamefont {Wang}\ \emph
  {et~al.}(2017{\natexlab{a}})\citenamefont {Wang}, \citenamefont {Chen},
  \citenamefont {Hang}, \citenamefont {Kee},\ and\ \citenamefont
  {Jiang}}]{Wang:2017arXiv}%
  \BibitemOpen
  \bibfield  {author} {\bibinfo {author} {\bibnamefont {Wang}, \bibfnamefont
  {Hai-Xiao}}, \bibinfo {author} {\bibfnamefont {Yige}\ \bibnamefont {Chen}},
  \bibinfo {author} {\bibfnamefont {Zhi~Hong}\ \bibnamefont {Hang}}, \bibinfo
  {author} {\bibfnamefont {Hae-Young}\ \bibnamefont {Kee}}, \ and\ \bibinfo
  {author} {\bibfnamefont {Jian-Hua}\ \bibnamefont {Jiang}}} (\bibinfo {year}
  {2017}{\natexlab{a}}),\ \bibfield  {title} {\enquote {\bibinfo {title}
  {Type-{II} {D}irac photons},}\ }\href
  {https://www.nature.com/articles/s41535-017-0058-z} {\bibfield  {journal}
  {\bibinfo  {journal} {npj Quantum Mater.}\ }\textbf {\bibinfo {volume}
  {2}}~(\bibinfo {number} {1}),\ \bibinfo {pages} {54}}\BibitemShut {NoStop}%
\bibitem [{\citenamefont {Wang}\ \emph
  {et~al.}(2016{\natexlab{a}})\citenamefont {Wang}, \citenamefont {Zhou},\ and\
  \citenamefont {Chong}}]{Wang:2016PRBa}%
  \BibitemOpen
  \bibfield  {author} {\bibinfo {author} {\bibnamefont {Wang}, \bibfnamefont
  {Hailong}}, \bibinfo {author} {\bibfnamefont {Longwen}\ \bibnamefont {Zhou}},
  \ and\ \bibinfo {author} {\bibfnamefont {YD}~\bibnamefont {Chong}}} (\bibinfo
  {year} {2016}{\natexlab{a}}),\ \bibfield  {title} {\enquote {\bibinfo {title}
  {Floquet {W}eyl phases in a three-dimensional network model},}\ }\href
  {https://journals.aps.org/prb/abstract/10.1103/PhysRevB.93.144114} {\bibfield
   {journal} {\bibinfo  {journal} {Phys. Rev. B}\ }\textbf {\bibinfo {volume}
  {93}}~(\bibinfo {number} {14}),\ \bibinfo {pages} {144114}}\BibitemShut
  {NoStop}%
\bibitem [{\citenamefont {Wang}\ \emph
  {et~al.}(2016{\natexlab{b}})\citenamefont {Wang}, \citenamefont {Xu},
  \citenamefont {Chen},\ and\ \citenamefont {Jiang}}]{Wang:2016PRBb}%
  \BibitemOpen
  \bibfield  {author} {\bibinfo {author} {\bibnamefont {Wang}, \bibfnamefont
  {HaiXiao}}, \bibinfo {author} {\bibfnamefont {Lin}\ \bibnamefont {Xu}},
  \bibinfo {author} {\bibfnamefont {HuanYang}\ \bibnamefont {Chen}}, \ and\
  \bibinfo {author} {\bibfnamefont {Jian-Hua}\ \bibnamefont {Jiang}}} (\bibinfo
  {year} {2016}{\natexlab{b}}),\ \bibfield  {title} {\enquote {\bibinfo {title}
  {Three-dimensional photonic {D}irac points stabilized by point group
  symmetry},}\ }\href {https://link.aps.org/doi/10.1103/PhysRevB.93.235155}
  {\bibfield  {journal} {\bibinfo  {journal} {Phys. Rev. B}\ }\textbf {\bibinfo
  {volume} {93}},\ \bibinfo {pages} {235155}}\BibitemShut {NoStop}%
\bibitem [{\citenamefont {Wang}\ \emph
  {et~al.}(2016{\natexlab{c}})\citenamefont {Wang}, \citenamefont {Jian},\ and\
  \citenamefont {Yao}}]{Wang:2016PRA}%
  \BibitemOpen
  \bibfield  {author} {\bibinfo {author} {\bibnamefont {Wang}, \bibfnamefont
  {Luyang}}, \bibinfo {author} {\bibfnamefont {Shao-Kai}\ \bibnamefont {Jian}},
  \ and\ \bibinfo {author} {\bibfnamefont {Hong}\ \bibnamefont {Yao}}}
  (\bibinfo {year} {2016}{\natexlab{c}}),\ \bibfield  {title} {\enquote
  {\bibinfo {title} {Topological photonic crystal with equifrequency {W}eyl
  points},}\ }\href@noop {} {\bibfield  {journal} {\bibinfo  {journal} {Phys.
  Rev. A}\ }\textbf {\bibinfo {volume} {93}}~(\bibinfo {number} {6}),\ \bibinfo
  {pages} {061801}}\BibitemShut {NoStop}%
\bibitem [{\citenamefont {Wang}\ \emph
  {et~al.}(2017{\natexlab{b}})\citenamefont {Wang}, \citenamefont {Xiao},
  \citenamefont {Liu}, \citenamefont {Zhu},\ and\ \citenamefont
  {Chan}}]{Wang:2017PRX}%
  \BibitemOpen
  \bibfield  {author} {\bibinfo {author} {\bibnamefont {Wang}, \bibfnamefont
  {Qiang}}, \bibinfo {author} {\bibfnamefont {Meng}\ \bibnamefont {Xiao}},
  \bibinfo {author} {\bibfnamefont {Hui}\ \bibnamefont {Liu}}, \bibinfo
  {author} {\bibfnamefont {Shining}\ \bibnamefont {Zhu}}, \ and\ \bibinfo
  {author} {\bibfnamefont {CT}~\bibnamefont {Chan}}} (\bibinfo {year}
  {2017}{\natexlab{b}}),\ \bibfield  {title} {\enquote {\bibinfo {title}
  {Optical interface states protected by synthetic {W}eyl points},}\ }\href
  {https://journals.aps.org/prx/abstract/10.1103/PhysRevX.7.031032} {\bibfield
  {journal} {\bibinfo  {journal} {Phys. Rev. X}\ }\textbf {\bibinfo {volume}
  {7}}~(\bibinfo {number} {3}),\ \bibinfo {pages} {031032}}\BibitemShut
  {NoStop}%
\bibitem [{\citenamefont {Wang}\ \emph
  {et~al.}(2015{\natexlab{b}})\citenamefont {Wang}, \citenamefont {Wang},
  \citenamefont {Xue}, \citenamefont {Yang}, \citenamefont {Hu},\ and\
  \citenamefont {Wu}}]{Wang:2015Scientificreports}%
  \BibitemOpen
  \bibfield  {author} {\bibinfo {author} {\bibnamefont {Wang}, \bibfnamefont
  {Yan-Pu}}, \bibinfo {author} {\bibfnamefont {Wei}\ \bibnamefont {Wang}},
  \bibinfo {author} {\bibfnamefont {Zheng-Yuan}\ \bibnamefont {Xue}}, \bibinfo
  {author} {\bibfnamefont {Wan-Li}\ \bibnamefont {Yang}}, \bibinfo {author}
  {\bibfnamefont {Yong}\ \bibnamefont {Hu}}, \ and\ \bibinfo {author}
  {\bibfnamefont {Ying}\ \bibnamefont {Wu}}} (\bibinfo {year}
  {2015}{\natexlab{b}}),\ \bibfield  {title} {\enquote {\bibinfo {title}
  {Realizing and characterizing chiral photon flow in a circuit quantum
  electrodynamics necklace},}\ }\href
  {https://www.nature.com/articles/srep08352} {\bibfield  {journal} {\bibinfo
  {journal} {Sci. Rep.}\ }\textbf {\bibinfo {volume} {5}},\ \bibinfo {pages}
  {8352}}\BibitemShut {NoStop}%
\bibitem [{\citenamefont {Wang}\ \emph
  {et~al.}(2016{\natexlab{d}})\citenamefont {Wang}, \citenamefont {Yang},
  \citenamefont {Hu}, \citenamefont {Xue},\ and\ \citenamefont
  {Wu}}]{Wang:2016NPJ}%
  \BibitemOpen
  \bibfield  {author} {\bibinfo {author} {\bibnamefont {Wang}, \bibfnamefont
  {Yan-pu}}, \bibinfo {author} {\bibfnamefont {Wan-li}\ \bibnamefont {Yang}},
  \bibinfo {author} {\bibfnamefont {Yong}\ \bibnamefont {Hu}}, \bibinfo
  {author} {\bibfnamefont {Zheng-yuan}\ \bibnamefont {Xue}}, \ and\ \bibinfo
  {author} {\bibfnamefont {Ying}\ \bibnamefont {Wu}}} (\bibinfo {year}
  {2016}{\natexlab{d}}),\ \bibfield  {title} {\enquote {\bibinfo {title}
  {Detecting topological phases of microwave photons in a circuit quantum
  electrodynamics lattice},}\ }\href
  {https://www.nature.com/articles/npjqi201615} {\bibfield  {journal} {\bibinfo
   {journal} {NPJ Quantum Inf.}\ }\textbf {\bibinfo {volume} {2}},\ \bibinfo
  {pages} {16015}}\BibitemShut {NoStop}%
\bibitem [{\citenamefont {Wang}\ \emph {et~al.}(2008)\citenamefont {Wang},
  \citenamefont {Chong}, \citenamefont {Joannopoulos},\ and\ \citenamefont
  {Solja\ifmmode \check{c}\else \v{c}\fi{}i\ifmmode~\acute{c}\else
  \'{c}\fi{}}}]{Wang:2008PRL}%
  \BibitemOpen
  \bibfield  {author} {\bibinfo {author} {\bibnamefont {Wang}, \bibfnamefont
  {Zheng}}, \bibinfo {author} {\bibfnamefont {Y.~D.}\ \bibnamefont {Chong}},
  \bibinfo {author} {\bibfnamefont {John~D.}\ \bibnamefont {Joannopoulos}}, \
  and\ \bibinfo {author} {\bibfnamefont {Marin}\ \bibnamefont {Solja\ifmmode
  \check{c}\else \v{c}\fi{}i\ifmmode~\acute{c}\else \'{c}\fi{}}}} (\bibinfo
  {year} {2008}),\ \bibfield  {title} {\enquote {\bibinfo {title}
  {Reflection-free one-way edge modes in a gyromagnetic photonic crystal},}\
  }\href {http://link.aps.org/doi/10.1103/PhysRevLett.100.013905} {\bibfield
  {journal} {\bibinfo  {journal} {Phys. Rev. Lett.}\ }\textbf {\bibinfo
  {volume} {100}},\ \bibinfo {pages} {013905}}\BibitemShut {NoStop}%
\bibitem [{\citenamefont {Wang}\ \emph {et~al.}(2009)\citenamefont {Wang},
  \citenamefont {Chong}, \citenamefont {Joannopoulos},\ and\ \citenamefont
  {Solja{\v{c}}i{\'c}}}]{Wang:2009Nature}%
  \BibitemOpen
  \bibfield  {author} {\bibinfo {author} {\bibnamefont {Wang}, \bibfnamefont
  {Zheng}}, \bibinfo {author} {\bibfnamefont {Yidong}\ \bibnamefont {Chong}},
  \bibinfo {author} {\bibfnamefont {JD}~\bibnamefont {Joannopoulos}}, \ and\
  \bibinfo {author} {\bibfnamefont {Marin}\ \bibnamefont {Solja{\v{c}}i{\'c}}}}
  (\bibinfo {year} {2009}),\ \bibfield  {title} {\enquote {\bibinfo {title}
  {Observation of unidirectional backscattering-immune topological
  electromagnetic states},}\ }\href
  {http://www.nature.com/nature/journal/v461/n7265/full/nature08293.html}
  {\bibfield  {journal} {\bibinfo  {journal} {Nature}\ }\textbf {\bibinfo
  {volume} {461}}~(\bibinfo {number} {7265}),\ \bibinfo {pages}
  {772--775}}\BibitemShut {NoStop}%
\bibitem [{\citenamefont {Wang}\ \emph {et~al.}(2013)\citenamefont {Wang},
  \citenamefont {Shen}, \citenamefont {Yu}, \citenamefont {Zhang},\ and\
  \citenamefont {Zheng}}]{Wang:2013JOSAB}%
  \BibitemOpen
  \bibfield  {author} {\bibinfo {author} {\bibnamefont {Wang}, \bibfnamefont
  {Zhuoyuan}}, \bibinfo {author} {\bibfnamefont {Linfang}\ \bibnamefont
  {Shen}}, \bibinfo {author} {\bibfnamefont {Zaihe}\ \bibnamefont {Yu}},
  \bibinfo {author} {\bibfnamefont {Xianmin}\ \bibnamefont {Zhang}}, \ and\
  \bibinfo {author} {\bibfnamefont {Xiaodong}\ \bibnamefont {Zheng}}} (\bibinfo
  {year} {2013}),\ \bibfield  {title} {\enquote {\bibinfo {title} {Highly
  efficient photonic-crystal splitters based on one-way waveguiding},}\ }\href
  {https://www.osapublishing.org/josab/abstract.cfm?uri=josab-30-1-173}
  {\bibfield  {journal} {\bibinfo  {journal} {J. Opt. Soc. Am. B}\ }\textbf
  {\bibinfo {volume} {30}}~(\bibinfo {number} {1}),\ \bibinfo {pages}
  {173--176}}\BibitemShut {NoStop}%
\bibitem [{\citenamefont {Wang}\ \emph {et~al.}(2011)\citenamefont {Wang},
  \citenamefont {Shen}, \citenamefont {Zhang}, \citenamefont {Wang},
  \citenamefont {Yu},\ and\ \citenamefont {Zheng}}]{Wang:2011JAP}%
  \BibitemOpen
  \bibfield  {author} {\bibinfo {author} {\bibnamefont {Wang}, \bibfnamefont
  {Zhuoyuan}}, \bibinfo {author} {\bibfnamefont {Linfang}\ \bibnamefont
  {Shen}}, \bibinfo {author} {\bibfnamefont {Xianmin}\ \bibnamefont {Zhang}},
  \bibinfo {author} {\bibfnamefont {Yigang}\ \bibnamefont {Wang}}, \bibinfo
  {author} {\bibfnamefont {Zaihe}\ \bibnamefont {Yu}}, \ and\ \bibinfo {author}
  {\bibfnamefont {Xiaodong}\ \bibnamefont {Zheng}}} (\bibinfo {year} {2011}),\
  \bibfield  {title} {\enquote {\bibinfo {title} {Photonic crystal cavity with
  one-way rotating state and its coupling with photonic crystal waveguide},}\
  }\href@noop {} {\bibfield  {journal} {\bibinfo  {journal} {Journal of Applied
  Physics}\ }\textbf {\bibinfo {volume} {110}}~(\bibinfo {number} {4}),\
  \bibinfo {pages} {043106}}\BibitemShut {NoStop}%
\bibitem [{\citenamefont {Watanabe}\ and\ \citenamefont
  {Lu}(2018)}]{Watanabe:2018PRL}%
  \BibitemOpen
  \bibfield  {author} {\bibinfo {author} {\bibnamefont {Watanabe},
  \bibfnamefont {Haruki}}, \ and\ \bibinfo {author} {\bibfnamefont {Ling}\
  \bibnamefont {Lu}}} (\bibinfo {year} {2018}),\ \bibfield  {title} {\enquote
  {\bibinfo {title} {Space group theory of photonic bands},}\ }\href
  {https://link.aps.org/doi/10.1103/PhysRevLett.121.263903} {\bibfield
  {journal} {\bibinfo  {journal} {Phys. Rev. Lett.}\ }\textbf {\bibinfo
  {volume} {121}},\ \bibinfo {pages} {263903}}\BibitemShut {NoStop}%
\bibitem [{\citenamefont {Waterstraat}(2016)}]{Waterstraat}%
  \BibitemOpen
  \bibfield  {author} {\bibinfo {author} {\bibnamefont {Waterstraat},
  \bibfnamefont {Nils}}} (\bibinfo {year} {2016}),\ \href@noop {} {\enquote
  {\bibinfo {title} {Fredholm operators and spectral flow},}\ }\Eprint
  {http://arxiv.org/abs/1603.02009} {arXiv:1603.02009} \BibitemShut {NoStop}%
\bibitem [{\citenamefont {Weimann}\ \emph {et~al.}(2017)\citenamefont
  {Weimann}, \citenamefont {Kremer}, \citenamefont {Plotnik}, \citenamefont
  {Lumer}, \citenamefont {Nolte}, \citenamefont {Makris}, \citenamefont
  {Segev}, \citenamefont {Rechtsman},\ and\ \citenamefont
  {Szameit}}]{Weimann:2017NatMat}%
  \BibitemOpen
  \bibfield  {author} {\bibinfo {author} {\bibnamefont {Weimann}, \bibfnamefont
  {S}}, \bibinfo {author} {\bibfnamefont {M}~\bibnamefont {Kremer}}, \bibinfo
  {author} {\bibfnamefont {Y}~\bibnamefont {Plotnik}}, \bibinfo {author}
  {\bibfnamefont {Y}~\bibnamefont {Lumer}}, \bibinfo {author} {\bibfnamefont
  {S}~\bibnamefont {Nolte}}, \bibinfo {author} {\bibfnamefont {KG}~\bibnamefont
  {Makris}}, \bibinfo {author} {\bibfnamefont {M}~\bibnamefont {Segev}},
  \bibinfo {author} {\bibfnamefont {MC}~\bibnamefont {Rechtsman}}, \ and\
  \bibinfo {author} {\bibfnamefont {A}~\bibnamefont {Szameit}}} (\bibinfo
  {year} {2017}),\ \bibfield  {title} {\enquote {\bibinfo {title}
  {Topologically protected bound states in photonic parity-time-symmetric
  crystals},}\ }\href
  {http://www.nature.com/nmat/journal/v16/n4/abs/nmat4811.html} {\bibfield
  {journal} {\bibinfo  {journal} {Nat. Mater.}\ }\textbf {\bibinfo {volume}
  {16}}~(\bibinfo {number} {4}),\ \bibinfo {pages} {433--438}}\BibitemShut
  {NoStop}%
\bibitem [{\citenamefont {Wen}\ and\ \citenamefont {Zee}(1992)}]{wen1992shift}%
  \BibitemOpen
  \bibfield  {author} {\bibinfo {author} {\bibnamefont {Wen}, \bibfnamefont
  {XG}}, \ and\ \bibinfo {author} {\bibfnamefont {A}~\bibnamefont {Zee}}}
  (\bibinfo {year} {1992}),\ \bibfield  {title} {\enquote {\bibinfo {title}
  {Shift and spin vector: {N}ew topological quantum numbers for the {H}all
  fluids},}\ }\href
  {https://journals.aps.org/prl/abstract/10.1103/PhysRevLett.69.953} {\bibfield
   {journal} {\bibinfo  {journal} {Phys. Rev. Lett.}\ }\textbf {\bibinfo
  {volume} {69}}~(\bibinfo {number} {6}),\ \bibinfo {pages} {953}}\BibitemShut
  {NoStop}%
\bibitem [{\citenamefont {Wertz}\ \emph {et~al.}(2010)\citenamefont {Wertz},
  \citenamefont {Ferrier}, \citenamefont {Solnyshkov}, \citenamefont {Johne},
  \citenamefont {Sanvitto}, \citenamefont {Lemaitre}, \citenamefont {Sagnes},
  \citenamefont {Grousson}, \citenamefont {Kavokin}, \citenamefont {Senellart},
  \citenamefont {Malpuech},\ and\ \citenamefont {Bloch}}]{Wertz:NatPhys2010}%
  \BibitemOpen
  \bibfield  {author} {\bibinfo {author} {\bibnamefont {Wertz}, \bibfnamefont
  {E}}, \bibinfo {author} {\bibfnamefont {L.}~\bibnamefont {Ferrier}}, \bibinfo
  {author} {\bibfnamefont {D.~D.}\ \bibnamefont {Solnyshkov}}, \bibinfo
  {author} {\bibfnamefont {R.}~\bibnamefont {Johne}}, \bibinfo {author}
  {\bibfnamefont {D.}~\bibnamefont {Sanvitto}}, \bibinfo {author}
  {\bibfnamefont {A.}~\bibnamefont {Lemaitre}}, \bibinfo {author}
  {\bibfnamefont {I.}~\bibnamefont {Sagnes}}, \bibinfo {author} {\bibfnamefont
  {R.}~\bibnamefont {Grousson}}, \bibinfo {author} {\bibfnamefont {A.~V.}\
  \bibnamefont {Kavokin}}, \bibinfo {author} {\bibfnamefont {P.}~\bibnamefont
  {Senellart}}, \bibinfo {author} {\bibfnamefont {G.}~\bibnamefont {Malpuech}},
  \ and\ \bibinfo {author} {\bibfnamefont {J.}~\bibnamefont {Bloch}}} (\bibinfo
  {year} {2010}),\ \bibfield  {title} {\enquote {\bibinfo {title} {Spontaneous
  formation and optical manipulation of extended polariton condensates},}\
  }\href {https://www.nature.com/articles/nphys1750} {\bibfield  {journal}
  {\bibinfo  {journal} {Nat. Phys.}\ }\textbf {\bibinfo {volume} {6}}~(\bibinfo
  {number} {11}),\ \bibinfo {pages} {860--864}}\BibitemShut {NoStop}%
\bibitem [{\citenamefont {Wimmer}\ \emph {et~al.}(2017)\citenamefont {Wimmer},
  \citenamefont {Price}, \citenamefont {Carusotto},\ and\ \citenamefont
  {Peschel}}]{Wimmer:2017NatPhys}%
  \BibitemOpen
  \bibfield  {author} {\bibinfo {author} {\bibnamefont {Wimmer}, \bibfnamefont
  {Martin}}, \bibinfo {author} {\bibfnamefont {Hannah~M}\ \bibnamefont
  {Price}}, \bibinfo {author} {\bibfnamefont {Iacopo}\ \bibnamefont
  {Carusotto}}, \ and\ \bibinfo {author} {\bibfnamefont {Ulf}\ \bibnamefont
  {Peschel}}} (\bibinfo {year} {2017}),\ \bibfield  {title} {\enquote {\bibinfo
  {title} {Experimental measurement of the {B}erry curvature from anomalous
  transport},}\ }\href {https://www.nature.com/articles/nphys4050} {\bibfield
  {journal} {\bibinfo  {journal} {Nat. Phys.}\ }\textbf {\bibinfo {volume}
  {13}}~(\bibinfo {number} {6}),\ \bibinfo {pages} {545--550}}\BibitemShut
  {NoStop}%
\bibitem [{\citenamefont {Wouters}(2007)}]{Wouters:PRB2007}%
  \BibitemOpen
  \bibfield  {author} {\bibinfo {author} {\bibnamefont {Wouters}, \bibfnamefont
  {Michiel}}} (\bibinfo {year} {2007}),\ \bibfield  {title} {\enquote {\bibinfo
  {title} {Resonant polariton-polariton scattering in semiconductor
  microcavities},}\ }\href
  {https://journals.aps.org/prb/abstract/10.1103/PhysRevB.76.045319} {\bibfield
   {journal} {\bibinfo  {journal} {Phys. Rev. B}\ }\textbf {\bibinfo {volume}
  {76}}~(\bibinfo {number} {4})}\BibitemShut {NoStop}%
\bibitem [{\citenamefont {Wouters}\ \emph {et~al.}(2008)\citenamefont
  {Wouters}, \citenamefont {Carusotto},\ and\ \citenamefont
  {Ciuti}}]{Wouters:PRB2008}%
  \BibitemOpen
  \bibfield  {author} {\bibinfo {author} {\bibnamefont {Wouters}, \bibfnamefont
  {Michiel}}, \bibinfo {author} {\bibfnamefont {Iacopo}\ \bibnamefont
  {Carusotto}}, \ and\ \bibinfo {author} {\bibfnamefont {Cristiano}\
  \bibnamefont {Ciuti}}} (\bibinfo {year} {2008}),\ \bibfield  {title}
  {\enquote {\bibinfo {title} {Spatial and spectral shape of inhomogeneous
  nonequilibrium exciton-polariton condensates},}\ }\href
  {http://link.aps.org/doi/10.1103/PhysRevB.77.115340} {\bibfield  {journal}
  {\bibinfo  {journal} {Phys. Rev. B}\ }\textbf {\bibinfo {volume} {77}},\
  \bibinfo {pages} {115340}}\BibitemShut {NoStop}%
\bibitem [{\citenamefont {Wu}\ \emph {et~al.}(2007)\citenamefont {Wu},
  \citenamefont {Bergman}, \citenamefont {Balents},\ and\ \citenamefont {{Das
  Sarma}}}]{Wu2007:PRL}%
  \BibitemOpen
  \bibfield  {author} {\bibinfo {author} {\bibnamefont {Wu}, \bibfnamefont
  {Congjun}}, \bibinfo {author} {\bibfnamefont {Doron}\ \bibnamefont
  {Bergman}}, \bibinfo {author} {\bibfnamefont {Leon}\ \bibnamefont {Balents}},
  \ and\ \bibinfo {author} {\bibfnamefont {S.}~\bibnamefont {{Das Sarma}}}}
  (\bibinfo {year} {2007}),\ \bibfield  {title} {\enquote {\bibinfo {title}
  {{Flat bands and Wigner crystallization in the honeycomb optical lattice}},}\
  }\href {http://link.aps.org/doi/10.1103/PhysRevLett.99.070401} {\bibfield
  {journal} {\bibinfo  {journal} {Phys. Rev. Lett.}\ }\textbf {\bibinfo
  {volume} {99}}~(\bibinfo {number} {7}),\ \bibinfo {pages}
  {070401}}\BibitemShut {NoStop}%
\bibitem [{\citenamefont {Wu}\ and\ \citenamefont {Hu}(2015)}]{Wu:2015PRL}%
  \BibitemOpen
  \bibfield  {author} {\bibinfo {author} {\bibnamefont {Wu}, \bibfnamefont
  {Long-Hua}}, \ and\ \bibinfo {author} {\bibfnamefont {Xiao}\ \bibnamefont
  {Hu}}} (\bibinfo {year} {2015}),\ \bibfield  {title} {\enquote {\bibinfo
  {title} {Scheme for achieving a topological photonic crystal by using
  dielectric material},}\ }\href
  {http://link.aps.org/doi/10.1103/PhysRevLett.114.223901} {\bibfield
  {journal} {\bibinfo  {journal} {Phys. Rev. Lett.}\ }\textbf {\bibinfo
  {volume} {114}},\ \bibinfo {pages} {223901}}\BibitemShut {NoStop}%
\bibitem [{\citenamefont {Wu}\ \emph {et~al.}(2017{\natexlab{a}})\citenamefont
  {Wu}, \citenamefont {Meng}, \citenamefont {Tian}, \citenamefont {Huang},
  \citenamefont {Xiang}, \citenamefont {Han},\ and\ \citenamefont
  {Wen}}]{Wu:arx2017}%
  \BibitemOpen
  \bibfield  {author} {\bibinfo {author} {\bibnamefont {Wu}, \bibfnamefont
  {Xiaoxiao}}, \bibinfo {author} {\bibfnamefont {Yan}\ \bibnamefont {Meng}},
  \bibinfo {author} {\bibfnamefont {Jingxuan}\ \bibnamefont {Tian}}, \bibinfo
  {author} {\bibfnamefont {Yingzhou}\ \bibnamefont {Huang}}, \bibinfo {author}
  {\bibfnamefont {Hong}\ \bibnamefont {Xiang}}, \bibinfo {author}
  {\bibfnamefont {Dezhuan}\ \bibnamefont {Han}}, \ and\ \bibinfo {author}
  {\bibfnamefont {Weijia}\ \bibnamefont {Wen}}} (\bibinfo {year}
  {2017}{\natexlab{a}}),\ \bibfield  {title} {\enquote {\bibinfo {title}
  {Direct observation of valley-polarized topological edge states in designer
  surface plasmon crystals},}\ }\href
  {https://www.nature.com/articles/s41467-017-01515-2} {\bibfield  {journal}
  {\bibinfo  {journal} {Nat. Commun.}\ }\textbf {\bibinfo {volume}
  {8}}~(\bibinfo {number} {1}),\ \bibinfo {pages} {1304}}\BibitemShut {NoStop}%
\bibitem [{\citenamefont {Wu}\ \emph {et~al.}(2017{\natexlab{b}})\citenamefont
  {Wu}, \citenamefont {Li}, \citenamefont {Hu}, \citenamefont {Ao},
  \citenamefont {Zhao},\ and\ \citenamefont {Gong}}]{Wu:2017AOM}%
  \BibitemOpen
  \bibfield  {author} {\bibinfo {author} {\bibnamefont {Wu}, \bibfnamefont
  {You}}, \bibinfo {author} {\bibfnamefont {Chong}\ \bibnamefont {Li}},
  \bibinfo {author} {\bibfnamefont {Xiaoyong}\ \bibnamefont {Hu}}, \bibinfo
  {author} {\bibfnamefont {Yutian}\ \bibnamefont {Ao}}, \bibinfo {author}
  {\bibfnamefont {Yifan}\ \bibnamefont {Zhao}}, \ and\ \bibinfo {author}
  {\bibfnamefont {Qihuang}\ \bibnamefont {Gong}}} (\bibinfo {year}
  {2017}{\natexlab{b}}),\ \bibfield  {title} {\enquote {\bibinfo {title}
  {Applications of topological photonics in integrated photonic devices},}\
  }\href {http://onlinelibrary.wiley.com/doi/10.1002/adom.201700357/abstract}
  {\bibfield  {journal} {\bibinfo  {journal} {Advanced Optical Materials}\
  }\textbf {\bibinfo {volume} {5}},\ \bibinfo {pages} {1700357}}\BibitemShut
  {NoStop}%
\bibitem [{\citenamefont {Xia}\ \emph {et~al.}(2017)\citenamefont {Xia},
  \citenamefont {Liu}, \citenamefont {Huang}, \citenamefont {Dai},
  \citenamefont {Jiao}, \citenamefont {Zang}, \citenamefont {Yu}, \citenamefont
  {Zheng},\ and\ \citenamefont {Liu}}]{Xia:PRB2017}%
  \BibitemOpen
  \bibfield  {author} {\bibinfo {author} {\bibnamefont {Xia}, \bibfnamefont
  {Bai-Zhan}}, \bibinfo {author} {\bibfnamefont {Ting-Ting}\ \bibnamefont
  {Liu}}, \bibinfo {author} {\bibfnamefont {Guo-Liang}\ \bibnamefont {Huang}},
  \bibinfo {author} {\bibfnamefont {Hong-Qing}\ \bibnamefont {Dai}}, \bibinfo
  {author} {\bibfnamefont {Jun-Rui}\ \bibnamefont {Jiao}}, \bibinfo {author}
  {\bibfnamefont {Xian-Guo}\ \bibnamefont {Zang}}, \bibinfo {author}
  {\bibfnamefont {De-Jie}\ \bibnamefont {Yu}}, \bibinfo {author} {\bibfnamefont
  {Sheng-Jie}\ \bibnamefont {Zheng}}, \ and\ \bibinfo {author} {\bibfnamefont
  {Jian}\ \bibnamefont {Liu}}} (\bibinfo {year} {2017}),\ \bibfield  {title}
  {\enquote {\bibinfo {title} {{Topological phononic insulator with robust
  pseudospin-dependent transport}},}\ }\href
  {https://link.aps.org/doi/10.1103/PhysRevB.96.094106} {\bibfield  {journal}
  {\bibinfo  {journal} {Phys. Rev. B}\ }\textbf {\bibinfo {volume}
  {96}}~(\bibinfo {number} {9}),\ \bibinfo {pages} {094106}}\BibitemShut
  {NoStop}%
\bibitem [{\citenamefont {Xiao}\ \emph
  {et~al.}(2016{\natexlab{a}})\citenamefont {Xiao}, \citenamefont {Lai},
  \citenamefont {Yu}, \citenamefont {Ma}, \citenamefont {Shvets},\ and\
  \citenamefont {Anlage}}]{Xiao2016prb}%
  \BibitemOpen
  \bibfield  {author} {\bibinfo {author} {\bibnamefont {Xiao}, \bibfnamefont
  {Bo}}, \bibinfo {author} {\bibfnamefont {Kueifu}\ \bibnamefont {Lai}},
  \bibinfo {author} {\bibfnamefont {Yang}\ \bibnamefont {Yu}}, \bibinfo
  {author} {\bibfnamefont {Tzuhsuan}\ \bibnamefont {Ma}}, \bibinfo {author}
  {\bibfnamefont {Gennady}\ \bibnamefont {Shvets}}, \ and\ \bibinfo {author}
  {\bibfnamefont {Steven~M.}\ \bibnamefont {Anlage}}} (\bibinfo {year}
  {2016}{\natexlab{a}}),\ \bibfield  {title} {\enquote {\bibinfo {title}
  {Exciting reflectionless unidirectional edge modes in a reciprocal photonic
  topological insulator medium},}\ }\href
  {https://link.aps.org/doi/10.1103/PhysRevB.94.195427} {\bibfield  {journal}
  {\bibinfo  {journal} {Phys. Rev. B}\ }\textbf {\bibinfo {volume} {94}},\
  \bibinfo {pages} {195427}}\BibitemShut {NoStop}%
\bibitem [{\citenamefont {Xiao}\ \emph {et~al.}(2010)\citenamefont {Xiao},
  \citenamefont {Chang},\ and\ \citenamefont {Niu}}]{Xiao:2010RMP}%
  \BibitemOpen
  \bibfield  {author} {\bibinfo {author} {\bibnamefont {Xiao}, \bibfnamefont
  {Di}}, \bibinfo {author} {\bibfnamefont {Ming-Che}\ \bibnamefont {Chang}}, \
  and\ \bibinfo {author} {\bibfnamefont {Qian}\ \bibnamefont {Niu}}} (\bibinfo
  {year} {2010}),\ \bibfield  {title} {\enquote {\bibinfo {title} {Berry phase
  effects on electronic properties},}\ }\href
  {https://link.aps.org/doi/10.1103/RevModPhys.82.1959} {\bibfield  {journal}
  {\bibinfo  {journal} {Rev. Mod. Phys.}\ }\textbf {\bibinfo {volume} {82}},\
  \bibinfo {pages} {1959--2007}}\BibitemShut {NoStop}%
\bibitem [{\citenamefont {Xiao}\ \emph {et~al.}(2017)\citenamefont {Xiao},
  \citenamefont {Zhan}, \citenamefont {Bian}, \citenamefont {Wang},
  \citenamefont {Zhang}, \citenamefont {Wang}, \citenamefont {Li},
  \citenamefont {Mochizuki}, \citenamefont {Kim}, \citenamefont {Kawakami}
  \emph {et~al.}}]{Xiao:2017NatPhys}%
  \BibitemOpen
  \bibfield  {author} {\bibinfo {author} {\bibnamefont {Xiao}, \bibfnamefont
  {L}}, \bibinfo {author} {\bibfnamefont {X}~\bibnamefont {Zhan}}, \bibinfo
  {author} {\bibfnamefont {ZH}~\bibnamefont {Bian}}, \bibinfo {author}
  {\bibfnamefont {KK}~\bibnamefont {Wang}}, \bibinfo {author} {\bibfnamefont
  {X}~\bibnamefont {Zhang}}, \bibinfo {author} {\bibfnamefont {XP}~\bibnamefont
  {Wang}}, \bibinfo {author} {\bibfnamefont {J}~\bibnamefont {Li}}, \bibinfo
  {author} {\bibfnamefont {K}~\bibnamefont {Mochizuki}}, \bibinfo {author}
  {\bibfnamefont {D}~\bibnamefont {Kim}}, \bibinfo {author} {\bibfnamefont
  {N}~\bibnamefont {Kawakami}},  \emph {et~al.}} (\bibinfo {year} {2017}),\
  \bibfield  {title} {\enquote {\bibinfo {title} {Observation of topological
  edge states in parity--time-symmetric quantum walks},}\ }\href
  {https://www.nature.com/articles/nphys4204} {\bibfield  {journal} {\bibinfo
  {journal} {Nat. Phys.}\ }\textbf {\bibinfo {volume} {13}}~(\bibinfo {number}
  {11}),\ \bibinfo {pages} {1117}}\BibitemShut {NoStop}%
\bibitem [{\citenamefont {Xiao}\ and\ \citenamefont
  {Fan}(2017{\natexlab{a}})}]{Xiao:2017arXiv}%
  \BibitemOpen
  \bibfield  {author} {\bibinfo {author} {\bibnamefont {Xiao}, \bibfnamefont
  {Meng}}, \ and\ \bibinfo {author} {\bibfnamefont {Shanhui}\ \bibnamefont
  {Fan}}} (\bibinfo {year} {2017}{\natexlab{a}}),\ \bibfield  {title} {\enquote
  {\bibinfo {title} {Photonic {C}hern insulator through homogenization of an
  array of particles},}\ }\href
  {https://link.aps.org/doi/10.1103/PhysRevB.96.100202} {\bibfield  {journal}
  {\bibinfo  {journal} {Phys. Rev. B}\ }\textbf {\bibinfo {volume} {96}},\
  \bibinfo {pages} {100202}}\BibitemShut {NoStop}%
\bibitem [{\citenamefont {Xiao}\ and\ \citenamefont
  {Fan}(2017{\natexlab{b}})}]{Xiao:2017arXiv2}%
  \BibitemOpen
  \bibfield  {author} {\bibinfo {author} {\bibnamefont {Xiao}, \bibfnamefont
  {Meng}}, \ and\ \bibinfo {author} {\bibfnamefont {Shanhui}\ \bibnamefont
  {Fan}}} (\bibinfo {year} {2017}{\natexlab{b}}),\ \bibfield  {title} {\enquote
  {\bibinfo {title} {Topologically charged nodal surface},}\ }\href
  {https://arxiv.org/abs/1709.02363} {\bibinfo  {journal} {arXiv:1709.02363}\
  }\BibitemShut {NoStop}%
\bibitem [{\citenamefont {Xiao}\ \emph
  {et~al.}(2016{\natexlab{b}})\citenamefont {Xiao}, \citenamefont {Lin},\ and\
  \citenamefont {Fan}}]{Xiao:2016PRL}%
  \BibitemOpen
\bibfield  {journal} {  }\bibfield  {author} {\bibinfo {author} {\bibnamefont
  {Xiao}, \bibfnamefont {Meng}}, \bibinfo {author} {\bibfnamefont {Qian}\
  \bibnamefont {Lin}}, \ and\ \bibinfo {author} {\bibfnamefont {Shanhui}\
  \bibnamefont {Fan}}} (\bibinfo {year} {2016}{\natexlab{b}}),\ \bibfield
  {title} {\enquote {\bibinfo {title} {Hyperbolic {W}eyl point in reciprocal
  chiral metamaterials},}\ }\href
  {https://journals.aps.org/prl/abstract/10.1103/PhysRevLett.117.057401}
  {\bibfield  {journal} {\bibinfo  {journal} {Phys. Rev. Lett.}\ }\textbf
  {\bibinfo {volume} {117}}~(\bibinfo {number} {5}),\ \bibinfo {pages}
  {057401}}\BibitemShut {NoStop}%
\bibitem [{\citenamefont {Xiao}\ \emph {et~al.}(2014)\citenamefont {Xiao},
  \citenamefont {Zhang},\ and\ \citenamefont {Chan}}]{Xiao:2014PRX}%
  \BibitemOpen
  \bibfield  {author} {\bibinfo {author} {\bibnamefont {Xiao}, \bibfnamefont
  {Meng}}, \bibinfo {author} {\bibfnamefont {Z.~Q.}\ \bibnamefont {Zhang}}, \
  and\ \bibinfo {author} {\bibfnamefont {C.~T.}\ \bibnamefont {Chan}}}
  (\bibinfo {year} {2014}),\ \bibfield  {title} {\enquote {\bibinfo {title}
  {Surface impedance and bulk band geometric phases in one-dimensional
  systems},}\ }\href {https://link.aps.org/doi/10.1103/PhysRevX.4.021017}
  {\bibfield  {journal} {\bibinfo  {journal} {Phys. Rev. X}\ }\textbf {\bibinfo
  {volume} {4}},\ \bibinfo {pages} {021017}}\BibitemShut {NoStop}%
\bibitem [{\citenamefont {Xu}\ \emph {et~al.}(2011)\citenamefont {Xu},
  \citenamefont {Weng}, \citenamefont {Wang}, \citenamefont {Dai},\ and\
  \citenamefont {Fang}}]{Xu:2011PRL}%
  \BibitemOpen
  \bibfield  {author} {\bibinfo {author} {\bibnamefont {Xu}, \bibfnamefont
  {Gang}}, \bibinfo {author} {\bibfnamefont {Hongming}\ \bibnamefont {Weng}},
  \bibinfo {author} {\bibfnamefont {Zhijun}\ \bibnamefont {Wang}}, \bibinfo
  {author} {\bibfnamefont {Xi}~\bibnamefont {Dai}}, \ and\ \bibinfo {author}
  {\bibfnamefont {Zhong}\ \bibnamefont {Fang}}} (\bibinfo {year} {2011}),\
  \bibfield  {title} {\enquote {\bibinfo {title} {Chern semimetal and the
  quantized anomalous {H}all effect in
  {${\mathrm{HgCr}}_{2}{\mathrm{Se}}_{4}$}},}\ }\href
  {https://link.aps.org/doi/10.1103/PhysRevLett.107.186806} {\bibfield
  {journal} {\bibinfo  {journal} {Phys. Rev. Lett.}\ }\textbf {\bibinfo
  {volume} {107}},\ \bibinfo {pages} {186806}}\BibitemShut {NoStop}%
\bibitem [{\citenamefont {Xu}\ \emph {et~al.}(2016)\citenamefont {Xu},
  \citenamefont {Wang}, \citenamefont {Xu}, \citenamefont {Chen},\ and\
  \citenamefont {Jiang}}]{Xu:OptExp2016}%
  \BibitemOpen
  \bibfield  {author} {\bibinfo {author} {\bibnamefont {Xu}, \bibfnamefont
  {Lin}}, \bibinfo {author} {\bibfnamefont {Hai-Xiao}\ \bibnamefont {Wang}},
  \bibinfo {author} {\bibfnamefont {Ya-Dong}\ \bibnamefont {Xu}}, \bibinfo
  {author} {\bibfnamefont {Huan-Yang}\ \bibnamefont {Chen}}, \ and\ \bibinfo
  {author} {\bibfnamefont {Jian-Hua}\ \bibnamefont {Jiang}}} (\bibinfo {year}
  {2016}),\ \bibfield  {title} {\enquote {\bibinfo {title} {{Accidental
  degeneracy in photonic bands and topological phase transitions in
  two-dimensional core-shell dielectric photonic crystals}},}\ }\href
  {https://www.osapublishing.org/abstract.cfm?URI=oe-24-16-18059} {\bibfield
  {journal} {\bibinfo  {journal} {Opt. Express}\ }\textbf {\bibinfo {volume}
  {24}}~(\bibinfo {number} {16}),\ \bibinfo {pages} {18059}}\BibitemShut
  {NoStop}%
\bibitem [{\citenamefont {Xu}\ \emph {et~al.}(2017)\citenamefont {Xu},
  \citenamefont {Wang},\ and\ \citenamefont {Duan}}]{Xu:2017PRL}%
  \BibitemOpen
  \bibfield  {author} {\bibinfo {author} {\bibnamefont {Xu}, \bibfnamefont
  {Yong}}, \bibinfo {author} {\bibfnamefont {Sheng-Tao}\ \bibnamefont {Wang}},
  \ and\ \bibinfo {author} {\bibfnamefont {L-M}\ \bibnamefont {Duan}}}
  (\bibinfo {year} {2017}),\ \bibfield  {title} {\enquote {\bibinfo {title}
  {Weyl exceptional rings in a three-dimensional dissipative cold atomic
  gas},}\ }\href
  {https://journals.aps.org/prl/abstract/10.1103/PhysRevLett.118.045701}
  {\bibfield  {journal} {\bibinfo  {journal} {Phys. Rev. Lett.}\ }\textbf
  {\bibinfo {volume} {118}}~(\bibinfo {number} {4}),\ \bibinfo {pages}
  {045701}}\BibitemShut {NoStop}%
\bibitem [{\citenamefont {Yablonovitch}(1987)}]{Yablonovitch:1987PRL}%
  \BibitemOpen
  \bibfield  {author} {\bibinfo {author} {\bibnamefont {Yablonovitch},
  \bibfnamefont {Eli}}} (\bibinfo {year} {1987}),\ \bibfield  {title} {\enquote
  {\bibinfo {title} {Inhibited spontaneous emission in solid-state physics and
  electronics},}\ }\href
  {https://journals.aps.org/prl/abstract/10.1103/PhysRevLett.58.2059}
  {\bibfield  {journal} {\bibinfo  {journal} {Phys. Rev. Lett.}\ }\textbf
  {\bibinfo {volume} {58}},\ \bibinfo {pages} {2059--2062}}\BibitemShut
  {NoStop}%
\bibitem [{\citenamefont {Yan}\ \emph {et~al.}(2018)\citenamefont {Yan},
  \citenamefont {Liu}, \citenamefont {Yan}, \citenamefont {Liu}, \citenamefont
  {Chen}, \citenamefont {Wang},\ and\ \citenamefont {Lu}}]{Yan:2018NatPhys}%
  \BibitemOpen
  \bibfield  {author} {\bibinfo {author} {\bibnamefont {Yan}, \bibfnamefont
  {Qinghui}}, \bibinfo {author} {\bibfnamefont {Rongjuan}\ \bibnamefont {Liu}},
  \bibinfo {author} {\bibfnamefont {Zhongbo}\ \bibnamefont {Yan}}, \bibinfo
  {author} {\bibfnamefont {Boyuan}\ \bibnamefont {Liu}}, \bibinfo {author}
  {\bibfnamefont {Hongsheng}\ \bibnamefont {Chen}}, \bibinfo {author}
  {\bibfnamefont {Zhong}\ \bibnamefont {Wang}}, \ and\ \bibinfo {author}
  {\bibfnamefont {Ling}\ \bibnamefont {Lu}}} (\bibinfo {year} {2018}),\
  \bibfield  {title} {\enquote {\bibinfo {title} {Experimental discovery of
  nodal chains},}\ }\href {https://www.nature.com/articles/s41567-017-0041-4}
  {\bibfield  {journal} {\bibinfo  {journal} {Nat. Phys.}\ }\textbf {\bibinfo
  {volume} {14}}~(\bibinfo {number} {5}),\ \bibinfo {pages} {461}}\BibitemShut
  {NoStop}%
\bibitem [{\citenamefont {Yan}\ \emph {et~al.}(2017)\citenamefont {Yan},
  \citenamefont {Bi}, \citenamefont {Shen}, \citenamefont {Lu}, \citenamefont
  {Zhang},\ and\ \citenamefont {Wang}}]{Yan:2017PRB}%
  \BibitemOpen
  \bibfield  {author} {\bibinfo {author} {\bibnamefont {Yan}, \bibfnamefont
  {Zhongbo}}, \bibinfo {author} {\bibfnamefont {Ren}\ \bibnamefont {Bi}},
  \bibinfo {author} {\bibfnamefont {Huitao}\ \bibnamefont {Shen}}, \bibinfo
  {author} {\bibfnamefont {Ling}\ \bibnamefont {Lu}}, \bibinfo {author}
  {\bibfnamefont {Shou-Cheng}\ \bibnamefont {Zhang}}, \ and\ \bibinfo {author}
  {\bibfnamefont {Zhong}\ \bibnamefont {Wang}}} (\bibinfo {year} {2017}),\
  \bibfield  {title} {\enquote {\bibinfo {title} {Nodal-link semimetals},}\
  }\href {https://link.aps.org/doi/10.1103/PhysRevB.96.041103} {\bibfield
  {journal} {\bibinfo  {journal} {Phys. Rev. B}\ }\textbf {\bibinfo {volume}
  {96}},\ \bibinfo {pages} {041103}}\BibitemShut {NoStop}%
\bibitem [{\citenamefont {Yang}\ \emph
  {et~al.}(2017{\natexlab{a}})\citenamefont {Yang}, \citenamefont {Guo},
  \citenamefont {Tremain}, \citenamefont {Barr}, \citenamefont {Gao},
  \citenamefont {Liu}, \citenamefont {B{\'e}ri}, \citenamefont {Xiang},
  \citenamefont {Fan}, \citenamefont {Hibbins} \emph
  {et~al.}}]{Yang:2017NatComm}%
  \BibitemOpen
  \bibfield  {author} {\bibinfo {author} {\bibnamefont {Yang}, \bibfnamefont
  {Biao}}, \bibinfo {author} {\bibfnamefont {Qinghua}\ \bibnamefont {Guo}},
  \bibinfo {author} {\bibfnamefont {Ben}\ \bibnamefont {Tremain}}, \bibinfo
  {author} {\bibfnamefont {Lauren~E}\ \bibnamefont {Barr}}, \bibinfo {author}
  {\bibfnamefont {Wenlong}\ \bibnamefont {Gao}}, \bibinfo {author}
  {\bibfnamefont {Hongchao}\ \bibnamefont {Liu}}, \bibinfo {author}
  {\bibfnamefont {Benjamin}\ \bibnamefont {B{\'e}ri}}, \bibinfo {author}
  {\bibfnamefont {Yuanjiang}\ \bibnamefont {Xiang}}, \bibinfo {author}
  {\bibfnamefont {Dianyuan}\ \bibnamefont {Fan}}, \bibinfo {author}
  {\bibfnamefont {Alastair~P}\ \bibnamefont {Hibbins}},  \emph {et~al.}}
  (\bibinfo {year} {2017}{\natexlab{a}}),\ \bibfield  {title} {\enquote
  {\bibinfo {title} {Direct observation of topological surface-state arcs in
  photonic metamaterials},}\ }\href
  {https://www.nature.com/articles/s41467-017-00134-1} {\bibfield  {journal}
  {\bibinfo  {journal} {Nat. Commun.}\ }\textbf {\bibinfo {volume}
  {8}}~(\bibinfo {number} {1}),\ \bibinfo {pages} {97}}\BibitemShut {NoStop}%
\bibitem [{\citenamefont {Yang}\ \emph
  {et~al.}(2018{\natexlab{a}})\citenamefont {Yang}, \citenamefont {Guo},
  \citenamefont {Tremain}, \citenamefont {Liu}, \citenamefont {Barr},
  \citenamefont {Yan}, \citenamefont {Gao}, \citenamefont {Liu}, \citenamefont
  {Xiang}, \citenamefont {Chen} \emph {et~al.}}]{yang2018ideal}%
  \BibitemOpen
  \bibfield  {author} {\bibinfo {author} {\bibnamefont {Yang}, \bibfnamefont
  {Biao}}, \bibinfo {author} {\bibfnamefont {Qinghua}\ \bibnamefont {Guo}},
  \bibinfo {author} {\bibfnamefont {Ben}\ \bibnamefont {Tremain}}, \bibinfo
  {author} {\bibfnamefont {Rongjuan}\ \bibnamefont {Liu}}, \bibinfo {author}
  {\bibfnamefont {Lauren~E}\ \bibnamefont {Barr}}, \bibinfo {author}
  {\bibfnamefont {Qinghui}\ \bibnamefont {Yan}}, \bibinfo {author}
  {\bibfnamefont {Wenlong}\ \bibnamefont {Gao}}, \bibinfo {author}
  {\bibfnamefont {Hongchao}\ \bibnamefont {Liu}}, \bibinfo {author}
  {\bibfnamefont {Yuanjiang}\ \bibnamefont {Xiang}}, \bibinfo {author}
  {\bibfnamefont {Jing}\ \bibnamefont {Chen}},  \emph {et~al.}} (\bibinfo
  {year} {2018}{\natexlab{a}}),\ \bibfield  {title} {\enquote {\bibinfo {title}
  {Ideal {W}eyl points and helicoid surface states in artificial photonic
  crystal structures},}\ }\href
  {http://science.sciencemag.org/content/359/6379/1013} {\bibfield  {journal}
  {\bibinfo  {journal} {Science}\ }\textbf {\bibinfo {volume} {359}},\ \bibinfo
  {pages} {eaaq1221}}\BibitemShut {NoStop}%
\bibitem [{\citenamefont {Yang}\ \emph
  {et~al.}(2017{\natexlab{b}})\citenamefont {Yang}, \citenamefont {Wu},\ and\
  \citenamefont {Zhang}}]{Yang:2017APL}%
  \BibitemOpen
  \bibfield  {author} {\bibinfo {author} {\bibnamefont {Yang}, \bibfnamefont
  {Bing}}, \bibinfo {author} {\bibfnamefont {Tong}\ \bibnamefont {Wu}}, \ and\
  \bibinfo {author} {\bibfnamefont {Xiangdong}\ \bibnamefont {Zhang}}}
  (\bibinfo {year} {2017}{\natexlab{b}}),\ \bibfield  {title} {\enquote
  {\bibinfo {title} {Engineering topological edge states in two dimensional
  magnetic photonic crystal},}\ }\href
  {https://aip.scitation.org/doi/full/10.1063/1.4973990} {\bibfield  {journal}
  {\bibinfo  {journal} {Appl. Phys. Lett.}\ }\textbf {\bibinfo {volume}
  {110}}~(\bibinfo {number} {2}),\ \bibinfo {pages} {021109}}\BibitemShut
  {NoStop}%
\bibitem [{\citenamefont {Yang}\ \emph
  {et~al.}(2017{\natexlab{c}})\citenamefont {Yang}, \citenamefont {Wu},\ and\
  \citenamefont {Zhang}}]{Yang:2017JOSAB}%
  \BibitemOpen
  \bibfield  {author} {\bibinfo {author} {\bibnamefont {Yang}, \bibfnamefont
  {Bing}}, \bibinfo {author} {\bibfnamefont {Tong}\ \bibnamefont {Wu}}, \ and\
  \bibinfo {author} {\bibfnamefont {Xiangdong}\ \bibnamefont {Zhang}}}
  (\bibinfo {year} {2017}{\natexlab{c}}),\ \bibfield  {title} {\enquote
  {\bibinfo {title} {Topological properties of nearly flat bands in
  two-dimensional photonic crystals},}\ }\href
  {https://www.osapublishing.org/josab/abstract.cfm?uri=josab-34-4-831}
  {\bibfield  {journal} {\bibinfo  {journal} {J. Opt. Soc. Am. B}\ }\textbf
  {\bibinfo {volume} {34}}~(\bibinfo {number} {4}),\ \bibinfo {pages}
  {831--836}}\BibitemShut {NoStop}%
\bibitem [{\citenamefont {Yang}\ \emph {et~al.}(2012)\citenamefont {Yang},
  \citenamefont {Yin}, \citenamefont {Chen}, \citenamefont {Kou}, \citenamefont
  {Feng},\ and\ \citenamefont {Oh}}]{Yang:2012PRA}%
  \BibitemOpen
  \bibfield  {author} {\bibinfo {author} {\bibnamefont {Yang}, \bibfnamefont
  {W~L}}, \bibinfo {author} {\bibfnamefont {Zhang-qi}\ \bibnamefont {Yin}},
  \bibinfo {author} {\bibfnamefont {Z.~X.}\ \bibnamefont {Chen}}, \bibinfo
  {author} {\bibfnamefont {Su-Peng}\ \bibnamefont {Kou}}, \bibinfo {author}
  {\bibfnamefont {M.}~\bibnamefont {Feng}}, \ and\ \bibinfo {author}
  {\bibfnamefont {C.~H.}\ \bibnamefont {Oh}}} (\bibinfo {year} {2012}),\
  \bibfield  {title} {\enquote {\bibinfo {title} {Quantum simulation of an
  artificial {A}belian gauge field using nitrogen-vacancy-center ensembles
  coupled to superconducting resonators},}\ }\href
  {https://link.aps.org/doi/10.1103/PhysRevA.86.012307} {\bibfield  {journal}
  {\bibinfo  {journal} {Phys. Rev. A}\ }\textbf {\bibinfo {volume} {86}},\
  \bibinfo {pages} {012307}}\BibitemShut {NoStop}%
\bibitem [{\citenamefont {Yang}\ \emph {et~al.}(2013)\citenamefont {Yang},
  \citenamefont {Poo}, \citenamefont {Wu}, \citenamefont {Gu},\ and\
  \citenamefont {Chen}}]{Yang:2013APL}%
  \BibitemOpen
  \bibfield  {author} {\bibinfo {author} {\bibnamefont {Yang}, \bibfnamefont
  {Yan}}, \bibinfo {author} {\bibfnamefont {Yin}\ \bibnamefont {Poo}}, \bibinfo
  {author} {\bibfnamefont {Rui-xin}\ \bibnamefont {Wu}}, \bibinfo {author}
  {\bibfnamefont {Yan}\ \bibnamefont {Gu}}, \ and\ \bibinfo {author}
  {\bibfnamefont {Ping}\ \bibnamefont {Chen}}} (\bibinfo {year} {2013}),\
  \bibfield  {title} {\enquote {\bibinfo {title} {Experimental demonstration of
  one-way slow wave in waveguide involving gyromagnetic photonic crystals},}\
  }\href {https://aip.scitation.org/doi/10.1063/1.4809956} {\bibfield
  {journal} {\bibinfo  {journal} {Appl. Phys. Lett.}\ }\textbf {\bibinfo
  {volume} {102}},\ \bibinfo {pages} {231113}}\BibitemShut {NoStop}%
\bibitem [{\citenamefont {Yang}\ \emph {et~al.}(2019)\citenamefont {Yang},
  \citenamefont {Gao}, \citenamefont {Xue}, \citenamefont {Zhang},
  \citenamefont {He}, \citenamefont {Yang}, \citenamefont {Singh},
  \citenamefont {Chong}, \citenamefont {Zhang},\ and\ \citenamefont
  {Chen}}]{yang:2018arxiv}%
  \BibitemOpen
  \bibfield  {author} {\bibinfo {author} {\bibnamefont {Yang}, \bibfnamefont
  {Yihao}}, \bibinfo {author} {\bibfnamefont {Zhen}\ \bibnamefont {Gao}},
  \bibinfo {author} {\bibfnamefont {Haoran}\ \bibnamefont {Xue}}, \bibinfo
  {author} {\bibfnamefont {Li}~\bibnamefont {Zhang}}, \bibinfo {author}
  {\bibfnamefont {Mengjia}\ \bibnamefont {He}}, \bibinfo {author}
  {\bibfnamefont {Zhaoju}\ \bibnamefont {Yang}}, \bibinfo {author}
  {\bibfnamefont {Ranjan}\ \bibnamefont {Singh}}, \bibinfo {author}
  {\bibfnamefont {Yidong}\ \bibnamefont {Chong}}, \bibinfo {author}
  {\bibfnamefont {Baile}\ \bibnamefont {Zhang}}, \ and\ \bibinfo {author}
  {\bibfnamefont {Hongsheng}\ \bibnamefont {Chen}}} (\bibinfo {year} {2019}),\
  \bibfield  {title} {\enquote {\bibinfo {title} {Realization of a
  three-dimensional photonic topological insulator},}\ }\href
  {https://www.nature.com/articles/s41586-018-0829-0} {\bibfield  {journal}
  {\bibinfo  {journal} {Nature}\ }\textbf {\bibinfo {volume} {565}},\ \bibinfo
  {pages} {622 -- 626}}\BibitemShut {NoStop}%
\bibitem [{\citenamefont {Yang}\ \emph
  {et~al.}(2018{\natexlab{b}})\citenamefont {Yang}, \citenamefont {Xu},
  \citenamefont {Xu}, \citenamefont {Wang}, \citenamefont {Jiang},
  \citenamefont {Hu},\ and\ \citenamefont {Hang}}]{Yang:2018PRL}%
  \BibitemOpen
  \bibfield  {author} {\bibinfo {author} {\bibnamefont {Yang}, \bibfnamefont
  {Yuting}}, \bibinfo {author} {\bibfnamefont {Yun~Fei}\ \bibnamefont {Xu}},
  \bibinfo {author} {\bibfnamefont {Tao}\ \bibnamefont {Xu}}, \bibinfo {author}
  {\bibfnamefont {Hai-Xiao}\ \bibnamefont {Wang}}, \bibinfo {author}
  {\bibfnamefont {Jian-Hua}\ \bibnamefont {Jiang}}, \bibinfo {author}
  {\bibfnamefont {Xiao}\ \bibnamefont {Hu}}, \ and\ \bibinfo {author}
  {\bibfnamefont {Zhi~Hong}\ \bibnamefont {Hang}}} (\bibinfo {year}
  {2018}{\natexlab{b}}),\ \bibfield  {title} {\enquote {\bibinfo {title}
  {Visualization of a unidirectional electromagnetic waveguide using
  topological photonic crystals made of dielectric materials},}\ }\href
  {https://link.aps.org/doi/10.1103/PhysRevLett.120.217401} {\bibfield
  {journal} {\bibinfo  {journal} {Phys. Rev. Lett.}\ }\textbf {\bibinfo
  {volume} {120}}~(\bibinfo {number} {21}),\ \bibinfo {pages}
  {217401}}\BibitemShut {NoStop}%
\bibitem [{\citenamefont {Yang}\ \emph
  {et~al.}(2017{\natexlab{d}})\citenamefont {Yang}, \citenamefont {Xiao},
  \citenamefont {Gao}, \citenamefont {Lu}, \citenamefont {Chong},\ and\
  \citenamefont {Zhang}}]{Yang:2017OE}%
  \BibitemOpen
  \bibfield  {author} {\bibinfo {author} {\bibnamefont {Yang}, \bibfnamefont
  {Zhaoju}}, \bibinfo {author} {\bibfnamefont {Meng}\ \bibnamefont {Xiao}},
  \bibinfo {author} {\bibfnamefont {Fei}\ \bibnamefont {Gao}}, \bibinfo
  {author} {\bibfnamefont {Ling}\ \bibnamefont {Lu}}, \bibinfo {author}
  {\bibfnamefont {Yidong}\ \bibnamefont {Chong}}, \ and\ \bibinfo {author}
  {\bibfnamefont {Baile}\ \bibnamefont {Zhang}}} (\bibinfo {year}
  {2017}{\natexlab{d}}),\ \bibfield  {title} {\enquote {\bibinfo {title} {Weyl
  points in a magnetic tetrahedral photonic crystal},}\ }\href
  {https://www.osapublishing.org/oe/abstract.cfm?uri=oe-25-14-15772} {\bibfield
   {journal} {\bibinfo  {journal} {Opt. Express}\ }\textbf {\bibinfo {volume}
  {25}}~(\bibinfo {number} {14}),\ \bibinfo {pages} {15772--15777}}\BibitemShut
  {NoStop}%
\bibitem [{\citenamefont {Yannopapas}(2011)}]{Yannopapas:2011PRB}%
  \BibitemOpen
  \bibfield  {author} {\bibinfo {author} {\bibnamefont {Yannopapas},
  \bibfnamefont {Vassilios}}} (\bibinfo {year} {2011}),\ \bibfield  {title}
  {\enquote {\bibinfo {title} {Gapless surface states in a lattice of coupled
  cavities: {A} photonic analog of topological crystalline insulators},}\
  }\href {https://journals.aps.org/prb/abstract/10.1103/PhysRevB.84.195126}
  {\bibfield  {journal} {\bibinfo  {journal} {Phys. Rev. B}\ }\textbf {\bibinfo
  {volume} {84}}~(\bibinfo {number} {19}),\ \bibinfo {pages}
  {195126}}\BibitemShut {NoStop}%
\bibitem [{\citenamefont {Yannopapas}(2014)}]{Yannopapas:2014IJMPB}%
  \BibitemOpen
  \bibfield  {author} {\bibinfo {author} {\bibnamefont {Yannopapas},
  \bibfnamefont {Vassilios}}} (\bibinfo {year} {2014}),\ \bibfield  {title}
  {\enquote {\bibinfo {title} {Dirac points, topological edge modes and
  nonreciprocal transmission in one-dimensional metamaterial-based
  coupled-cavity arrays},}\ }\href
  {http://www.worldscientific.com/doi/abs/10.1142/S0217979214410069} {\bibfield
   {journal} {\bibinfo  {journal} {Int. J. Mod. Phys. B}\ }\textbf {\bibinfo
  {volume} {28}}~(\bibinfo {number} {02}),\ \bibinfo {pages}
  {1441006}}\BibitemShut {NoStop}%
\bibitem [{\citenamefont {Yao}\ \emph {et~al.}(2018)\citenamefont {Yao},
  \citenamefont {Song},\ and\ \citenamefont {Wang}}]{yao2018non}%
  \BibitemOpen
  \bibfield  {author} {\bibinfo {author} {\bibnamefont {Yao}, \bibfnamefont
  {Shunyu}}, \bibinfo {author} {\bibfnamefont {Fei}\ \bibnamefont {Song}}, \
  and\ \bibinfo {author} {\bibfnamefont {Zhong}\ \bibnamefont {Wang}}}
  (\bibinfo {year} {2018}),\ \bibfield  {title} {\enquote {\bibinfo {title}
  {Non-{H}ermitian {C}hern bands},}\ }\href
  {https://link.aps.org/doi/10.1103/PhysRevLett.121.136802} {\bibfield
  {journal} {\bibinfo  {journal} {Phys. Rev. Lett.}\ }\textbf {\bibinfo
  {volume} {121}},\ \bibinfo {pages} {136802}}\BibitemShut {NoStop}%
\bibitem [{\citenamefont {Yao}\ and\ \citenamefont {Wang}(2018)}]{yao2018edge}%
  \BibitemOpen
  \bibfield  {author} {\bibinfo {author} {\bibnamefont {Yao}, \bibfnamefont
  {Shunyu}}, \ and\ \bibinfo {author} {\bibfnamefont {Zhong}\ \bibnamefont
  {Wang}}} (\bibinfo {year} {2018}),\ \bibfield  {title} {\enquote {\bibinfo
  {title} {Edge states and topological invariants of non-{H}ermitian
  systems},}\ }\href {https://link.aps.org/doi/10.1103/PhysRevLett.121.086803}
  {\bibfield  {journal} {\bibinfo  {journal} {Phys. Rev. Lett.}\ }\textbf
  {\bibinfo {volume} {121}},\ \bibinfo {pages} {086803}}\BibitemShut {NoStop}%
\bibitem [{\citenamefont {Yao}\ \emph {et~al.}(2017)\citenamefont {Yao},
  \citenamefont {Yan},\ and\ \citenamefont {Wang}}]{Yao:2017PRB}%
  \BibitemOpen
  \bibfield  {author} {\bibinfo {author} {\bibnamefont {Yao}, \bibfnamefont
  {Shunyu}}, \bibinfo {author} {\bibfnamefont {Zhongbo}\ \bibnamefont {Yan}}, \
  and\ \bibinfo {author} {\bibfnamefont {Zhong}\ \bibnamefont {Wang}}}
  (\bibinfo {year} {2017}),\ \bibfield  {title} {\enquote {\bibinfo {title}
  {Topological invariants of {F}loquet systems: {G}eneral formulation, special
  properties, and {F}loquet topological defects},}\ }\href
  {https://link.aps.org/doi/10.1103/PhysRevB.96.195303} {\bibfield  {journal}
  {\bibinfo  {journal} {Phys. Rev. B}\ }\textbf {\bibinfo {volume} {96}},\
  \bibinfo {pages} {195303}}\BibitemShut {NoStop}%
\bibitem [{\citenamefont {Yariv}(1976)}]{yariv1976introduction}%
  \BibitemOpen
  \bibfield  {author} {\bibinfo {author} {\bibnamefont {Yariv}, \bibfnamefont
  {Amnon}}} (\bibinfo {year} {1976}),\ \href@noop {} {\emph {\bibinfo {title}
  {Introduction to optical electronics}}}\ (\bibinfo  {publisher} {Holt,
  Rinehart and Winston, Inc., New York, NY})\BibitemShut {NoStop}%
\bibitem [{\citenamefont {Yariv}\ \emph {et~al.}(1999)\citenamefont {Yariv},
  \citenamefont {Xu}, \citenamefont {Lee},\ and\ \citenamefont
  {Scherer}}]{Yariv:OptLett1999}%
  \BibitemOpen
  \bibfield  {author} {\bibinfo {author} {\bibnamefont {Yariv}, \bibfnamefont
  {Amnon}}, \bibinfo {author} {\bibfnamefont {Yong}\ \bibnamefont {Xu}},
  \bibinfo {author} {\bibfnamefont {Reginald~K}\ \bibnamefont {Lee}}, \ and\
  \bibinfo {author} {\bibfnamefont {Axel}\ \bibnamefont {Scherer}}} (\bibinfo
  {year} {1999}),\ \bibfield  {title} {\enquote {\bibinfo {title}
  {Coupled-resonator optical waveguide: {A} proposal and analysis},}\ }\href
  {https://www.osapublishing.org/ol/abstract.cfm?uri=ol-24-11-711} {\bibfield
  {journal} {\bibinfo  {journal} {Opt. Lett.}\ }\textbf {\bibinfo {volume}
  {24}}~(\bibinfo {number} {11}),\ \bibinfo {pages} {711--713}}\BibitemShut
  {NoStop}%
\bibitem [{\citenamefont {Yi}\ and\ \citenamefont {Karzig}(2016)}]{Yi:2016PRB}%
  \BibitemOpen
  \bibfield  {author} {\bibinfo {author} {\bibnamefont {Yi}, \bibfnamefont
  {Kexin}}, \ and\ \bibinfo {author} {\bibfnamefont {Torsten}\ \bibnamefont
  {Karzig}}} (\bibinfo {year} {2016}),\ \bibfield  {title} {\enquote {\bibinfo
  {title} {{Topological polaritons from photonic Dirac cones coupled to
  excitons in a magnetic field}},}\ }\href
  {https://link.aps.org/doi/10.1103/PhysRevB.93.104303} {\bibfield  {journal}
  {\bibinfo  {journal} {Phys. Rev. B}\ }\textbf {\bibinfo {volume}
  {93}}~(\bibinfo {number} {10}),\ \bibinfo {pages} {104303}}\BibitemShut
  {NoStop}%
\bibitem [{\citenamefont {Yin}\ \emph {et~al.}(2018)\citenamefont {Yin},
  \citenamefont {Jiang}, \citenamefont {Li}, \citenamefont {L{\"u}},\ and\
  \citenamefont {Chen}}]{yin2018geometrical}%
  \BibitemOpen
  \bibfield  {author} {\bibinfo {author} {\bibnamefont {Yin}, \bibfnamefont
  {Chuanhao}}, \bibinfo {author} {\bibfnamefont {Hui}\ \bibnamefont {Jiang}},
  \bibinfo {author} {\bibfnamefont {Linhu}\ \bibnamefont {Li}}, \bibinfo
  {author} {\bibfnamefont {Rong}\ \bibnamefont {L{\"u}}}, \ and\ \bibinfo
  {author} {\bibfnamefont {Shu}\ \bibnamefont {Chen}}} (\bibinfo {year}
  {2018}),\ \bibfield  {title} {\enquote {\bibinfo {title} {Geometrical meaning
  of winding number and its characterization of topological phases in
  one-dimensional chiral non-{H}ermitian systems},}\ }\href
  {https://journals.aps.org/pra/abstract/10.1103/PhysRevA.97.052115} {\bibfield
   {journal} {\bibinfo  {journal} {Phys. Rev. A}\ }\textbf {\bibinfo {volume}
  {97}}~(\bibinfo {number} {5}),\ \bibinfo {pages} {052115}}\BibitemShut
  {NoStop}%
\bibitem [{\citenamefont {Yoshida}\ \emph {et~al.}(2018)\citenamefont
  {Yoshida}, \citenamefont {Peters},\ and\ \citenamefont
  {Kawakami}}]{yoshida2018non}%
  \BibitemOpen
  \bibfield  {author} {\bibinfo {author} {\bibnamefont {Yoshida}, \bibfnamefont
  {Tsuneya}}, \bibinfo {author} {\bibfnamefont {Robert}\ \bibnamefont
  {Peters}}, \ and\ \bibinfo {author} {\bibfnamefont {Norio}\ \bibnamefont
  {Kawakami}}} (\bibinfo {year} {2018}),\ \bibfield  {title} {\enquote
  {\bibinfo {title} {Non-{H}ermitian perspective of the band structure in
  heavy-fermion systems},}\ }\href
  {https://link.aps.org/doi/10.1103/PhysRevB.98.035141} {\bibfield  {journal}
  {\bibinfo  {journal} {Phys. Rev. B}\ }\textbf {\bibinfo {volume} {98}},\
  \bibinfo {pages} {035141}}\BibitemShut {NoStop}%
\bibitem [{\citenamefont {Yoshioka}(2002)}]{YoshiokaBook}%
  \BibitemOpen
  \bibfield  {author} {\bibinfo {author} {\bibnamefont {Yoshioka},
  \bibfnamefont {Daijiro}}} (\bibinfo {year} {2002}),\ \href@noop {} {\emph
  {\bibinfo {title} {The quantum {H}all effect}}}\ (\bibinfo  {publisher}
  {Springer},\ \bibinfo {address} {New York})\BibitemShut {NoStop}%
\bibitem [{\citenamefont {You}\ and\ \citenamefont
  {Nori}(2011)}]{You:2011Nature}%
  \BibitemOpen
  \bibfield  {author} {\bibinfo {author} {\bibnamefont {You}, \bibfnamefont
  {JQ}}, \ and\ \bibinfo {author} {\bibfnamefont {Franco}\ \bibnamefont
  {Nori}}} (\bibinfo {year} {2011}),\ \bibfield  {title} {\enquote {\bibinfo
  {title} {Atomic physics and quantum optics using superconducting circuits},}\
  }\href
  {http://www.nature.com/nature/journal/v474/n7353/full/nature10122.html}
  {\bibfield  {journal} {\bibinfo  {journal} {Nature}\ }\textbf {\bibinfo
  {volume} {474}}~(\bibinfo {number} {7353}),\ \bibinfo {pages}
  {589--597}}\BibitemShut {NoStop}%
\bibitem [{\citenamefont {Yu}\ \emph {et~al.}(2014)\citenamefont {Yu},
  \citenamefont {Wang}, \citenamefont {Shen},\ and\ \citenamefont
  {Deng}}]{Yu:2014EML}%
  \BibitemOpen
  \bibfield  {author} {\bibinfo {author} {\bibnamefont {Yu}, \bibfnamefont
  {Zaihe}}, \bibinfo {author} {\bibfnamefont {Zhuoyuan}\ \bibnamefont {Wang}},
  \bibinfo {author} {\bibfnamefont {Linfang}\ \bibnamefont {Shen}}, \ and\
  \bibinfo {author} {\bibfnamefont {Xiaohua}\ \bibnamefont {Deng}}} (\bibinfo
  {year} {2014}),\ \bibfield  {title} {\enquote {\bibinfo {title} {One-way
  electromagnetic mode at the surface of a magnetized gyromagnetic medium},}\
  }\href {https://link.springer.com/article/10.1007/s13391-014-3204-9}
  {\bibfield  {journal} {\bibinfo  {journal} {Electron. Mater. Lett.}\ }\textbf
  {\bibinfo {volume} {10}}~(\bibinfo {number} {5}),\ \bibinfo {pages}
  {969--973}}\BibitemShut {NoStop}%
\bibitem [{\citenamefont {Yu}\ and\ \citenamefont
  {Fan}(2009)}]{Yu:NatPhot2009}%
  \BibitemOpen
  \bibfield  {author} {\bibinfo {author} {\bibnamefont {Yu}, \bibfnamefont
  {Zongfu}}, \ and\ \bibinfo {author} {\bibfnamefont {Shanhui}\ \bibnamefont
  {Fan}}} (\bibinfo {year} {2009}),\ \bibfield  {title} {\enquote {\bibinfo
  {title} {Complete optical isolation created by indirect interband photonic
  transitions},}\ }\href {https://www.nature.com/articles/nphoton.2008.273}
  {\bibfield  {journal} {\bibinfo  {journal} {Nat. Photonics}\ }\textbf
  {\bibinfo {volume} {3}}~(\bibinfo {number} {2}),\ \bibinfo {pages}
  {91--94}}\BibitemShut {NoStop}%
\bibitem [{\citenamefont {Yu}\ \emph {et~al.}(2008)\citenamefont {Yu},
  \citenamefont {Veronis}, \citenamefont {Wang},\ and\ \citenamefont
  {Fan}}]{Yu:2008PRL}%
  \BibitemOpen
  \bibfield  {author} {\bibinfo {author} {\bibnamefont {Yu}, \bibfnamefont
  {Zongfu}}, \bibinfo {author} {\bibfnamefont {Georgios}\ \bibnamefont
  {Veronis}}, \bibinfo {author} {\bibfnamefont {Zheng}\ \bibnamefont {Wang}}, \
  and\ \bibinfo {author} {\bibfnamefont {Shanhui}\ \bibnamefont {Fan}}}
  (\bibinfo {year} {2008}),\ \bibfield  {title} {\enquote {\bibinfo {title}
  {One-way electromagnetic waveguide formed at the interface between a
  plasmonic metal under a static magnetic field and a photonic crystal},}\
  }\href {https://journals.aps.org/prl/abstract/10.1103/PhysRevLett.100.023902}
  {\bibfield  {journal} {\bibinfo  {journal} {Phys. Rev. Lett.}\ }\textbf
  {\bibinfo {volume} {100}}~(\bibinfo {number} {2}),\ \bibinfo {pages}
  {023902}}\BibitemShut {NoStop}%
\bibitem [{\citenamefont {Yuan}\ and\ \citenamefont
  {Fan}(2015{\natexlab{a}})}]{Yuan:2015PRL}%
  \BibitemOpen
  \bibfield  {author} {\bibinfo {author} {\bibnamefont {Yuan}, \bibfnamefont
  {Luqi}}, \ and\ \bibinfo {author} {\bibfnamefont {Shanhui}\ \bibnamefont
  {Fan}}} (\bibinfo {year} {2015}{\natexlab{a}}),\ \bibfield  {title} {\enquote
  {\bibinfo {title} {Three-dimensional dynamic localization of light from a
  time-dependent effective gauge field for photons},}\ }\href
  {https://link.aps.org/doi/10.1103/PhysRevLett.114.243901} {\bibfield
  {journal} {\bibinfo  {journal} {Phys. Rev. Lett.}\ }\textbf {\bibinfo
  {volume} {114}},\ \bibinfo {pages} {243901}}\BibitemShut {NoStop}%
\bibitem [{\citenamefont {Yuan}\ and\ \citenamefont
  {Fan}(2015{\natexlab{b}})}]{Yuan:2015PRA}%
  \BibitemOpen
  \bibfield  {author} {\bibinfo {author} {\bibnamefont {Yuan}, \bibfnamefont
  {Luqi}}, \ and\ \bibinfo {author} {\bibfnamefont {Shanhui}\ \bibnamefont
  {Fan}}} (\bibinfo {year} {2015}{\natexlab{b}}),\ \bibfield  {title} {\enquote
  {\bibinfo {title} {Topologically nontrivial {F}loquet band structure in a
  system undergoing photonic transitions in the ultrastrong-coupling regime},}\
  }\href {https://link.aps.org/doi/10.1103/PhysRevA.92.053822} {\bibfield
  {journal} {\bibinfo  {journal} {Phys. Rev. A}\ }\textbf {\bibinfo {volume}
  {92}},\ \bibinfo {pages} {053822}}\BibitemShut {NoStop}%
\bibitem [{\citenamefont {Yuan}\ and\ \citenamefont
  {Fan}(2016)}]{Yuan:2016Optica}%
  \BibitemOpen
  \bibfield  {author} {\bibinfo {author} {\bibnamefont {Yuan}, \bibfnamefont
  {Luqi}}, \ and\ \bibinfo {author} {\bibfnamefont {Shanhui}\ \bibnamefont
  {Fan}}} (\bibinfo {year} {2016}),\ \bibfield  {title} {\enquote {\bibinfo
  {title} {Bloch oscillation and unidirectional translation of frequency in a
  dynamically modulated ring resonator},}\ }\href
  {https://www.osapublishing.org/optica/abstract.cfm?uri=optica-3-9-1014}
  {\bibfield  {journal} {\bibinfo  {journal} {Optica}\ }\textbf {\bibinfo
  {volume} {3}}~(\bibinfo {number} {9}),\ \bibinfo {pages}
  {1014--1018}}\BibitemShut {NoStop}%
\bibitem [{\citenamefont {Yuan}\ \emph
  {et~al.}(2016{\natexlab{a}})\citenamefont {Yuan}, \citenamefont {Shi},\ and\
  \citenamefont {Fan}}]{Yuan:2016OptLett}%
  \BibitemOpen
  \bibfield  {author} {\bibinfo {author} {\bibnamefont {Yuan}, \bibfnamefont
  {Luqi}}, \bibinfo {author} {\bibfnamefont {Yu}~\bibnamefont {Shi}}, \ and\
  \bibinfo {author} {\bibfnamefont {Shanhui}\ \bibnamefont {Fan}}} (\bibinfo
  {year} {2016}{\natexlab{a}}),\ \bibfield  {title} {\enquote {\bibinfo {title}
  {Photonic gauge potential in a system with a synthetic frequency
  dimension},}\ }\href
  {https://www.osapublishing.org/ol/abstract.cfm?uri=ol-41-4-741} {\bibfield
  {journal} {\bibinfo  {journal} {Opt. Lett.}\ }\textbf {\bibinfo {volume}
  {41}}~(\bibinfo {number} {4}),\ \bibinfo {pages} {741--744}}\BibitemShut
  {NoStop}%
\bibitem [{\citenamefont {Yuan}\ \emph {et~al.}(2017)\citenamefont {Yuan},
  \citenamefont {Wang},\ and\ \citenamefont {Fan}}]{Yuan:2017PRA}%
  \BibitemOpen
  \bibfield  {author} {\bibinfo {author} {\bibnamefont {Yuan}, \bibfnamefont
  {Luqi}}, \bibinfo {author} {\bibfnamefont {Da-wei}\ \bibnamefont {Wang}}, \
  and\ \bibinfo {author} {\bibfnamefont {Shanhui}\ \bibnamefont {Fan}}}
  (\bibinfo {year} {2017}),\ \bibfield  {title} {\enquote {\bibinfo {title}
  {Synthetic gauge potential and effective magnetic field in a {R}aman medium
  undergoing molecular modulation},}\ }\href
  {https://link.aps.org/doi/10.1103/PhysRevA.95.033801} {\bibfield  {journal}
  {\bibinfo  {journal} {Phys. Rev. A}\ }\textbf {\bibinfo {volume} {95}},\
  \bibinfo {pages} {033801}}\BibitemShut {NoStop}%
\bibitem [{\citenamefont {Yuan}\ \emph
  {et~al.}(2016{\natexlab{b}})\citenamefont {Yuan}, \citenamefont {Xiao},\ and\
  \citenamefont {Fan}}]{Yuan:2016PRB}%
  \BibitemOpen
  \bibfield  {author} {\bibinfo {author} {\bibnamefont {Yuan}, \bibfnamefont
  {Luqi}}, \bibinfo {author} {\bibfnamefont {Meng}\ \bibnamefont {Xiao}}, \
  and\ \bibinfo {author} {\bibfnamefont {Shanhui}\ \bibnamefont {Fan}}}
  (\bibinfo {year} {2016}{\natexlab{b}}),\ \bibfield  {title} {\enquote
  {\bibinfo {title} {Time reversal of a wave packet with temporal modulation of
  gauge potential},}\ }\href
  {https://link.aps.org/doi/10.1103/PhysRevB.94.140303} {\bibfield  {journal}
  {\bibinfo  {journal} {Phys. Rev. B}\ }\textbf {\bibinfo {volume} {94}},\
  \bibinfo {pages} {140303}}\BibitemShut {NoStop}%
\bibitem [{\citenamefont {Yuan}\ \emph {et~al.}(2018)\citenamefont {Yuan},
  \citenamefont {Xiao}, \citenamefont {Lin},\ and\ \citenamefont
  {Fan}}]{Yuan:2017arXiv}%
  \BibitemOpen
  \bibfield  {author} {\bibinfo {author} {\bibnamefont {Yuan}, \bibfnamefont
  {Luqi}}, \bibinfo {author} {\bibfnamefont {Meng}\ \bibnamefont {Xiao}},
  \bibinfo {author} {\bibfnamefont {Qian}\ \bibnamefont {Lin}}, \ and\ \bibinfo
  {author} {\bibfnamefont {Shanhui}\ \bibnamefont {Fan}}} (\bibinfo {year}
  {2018}),\ \bibfield  {title} {\enquote {\bibinfo {title} {Synthetic space
  with arbitrary dimensions in a few rings undergoing dynamic modulation},}\
  }\href {https://link.aps.org/doi/10.1103/PhysRevB.97.104105} {\bibfield
  {journal} {\bibinfo  {journal} {Phys. Rev. B}\ }\textbf {\bibinfo {volume}
  {97}},\ \bibinfo {pages} {104105}}\BibitemShut {NoStop}%
\bibitem [{\citenamefont {Yves}\ \emph
  {et~al.}(2017{\natexlab{a}})\citenamefont {Yves}, \citenamefont {Fleury},
  \citenamefont {Berthelot}, \citenamefont {Fink}, \citenamefont {Lemoult},\
  and\ \citenamefont {Lerosey}}]{Yves:NatComm2017}%
  \BibitemOpen
  \bibfield  {author} {\bibinfo {author} {\bibnamefont {Yves}, \bibfnamefont
  {Simon}}, \bibinfo {author} {\bibfnamefont {Romain}\ \bibnamefont {Fleury}},
  \bibinfo {author} {\bibfnamefont {Thomas}\ \bibnamefont {Berthelot}},
  \bibinfo {author} {\bibfnamefont {Mathias}\ \bibnamefont {Fink}}, \bibinfo
  {author} {\bibfnamefont {Fabrice}\ \bibnamefont {Lemoult}}, \ and\ \bibinfo
  {author} {\bibfnamefont {Geoffroy}\ \bibnamefont {Lerosey}}} (\bibinfo {year}
  {2017}{\natexlab{a}}),\ \bibfield  {title} {\enquote {\bibinfo {title}
  {{Crystalline metamaterials for topological properties at subwavelength
  scales}},}\ }\href {http://www.nature.com/doifinder/10.1038/ncomms16023}
  {\bibfield  {journal} {\bibinfo  {journal} {Nat. Commun.}\ }\textbf {\bibinfo
  {volume} {8}},\ \bibinfo {pages} {16023}}\BibitemShut {NoStop}%
\bibitem [{\citenamefont {Yves}\ \emph
  {et~al.}(2017{\natexlab{b}})\citenamefont {Yves}, \citenamefont {Fleury},
  \citenamefont {Lemoult}, \citenamefont {Fink},\ and\ \citenamefont
  {Lerosey}}]{Simon:NJP2017}%
  \BibitemOpen
  \bibfield  {author} {\bibinfo {author} {\bibnamefont {Yves}, \bibfnamefont
  {Simon}}, \bibinfo {author} {\bibfnamefont {Romain}\ \bibnamefont {Fleury}},
  \bibinfo {author} {\bibfnamefont {Fabrice}\ \bibnamefont {Lemoult}}, \bibinfo
  {author} {\bibfnamefont {Mathias}\ \bibnamefont {Fink}}, \ and\ \bibinfo
  {author} {\bibfnamefont {Geoffroy}\ \bibnamefont {Lerosey}}} (\bibinfo {year}
  {2017}{\natexlab{b}}),\ \bibfield  {title} {\enquote {\bibinfo {title}
  {{Topological acoustic polaritons: Robust sound manipulation at the
  subwavelength scale}},}\ }\href
  {https://iopscience.iop.org/article/10.1088/1367-2630/aa66f8/meta} {\bibfield
   {journal} {\bibinfo  {journal} {New J. Phys.}\ }\textbf {\bibinfo {volume}
  {19}}~(\bibinfo {number} {7}),\ \bibinfo {pages} {075003}}\BibitemShut
  {NoStop}%
\bibitem [{\citenamefont {Zak}(1964)}]{Zak:1964PR}%
  \BibitemOpen
  \bibfield  {author} {\bibinfo {author} {\bibnamefont {Zak}, \bibfnamefont
  {J}}} (\bibinfo {year} {1964}),\ \bibfield  {title} {\enquote {\bibinfo
  {title} {Magnetic translation group},}\ }\href
  {https://link.aps.org/doi/10.1103/PhysRev.134.A1602} {\bibfield  {journal}
  {\bibinfo  {journal} {Phys. Rev.}\ }\textbf {\bibinfo {volume} {134}},\
  \bibinfo {pages} {A1602--A1606}}\BibitemShut {NoStop}%
\bibitem [{\citenamefont {Zak}(1989)}]{Zak:1989PRL}%
  \BibitemOpen
  \bibfield  {author} {\bibinfo {author} {\bibnamefont {Zak}, \bibfnamefont
  {J}}} (\bibinfo {year} {1989}),\ \bibfield  {title} {\enquote {\bibinfo
  {title} {Berry's phase for energy bands in solids},}\ }\href
  {https://link.aps.org/doi/10.1103/PhysRevLett.62.2747} {\bibfield  {journal}
  {\bibinfo  {journal} {Phys. Rev. Lett.}\ }\textbf {\bibinfo {volume} {62}},\
  \bibinfo {pages} {2747--2750}}\BibitemShut {NoStop}%
\bibitem [{\citenamefont {Zang}\ and\ \citenamefont
  {Jiang}(2011)}]{Zang:2011JOSAB}%
  \BibitemOpen
  \bibfield  {author} {\bibinfo {author} {\bibnamefont {Zang}, \bibfnamefont
  {Xiaofei}}, \ and\ \bibinfo {author} {\bibfnamefont {Chun}\ \bibnamefont
  {Jiang}}} (\bibinfo {year} {2011}),\ \bibfield  {title} {\enquote {\bibinfo
  {title} {Edge mode in nonreciprocal photonic crystal waveguide: manipulating
  the unidirectional electromagnetic pulse dynamically},}\ }\href
  {https://www.osapublishing.org/josab/abstract.cfm?uri=josab-28-3-554}
  {\bibfield  {journal} {\bibinfo  {journal} {J. Opt. Soc. Am. B}\ }\textbf
  {\bibinfo {volume} {28}}~(\bibinfo {number} {3}),\ \bibinfo {pages}
  {554--557}}\BibitemShut {NoStop}%
\bibitem [{\citenamefont {Zeng}\ \emph {et~al.}(2015)\citenamefont {Zeng},
  \citenamefont {Wang},\ and\ \citenamefont {Zhai}}]{Zeng:2015PRL}%
  \BibitemOpen
  \bibfield  {author} {\bibinfo {author} {\bibnamefont {Zeng}, \bibfnamefont
  {Tian-Sheng}}, \bibinfo {author} {\bibfnamefont {Ce}~\bibnamefont {Wang}}, \
  and\ \bibinfo {author} {\bibfnamefont {Hui}\ \bibnamefont {Zhai}}} (\bibinfo
  {year} {2015}),\ \bibfield  {title} {\enquote {\bibinfo {title} {Charge
  pumping of interacting fermion atoms in the synthetic dimension},}\ }\href
  {https://link.aps.org/doi/10.1103/PhysRevLett.115.095302} {\bibfield
  {journal} {\bibinfo  {journal} {Phys. Rev. Lett.}\ }\textbf {\bibinfo
  {volume} {115}},\ \bibinfo {pages} {095302}}\BibitemShut {NoStop}%
\bibitem [{\citenamefont {Zeuner}\ \emph {et~al.}(2015)\citenamefont {Zeuner},
  \citenamefont {Rechtsman}, \citenamefont {Plotnik}, \citenamefont {Lumer},
  \citenamefont {Nolte}, \citenamefont {Rudner}, \citenamefont {Segev},\ and\
  \citenamefont {Szameit}}]{Zeuner:2015PRL}%
  \BibitemOpen
  \bibfield  {author} {\bibinfo {author} {\bibnamefont {Zeuner}, \bibfnamefont
  {Julia~M}}, \bibinfo {author} {\bibfnamefont {Mikael~C.}\ \bibnamefont
  {Rechtsman}}, \bibinfo {author} {\bibfnamefont {Yonatan}\ \bibnamefont
  {Plotnik}}, \bibinfo {author} {\bibfnamefont {Yaakov}\ \bibnamefont {Lumer}},
  \bibinfo {author} {\bibfnamefont {Stefan}\ \bibnamefont {Nolte}}, \bibinfo
  {author} {\bibfnamefont {Mark~S.}\ \bibnamefont {Rudner}}, \bibinfo {author}
  {\bibfnamefont {Mordechai}\ \bibnamefont {Segev}}, \ and\ \bibinfo {author}
  {\bibfnamefont {Alexander}\ \bibnamefont {Szameit}}} (\bibinfo {year}
  {2015}),\ \bibfield  {title} {\enquote {\bibinfo {title} {Observation of a
  topological transition in the bulk of a non-{H}ermitian system},}\ }\href
  {http://link.aps.org/doi/10.1103/PhysRevLett.115.040402} {\bibfield
  {journal} {\bibinfo  {journal} {Phys. Rev. Lett.}\ }\textbf {\bibinfo
  {volume} {115}},\ \bibinfo {pages} {040402}}\BibitemShut {NoStop}%
\bibitem [{\citenamefont {Zhang}\ \emph {et~al.}(2011)\citenamefont {Zhang},
  \citenamefont {Jung}, \citenamefont {Fiete}, \citenamefont {Niu},\ and\
  \citenamefont {MacDonald}}]{Zhang:PRL2011}%
  \BibitemOpen
  \bibfield  {author} {\bibinfo {author} {\bibnamefont {Zhang}, \bibfnamefont
  {Fan}}, \bibinfo {author} {\bibfnamefont {Jeil}\ \bibnamefont {Jung}},
  \bibinfo {author} {\bibfnamefont {Gregory~A}\ \bibnamefont {Fiete}}, \bibinfo
  {author} {\bibfnamefont {Qian}\ \bibnamefont {Niu}}, \ and\ \bibinfo {author}
  {\bibfnamefont {Allan~H}\ \bibnamefont {MacDonald}}} (\bibinfo {year}
  {2011}),\ \bibfield  {title} {\enquote {\bibinfo {title} {{Spontaneous
  quantum Hall states in chirally stacked few-layer graphene systems}},}\
  }\href {https://link.aps.org/doi/10.1103/PhysRevLett.106.156801} {\bibfield
  {journal} {\bibinfo  {journal} {Phys. Rev. Lett.}\ }\textbf {\bibinfo
  {volume} {106}}~(\bibinfo {number} {15}),\ \bibinfo {pages}
  {156801}}\BibitemShut {NoStop}%
\bibitem [{\citenamefont {Zhang}\ \emph {et~al.}(2013)\citenamefont {Zhang},
  \citenamefont {Yang}, \citenamefont {Chen}, \citenamefont {Li},\ and\
  \citenamefont {Xia}}]{Zhang:2013OLT}%
  \BibitemOpen
  \bibfield  {author} {\bibinfo {author} {\bibnamefont {Zhang}, \bibfnamefont
  {Le}}, \bibinfo {author} {\bibfnamefont {Dongxiao}\ \bibnamefont {Yang}},
  \bibinfo {author} {\bibfnamefont {Kan}\ \bibnamefont {Chen}}, \bibinfo
  {author} {\bibfnamefont {Tao}\ \bibnamefont {Li}}, \ and\ \bibinfo {author}
  {\bibfnamefont {Song}\ \bibnamefont {Xia}}} (\bibinfo {year} {2013}),\
  \bibfield  {title} {\enquote {\bibinfo {title} {Design of nonreciprocal
  waveguide devices based on two-dimensional magneto-optical photonic
  crystals},}\ }\href
  {https://www.sciencedirect.com/science/article/pii/S003039921300056X}
  {\bibfield  {journal} {\bibinfo  {journal} {Opt. Laser Technol.}\ }\textbf
  {\bibinfo {volume} {50}},\ \bibinfo {pages} {195--201}}\BibitemShut {NoStop}%
\bibitem [{\citenamefont {Zhang}\ and\ \citenamefont
  {Hu}(2001)}]{Zhang:2001Science}%
  \BibitemOpen
  \bibfield  {author} {\bibinfo {author} {\bibnamefont {Zhang}, \bibfnamefont
  {Shou-Cheng}}, \ and\ \bibinfo {author} {\bibfnamefont {Jiangping}\
  \bibnamefont {Hu}}} (\bibinfo {year} {2001}),\ \bibfield  {title} {\enquote
  {\bibinfo {title} {A four-dimensional generalization of the quantum {H}all
  effect},}\ }\href {http://science.sciencemag.org/content/294/5543/823}
  {\bibfield  {journal} {\bibinfo  {journal} {Science}\ }\textbf {\bibinfo
  {volume} {294}}~(\bibinfo {number} {5543}),\ \bibinfo {pages}
  {823--828}}\BibitemShut {NoStop}%
\bibitem [{\citenamefont {Zhang}\ \emph {et~al.}(2018)\citenamefont {Zhang},
  \citenamefont {Song}, \citenamefont {Alexandradinata}, \citenamefont {Weng},
  \citenamefont {Fang}, \citenamefont {Lu},\ and\ \citenamefont
  {Fang}}]{Zhang:2018PRL}%
  \BibitemOpen
  \bibfield  {author} {\bibinfo {author} {\bibnamefont {Zhang}, \bibfnamefont
  {Tiantian}}, \bibinfo {author} {\bibfnamefont {Zhida}\ \bibnamefont {Song}},
  \bibinfo {author} {\bibfnamefont {A.}~\bibnamefont {Alexandradinata}},
  \bibinfo {author} {\bibfnamefont {Hongming}\ \bibnamefont {Weng}}, \bibinfo
  {author} {\bibfnamefont {Chen}\ \bibnamefont {Fang}}, \bibinfo {author}
  {\bibfnamefont {Ling}\ \bibnamefont {Lu}}, \ and\ \bibinfo {author}
  {\bibfnamefont {Zhong}\ \bibnamefont {Fang}}} (\bibinfo {year} {2018}),\
  \bibfield  {title} {\enquote {\bibinfo {title} {Double-{W}eyl phonons in
  transition-metal monosilicides},}\ }\href
  {https://link.aps.org/doi/10.1103/PhysRevLett.120.016401} {\bibfield
  {journal} {\bibinfo  {journal} {Phys. Rev. Lett.}\ }\textbf {\bibinfo
  {volume} {120}},\ \bibinfo {pages} {016401}}\BibitemShut {NoStop}%
\bibitem [{\citenamefont {Zhang}\ \emph {et~al.}(2012)\citenamefont {Zhang},
  \citenamefont {Li},\ and\ \citenamefont {Jiang}}]{Zhang:2012APL}%
  \BibitemOpen
  \bibfield  {author} {\bibinfo {author} {\bibnamefont {Zhang}, \bibfnamefont
  {Xiaogang}}, \bibinfo {author} {\bibfnamefont {Wei}\ \bibnamefont {Li}}, \
  and\ \bibinfo {author} {\bibfnamefont {Xunya}\ \bibnamefont {Jiang}}}
  (\bibinfo {year} {2012}),\ \bibfield  {title} {\enquote {\bibinfo {title}
  {Confined one-way mode at magnetic domain wall for broadband high-efficiency
  one-way waveguide, splitter and bender},}\ }\href
  {https://aip.scitation.org/doi/full/10.1063/1.3679172} {\bibfield  {journal}
  {\bibinfo  {journal} {Appl. Phys. Lett.}\ }\textbf {\bibinfo {volume}
  {100}}~(\bibinfo {number} {4}),\ \bibinfo {pages} {041108}}\BibitemShut
  {NoStop}%
\bibitem [{\citenamefont {{Zhang}}\ \emph {et~al.}(2018)\citenamefont
  {{Zhang}}, \citenamefont {{Li}}, \citenamefont {{Malpuech}}, \citenamefont
  {{Zhang}}, \citenamefont {{Bleu}}, \citenamefont {{Koniakhin}}, \citenamefont
  {{Li}}, \citenamefont {{Zhang}}, \citenamefont {{Xiao}},\ and\ \citenamefont
  {{Solnyshkov}}}]{zhang:2018arxiv}%
  \BibitemOpen
  \bibfield  {author} {\bibinfo {author} {\bibnamefont {{Zhang}}, \bibfnamefont
  {Z}}, \bibinfo {author} {\bibfnamefont {F.}~\bibnamefont {{Li}}}, \bibinfo
  {author} {\bibfnamefont {G.}~\bibnamefont {{Malpuech}}}, \bibinfo {author}
  {\bibfnamefont {Y.}~\bibnamefont {{Zhang}}}, \bibinfo {author} {\bibfnamefont
  {O.}~\bibnamefont {{Bleu}}}, \bibinfo {author} {\bibfnamefont
  {S.}~\bibnamefont {{Koniakhin}}}, \bibinfo {author} {\bibfnamefont
  {C.}~\bibnamefont {{Li}}}, \bibinfo {author} {\bibfnamefont {Y.}~\bibnamefont
  {{Zhang}}}, \bibinfo {author} {\bibfnamefont {M.}~\bibnamefont {{Xiao}}}, \
  and\ \bibinfo {author} {\bibfnamefont {D.}~\bibnamefont {{Solnyshkov}}}}
  (\bibinfo {year} {2018}),\ \bibfield  {title} {\enquote {\bibinfo {title}
  {{Particle-like behavior of topological defects in linear wave packets in
  photonic graphene}},}\ }\href {https://arxiv.org/abs/1806.05540} {\bibinfo
  {journal} {arXiv:1806.05540}\ }\BibitemShut {NoStop}%
\bibitem [{\citenamefont {Zhao}\ \emph {et~al.}(2018)\citenamefont {Zhao},
  \citenamefont {Miao}, \citenamefont {Teimourpour}, \citenamefont {Malzard},
  \citenamefont {El-Ganainy}, \citenamefont {Schomerus},\ and\ \citenamefont
  {Feng}}]{Zhao:2018NatComm}%
  \BibitemOpen
\bibfield  {journal} {  }\bibfield  {author} {\bibinfo {author} {\bibnamefont
  {Zhao}, \bibfnamefont {Han}}, \bibinfo {author} {\bibfnamefont {Pei}\
  \bibnamefont {Miao}}, \bibinfo {author} {\bibfnamefont {Mohammad~H}\
  \bibnamefont {Teimourpour}}, \bibinfo {author} {\bibfnamefont {Simon}\
  \bibnamefont {Malzard}}, \bibinfo {author} {\bibfnamefont {Ramy}\
  \bibnamefont {El-Ganainy}}, \bibinfo {author} {\bibfnamefont {Henning}\
  \bibnamefont {Schomerus}}, \ and\ \bibinfo {author} {\bibfnamefont {Liang}\
  \bibnamefont {Feng}}} (\bibinfo {year} {2018}),\ \bibfield  {title} {\enquote
  {\bibinfo {title} {{Topological hybrid silicon microlasers}},}\ }\href
  {https://doi.org/10.1038/s41467-018-03434-2} {\bibfield  {journal} {\bibinfo
  {journal} {Nat. Commun.}\ }\textbf {\bibinfo {volume} {9}}~(\bibinfo {number}
  {1}),\ \bibinfo {pages} {981}}\BibitemShut {NoStop}%
\bibitem [{\citenamefont {Zhen}\ \emph {et~al.}(2015)\citenamefont {Zhen},
  \citenamefont {Hsu}, \citenamefont {Igarashi}, \citenamefont {Lu},
  \citenamefont {Kaminer}, \citenamefont {Pick}, \citenamefont {Chua},
  \citenamefont {Joannopoulos},\ and\ \citenamefont
  {Solja{\v{c}}i{\'c}}}]{zhen2015spawning}%
  \BibitemOpen
  \bibfield  {author} {\bibinfo {author} {\bibnamefont {Zhen}, \bibfnamefont
  {Bo}}, \bibinfo {author} {\bibfnamefont {Chia~Wei}\ \bibnamefont {Hsu}},
  \bibinfo {author} {\bibfnamefont {Yuichi}\ \bibnamefont {Igarashi}}, \bibinfo
  {author} {\bibfnamefont {Ling}\ \bibnamefont {Lu}}, \bibinfo {author}
  {\bibfnamefont {Ido}\ \bibnamefont {Kaminer}}, \bibinfo {author}
  {\bibfnamefont {Adi}\ \bibnamefont {Pick}}, \bibinfo {author} {\bibfnamefont
  {Song-Liang}\ \bibnamefont {Chua}}, \bibinfo {author} {\bibfnamefont
  {John~D}\ \bibnamefont {Joannopoulos}}, \ and\ \bibinfo {author}
  {\bibfnamefont {Marin}\ \bibnamefont {Solja{\v{c}}i{\'c}}}} (\bibinfo {year}
  {2015}),\ \bibfield  {title} {\enquote {\bibinfo {title} {Spawning rings of
  exceptional points out of {D}irac cones},}\ }\href
  {https://www.nature.com/articles/nature14889} {\bibfield  {journal} {\bibinfo
   {journal} {Nature}\ }\textbf {\bibinfo {volume} {525}}~(\bibinfo {number}
  {7569}),\ \bibinfo {pages} {354--358}}\BibitemShut {NoStop}%
\bibitem [{\citenamefont {Zheng}\ and\ \citenamefont
  {Zhai}(2014)}]{Zheng:2014PRA}%
  \BibitemOpen
  \bibfield  {author} {\bibinfo {author} {\bibnamefont {Zheng}, \bibfnamefont
  {Wei}}, \ and\ \bibinfo {author} {\bibfnamefont {Hui}\ \bibnamefont {Zhai}}}
  (\bibinfo {year} {2014}),\ \bibfield  {title} {\enquote {\bibinfo {title}
  {{Floquet topological states in shaking optical lattices}},}\ }\href
  {https://doi.org/10.1103/PhysRevA.89.061603} {\bibfield  {journal} {\bibinfo
  {journal} {Phys. Rev. A}\ }\textbf {\bibinfo {volume} {89}}~(\bibinfo
  {number} {6}),\ \bibinfo {pages} {061603}}\BibitemShut {NoStop}%
\bibitem [{\citenamefont {Zhou}\ \emph {et~al.}(1999)\citenamefont {Zhou},
  \citenamefont {Spivak},\ and\ \citenamefont {Altshuler}}]{Zhou:1999}%
  \BibitemOpen
  \bibfield  {author} {\bibinfo {author} {\bibnamefont {Zhou}, \bibfnamefont
  {F}}, \bibinfo {author} {\bibfnamefont {B.}~\bibnamefont {Spivak}}, \ and\
  \bibinfo {author} {\bibfnamefont {B.}~\bibnamefont {Altshuler}}} (\bibinfo
  {year} {1999}),\ \bibfield  {title} {\enquote {\bibinfo {title} {Mesoscopic
  mechanism of adiabatic charge transport},}\ }\href
  {https://link.aps.org/doi/10.1103/PhysRevLett.82.608} {\bibfield  {journal}
  {\bibinfo  {journal} {Phys. Rev. Lett.}\ }\textbf {\bibinfo {volume} {82}},\
  \bibinfo {pages} {608--611}}\BibitemShut {NoStop}%
\bibitem [{\citenamefont {Zhou}\ \emph {et~al.}(2018)\citenamefont {Zhou},
  \citenamefont {Peng}, \citenamefont {Yoon}, \citenamefont {Hsu},
  \citenamefont {Nelson}, \citenamefont {Fu}, \citenamefont {Joannopoulos},
  \citenamefont {Solja{\v{c}}i{\'c}},\ and\ \citenamefont
  {Zhen}}]{Zhou:2018Science}%
  \BibitemOpen
  \bibfield  {author} {\bibinfo {author} {\bibnamefont {Zhou}, \bibfnamefont
  {Hengyun}}, \bibinfo {author} {\bibfnamefont {Chao}\ \bibnamefont {Peng}},
  \bibinfo {author} {\bibfnamefont {Yoseob}\ \bibnamefont {Yoon}}, \bibinfo
  {author} {\bibfnamefont {Chia~Wei}\ \bibnamefont {Hsu}}, \bibinfo {author}
  {\bibfnamefont {Keith~A}\ \bibnamefont {Nelson}}, \bibinfo {author}
  {\bibfnamefont {Liang}\ \bibnamefont {Fu}}, \bibinfo {author} {\bibfnamefont
  {John~D}\ \bibnamefont {Joannopoulos}}, \bibinfo {author} {\bibfnamefont
  {Marin}\ \bibnamefont {Solja{\v{c}}i{\'c}}}, \ and\ \bibinfo {author}
  {\bibfnamefont {Bo}~\bibnamefont {Zhen}}} (\bibinfo {year} {2018}),\
  \bibfield  {title} {\enquote {\bibinfo {title} {Observation of bulk {F}ermi
  arc and polarization half charge from paired exceptional points},}\ }\href
  {http://science.sciencemag.org/content/early/2018/01/10/science.aap9859}
  {\bibfield  {journal} {\bibinfo  {journal} {Science}\ }\textbf {\bibinfo
  {volume} {359}},\ \bibinfo {pages} {eaap9859}}\BibitemShut {NoStop}%
\bibitem [{\citenamefont {Zhou}\ \emph
  {et~al.}(2017{\natexlab{a}})\citenamefont {Zhou}, \citenamefont {Ying},
  \citenamefont {Lu}, \citenamefont {Shi}, \citenamefont {Zi},\ and\
  \citenamefont {Yu}}]{Zhou:2017arXiv}%
  \BibitemOpen
  \bibfield  {author} {\bibinfo {author} {\bibnamefont {Zhou}, \bibfnamefont
  {Ming}}, \bibinfo {author} {\bibfnamefont {Lei}\ \bibnamefont {Ying}},
  \bibinfo {author} {\bibfnamefont {Ling}\ \bibnamefont {Lu}}, \bibinfo
  {author} {\bibfnamefont {Lei}\ \bibnamefont {Shi}}, \bibinfo {author}
  {\bibfnamefont {Jian}\ \bibnamefont {Zi}}, \ and\ \bibinfo {author}
  {\bibfnamefont {Zongfu}\ \bibnamefont {Yu}}} (\bibinfo {year}
  {2017}{\natexlab{a}}),\ \bibfield  {title} {\enquote {\bibinfo {title}
  {Electromagnetic scattering laws in {W}eyl systems},}\ }\href
  {https://www.nature.com/articles/s41467-017-01533-0} {\bibfield  {journal}
  {\bibinfo  {journal} {Nat. Commun.}\ }\textbf {\bibinfo {volume}
  {8}}~(\bibinfo {number} {1}),\ \bibinfo {pages} {1388}}\BibitemShut {NoStop}%
\bibitem [{\citenamefont {Zhou}\ \emph
  {et~al.}(2017{\natexlab{b}})\citenamefont {Zhou}, \citenamefont {Luo},
  \citenamefont {Wang}, \citenamefont {Guo}, \citenamefont {Zhou},
  \citenamefont {Pu},\ and\ \citenamefont {Zhou}}]{Zhou:2017PRL}%
  \BibitemOpen
  \bibfield  {author} {\bibinfo {author} {\bibnamefont {Zhou}, \bibfnamefont
  {Xiang-Fa}}, \bibinfo {author} {\bibfnamefont {Xi-Wang}\ \bibnamefont {Luo}},
  \bibinfo {author} {\bibfnamefont {Su}~\bibnamefont {Wang}}, \bibinfo {author}
  {\bibfnamefont {Guang-Can}\ \bibnamefont {Guo}}, \bibinfo {author}
  {\bibfnamefont {Xingxiang}\ \bibnamefont {Zhou}}, \bibinfo {author}
  {\bibfnamefont {Han}\ \bibnamefont {Pu}}, \ and\ \bibinfo {author}
  {\bibfnamefont {Zheng-Wei}\ \bibnamefont {Zhou}}} (\bibinfo {year}
  {2017}{\natexlab{b}}),\ \bibfield  {title} {\enquote {\bibinfo {title}
  {Dynamically manipulating topological physics and edge modes in a single
  degenerate optical cavity},}\ }\href
  {https://link.aps.org/doi/10.1103/PhysRevLett.118.083603} {\bibfield
  {journal} {\bibinfo  {journal} {Phys. Rev. Lett.}\ }\textbf {\bibinfo
  {volume} {118}},\ \bibinfo {pages} {083603}}\BibitemShut {NoStop}%
\bibitem [{\citenamefont {Zhou}\ \emph
  {et~al.}(2017{\natexlab{c}})\citenamefont {Zhou}, \citenamefont {Wang},
  \citenamefont {Leykam},\ and\ \citenamefont {Chong}}]{Xin:NJP2017}%
  \BibitemOpen
  \bibfield  {author} {\bibinfo {author} {\bibnamefont {Zhou}, \bibfnamefont
  {Xin}}, \bibinfo {author} {\bibfnamefont {You}\ \bibnamefont {Wang}},
  \bibinfo {author} {\bibfnamefont {Daniel}\ \bibnamefont {Leykam}}, \ and\
  \bibinfo {author} {\bibfnamefont {Y~D}\ \bibnamefont {Chong}}} (\bibinfo
  {year} {2017}{\natexlab{c}}),\ \bibfield  {title} {\enquote {\bibinfo {title}
  {Optical isolation with nonlinear topological photonics},}\ }\href
  {http://stacks.iop.org/1367-2630/19/i=9/a=095002} {\bibfield  {journal}
  {\bibinfo  {journal} {New J. Phys.}\ }\textbf {\bibinfo {volume}
  {19}}~(\bibinfo {number} {9}),\ \bibinfo {pages} {095002}}\BibitemShut
  {NoStop}%
\bibitem [{\citenamefont {Zhu}\ \emph {et~al.}(2018{\natexlab{a}})\citenamefont
  {Zhu}, \citenamefont {Zhong}, \citenamefont {Ke}, \citenamefont {Qin},
  \citenamefont {Sukhorukov}, \citenamefont {Kivshar},\ and\ \citenamefont
  {Lee}}]{Zhu:2018PRA}%
  \BibitemOpen
  \bibfield  {author} {\bibinfo {author} {\bibnamefont {Zhu}, \bibfnamefont
  {Bo}}, \bibinfo {author} {\bibfnamefont {Honghua}\ \bibnamefont {Zhong}},
  \bibinfo {author} {\bibfnamefont {Yongguan}\ \bibnamefont {Ke}}, \bibinfo
  {author} {\bibfnamefont {Xizhou}\ \bibnamefont {Qin}}, \bibinfo {author}
  {\bibfnamefont {Andrey~A.}\ \bibnamefont {Sukhorukov}}, \bibinfo {author}
  {\bibfnamefont {Yuri~S.}\ \bibnamefont {Kivshar}}, \ and\ \bibinfo {author}
  {\bibfnamefont {Chaohong}\ \bibnamefont {Lee}}} (\bibinfo {year}
  {2018}{\natexlab{a}}),\ \bibfield  {title} {\enquote {\bibinfo {title}
  {Topological {F}loquet edge states in periodically curved waveguides},}\
  }\href {https://link.aps.org/doi/10.1103/PhysRevA.98.013855} {\bibfield
  {journal} {\bibinfo  {journal} {Phys. Rev. A}\ }\textbf {\bibinfo {volume}
  {98}},\ \bibinfo {pages} {013855}}\BibitemShut {NoStop}%
\bibitem [{\citenamefont {Zhu}\ and\ \citenamefont
  {Jiang}(2011)}]{Zhu:2011JLT}%
  \BibitemOpen
  \bibfield  {author} {\bibinfo {author} {\bibnamefont {Zhu}, \bibfnamefont
  {Haibin}}, \ and\ \bibinfo {author} {\bibfnamefont {Chun}\ \bibnamefont
  {Jiang}}} (\bibinfo {year} {2011}),\ \bibfield  {title} {\enquote {\bibinfo
  {title} {Extraordinary coupling into one-way magneto-optical photonic crystal
  waveguide},}\ }\href
  {https://www.osapublishing.org/jlt/abstract.cfm?uri=jlt-29-5-708} {\bibfield
  {journal} {\bibinfo  {journal} {J. of Lightwave Technol.}\ }\textbf {\bibinfo
  {volume} {29}}~(\bibinfo {number} {5}),\ \bibinfo {pages}
  {708--713}}\BibitemShut {NoStop}%
\bibitem [{\citenamefont {Zhu}\ \emph {et~al.}(2018{\natexlab{b}})\citenamefont
  {Zhu}, \citenamefont {Wang}, \citenamefont {Xu}, \citenamefont {Lai},
  \citenamefont {Jiang},\ and\ \citenamefont {John}}]{Zhu:PRB2018}%
  \BibitemOpen
  \bibfield  {author} {\bibinfo {author} {\bibnamefont {Zhu}, \bibfnamefont
  {Xuan}}, \bibinfo {author} {\bibfnamefont {Hai-Xiao}\ \bibnamefont {Wang}},
  \bibinfo {author} {\bibfnamefont {Changqing}\ \bibnamefont {Xu}}, \bibinfo
  {author} {\bibfnamefont {Yun}\ \bibnamefont {Lai}}, \bibinfo {author}
  {\bibfnamefont {Jian-Hua}\ \bibnamefont {Jiang}}, \ and\ \bibinfo {author}
  {\bibfnamefont {Sajeev}\ \bibnamefont {John}}} (\bibinfo {year}
  {2018}{\natexlab{b}}),\ \bibfield  {title} {\enquote {\bibinfo {title}
  {{Topological transitions in continuously deformed photonic crystals}},}\
  }\href {https://link.aps.org/doi/10.1103/PhysRevB.97.085148} {\bibfield
  {journal} {\bibinfo  {journal} {Phys. Rev. B}\ }\textbf {\bibinfo {volume}
  {97}}~(\bibinfo {number} {8}),\ \bibinfo {pages} {85148}}\BibitemShut
  {NoStop}%
\bibitem [{\citenamefont {Zilberberg}\ \emph {et~al.}(2018)\citenamefont
  {Zilberberg}, \citenamefont {Huang}, \citenamefont {Guglielmon},
  \citenamefont {Wang}, \citenamefont {Chen}, \citenamefont {Kraus},\ and\
  \citenamefont {Rechtsman}}]{Zilberberg:2018Nature}%
  \BibitemOpen
  \bibfield  {author} {\bibinfo {author} {\bibnamefont {Zilberberg},
  \bibfnamefont {Oded}}, \bibinfo {author} {\bibfnamefont {Sheng}\ \bibnamefont
  {Huang}}, \bibinfo {author} {\bibfnamefont {Jonathan}\ \bibnamefont
  {Guglielmon}}, \bibinfo {author} {\bibfnamefont {Mohan}\ \bibnamefont
  {Wang}}, \bibinfo {author} {\bibfnamefont {Kevin~P}\ \bibnamefont {Chen}},
  \bibinfo {author} {\bibfnamefont {Yaacov~E}\ \bibnamefont {Kraus}}, \ and\
  \bibinfo {author} {\bibfnamefont {Mikael~C}\ \bibnamefont {Rechtsman}}}
  (\bibinfo {year} {2018}),\ \bibfield  {title} {\enquote {\bibinfo {title}
  {Photonic topological boundary pumping as a probe of {4D} quantum {H}all
  physics},}\ }\href {https://www.nature.com/articles/nature25011} {\bibfield
  {journal} {\bibinfo  {journal} {Nature}\ }\textbf {\bibinfo {volume}
  {553}}~(\bibinfo {number} {7686}),\ \bibinfo {pages} {59}}\BibitemShut
  {NoStop}%
\end{thebibliography}%


%
